\documentclass[
10pt, 
english, 
singlespacing, 
headsepline, 
]{MastersDoctoralThesis} 

\usepackage{enumitem}

\usepackage[utf8]{inputenc} 
\usepackage[T1]{fontenc} 

\usepackage[acronym]{glossaries} 

\usepackage[sc]{mathpazo} 

\usepackage{todonotes}

\usepackage[backend=biber,style=phys,natbib=true]{biblatex} 

\usepackage{comment}

\usepackage{subcaption}

\usepackage{epigraph}
\usepackage[autostyle=true]{csquotes} 
\usepackage{physics}
\usepackage{svg}
\usepackage[]{hyperref}
\usepackage[noabbrev]{cleveref} 
\usepackage{pdfpages} 
\usepackage[pagestyles]{titlesec} 
\titleformat{\chapter}[display]{\normalfont\bfseries}{}{0pt}{\Huge}
\newpagestyle{mystyle}
{\sethead[\thepage][][\chaptertitle]{}{}{\thepage}}
\usepackage{url}

\usepackage{relsize}

\usepackage{enumitem}

\usepackage{float}
\usepackage{wrapfig}
\restylefloat{figure}

\usepackage{MnSymbol}
\usetikzlibrary{shapes.geometric}

\usepackage{tikz}
\usetikzlibrary{calc}

\usepackage{feynmp-auto}

\usepackage{xfrac}

\usepackage{slashed}

\usepackage{bm}
\usepackage{mathrsfs}  

\usepackage{mathtools}

\newcommand*{\del}{\mathop{\mathrm{{}\partial}}\mathopen{}}

\newcommand{\myarrow}[1]{\raisebox{-1pt}{$\stackrel{\tikz[baseline]{\draw[->] (0, 0) arc (260:310:50pt);}}{#1}$}}

\newcommand\myatop[2]{\genfrac{}{}{0pt}{1}{#1}{#2}}

\newcommand*{\saddle}{\includegraphics[width=.8em]{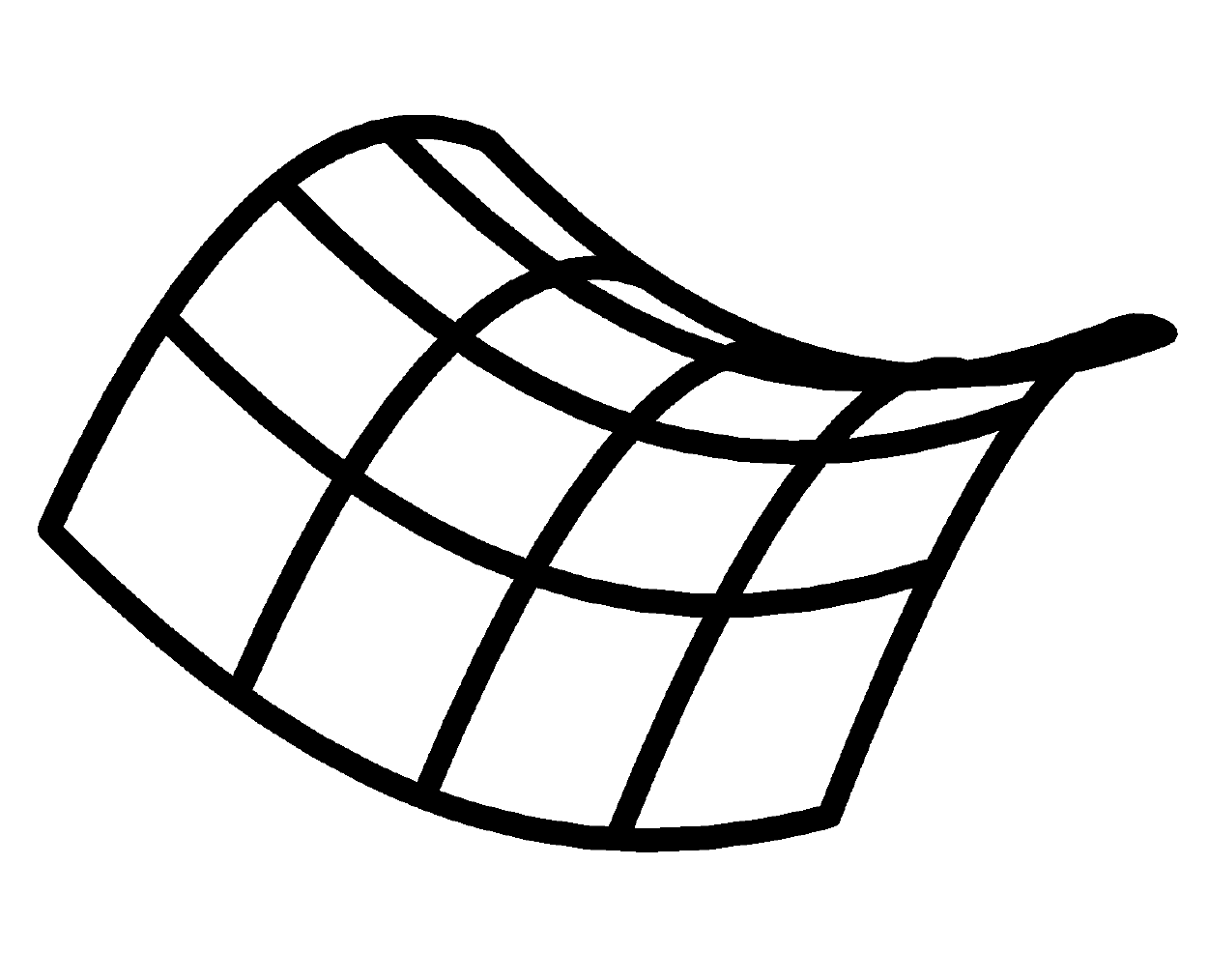}\!\!}
\newcommand*{\Qop}{\mathscr{Q}}

\newcommand*{\Qb}{\mathbb{Q}}

\newcommand*{\n}{\boldsymbol{n}}

\newcommand*{\setN}{\ensuremath{\mathbb{N}}}
\newcommand*{\setZ}{\ensuremath{\mathbb{Z}}}

\newcommand*{\setR}{\ensuremath{\mathbb{R}}}

\newcommand*{\setS}{\ensuremath{\mathbb{S}}}
\newcommand*{\setT}{\ensuremath{\mathbb{T}}}

\newcommand{\curly}{\mathrel{\leadsto}}

\newcommand*{\Qp}{\mathcal{Q}}

\newcommand*{\Sum}{\mathlarger{\mathlarger{\sum}}}

\newcommand*{\Int}{\mathlarger{\int}}

\newcommand*{\Lp}{\mathscr{L}}

\newcommand*{\Laplacian}{\mathop{{}\bigtriangleup}\nolimits}

\NewDocumentCommand{\Opp}{sO{\mathcal{Q}}oo}{%
  \IfBooleanTF{#1}
    {\IfNoValueTF{#3}
      {\mathscr{O}^{-#2}}%
      {\IfNoValueTF{#4}
         {\mathscr{O}^{-#2;#3}}%
         {\mathscr{O}_{#4}^{-#2;#3}}%
      }}%
    {\IfNoValueTF{#3}
      {\mathscr{O}^{#2}}%
      {\IfNoValueTF{#4}
         {\mathscr{O}^{#2;#3}}%
         {\mathscr{O}_{#4}^{#2;#3}}%
      }}%
}
\NewDocumentCommand{\Vpp}{sO{\mathcal{Q}}o}{%
  \IfBooleanTF{#1}
    {\IfNoValueTF{#3}
      {\mathscr{O}^{-#2}_{\ell_2 m_2}}%
      {\mathscr{O}^{-#2}_{#3}}%
    }%
    {\IfNoValueTF{#3}
      {\mathscr{O}^{#2}_{\ell_1 m_1}}%
      {\mathscr{O}^{#2}_{#3}}%
    }%
}

\DeclareMathOperator{\Proj}{U}

\newcommand{\normord}[1]{\vcentcolon\mathrel{\mkern1mu#1\mkern1mu}\vcentcolon}

\def\scalar#1{\mathinner{({#1})}}

\DeclarePairedDelimiter\ceil{\lceil}{\rceil}
\DeclarePairedDelimiter\floor{\lfloor}{\rfloor}
\DeclareMathOperator{\Id}{1}

\DeclareMathOperator{\A}{\mathscr{A}}

\NewDocumentCommand{\Anew}{ooo}{\mathinner{\A\IfValueT{#1}{_{#1}}\IfValueT{#2}{^{#2}}(\tau_\text{in}, \tau_\text{out} \IfValueT{#3}{ \mid #3})}}

\DeclareDocumentCommand\DD{ e{^} m }{
  \!\mathinner{\mathcal{D}\IfValueT{#1}{^{#1}}{#2}}
}

\DeclareMathOperator{\spec}{spec}

\DeclareMathOperator*{\SumInt}{%
	\mathchoice%
	{\ooalign{$\displaystyle\Sum$\cr\hidewidth$\displaystyle\Int$\hidewidth\cr}}
	{\ooalign{\raisebox{.14\height}{\scalebox{.7}{$\textstyle\Sum$}}\cr\hidewidth$\textstyle\Int$\hidewidth\cr}}
	{\ooalign{\raisebox{.2\height}{\scalebox{.6}{$\scriptstyle\Sum$}}\cr$\scriptstyle\Int$\cr}}
	{\ooalign{\raisebox{.2\height}{\scalebox{.6}{$\scriptstyle\Sum$}}\cr$\scriptstyle\Int$\cr}}
}

\DeclareMathOperator\arctanh{arctanh}

\DeclareMathOperator\arccoth{arccoth}

\DeclareMathOperator{\Disc}{Disc}

\thesistitle{A study of Bosonic and Fermionic Theories at Large Charge} 
\supervisor{Prof. Dr. Susanne Reffert} 

\firstname{Ioannis} 
\lastname{Kalogerakis} 
\author{\fname~ \lname} 
\matriculationnumber{19-120-955} 
\degree{Doctor of Science} 
\institute{\href{https://www.itp.unibe.ch/}{Institute for Theoretical Physics}} 
\faculty{\href{https://www.philnat.unibe.ch}{Faculty of Science}} 
\university{\href{https://www.unibe.ch/index_eng.html}{University of Bern}} 

\department{\href{https://www.izb.unibe.ch/}{Department of Biology}} 
\group{\href{https://www.GROUP_NAME.net/}{Group name}} 
\subject{Cell biology} 

\usepackage{float}

\usepackage{titlesec}

\titleformat{\section}
  {\normalfont\Large\bfseries}{\thesection}{1em}{}[{\titlerule[2pt]}]

\titleformat{\subsection}
  {\normalfont\Large\bfseries}{\thesubsection}{1em}{}[{\titlerule[0.8pt]}]

\titleformat{\subsubsection}
  {\normalfont\Large\bfseries}{\thesubsubsection}{1em}{}[{\titlerule[0.3pt]}]

\AtBeginDocument{
\hypersetup{pdftitle=\ttitle} 
\hypersetup{pdfauthor=\authorname} 
\hypersetup{pdfkeywords=\keywordnames} 
}

\newacronym{qft}{QFT}{Quantum Field Theory}
\newacronym{qfts}{QFTs}{Quantum Field Theories}
\newacronym{cft}{CFT}{Conformal Field Theory}
\newacronym{cfts}{CFTs}{Conformal Field Theories}
\newacronym{eft}{EFT}{Effective Field Theory}
\newacronym{lce}{LCE}{Large Charge Expansion}
\newacronym{wf}{WF}{Wilson-Fisher}
\newacronym{ir}{IR}{Infrared}
\newacronym{uv}{UV}{Ultraviolet}
\newacronym{ssb}{SSB}{Spontaneous Symmetry Breaking}
\newacronym{scts}{SCT}{special conformal transformation}
\newacronym{ope}{OPE}{operator product expansion}
\newacronym{rg}{RG}{renormalisation group}
\newacronym{eom}{eom}{equation of motion}
\newacronym{1pi}{1PI}{one-particle irreducible}
\newacronym{ms}{MS}{minimal subtraction}
\newacronym{rgf}{RG functions}{renormalisation group functions}
\newacronym{rge}{RGE}{renormalisation group equation}
\newacronym{pde}{PDE}{partial differential equation}
\newacronym{ccwz}{CCWZ}{Coleman-Callan-Wess-Zumino}
\newacronym{vev}{VEV}{vacuum expectation value}
\newacronym{lsm}{LSM}{linear sigma model}
\newacronym{nlsm}{NLSM}{non-linear sigma model}
\newacronym{dof}{DOF}{degrees of freedom}
\newacronym{ode}{ODE}{ordinary differential equation}
\newacronym{wl}{WL}{worldline}
\newacronym{gn}{GN}{Gross-Neveu}
\newacronym{gny}{GNY}{Gross-Neveu-Yukawa}
\newacronym{njl}{NJL}{Nambu–Jona–Lasinio}
\newacronym{pg}{PG}{Pauli–Gürsey}
\newacronym{bcs}{BCS}{Bardeen--Cooper--Schrieffer}
\newacronym{nlo}{NLO}{next-to-leading-order}

\begin{document}

\frontmatter 

\pagestyle{plain} 

\begin{titlepage}

 \begin{center}
\hypersetup{hidelinks} 
 \vspace*{.06\textheight}

 {\huge \bfseries \ttitle\par}\vspace{1.5cm} 
  {\Large Inaugural dissertation}\break
  {\Large of the \facname,}\break
  {\Large \univname}
  \vspace{1.5cm}
 
  {\Large presented by}
 
  \vspace{1cm}
 
  {\LARGE \bfseries \authorname \par}\vspace{0.4cm} 
  {\Large from Greece}\\[0.5cm]

   \vfill

 \emph{Supervisor of the doctoral thesis:}\\[1mm]
 {\Large \supname}\\ 
 {\Large \instname}\\ 
 {\Large  \facname ~of the \univname} \\

 \vfill
 \end{center}
\end{titlepage}

\begin{titlepage}
{
\vspace*{-3cm}
\hfill
}
 \begin{center}
\hypersetup{hidelinks} 
 \vspace*{.06\textheight}

 {\huge \bfseries \ttitle\par}\vspace{1.5cm} 
  {\Large Inaugural dissertation}\break
  {\Large of the \facname,}\break
  {\Large \univname}
  \vspace{1.5cm}
 
  {\Large presented by}
 
  \vfill
 
  {\LARGE \bfseries \authorname \par}\vspace{0.4cm} 
  {\Large from Greece}\\[0.5cm]

   \vfill

 \emph{Supervisor of the doctoral thesis:}\\[1mm]
 {\Large \supname}\\ 
 {\Large \instname}\\ 
 {\Large  \facname ~of the \univname} \\
 
\vfill

 {\Large Accepted by the \facname.}
 
\vfill
 
  \end{center}
 
 Bern, \ \today \hfill  The Dean \hspace{2.3cm} \break
\raggedleft	 Prof. Dr. Marco Herwegh

\end{titlepage}


\cleardoublepage
\vspace*{0.2\textheight}

\hfill \textit{To the love of my life, Helena, for always being there for me, and to my parents for all their sacrifices.}


\cleardoublepage
\vspace*{0.2\textheight}

\noindent\enquote{\itshape Whereof one cannot speak, thereof one must be silent.}\bigbreak

\hfill Ludwig Wittgenstein


\begin{abstract}
\addchaptertocentry{\abstractname} 

The aim of this thesis is to systematically and consistently study strongly coupled bosonic and fermionic conformal field theories using the large quantum number expansion. The idea behind it is to study sectors of conformal field theories that are characterised by large quantum numbers under global symmetries. In this limit, the conformal field theories, even if they initially were strongly coupled and interacting, can now be written in terms of an effective field theory that is weakly coupled. Some common effective field theories that appear in the literature are the bosonic conformal superfluid and the Fermi sphere, condensed matter systems characterised by a high particle density, making the study of such systems a cross-disciplinary matter.

In the first part of the thesis, we start by reviewing concepts of quantum field theories, conformal field theories and of the large charge expansion that are essential for the subsequent chapters. 

The second part of the thesis is devoted to the analysis of the \(O(2)\) vector model in the large charge expansion, where we start by reviewing the classical treatment of the model, and we continue by examining its quantum behaviour. Following that, we compute three and four-point correlation functions of large charge primaries with insertions of the conserved current \(\mathcal{J}^\mu\) and/or the energy momentum tensor \(T^{\mu \nu}\) and in the final part of the chapter, we compute three and four-point correlation functions of spinning charged primary operators \(\Vpp[\Qp][\ell m]\) with the insertion of light charged spinning primary operators \(\mathscr{O}^q\) with \(q \ll \Qp\) in the middle. 

In the third part of the thesis, we use the resurgence methodology to study the asymptotics of the \(O(N)\) vector model. We start by introducing the \(O(N)\) model at large charge and large-\(N\) and we study its asymptotics on the torus \(\setT^2\) and the sphere \(\setS^2\) using the resurgence framework. Then, we derive a worldline interpretation of the heat-kernel trace that replicates the previously computed results, and we deduce that its geometric nature is robust enough to allow us to extrapolate the results to finite \(N\). Finally, we compare the resurgent results with the small-charge regime which is accessible at the doubling-scaling limit, while we theorise that our conjunctions hold for finite values of \(N\), and we find the value of optimal truncation. 

In the last part of the thesis, we study various fermionic models with a four-fermion interaction at large charge and large-\(N\). We start by presenting the models that we will examine in the subsequent sections and afterwards, we discuss the symmetry breaking pattern of the models, and we investigate the appearance of a condensate. We continue by studying the spectrum of the fluctuations for the models, and we compute the scaling dimension of the \acrlong{gn} model at large charge and of the \acrlong{njl} model both at large and small-charge regime.

\end{abstract}


\begin{acknowledgements}
\addchaptertocentry{\acknowledgementname} 
My personal belief has always been that for every problem, there exists a rational
answer. Every human being is born with some sort of curiosity about their existence and
place in the universe. Growing up, I had my set of philosophical questions as
well, and for me, the scientific method has always been an ideal way to tackle such
questions. My passion for physics and especially theoretical physics is motivated
strongly by the fact that it not only explains every day phenomena but also allows us
to imagine things unseen through thought experiments and mathematical reasoning. 

This line of questioning eventually led me to do my PhD at the~\instname~at the~\univname~under the supervision of Susanne Reffert. Therefore, the first obvious, but sincere, acknowledgements are to her. I am truly grateful for believing in me, choosing me as a PhD student, and giving me the chance to (hopefully) attain the highest academic degree. During these three and a half years, Susanne has been an immense help in the pandemic and post-pandemic period where the lives of all of us drastically changed, and we managed to overcome the difficulties that arose and have a meaningful and fruitful collaboration. Her broad knowledge of physics, the clarity of her thoughts, and her incredible ability to accurately express complex concepts and ideas are inspirational. I am grateful for helping me be a better physicist and being part of a great team in such a prestigious institute. For what matters, sincerely, thanks!

At this point, I want to take the opportunity to personally thank every member of the team, starting with Domenico Orlando. His innovative ideas and his passion for physics have been stimulating. I also want to thank him for being his teaching assistant for the last three years, during which we had an excellent collaboration. He was available at all times to help and answer all possible questions, and he was always a joy to be around. 

\medskip

Next, I want to thank my collaborators, Nicola Dondi, Rafael Moser, Vito Pellizzani, Fabio Apruzzi and Donald Youmans, who besides co-workers have also become friends. Nicola, thank you for sharing the office with me, and working with me on all of my projects. I learned so much from you, and I wish you the best in your future academic career, which you truly deserve. Donald, thank you for our conversations, for unofficially sharing the office with me and Nicola and all the great times we had at the office and outside. Vito and Rafael, thank you for sharing the demanding experience of doing a PhD, and for our conversations about physics and more. I wish you both the best to your future career plans. Especially Rafael, I took great pleasure in our collaboration on most of our projects, I admired your talent and brightness from the time that you were my student, and it was an honour working with you. Last but not least, I want to thank Thiago Araujo for our short time together, I am genuinely sorry that the pandemic did not let me get to know you better, and we did not get the opportunity to write a paper together that we both wanted so much. 

Finally, I want to thank three people I worked with during these years, Matthias Blau, Simeon Hellerman and Gilberto Colangelo, I learned a lot from you, and it was an honour working together. 

\medskip

Ultimately, it has been six years since I left Greece, but in all those years my friends from back home were always there for me and for this, I am truly grateful. Doing a PhD would have been impossible if I did not have the help and support of my family and especially my parents, who sacrificed so much for me to be here. Words cannot describe my gratitude and appreciation for all that you have offered me and how you raised me. The last person I want to thank is the one closest to my heart. Helena, thank you for all your sacrifices all these years and all the sleepless nights together during the writing of this thesis. Thank you for being there for me and helping me overcome all the adversities that we have faced together. Thank you for supporting me in spirit and in body during the pandemic and after it. If it wasn't for you, I do not know if I would have been able to reach thus far. You are my anchor and the love of my life, and for that, I feel truly blessed.

\end{acknowledgements}


{
  \hypersetup{linkcolor=black} 
  \tableofcontents 
}


\begin{abbreviations}{ll} 

\textbf{QFT} & \textbf{Q}uantum \textbf{F}ield \textbf{T}heory\\
\textbf{CFT} & \textbf{C}onformal \textbf{F}ield \textbf{T}heory\\
\textbf{EFT} & \textbf{E}ffective \textbf{F}ield \textbf{T}heory\\
\textbf{LCE} & \textbf{L}arge \textbf{C}harge \textbf{E}xpansion\\
\textbf{SSB} & \textbf{S}pontaneous \textbf{S}ymmetry \textbf{B}reaking\\
\textbf{WF} & \textbf{W}ilson-\textbf{F}isher\\
\textbf{RG} & \textbf{R}enormalisation \textbf{G}roup\\
\textbf{RGE} & \textbf{R}enormalisation \textbf{G}roup \textbf{E}quation\\
\textbf{IR} & \textbf{I}nfrared\\
\textbf{UV} & \textbf{U}ltraviolet\\
\textbf{SCTs} & \textbf{S}pecial \textbf{C}onformal \textbf{T}ransformations\\
\textbf{OPE} & \textbf{O}perator \textbf{P}roduct \textbf{E}xpansion\\
\textbf{EOM} & \textbf{E}quation \textbf{O}f \textbf{M}otion\\
\textbf{1PI} & \textbf{O}ne \textbf{P}article \textbf{I}rreducible\\
\textbf{MS} & \textbf{M}inimal-\textbf{S}ubtraction\\
\textbf{PDE} & \textbf{P}artial \textbf{D}ifferential \textbf{E}quation\\
\textbf{CCWZ} & \textbf{C}oleman-\textbf{C}allan-\textbf{W}ess-\textbf{Z}umino\\
\textbf{BCS} & \textbf{B}ardeen-\textbf{C}ooper \textbf{S}chrieffer\\
\textbf{VEV} & \textbf{V}acuum\textbf{E}xpectation \textbf{V}alue\\
\textbf{LSM} & \textbf{L}inear \textbf{S}igma \textbf{M}odel\\
\textbf{NLSM} & \textbf{N}on \textbf{L}inear \textbf{S}igma \textbf{M}odel\\
\textbf{DOF} & \textbf{D}egrees \textbf{O}f \textbf{F}reedom\\
\textbf{PDE} & \textbf{P}artial \textbf{D}ifferential \textbf{E}quation\\
\textbf{ODE} & \textbf{O}rdinary \textbf{D}ifferential \textbf{E}quation\\
\textbf{WL} & \textbf{W}orldline\\
\textbf{NJL} & \textbf{N}ambu–\textbf{J}ona–\textbf{L}asinio \\
\textbf{GN} & \textbf{G}ross-\textbf{N}eveu\\
\textbf{GNY} & \textbf{G}ross-\textbf{N}eveu-\textbf{Y}ukawa\\
\textbf{PG} & \textbf{P}auli-\textbf{G}ürsey  \\
\textbf{BCS} & \textbf{B}ardeen-\textbf{C}ooper-\textbf{S}chrieffer \\
\textbf{NLO} & \textbf{N}ext-to-\textbf{L}eading-\textbf{O}rder

\end{abbreviations}


\mainmatter 

\pagestyle{thesis} 



\chapter{Introduction} 

\label{Chapter1} 

\epigraph{\itshape “I am just a child who has never grown up. I still keep asking these 'how' and 'why' questions. Occasionally, I find an answer.”}{Stephen Hawking}

\acrfull{qft} is the theoretical framework that unifies classical field theory, special relativity, and, of course, quantum mechanics and is suitable for describing a wide-ranging collection of physical systems, extending from condensed matter theory up to high-energy physics. The study of \acrlong{qfts} has strong implications in a variety of related subjects like mathematics, cosmology, black holes, quantum gravity and its best realisation which is string theory. One of the most prominent methods of studying the properties of \acrshort{qfts} is examining the Wilsonian \acrfull{rg} flow \cite{PhysRevB.4.3174,PhysRevB.4.3184,komargodski2012constraints}. There is strong evidence that the end points of the \acrshort{rg} flow of local, unitary and relativistic \acrlong{qfts} are described by fixed points so that the \acrshort{rg} flow ranges between the \acrfull{uv} limit found at small distances and/or high energies and the \acrfull{ir} limit that is found at long distances and/or low energies. At the fixed points live \acrfull{cfts}, which are a sub-branch of \acrshort{qfts} that exhibit symmetry under rescaling of physical lengths, and they are like beacons of light in the uncharted territory of \acrlong{qfts}. On that account, in accordance with the Wilsonian universality principle, every \acrshort{qft} can be categorised using \acrshort{cfts} and their relevant deformations. 

\acrshort{cfts} play an important role in theoretical physics, and a short but in-exhaustive list of uses consists of the following: they represent critical points in statistical mechanics, they describe the world-sheet of string theory, and they are connected to quantum gravity through the AdS/CFT correspondence. Their fairly constrained structure permits us to express all observables in terms of a set of two numbers, the scaling dimension \(\Delta\) of every primary operator of the theory, and the \acrfull{ope} coefficients, collectively known as the \acrshort{cft} data. Alas, most \acrshort{cfts} are strongly coupled, and the related data is very hard to collect without any further simplifying assumptions that would permit any form of semiclassical approximation. The means at our disposal are very limited, and they consist mainly of numerical approaches like the Monte Carlo method \cite{campostrini2001critical,campostrini2002critical}, non-perturbative approaches like the conformal bootstrap \cite{polyakov1970conformal,ferrara1972covariant,belavin1984infinite,dolan2001conformal,dolan2004conformal,RiccardoRattazzi_2008,https://doi.org/10.48550/arxiv.1602.07982} and a few perturbative approaches like large-\(N\) expansion~\cite{Moshe_2003,zinn2021quantum} and the small-\(\varepsilon\) approximation~\cite{WILSON197475}. Therefore, we should examine in depth any new technique that aims to deal with this issue.

The approach we are examining in this thesis is the large quantum number expansion. It has been observed that the conformal data of operators related to \acrshort{cft} sectors of large quantum numbers, like large spin \(J\)~\cite{alday2007comments,komargodski2013convexity,fitzpatrick2013analytic} exhibit considerable simplifications. Very recently, it was realised that similar simplifications arise in sectors of large global charge \cite{hellerman2015cft,monin2017semiclassics,Gaume:2020bmp} and this new methodology is called the \acrfull{lce}. The idea behind it is to study sectors of \acrlong{cfts} that are characterised by large quantum numbers under global symmetries. In this limit, the \acrshort{cfts}, even if they initially were strongly coupled and interacting, can now be written in terms of an \acrfull{eft} that is weakly coupled. Some common \acrshort{eft}s that appear are the \textit{bosonic conformal superfluid} and the \textit{Fermi sphere}, condensed matter systems characterised by a high particle density, making the study of such systems a cross-disciplinary matter.



Generally, the \acrshort{lce} was applied with great success to the case of the \(O(N)\) model and similar bosonic models as a means of analytically accessing these strongly coupled \acrshort{cfts}. In the large-charge sector, we are allowed to write an \acrshort{eft}, known as bosonic conformal superfluid, as an expansion in inverse powers of the large charge \(\Qp\) and calculate the relevant \acrshort{cft} data encoding all the information about the testable predictions of the theory~\cite{badel2019epsilon,giombi2021large,antipin2020charging,antipin2020charging2,jack2021anomalous,jack2022scaling,monin2017semiclassics,jafferis2018conformal,arias2020correlation,cuomo2021note,Cuomo:2020thesis,komargodski2021spontaneously,banerjee2018conformal,banerjee2019conformal,banerjee2022subleading,dondi2022spinning,alvarez2017compensating,alvarez2019large,moser2022convexity}. Therefore, the idea of fixing the charge and restricting the system to the large-charge Hilbert space sector, results in a ground state that spontaneously breaks boost invariance, and the combination of simultaneously breaking boost and charge invariance gives rise to a condensate and a number of Goldstone modes in terms of which we can now express the \acrshort{eft}. 

For the class of bosonic superfluids in \(d=3\) spacetime dimensions, the scaling dimension of the lowest charged primary operator \(\Opp\) reads~\cite{hellerman2015cft,monin2017semiclassics}
\begin{equation}\label{eq:scalingON}
	\Delta_\Qp = d_{3/2} \Qp^{3/2} + d_{1/2} \Qp^{1/2} - 0.0937 +\order{\Qp^{-1/2}},
\end{equation}
where with \(d_{i}\) we denote the Wilsonian parameters that are inaccessible within the validity of the \acrshort{eft} but can be computed if another controlling parameter is added to the system, like large-\(N\)~\cite{alvarez2019large}. 
The first two terms in the above expression are originating from the large-charge ground state, but the \(\Qp^0\) contribution is the Casimir energy of the fluctuations around the classical saddle, and it gets no other corrections from any terms of the theory; hence, it is a universal prediction and a feature of all conformal superfluid \acrshort{eft}s.

For the case of the \(O(N)\) model, the massless spectrum on top of the fixed ground state can be computed~\cite{alvarez2017compensating} and is made up by one conformal or type I Goldstone mode and \(N-1\) Goldstone modes of type II that exhibit a quadratic dispersion relation
\begin{align}
	\omega_{\text{I}} &= \frac{1}{\sqrt{2}}p + \dots, &  \omega_{\text{II}} &= \frac{p^2}{2\mu}+ \dots,
\end{align}
and every one of them comes along with a massive mode of mass \(\mu \sim \sqrt{\Qp}\).



So far in the literature, the scaling dimension \(\Delta_\Qp\) of the lowest primary large-charge operators \(\Opp\) have been computed~\cite{hellerman2015cft,monin2017semiclassics,Gaume:2020bmp,alvarez2019large,badel2019epsilon,giombi2021large,antipin2020charging,antipin2020charging2,jack2021anomalous,jack2022scaling}, with the results being separately verified through lattice computations~\cite{banerjee2018conformal,banerjee2019conformal,banerjee2022subleading,singh2022large}, and only few \(n\)-point functions have dispersedly turned up that compute the \acrshort{ope} coefficients~\cite{monin2017semiclassics,jafferis2018conformal,arias2020correlation,cuomo2021note,Cuomo:2020thesis,komargodski2021spontaneously} so that we can have complete knowledge of the \acrshort{cft} data. Therefore, there has been a dire need to systematically collect and study in a unified language all possible three and four-point correlation functions with current insertions for a general \acrshort{cft} in \(d\) spacetime dimensions that exhibits a global \(O(2)\) phase symmetry as a subgroup of a larger symmetry group like \(O(N)\) and that has a low-energy description as a superfluid \acrshort{eft}. This is the subject of \Cref{Chapter3} following closely \cite{dondi2022spinning}. 

The aim is not only to review known results, but to go beyond the state of the art by computing for the first time correlation functions between spinning phonon states, which are excitations over the homogeneous scalar ground state. The existence of these spinning states which are known as \emph{superfluid phonon} states was already predicted by the very first paper~\cite{hellerman2015cft}, and correspond to spinning primaries \(\Opp_{\ell m}\) labelled by different quantum numbers, \emph{e.g.} spin \(\ell\), but same charge \(\Qp\) as the scalar primary \(\Opp\). From a technical perspective, we use the fact the superfluid \acrshort{eft} is a weakly coupled system, which at leading order can be perceived as a free theory, and we can apply the usual prescriptions, like canonical quantisation to compute correlators using the standard operator algebra. The form of the superfluid \acrshort{eft} is constrained by conformal invariance and the symmetry breaking pattern with the leading term scaling as \(\Qp^{d/\pqty{d-1}}\), and there are in total \(\ceil{(d+1)/2} \) terms that scale positively in \(\Qp \) that originate from curvature terms in the action, with the last one scaling as \( \Qp^{1/(d-1)}\) while the Casimir energy of the fluctuations scales as \(\Qp^0\). Hence, we can express correlation functions as a series of semiclassical terms with positive \(\Qp\) scaling and a one-loop quantum contribution of order \(\Qp^0\), while neglecting all the terms that have negative scaling in the charge \(\Qp\). Moreover, since the saddle point of the theory corresponds to a classically homogeneous ground state, the only position dependence in the form of the correlation functions has to come from quantum fluctuations. In our computations, we consider only non-supersymmetric theories, while theories with supersymmetry have to be studied separately due to their characteristics and have been examined in~\cite{hellerman2015cft,hellerman2017operator,hellerman2017large,bourget2018limit,hellerman2019universal,Hellerman_2021S,beccaria1809large,hellerman2021large,hellerman2021exponentially,giombi2022large,giombi2022large2}.



One of the most impressive results in the large-charge analysis of the \(O(N)\) vector model in \(d=3\) spacetime dimensions at the conformal point, is that the \acrshort{lce} seems to work even for relatively small charges, a feature that is quite remarkable and surprising at the same time, since we presumed that the semiclassical expansion would only function in systems with a considerable number of \acrfull{dof}. This characteristic of the superfluid \acrshort{eft} was first discovered in a series of papers by Banerjee et al.~\cite{banerjee2018conformal,banerjee2019conformal} when comparing the conformal dimension of the lowest charged operator \(\Opp\) with lattice computations for the cases of the \(O(2)\) and the \(O(4)\) model. Moreover, it was discovered that only a few terms in the effective action were enough to replicate the results of the lattice computations with high precision.

From the standpoint of the superfluid \acrshort{eft}, it is not clear why the predictions made in the large-charge regime can be extrapolated to the small-charge limit. However, with the inclusion of an additional controlling parameter to the system, it is possible to move past the \acrshort{eft} and try to understand this phenomenon. For instance, this is possible if we examine the large-\(N\) limit of the \(O(N)\) vector model at large charge and large \(N\)~\cite{alvarez2019large,giombi2021large}. In the newly derived double-scaling limit, defined as the limit where \(\Qp\to \infty, \, \,  N\to \infty\), with their ratio \(\Qp/(2N) = \Qb \) being constant, it is feasible to make exact predictions at leading order in \(N\) for every value of the charge \(\Qb\). 

Inspired by that, in \Cref{Chapter4} we take up where Álvarez-Gaumé et al.~\cite{alvarez2019large} left off, and we demonstrate that the \acrshort{lce} in the double-scaling limit is actually asymptotic and this can be attributed to the asymptotic  character  of the Seeley--DeWitt expansion \cite{dewitt1957dynamical,seeley1967complex} of the heat-kernel trace and the associated zeta function on the two-sphere \(\setS^2\)~\cite{CANDELAS1984397,dowker2005barnes}. Dyson~\cite{PhysRev.85.631} was the first to assert that the existence of asymptotic series is a common characteristic of perturbative solutions of quantum mechanical problems. However, this asymptotic nature of the solutions indicates the existence of non-perturbative phenomena in the corresponding theory, a fact that was initially examined in a quantitative manner in the framework of anharmonic oscillators 
in a succession of papers by Bender and Wu with the initial being~\cite{PhysRevLett.27.461}. Translated to modern-day language, the topic has been renamed to \emph{resurgent asymptotics}, or for simplicity \emph{resurgence}, and was originally developed by Écalle~\cite{ecalle1981fonctions}.

Therefore, in the present work, we will utilise the resurgence methodology to demonstrate which non-perturbative corrections appear in the double-scaling limit of the \acrshort{lce} and in what way non-perturbative ambiguities are lifted. In the case of the sphere \(\setS^2\) resurgence analysis alone fails to provide an unambiguous result for the non-perturbative corrections. This ambiguity can be resolved in two ways, either by utilising the resurgence methodology for the Dawson's function, or using a geometric interpretation in terms of worldline instantons. 

The second procedure does not a priori depend on large \(N\) given the fact that the final result is a finite-volume effect that is connected to the geometric structure of the compactification manifold. Therefore, we obtain a nice geometric understanding of both the non-perturbative contributions and of the Borel ambiguities and the picture is robust enough to let us go beyond the double-scaling limit and suggest a precise form for the grand potential \(\omega\) that holds true for all values of the charge \(\Qb\). We are able to verify our computation numerically with excellent precision, and we are in a position to conjecture that the \acrshort{lce} is always asymptotic, even in finite \(N\), with an optimal truncation \(N^* = \order*{\sqrt{\Qp}}\) and an error of order \(\epsilon(\Qp) = \order*{e^{- \sqrt{\Qp}}}\) which is in agreement with the lattice simulations~\cite{banerjee2018conformal,banerjee2019conformal}. 

The non-perturbative corrections that we compute can be seen as “classical” and originate from the fact that the \acrshort{eft} is an asymptotic expansion on its own with a \((2n)!\) factorial growth. Their scaling of \(\order*{e^{- \sqrt{\Qp}}}\) has to be compared with the usual instantons that originate in the proliferation of Feynman diagrams, which scale as \(\order*{e^{-\Qp^{3/2}}}\) and are therefore subleading. This \((2n)!\) factorial growth is also present in calculations of the effective action of the Euler--Heisenberg type of theories~\cite{heisenberg2006consequences,dunne2005heisenberg}. In those cases, it has been demonstrated \cite{dunne2005worldline,dunne2006worldline} that this \((2n)!\) growth is actuated by worldline instantons, which is precisely what happens in our problem. 



Meanwhile, while the majority of the \acrshort{lce} literature is concentrated around bosonic systems, there have been some attempts to study relativistic fermionic systems at large charge~\cite{delacretaz2022thermalization,komargodski2021spontaneously,antipin2022yukawa} \footnote{The case of the unitary Fermi gas, which is described in terms of a non-relativistic \acrshort{cft}, has been examined in~\cite{favrod2018large,kravec2019nonrelativistic,kravec2019spinful,orlando2021near,hellerman2020droplet,pellizzani2022operator,hellerman2022nonrelativistic} using the large charge approach.}.
Interestingly, in~\cite{komargodski2021spontaneously} the authors showed that the superfluid pattern is inapplicable for the free fermion case, where the fixed charge ground state is described by a Fermi surface.

Therefore, in \Cref{Chapter5} we aim to fill this gap in the literature and consistently investigate various fermionic models with a four-fermion interaction term in \(d=3\) spacetime dimensions, like the \acrfull{gn} model~\cite{gross1974dynamical}, the chiral \acrshort{gn} or \acrfull{njl} model~\cite{nambu1961dynamical,nambu1961dynamical2}, and finally, its \(SU(2) \times SU(2)\) generalisation. Since we want to have extra control over our system, we will use an additional controlling parameter, which is the large-\(N\) limit~\cite{alvarez2019large} of fermion flavours.

We discover that two types of qualitative behaviour are possible according to our findings:
\begin{enumerate}[left= 0pt]

    \item For \acrlong{gn} types of models we see that there is no \acrfull{ssb} taking place in sectors of large baryon number, and the large-\(N\) physics is described in terms of an approximate Fermi surface strictly in the \(N \to \infty\) limit. It is not yet clear if the Fermi surface remains when subleading corrections in \(N\) are considered. The same Fermi surface behaviour has been found for the case of the free fermion~\cite{komargodski2021spontaneously,dondi2022fermionic}. The ground state is related to the scalar primary \(\Opp_{\text{FS}}\) that first appeared in~\cite{komargodski2021spontaneously} and the scaling dimension \(\Delta_{\text{FS}}\) reads
\begin{equation}
	\frac{\Delta_{\text{FS}}(\Qp)}{N} =
\frac{2}{3} \pqty{\frac{\Qp}{2N}}^{3/2} + \frac{1}{12} \pqty{\frac{\Qp}{2N}}^{1/2} - \frac{1}{192} \pqty{\frac{\Qp}{2N}}^{-1/2} + \order{\pqty{\frac{\Qp}{2N}}^{-3/2}}.
\end{equation}
We notice that this large-charge sector has no \(\Qp^0\) universal contribution corresponding to the Casimir energy of fluctuations, since there are no Goldstone modes.

\item For \acrfull{njl} types of models, the “chiral” four-fermion interaction term permits \acrshort{ssb} to take place and the large-\(N\) physics is described by a conformal superfluid \acrshort{eft} in a different universality class than the bosonic case~\cite{hellerman2015cft,monin2017semiclassics,Gaume:2020bmp}. The ground state corresponds to the primary operator \(\Opp_{\text{SF}}\) and in the limit that \(\Qp/N \gg 1\) the conformal dimension reads
\begin{equation}
  \frac{\Delta_{\text{SF}}(\Qp)}{2N} = \frac{2}{3} \pqty{\frac{\Qp}{2N \kappa_0}}^{3/2} + \frac{1}{6} \pqty{\frac{\Qp}{2N \kappa_0}}^{1/2} - 0.0937 +  \dots \, ,
\end{equation}
where the constant parameter \(\kappa_0\) is given by \(\kappa_0 \tanh \kappa_0 =1\).
The result is similar to the one derived in the case of the bosonic superfluid \acrshort{eft} of the \(O(N)\) model~\cite{alvarez2019large} but the Wilsonian coefficients are different, indicating that the fermionic \acrshort{cft} is in a different universality class. Moreover, working in the large-\(N\) limit we have access to the small-charge regime \(\Qp/N \ll 1\), where the scaling dimension reads
\begin{equation}
	\frac{\Delta(\Qp)}{2N}  = \frac{1}{2}\left(\frac{\Qp}{2N}\right) +\frac{2}{\pi^2}\left(\frac{\Qp}{2N} \right)^2+ \dots \, ,
\end{equation}
which is in accordance with the usual perturbative result for the free bosonic scalar operator of mass dimension one and charge two. These results also hold true for generalisations of the \acrshort{njl} model. 
\end{enumerate} 

As a final note, for every fermionic model that supports a large charge superfluid ground state, there is a physically intuitive way to comprehend the existence of a bosonic condensate. For instance, for the \acrshort{njl} model we can carry out a \acrfull{pg} transformation~\cite{pauli1957conservation,gursey1958relation}, defined as
\begin{align}
        \Psi &\mapsto \frac{ 1}{2} \left[ (1-\Gamma_5) \Psi + (1+ \Gamma_5) C_4 \bar\Psi^T  \right] ,&
        \bar\Psi &\mapsto \frac{ 1}{2} \left[ \bar\Psi (1+\Gamma_5) - \Psi^T C_4 (1- \Gamma_5) \right] ,
\end{align}
to derive a model that exhibits a Cooper-type interaction~\cite{kleinert1998two,ebert2016competition}.
Every computation can be reduplicated in the context of the Cooper model, with the same results as before. In the Cooper-pair context, it is evident that the nature of the condensate is that of Cooper pairs that describe a superconductor. The attractive interaction results in a Cooper instability, and we now have a system described by condensing bosons at large charge, and that is why the results are so similar to the \( O(N) \) vector model.

\medskip

The plan of the thesis is as follows: in \Cref{Chapter2} we start by briefly reviewing concepts of \acrshort{qfts} in \cref{sec:qft}, \acrshort{cfts} in \cref{sec.CFT} and the \acrshort{lce} in \cref{sec.largecharge} that are important for the subsequent chapters. 
In \Cref{Chapter3} we examine the \(O(2)\) model at large charge as a working example. In \cref{sec:O(2)_review} we review the classical treatment of the model, while in \cref{sec.canonicalandpathintegralquantisation} we now examine its quantum behaviour. Following that, in \cref{sec:ConformalAlgebraAndChargeCorrelators} we compute three and four-point correlation functions of large charge primaries with insertions of the conserved current \(\mathcal{J}^\mu\) and/or the energy momentum tensor \(T^{\mu \nu}\) and in the final part of the chapter, in \cref{sec:HLH} we compute three and four-point correlation functions of spinning charged primary operators \(\Vpp[\Qp][\ell m]\) with the insertion of light charged spinning primary operators \(\mathscr{O}^q\) with \(q \ll \Qp\) in the middle. 
In \Cref{Chapter4} we use the resurgence methodology to study the asymptotics of the \(O(N)\) vector model. In \cref{sec:O2N} we introduce the \(O(N)\) model at large charge and large-\(N\) and we study its asymptotics on the torus \(\setT^2\) and the sphere \(\setS^2\) using the resurgence framework. In \cref{sec:worldline} we derive a worldline interpretation of the heat-kernel trace that replicates the previously computed results, and we deduce that its geometric nature is robust enough to allow us to extrapolate the results to finite \(N\). In \cref{sec:resurgence} we compare the resurgent results with the small-charge regime which is accessible at the doubling-scaling limit, while in \cref{sec:lessons-from-large-N} we theorise that our conjunctions hold for finite \(N\), and we find the value of optimal truncation. 
In \Cref{Chapter5} we study various fermionic models with a four-fermion interaction at large charge and large-\(N\). In \cref{sec:models} we start by presenting the models that we will examine in the subsequent sections. Afterwards, in \cref{sec:symmetry} we discuss the symmetry breaking pattern of the models, and we investigate the appearance of a condensate. In \cref{sec:fluctuations} we study the spectrum of the fluctuations for the models, while in \cref{sec:conformalDim} we compute the scaling dimension of the \acrshort{gn} model at large charge and of the \acrshort{njl} model both at the large and at the small-charge regime. 
%

\chapter{Quantum, Conformal Field Theories and the Large-Charge Expansion} 

\label{Chapter2} 

\epigraph{\itshape “As far as laws of mathematics refer to reality, they are not certain; and
as far as they are certain, they do not refer to reality.” }{Albert Einstein }

The aim of this chapter is to give a pedagogical introduction to \acrlong{qfts}, \acrlong{cfts} and to the \acrlong{lce} to make the reading of the present thesis as self-contained as possible for the interested reader.

Therefore, the first part of this chapter is a brief introduction to \acrlong{qfts}, starting from classical field theory and conserved charges and moving to quantisation, regularisation, and the \acrlong{rg} flow. Since these are all well-known results for anyone with basic knowledge of \acrshort{qft}, this section will be kept short and used as a motivation for \acrshort{cfts}. There are many excellent and thorough books and lectures notes on the subject, some of the most exhaustive are the ones from Weinberg \cite{weinberg2005quantum1,weinberg2005quantum2,weinberg2005quantum3} while others include \cite{Peskin:1995ev,zinn2021quantum,Srednicki:1019751}.

The second part of the chapter is a basic introduction to \acrlong{cfts} and is accompanied by extensive derivations of the results presented in detail in \Cref{AppendixA}. The structure and the derivations are based on the \acrshort{cft} graduate course given at the \univname \ where the author was a tutor to the homonymous course. This section is mainly inspired by the lectures of Qualls \cite{Qualls:2015qjb}, the book of Ammon and Erdmenger \cite{ammon_erdmenger_2015} and the notes of Gillioz \cite{https://doi.org/10.48550/arxiv.2207.09474}.

The third and last part of the chapter is a terse introduction to the \acrlong{lce}. On general grounds, a systematic methodology is well-defined for the case that the system at hand has a description in terms of a conformal superfluid \acrshort{eft} with an \(O(2)\) Abelian symmetry subgroup.
Thus, the point of this section is to concisely review the most essential tools required to understand the \acrshort{lce}, like \acrlong{ssb} and the \textit{Goldstone theorem}. Any supplementary material will be presented on the spot in the relevant chapter.


\section{Quantum Field Theory}\label{sec:qft}

\acrlong{qft} is the theoretical framework that unifies classical field theory, special relativity, and of course quantum mechanics. A full analytical and in depth development of \acrshort{qft} is far beyond the scope of this chapter and of this thesis for that matter. Nevertheless, we will try to introduce crucial ingredients that will be used in the following sections. Therefore, by restricting to scalar fields \(\phi\), and working solely in flat space --- a great classical curved space \acrshort{qft} textbook is by Birrell and Davies \cite{birrell_davies_1982} while an explanatory recent paper is by Witten \cite{https://doi.org/10.48550/arxiv.2112.11614} --- we are able to tackle the most important issues. Starting with classical field theory, symmetries and charge conservation, we will move on to the quantisation of the fields, introducing the notation for both canonical and path integral quantisation. We will then briefly comment on the notion of Wick rotation and the connection of the path integral to statistical mechanics, and finish this review by addressing regularisation, renormalisation and the Wilsonian \acrlong{rg}. 


\subsection{Classical field theory and symmetries} \label{sec.classicalfield}

In all our analysis from this point on we always assume the existence of a \textit{globally hyperbolic spacetime}, in other words a \textit{pseudo-Riemannian manifold} $\left( \mathcal{M}, g_{\mu \nu} \right)$ \footnote{For a mathematically rigorous definition of manifolds see \Cref{sec.differentialgeometry}} along with a complete \textit{Cauchy hyper-surface} $\Sigma$ \footnote{A hyper-surface is a $(d-1)$-dimensional submanifold $\Sigma$ of a $d$-dimensional manifold $(\mathcal{M}, g_{\mu\nu})$. For a proper definition and discussion \cite[see][]{Hawking:1973uf}.} where initial conditions of the fields can be properly formulated in order to solve the initial value problem \footnote{For a globally hyperbolic spacetime, knowing a set of past or future initial conditions, allows us to determine the whole set of events in that
spacetime.}. As this chapter is purely introductory, for a thorough analysis of classical field theory, see \cite{arnold2013mathematical}.
\par
We start by defining a scalar field \(\phi(x) \in \setR^{d-1,1}\), which is the \(d\)-dimensional Minkowski space with one temporal and $d-1$ spatial directions. In the usual field theory language, we assume that $x$ indicates points in Minkowski space and consists of vector components $x^\mu$, where $ \mu= 0, \dots, d-1$, with $x^0 =c t$  \footnote{For the rest of the thesis and unless specified otherwise, we will set the speed of light to one so that we can use the same units for all spacetime components. } while $x^i = 1, \dots, d-1$. 

The infinitesimal line element for Minkowski space is specified, as
\begin{equation}\label{eq.Minklinelement}
    {\dd s}^2 \equiv - \left(\dd x^0\right)^2 + \Sum\limits_{i=1}^{d-1} \left(\dd x^i \right)^2 \equiv \eta_{\mu \nu} \dd x^\mu \dd x^\nu.
\end{equation}
It is obvious from the above expression \(\pqty{\ref{eq.Minklinelement}}\) that the Minkowski metric is a diagonal matrix 
\begin{equation}
    \eta_{\mu \nu} = \textrm{diag} (-1, \underbrace{1,\dots,1}_{\left(d-1\right) \ \textrm{times}} ),
\end{equation}
that satisfies $\eta^{\mu \nu} \eta_{\nu \sigma} = \delta^{\mu}_{\ \sigma}$ and can be used to lower and raise indices in a $d$-vector as $V_\mu = \eta_{\mu \nu} V^{\nu}$. 

The most elementary principle in physics is that the fundamental physical laws are independent of the reference system. In other words, observers doing experiments in different reference frames may have different outlooks, but the underlying physical laws are identical. In mathematical terms, this can be interpreted in the following manner. Having a map \footnote{In physical language  a map is translated to coordinate system.}, there is a set of transformations that leaves the infinitesimal line element invariant
\begin{equation}\label{eq.metricinvariance}
 \eta_{\mu \nu} \dd x^\mu \dd x^\nu = \eta_{\mu \nu} \dd x'^\mu \dd x'^\nu.
\end{equation}
The above principle is true for all maps that are \textit{invertible} --- isomorphisms --- and \textit{differentiable} --- smooth transformations --- so it is known as \textit{diffeomorphism invariance}. 

\Cref{eq.metricinvariance} is satisfied by a set of transformations, namely \textit{translations} by a constant vector
\begin{equation}\label{eq.translations}
    x^\mu \to x'^\mu = x^\mu + a^\mu,
\end{equation}
and \textit{Lorentz transformations} 
\begin{equation}\label{eq.Lorentztransformations}
    x^\mu \xrightarrow{\Lambda} x'^\mu = \Lambda^{\mu}_{\ \nu} x^\nu,
\end{equation}
where the matrix \(\Lambda\) satisfies
\begin{equation}
    \Lambda^{\mu}_{\ \rho} \Lambda^{\nu}_{\ \sigma} \eta_{\mu \nu} = \eta_{\rho \sigma}.
\end{equation}
Lorentz transformations \( \Lambda\) consist of rotations in spatial directions and boosts. Altogether, they form the Lorentz group \( SO(d-1,1)\). 

The combination of Lorentz transformations \( \Lambda\) and translations by \( a \) composes the Poincaré group \( ISO (d-1,1) \). These transformations can be gathered in a set \( \left( \Lambda, a \right) \) and their impact on \( x\) is
\begin{equation}
    x^\mu \xrightarrow{\left( \Lambda, \ a \right)} x'^\mu = \Lambda^{\mu}_{\ \nu} x^\nu + a^\mu.
\end{equation}
Two consecutive Poincaré transformations  \( \left( \Lambda_1, a_1 \right) \) and \( \left( \Lambda_2, a_2 \right) \) act as
\begin{align}
 \left( \Lambda_1, a_1 \right) \circ \left( \Lambda_2, a_2 \right) & = \left(\Lambda_1 \Lambda_2, a_1+ \Lambda_1 a_2 \right) & \in  ISO(d-1,1),
\end{align} 
or in component form
\begin{equation}
    x^\mu  \xrightarrow{\left(\Lambda_1 \Lambda_2, \ a_1+ \Lambda_1 a_2 \right)} x'^\mu = \Lambda^\mu_{\ \sigma} \Lambda^{\sigma}_{\ \rho} x^\rho + \Lambda^\mu_{\ \rho} a^\rho +a^\mu. 
\end{equation}
This is the underlying symmetry of space-time, manifest to all relativistic \acrlong{qfts} and it and the most accurate interpretation of the physical world, at least up to energy
levels where quantum gravity comes into play.

From now on, we will concentrate on the case of a real scalar field $\phi$ defined in \(d\)-dimensional spacetime. This field is a map that acts on spacetimes points $x$ and appoints them a real value \( \phi(x)\) and the change of the system can be described by an action \(S [\phi]\). In a more precise mathematical language we can express this as follows : assuming  \(\mathcal{M}\) to be a differentiable manifold, \( \mathcal{T}_p(\mathcal{M})\) \footnote{See \Cref{sec.differentialgeometry}} is the tangent space at a point \(p\)  and \(\mathscr{L} : \mathcal{T}_p(\mathcal{M}) \to \mathbb{R}\) is a differentiable functional, then \(\phi : \mathbb{R} \to \mathcal{M}\) is known as \textit{motion of the Lagrangian system with configuration manifold} \(\mathcal{M}\) \textit{and Lagrangian density} \(\mathscr{L}\) \footnote{In a shorthand notation it is usually just called \textit{Lagrangian}.} as long as $\phi$ is an extremum of
\begin{equation}\label{eq.definitionaction}
    S[\phi] = \Int \dd^d x \ \mathscr{L} \left( \phi, \del_\mu \phi \right).
\end{equation}
Both \(\mathscr{L}\) and consequently \(S\) depend on the field \( \phi(x)\) and the derivative of it \(\del_\mu \phi\), where \(\del_\mu = {\del}/{\del x^\mu}\) and in the usual field theory prescription only first derivatives of the fields are contained in the action and only local terms.

For a scalar theory to exhibit Poincaré invariance the Lagrangian can only be expressed regarding the field \( \phi \) in some power and on
\begin{equation}\label{eq.kinetictermscalar}
    - \left( \del_t \phi(t, x) \right)^2 + \left( \nabla \phi( t,x) \right)^2 \equiv \eta^{\mu \nu} \del_\mu \phi(x) \del_\nu \phi(x). 
\end{equation}
Thus from dimensional analysis \footnote{Remember --- \cite[see][]{deligne1999quantum1,deligne1999quantum2} --- that the mass dimensions 
\begin{align*}
    [S]&=0, & [\del_\mu] &= + 1, & [\dd^d x] = -d,
\end{align*} 
so that 
\begin{equation*}
    [\phi] = \frac{d-2}{2}.
\end{equation*}} 
and \cref{eq.kinetictermscalar}, we can deduce that the simplest form that a Lagrangian of a scalar field \(\phi \) can take is
\begin{equation}\label{eq.freelagragian}
    \mathscr{L}_{\textrm{free}} (\phi, \del_\mu \phi) = - \frac{1}{2}\eta^{\mu\nu} \del_{\mu}\phi (x) \del_{\nu}\phi(x) -\frac{1}{2} m^2 {\phi(x)}^2,
\end{equation}
where $m$ is the mass of the field and there are no interaction terms present. Hence, using \cref{eq.definitionaction,eq.freelagragian} the free action of a scalar field $\phi(x)$ is
\begin{equation}
     S[\phi]= -\frac{1}{2} {\Int} \dd^d x \left( \eta^{\mu\nu} \del_{\mu}\phi (x) \del_{\nu}\phi(x) +m^2 {\phi(x)}^2  \right).  
\end{equation}
The \textit{Euler-Lagrange} equation or \acrfull{eom} is derived by taking the derivative of the functional \( S[\phi]\) and demanding that \( \phi(x)\) is an extremum of it, i.e. \({\delta S}/{\delta \phi} =0\) so that
\begin{align}
    \frac{\delta S}{\delta \phi} &= \frac{\del \mathscr{L}}{\del \phi} - \del_\mu \left( \frac{\del \mathscr{L}}{\del \left( \del_\mu \phi \right)}\right), \\
    \frac{\delta S}{\delta \phi} &=0 \implies  \del_\mu \left( \frac{\del \mathscr{L}}{\del \left( \del_\mu \phi \right)}\right) = \frac{\del \mathscr{L}}{\del \phi}. \label{eq.extremalofLangrangian}
\end{align}
If now we use \cref{eq.freelagragian} the free scalar \acrshort{eom} takes the form
\begin{equation}\label{eq.KleinGordon}
    \left(- \Box + m^2 \right) \phi(x) =0,
\end{equation}
where \(\Box \equiv \del_\mu \del^\mu = - {\del_t}^2 + \nabla^2\) is known as \textit{D'Alembertian} or \textit{D' Alembert operator} and \cref{eq.KleinGordon} is known in the literature as \textit{Klein-Gordon} equation. It can be shown that the field \(\phi(x)\) that satisfies \cref{eq.KleinGordon} can be decomposed into positive and negative frequency modes as
\begin{equation}\label{eq.modedecomposition}
    \phi(x) = \frac{1}{\left( 2 \pi \right)^{d-1}} {\Int} \frac{\dd^{d-1} \vec{k}}{2 \omega_k} \left[ a(\vec{k}) e^{ikx} + a^* (\vec{k}) e^{-ikx} \right] \Big|_{k^0 = \omega_k},
\end{equation}
where \( \omega_k =  \sqrt{\vec{k}^2 + m^2} \) and also \(k x = - k^0x^0 + k^ix^i\).

Besides the free theory, we can also include interaction terms, which will be denoted as \(\mathscr{L}_{\textrm{int}}\). Frequently, the interaction term is realized as a polynomial of the form
\begin{equation}\label{eq.interactionterms}
\mathscr{L}_{\textrm{int}} (\phi) = - \frac{g_p}{p!} {\phi(x)}^p,
\end{equation}
where normally \( p \geq 3, n \in \mathbb{N}\) and \( g_p \in \mathbb{R}\) is known as the \textit{coupling constant} and dictates the strength of the interaction. 

Finally, a side comment that will prove useful for later discussions. If we imagine that we have a system that consists of two real fields \(\phi_1\) and \(\phi_2\) that do not interact with each other --- this system exhibits an internal \(O(2)\) symmetry --- and share the same mass \(m\), then it is possible to write the combined Lagrangian of the system
\begin{equation}
    \mathscr{L} (\phi_i, \del_\mu \phi_i) = - \frac{1}{2} \del_\mu \phi_i \del^\mu \phi_i - \frac{1}{2} m^2 {\phi_i}^2,
\end{equation}
where \(i=1,2\), in terms of one field by defining
\begin{align}\label{eq.O2toU1transformationfields}
    \varphi & = \frac{1}{\sqrt{2}} \left( \phi_1 + i \phi_2 \right), & \varphi^{*} & = \frac{1}{\sqrt{2}} \left( \phi_1 - i \phi_2 \right),
\end{align}
where \( \varphi\) is a \textit{complex scalar} and \( \varphi^*\) is its complex conjugate. Then the Lagrangian takes the form 
\begin{equation}\label{eq.U(1)system}
    \mathscr{L} (\varphi, \del_\mu \varphi) = - \del_\mu \varphi^*\del^\mu \varphi -  m^2 \varphi^* \varphi.
\end{equation}
This analysis can be extended beyond the free theory to include also interaction terms and to systems that have more than two fields, so they exhibit a larger internal symmetry than \(O(2)\). 

A crucial element of theoretical physics and variational calculus is the relationship between \textit{continuous symmetries} and \textit{conservation laws}. In fact, one of the most important theorems is called after Noether \cite{Noether_1971} and in accordance with it, continuous symmetries generate \textit{conserved currents} and \textit{charges}.

The starting point is to consider the transformation
\begin{equation}\label{eq.deformationfield}
    \phi(x) \to {\phi'}(x) = \phi(x) + \delta\alpha \delta \phi(x),
\end{equation}
where \( \delta\alpha \) is a random infinitesimal variable that is related to the deformation \(\delta \phi (x)\). We assume that the action remains invariant under \cref{eq.deformationfield} in the sense that
\begin{equation}
    S[\phi] = S[ {\phi'}],
\end{equation}
which holds true as long as the Lagrangian is invariant under \(\delta \phi (x)\) too, up to a total derivative term expressed regarding an arbitrary vector \(X^\mu\) as
\begin{equation}\label{eq.deformationLagrangian}
    \mathscr{L} \left(  {\phi'}, \del_\mu  {\phi'} \right) = \mathscr{L} \left( \phi, \del_\mu \phi \right) + \delta\alpha \del_\nu X^\nu.
\end{equation}
Actually \cref{eq.deformationLagrangian} indicates that
\begin{align}
    \delta\alpha \del_\nu X^\nu & \overset{!}{=}  \mathscr{L} \left(  {\phi'}, \del_\mu  {\phi'} \right) - \mathscr{L} \left( \phi, \del_\mu \phi \right) = \mathscr{L} \left( \phi + \delta\alpha \delta \phi, \del_\mu \phi + \delta\alpha \del_\mu \delta \phi \right) - \mathscr{L} \left( \phi, \del_\mu \phi \right) \nonumber\\
    & = \delta\alpha \left\{ \frac{\del \mathscr{L}}{\del \phi} \delta \phi + \frac{\del \mathscr{L}}{\del \left( \del_\mu \phi \right)} \del_\mu \delta\phi\right\} + \order{\alpha^2}\\
    & = \delta\alpha \Bigg\{ \underbrace{\left( \frac{\del \mathscr{L}}{\del \phi} - \del_\mu \left( \frac{\del \mathscr{L}}{\del \left( \del_\mu \phi \right)} \right)\right)}_{= \ 0 \ \textrm{due to \cref{eq.extremalofLangrangian}}} \delta\phi + \del_\mu \left( \frac{\del \mathscr{L}}{\del \left( \del_\mu \phi \right)} \delta\phi \right) \Bigg\} + \order{\alpha^2}, \nonumber
\end{align}
or to put it in another way
\begin{equation}
    \delta\alpha \del_\mu \left(  \frac{\del \mathscr{L}}{\del \left( \del_\mu \phi \right)} \delta\phi  -  X^\mu \right) = 0.
\end{equation}
This quantity is precisely the conserved current \(\mathcal{J}^\mu\) that corresponds to the symmetry \(\delta \phi\) of \(\phi \) as
\begin{align}\label{eq.currentconservation}
    \mathcal{J}^\mu &= \frac{\del \mathscr{L}}{\del \left( \del_\mu \phi \right)} \frac{\delta\phi}{\delta \alpha}- \frac{X^\mu}{\delta \alpha}, & \del_\mu \mathcal{J}^\mu & = 0.
\end{align}
With this in mind we can define a conserved charge \( \mathcal{Q}\) for the current \(\mathcal{J}^\mu\) by integrating the temporal unit \(\mathcal{J}^{t}\) in spatial hypersurfaces \(\mathbb{R}^{d-1}\) of constant time \(t\) as
\begin{equation}\label{eq.charge}
    \mathcal{Q} = \Int\limits_{\mathbb{R}^{d-1}} \dd^{d-1} x \ \mathcal{J}^{t}.
\end{equation}
The Noether charge \( \mathcal{Q}\) is time independent and using Gauss's law this can be proven as follows. By current conservation \(\pqty{\ref{eq.currentconservation}}\) we see that
\begin{equation}\label{eq.currentcompenents}
    \del_t \mathcal{J}^t = - \del_i \mathcal{J}^i,
\end{equation}
where \(i = 1, \dots, d-1\).
Differentiating the charge regarding time and applying \cref{eq.currentcompenents} gives us
\begin{equation}\label{eq.chargeconservationproof}
    \frac{\dd}{\dd t} \mathcal{Q} = \Int\limits_{\mathbb{R}^{d-1}} \dd ^{d-1} x \ \del_t \mathcal{J}^{t} = -\Int\limits_{\mathbb{R}^{d-1}} \dd ^{d-1} x \ \del_i \mathcal{J}^{i} = \Int\limits_{\del \mathbb{R}^{d-1}} \dd ^{d-2}x \ \eta_i \mathcal{J}^{i} = 0 \, ,
\end{equation}
where \( \eta_i\) is an outward-pointing normal vector. It is obvious written like this that the last term is a boundary term and by Stoke’s theorem it is assumed to vanish \footnote{
All reasonable functions in physics have compact support.}.

At this point, it would be fruitful and advantageous for future reference to specifically examine the Noether currents and charges related to the Poincaré symmetries and particularly translations and Lorentz transformations.
\begin{enumerate}
    \item \underline{\textbf{\emph{Translations}}}

    The first thing we want to investigate are translations given by \cref{eq.translations}. This kind of transformations acts on the field as
    \begin{equation}
        \phi(x) \overset{P}{\to} \phi'(x') = \phi(x - a) = \phi(x) - a^\mu \del_\mu \phi(x) + \order{a^2}.
    \end{equation}
    Given this transformation of the field, the Lagrangian is also affected 
    \begin{equation}
        \mathscr{L} \overset{P}{\to} \mathscr{L'} = \mathscr{L} - a^\nu \del_\mu \left( \delta^\mu_{\ \nu} \mathscr{L} \right) + \order{a^2}.
    \end{equation}
    Hence, using \cref{eq.deformationfield,eq.deformationLagrangian} we can read off the values of \(\delta \phi \) and \(X^\mu\) to be
    \begin{align}
        \frac{\delta \phi}{\delta a^\nu} &= -  \del_\nu \phi, & \frac{X^\mu}{\delta a^\nu} & = - \delta^\mu_{\ \nu}  \mathscr{L}.
    \end{align}
    Thus, by applying the Noether's theorem, and \cref{eq.currentconservation} we find the conserved current to be
    \begin{equation}\label{eq.ThetaEM}
         \Theta^{\mu \nu}  = -\frac{\del \mathscr{L}}{\del \left( \del_\mu \phi \right)} \del^\nu \phi + \eta^{\mu \nu} \mathscr{L} .
    \end{equation}
    Now, we can find the conserved charges. There are two kinds, one associated with time translations and the other is related to space translations. 
    \begin{enumerate}[label=\roman*.]
   \item \underline{\textbf{\emph{Time Translations}}}
    
  \begin{align}
      H & \equiv \Int\limits_{\mathbb{R}^{d-1}} \dd^{d-1}x \ \mathcal{H} =  \Int\limits_{\mathbb{R}^{d-1}} \dd^{d-1}x \ \Theta^{tt} \nonumber \\ 
      & = \Int\limits_{\mathbb{R}^{d-1}} \dd^{d-1}x\left( \Pi \del_t \phi - \mathscr{L} \right)
      = \Int\limits_{\mathbb{R}^{d-1}} \dd^{d-1}x\left( \frac{1}{2} \Pi^2 + \frac{1}{2} \left( \nabla \phi \right)^2 + \frac{1}{2} m^2 \phi^2 \right). \label{eq.Hamiltoniandensity}
  \end{align}
  \item \underline{\textbf{\emph{Space Translations}}}
     
  \begin{equation}\label{eq.physicalmomentum}
      P^i =  \Int\limits_{\mathbb{R}^{d-1}} \dd^{d-1}x \ \Theta^{ti} = - \Int\limits_{\mathbb{R}^{d-1}} \dd^{d-1}x \ \Pi \del^i \phi,
  \end{equation}
  \end{enumerate}
   where as usual \(H\) stands for the \textit{Hamiltonian}, \(\mathcal{H}\) is the \textit{Hamiltonian density}, \(P^i\) is the \textit{physical momentum} and finally \( \Pi \) is the conjugate momentum of the field \( \phi\) defined as
\begin{equation}\label{eq.conjugatemomentum}
       \Pi =  \frac{\del \mathscr{L}}{\del \left( \del_t \phi \right)} = - \eta^{\mu 0}\del_{\mu}\phi.
   \end{equation}
   \item \underline{\textbf{\emph{Lorentz Transformations}}}
    
   Next in order we want to examine Lorentz transformations derived in \cref{eq.Lorentztransformations}. The infinitesimal form of such a transformation is well known in the literature --- \cite[see][]{weinberg2005quantum1} --- and it is 
\begin{align}\label{eq.infinitesimalLorentz}
        \Lambda^{\mu}_{\ \nu}  & \equiv \delta^\mu_{\ \nu} + \omega^{\mu}_{\ \nu}, & \omega_{\mu \nu} & = - \omega_{\nu \mu}.
    \end{align}\
    Under \cref{eq.infinitesimalLorentz} the field \(\phi\) changes as 

    \begin{equation}
        \phi(x^\mu) \overset{\Lambda}{\to} \phi'(x'^\mu) = \phi\left( x^\mu - \omega^\mu_{\ \rho} x^\rho \right) = \phi(x) - \omega^\mu_{\ \rho} x^\rho \del_\mu \phi(x),
    \end{equation}
    where now \( a^\mu = \omega^\mu_{\ \rho} x^\rho\). Repeating the same procedure as in the translations, we find that the conserved current is
    \begin{align}\label{eq.lorentzcurrent}
      N^{\mu \nu \rho} & = x^\nu\Theta^{\mu \rho} - x^\rho \Theta^{\mu \nu}, &  \del_\mu N^{\mu \nu \rho} & =0.
    \end{align}
    The Noether charge related to the Lorentz transformations is
    \begin{equation}\label{eq.lorentzcharge}
        M^{\nu \rho} = \Int\limits_{\mathbb{R}^{d-1}} \dd^{d-1}x \ N^{t \nu \rho}.
    \end{equation}
    We see that \(\Theta_{\mu \nu}\) appears in the conserved current for both translations and Lorentz transformations. We can deduce that this is the familiar \textit{energy-momentum tensor} and in the above derivation it is not a priori symmetric. Nevertheless, it can be shown --- see \cite{Qualls:2015qjb} \S 2 for details of the computation and \cite{feynman2018feynman} for a broader discussion, and an alternative derivation more in the spirit of General Relativity --- that if the system features Poincaré invariance the energy-momentum tensor can be made symmetric and this is called the \textit{Belinfante energy-momentum tensor} defined as
    \begin{equation}\label{eq.BelinfanteEM}
        T^{\mu \nu} = \Theta^{\mu \nu} + \del_\rho B^{\rho \mu \nu},
    \end{equation}
    where \( B^{\rho \mu \nu}\) is the tensor that is antisymmetric in the exchange of the two first indices.
\end{enumerate}
Noether's theorem is also applicable to continuous symmetries that go beyond the usual spacetime ones. For example, if we consider the Lagrangian of \cref{eq.U(1)system}
\begin{equation*}
    \mathscr{L} (\varphi, \del_\mu \varphi) = - \del_\mu \varphi^*\del^\mu \varphi -  m^2 \varphi^* \varphi,
\end{equation*}
it possesses a \(U(1) \) symmetry that is a subset of the original \(O(2) \supseteq U(1)\) symmetry. The system is invariant under the transformation
\begin{align}\label{eq.U(1)symmetry}
    \varphi(x) & \to \varphi' (x) = e^{ia} \varphi (x), & \varphi^* (x) & \to {\varphi^*}'(x)  = e^{-ia} \varphi^* (x),
\end{align}
with \(a\) being an arbitrary, not space dependent parameter of the symmetry. The conserved current and charge are well known in the literature --- for example \cite[see][]{Beisert14} --- and \(\mathcal{J}^\mu\) can be found using \cref{eq.currentconservation} and 
\begin{align}
    \delta \varphi &= i \varphi \ \delta a , & \delta \varphi^* & = - i  \varphi^* \ \delta a, & X^\mu & =0.
\end{align}
to be
\begin{align}
    \mathcal{J}^\mu & =  \frac{\del \mathscr{L}}{\del \left( \del_\mu \varphi \right)} \frac{\delta\varphi}{\delta \alpha} +  \frac{\del \mathscr{L}}{\del \left( \del_\mu \varphi^* \right)} \frac{\delta\varphi^*}{\delta \alpha}  \nonumber \\
    & = -i \left( \varphi \del^\mu \varphi^* - \varphi^* \del^\mu \varphi \right).
\end{align}
Then employing \cref{eq.charge} the charge \(\mathcal{Q}\) is determined to be
\begin{align}
    \mathcal{Q} = \Int\limits_{\mathbb{R}^{d-1}} \dd^{d-1}x \ \mathcal{J}^t & =   - i \Int\limits_{\mathbb{R}^{d-1}} \dd^{d-1}x    \left( \varphi \del^t \varphi^* - \varphi^* \del^t \varphi \right)          \nonumber \\
    &=  i \Int\limits_{\mathbb{R}^{d-1}} \dd^{d-1}x \left( \varphi \dot{\varphi}^*  - \varphi^* \dot{\varphi}  \right).
\end{align}
The above transformation is known as a \textit{global} transformation, and the symmetry at hand is a perfect example of an \textit{internal} symmetry that we will utilize in the following chapters and are crucial in the context of the \acrlong{lce}.

\subsection{Quantisation}\label{subsec.Quantisation}

At this point, we want to quantise our scalar field \( \phi\) and to do so we will utilise two distinct but formally equivalent methods: the \textit{canonical} quantisation procedure and the \textit{path integral} formulation of \acrshort{qfts}.

The first approach to examine is the canonical quantisation. The word canonical originates from the Hamiltonian formulation of classical field theory. In general, the idea behind canonical quantisation is to promote the classical fields to quantum operators and at the same time try to maintain the underlying formal structure like the symmetries of the classical system to the maximum possible amount. It is known that in the Hamiltonian interpretation of classical field theory, the dynamics of a system are encoded in the Poisson brackets. The canonical quantisation replaces the Poisson brackets with commutation relations between the new quantum operators, a procedure which moderately conserves the overall structure of the theory.

The second quantisation methodology that we will examine is the Feynman's path integral approach, which utilises the action principle of the classical field theory by replacing the classical unique trajectory by a functional integral with integration measure \(\mathcal{D}\phi\) over all possible field configurations \(\phi\) to calculate a quantum transition amplitude.
The original idea of using path integrals to solve issues in diffusion and Brownian motion was proposed by Norbert Wiener --- see \cite{Chaichian2001PathII} for further references--- who introduced the \textit{Wiener integral}. The notion of path integrals was furthered expanded by Dirac~\cite{Dirac:1933xn} while the modern approach to the subject was developed by Feynman in his doctoral thesis~\cite{feynman2005} and completed in \cite{feynman2010quantum}. 

On the whole, there are many advantages to the use of the path integral approach. Foremost, \textit{Lorentz covariance} is clearer than in the operator language of the canonical formalism. Moreover, it is far less complicated to make a coordinate change in the path integral system than in the operators. Furthermore, path integrals exhibit a very close relation to stochastic processes and thermodynamics, a feat that is further examined in \cref{sec.thermalQFT}. Finally, a priori, it is more straightforward to find the right form of a system in the Lagrangian than in the Hamiltonian language. On the other hand, a disadvantage of the path integral approach is that unitarity is not immediately apparent. In any case, the equivalence between the formalisms has been proven, \emph{e.g.} see \cite{weinberg2005quantum1}, hence any underlying downsides of one procedure can be solved in the other.


\subsubsection{Canonical quantisation}\label{sec.canonicalquantisation}

We start our analysis by examining a field \(\phi\) satisfying \cref{eq.KleinGordon} which can be expanded in the usual mode decomposition of positive and negative frequencies like in \cref{eq.modedecomposition} \footnote{For an analysis for a general curved spacetime see \Cref{sec.curvedQFT}}. Our task is to turn the previously classical field and its conjugate momentum into operators  \footnote{To emphasize the difference between classical fields and quantum operators, in these sections only we will use the hat notation for the operators.} . Using \cref{eq.conjugatemomentum} we get
\begin{equation}
    \Pi(x) = \del_t \phi(x).
\end{equation}
At which point we promote the field \(\phi(x)\) and its conjugate field \(\Pi(x)\) to operators
\begin{align}
    \phi(x) & \to \Hat{\phi}(x), & \Pi(x) & \to \Hat{\Pi}(x).
\end{align}
Thereupon, following the classical notion of Poisson brackets
\begin{equation}
    \left\{ \phi(x), \Pi(y) \right\} = \delta^{d-1}(x-y),
\end{equation}
we impose canonical commutation relations such that
\begin{align}\label{eq.commutationfields}
         \left[ \hat\phi(x), \hat\Pi(y)  \right] &= i \delta^{d-1}(x-y), \nonumber \\
         \left[ \hat\phi(x),\hat\phi(y) \right] &= \left[ \hat\Pi(x) , \hat\Pi(y) \right] =0.
\end{align}
Accordingly, the same thing can be done for the previously classical modes \(a(\Vec{k})\) and \(a^*(\Vec{k})\) which now become quantum operators in momentum space
\begin{align}
    a(\Vec{k}) & \to \hat{a} (\Vec{k}), & a^*(\Vec{k}) & \to \Hat{a}^\dagger  (\Vec{k}).
\end{align}
Therefore, the field \( \hat\phi(x)\) is written as
\begin{equation}\label{eq.quantumfielddecomp}
   \hat\phi(x) = \frac{1}{\left( 2 \pi \right)^{d-1}} {\Int} \frac{\dd^{d-1} \vec{k}}{2 \omega_k} \left[ \hat a(\vec{k}) e^{ikx} + \hat a^\dagger (\vec{k}) e^{-ikx} \right] \Big|_{k^0 = \omega_k},
\end{equation}
whereas before \( \omega_{\Vec{k}} =  \sqrt{\vec{k}^2 + m^2} \) and again \(k x = -k^0 x^0 + k^ix^i =- \omega_k t + \Vec{k}x^i\). Consequently, from \cref{eq.quantumfielddecomp} we can express the Fourier coefficient \( \hat{a} (\Vec{k})\) as 
\begin{equation}\label{eq.Fouriermodes}
    \hat{a}(k) = {\Int} \dd^{d-1}x \ e^{-i k x} \left[ 
 \omega_k \hat{\phi}(x) + i \hat{\Pi}(x) \right].
\end{equation}
From \cref{eq.commutationfields,eq.Fouriermodes} we can see that the Fourier modes \( \hat{a} (\Vec{k})\) and \(\hat a^\dagger(\Vec{k})\) satisfy the subsequent commutation relations 
\begin{align}\label{eq.fouriercommutators}
    \comm {\hat{a} (\Vec{k})}{\hat a^\dagger(\Vec{k'})} & = 2 \omega_k (2 \pi)^{d-1} \delta^{d-1} ( \Vec{k} - \Vec{k'}), \\
    \comm{\hat{a} (\Vec{k})}{\hat{a} (\Vec{k'})} & =\comm{ \hat a^\dagger(\Vec{k})}{\hat a^\dagger(\Vec{k'})}  =0.
\end{align}
But now we observe that these are precisely the commutation relations of a \textit{quantum harmonic oscillator}. Thus, we can understand \( \hat{a} (\Vec{k})\) and \(\hat a^\dagger(\Vec{k})\) as ladder operators that create or annihilate particles of momentum \(\Vec{k}\).

It is also possible to define the \textit{vacuum state} of the theory \( \ket{0} \) to be the one that is eradicated by all the annihilation operators
\begin{equation}
    \hat{a} (\Vec{k}) \ket{0} = 0.
\end{equation}
We also assert that the vacuum state is properly normalised in the sense that
\begin{equation}\label{eq.vacuumnormalised}
    \bra{0}\ket{0} = 1.
\end{equation}
For this vacuum state we can construct a Fock space by repeated action of creation operators. With a single excitation, we can construct a single-particle state \( \Vec{\ket{k}} \) as
\begin{equation}
   \Vec{\ket{k}}  \coloneqq \hat a^\dagger(\Vec{k}) \ket{0}.
\end{equation}
In the same spirit, multiple harmonic oscillators can be excited at the same time to produce a multi-particle state as
\begin{equation}\label{eq.genericstate}
    \vert \Vec{k}_1, \Vec{k}_2,\dots, \Vec{k}_n \rangle  \coloneqq \hat a^\dagger(\Vec{k}_1) a^\dagger(\Vec{k}_2) \dots a^\dagger(\Vec{k}_n) \ket{0}.
\end{equation}
As we can see from \cref{eq.genericstate} for a \acrlong{qft}, a generic state is defined upon acting on the vacuum \( \ket{0}\) and is a linear combination of multiple identical particle states each of which has its Hilbert space \( \mathscr{H}\). This generic vector space is named after \cite{Fock1932KonfigurationsraumUZ}.

Casually, the Fock space is described as the combination of a series of Hilbert spaces \( \mathscr{H} \) made up of the vacuum state \( \mathscr{C} \), a single-particle state \( \mathscr{H}\), a two-particle state, et cetera. In a more mathematical language, the Fock space is defined as the Hilbert space realization of the direct sum of the symmetrical or anti-symmetrical tensor in a tensorial power of a single-particle Hilbert space \(\mathscr{H}\)
\begin{equation}
    F_\nu (\mathscr{H}) = \overline{\bigoplus\limits_{n=0}^{\infty} S_\nu \mathscr{H}^{\bigoplus n}},
\end{equation}
where \( S_{\nu }\) is characterised as the operator that either symmetrises or anti-symmetrises a tensor, contingent on the nature of the Hilbert space \(\mathscr{H}\) that either denotes particles obeying Bose-Einstein \((\nu = +)\) or Fermi-Dirac \((\nu = -)\) statistics.


\subsubsection{Path Integral quantisation}\label{sec.pathintegral}
We now turn our attention to the path integral quantisation. Beyond the advantages that were referenced in \cref{subsec.Quantisation}, this methodology is better suited when we study changes in the spectrum due to interactions, since we can utilise standard loop expansions methods. Thus, in this section, we can move beyond the free Lagrangian and add interaction terms.

As mentioned before in \cref{subsec.Quantisation}, the idea of the path integral is to sum over all possible field configurations \( \phi\) between an initial state \(\ket{\phi_i}\) at a time \(t_i\) and a final state \(\ket{\phi_f}\) at a time \(t_f\) denoted as
\begin{align}
    \hat{\phi}(t_i,x) \ket{\phi_i, t_i} &= \phi_i(x) \ket{\phi_i, t_i}, & \hat{\phi}(t_f,x) \ket{\phi_f, t_f} &= \phi_f(x) \ket{\phi_f, t_f},
\end{align}
with an integration measure \( \mathcal{D} \phi\) that formally takes the form
\begin{equation}
    \mathcal{D}\phi \propto \mathlarger{\prod}_{t_i 	\leq t 	\leq t_f} \mathlarger{\prod}_{x \in \mathbb{R}^{d-1}} \dd{\phi(t, x)}.
\end{equation}
Therefore the transition amplitude becomes
\begin{equation}
    \bra{\phi_f, t_f}\ket{\phi_i, t_i} = N {\int} \mathcal{D}\phi \  \exp\bqty{ i {\int}_{t_i}^{t_f} \dd{t} {\int}_{\mathbb{R}^{d-1}} \dd[d-1]{x}\mathscr{L}_{\textrm{free}} ( \phi, \del \phi) }, 
\end{equation}
where \( N\) is a normalisation parameter. Normally, when we examine a \acrshort{qft} we are interested in the vacuum to vacuum amplitude. To compute this, we send 
\begin{align}
    t_i & \to - \infty, & t_f & \to + \infty,
\end{align}
and at the same time we assume that
\begin{equation}
    \phi_i(x) = \phi_f(x) =0.
\end{equation}
Given the above suppositions, the vacuum-to-vacuum transition becomes
\begin{equation}
    \bra{0, + \infty}\ket{0, -\infty} \equiv \bra{0}\ket{0}  = N {\int} \mathcal{D}\phi \  \exp\left[ i {\int} \dd[d]{x} \mathscr{L}_{\textrm{free}} ( \phi, \del \phi) \right],
\end{equation}
where \(N\) is chosen in such a way that \cref{eq.vacuumnormalised} is satisfied.

Beyond the transition amplitude, we also want to compute \textit{correlation functions} 
\begin{align}\label{eq.pathintegraldefinition}
    \mel{0}{\mathcal{T} \hat{\phi}(x_1) \hat{\phi}(x_2) \dots \hat{\phi}(x_n)}{0} & \equiv \expval{\phi(x_1)\phi(x_2)\dots\phi(x_n)} \equiv G^{(n)} (x_1, \dots, x_n)  \\
    & = N {\int} \mathcal{D}{\phi} \ \phi(x_1)\dots\phi(x_n) \exp\left[ i {\int} \dd[d]{x} \mathscr{L}_{\textrm{free}} \left( \phi, \del{\phi} \right) \right], \nonumber
\end{align}
where in the first line \(\mathcal{T}\) expresses the \textit{time ordering prescription} so that operators that enter the path integral at a later time appear on the left of these operators that entered at an earlier stage \footnote{Mathematically this is denoted as
\begin{equation}
    \mathcal{T} \hat{\phi}(x) \hat{\phi}(y) \equiv \Theta\pqty{x^0-y^0} \hat{\phi}(x) \hat{\phi}(y) + \Theta\pqty{y^0-x^0}
    \hat{\phi}(y) \hat{\phi}(x),
\end{equation}
where \( \Theta\) is the Heaviside step function.}.

It is not always easy to compute correlation functions of the form of \cref{eq.pathintegraldefinition}. For our convenience it is easier to define the \textit{generating functional} \( \mathcal{Z}_0[J]\) which is formally expressed as
\begin{equation}\label{eq.generatingfunctional}
    \mathcal{Z}_0[J] \equiv \expval{\exp\left[ i {\int} \dd[d]{x} J(x) \phi(x)\right]}.
\end{equation}
where \(J{(x)}\) is the classical source \footnote{We denote \(\mathcal{Z}_0[J]\) the generating functional of the free theory contrary to the interacting theory where the generating function is without subscript.}. Now the correlation function can be computed regarding the generating functional \(\mathcal{Z}_0[J]\) as
\begin{equation}\label{eq.correlationfunctiongeneratingfunctional}
\expval{\phi(x_1)\phi(x_2)\dots\phi(x_n)} = \left( -i \right)^n \eval{\frac{\delta^n \mathcal{Z}_0[J]}{ 
 \delta{J(x_1)}\dots \delta{J(x_n)} 
 }}_{J=0},
\end{equation}
and the actual form of the generating functional that reproduces \cref{eq.pathintegraldefinition} via \cref{eq.correlationfunctiongeneratingfunctional} is 
\begin{equation}
     \mathcal{Z}_0[J] = N {\int} \mathcal{D}{\phi} \exp\left[ i {\int}\dd[d]{x} \left( \mathscr{L}_{\textrm{free}} (\phi, \del{\phi}) + J(x) \phi(x) \right)    \right].
\end{equation}
For the case of the free Lagrangian of \cref{eq.freelagragian} the generating functional takes the form
\begin{align}\label{eq.generatingfunctionalfreefield}
    \mathcal{Z}_0[J] & = N {\int}\mathcal{D}{\phi} \exp\left[ i {\int}\dd[d]{x} \left( -\frac{1}{2} \phi \left( - \Box + m^2\right)\phi +J\phi \right) \right] \nonumber \\ 
    & \simeq \lim_{\epsilon \to 0}{ N {\int}\mathcal{D}{\phi} \exp\left[ i {\int}\dd[d]{x} \left( -\frac{1}{2} \phi \left( - \Box + m^2 - i \epsilon \right)\phi +J\phi \right) \right]}.
\end{align}
In the first line we have integrated by parts and in the usual field theory approach we assumed that the boundary term is vanishing. In the second line we introduced the small \(\epsilon\) trick called the \textit{Feynman prescription}, \emph{e.g.} see~\cite{bjorken1965relativistic},
to avoid the poles that appear in the real line of the integrand and ensure the convergence properties of the path integral.

Looking closely at \cref{eq.generatingfunctionalfreefield} we observe that by integrating by parts, the integrand is Gaussian in the fields \(\phi\), thus we can perform the Gaussian integration to obtain
\begin{equation}\label{eq.generatingfunctionalfeynmanprop}
    \mathcal{Z}_0[J] = \exp\bqty{  \frac{i}{2} {\int} \dd[d]{x} \dd[d]{y} J(x) \Delta_{F}(x-y) J(y) },
\end{equation}
where the quantity \( \Delta_{F} \) is known as the \textit{Feynman propagator} and for the case of a real scalar field it reads
\begin{equation}\label{eq.Feynmanpropagator}
    \Delta_{F}(x-y) ={\int} \frac{\dd[d]{k}}{(2 \pi)^d} \frac{e^{i k (x-y)}}{k^2 + m^2 - i \epsilon}.
\end{equation}
It can be shown that \cref{eq.Feynmanpropagator} satisfies the equation
\begin{equation}
    \left( - \Box + m^2 \right) \Delta_{F}(x-y) = \delta^{d}(x-y),
\end{equation}
and thus we can deduce that the Feynman propagator of \cref{eq.Feynmanpropagator} is nothing less than the Green's function of Klein-Gordon \cref{eq.KleinGordon}.

Finally, using \cref{eq.pathintegraldefinition,eq.correlationfunctiongeneratingfunctional,eq.generatingfunctionalfeynmanprop} we find that
\begin{equation}\label{eq:Greenfunction2}
    G^{(2)} (x_1,x_2) \equiv \expval{\phi{(x_1)} \phi{(x_2)}} = -i \Delta_{F}(x_1-x_2).
\end{equation}
Things get more complicated for a \acrlong{qft} when interactions are included. For the purpose of this section, the interaction terms that we will consider are in the same spirit as these of \cref{eq.interactionterms}. Given such a term, the generating functional now takes the more general form
\begin{equation}\label{eq.generatingfunctionalgeneral}
    \mathcal{Z}\bqty{J} = N {\int} \mathcal{D}{\phi} \exp\bqty{i {\int} \dd[d]{x} \pqty{\mathscr{L}_{\textrm{free}} + \mathscr{L}_{\textrm{int}} + J\phi}},
\end{equation}
which is not Gaussian any more, so the integral cannot be exactly computed. Nevertheless, there is a way out of this predicament. If the coupling constant \(g_n\) of \cref{eq.interactionterms} is small so that the theory is weakly coupled, then we are in an ideal place to take advantage of the strongest tool in theoretical physics, which is \textit{perturbation theory} --- for a review see~\cite{holmes1998introduction,bender1999advanced}. Then it is possible to manipulate \cref{eq.generatingfunctionalgeneral} in the following manner
\begin{align}
    \mathcal{Z}\bqty{J} & = N \exp \bqty{i {\int} \dd[d]{x} \mathscr{L}_{\textrm{int}} \ \pqty{ \frac{1}{i} \fdv{J(x)}}} {\int} \mathcal{D}{\phi} \exp \bqty{i {\int} \dd[d]{x} \pqty{\mathscr{L}_{\textrm{free}} + J \phi}} \nonumber \\
    & = \exp \bqty{ i {\int} \dd[d]{x} \mathscr{L}_{\textrm{int}} \ \pqty{ \frac{1}{i} \fdv{J(x)}}} \ \mathcal{Z}_0\bqty{J}.
\end{align}
It is now clear that knowing the generating functional of the free theory along with the interacting Lagrangian is enough to work out the more general generating functional \(\mathcal{Z}\bqty{J}\).

Actually, this approach has laid the groundwork to use \textit{Feynman diagrams}. In this graphical approach, propagators are depicted as lines going from one point to the other, while interactions terms are portrayed as vertices. For example, in this notation, the generating functional \( \mathcal{Z}_0 \bqty{J}\) encodes the propagator of the free field theory. 

At this point, we will formulate the \textit{Feynman rules} for the \(\phi^4\) theory, which is the simplest possible case and more complicated cases will be examined in the subsequent chapters. Using \cref{eq.interactionterms}, the interaction Lagrangian for the \(\phi^4\) is 
\begin{equation}
    \mathscr{L}_{\textrm{int}} = - \frac{g}{4!}\phi^4,
\end{equation}
and the Lagrangian of the system becomes
\begin{equation}\label{eq.phi^4langragian}
    \mathscr{L} = -\frac{1}{2} \del^\mu \phi \del_\mu \phi - \frac{1}{2} m^2 \phi^2 - \frac{g}{4!} \phi^4.
\end{equation}
Therefore, given the above Lagrangian the propagator \( -i \Delta_F (x_i-y_i) \) is pictured as
\begin{center}
\begin{fmffile}{Propagator}
 \begin{fmfgraph*}(60,10)\label{xspacepropagator}
  \fmfstraight
   \fmfleft{i1}
   \fmfright{o1}
   \fmf{plain}{i1,o1}
   \fmflabel{\(x_i\)}{i1}
   \fmflabel{\(y_i\)}{o1}
 \end{fmfgraph*}
\end{fmffile}
\end{center}
Additionally, the interaction term is delineated as a vertex with several legs equal to the power of the field in \cref{eq.interactionterms} and the whole term is weighted by a factor of \(ig\). For \(\phi^4\) this has four legs as
\begin{center}
    \begin{fmffile}{vertex}
 \begin{fmfgraph*}(80,40)
   \fmfleft{i1,i2}
   \fmfright{o1,o2}
   \fmf{plain}{i1,v1}
   \fmf{plain}{i2,v1}
   \fmfv{lab=\(y_i\),lab.dist=-.2w}{v1}
   \fmf{plain}{o1,v1}
   \fmf{plain}{o2,v1}
 \end{fmfgraph*}
\end{fmffile}
\end{center}
After connecting all the legs appropriately, the last step is to integrate over all the \(y_i\) coordinates of every vertex in the position space, and also contemplate and evaluate the symmetry factors.

This can be generalized for any power of the field \( \phi\) in the interaction term. In the same spirit as above, the Feynman rules for calculating \( G^\pqty{n}\pqty{x_1,\dots,x_n}\) to order \(n\) in \(g\) for \( \mathscr{L}_{\textrm{int}} = -\frac{g}{p!} \phi^p\) in \(x\)-space are
\begin{enumerate}[label=\textbf{R.\arabic*},ref=R.\arabic*]
    \item\label{R1}  Draw the basic elements :
    \begin{itemize}
        \item \underline{\textbf{\emph{The external points \(x_1, \dots,x_n\)}}}.
\begin{equation*}
    \underset{x_1}{\bullet},\ \underset{x_2}{\bullet}, \dots\dots\dots\dots\dots\dots\dots\dots, \underset{x_n}{\bullet}
\end{equation*}
        \item \underline{\textbf{\emph{\(n\) vertices of order \(p\)}}}.
\begin{equation*}
\vcenter{\hbox{\includegraphics[scale=0.3]{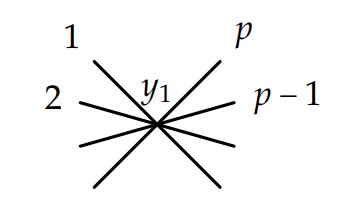}}} \underbrace{\dots\dots\dots}_{n-2} \vcenter{\hbox{\includegraphics[scale=0.3]{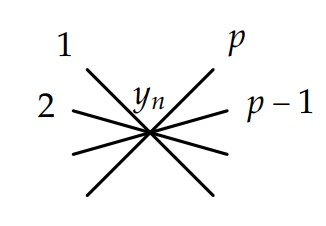}}}
\end{equation*}
\end{itemize}
\item\label{R2} Connect the basic elements in all possible ways, remembering that legs are different for symmetry reasons. 
\item\label{R3} Each different drawing corresponds to a particular diagram \(\mathcal{I}_D\) weighted by its symmetry factor, and each line corresponds to \( -i \Delta_F  \). Evaluate every distinct diagram \(\mathcal{I}_D\)
\begin{equation}
    \mathcal{I}_D = \pqty{\frac{ig}{p!}}^n {\int} \dd[d]{y_1} \ \dots {\int} \dd[d]{y_n} \ \mathlarger{\prod} \pqty{-i} \Delta_F (a - b),
\end{equation}
where \( a, b \in \bqty{x_1,\dots,x_n, y_1, \dots, y_n} \).
\item\label{R4} Compute the correlation function  \( G^\pqty{n}\pqty{x_1,\dots,x_n}\) as
\begin{equation*}
    G^\pqty{n}\pqty{x_1,\dots,x_n} = \frac{1}{n!}\mathlarger{\sum}\limits_D  \mathcal{I}_D.
\end{equation*}
\end{enumerate}
Ordinarily, it is more useful to compute the Feynman diagrams in momentum instead of coordinate space. This comes with the clear advantage that due to translation invariance, the overall momentum in a vertex shall be zero. Thus, the goal is to transform the Green functions from the real space to momentum space as
\begin{align}
    G^\pqty{n} \pqty{p_1,\dots,p_n}& = {\int} \dd[d]{x_1} \ \dots {\int} \dd[d]{x_n} G^\pqty{n} \pqty{x_1,\dots,x_n} e^{-i \pqty{p_1 x_1 + \dots + p_n x_n}},\\
    G^\pqty{n} \pqty{x_1,\dots,x_n}& = {\int} \frac{\dd[d]{p_1}}{\pqty{2 \pi}^d}\  \dots {\int} \frac{\dd[d]{p_n}}{\pqty{2 \pi}^d} G^\pqty{n} \pqty{p_1,\dots,p_n} e^{i \pqty{p_1 x_1 + \dots + p_n x_n}},
\end{align}
and we assume that all the momenta are incoming. Then the Feynman rules in \(p\)-space are as following. Rules \ref{R1}, \ref{R2} and \ref{R4} remain unchanged, while \ref{R3} changes in the following manner. Now every external line is characterised by its momentum \(p\) and is depicted as
\begin{center}
    \begin{fmffile}{Propagator_momentumexternal}
 \begin{fmfgraph*}(60,60)
  \fmfstraight
   \fmfleft{i1}
   \fmfright{o1}
   \fmf{fermion,label=\(p\)}{i1,o1}
 \end{fmfgraph*}
\end{fmffile}
\end{center}
and also comes with a component \( {\pqty{p^2 + m^2}^{-1}} \). Moreover, every internal line of momentum \(k\) is drawn as
\begin{center}
    \begin{fmffile}{Propagator_momentuminternal}
 \begin{fmfgraph*}(60,60)
  \fmfstraight
   \fmfleft{i1}
   \fmfright{o1}
   \fmf{fermion,label=\(k\)}{i1,o1}
 \end{fmfgraph*}
\end{fmffile}
\end{center}
and comes with a component \( { \pqty{k^2 + m^2-i \epsilon}^{-1}} \).\newline
Now for every vertex we shall add an element \( i g \pqty{2 \pi}^d \delta^d \pqty{\Sum_i p_i} \) while the presence of the delta function secures that the momentum is conserved in every vertex

\begin{center}
    \begin{fmffile}{vertex_momentum}
 \begin{fmfgraph*}(80,40)
   \fmfleft{i1,i2}
   \fmfright{o1,o2}
   \fmf{fermion}{i1,v1}
   \fmf{fermion}{i2,v1}
   \fmf{fermion}{o1,v1}
   \fmf{fermion}{o2,v1}
   \fmflabel{\(p_2\)}{i1}
   \fmflabel{\(p_1\)}{i2}
   \fmflabel{\(p_4\)}{o1}
   \fmflabel{\(p_3\)}{o2}
 \end{fmfgraph*}
\end{fmffile}
\end{center}
Finally the integration is performed over all remaining independent momenta with a integration parameter \( {\Int} \frac{\dd[d]{k}}{\pqty{2 \pi}^d}\) and all symmetry factors shall be accounted for. 

Besides the usual generating functional \(\mathcal{Z}\bqty{J}\), it is possible and useful for us to define two additional ones, \( W\bqty{J}\) and \( \Gamma\bqty{\varphi}\). A priori these new generating functional contain the same amount of physical data, but it is easier to compute them. The first one, \( W\bqty{J}\), is the generating functional of \textit{connected diagrams} and is defined as
\begin{equation}\label{eq.connectedgeneratingfunctional}
     \mathcal{Z}\bqty{J} \equiv e^{i  W\bqty{J}}.
\end{equation}
These are the Feynman diagrams that do not have any disconnected lines. A connected \(n\)-point function is expressed as
\begin{equation}\label{eq.connectednpointfunction}
    \expval{\phi(x_1)\phi(x_2)\dots\phi(x_n)}_c = \pqty{ -i }^{n-1} \eval{\frac{\delta^n W\bqty{J}}{ 
    \delta{J(x_1)} \dots \delta{J(x_n)} 
    }}_{J=0}.
\end{equation}
There is a noteworthy subgroup for the connected diagrams. These are the \acrfull{1pi} diagrams which are defined as the ones that cannot become two non-trivial separate diagrams by cutting a singular inner line. Every diagram that is not \acrshort{1pi} possess such a line and is called \textit{reducible}. As a matter of fact, the generator of the \acrshort{1pi} correlation functions is the \textit{quantum effective action} \( \Gamma\bqty{\varphi}\), first defined in \cite{PhysRev.127.965} using perturbative methods and then in \cite{university1964relativity} and separately in \cite{Jona-Lasinio:1964zvf} using non-perturbative methodology. Since the effective action is an important physical tool, it calls for a more thorough examination.

From the \acrshort{qft} scope \( \Gamma\bqty{\varphi}\) stands as the adjusted definition of the classical action \( S\bqty{\phi}\) that also considers quantum corrections and at the same time ensures that the stationary-action principle holds true. In other words, extremising \( \Gamma\bqty{\varphi}\) produces the \acrshort{eom} but instead of the classical fields, now it considers the vacuum expectation value of the aforementioned quantum fields. \( \Gamma\bqty{\varphi}\) is determined by making use of the Legendre transformation of \(W\bqty{J}\) as
\begin{equation}\label{eq.effectiveaction}
    \Gamma\bqty{\phi} \equiv W\bqty{J} - {\int} \dd[d]{x} J\pqty{x} \varphi\pqty{x},
\end{equation}
where \(J\pqty{x}\) is the non-zero source that ensures that the classical scalar field has the expectation value \(\varphi\pqty{x}\) expressed as the solution of
\begin{equation}
    \varphi\pqty{x} = \expval{\hat\phi\pqty{x}}_J = \fdv{W \bqty{J}}{J\pqty{x}}.
\end{equation}
In other words, being an expectation value the classical field \(\phi\pqty{x}\) can be seen as the weighted mean over quantum fluctuations in the presence of the sources \(J\pqty{x}\). Applying the functional derivative with respect to \(\varphi\pqty{x}\) to \(\Gamma\bqty{\varphi}\) and using \cref{eq.effectiveaction} generates the quantum \acrshort{eom}
\begin{equation}\label{eq.quantumsource}
    J\pqty{x} = \fdv{\Gamma\bqty{\varphi}}{\varphi\pqty{x}}.
\end{equation}
Hence, lacking a source, \emph{i.e.} \(J\pqty{x}=0\), \cref{eq.quantumsource} indicates that the vacuum expectation value of the fields \(\phi\) actually extremise the quantum effective action \(\Gamma\bqty{\varphi}\) instead of the classical action \( S\bqty{\phi}\). This is precisely the principle of least action, but in the full quantum theory. This alteration from the classical theory originates in the path integral approach, where every possible field configuration is accounted for in the path integral, in contrast to the classical field theory where solely the classical configurations play a part.

As mentioned before, \(\Gamma\bqty{\varphi}\) is also the generating functional for the \acrshort{1pi} correlation functions. Consequently, the \acrlong{1pi} \(n\)-point functions are 
\begin{align}\label{eq.n-point1pi}
    \expval{\hat{\phi}\pqty{x_1} \dots \hat{\phi}\pqty{x_n}}_{\acrshort{1pi}} & = \eval{\Gamma^{\pqty{n}} \pqty{x_1, \dots, x_n}}_{J=0}, \nonumber \\
    & = \eval{\frac{\delta^n \Gamma\bqty{\varphi}}{\delta \varphi\pqty{x_1} \dots \delta \varphi\pqty{x_n}} }_{J=0}.
\end{align}
In \cref{eq.n-point1pi} \(\Gamma\bqty{\varphi}\) is the sum of all \acrlong{1pi} Feynman graphs. The relationship between \(W\bqty{J}\) and \(\Gamma\bqty{\varphi}\) also indicates that there should be some beneficial correlations between their Green's functions. As a matter of fact, by applying the chain rule in the above equations we can show that
\begin{equation}
    \fdv{J\pqty{x}} = {\int} \dd[d]{y} \frac{\delta^2 \ W\bqty{J}}{\delta J\pqty{x} \delta J\pqty{y}} \fdv{\varphi \pqty{y}}, 
\end{equation}
and, moreover, this leads to
\begin{equation}\label{eq.inverseexactprop}
    \Gamma^{\pqty{2}}\pqty{x,y} = \pqty{\frac{\delta^2 \ W\bqty{J}}{\delta J\pqty{x} \delta J\pqty{y}}}^{-1}.
\end{equation}
Thus \( \eval{\Gamma^{\pqty{2}}}_{J=0}\) is nothing less than the inverse of the exact propagator. 

Finally, it is useful to have an expression for \(\Gamma^{\pqty{n}} \pqty{x_1, \dots, x_n}\) in momentum space. To achieve that we should Fourier transform \( \Gamma^{\pqty{n}} \pqty{x_1, \dots, x_n}\) to obtain the \textit{vertex functions} \( \Gamma^\pqty{n} \pqty{p_1, \dots, p_n}\) as
\begin{equation}\label{eq.vertexfunctions}
    \pqty{2 \pi}^d \delta^d \pqty{\Sum_{i=1}^n p_i} \Gamma^\pqty{n} \pqty{p_1, \dots, p_n} = \mathlarger{\mathlarger{\prod}}_{k=1}^n {\int} \dd[d]{x_k} e^{-i x_k p_k} \Gamma^{\pqty{n}} \pqty{x_1, \dots, x_n},
\end{equation}
where all the momenta are incoming. With that in mind \cref{eq.effectiveaction} may be expressed as
\begin{align}
    \Gamma\bqty{\varphi} & = \frac{1}{2} {\Int} \frac{\dd[d]{p}}{\pqty{2 \pi}^d} \varphi\pqty{-p} \pqty{p^2 + m^2 - \Pi\pqty{p^2}} \varphi\pqty{p}  \\
    & +  \Sum_{n=3}^{\infty} \ \frac{1}{n!} {\Int} \frac{\dd[d]{p_1}}{\pqty{2 \pi}^d} \ \dots {\Int} \frac{\dd[d]{p_n}}{\pqty{2 \pi}^d}  \pqty{2 \pi}^d \delta^d \pqty{\Sum_{i=1}^n p_i} \Gamma^\pqty{n} \pqty{p_1, \dots, p_n} \varphi\pqty{p_1}  \dots \varphi\pqty{p_n}, \nonumber
\end{align}
whereas in a matter of fact by \cref{eq.inverseexactprop} the exact propagator is nothing but
\begin{equation}\label{eq.exactpropagator}
    G^\pqty{2}\pqty{p,-p} = \pqty{\Gamma^\pqty{2}\pqty{p,-p}}^{-1} = \frac{1}{\pqty{p^2 + m^2 - \Pi \pqty{p^2}}},
\end{equation}
and furthermore \( \Pi \pqty{p^2} \) is the self-energy of the system which consists of all the \acrshort{1pi} corrections to the two point function.


\subsection{Wick rotation and thermal QFT}\label{sec.thermalQFT}

At this point and having introduced canonical and path integral quantisation, what we want to do is to introduce a mathematical technique to simplify the computation of path integrals and also lay down the groundwork for thermal \acrshort{qft} \footnote{For thermal operators and the \textit{KMS} condition see \Cref{sec.KMS}}. Thus, this section is mainly inspired by Hartman~\cite{Hartman} and follows closely \cite{Kalogerakis2019}. 

A priori path integrals in Lorentzian signature, similar to the ones in \cref{eq.pathintegraldefinition} and after, are hard to compute, and this is because the exponential factor \( e^{i S}\nsucc 0\) and thus it describes fast oscillation. The remedy out of this is to perform a trick introduced by Wick~\cite{PhysRev.96.1124}, which is by then known as \textit{Wick rotation}. What we actually do is that we analytically continue in the temporal direction from real Lorentzian to imaginary Euclidean time as
\begin{equation}\label{eq.Wickrotation}
    t \to - i \tau.
\end{equation}
Doing so, the generating functional of \cref{eq.generatingfunctionalgeneral} becomes
\begin{equation}\label{eq.Euclideangeneratingfunctional}
    \mathcal{Z}\bqty{J} = {\int} \mathcal{D} \phi \, e^{-S_E + {\int} \dd[d]{x} J(x) \phi (x)},
\end{equation}
where \footnote{Obviously, \cref{eq.EuclideanLangragian} is only valid for the case of the scalar field \(\phi(x)\)  that we are examining thus far.}
\begin{align}
   S_E &=  {\int} \dd[d-1]{x}  \dd{\tau} \ \mathscr{L}_E,\\
   \mathscr{L}_E & = \frac{1}{2} \del^\mu \phi \del_\mu \phi + \frac{1}{2} m^2 \phi^2 - \mathscr{L}_{\textrm{int}}[\phi].
    \, \label{eq.EuclideanLangragian} 
\end{align}
From the above definition of the generating functional and the path integral, it is understandable that the convergence properties in the Euclidean formulation \footnote{Going from Lorentzian to Euclidean signature the metric \( \eta_{\mu \nu} \) for  raising and lowering indices is replaced by the Kronecker delta \( \delta_{\mu\nu}\). Furthermore, Einstein summation in the indices is always presumed so that \( \del^\mu \phi \del_\mu \phi = \left( \frac{\del \phi}{\del \tau} \right)^2 + \left( \nabla \phi \right)^2. \) }  are more apparent since the exponential \( e^{-S_E} \succ 0 \) hence it is heavily damped. A more rigorous proof of this connection has been given in \cite{Osterwalder:1973dx,Osterwalder:1974tc} \footnote{The Osterwalder–Schrader theorem states that under a number of axioms, it is possible to analytically continue \textit{Euclidean Schwinger functions} to \textit{Lorentzian Wightman distributions} that fulfil Wightman axioms, and therefore they properly define a \acrshort{qft}.}.

But there is more to the story. There is a connection between Euclidean path integrals and thermodynamics. By construction, path integrals characterise transition amplitudes and for instance, for Euclidean time \( \tau \) and given two fields \(\phi_1\) and \(\phi_2\), the transition under the evolution \(e^{-\beta H}\) can be described by the subsequent path integral
\begin{equation}\label{5.2.1}
   \mel{\phi_2}{e^{-\beta H}}{\phi_1} = \Int\limits_{\phi(\tau=0) = \phi_1}^{\phi(\tau=\beta) = \phi_2} \mathcal{D} \phi \ e^{-S_E [\phi]},
\end{equation}
where \(\phi_1\) and \(\phi_2\) are the boundaries of integration and the path integral assumes a 
foliation in spatial and temporal spacetime dimensions. 

The exact form of the path integral is obviously dependent on the topology of spacetime, so for instance if the manifold is \(\setR \times \setS^{d-1}\) it can be depicted as
\begin{equation}\label{eq.transitionamplitude}
\mel{\phi_2}{e^{-\beta H}}{\phi_1} \rangle=
\overset{\phi_2}{\underset{\phi_1}{\vcenter{\hbox{\includegraphics[width=2.5cm,height=3.cm]{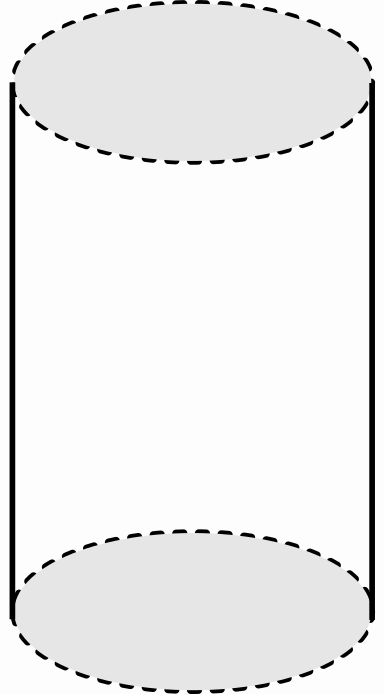}  }}}}\overset{\Bigg\uparrow}{\underset{\Bigg\downarrow}{\beta}}
\end{equation}
where the integration is performed over all the field configurations \( \phi\). 

Based on the above logic, we can define a \textit{cut} to be a \textit{Cauchy slice} for a fixed value of \( \tau\) that is a \textit{codimension}-\(1\) surface \(\Sigma\). Therefore, the transition amplitude of \cref{eq.transitionamplitude} admits two cuts, one for \(\tau=0\) and the other for \(\tau=\beta\), where \( \beta \) is just a time value for the moment. Thus, it is also possible to define a path integral with a single fixed boundary along with an open cut. This configuration resembles a quantum state 
\begin{equation}
    \ket{\psi}
 = e^{-\beta H} \ket{\phi_1},
\end{equation}
and looks like
\begin{equation}
\ket{\psi} =
\underset{\phi_1}{\vcenter{\hbox{\includegraphics[width=2.5cm,height=3.0cm]{Figures/cylinder.png}  }}} \overset{\Bigg\uparrow}{\underset{\Bigg\downarrow}{\beta}}
\end{equation}
Consequently, we can deduce that a general quantum state on $\Sigma$ is the path integral that admits any open cut $\Sigma$. Path integrals with open cuts can be sewed together to construct a transition amplitude, so for example, for a theory defined on a line, the vacuum to vacuum amplitude is
\begin{align}
\bra{0}\ket{0} & = \Sum_{\phi_1} \bra{0} \ket{\phi_1}\bra{\phi_1}\ket{0}\nonumber \\
& = \Int\mathcal{D} \phi e^{- S_E[\phi] } =
\overset{\infty}{\underset{-\infty}{\vcenter{\hbox{\includegraphics[width=2.5cm,height=2.5cm]{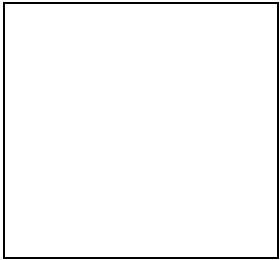}}}}}
\end{align}
where two path integrals have been glued together. We can see that the first one is \( \ket{0}  \in (-\infty \ < \tau \leq 0] \), while \( \bra{0} \in [0\leq \ \tau < \ \infty) \). The summation runs over all the intermediate states at \( \tau = 0\) and the path integral should be smooth and connected.

This methodology can be easily extended to include a local operator \(\mathscr{O}\pqty{x}\), by inserting them in the path integral, as
\begin{equation}
   \mel{\phi_i}{\mathscr{O}_1(x_1) \mathscr{O}_2 (x_2)}{\phi_j}  = \int\limits_{C(\phi)} \mathcal{D}\phi \ \mathscr{O}_1(x_1) \mathscr{O}_2 (x_2) \ e^{-S_E[\phi]} ,
\end{equation}
where $C(\phi)$ are the fields that have to be integrated over.

Hence, we can extend this discussion to also include density matrices, which are just operators where both ends admit open cuts. For example
\begin{equation}\label{5.2.4}
    \rho = e^{-\beta H},
\end{equation}
is the density matrix in thermal equilibrium with a temperature \(T = {1} / {\beta}\) and corresponds to the transition 
\begin{equation}
\rho =
\vcenter{\hbox{\includegraphics[width=2.5cm,height=3.0cm]{Figures/cylinder.png}}} \overset{\Bigg\uparrow}{\underset{\Bigg\downarrow}{\beta}}
\end{equation}
Thus, \( \rho \) can be thought of as a time evolution operator in the Heisenberg picture. This definition of the density matrix can be very insightful in the sense that the thermal partition function, $\mathcal{Z}\pqty{\beta}$ is defined as
\begin{align}\label{eq.partitionfunctionquantum}
     \mathcal{Z}\pqty{\beta} & \equiv \Tr\bqty{e^{-\beta H}} \\ \nonumber
     & = \Sum\limits_{\phi_i} \mel{\phi_i}{e^{-\beta H}}{\phi_i} = \Sum_{\phi_i} \overset{\phi_i}{\underset{\phi_i}{\vcenter{\hbox{\includegraphics[width=2.5cm,height=3cm]{Figures/cylinder.png}  }}}} \overset{\bigg\uparrow}{\underset{\bigg\downarrow}{\beta}} 
\end{align}
where due to the definition of the trace the summation is over the same initial and final field configuration. The path integral is constrained to have field configurations with periodic boundary conditions on the temporal coordinate
\begin{align}\label{eq.partitionfunctionintegral}
     \mathcal{Z}\pqty{\beta} &=\Int\limits_{\phi(\tau,x)}^{\phi(\tau + \beta, x)} \mathcal{D} \phi \ e^{-S_E [\phi]} \nonumber \\
     & =\underset{ {} \qquad \beta \ \myarrow{}}{\vcenter{\hbox{\includegraphics{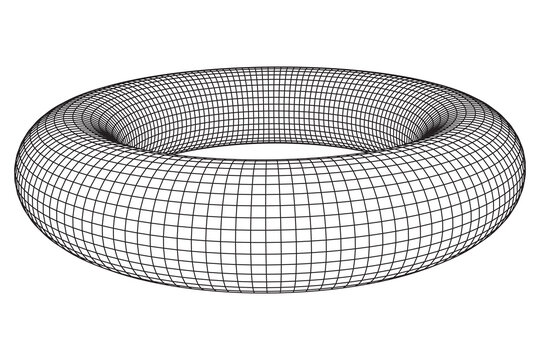}  }}}
\end{align}
What happened is that the path integral of a \acrshort{qft} in Euclidean metric that was defined in a \(d\)-dimensional space with cylindrical topology of circumference \(\beta\) can be interpreted as a thermal average living in \(d-1\) spatial dimensions of a quantum statistical system. \footnote{This result can be generalized. We can formally state, that if the initial spatial coordinates that the \acrshort{qft} was formulated were \(\mathbb{R}^{d-1}\), then \(\mathcal{Z}\pqty{\beta}\) would be the path integral evaluated on \(\mathbb{R}^{d-1} \times \mathbb{S}^{1}\).} Actually, the generating functional (\ref{eq.Euclideangeneratingfunctional}) of the original Euclidean \(d\)-dimensional \acrshort{qft} coincides with the partition function (\ref{eq.partitionfunctionquantum}) of the quantised statistical system
\begin{equation}\label{eq.partitionfunctionandgenerating}
    \underbrace{\mathcal{Z}[J = 0]}_{\mathbb{R} \times \mathbb{S}^{d-1}} = \underbrace{\mathcal{Z}\pqty{\beta}}_{\mathbb{S}^1 \times \mathbb{S}^{d-1}}.
\end{equation}
Consequently, using standard thermodynamical arguments we can use the partition function of \cref{eq.partitionfunctionintegral} to compute 
\begin{align}
  &\textrm{The free energy} & F &= - \frac{1}{\beta } \log{\mathcal{Z}}, \label{eq.5.2.6} \\
& \textrm{The energy} & \langle E \rangle &= \frac{\del(\beta F)}{\del \beta}, \\
& \textrm{The entropy} & S &= \pqty{\beta \pdv{\beta} - 1} \pqty{\beta F}.
\end{align}
As a matter of fact
 \cref{eq.connectedgeneratingfunctional} also has an interesting interpretation in Euclidean signature as it takes the form
\begin{equation}
    e^{- W \bqty{J}} = \mathcal{Z}\bqty{J}.
\end{equation}
We can see that this is nothing short of the free energy of the statistical mechanical system. Similarly, \( \Gamma\bqty{\varphi}\) as a Legendre transform of \(W \bqty{J} \) is associated with the \textit{Gibbs free energy}.


\subsection{Regularisation, renormalisation and fixed points}\label{sec.RGflow}

Generally, local \acrshort{qfts} are intrinsically encumbered by divergences that impinge on the perturbation theory. The most prominent of these divergences are infinities that occur at very short distances and/or very high momenta, a region that is known as the\acrshort{uv} limit in contrast to the region of long distances and/or low momenta which is known as the \acrshort{ir} limit.

The independent mathematical techniques that are employed to resolve and regulate these divergences are called \textit{regularisation} and \textit{renormalisation}. 
\begin{itemize}
    \item \underline{\textbf{\emph{Regularisation}}}
    
    This method consists of removing the divergences by an appropriate regularisation scheme. Since a priori no physical quantity depends on the specific choice of scheme, all that we have to decide is which procedure is more appropriate for the problem at hand. Specifically, regularisations that preserve the symmetries of a system are often preferred to non-preserving ones, but nevertheless any choice properly executed leads to the same physical predictions. In most of the introductory books to \acrshort{qft}, \emph{e.g.} see~\cite{Peskin:1995ev,Srednicki:1019751}, the techniques that are employed are either \textit{cut-off regularization} or \textit{dimensional regularisation} developed by ’t Hooft and Veltman \cite{tHooft:1972tcz}. In cut-off regularisation a --- most of the time smooth --- cut-off regulator \(\Lambda\) is introduced in momentum space to regularise a divergent integral and then the limit \(\Lambda \to \infty\) is taken at the end of the computation. The problem with this technique is that it is not Lorentz invariant. On the other hand, in dimensional regularisation any systematic divergences are turning up in the form of poles in the physical value of the spacetime dimension \(d\). For example, in four spacetime dimensions writing \(d = 4- \varepsilon\), divergences appear when the regulator \(\varepsilon \to 0\). These are usually cancelled by applying renormalisation techniques to gain meaningful physical results. In the subsequent chapters, the scheme most used is the \textit{zeta function regularisation}, for which details will be given later, but for a more thorough definition see~\cite{elizalde1994zeta,Dunne_2008} and references therein.
    \item \underline{\textbf{\emph{Renormalisation}}}
    
    In the renormalisation scheme, all infinities lie in the \textit{bare parameters} while the new \textit{renormalised parameters} are finite but depend on the energy scale. This inevitably leads to the famous \textit{running of the coupling} since renormalised observable parameters run when the energy scale is changed. How they change is the subject of the \textit{renormalisation group equation} which we will study in the upcoming sections. 
\end{itemize}

\subsubsection{Regularisation and the renormalisation group}

We will review regularisation and renormalisation in the context of  \(\phi^4\) theory to one loop in perturbation theory in \(d=4\) spacetime dimensions. The Lagrangian of the system is defined in \cref{eq.phi^4langragian} and by making use of Feynman graphs we want to compute the \acrlong{1pi} vertex functions  \( \Gamma^\pqty{2}_{1-\textrm{loop}}\) and \( \Gamma^\pqty{4}_{1-\textrm{loop}}\) defined in \cref{eq.vertexfunctions}.

Hence, let us examine \( \Gamma^\pqty{2}_{1-\textrm{loop}} \pqty{p,-p}\) in detail. Following the Feynman rules of \cref{sec.pathintegral} and ignoring factors of \( {\pqty{p^2 + m^2}^{-1}} \) that arise from external legs, the \( \Gamma^{\pqty{2}}_{1-\textrm{loop}} \pqty{p,-p}\) diagram is 
\begin{center}
\begin{fmffile}{feyngraph}
 \begin{align*}
 \Gamma^{\pqty{2}}_{1-\textrm{loop}} \pqty{p,-p}
   & =  \quad\parbox{100pt}
  {
    \begin{fmfgraph*}(100,80)
       \fmfleft{i}
       \fmfright{o}
       \fmftop{m}
       \fmfv{label=p,l.a=60}{i}
       \fmfv{label=p,l.a=120}{o}
       \fmflabel{k}{m}
       \fmf{fermion}{i,v1}
       \fmf{fermion}{v1,o}
       \fmf{fermion,right,tension=0}{v1,m,v1}
    \end{fmfgraph*}}
    \\
       & =\frac{ig}{2} {\Int} \frac{\dd[d]{k}}{\pqty{2 \pi}^d} \frac{1}{k^2+ m^2 - i \epsilon}.
    \end{align*} 
\end{fmffile}
\end{center}
Using the Wick rotation technique of \cref{sec.thermalQFT}, we send \(k_0 \to ik_0\) and now it is possible to send \(\epsilon \to 0\). The above integral can be computed using standard techniques, and it gives
\begin{equation}\label{eq.1loopgamma2}
    \Gamma^\pqty{2}_{1-\textrm{loop}} \pqty{p,-p} = - \frac{g}{2} \frac{\Gamma \pqty{1-\frac{d}{2}}}{\pqty{4 \pi}^{d/2}} m^{\pqty{d-2}/2},
\end{equation}
where \(\Gamma \pqty{1-\frac{d}{2}} \) is the \textit{Gamma function} defined for positive integers as
\begin{equation}
    \Gamma\pqty{n} = \pqty{n-1}!.
\end{equation}
Looking at the properties of the Gamma function, it is not hard to find that \(\Gamma\pqty{-1}\) is ill-defined, thus for \(d=4\) \cref{eq.1loopgamma2} is divergent. At this point, we have to choose a regularisation scheme, and we pick the dimensional regularisation by going to a non-integer number of spacetime dimensions \(d = 4- \varepsilon\). Furthermore, we are going to use the analytic expansion of the Gamma function
\begin{equation}\label{eq.analyticexpansionofgammafunction}
    \Gamma(-n + \delta) = \frac{\pqty{-1}^n}{n!} \Bqty{\frac{1}{\delta} + \psi\pqty{n+1} + \order{\delta}},
\end{equation}
where \(\psi\pqty{n+1} \) is the digamma function \footnote{For more details see~\cite{abramowitz1965handbook}.} that has an expansion in terms of Harmonic numbers as
\begin{equation}\label{eq.digamma}
    \psi\pqty{n+1} = \Sum\limits_{k=1}^n \frac{1}{k} - \gamma,
\end{equation}
where \( \gamma \) is the Euler–Mascheroni constant. \newline
For \(d=4 - \varepsilon\) we see that we have to set \(n=1\) and \(\delta={\varepsilon}/{2}\) in \cref{eq.analyticexpansionofgammafunction,eq.digamma} and now it is possible to isolate the divergence of the vertex function \(\Gamma^{\pqty{2}}_{1-\textrm{loop}}\) in \cref{eq.1loopgamma2} and evaluate it as
\begin{equation}\label{eq.selfenergyresult}
    \Gamma^{\pqty{2}}_{1-\textrm{loop}} \pqty{p,-p} \sim \frac{1}{2} \frac{g}{16 \pi^2} m^2 \pqty{\frac{2}{\varepsilon} + 1 - \ln{m^2}} \pqty{e^{- \gamma} 4 \pi}^{\varepsilon/2}.
\end{equation}
Thus the divergent part of \(\Gamma^{\pqty{2}}_{1-\textrm{loop}}\) is 
\begin{equation}\label{eq.1loop2pointdivergence}
    \Gamma^{\pqty{2}}_{1-\textrm{loop}, \textrm{div}} \pqty{p,-p} = \frac{gm^2}{16 \pi^2 \varepsilon}.
\end{equation}
Similarly it is possible to compute the divergent part of the vertex function \(\Gamma^{\pqty{4}}_{1-\textrm{loop}}\) which is
\begin{equation}\label{eq.1loop4pointdivergence}
    \Gamma^{\pqty{4}}_{1-\textrm{loop}, \textrm{div}} \pqty{p_1,p_2,p_3,p_4} = \frac{3g^2}{16 \pi^2 \varepsilon}.
\end{equation}
As a side comment it is useful to note that while using a cut-off regularisation scheme the \( 1/\varepsilon\) regulator would be replaced by an \(\ln{\Lambda}\) regulator where \(\Lambda\) is the value of the cut-off.

Now the goal is to find a way to systematically absorb divergences like the ones encountered in \cref{eq.1loop2pointdivergence,eq.1loop4pointdivergence}. To do so, we take advantage of the fact that the theory at hand is renormalisable \footnote{The criteria for renormalisability of \acrshort{qfts} can be found in \Cref{sec.renormalisable}.}. In renormalisable \acrlong{qfts}, it is possible to remove divergences by including additional terms in the initial Lagrangian that have a form that is similar to the already existing terms, but their coefficients are now divergent and are known as \textit{counterterms}. For the specific example of \cref{eq.phi^4langragian} these terms are
\begin{equation}\label{eq.countertermlagrangian}
    \mathscr{L}_{\textrm{ct}} = - \frac{A}{2}\del^\mu \phi \del_\mu \phi - \frac{B}{2} \phi^2 - \frac{C}{4!} \phi^4, 
\end{equation}
where \(A,B,C\) are the now unknown coefficients that are included to cancel the divergences. The presence of the above terms generates new extra vertices that impact the vertex functions at leading order as
\begin{align}
    \Gamma^\pqty{2}_{\textrm{ct},\textrm{tree}} \pqty{p,-p} & = - A p^2-B, \nonumber \\
    \Gamma^\pqty{4}_{\textrm{ct},\textrm{tree}} & = - C.
\end{align}
Thus, to cancel the divergences in the vertex functions the coefficients \(A,B,C\) get their values fixed 
\begin{align}\label{eq.valuesABC}
    A & =0, & B & = \frac{gm^2}{16 \pi^2 \varepsilon}, & C & = \frac{3g^2}{16 \pi^2 \varepsilon},
\end{align}
and this selection leads to 
\begin{equation}
   \lim_{\varepsilon \to 0} \Bqty{ \Gamma^\pqty{n}_{1-\textrm{loop}} + \Gamma^\pqty{n}_{\textrm{ct},\textrm{tree}} }= \textrm{finite}.
\end{equation}
Hence there are no poles any more in the vanishing limit of the regulator and setting \(\varepsilon=0\) is an appropriate action. As this was only a one-loop order result, the coefficient \(A\) was trivially zero, but including higher-loop corrections in the original perturbative expansion would lead to new divergences, thus non-vanishing values of \(A\) would be necessary to absorb these.

One of the key elements of renormalisable \acrshort{qfts} is that all divergences can be cancelled by a finite number of counterterms and a limited number of vertex functions contain infinitely many divergences that can be absorbed into a limited number of masses and couplings. Therefore, in order for this to happen, the counterterm Lagrangian has to be of a form similar to the initial Lagrangian. On that account, we can introduce the \textit{renormalised perturbation theory} and, as a consequence, \(A,B,C\) do not need to be single numbers like in \cref{eq.valuesABC} but can be generic to all order coefficients that are expressed in a power series form in terms of the coupling \(g\) and the chosen regulator. 

Along these lines, we can combine \cref{eq.phi^4langragian,eq.countertermlagrangian} and define the \textit{bare Lagrangian} as
\begin{align}\label{eq.barelagrangian}
    \mathscr{L}_{\textrm{bare}} & = \mathscr{L} + \mathscr{L}_{\textrm{ct}} \nonumber \\
    & = -\frac{1}{2} \del^\mu \phi \del_\mu \phi \pqty{1+A} -\frac{1}{2}m^2 \phi^2 \pqty{1+B} - \frac{1}{4!} g \phi^4 \pqty{1+C} \\
    & =  -\frac{1}{2} \del^\mu \phi_0 \del_\mu \phi_0 -\frac{1}{2} m_0^2 \phi_0^2 - \frac{1}{4!} g_0 \phi_0^4, \nonumber
\end{align}
where
\begin{align}
    Z_\phi & = 1 + A, & \phi_0 &=  Z_\phi^{1/2} \phi, \label{eq.fieldrenorm} \\
    Z_{m^2} & = 1+B, &  m_0^2 & = m^2 \frac{\pqty{1+B}}{\pqty{1+A}} \equiv m^2 \frac{Z_{m^2}}{ Z_\phi}, \label{eq.massrenorm} \\
    Z_g & = 1+C, & g_0 &= \frac{g+C}{\pqty{1+A}^2} \equiv \frac{gZ_g}{Z_{\phi}^2}. \label{eq.couplingrenorm}
\end{align}
In this notation \( \phi_0, m_0, g_0\) are the bare quantities while \(\phi, m, g\) are the renormalised ones. The relation between them is expressed by the multiplicative renormalisation terms like the field renormalisation \(Z_\phi\). In this fashion, \( Z_g / Z_\phi^2 \) denotes the coupling renormalisation and \(Z_{m^2}\) the mass renormalisation. It is the bare parameters \( \phi_0, m_0, g_0\) that can be described in a power series form in \(g\) with ever-higher poles in the regulator \(\varepsilon\).

In retrospect, the values of \(A,B,C\) in \cref{eq.valuesABC} are not uniquely defined, as they can be chosen in such a way that they also include finite terms. The choice to pick such values that only subtract the divergences is known as \textit{\acrfull{ms} scheme} and for which we get
\begin{align}\label{eq.msscheme}
    m_0^2 & = m^2 \pqty{1 + \frac{g}{16 \pi^2 \varepsilon}}, & g_0 &= g \pqty{1 + \frac{3}{16 \pi^2 \varepsilon}}, & Z_\phi & = 1 + \order{g^2}.
\end{align}
Now, looking at \cref{eq.massrenorm,eq.couplingrenorm} it is clear that as \(\varepsilon \to 0\) then \(A,B \to \infty \). Nevertheless, this is not a problem since these are not physical observables.

Moreover, an important issue is the following : for the \( \phi^4\) theory, the initial coupling constant \(g\) had a zero mass-dimension, \emph{i.e.} \(\bqty{g}=0\), which, as we can see in \Cref{sec.renormalisable} is an important criterion for the theory to be renormalisable. But when we choose the dimensional regularisation scheme, and we go away from \(d=4\), then \(g\) is no longer dimensionless and using \cref{eq:massdimensioncoupling} we see that it has a mass dimension of
\begin{equation}
    \bqty{g} = \varepsilon.
\end{equation}
To keep it dimensionless we have to introduce a mass parameter \(\mu\) upon which physical quantities shall not depend and this alters \(g\) as
\begin{equation}\label{eq.gwithmu}
    g \to g \mu^\varepsilon.
\end{equation}
This alteration also affects the value of \(C\) in \cref{eq.valuesABC} which now is 
\begin{equation}
    C \to C \mu^\varepsilon,
\end{equation}
and as a result \cref{eq.couplingrenorm} becomes
\begin{equation}
    g_0 = \mu^\varepsilon \frac{g Z_g}{Z_{\phi}^2}.
\end{equation}
The bare Lagrangian of \cref{eq.barelagrangian} gives by construction finite, \(\mu\)-independent results for physical observables. So, as a general rule, bare parameters shall not depend on \(\mu\) since this is not a variable of the initial theory. 

To summarise, the bare parameters and the bare fields are \(\mu\) independent, but they are infinite in the limit that the regulator vanishes, hence they diverge. On the other hand, the renormalised fields and parameters are finite but now depend on the arbitrary mass parameter \(\mu\) which unavoidably leads to the running of couplings as we will see in a bit.

But first through \cref{eq.barelagrangian} we can work out the \acrlong{1pi} renormalised vertex functions \( \Gamma^\pqty{n} \pqty{p_1,\dots,p_n; m,g}\) regarding the bare \acrlong{1pi} proper vertex function \( \Gamma_0^\pqty{n} \pqty{p_1,\dots,p_n; m_0,g_0}\) as
\begin{equation}
    \Gamma_0^\pqty{n} \pqty{p_1,\dots,p_n} = \Gamma^\pqty{n} \pqty{p_1,\dots,p_n} + \Gamma_{\textrm{ct}}^\pqty{n} \pqty{p_1,\dots,p_n}.
\end{equation}
As an outcome, we get a well-defined result for the vertex function \( \Gamma^\pqty{n} \pqty{p_1,\dots,p_n; m,g}\), independent of the regulator.

Now, going back to our previous point, the connected \(n\)-point function is defined in \cref{eq.connectednpointfunction}. We want to Fourier transform it to momentum space and look at the bare version of it as expressed by the bare field \( \phi_0\) and the bare couplings \(m_0,g_0\)
\begin{equation}\label{eq.npointbare}
     G_{0; c}^\pqty{n} \pqty{p_1, \dots, p_n}  = \expval{\phi_0 \pqty{p_1} \dots \phi_0 \pqty{p_n}}_c .
\end{equation}
By applying \cref{eq.fieldrenorm} the expression becomes
\begin{align}\label{eq.connectedandbaregreens}
    G_{0; c}^\pqty{n} \pqty{p_1, \dots, p_n} & = Z_{\phi}^{n/2} \expval{\phi \pqty{p_1} \dots \phi \pqty{p_n}}_c \nonumber \\
    & = Z_{\phi}^{n/2} G_{ c}^\pqty{n} \pqty{p_1, \dots, p_n}, & n & \geq 1,
\end{align}
where now \( G_{ c}^\pqty{n} \) stands for the renormalised \(n\)-point correlation function, written with respect to the renormalised parameters \(\phi, m,g\).

We can express the proper, bare vertex functions \( \Gamma_0^\pqty{n} \pqty{p_1,\dots,p_n;m_0,g_0} \), which are derived from the \acrlong{1pi}
elements of the bare connected \(n\)-point correlation functions of \cref{eq.npointbare} by cutting off the external legs, and the renormalised quantities by using \cref{eq.exactpropagator,eq.connectedandbaregreens} as
\begin{align}\label{eq.vertexfunctionbareandrenorm}
    \Gamma^\pqty{n} \pqty{p_1,\dots,p_n} & = Z_{\phi}^{n/2}  \Gamma_0^\pqty{n} \pqty{p_1,\dots,p_n}, & n & \geq 1.
\end{align}
Now we observe that \cref{eq.fieldrenorm,eq.massrenorm,eq.couplingrenorm} do not solely depend on the bare parameters but also on the arbitrary mass scale \( \mu\) as
\begin{align}
    \phi^2 & = Z_\phi^{-1} \pqty{g\pqty{\mu}} \phi_0^2, \label{eq.phiwithmu} \\
    m^2 \pqty{\mu} & = \frac{Z_\phi \pqty{g\pqty{\mu}}}{Z_{m^2} \pqty{g\pqty{\mu}}} m_0^2, \label{eq.mwithmu}\\
    g\pqty{\mu} &= \mu^{- \varepsilon} \frac{Z_\phi^2 \pqty{g\pqty{\mu}}}{Z_g \pqty{g\pqty{\mu}}}g_0. \label{eq.gwithmu2}
\end{align}
As a consequence, the renormalised vertex functions \( \Gamma^\pqty{n} \pqty{p_1,\dots,p_n;m,g,\mu} \) depend on the mass scale \(\mu\) in two separate ways: both directly from factors of \(\mu^\varepsilon\) that are produced by replacing \(g\) with \(g\mu^\varepsilon\) in \cref{eq.gwithmu} and indirectly through \cref{eq.mwithmu,eq.gwithmu2}. On the contrary, the bare vertex functions \( \Gamma_0^\pqty{n} \pqty{p_1,\dots,p_n;m_0,g_0} \) are independent of \(\mu\). Hence, we see that at the right-hand side of \cref{eq.vertexfunctionbareandrenorm} the only dependence on \(\mu\) comes from \(Z_\phi\) via \cref{eq.phiwithmu}.

As stated before, any bare quantity is unquestionably independent of the arbitrarily imported mass scale \(\mu\) thus rewriting \cref{eq.vertexfunctionbareandrenorm} as
\begin{align}\label{eq.barerenormalisedpropervertex}
     \Gamma_0^\pqty{n} \pqty{p_1,\dots,p_n} & = Z_{\phi}^{-n/2}  \Gamma^\pqty{n} \pqty{p_1,\dots,p_n}, & n & \geq 1,
\end{align}
indicates a non-trivial behaviour of the renormalised vertex \( \Gamma^\pqty{n} \pqty{p_1,\dots,p_n;m,g,\mu}\) under alterations of \(\mu\). To put it another way, the combined alterations of both \( \Gamma^\pqty{n} \pqty{p_1,\dots,p_n;m,g,\mu}\) and of the rest of the renormalised functions have to be related in such a way to ensure that the physical data contained in the renormalized functions is unchanged under alterations of \(\mu\).

We can compute these alterations by applying the --- by construction dimensionless --- operator \( \mu \pdv{\mu} \) to \cref{eq.barerenormalisedpropervertex} presuming that the bare parameters remain fixed and remembering that the bare vertex function does not depend on \(\mu\)
\begin{equation}\label{eq.invarianceofbarevertex}
    0 =\mu \pdv{\mu} \Gamma_0^\pqty{n} \pqty{p_1,\dots,p_n} =\mu \pdv{\mu} \pqty{Z_{\phi}^{-n/2} \ \Gamma^\pqty{n} \pqty{p_1,\dots,p_n}}.
\end{equation}
At this point we can apply the chain rule for which \cref{eq.invarianceofbarevertex} for \(n \geq 1\) becomes
\begin{equation}\label{eq.RGE1}
    \bqty{-n \mu \pdv{\mu} \eval{\log{Z_\phi^{1/2}}}_0  + \mu \eval{\pdv{g}{\mu}}_0 \pdv{g} + \mu \eval{\pdv{m}{\mu}}_0 \pdv{\mu} + \mu \pdv{\mu}} \Gamma^\pqty{n} \pqty{p_1,\dots,p_n;m,g,\mu}=0,
\end{equation}
where the notation \( \eval_0\) is used as a reminder that the bare variables \(m_0, g_0 \) are fixed. Thus, \cref{eq.RGE1} signifies the invariance of \( \Gamma^\pqty{n} \pqty{p_1,\dots,p_n;m,g,\mu}\)  under a change from \( \Bqty{\mu, m\pqty{\mu},g\pqty{g}} \to \Bqty{ \mu', m\pqty{\mu'}, g\pqty{\mu'}} \). The physical observables of the system are unchanged under a transformation of the arbitrary mass parameter \( \mu \to \mu'\) as long as the coupling \(g\pqty{\mu}\) and the mass \( m \pqty{\mu}\) are transformed accordingly. It is important to note that the mass parameter \(\mu\) is not an independent factor. 

Thus, we can define three equations, the \textit{\acrfull{rgf}} that encapsulate the relationship of \(g, m, Z_\phi\) with the mass scale \(\mu\)
\begin{align}
   & \textrm{\textit{The anomalous field dimension :}}  & \gamma\pqty{m, g, \mu} & \equiv \mu \pdv{\mu} \eval{\log{Z_\phi^{1/2}}}_0, \label{eq.gammafunction} \\
    & \textrm{\textit{The anomalous mass dimension :}} & \gamma_m \pqty{m, g, \mu} & \equiv \frac{\mu}{m} \eval{\pdv{m}{\mu}}_0,  \label{eq.gammamfunction}\\
    & \textrm{\textit{The beta function :}} & \beta\pqty{m,g,\mu} & \equiv \mu \eval{\pdv{g}{\mu}}_0. \label{eq.betafunction} 
\end{align}
Using \cref{eq.gammafunction,eq.gammamfunction,eq.betafunction} we can express \cref{eq.RGE1} as
\begin{equation}\label{eq.RGE2}
    \bqty{ \mu \pdv{\mu} + \beta\pqty{m,g,\mu} \pdv{g}   - n \gamma\pqty{m, g, \mu} + m \gamma_m \pqty{m, g, \mu}  \pdv{m}} \Gamma^\pqty{n} \pqty{p_1,\dots,p_n;m,g,\mu}=0,
\end{equation}
which is the \textit{\acrfull{rge}} \footnote{There is an alternative way to see the running of the mass, see \cref{sec.renormalisable}.}.

Now, solving a \acrshort{pde} similar to \cref{eq.RGE2} is a priori challenging, given that \cref{eq.gammafunction,eq.gammamfunction,eq.betafunction} can depend on \(m, g\) and \( \mu \) simultaneously. Thankfully there is a remedy and this has been given by 't Hooft~\cite{THOOFT1973455} and Collins and Macfarlane~\cite{PhysRevD.10.1201} and that is that in the \acrshort{ms} scheme any counterterms are mass independent, and they
solely depend on the coupling \(g\), apart from the regulator. Therefore, the \acrshort{rg}
\cref{eq.gammafunction,eq.gammamfunction,eq.betafunction} are independent of the mass \(m\) and arbitrary mass scale \(\mu\), and depend uniquely upon \(g\)
\begin{align}
    \gamma\pqty{ g} & \overset{MS}{=} \mu \pdv{\mu} \eval{\log{Z_\phi^{1/2}}}_0, \label{eq.gammafunctiong} \\
     \gamma_m \pqty{g} & \overset{MS}{=} \frac{\mu}{m} \eval{\pdv{m}{\mu}}_0,  \label{eq.gammamfunctiong}\\
    \beta\pqty{g} & \overset{MS}{=} \mu \eval{\pdv{g}{\mu}}_0. \label{eq.betafunctiong} 
\end{align}
Applying \cref{eq.gammafunctiong,eq.gammamfunctiong,eq.betafunctiong} on \cref{eq.RGE2}, the expression of \acrshort{rge} becomes
\begin{equation}\label{eq.RGE3}
    \bqty{ \mu \pdv{\mu} + \beta\pqty{g} \pdv{g}   - n \gamma\pqty{g} + m \gamma_m \pqty{ g}  \pdv{m}} \Gamma^\pqty{n} \pqty{p_1,\dots,p_n;m,g,\mu}=0,
\end{equation}
and now the solution of this \acrlong{pde} is considerably easier than \cref{eq.RGE2}.

Let us now try to compute the \acrshort{rgf} using the fact that via the \acrshort{ms} scheme the renormalisation functions revolve around \( \mu\) only through the renormalised coupling \( g \pqty{\mu}\). First we will examine the beta function, and we will insert \cref{eq.gwithmu2} into \cref{eq.betafunctiong} for which we find 
\begin{equation}\label{eq.betafunctionlog}
    \beta \pqty{g} = - \varepsilon \bqty{\dv{g} \log{\pqty{g Z_g Z_\phi^{-2}}}}^{-1}.
\end{equation}
We can further relate the beta function with the anomalous field dimension \(\gamma\pqty{g}\). To do so, we use chain rule in \cref{eq.gammafunctiong} to write it in the form
\begin{equation}\label{eq.anomalousfielddimensionlog}
    \gamma\pqty{g} = \mu \eval{\pdv{g}{\mu}}_{0} \dv{g} \log{Z_\phi^{1/2}} = \beta \pqty{g} \dv{g} \log{Z_\phi^{1/2}},
\end{equation}
where in the last expression we used the definition of the beta function (\ref{eq.betafunctiong}). Substituting this back to \cref{eq.betafunctionlog} it becomes
\begin{equation}\label{eq.betafunctiongammaform}
    \beta\pqty{g} = \frac{-\varepsilon + 4 \gamma\pqty{g}}{\dd{\log{\bqty{g Z_g\pqty{g}}}} / \dd{g}}.
\end{equation}
Similarly, using \cref{eq.mwithmu} and using the chain rule as before, we can derive the following relation for the anomalous mass dimension \( \gamma_m \pqty{g}\)
\begin{equation}\label{eq.gammambetaform}
    \gamma_m\pqty{g} = -\frac{\beta\pqty{g}}{2} \bqty{\dv{g} \log{Z_{m^2}}- \dv{g}\log{Z_\phi}} = -\frac{\beta\pqty{g}}{2} \dv{g} \log{Z_{m^2}} + \gamma\pqty{g},
\end{equation}
and in the last expression we used the definition of \cref{eq.anomalousfielddimensionlog}.

Theoretically, all three \cref{eq.anomalousfielddimensionlog,eq.betafunctiongammaform,eq.gammambetaform} depend on the regulator \(\varepsilon\) so in principle the \acrlong{rgf} should be denoted as
\begin{align}
    \beta & = \beta\pqty{g, \varepsilon}, & \gamma & = \gamma\pqty{g, \varepsilon}, & \gamma_m & = \gamma_m \pqty{g, \varepsilon}.
\end{align}
Nevertheless, the dependence on the regulator is constrained by the fact that the theory is renormalisable and hence the \acrshort{rgf} are finite when the regulator vanishes, \emph{e.g.} \(\varepsilon \to 0\), so there should be no poles in \(\varepsilon\). Actually, it can be shown that in a direct computation of the right-hand part of \cref{eq.anomalousfielddimensionlog,eq.betafunctiongammaform,eq.gammambetaform} all poles in the regulator \(\varepsilon\) cancel out. Therefore, we can write the \acrshort{rgf} in a power series form in \(\varepsilon\) that contains non-negative powers of the regulator \(\varepsilon^n\), for example, for the beta function we have 
\begin{equation}\label{eq.betaexpansioninepsilongeneral}
    \beta\pqty{g,\varepsilon} = \beta_0\pqty{g} + \varepsilon \beta_1\pqty{g} + \varepsilon^2 \beta_2\pqty{g} + \order{\varepsilon^3}.
\end{equation}
Furthermore, things get even simpler, since of all the non-negative factors \(\varepsilon^n\), only the beta function from above contains such a term, and this is the very first order in \(\varepsilon\) which is \(\beta_1\pqty{g}\).

This can be shown in the following way: using \cref{eq.valuesABC,eq.fieldrenorm,eq.couplingrenorm,eq.massrenorm} it is easy to deduce the most general form of the expansion in inverse powers of the regulator \(\varepsilon\) for the multiplicative renormalisation terms in the \acrshort{ms} scheme, and these are
\begin{align}
    Z_\phi \pqty{g, \varepsilon} & = 1 + \Sum_{n=1}^{\infty} Z_{\phi,n} \pqty{g} \frac{1}{\varepsilon^n},\label{eq.generalformzphi} \\
     Z_{m^2} \pqty{g, \varepsilon} & = 1 + \Sum_{n=1}^{\infty} Z_{m^2,n} \pqty{g} \frac{1}{\varepsilon^n},\label{eq.generalformzm} \\
      Z_g \pqty{g, \varepsilon} & = 1 + \Sum_{n=1}^{\infty} Z_{g,n} \pqty{g} \frac{1}{\varepsilon^n}.\label{eq.generalformzg}
\end{align}
Then we can insert the general form of the multiplicative renormalisation constants as written in \cref{eq.generalformzphi,eq.generalformzm,eq.generalformzg} into \cref{eq.anomalousfielddimensionlog,eq.betafunctiongammaform,eq.gammambetaform} to obtain after some formal manipulations
\begin{align}
    & \gamma\pqty{g, \varepsilon} \bqty{1 + \Sum_{n=1}^{\infty} Z_{\phi,n} \pqty{g} \frac{1}{\varepsilon^n}}  = \frac{1}{2}\beta\pqty{g, \varepsilon} \Sum_{n=1}^{\infty} Z'_{\phi,n} \pqty{g} \varepsilon^{-n}, \label{eq.generalisedgamma}\\
   & \beta\pqty{g, \varepsilon} \Bqty{1 + \Sum_{n=1}^{\infty}\bqty{g Z_{g,n} \pqty{g}}' \frac{1}{\varepsilon^n}}  = \bqty{-\varepsilon + 4 \gamma\pqty{g,\varepsilon}}g \bqty{1 + \Sum_{n=1}^{\infty} Z_{g,n} \pqty{g} \varepsilon^{-n}}, \label{eq.generalisedbeta}\\
   & \bqty{- \gamma_m \pqty{g,\varepsilon} + \gamma\pqty{g,\varepsilon}}  \pqty{1 +  \Sum_{n=1}^{\infty} Z_{m^2,n} \pqty{g} \frac{1}{\varepsilon^n}} = \frac{\beta\pqty{g,\varepsilon}}{2} \Sum_{n=1}^{\infty} Z'_{m^2,n} \pqty{g} \frac{1}{\varepsilon^{n}}. \label{eq.generalisedgammam}
\end{align}
At this time we can straightforwardly insert \cref{eq.generalisedgamma} in \cref{eq.generalisedbeta} eliminating the anomalous field dimension \(\gamma\pqty{g,\varepsilon}\) and then expand in powers of the regulator. By doing so, we see that the beta function's expansion in powers of \(\varepsilon\) given in \cref{eq.betaexpansioninepsilongeneral} cannot contain any term in \(\varepsilon\) beyond order one
\begin{equation}\label{eq.betafunctionexpansion}
    \beta\pqty{g,\varepsilon} = \beta_0 \pqty{g} + \varepsilon \beta_1\pqty{g}.
\end{equation}
Thus, we can use the result of \cref{eq.betafunctionexpansion} to remove the beta function from \cref{eq.generalisedgamma,eq.generalisedgammam} and have an expression for the functions of the anomalous mass dimension \(\gamma_m\pqty{g,\varepsilon}\) and the anomalous field dimension \(\gamma \pqty{g,\varepsilon}\) in terms of \(\varepsilon\). Currently, we have a system of three equations that we can equate the finite terms of the same order to find
\begin{align}
    \beta_0 + \varepsilon\beta_1 + \beta_1 \pqty{Z_{g,1} + gZ'_{g,1}} & = \pqty{-\varepsilon + 4 \gamma} g - gZ_{g,1}, \\
    \gamma & = \frac{1}{2} \beta_1 Z'_{\phi,1}, \\
    \gamma_m - \gamma & = - \frac{1}{2} \beta_1 Z'_{m^2, 1},
\end{align}
where the \('\) denotes differentiation regarding \(g\) and the system at hand can be easily solved to generate the following solutions
\begin{align}
    \beta_1\pqty{g} & = -g,\\
    \beta_0 \pqty{g} & = g^2 Z'_{g,1} \pqty{g} + 4g \gamma\pqty{g},\\
    \gamma \pqty{g} & = - \frac{1}{2} g Z'_{\phi,1} \pqty{g},\\
    \gamma_m\pqty{g} & = \gamma\pqty{g} + \frac{1}{2}gZ'_{m^2, 1} \pqty{g}.
\end{align}
In this way, all the \acrlong{rgf} are written in terms of the first  derivatives of the three multiplicative renormalisation terms and more concretely as the derivatives of residues of a simple pole. Moreover, as promised, the regulator \(\varepsilon\) only appears at the beta function
\begin{equation}
    \beta\pqty{g} = -\varepsilon g + g^2 Z'_{g,1} + 4 g \gamma\pqty{g}.
\end{equation}
In order for the \acrshort{rgf} to be finite physical observables at the point when the regulator \(\varepsilon\) vanishes, i.e. \(\varepsilon \to 0\), it is required that no higher pole from \cref{eq.generalisedgamma,eq.generalisedbeta,eq.generalisedgammam} contributes. This can be verified --- see \cite{kleinert2001critical} for details --- and the residues can be computed from the one-loop \acrshort{1pi} diagrams so that at the end
\begin{align}
    \beta\pqty{g} & = - \epsilon g + \frac{3g^2}{\pqty{4\pi}^2} -\frac{17}{3} \frac{g^3}{\pqty{4 \pi}^4} \label{eq.betafunctionfinalform},\\
    \gamma\pqty{g} & = \frac{1}{12} \frac{g^2}{\pqty{4\pi}^4},\\
    \gamma_m \pqty{g} & = \frac{1}{2} \frac{g}{\pqty{4 \pi}^2 } - \frac{5}{12} \frac{g^2}{\pqty{4 \pi}^4}.
\end{align}

\subsubsection{The Wilsonian renormalisation group and fixed points}\label{sec.Wilsonianrg}

A different interpretation to the \acrlong{rg} came from Wilson~\cite{RevModPhys.47.773,PhysRevB.4.3174,PhysRevB.4.3184} and Wilson and Fisher~\cite{PhysRevLett.28.240}. As stated at the beginning of \cref{sec.RGflow}, there are plenty of methods to regularise a divergent system. One of the most prominent is the cut-off regularisation and in this approach, the cut-off \( \Lambda\) is but a mere tool of the system without any physical consequence that has to go away at the end of the computation. But in the Wilsonian approach to renormalisation, the cut-off \(\Lambda\) is not just a regulator, but now it is an important physical scale that affects the behaviour of the system. Wilson's idea was motivated and inspired by statistical and condensed matter physics and the work of Kadanoff~\cite{PhysicsPhysiqueFizika.2.263}.

In the Wilsonian approach we start again by considering the generating functional in Euclidean signature of \cref{eq.Euclideangeneratingfunctional} which is
\begin{equation*}
     \mathcal{Z}\bqty{J} = \mathlarger{\int} \mathcal{D} \phi \, e^{-S_E + {\int} \dd[d]{x} J(x) \phi (x)}.
\end{equation*}
Nevertheless, now we introduce a \acrshort{uv} cut-off \(\Lambda\) that limits the integration units in the sense that now the path integral is performed merely over field configurations \(\phi \pqty{k}\) where \( \abs{k} < \Lambda\) as
\begin{equation}\label{eq.wilsoniangeneratingfunctional}
    \mathcal{Z}\bqty{J} = \mathlarger{\int} \mathcal{D} \phi_{\abs{k}< \Lambda} \, e^{-S^{\textrm{eff}}_E \bqty{\phi;\Lambda} + {\int} \dd[d]{x} J(x) \phi (x)},
\end{equation}
and we demand that \(J\pqty{k}=0\) for \(k > \Lambda\). 
\( S^{\textrm{eff}}_E  \) is the \textit{Wilsonian effective Euclidean action} and the integration measure \(\mathcal{D} \phi_{\abs{k}< \Lambda}\) takes the form
\begin{equation}
    \mathcal{D} \phi_{\abs{k}< \Lambda} = \underset{\abs{k}< \Lambda}{\mathlarger{\prod}} \dd{\phi\pqty{k}}.
\end{equation}
The effective action depends on the large-\(k\) modes that were integrated out and is given by
\begin{equation}
    S^{\textrm{eff}}_E \bqty{\phi;\Lambda} = \mathlarger{\int} \mathcal{D}{\phi_{\abs{k}>\Lambda}} e^{-S_E\bqty{\phi}}. 
\end{equation}
As a matter of fact, the effective action is non-local for scales of \(x^\mu 	\sim 1/\Lambda\) and that happens since high-frequency modes are eliminated from the system. Nonetheless, it is possible to rewrite \( S^{\textrm{eff}}_E  \) as an expansion of local operators made of light fields. This particular procedure is known as \textit{Wilsonian \acrfull{ope}}.  Then the Wilsonian effective action can be expressed in terms of an object, known as the \textit{effective Lagrangian}, see \cite{NEUBERT_2006} for details. This item now is an infinite sum that in principle contains all possible local operators that are permitted by the symmetries of the system, and these operators are multiplied by some coupling coefficients that are finite functions of the cut-off scale \(\Lambda\) and are known as \textit{Wilsonian coefficients}.

In the specific case of the \(\phi^4\) theory, the effective Lagrangian in coordinate space is given by 
\begin{equation}
    \mathscr{L}_{E}^{\textrm{eff}} = \frac{Z\pqty{\Lambda}}{2} \del_\mu \phi \del^\mu \phi + \frac{m^2\pqty{\Lambda}}{\phi^2} + \frac{g\pqty{\Lambda}}{4!} \phi^4 + \order{\frac{1}{\Lambda^2}},
\end{equation}
and \( Z\pqty{\Lambda},m^2\pqty{\Lambda}\) and \(g\pqty{\Lambda}\) are precisely the finite functions of the cut-off scale \(\Lambda\) that we mentioned before. Furthermore, the term \( \order{\frac{1}{\Lambda^2}}\) in the effective Lagrangian is there to denote higher-order terms like \(\phi^6\) and so on, or terms that also include derivatives. It is these terms that are created from one loop quantum corrections, and they compensate for the absence of the large frequency Fourier modes in \cref{eq.wilsoniangeneratingfunctional} by generating additional interactions amid the leftover modes \(\phi\pqty{k}\), that were controlled before by the quantum fluctuations of the high-frequency \(k\) modes.

The question at hand now is what will happen if we decrease the cut-off \(\Lambda\) even lower to a value \(b \Lambda\), where \(b <1\). This is actually addressed by the Wilsonian approach and the answer is that field configurations with momenta spanning in the range \(\Lambda\) and \(b \Lambda\) are being integrated out in the sense that
\begin{equation}\label{eq.partitionfunctionwilsonian2}
    \mathcal{Z}\bqty{J} = \mathlarger{\int} \mathcal{D}{\phi_{\abs{k}< b \Lambda}} e^{- S_E^{\textrm{eff}} \bqty{\phi; b\Lambda}+ {\int}\dd[d]{x} J\pqty{x} \phi \pqty{x}   },
\end{equation}
and the new effective action contains only field configurations \(\phi \pqty{k}\) where \( \abs{k} < b\Lambda\).

We observe that Fourier components \(\phi \pqty{k}\) with momenta between \(b\Lambda < \abs{k} < \Lambda\) do not any more appear in \cref{eq.partitionfunctionwilsonian2} and they are also no longer present in the effective Lagrangian that describes the system at the current lower energy scale \footnote{Actually, the method of integrating out high-frequency field configurations is associative, so in reality it does not really matter if we integrate out modes with momenta in the range \(\Lambda > \abs{k} > b\Lambda\) and then additional modes with momenta between \(b\Lambda > \abs{k} > b'\Lambda\) with \(b>b'\) or we integrate out all modes with momenta \(\Lambda > \abs{k} > b'\Lambda\) from the start. In any case, integrating
out high-frequency Fourier modes is an irreversible procedure, and it is impossible to go back up.}. Notwithstanding, their physical data is still encrypted in changes of the Wilsonian physical variables --- like \(Z\) and \(g\) --- at the latest energy scale.  This course of action that integrates out high-frequency modes results in coarse-graining and reduces the total number of degrees of freedom of the system. 

Through this procedure, the running coupling constant \(g\pqty{\Lambda}\) satisfies the following differential equation 
\begin{equation}\label{eq.betafunctionrunning}
    \beta\pqty{g\pqty{\Lambda}} = \Lambda \dv{g\pqty{\Lambda}}{\Lambda} = \dv{g\pqty{\Lambda}}{\log{\Lambda}}.
\end{equation}
Principally, the Wilsonian approach to \acrlong{rg} provides us with a new meaning for the beta function. What the beta function truly assesses is how the coupling \(g\pqty{\Lambda}\) changes by integrating out high-frequency modes.  

Back to \cref{eq.betafunctionrunning}, the solution to this differential equation is easy to derive and is
\begin{equation}\label{eq.betafunctionandlambda}
    \log{\Lambda} = \mathlarger{\int}_g^{g\pqty{\Lambda}} \frac{\dd{g'}}{\beta\pqty{g'}}.
\end{equation}

There is an important observation to be made. Any point in the configuration space created by the coupling constants where the beta function \(\beta\pqty{g\pqty{\Lambda}}\) 
disappears, i.e. \(\beta\pqty{g^*} =0\), is known as a \textit{fixed point} \(g^*\). Fixed points have the property that the running coupling \(g\pqty{\Lambda}\) turns out to be independent of \( \Lambda\) and satisfy
\begin{equation}\label{eq.propertyfixedpoint}
    \bqty{\dv{g\pqty{\Lambda}}{\Lambda}}_{g= g^*} =0.
\end{equation}
We want to examine how a fixed point that satisfies \cref{eq.propertyfixedpoint} is obtained from \cref{eq.betafunctionandlambda}. We know from \cref{eq.betafunctionfinalform} that the beta function at leading order in perturbation theory has the form
\begin{align}
    \beta\pqty{g} & = - \varepsilon g + \frac{3g^2}{16 \pi^2} + \order{g^3},\nonumber \\
    &= - \varepsilon g + b g^2  + \order{g^3}. \label{eq.betafunctionb}
\end{align}
We notice that the beta function begins with a negative slope and vanishes at the values
\begin{align}\label{eq.fixedpointIR}
  g_{\textrm{UV}}^* &=0, &  g_{\textrm{IR}}^* & = \frac{\varepsilon}{b}. 
\end{align}
The above analysis holds true whether we know the value of the beta function at best to \(\order{g^2}\) in our perturbative expansion as long as we consider small enough values for \(\varepsilon\). The points \(g^*\) are the fixed points in the \acrshort{rg} flow. In general, there are two scenarios
\begin{enumerate}[label=\textbf{I.\arabic*},ref=I.\arabic*]
    \item\label{I1} If \(\beta\pqty{g\pqty{\Lambda}} <0\) for small \(g\pqty{\Lambda}\) , but \( \exists \ g^*\) for which \(\beta\pqty{g^*}=0\) then \(g^*\) is an \textit{\acrshort{ir} fixed point}.
    \item\label{I2} If \(\beta\pqty{g\pqty{\Lambda}} >0\) for small \(g\pqty{\Lambda}\) , but \( \exists \ g^*\) for which \(\beta\pqty{g^*}=0\) then \(g^*\) is an \textit{\acrshort{uv} fixed point}.
\end{enumerate}
From the arguments above, it is clear that we are in the case \ref{I1}, but we can further strengthen our analysis in the following way. We can add \cref{eq.betafunctionb} into \cref{eq.betafunctionandlambda} to derive
\begin{equation}\label{eq.loglambda}
    \log{\Lambda} = \mathlarger{\int}_g^{g\pqty{\Lambda}} \frac{\dd{g'}}{- \varepsilon g' + b g'^2}.
\end{equation}
From \cref{eq.loglambda} we see that when \(g'=g^*\) then \(\Lambda \to 0\). This is the low-energy limit of the theory, and hence we speak of an \acrlong{ir} fixed point.

At this point, we can even examine the stability of the fixed point. We start by Taylor expanding the beta function \(\beta\pqty{g}\) around the non-trivial zero \(g^*_{\textrm{IR}}\) \footnote{From now on and unless specified otherwise, to lighten the notation \(g^*\) denotes the \acrshort{ir} fixed point.}
\begin{equation}
    \beta\pqty{g} \sim \beta'\pqty{g^*} \pqty{g-g^*} + \dots \equiv \omega \pqty{g-g^*}+ \dots,
\end{equation}
where again \('\) signifies differentiation regarding \(g\). Omega is the slope of the beta function evaluated at the fixed point \(g^*\)
\begin{equation}
    \omega \equiv \beta'\pqty{g^*},
\end{equation}
and it controls the primary corrections of the scaling laws. By construction, the sign of omega determines whether the fixed point is stable or not. For an \acrshort{ir} fixed point to be stable, omega must be positive. Hence, going back to \cref{eq.betafunctionb} and inserting \(g^*\) from \cref{eq.fixedpointIR}
we compute the omega to be
\begin{equation}
    \omega = \varepsilon.
\end{equation}
Thus \(g^*\) is a stable \acrshort{ir} fixed point. As a general rule, the behaviour of the beta function may vary for bigger coupling. For example, it is possible that more zeros of the beta function exist on the right of \(g^*\). In the present case, we notice from \cref{eq.loglambda} that for positive values of the beta function, the coupling \(g\pqty{\Lambda}\) always heads toward zero from the right side. On the other hand, for negative values of the beta function, it goes away from zero towards the right side.

We also want to examine what happens for \(g^*_{\textrm{\acrshort{uv}}}\) and to do so we go back to \cref{eq.loglambda} and we make a variable change from \(g \to x =1/g\), for which the equation takes the form
\begin{equation}
    \log{\Lambda} = - \frac{1}{\varepsilon} \mathlarger{\int}_{1/g}^{1/g\pqty{\Lambda}} \frac{\dd{x}}{1/g^* - x},
\end{equation}
which can be straightforwardly integrated to
\begin{equation}
    \Lambda = \frac{\abs{1/g^* - 1/g\pqty{\Lambda}}^{1/\varepsilon}}{\abs{1/g^* - 1/g}^{1/\varepsilon}},
\end{equation}
and this in turn yields
\begin{equation}\label{eq.couplingwithlambda}
    g\pqty{\Lambda} = \frac{g^*}{1 + \Lambda^\varepsilon \pqty{g^*/g -1}}.
\end{equation}
If we study \cref{eq.couplingwithlambda}, we observe that in the limit that \(\Lambda \to \infty\), the coupling constant \(g\pqty{\Lambda} \to 0\) and this is but the trivial zero of the beta function (\ref{eq.betafunctionb}). This limit corresponds to the high-energy limit of the theory, and is known as an \acrfull{uv} fixed point. For \(g=0\) it is obvious that correlation functions act as those that belong to a free theory that, as we have seen in \cref{sec.pathintegral}, the relation between the fields is Gaussian. In consequence, this zero of the beta function is known as \textit{Gaussian} or \textit{trivial fixed point}. For the case of the \(\phi^4\) the trivial fixed point is \acrshort{uv} stable. All of this analysis can be depicted in the following figure:
\begin{equation}
\underset{\textrm{The flow of the coupling } \ g\pqty{\Lambda} \ \textrm{when} \ \Lambda \to 0 \ \textrm{is delineated by arrows.} }{\overset{\beta(g)}{\underset{g^*_{\textrm{UV}} \qquad \qquad \qquad  \quad g^*_{\textrm{IR}}\hspace*{\fill}}{\vcenter{\hbox{\includegraphics[scale=0.27]{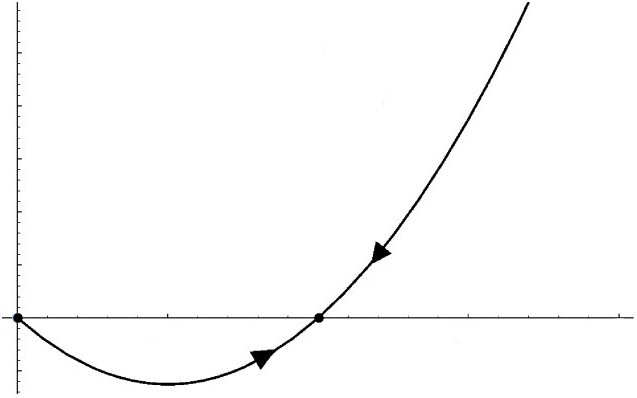}}}}}_{ \begin{matrix}
 & g/g^* & \\
 &  &  \\
  &  & \\
  & & \\
  & & 
\end{matrix}}} \nonumber
\end{equation}
At the end, we can see that in the case of \(d =4\) which is for \(\varepsilon =0\), the beta function admits a single fixed point, which is the trivial
Gaussian fixed point. 

On the other hand, for \(d=3\) where \(\varepsilon =1\) the beta function acquires two fixed points. One is still the trivial Gaussian one in the \acrshort{uv}, but there is also an \acrshort{ir} stable fixed point, the famous \acrfull{wf} fixed point.

The reason that we were so interested in fixed points is the following: obviously, if the beta function is zero, then the coupling \(g\) is a constant, there is no running of the coupling, it exhibits scale invariance, and it is unchanged under different energy scales. Therefore, any fixed point \(g^*\) of the \acrlong{rg} flow corresponds to a scale invariant --- and as we will see a conformal invariant --- \acrlong{qft}, which is the subject of \cref{sec.CFT}. In fact, the whole \acrshort{rg} flow can be expressed in the following manner: we start from the \acrshort{uv} fixed point \(g^*_{\textrm{UV}}\), where a \acrlong{cft} lives, and we add a \textit{relevant deformation}. This operator breaks conformal invariance, so we obtain a \acrshort{qft} that is not scale invariant. Then, if the theory exhibits a second attractive fixed point at the \acrlong{rg} flow, the system runs until it hits the \acrshort{ir} fixed point \(g^*_{\textrm{IR}}\), where again a \acrshort{cft} lies. Looking at \acrshort{qfts} from this perspective, known as the \textit{principle of Wilsonian universality}, every known \acrlong{qft} is classified by a \acrlong{cft} and its relevant deformations.

\section{Conformal Field Theory}\label{sec.CFT}

A \acrshort{cft} is a \acrshort{qft} which possesses extra symmetry under \textit{conformal transformations}. Therefore, besides the usual symmetries under translations, rotations and boosts, $d+1$ additional symmetries appear. More specifically, invariance under \textit{dilatations} and under \textit{\acrfull{scts}}. Combined, these symmetries generate the \textit{conformal group}. 

As we have seen in \cref{sec.Wilsonianrg}, \acrshort{cfts} exist in fixed points of the \acrshort{rg} flow, where the beta function vanishes. Actually, \acrshort{cfts} play an important role in theoretical physics and a short but in-exhaustive list of uses consists of the following:

\begin{itemize}[left= 0pt]
  \item  They are used to describing critical points in statistical mechanics. These are the points where a \textit{continuous phase transition} takes place. A phase transition is a point in the configuration space where \cref{eq.5.2.6} becomes a nonanalytic function of at least one of its parameters in the macroscopic limit. As a matter of fact, one phenomenon that is described by an interacting \acrshort{cft} is the ferromagnetic-paramagnetic second-order phase transition that takes place in magnetic materials. Such a system is commonly represented as a \(3\)-dimensional lattice, and every unit of the lattice is characterised by a spin variable. The nearest neighbour spins interact in such a manner that the energy of the lattice system is lowest in the case that the spins are aligned. So, when the system is at a very low temperature, all the spins are parallel to each other facing the same spatial direction, the system exhibits a net magnetic moment and is in its \textit{ferromagnetic phase}. Contrarily, at high temperature there is no organized alignment of the spins due to thermal fluctuations, and hence the system does not exhibit a net magnetic moment and this phase is called a \textit{paramagnetic phase}. So obviously, there is a temperature between these two phases, called the \textit{Curie temperature} \(T_C\), where the system changes phase and this is the critical point of this statistical system. If now we restrict the original system a bit more, and we demand that the spins in every unit can only acquire an up or down orientation, the \acrshort{cft} that portrays the critical point is the \(3\)-dimensional Ising model \cite{ising1924beitrag}. Captivatingly, it is the same \acrshort{cft} model that describes the liquid to gas phase transition in water. This is a pure manifestation of the Wilsonian Universality principle that we discussed in \cref{sec.Wilsonianrg}, that indistinguishable \acrlong{cfts} describe phase transitions of completely dissimilar macroscopical physical systems. These systems are described by a different \acrshort{qft} but the fixed point where the phase transition occurs is characterised by a similar \acrshort{cft}.

\item They describe the \textit{world-sheet} of \textit{string theory}. Although no more information about that will be given in this thesis, there is an abundance of books that the interested reader may refer to. A classical masterpiece that is somewhat obsolete in certain areas but ever useful is by Green, Schwarz, and Witten~\cite{green1988superstring1,green1988superstring2} while distinguished modern approaches consist of~\cite{polchinski2005string1,polchinski2005string2,kiritsis2019string}.

\item They are connected to quantum gravity through the \textit{AdS/CFT correspondence}. The \textit{gauge/gravity} duality was initially formulated by Maldacena~\cite{Maldacena:1997re} and then by Witten~\cite{https://doi.org/10.48550/arxiv.hep-th/9802150} and Gubser et al.~\cite{Gubser_1998}, and is the finest fulfilment of the idea of holography of 't Hooft~\cite{https://doi.org/10.48550/arxiv.gr-qc/9310026} and Susskind~\cite{doi:10.1063/1.531249} up to now. For an enlightening introduction, see Aharony et al.~\cite{Aharony:1999ti}.
\end{itemize}
\acrshort{cfts} are a vast subject, hence in this thesis, we will focus on properties of \acrshort{cfts} in $d>3$ spacetime dimensions.

\subsection{Conformal transformation and conformal algebra}\label{Conformaltransformations}

We start by considering the metric tensor $g_{\mu \nu}$ of the \textit{Pseudo-Riemannian} manifold $\mathcal{M}$ with line element $\dd s^2 = g_{\mu \nu} \dd x^\mu \dd x^\nu$. A conformal transformation is the differentiable map $\phi_\alpha$ that preserves the angles and re-scales the lengths by a spacetime dependent scale factor $\Omega(x) > 0$ as
\begin{equation} \label{definitionconformal}
    g_{\mu \nu} \xrightarrow{\phi_\alpha} g'_{\mu \nu} (x') = \Omega(x) g_{\mu \nu} (x).
\end{equation}
Moreover, given the transformation $x^\mu \to x'^\mu$, the relevant metric transformation is
\begin{equation} \label{metrictransformation}
g_{\rho \sigma}(x) \to g'_{\rho \sigma}(x') = \del_{\rho} x'^{\mu} \del_\sigma x'^{\nu} g_{\mu \nu}(x). \footnote{It is clear from the transformation of the line element $\dd s^2$ that at least locally the angles remain invariant and furthermore the causal structure is preserved.}
\end{equation}
For the case of the $d$-dimensional space $\mathbb{R}^{d-1,1}$ with flat metric $\eta_{\mu \nu}$ we can write the scale factor $\Omega(x)$ using the definition of the conformal transformation (\ref{definitionconformal}) and the transformation of the metric (\ref{metrictransformation}) as
\begin{equation} \label{scalefactordefinition}
    \eta_{\rho \sigma} \del_\mu x'^{\rho} \del_\nu x'^{\sigma} = \Omega(x) \eta_{\mu \nu}.
\end{equation}
It is well known that for $\Omega(x) =1$ we gain the Poincaré group that was discussed briefly in \cref{sec.classicalfield}. Additionally, one can clearly see that in the case that $\Omega \neq 1$ but still a constant, this is a global rescaling transformation.

To work out the most general form of the conformal transformations, we start by taking an \textit{infinitesimal coordinate transformation} of $x^\mu$ up to first order in $\epsilon(x) \ll 1$ as
\begin{equation}\label{transformationofx}
    x^\mu \to x'^\mu = x^\mu + \epsilon^\mu (x) + \order{\epsilon^2},
\end{equation}
where we used the fact that being a differentiable transformation, we can always Taylor-expand it about an infinitesimal vector $\epsilon^\mu$ and be able to ignore all terms of quadratic order, \emph{i.e.}~\(\order{\epsilon^2}\) and beyond. We can use the above result to deduce how the metric tensor transforms as:
\begin{align}\label{conformalmetrictransformation}
    \eta_{\rho \sigma} \del_\mu x'^{\rho} \del_\nu x'^{\sigma} & = \eta_{\rho \sigma} \big( \delta^\rho_\mu  + \del_\mu \epsilon^\rho + \order{\epsilon^2} \big)  \big( \delta^\sigma_\nu + \del_\nu \epsilon^\sigma + \order{\epsilon^2} \big) \nonumber \\
    & = \eta_{\mu \nu} + \del_\nu \epsilon_\mu + \del_\mu \epsilon_\nu + \order{\epsilon^2}.
\end{align}
By plugging back the result of~\cref{conformalmetrictransformation} in to the definition of the conformal transformation, i.e.~\cref{definitionconformal}, we observe that the conformality condition is only satisfied if and only if 
\begin{equation} \label{infinitesimalconfcond}
    \del_\mu \epsilon_\nu  + \del_\nu \epsilon_\mu = f(x) \eta_{\mu \nu},
\end{equation}
where $f(x)$ is an arbitrary function that we have to determine. To do that, we multiply both sides of~\cref{infinitesimalconfcond} by the inverse of the metric tensor $\eta^{\mu \nu}$ and then we perform the necessary index contractions \footnote{Remember $\eta_{\mu \nu} \eta^{\mu \nu} = d $.}
\begin{equation}
    2 \del_\mu \epsilon^\mu = d  \cdot  f(x) \implies f(x) = \frac{2}{d} \del_\mu \epsilon^\mu = \frac{2}{d} \del  \ \cdot \ \epsilon.
\end{equation}
At this point, using the result for $f(x)$ we can rewrite~\cref{infinitesimalconfcond} as
\begin{equation} \label{Usefulconditions}
    \del_\mu \epsilon_\nu  + \del_\nu \epsilon_\mu = \frac{2}{d} ( \del  \ \cdot \ \epsilon ) \eta_{\mu \nu}.
\end{equation}
Meanwhile, we can also reason out the scale factor $\Omega(x)$ using all the above results as
\begin{equation}
    \Omega(x) = 1 + \frac{2}{d} ( \del  \ \cdot \ \epsilon ) + \order{\epsilon^2}.
\end{equation}
Our final goal is to deduce the conformal Killing equation. To achieve that, we start by multiplying~\cref{Usefulconditions} with $\del^\nu$, which leads to
\begin{equation} \label{1ststep}
    \del_\mu ( \del \ \cdot \ \epsilon) + \Box \epsilon_\mu = \frac{2}{d} \del_\mu ( \del \ \cdot \ \epsilon).
\end{equation}
Then we act again in~\cref{1ststep} with \(\del_\nu\), which in turn gives
\begin{equation} \label{2ndstep}
    \del_\mu \del_\nu ( \del \ \cdot \ \epsilon) + \Box \del_\nu \epsilon_\mu = \frac{2}{d} \del_\mu \del_\nu ( \del \ \cdot \ \epsilon).
\end{equation}
Finally, if we swap \(\mu \leftrightarrow \nu\) in~\cref{2ndstep} then add the result back into~\cref{2ndstep} and we use the result of~\cref{Usefulconditions}, we finally arrive to the desired result
\begin{equation}\label{ConformalKilling}
    ( \eta_{\mu \nu} \Box + (d-2) \del_\mu \del_\nu ) ( \del \ \cdot \ \epsilon) = 0,
\end{equation}
which is the conformal Killing equation.

It is obvious that for $d=2$, the whole expression simplifies significantly. For the purpose of this thesis, we only care about the $d>2$ case. As a side-note, also $d=1$ is a special case, since all manifolds for $d=1$ are trivially conformally flat. This can be seen by multiplying \cref{ConformalKilling} by \(\eta^{\mu \nu}\), in which case the outcome is
\begin{equation}
    \left( d-1 \right) \Box \left( \del \ \cdot \ \epsilon \right) = 0,
\end{equation}
which is trivially realized for $d=1$.

Getting back to the conformal Killing~\cref{ConformalKilling}, it is obvious by looking at the derivatives that $\epsilon$ has at best a quadratic dependence in $x^\mu$. Hence, it can be written in the following manner
\begin{equation}\label{infinitesimaldependence}
    \epsilon_\mu (x) = a_\mu + b_{\mu \nu} x^\nu + c_{ \mu \nu \rho} x^\nu x^\rho,
\end{equation}
where $a_\mu, \ b_{\mu \nu}, \ c_{\mu \nu \rho}$ are infinitesimal constant parameters. Furthermore, by exchanging ~$x^\nu \leftrightarrow x^\rho$ in the quadratic coefficient, it is obvious that $c_{\mu \nu \rho}$ should be symmetric in the relative indices; otherwise the expression would be zero.  

At this point, it is obvious how to proceed. Since every component of the expression (\ref{infinitesimaldependence}) should be satisfied individually and independently of the position $x^\mu$, we can analyse them separately to find the corresponding generators \footnote{see~\cref{AppendixA} for the relation of generator and infinitesimal transformations.}.
\begin{itemize}[left= 0pt]
    \item The easiest term we can start with is the constant $a_\mu$. This is well known from any introduction to \acrshort{qft} --- \emph{e.g.} see~\cite{weinberg2005quantum1,Peskin:1995ev} --- to be related to the following infinitesimal transformation 
    \begin{equation}
       x^\mu \to  x'^\mu = x^\mu  + a^\mu,
    \end{equation}
    which corresponds to the translation of \cref{eq.translations} that we have examined in \cref{sec.classicalfield}. Using~\cref{Generator} and assuming that the last term is zero \footnote{This is to assume that a field $\Phi$ is not affected by the transformation, which is not entirely true as we have seen in \cref{sec.classicalfield}, and we will come back to this later.} then it is easy to see that the generator of this infinitesimal translation is the momentum operator
    \begin{equation}
        P_\mu = -i \del_\mu.
    \end{equation}
    \item The next term we are interested in is the $b_{\mu \nu}$ term, which is linear in $x^\nu$ in the expression~(\ref{infinitesimaldependence}). By using~\cref{Usefulconditions} we can work the following relation for this coefficient
\begin{equation}\label{equationforb}
        b_{ \mu \nu} + b_{\nu \mu} = \frac{2}{d} \left( \eta^{\rho \sigma} b_{\rho \sigma} \right) \eta_{\mu \nu}.
    \end{equation}
    Assuming that $b_{\mu \nu}$ is a generic function that can be decomposed into a symmetric and an antisymmetric part, then it is obvious from~\cref{equationforb} that the symmetric part should be analogous to the metric $\eta_{\mu \nu}$ times a constant~$\alpha$ for the above expression to hold. Thus, $b_{\mu \nu}$ can be broken down as
    \begin{equation}\label{splitb}
        b_{\mu \nu} = \alpha \eta_{\mu \nu} + \omega_{\mu \nu},
    \end{equation}
    where $\omega_{\mu \nu}$ is antisymmetric under the exchange of $\mu \leftrightarrow \nu$. 
    \begin{enumerate}
        \item First, we examine the antisymmetric term $\omega_{\mu \nu}$. Using~\cref{transformationofx} this corresponds to the subsequent infinitesimal transformation
        \begin{equation}
            x^\mu \to x'^\mu = \Lambda^{\mu}_{\ \nu} x^\nu \equiv \left(\delta^\mu_{\ \nu} + \omega^{\mu}_{\ \nu}   \right)x^\nu.
        \end{equation}
        This form is already known and corresponds to \cref{eq.Lorentztransformations,eq.infinitesimalLorentz} of \cref{sec.classicalfield}. Again, using the condition for the generators~(\ref{Generator}) we can associate the infinitesimal Lorentz rotations to the angular momentum operator $J_{\mu \nu}$ as
        \begin{equation}
            J_{\mu \nu} = i \left( x_\mu \del_\nu - x_\nu \del_\mu \right).
        \end{equation}
        \item On the other hand, the symmetric part of~\cref{splitb} is new, in the sense that this is not a standard part of the Poincaré algebra and is special to \acrshort{cfts}. The infinitesimal transformation corresponds to
        \begin{equation}\label{infinitesimaldilatation}
            x^\mu \to x'^\mu = x^\mu + \alpha x^\mu.
        \end{equation}
        The transformation at hand is easy to be seen to be a scale transformation, also called a dilatation. Obviously, scale symmetry cannot be a fundamental symmetry of the world, since it implies the absence of an underlying energy scale that is the same for any observer in a reference frame. Nonetheless, in certain systems it can be an almost accurate symmetry, and thus it is worth reviewing. This transformation is now associated with a new operator $D$ that is defined as
        \begin{equation}
            D = -i x^\mu \del_\mu,
        \end{equation}
        and it is known in  the literature as the \textit{dilatation} operator.
    \end{enumerate}
    \item The last term to examine is $c_{\mu \nu \rho}$. This term takes the form~\cite{Qualls:2015qjb}
    \begin{align}
        c_{\mu \nu \rho}& = \eta_{\mu \rho} b_{\nu} + \eta_{\mu \nu} b_{\rho} - \eta_{\nu \rho}b_\mu, & b_\mu & = \frac{1}{d} c^\nu_{\ \nu \mu}. 
    \end{align}
    Plugging this solution into \cref{transformationofx}, the infinitesimal transformation at hand is
    \begin{equation}
        x^\mu \to x'^\mu = x^\mu + 2 \left( x \cdot b \right) x^\mu - x^2 b^\mu,
    \end{equation}
    where $\left( x \cdot b \right) = b_\mu x^\mu$ and $x^2 = x^\mu x_\mu$. In the literature, this expression is also called \acrlong{scts}. The basic logic is that in a scale invariant world, where no underlying fundamental scale exists, observers may even be inclined to alter their definition of scale as they move around. To find the generator, all we have to do is use \cref{Generator} and obtain
    \begin{equation}
        K_\mu = -i \left(2 x_\mu x^\nu \del_\nu - x^2 \del_\mu \right).
    \end{equation}
\end{itemize}
At this point we can collect the generators of the infinitesimal transformations that we derived
\begin{align}\label{conformalgenerators}
& P_\mu  = -i \del_\mu  \nonumber  \\ 
& J_{\mu \nu}  = i \left( x_\mu \del_\nu - x_\nu \del_\mu \right), \\
 & D = -i x^\mu \del_\mu, \nonumber  \\
    & K_\mu  = -i \left(2 x_\mu x^\nu \del_\nu - x^2 \del_\mu \right). \nonumber 
\end{align}
Besides the usual ones --- remember that combining translation and Lorentz transformations generates the Poincaré group --- we have two extra that we have not seen before. 

It can be proven that the total of the conformal transformations generates a group, in the sense that the combination
of conformal transformations remains a conformal transformation. It is a well-known fact from \acrlong{qft}, that any group is defined by its generators and their associated
commutation relations, AKA the algebra. Using the above generators~(\ref{conformalgenerators}) --- for a detailed derivation of the commutation relations concerning the “new” generators $D$ and $K_\mu$ see \Cref{AppendixA} --- we can write down the conformal algebra
\begin{align}\label{eq.conformalalgebra}
    \left[ D, P_\mu  \right] & = i P_\mu, \nonumber \\
    \left[ D, K_\mu  \right] & = - i K_\mu, \nonumber \\ 
    \left[ K_\mu, P_\nu  \right] & = 2i \left( \eta_{\mu \nu} D - J_{\mu \nu} \right), \\
    \left[ K_\rho, J_{\mu \nu}  \right] & = i \left( \eta_{\rho \mu} K_\nu - \eta_{\rho \nu} K_\mu \right), \nonumber \\
    \left[ P_{\rho}, J_{\mu,\nu}  \right] & = i \left( \eta_{\rho \mu} P_\nu - \eta_{\rho \nu} P_\mu \right), \nonumber \\
    \left[ J_{\mu \nu}, J_{\rho \sigma}  \right] & = i \left( \eta_{\nu \rho }J_{\mu \sigma} + \eta_{ \mu \sigma}J_{ \nu \rho} - \eta_{ \mu \rho }J_{ \nu \sigma} - \eta_{\nu \sigma}J_{\mu \rho}  \right). \nonumber
\end{align}
More about the conformal group, its representations and the algebra will be discussed in the following section. Now, we will briefly discuss the finite form of the new transformations.

In general, any finite transformation has to follow from a series of infinitesimal ones \footnote{Notwithstanding, one has to remember that the order of applying conformal transformations is not commutative, a rotation and then a special conformal transformation will not correspond to the same outcome if performed in the opposite order.}. The finite transformation of translation and Lorentz rotations is well known. However, the finite transformation of the dilatation and the \acrshort{scts} is new. We start with the dilatation, which is easier both conceptually and mathematically. 

The finite form of the scale transformation is, 
\begin{equation}
  x^\mu \to x'^\mu = \alpha x^\mu.
\end{equation}
We can parameterize $\alpha = e^{a}, \ a \in \mathbb{R}$, so that 
 \begin{equation}
      x^\mu \to x'^\mu = e^{a} x^\mu.
 \end{equation}
Written like that it is easy to see the infinitesimal form of \cref{infinitesimaldilatation} by Taylor expanding to first order \(\order{a^2}\).

On the other hand, special conformal transformations do not exponentiate so easily. Actually, the simplest way to derive them needs some sidetracking. So, we will first state here the finite form and then try to interpret it. Without further ado, the finite form of \acrshort{scts} is

\begin{equation}\label{finiteSCTs}
	x^\mu \to x'^\mu = \frac{x^\mu - b^\mu x^2}{1- 2 \left(b\cdot x \right)+ b^2 x^2 }.
\end{equation}
And the related scale factor is
\begin{equation}
	\Omega(x) = \left( 1 - 2bx + b^2 x^2 \right)^2.
\end{equation}
A detailed computation of how to obtain the infinitesimal form, the scale factor $\Omega(x)$ and the generator $K_\mu$ of \acrshort{scts} from the finite form of \cref{finiteSCTs} can be found in \Cref{AppendixA}.

To comprehend \acrshort{scts} instinctively, we start by defining an inversion, which is a discrete transformation
\begin{equation}
         x^\mu \to x'^\mu = \frac{x^\mu}{x^2}.
\end{equation}
Although this particular transformation does not possess an infinitesimal form, it nonetheless shares the vital features of conformal transformations. Then, we can check, \emph{i.e.} see \Cref{AppendixA}, that infinitesimal \acrshort{scts} are derived by performing an inversion then a translation, followed by a final inversion

\begin{equation}\label{SCTprocedure}
    \frac{x'^\mu}{x'^2} = \frac{x^\mu}{x^2} - b^\mu.
\end{equation}
Now considering that this procedure contains two inversions, and given that the inversion is the inverse of itself, it makes no difference if the inversion is a real symmetry of the examined system or not. 
The benefit of this description is that it can be smoothly exponentiated since the combination of many
infinitesimal \acrshort{scts} can be expressed as in \cref{SCTprocedure}, which holds true for finite $b^\mu$.

We have to point out some global characteristics of \acrshort{scts} and inversions. Obviously, concerning the inversion, any point that satisfies $x^2 =0$ is mapped to $\infty$. This is no part of either flat Euclidean or Minkowski space. Something similar happens for \acrshort{scts}. Concentrating solely on flat Euclidean space we can see that there are some points of interest
\begin{itemize}
    \item The point $x=0$ which coincides with the origin is clearly mapped to itself.
    \item Considering the vector $b^\mu$, points that satisfy
    \begin{equation*}
        1-2  \left(b\cdot x \right)+ b^2 x^2 =0,
    \end{equation*}
    in other words, any point $x^\mu = \frac{b^\mu}{b^2}$ can be seen to be mapped to $\infty$.
    \item On the other hand, the opposite is true for $x \rightarrow \infty$ that is actually mapped to the vector $-\frac{b^\mu}{b^2}$.
\end{itemize}
These can all be explained by considering that \acrshort{scts} and translations are linked by inversion. From the above we can see that \acrshort{scts} do not affect the origin but move any other point, $\infty$ as well $\infty$. Contrary to that, translations are well known to move all points, but $\infty$. As for dilatations and Lorentz rotations, it is clear that they affect neither the origin nor $\infty$. All of these traits are essential when we will consider correlation functions, as the form of two and three point functions is much more constrained in a \acrshort{cft} due to these properties and actually, it is independent of position in space and depends solely on the form of operators, \emph{e.g.} scalar, spinning, etc.


\subsection{Conformal group and field transformations}\label{sec.Conformalgroup}

We now turn our attention back to the conformal group and its representations. To study the conformal group, we examine the related Lie algebra, which is called the conformal algebra. 

The first step is to count the number of generators. We can do that directly to find
\begin{align*}
    1 \ \textrm{dilatation} &+ d \ \textrm{translations} + d \ \acrshort{scts} \\
    &+ \frac{d \left( d-1 \right)}{2} \ \textrm{Lorentz} = \frac{\left( d+2 \right) \left( d+1 \right)}{2} \ \textrm{generators},
\end{align*}
which matches exactly the number of generators of a $\mathfrak{so}(d+2)$ kind of algebra. We can see that in a more clear way if we write different generators
\begin{align}
    \Tilde{J}_{\mu, \nu} & \equiv J_{\mu \nu}, \nonumber \\
    \Tilde{J}_{-1, \nu} & \equiv \frac{1}{2} \left( P_\mu - K_\mu \right), \nonumber \\
    \Tilde{J}_{0, \nu} & \equiv \frac{1}{2} \left( P_\mu + K_\mu \right), \\
    \Tilde{J}_{-1, 0} & \equiv D.\nonumber
\end{align}
Using the alternate generators, we can show that they satisfy the following commutation relation
\begin{equation}
    \left[ \Tilde{J}_{mn}, \Tilde{J}_{qp} \right] = i \left(    \eta_{mq} \Tilde{J}_{np} + \eta_{np} \Tilde{J}_{mq} - \eta_{ mp} \Tilde{J}_{nq} - \eta_{nq} \Tilde{J}_{mp} \right).
\end{equation}
To analyse the system more we have to specify the underlying metric. 
\begin{itemize}
    \item In Euclidean signature, the metric at hand is $\eta = \textrm{diag} \left(-1, 1, \dots,1 \right)$. Then the algebra is $\mathfrak{so}\left( d+1, 1 \right)$ while the Lie group is the $SO \left( d+1,1 \right)$.
    \item In the Minkowski signature, the metric at hand is replaced by $\Tilde\eta = \textrm{diag} \left(-1, 1, \dots,1, -1 \right)$. Then the algebra is $\mathfrak{so}\left( d, 2 \right)$ while the Lie group is the $SO \left( d,2 \right)$.
\end{itemize}

In \cref{Conformaltransformations} we assumed that an arbitrary local field $\Phi$ \footnote{In this section, we will be using big $\Phi$ for arbitrary fields that may or may not have spin, while the small $\phi$ of \cref{sec.classicalfield} will be used for spinless scalar fields as usual.} that from now on, we assume that exists in an irreducible representation of the conformal group, in other words a field $\Phi$ in a \acrshort{cft} that has to transform in an irreducible representation of the conformal algebra, was not affected at all by the infinitesimal conformal transformations. That statement is not true, since classical fields are indeed impacted by the generators \footnote{As usual, invariance at the classical level does not necessarily mean invariance at the quantum level, and this is true as well for conformal invariance.  Nevertheless, this is an entirely other subject.}  and our goal is to deduce the form of these generators. To do so we are going to use the method of induced representations, and we demand that given an infinitesimal conformal transformation $\epsilon_a$, a field $\Phi$ transforms as
\begin{equation}
    \Phi(x) \to \Phi' (x') = \left( 1 - i \epsilon_a T_a \right) \Phi(x),
\end{equation}
where $T_a$ is a matrix representation. 

Now, we observe that there is a subgroup within the conformal group for which the point $x=0$ remains invariant \footnote{Remember that in \cref{Conformaltransformations} we discussed that only translation affect the origin.}. We can then express the action of infinitesimal conformal transformations on the field $\Phi$ as

\begin{align}
    J_{\mu \nu} \Phi(0) &= S_{\mu \nu} \Phi(0), \\
    D \Phi(0) &= \Tilde{\Delta} \Phi(0), \\
    K_\mu \Phi(0) &= \kappa_\mu \Phi(0), 
\end{align}
where $S_{\mu \nu} , \ \Tilde{\Delta}, \ \kappa_\mu$ are the values of the generators $J_{\mu \nu}, \ D, \ K_\mu$ at the point $x=0$. We can check -—- see \Cref{AppendixA} for details --— that these values constitute a matrix representation, and hence we can sum up the algebra as

\begin{equation}\label{reducedalgebra}
		\begin{split}
		[\tilde{\Delta},S_{\mu \nu}] & = 0,\\
		[\tilde{\Delta},\kappa_\mu] & = -i \kappa_\mu,\\
		[\kappa_\nu, \kappa_\mu] & = 0,\\
		[\kappa_\mu,S_{\nu \rho}]& =i \Big( \eta_{\mu \nu} \kappa_\rho - \eta_{\mu \rho} \kappa_\nu \Big),\\
		[S_{\mu \nu}, S_{\rho \sigma}] & = i \Big( \eta_{\nu \rho} S_{\mu \sigma} + \eta_{\mu \sigma}S_{\nu \rho} - \eta_{\mu \rho} S_{\nu \sigma} - \eta_{\nu \sigma}S_{\mu \rho} \Big).
		\end{split}
\end{equation}
At this point, we can make use of Schur's lemma. In layman's terms, it states that a matrix commuting with the generator $S_{\mu \nu}$ should be proportional to the identity. Thus, from \cref{reducedalgebra} we can deduce that 
\begin{enumerate}[label=\textbf{P.\arabic*},ref=P.\arabic*]
    \item \label{l1} $\Tilde{\Delta}$ should be a number that is proportionate to the identity. 
    \item \label{l2}The matrix $\kappa_\mu$ shall vanish due to \ref{l1}.
\end{enumerate}

We will come back to this later. Now, we want to extend our result to any arbitrary spacetime point, and to do so we will make use of the \textit{Hausdorff} formula 
\begin{equation}\label{Hausdorff}
		e^{-A} B e^A = B + [B,A] + \frac{1}{2!} [[B,A],A]+ \dots,
\end{equation}
by which we can act with the operator $\mathcal{T}(x) = e^{-i x^\lambda P_{\lambda}}$ on the rest of the conformal generators --- see \Cref{AppendixA} for a analytical derivation --- as
\begin{align}\label{eq:eq:Hausedorf}
    \mathcal{T}(x)^{-1} J_{\mu \nu} \mathcal{T}(x) & = J_{\mu \nu} - x_\mu P_\nu + x_\nu P_\mu, \\
    \mathcal{T}(x)^{-1} D \mathcal{T}(x) & = D + x^\nu P_\nu , \\
    \mathcal{T}(x)^{-1} K_{\mu} \mathcal{T}(x) & = K_{\mu} +2 x_\mu D - 2 x^\nu J_{\mu \nu} + 2 x_\mu \left( x^\nu P_\nu \right) - x^2 P_\mu.
\end{align}
Using these results, we are halfway to inferring the commutation relations of the field $\Phi$
\begin{align}
    \left[ P_\mu, \Phi(x) \right] & = - i \del_\mu \Phi(x), \\
     \left[ D, \Phi(x) \right] & = \left(- i x^\nu \del_\nu + \Tilde{\Delta} \right) \Phi(x), \label{Dilatationfield} \\
     \left[ J_{\mu \nu}, \Phi(x) \right] & =S_{\mu \nu} \Phi(x) +i\left( x_\mu \del_\nu - x_\nu \del_\mu \right) \Phi(x), \\
      \left[  K_{\mu}, \Phi(x) \right] & =\left( \kappa_{\mu} +2 x_\mu \tilde{\Delta} - 2 x^\nu S_{\mu \nu} - 2 i x_\mu \left( x^\nu \del_\nu \right) + i x^2 \del_\mu \right)\Phi(x). 
\end{align}
Let's examine more closely \cref{Dilatationfield}. At the origin, $x=0$ this takes the form
\begin{equation}
    \left[ D, \Phi(0) \right]  =  \Tilde{\Delta} \Phi(0),
\end{equation}
and from postulate~\ref{l1} we know that $\tilde{\Delta}$ is a number, and also that $\tilde{\Delta}$ is non-Hermitian generator \footnote{One can see that representations of the dilatation are non-unitary. }. Hence, $\tilde{\Delta} = - i \Delta$, where the number $\Delta$ is called the \textit{scaling dimension} of the field \footnote{The factor of $i$ is there to make sure that the scaling dimension $\Delta$ is a real number when the field $\Phi$ is a
real.}. Under a dilatation, $x^\mu \xrightarrow{D} x'^\mu = \lambda x^\mu$ the field transforms as
\begin{equation}
    \Phi(x) \xrightarrow{D} \Phi'(x') = \lambda^{- \Delta} \Phi (x).
\end{equation}
Thus, we derive that the field $\Phi(x)$ --- that we assumed to transform in an irreducible representation of the conformal algebra --- exhibits a fixed scaling dimension and for that reason, it is an eigenstate of the dilatation $D$.

Finally, using postulate~\ref{l2}, we see that $\kappa_\mu =0$ and the commutation relations of the field $\Phi(x)$ are
\begin{align}
    \left[ P_\mu, \Phi(x) \right] & = - i \del_\mu \Phi(x) \equiv \mathcal{P}_\mu \Phi (x), \\
     \left[ D, \Phi(x) \right] & = -i \left( x^\nu \del_\nu  + \Delta \right) \Phi(x) \equiv \mathcal{D} \Phi(x), \\
     \left[ J_{\mu \nu}, \Phi(x) \right] & =S_{\mu \nu} \Phi(x) +i\left( x_\mu \del_\nu - x_\nu \del_\mu \right) \Phi(x) \equiv \mathcal{J}_{\mu \nu}, \\
      \left[  K_{\mu}, \Phi(x) \right] & =\left(  -2i x_\mu \Delta - 2 x^\nu S_{\mu \nu} - 2 i x_\mu \left( x^\nu \del_\nu \right) + i x^2 \del_\mu \right)\Phi(x) \equiv \mathcal{K}_{\mu}.
\end{align}


\subsection{Primary and descendants fields}\label{sec.primaryfields}

All the analysis of \cref{sec.Conformalgroup} can be utilized to derive the transformation rules of a field $\Phi(x)$ under conformal transformations. From this point on, we will be assuming that the examined field is scalar and will be denoted as $\phi(x)$ as in \cref{sec.classicalfield} and thereafter. 

Before examining how the transformation properties of the field $\phi$ constrain the form of correlation functions in a \acrshort{cft} we want to discuss some details about the spectrum and the nature of the fields. 

Given a generic state \(\ket{\Delta}\) that is an eigenstate of the dilatation in the sense that 

\begin{equation}\label{eq.scalingdimension}
    D \ket{\Delta} = i \Delta \ket{\Delta}, 
\end{equation}
we want to examine how it changes when acted upon by $K_\mu$ and $P_\mu$. To do so, we use some standard group theory techniques, for example see Tung~\cite{tung1985group}, \(\S \, 7\).

Acting with the special conformal generator $K_\mu$ we see that

\begin{equation}
   D K_\mu  \ket{\Delta} = \left[ \left[ D, K_\mu \right] + K_\mu D \right] \ket{\Delta} = \left( -i K_\mu +i \Delta K_\mu  \right) \ket{\Delta} = i \left( \Delta -1 \right) K_\mu \ket{\Delta},
\end{equation}
where \cref{eq.conformalalgebra} was used. Now we can see that \(K_\mu \ket{\Delta} \) is either an eigenstate of $D$ with a new eigenvalue $\Delta -1$ or otherwise it should be the null vector. Thus, this demonstrates that $K_\mu$ decreases the scaling dimension, and hence it is a lowering operator.

An analogous derivation can be made for $P_\mu$
\begin{equation}\label{eq.raisingthescaling}
     D P_\mu  \ket{\Delta} = \left[ \left[ D, P_\mu \right] + P_\mu D \right] \ket{\Delta} = \left( i P_\mu + i \Delta P_\mu \right) \ket{\Delta} = i \left( \Delta + 1 \right) P_\mu \ket{\Delta},
\end{equation}
where again \cref{eq.conformalalgebra} was used, and similarly as before, we can conclude that the vector \(P_\mu \ket{\Delta}\) is either the null vector or an eigenstate of $D$. This outcome is in accordance with the known result that the mass dimension of the derivative is 

\begin{equation*}
    [\del_\mu] = + 1.
\end{equation*} 
Assuming that in a unitary \acrshort{cft} conformal dimensions have to be bounded from below, the fact that $K_\mu$ lowers the scaling indefinitely seems to contradict this. The way out of this puzzle is to simply accept that for some general state, when acted upon by $K_\mu$ it will hit zero and will go no further down. It is equivalent to imagining that there are some local fields $\phi$ that vanish when acted upon by $K_\mu$. These fields exhibit the lowest possible scaling dimension in a specified irreducible multiplet of the algebra, and this non-zero, positive scaling is fixed by unitarity bounds \cite{ammon_erdmenger_2015} \(\S \, 3.2.2\). 

These fields are known as \textit{primary} fields, and they have to obey the following conditions
\begin{align}
    &\left[ D, \phi(0) \right] =-i \Delta \phi(0),\\
    &\left[ K_\mu, \phi(0) \right]\label{eq.stateannihilation} =0. 
\end{align}
Actually, in the literature one can find two different definitions for the primary field but it has been shown in~\cite{Campos_Delgado_2022} that these two are equivalent.

Now acting on a local conformal primary field with $P_\mu$ increases the scaling dimension by $+1$. Thus, $P_\mu$ is a raising operator and the fields constructed by the action of $P_\mu$ are called \textit{descendants}. Acting on a primary field with $P_\mu$ repeatedly can produce infinite descendants, and a primary field and all the descendants produced by it are called a \textit{conformal family}.


\subsection{Constraints on correlation functions of primary operators}

Going back to the transformation rules of the local primary field $\phi$, we first need to derive the transformation rules for $x \to x'$ for a conformal transformation with scale factor $\Omega(x)$. The relevant Jacobian is
\begin{equation}
    \abs{\frac{\del x'}{\del x}} = \frac{1}{\sqrt{\textrm{det}g'_{\mu \nu}}} = \Omega(x)^{-d/2}.
\end{equation}
Under this transformation, the scalar field $\phi(x)$ changes as
\begin{equation}
    \phi(x) \to \phi'(x') = \abs{\frac{\del x'}{\del x}}^{- \Delta/d} \phi(x).
\end{equation}
This is precisely the transformation rule that has to be satisfied by a local primary scalar field.

We are interested to seeing if and how conformal invariance constrains the form of the correlation functions of primary fields. It turns out that not only it imposes great constraints on the form of the Green's functions, but actually the two point function is uniquely defined up to the value of the scaling dimension, which is a result of the extra constraints coming from scaling invariance and special coordinate transformation.

We will use the transformation rules of a $n$-point function 
\begin{equation}
		\langle \phi_1(x_1) \phi_2(x_2) \dots \phi_n(x_n) \rangle = {\abs{\frac{\partial x'}{\partial x}}}_{x=x_1}^{\Delta_1 / d}{\abs{\frac{\partial x'}{\partial x}}}_{x=x_2}^{\Delta_2 / d} \dots  {\abs{\frac{\partial x'}{\partial x}}}_{x=x_n}^{\Delta_n / d} \langle \phi_1(x_1') \phi_2(x_2') \dots \phi_n(x_n') \rangle. 
\end{equation} 
We are interested in two, three and four-point functions. The full analytic derivation can be found in \Cref{AppendixA} and this chapter contains the most important results.

\subsubsection{Two-point functions}

From \acrshort{qft} it is well known that the two-point function can only depend on a quantity that is the difference between two points, and it is also Lorentz invariant. The only such known quantity is the absolute value of the interval between two different spacetime points, \emph{i.e.}
\begin{equation}
    \abs{x_1 -x_2} \equiv \sqrt{\eta_{\mu \nu}\big(x_1^\mu - x_2^\mu\big)\big(x_1^\nu-x_2^\nu\big)}.
\end{equation}
Using this expression, we can write that the most general form of the two-point function that is invariant under Poincaré transformations as
\begin{equation}\label{eq.generalform}
	\langle \phi_1(x_1) \phi_2(x_2) \rangle = c_{12} {\abs{x_1 -x_2}}^\alpha.
\end{equation}
Now we should apply the new symmetries that are manifest in the conformal algebra. 

From dilatation invariance we have that
\begin{equation}\label{eq.1}
	\langle \phi_1(x_1) \phi_2(x_2) \rangle = \lambda^{{\Delta_1}+{\Delta_2}}\langle \phi_1(\lambda x_1) \phi_2(\lambda x_2) \rangle.
\end{equation}
And combining these, we can write the two-point function as
\begin{equation}
			\langle \phi_1(x_1) \phi_2(x_2) \rangle = \frac{c_{12}}{{\abs{x_1 -x_2}}^{{\Delta_1}+{\Delta_2}}}.
\end{equation}
Using the \acrshort{scts} we can show that it if ${\Delta}_1 \neq {\Delta}_2$, the two-point function vanishes hence 

\begin{equation}
\langle \phi_1(x_1) \phi_2(x_2) \rangle =  \left\{ \begin{aligned}
  & \frac{c_{12}}{{\abs{x_1 - x_2}}^{2 \Delta_1}}  &\textrm{if}&  &{\Delta}_1 = {\Delta}_2,\\
  &0  &\textrm{if}& &{\Delta}_1 \neq {\Delta}_2.
\end{aligned}\right.
\end{equation}
There is one more possible simplification that can be done. The constant $c_{12}$ is real, and it is also symmetric under $\phi_1 \leftrightarrow \phi_2$ so that $c_{12} = c_{21}$. Using this fact, we can pick a basis of diagonal scalar primary operator $\mathscr{O}$ such that $c_{12}= \delta_{12}$. 

So finally, the two-point function takes the form
\begin{equation}\label{eq.conformaltwopointfunction}
	\left\langle \mathscr{O}_1(x_1) \mathscr{O}_2(x_2) \right\rangle =  \frac{1}{{\abs{x_1 - x_2}}^{2 \Delta}},
\end{equation}
where the only unknown is the scaling dimension $\Delta$ of the fields. We can see that this is a forceful constraint, which is the immense power of \acrshort{cfts}.
\subsubsection{Three-point functions}

In the same spirit as before, we start with the general form of \cref{eq.generalform}, with three fields, and follow the same methodology \footnote{From this point on we always assume that we are in the diagonal basis of scalar primary operator $\mathscr{O}$ that are already redefined for the two-point function}

\begin{enumerate}
    \item Translations and Lorentz rotations apply the same to all $n$-point functions, so they constrain the three-point function to be determined by the absolute value of the interval between the three points in some power,
    	\begin{equation}
		\expval{ \mathscr{O}_1(x_1) \mathscr{O}_2(x_2) \mathscr{O}_3(x_3) } = C_{\mathscr{O}_1\mathscr{O}_2\mathscr{O}_3} {\abs{x_1-x_2}}^\alpha {\abs{x_2-x_3}}^b {\abs{x_1-x_3}}^c,
		\end{equation}
    \item From dilatation we obtain the constraint
		\begin{equation}
			\Delta_1 + \Delta_2 + \Delta_3+\alpha+b+c=0.
	  \end{equation}
	\item From \acrshort{scts} we find the following three constraints
		\begin{gather}
			2 \Delta_1 + \alpha + c =0,\\
			2 \Delta_2 + \alpha  + b = 0,\\
			2 \Delta_3 + b +c =0,
		\end{gather}
		that we can solve for $\alpha, b, c$ to get
		\begin{gather}
			\alpha = \Delta_3 - \Delta_1 - \Delta_2,\\
			b = \Delta_1 - \Delta_2 -\Delta_3,\\
			c = \Delta_2 - \Delta_1 - \Delta_3.		
		\end{gather}
\end{enumerate}
Using the above conditions the final form of the three-point function is
\begin{equation}\label{eq.generalform3pointfunction}
    \expval{ \mathscr{O}_1(x_1) \mathscr{O}_2(x_2) \mathscr{O}_3(x_3) } = \frac{C_{\mathscr{O}_1\mathscr{O}_2\mathscr{O}_3}}{{\vert x_{12} \vert}^{{\Delta}_1 + {\Delta}_2-{\Delta}_3}{\vert x_{23} \vert}^{{\Delta}_2 + {\Delta}_3-{\Delta}_1}{\vert x_{31} \vert}^{{\Delta}_3 + {\Delta}_1-{\Delta}_2}}, 
\end{equation}
where \(x_{ij} \equiv x_i -x_j\) for shortness.

The constant \(C_{\mathscr{O}_1\mathscr{O}_2\mathscr{O}_3}\) cannot be normalized to unity, since the operators are already redefined and thus this number, which is known as  \textit{\acrlong{ope}} coefficient, is a characteristic of the theory, and along with the scaling dimension \(\Delta\) they constitute the \acrshort{cft} data that has to be computed to classify the \acrshort{cft}. 

In general, a generic CFT is described only by the conformal scaling dimensions and the OPE coefficients, since higher-point functions can be calculated in principle by connecting three-point functions appropriately. 

\subsubsection{Four-point functions}

Again, for four-point functions the starting point and the logic is similar as before. But for the four-point function there is a difference, since by having four points and more it is possible to create certain dimensionless invariant coefficients that preserve the CFT symmetries. These cross ratios, are
\begin{align}
		\chi_1 & = \left( \frac{\abs{x_{12}}\abs{x_{34}}}{\abs{x_{13}}\abs{x_{24}}}\right)^2, &  \chi_2 &=\left( \frac{\abs{x_{12}}\abs{x_{34}}}{\abs{x_{23}}\abs{x_{14}}}\right)^2.
\end{align}
The four-point function can be written in its most general form as a product of two such cross ratios in terms of \textit{conformal blocks}, as
\begin{equation}\label{eq.4pointfunction}
    \left\langle \mathscr{O}_1(x_1) \mathscr{O}_2(x_2) \mathscr{O}_3(x_3) \mathscr{O}_4(x_4) \right\rangle = \mathcal{F}\left(\frac{{x}_{12}{x}_{34}}{{x}_{13}{x}_{24}}, \frac{{x}_{12}{x}_{34}}{{x}_{23}{x}_{14}} \right) \mathlarger{\prod}\limits_{i<j}^{4} {x_{ij}}^{\Delta /3 - {\Delta}_i -{\Delta}_j }, 
\end{equation}
where $\Delta = \Sum\limits_{i=1}^{4}{\Delta}_i $ and $\mathcal{F}\left( \chi_1, \chi_2 \right)$ is a function of all possible cross ratios. It is important to note that if we manage to write the four-point function in the form of \cref{eq.4pointfunction} then we can read off the scaling dimensions of the scalar operators and the \acrshort{ope} coefficients.


\subsection{The energy-momentum tensor}\label{sec.Energymomentumtensor}

Going back to \cref{sec.classicalfield} we remind ourselves that according to Noether's theorem, continuous symmetries generate conserved currents and charges. The conserved current for spacetime translations is given by \cref{eq.ThetaEM} and is the energy-momentum tensor, while the conserved charges are given in \cref{eq.Hamiltoniandensity,eq.physicalmomentum} and in a shorthand notation they can be written as
\begin{equation}
    P_\nu = \mathlarger{\int} \dd[d-1]{x} T^0_\nu.
\end{equation}
Meanwhile, for Lorentz transformations the current is specified in \cref{eq.lorentzcurrent} and the conserved charge is stated in \cref{eq.lorentzcharge} as
\begin{equation}
    M_{\nu \rho} = \mathlarger{\int} \dd[d-1]{x} \pqty{x_\nu T^0_{ \ \rho} - x_\rho T^0_{\ \nu}  },
\end{equation}
where \(T_{\mu \nu}\) is the Belinfante energy-momentum tensor of \cref{eq.BelinfanteEM}
\begin{equation*}
        T^{\mu \nu} = \Theta^{\mu \nu} + \del_\rho B^{\rho \mu \nu}.
\end{equation*}
In the same spirit, we can deduce the conserved currents and charges for the rest of the conformal transformations.

For example, dilatation of \cref{infinitesimaldilatation} associated with the infinitesimal transformation
\begin{equation*}
    x^\mu \to x'^\mu = x^\mu + \alpha x^\mu,
\end{equation*}
generates the current \(\mathcal{J}_{\pqty{D} \mu}\) which is computed to be
\begin{equation} \label{eq.dilatationcurrent}
    \mathcal{J}_{\pqty{D} \mu} = x^\nu T_{\mu \nu},
\end{equation}
and the relevant charge is
\begin{equation}
    D = \mathlarger{\int} \dd[d-1]{x} x^\rho T^0_{\ \rho}.
\end{equation}
For special conformal transformations, the conserved current and charge are
\begin{align}
    \mathcal{J}_{\pqty{K} \mu \nu} & = x^2 T_{\mu \nu} - 2x_\nu x^\rho T_{\mu \rho}, \\
    K_\nu & = \mathlarger{\int} \dd[d-1]{x} \pqty{x^2 T^0_{\ \nu} - 2 x_\nu x^\rho T^0_{\ \rho}}.
\end{align}

In a similar vein as the Poincaré invariance imposed constraints on the stress-energy tensor and more specifically symmetrisation in its indices, conformal invariance inflicts new restrictions on it. From conservation of \cref{eq.dilatationcurrent} \footnote{For details, see \Cref{sec.conformalEM}.} we observe that
\begin{align}\label{eq.tracelessEM}
    0 & = \del^\nu \mathcal{J}_{\pqty{D} \nu} = \del^\nu \pqty{x^\rho T_{\nu \rho}} \nonumber \\ 
    & = \pqty{\del^\nu x^\rho} T_{\nu \rho} + x^\rho \del^\nu T_{\nu \rho} \\
    & = T^\rho_{\ \rho}. \nonumber
\end{align}
It is obvious now that the conformal energy-momentum tensor is traceless. Since the dilatation charge gives rise to scale transformations, tracelessness of the stress-energy tensor signifies scale invariance of the theory, or conversely, if the classical theory is scale invariant the stress-energy tensor should be traceless.

This property is an important criterion to examine to verify that a field theory truly is a \acrlong{cft}. 


\subsection{State-operator correspondence and radial quantisation}\label{sec.stateoperator}

If we want to solve equations of motion to describe the time evolution of a system, we need to impose an initial condition at one point in time, and then under certain assumptions the whole evolution of the system both forward and backwards can be determined. So, it is a usual practise while studying \acrlong{qfts} to foliate the spacetime in a specific way to facilitate such computations, with every foliated leaf carrying its exclusive Hilbert space. The manner of the foliation varies and actually is connected to the symmetries of the theory. 

\begin{wrapfigure}{R}{0.30\textwidth} 
    \centering
    \includegraphics[width=0.30\textwidth]{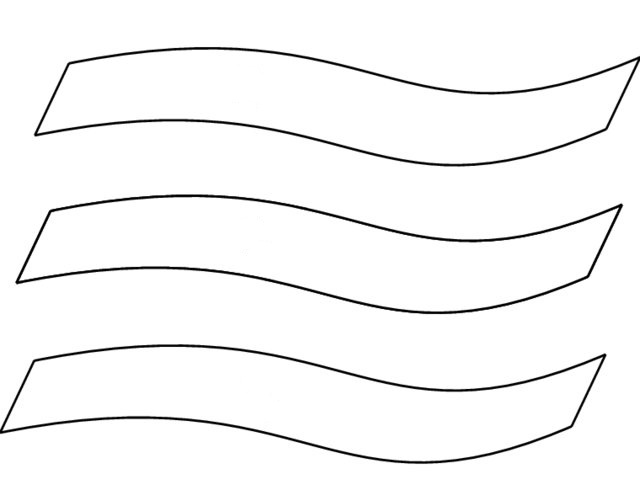}
    \caption{The \(d\)-dimensional spacetime is foliated in \(\pqty{d-1}\)- dimensional leafs.}
    \label{fig.spacetimefoliation}
\end{wrapfigure}

In the most common manner, when we examine a \acrlong{qft} that exhibits Poincaré invariance, we usually divide the \(d\)-dimensional spacetime manifold \(\pqty{\mathcal{M}, g_{\mu \nu}}\) into \(\pqty{d-1}\) - dimensional hypersurfaces of equal time like in \Cref{fig.spacetimefoliation}. Since in \cref{sec.classicalfield} we made the assumption that our manifold is globally hyperbolic, the conditions at any spatial slice at any point in time determine the system everywhere. Therefore, we express the spacetime as a series of spatial slices that evolve in time, and now we have a problem which amounts to finding a way of describing how those spatial slices evolve. Given the Poincaré symmetry, we already know of an operator that allows us to time translate between spatial slices, and this is the Hamiltonian given in \cref{eq.Hamiltoniandensity}. On top of these spatial slices live states that can be uniquely characterised by their momenta as
\begin{equation}
    P^\mu \ket{k} = k^\mu \ket{k}.
\end{equation}
Then, using the Hamiltonian operator, we can define a \textit{unitary evolution operator}
\begin{equation}
    U = e^{i H t},
\end{equation}
that connects these states. 

Although this is a perfectly fine foliation for Poincaré invariant theories, in the case of \acrlong{cfts} there is a more convenient way to foliate the system.

\begin{wrapfigure}{L}{0.4\textwidth}
    \centering
    \includegraphics[scale=0.3]{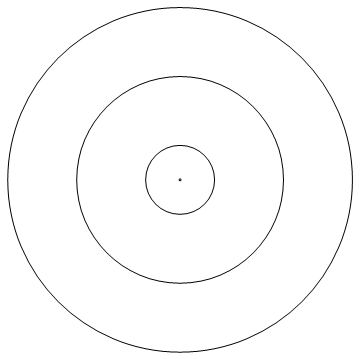}
    \caption{Spheres centred at the origin of Euclidean spacetime.}
    \label{fig.radialquantisation}
\end{wrapfigure}

For \acrshort{cfts} we can use the enhanced symmetry to find a different way to foliate our manifold. So, instead of time translations, we are going to use dilatations. We will split the spacetime manifold in \(\mathbb{S}^{d-1}\) spheres of differing radii that are centred around the origin of the \(d\)-dimensional Euclidean space as in \Cref{fig.radialquantisation}, with a metric element
\begin{equation}\label{eq.sphericalmetric}
    \dd{s^2} = \dd{r^2} + r^2 \dd{\Omega},
\end{equation}
where \(\dd{\Omega}\) is the volume element of the unit sphere. To pass from one sphere to another, we use the generator of the dilatation \(D\). In a manner similar as before, we want to distinguish states based on their eigenvalues, but instead of their momenta, now we are going to use the scaling dimension denoted in \cref{eq.scalingdimension} as
\begin{equation*}
    D \ket{\Delta} = i \Delta \ket{\Delta},
\end{equation*}
and their spin \(\ell\)
\begin{equation*}
    J_{\mu \nu} \ket{\Delta, \ell} = S_{\mu \nu} \ket{\Delta,\ell},
\end{equation*}
since as we can see from the algebra \ref{eq.conformalalgebra}, only \(S_{\mu \nu}\) commutes with the dilatation. 

At this point, to denote the evolution operator, we make the following coordinate change
\begin{equation}\label{eq.tautransformation}
    \tau \equiv R_0 \log{\pqty{r/R_0}},
\end{equation}
where \(r\) is the radial direction and \(R_0\) is the radius of the sphere which we normally set \(R_0=1\) but we reinstated here for dimensional reasons. Then the metric of \cref{eq.sphericalmetric} becomes
\begin{align}
    \dd{s^2_{\textrm{flat}}} & = \dd{r^2} + r^2 \dd{\Omega} = \frac{r^2}{{R_0}^2} \pqty{\frac{R_0^2}{r^2} \dd{r^2} + R_0^2 \dd{\Omega}} \nonumber \\
    & = e^{2 \tau/ R_0} \pqty{\dd{\tau^2} + R_0^2 \dd{\Omega}}\\
    & = e^{2 \tau /R_0} \dd{s^2_\textrm{cyl}}.
\end{align}
It is now obvious that the flat metric \(\mathbb{R}^d\) expressed in spherical coordinates in \cref{eq.sphericalmetric} is conformally equivalent to the cylinder \(\mathbb{R} \times \mathbb{S}^{d-1} \) through a Weyl transformation \footnote{From now on we set \(R_0 =1\)}. In other words, we found a metric that best encapsulates the physics of the system since the \(\mathbb{S}^{d-1} \) are the spheres of before, and now \(\tau\) is the proper time coordinate to express the evolution of the states. Hence, the evolution operator is now
\begin{equation}
    U = e^{i D \tau},
\end{equation}
and acting with it on an eigenstate of the dilatation \(\ket{\Delta}\) using \cref{eq.scalingdimension} it gives
\begin{equation}
    U \ket{\Delta} = e^{- \Delta \tau} \ket{\Delta} = r^{- \Delta} \ket{\Delta}.
\end{equation}
This whole procedure is called \textit{radial quantisation} and can be depicted as
\begin{align*}
&\vcenter{\hbox{\includegraphics[scale=0.35]{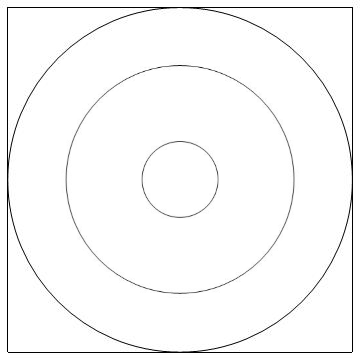}}}  &\mathlarger{\mathlarger{\mathlarger{\longleftrightarrow}}}& 
&\vcenter{\hbox{\includegraphics[scale=0.27]{Figures/cylinder.png}}} { \begin{matrix}
\infty &  & \\
 \bigg{\uparrow}&  &  \\
  \bigg{\vert} &  & \\
  \tau  &  & \\
   \bigg{\vert}  &  & \\
 \bigg{\vert} & & \\
  - \infty& & 
\end{matrix}}
\end{align*}
We can analyse the situation even more. We observe that the limit \(r \to 0\) in the flat case corresponds to the limit \(\tau \to -\infty\) in the cylinder case and also for \( r \to \infty\) after the Weyl transformation we get \( \tau \to \infty\).

In radial quantisation, the way to generate a state at a given time living on the sphere, is by inserting an operator in the sphere. We have the following cases
\begin{enumerate}[left= 0pt]
    \item The first case is that we do not make any operator insertion anywhere. Thus, this system coincides with the vacuum state \(\ket{0}\). This is a unique ground state that has the property that is invariant under conformal transformations. For this state, the dilatation eigenvalue which coincides with its radial quantization energy is \(0\).
    \item The second case is to insert an operator \(\mathscr{O}\) with scaling dimension \(\Delta\) at \(x=0\). The associated state is
    \begin{equation}
        \ket{\Delta} \equiv \mathscr{O}_{\Delta}\pqty{0}\ket{0},
    \end{equation}
    where we used the subscript to denote the scaling dimension of the operator. The energy of the state is equal to the scaling dimension of the operator. 
    \item Finally, we can insert an operator at a point different from the origin \(0\). The state produced is %
    \begin{equation}
        \ket{\chi} \equiv \mathscr{O}_{\Delta} \pqty{x} \ket{0}.
    \end{equation}
    Actually this state is not an eigenstate of the dilatation operator, since the dilatation moves the point of insertion. We can use the same trick as in \cref{sec.Conformalgroup} and move the point of insertion by using \( \mathcal{T}\pqty{x} = e^{-i x^\lambda P_\lambda}\) in which case we find that
    \begin{equation}
        \ket{\chi} = \mathcal{T}^{-1}\pqty{x} \mathscr{O}_{\Delta} \pqty{0} \mathcal{T}\pqty{x} \ket{0} = \mathlarger{\sum}_n \frac{x^n}{n!} (i P)^n \ket{\Delta}.
    \end{equation}
    Using the analysis of \cref{sec.primaryfields} we see that this state is actually a superposition of states, each of them having its eigenvalue since via \cref{eq.raisingthescaling} we notice that every time that we act with \(P\) on the state \(\ket{\Delta}\), it raises the conformal dimension of the state by one. 
\end{enumerate}
Then the statement is that if one picks a generic state \(\ket{\Delta}\) at \(\tau=- \infty\) on the cylinder \(\mathbb{R} \times \mathbb{S}^{d-1}\), by just making a conformal transformation, one can relate this state to a local operator \(\mathscr{O}_{\Delta}\pqty{0}\) with scaling dimension \(\Delta\) on the origin of \(\mathbb{R}^d\).
As a matter of fact, we can always map operators to states, by just acting with the operator on the vacuum. In other words, placing a primary operator at \(x=0\), we produce a state with a scaling
\(\Delta\) that due to \cref{eq.stateannihilation} is annihilated by the operator \(K_\mu\). But, the other way around, that starting from a state with scaling \(\Delta\) that is annihilated by applying \(K_\mu\), we can construct a
related local primary operator is a unique characteristic of \acrlong{cfts}. This property is the \textit{state-operator correspondence}, where states on the cylinder are
in a one-to-one correspondence with local operators in the plane.

As we saw earlier, the flat metric and the cylinder metric are related by a scale transformation, hence for a conformally invariant theory, correlation functions computed on flat space and correlation functions computed on the cylinder are related through a Weyl transformation. In short, correlation functions on the cylinder are simply computed in another geometry than the ones of flat space. Using radial quantization we can compute the two-point functions of two diagonal scalar primary operators \(\mathscr{O}\pqty{\tau}\) as
\begin{align}
    {\bra{0} \mathscr{O}_1 \pqty{\tau_\textrm{out}} \mathscr{O}_2\pqty{\tau_\textrm{in}} \ket{0}}_{\textrm{cyl}} & = \expval{\mathscr{O}_1 \pqty{\tau_\textrm{out}} \mathscr{O}_2\pqty{\tau_\textrm{in}}}_{\textrm{cyl}} \nonumber \\
    & = \expval{\mathscr{O}_1 \pqty{x_\textrm{out}} \mathscr{O}_2\pqty{x_\textrm{in}}}_{\textrm{flat}} \pqty{\frac{\abs{x_\textrm{out}}}{R_0}}^{\Delta} \pqty{\frac{\abs{x_\textrm{in}}}{R_0}}^{\Delta}  \nonumber \\
    & = \abs{x_\textrm{out} -x_\textrm{in} }^{-2 \Delta} \pqty{\frac{\abs{x_\textrm{out}}}{R_0}}^{\Delta} 
 \pqty{\frac{\abs{x_\textrm{in}}}{R_0}}^{\Delta} \label{eq.twopointtransform},
\end{align}
where in the last line we used the result of \cref{eq.conformaltwopointfunction}. By working in polar coordinates, we identify 
\begin{align}\label{eq.polartocylinder}
    x_\textrm{out} & = r_\textrm{out}, & x_\textrm{in}& = r_\textrm{in},
\end{align}
and also 
\begin{equation}\label{eq.differentabsvalue}
    \abs{ x_\textrm{out} - x_\textrm{in}} = \bqty{{r^2_\textrm{out}}+ {r^2_\textrm{in}} - 2 {r_\textrm{out}} {r_\textrm{in}} \n_{\textrm{out}}\n_{\textrm{in}}}^{1/2},
\end{equation}
where \( \n\) is the unit directional vector. Then combining \cref{eq.polartocylinder,eq.differentabsvalue}, the two-point function of \cref{eq.twopointtransform} becomes
\begin{align}
    \expval{\mathscr{O}_1 \pqty{\tau_\textrm{out}} \mathscr{O}_2\pqty{\tau_\textrm{in}}}_{\textrm{cyl}} &= \abs{{r^2_\textrm{out}}+ {r^2_\textrm{in}} - 2 {r_\textrm{out}} \ {r_\textrm{in}} \ \n_{\textrm{out}} \ \n_{\textrm{in}}}^{- \Delta} {r^{\Delta}_\textrm{out}} \ {r^{\Delta}_\textrm{in}} \nonumber\\
    & = \pqty{\frac{r_\textrm{in}}{r_\textrm{out}}} \pqty{1 + \pqty{\frac{r_\textrm{in}}{r_\textrm{out}}}^2 - 2 \frac{r_\textrm{in}}{r_\textrm{out}} \ \n_{\textrm{out}} \ \n_{\textrm{in}} }^{-\Delta}.
\end{align}
Finally, using \cref{eq.tautransformation} we find that
\begin{align}\label{eq.twopointfunctioncylinder}
     \expval{\mathscr{O}_1 \pqty{\tau_\textrm{out}} \mathscr{O}_2\pqty{\tau_\textrm{in}}}_{\textrm{cyl}} &= R_0^{-2 \Delta} e^{- \Delta \pqty{\tau_{\textrm{out}} - \tau_{\textrm{in}}}/R_0} \pqty{1 + e^{- 2\Delta \pqty{\tau_{\textrm{out}} - \tau_{\textrm{in}}}/R_0}  - 2 e^{- \Delta \pqty{\tau_{\textrm{out}} - \tau_{\textrm{in}}}/R_0} \ \n_{\textrm{out}} \ \n_{\textrm{in}} }^{-\Delta} \nonumber \\
     & \underset{\pqty{\tau_{\textrm{out}} - \tau_{\textrm{in}}} \to \infty}{\simeq} R_0^{-2 \Delta} \exp\Bqty{- \Delta \pqty{\tau_{\textrm{out}} - \tau_{\textrm{in}}}/R_0} \nonumber \\
     &= \Anew.
\end{align}
and we introduced the symbol \(\Anew\) for future benefit.

Hence, we see that in the limit of infinite separation in cylinder time, solely the leading term in the expansion counts. Thus, the ground state energy in the cylindrical frame can be straightforwardly identified with the scaling dimension
\begin{equation}\label{eq.energyandscaling}
    \Delta = R_0 E.
\end{equation}
It is important to point out that when we mention the energy of the ground state, this discussion only holds true for the cylindrical geometry. In contrast with flat space, the cylinder frame provides an intrinsic scale to the \acrlong{cft} ---  denoted as the radius of the sphere \(R_0\) --- thus the theory has a discrete spectrum. In the flat space geometry, a \acrlong{cft} has a continuous spectrum and does not acquire a mass gap. 

A discussion similar to the one for the two-point function can be repeated, but now for three scalar primaries. Setting \(R_0 =1\) for ease of notation, a three-point function of a \acrshort{cft} on the cylinder is related to the one on flat space as  
\begin{equation}\label{eq.3pointflatcylinder}
    \expval{ \mathscr{O}_1 (\tau_{\textrm{out}}) \mathscr{O}_2 (\tau) \mathscr{O}_3 (\tau_{\textrm{in}}) }_{\textrm{cyl}} =   \expval{\mathscr{O}_1 (x_{\text{out}}) \mathscr{O}_2 (x) \mathscr{O}_3 (x_{\text{in}}) }_{\text{flat}} \, {\abs{x_{\text{out}}}}^{\Delta_1} \, {\abs{x}}^{\Delta_2} {\abs{x_{\text{in}}}}^{\Delta_3}.
\end{equation}
From \cref{eq.generalform3pointfunction} we know that
\begin{equation}
    \expval{\mathscr{O}_1 (x_{\text{out}}) \mathscr{O}_2 (x) \mathscr{O}_3 (x_{\text{in}}) }_{\text{flat}} = \frac{C_{\mathscr{O}_1\mathscr{O}_2\mathscr{O}_3}}{\abs{ x_{12}}^{\Delta_1 + \Delta_2 - \Delta_3} \abs{ x_{23}}^{\Delta_2 + \Delta_3 - \Delta_1} \abs{ x_{13}}^{\Delta_1 + \Delta_3 - \Delta_2} }, \nonumber
\end{equation}
where \(x_{ij} \equiv x_i -x_j\) and
\begin{align}\label{eq.xouxxin}
    x_1 & \equiv x_{\text{out}}, &
    x_2 & \equiv x, &
    x_3 & \equiv x_{\text{in}}.
\end{align}
Hence combining \cref{eq.3pointflatcylinder,eq.generalform3pointfunction,eq.xouxxin} we get
\begin{align}\label{eq.threepointuptoapoint}
   & \expval{\mathscr{O}_1 \pqty{\tau_{\textrm{out}}} \mathscr{O}_2 \pqty{\tau} \mathscr{O}_3 \pqty{\tau_{\textrm{in}}}}_{\textrm{cyl}} \\
   & = C_{\mathscr{O}_1\mathscr{O}_2\mathscr{O}_3} \abs{ x_{\textrm{out}}-x}^{-\pqty{\Delta_1 + \Delta_2 - \Delta_3 }}   \abs{ x - x_{\textrm{in}}}^{- \pqty{\Delta_2 + \Delta_3 - \Delta_1 }} \abs{ x_{\textrm{out}} - x_{\textrm{in}}}^{- \pqty{\Delta_1 + \Delta_3 - \Delta_2 }} {\abs{x_{\textrm{out}}}}^{\Delta_1} {\abs{x}}^{\Delta_2} {\abs{x_{\text{in}}}}^{\Delta_3}. \nonumber 
\end{align}
Again working in polar coordinates we recognize 
\begin{align}
    x_{\text{out}}& = r_{\text{out}}, & 
    x& = r, &
    x_{\text{in}} &= r_{\text{in}},
\end{align} 
so that
\begin{align}
    \abs{x_{\text{out}}-x} & = \bqty{ {r_{\text{out}}^2 + r^2 - 2  \ r_{\text{out}} \ r \ \n_{\text{out}} \  \n } }^{1/2}, \\
    \abs{x_{\text{in}}-x} & = \bqty{ {r_{\text{in}}^2 + r^2 - 2  \ r_{\text{in}} \ r \ \n_{\text{in}} \ \n }  }^{1/2}, \\
    \abs{x_{\text{out}}-x_{\text{in}}} & = \bqty{ {r_{\text{out}}^2 + r_{\text{in}}^2 - 2 \  r_{\text{out}} \ r_{\text{in}} \ \n_{\text{out}} \ \n_{\text{in}}} }^{1/2}.
\end{align}
In the limit 
\begin{align}
    x_{\text{out}}& \rightarrow \infty, & x_{\text{in}}&  \rightarrow 0,  &\longrightarrow&  &r_{\text{out}} &\rightarrow \infty, & r_{\text{in}} & \rightarrow 0,
\end{align}
we get the following simplifications
\begin{align}
    \bqty{ {r_{\text{out}}^2 + r^2 - 2  \ r_{\text{out}} \ r \ \n_{\text{out}} \  \n } }^{1/2} & \cong r_\textrm{out}, \\
    \bqty{ {r_{\text{in}}^2 + r^2 - 2  \ r_{\text{in}} \ r \ \n_{\text{in}} \ \n }  }^{1/2} & \cong r, \\
    \bqty{ {r_{\text{out}}^2 + r_{\text{in}}^2 - 2  \ r_{\text{out}} \  r_{\text{in}} \ \n_{\text{out}} \ \n_{\text{in}}}   }^{1/2} & \cong r_\textrm{out}.
\end{align}
Then \cref{eq.threepointuptoapoint} takes the form
\begin{align}
\expval{ \mathscr{O}_1 (\tau_{\text{out}}) \mathscr{O}_2 (\tau) \mathscr{O}_3 (\tau_{\text{in}}) }_{\text{cyl}} & =  C_{\mathscr{O}_1\mathscr{O}_2\mathscr{O}_3} r_{\text{out}}^{-\pqty{\Delta_1 + \Delta_2 - \Delta_3}}  r^{- \pqty{\Delta_2 + \Delta_3 - \Delta_1 }} r_{\text{out}}^{- \pqty{\Delta_1 + \Delta_3 - \Delta_2 }} {r^{\Delta_1}_{\text{out}}} r^{\Delta_2} {r^{\Delta_3}_{\text{in}}} \nonumber \\
& = C_{\mathscr{O}_1\mathscr{O}_2\mathscr{O}_3} {\pqty{\frac{r}{r_{\text{out}}}}}^{\Delta_1} {\pqty{\frac{r_{\text{in}}}{r}}}^{\Delta_3}.
\end{align}
Finally, using \cref{eq.tautransformation} we get
\begin{align}
    \expval{ \mathscr{O}_1 (\tau_{\text{out}}) \mathscr{O}_2 (\tau) \mathscr{O}_3 (\tau_{\text{in}}) }_{\text{cyl}} & = C_{\mathscr{O}_1\mathscr{O}_2\mathscr{O}_3} e^{\Delta_1 ( \tau - \tau_{\text{out}})/ R_0} e^{\Delta_3 ( \tau_{\text{in}} - \tau )/ R_0} \nonumber \\
    & = C_{\mathscr{O}_1\mathscr{O}_2\mathscr{O}_3}  e^{\tau \pqty{\Delta_1 - \Delta_3 } /R_0 } e^{\tau_{\text{in}} \Delta_3 /R_0} e^{ - \tau_{\text{out}} \Delta_1 /R_0}  e^{ \tau_{\text{out}} \Delta_3 /R_0} e^{ - \tau_{\text{out}} \Delta_3 /R_0} \nonumber \\
    & = C_{\mathscr{O}_1\mathscr{O}_2\mathscr{O}_3} \, e^{- \Delta_3 (\tau_{\text{out}} -\tau_{\text{in}})  /R_0}  \, e^{   \pqty{\Delta_1 - \Delta_3 }\pqty{\tau - \tau_{\text{out}}} /R_0 }.\label{eq.formof3pointfunction}
\end{align}
It is obvious that in the infinite separation limit the scaling of the middle \( \mathscr{O}_2\) operator in \cref{eq.formof3pointfunction} does not affect the outcome of the computation. 

Furthermore, a special case is for \(\Delta_3 = \Delta_1 = \Delta\) since in this case, the dependence on \(\tau\) drops completely and

\begin{equation}\label{eq.threepointnotau}
    \expval{ \mathscr{O}^{\Delta}_1 (\tau_{\text{out}}) \mathscr{O}_2 (\tau) \mathscr{O}^{\Delta}_3 (\tau_{\text{in}}) }_{\text{cyl}} = C_{\mathscr{O}_1\mathscr{O}_2\mathscr{O}_3} \, e^{- \Delta(\tau_{\text{out}} -\tau_{\text{in}})  /R_0}.
\end{equation}
A generalisation to spinning correlators can be found in \cite{Rychkov_2017} and in \Cref{sec:constraints}. 

The derivations at hand will prove useful in \cref{Chapter3}, when we compute the three-point functions of scalar primary operators of large charge, and we want to study the \acrshort{cft} data.

\section{Large Charge}\label{sec.largecharge}

Having studied important aspects of \acrlong{cfts} in \cref{sec.CFT}, we saw in detail that the constraints that conformal symmetry imposes on the system restrict all the information that we need to know thoroughly to determine every prediction it can make down to two sets of numbers, known as the \textit{\acrshort{cft} data}: the scaling dimension \(\Delta\) of every primary operator of the theory, and the \acrlong{ope} coefficient \(C_{ \mathscr{O}_1 \mathscr{O}_2 \mathscr{O}_3}\).  Knowledge of this data can help us compute theoretical quantities that are physical observables, and hence we can compare their values with the ones coming from experiments. Moreover, we can use the \acrshort{cft} data to categorize \acrshort{qft} systems upon having or not critical phases that belong to the same universality class, as in the examples of \cref{sec.CFT}.

Sadly, computing the \acrshort{cft} data is generally demanding and usually additional simplifying assumptions have to be imposed. The problem lies in the fact that not all \acrlong{cfts} are weakly coupled, but in most cases they are interacting and also not easily accessible by exact methodologies similar to the \(3\)-dimensional Ising model. Furthermore, innately \acrshort{cfts} lack an intrinsic scale and usual perturbative approaches are not applicable but non-perturbative methods are necessary. This is the reason we should examine in depth any technique that aims to deal with this issue. Some of the most prevalent methods are
 \begin{enumerate}[label=\roman*,ref=\roman*,left= 0pt]
    \item \underline{\textbf{\emph{Monte Carlo methods}}}
    
    This classical method is a general class of numerical algorithms that depends on recurring random sampling to attain numeric results. In the case of \acrshort{cfts}, it is more suitable for spin lattice systems similar to the Ising model and similar. 
    \item \underline{\textbf{\emph{Conformal bootstrap}}}\label{m2}
    
    The method is a non-perturbative approach with the aim of constraining and numerically solving \acrshort{cfts}. Dissimilar to more conventional approaches of \acrshort{qfts}, this procedure does not make use of the system's Lagrangian, and it rather works with the generic axiomatic variables like the scaling dimension of the local operators and their \acrshort{ope} coefficients. Then it tries to solve a group of self-consistent restrictions of these variables to gain numerical restraints, with the final goal to increase the accuracy of these restraints. The modern manifestation of this approach emerged after the paper by Rattazzi et al.~\cite{RiccardoRattazzi_2008}. Tested to the Ising model \acrshort{cft}, it gave the most accurate predictions~\cite{PhysRevD.86.025022,simmons2015semidefinite}.
    \item \underline{\textbf{\emph{Perturbative approaches}}}
    
    There is more than one perturbative approach, so it is not exactly a unique methodology. The idea is that an additional controlling parameter exists, like \(N\) in the large-\(N\) expansion~\cite{Moshe_2003} or \(\varepsilon\) in the small \(\varepsilon\) approximation~\cite{WILSON197475}. Then the conformal data of any interacting \acrshort{cft} can be calculated in terms of the conformal data of the free \acrshort{cft} as an asymptotic expansion in the controlling parameter.
    \item \underline{\textbf{\emph{Large quantum number}}}
    
    It has been observed that the conformal data of operators related to \acrshort{cft} sectors of large quantum numbers like large spin \(J\)~\cite{komargodski2013convexity,fitzpatrick2013analytic} exhibit considerable simplifications. Very recently, it was appreciated that similar simplifications arise in sectors of large global charge. This new methodology is the \acrlong{lce} that we will study in detail.
\end{enumerate}
The idea behind it is to study sectors of \acrlong{cfts} that are characterised by large quantum numbers under the global symmetries. In this limit, \acrshort{cfts}, even if they initially were in a strong coupling regime, can now be written in terms of an \acrlong{eft} in the spirit of \cref{sec.Wilsonianrg}. This gives rise to a new class within the \acrshort{cfts} as now we are interested in finding global properties that are true for entire families of \acrshort{cfts} that exhibit an identical symmetry structure. For example, some common \acrshort{eft}s that appear are the \textit{bosonic conformal superfluid} and the \textit{Fermi sphere}, condensed matter systems characterised by a high particle density, making the study of such systems a cross-disciplinary matter.

The first paper concerning \acrshort{lce} came out on \(2015\) by Hellerman et al.~\cite{hellerman2015cft} and it concentrated on \(3\)-dimensional \acrlong{cfts} with an underlying \(O(2)\) global symmetry. There, the authors demonstrated that under certain conditions, it is possible to write an effective action in terms of the Goldstone modes only. Moreover, they showed that it is possible to compute the scaling dimension of the lowest scalar primary operator of large charge \(\Opp\) using radial quantization and \cref{eq.energyandscaling} and the outcome is a universal scaling law which goes like 
\begin{equation}\label{eq.largechargescale}
    \Delta_\Qp \sim \Qp^{3/2} + \dots,
\end{equation}
where \(\Delta_\mathcal{Q}\) is the scaling dimension of \(\Opp\). Furthermore, there are subleading terms that come in an expansion of inverse powers of the charge \(\mathcal{Q}\). Most importantly, the only term that scales as \(\mathcal{Q}^0\) is the \textit{Casimir energy} of the quantum fluctuations, since there is no classical term with such scaling. This number is thus a clear prediction of the theory \footnote{More details on that will be given in \Cref{Chapter3}.}. The same \acrshort{eft}, known as conformal superfluid, has been separately shown to be consistently derived by the application of the \acrfull{ccwz} construction in \cite{monin2017semiclassics}.

In a very interesting turn of events, the predictions of \cite{hellerman2015cft} have been confirmed independently using Monte Carlo simulations in \cite{banerjee2018conformal}, employing small-\(\varepsilon\) approximation in \cite{badel2019epsilon} and via large-\(N\) expansion in \cite{alvarez2019large}. 

In the original paper of Hellerman et al. \cite{hellerman2015cft} there have been worth mentioning predictions about the spectrum of large-charge operators that also carry a small spin \(J\). In the validity of the \acrlong{eft}, these operators are related to \textit{superfluid phonons}, Goldstone bosons that arise from the superfluid's \acrlong{ssb} pattern \footnote{A brief analysis on Goldstone bosons and \acrshort{ssb} is given in \cref{sec.Goldstonetheorem}.}. An overview of the story thus far with an extensive analysis of the global properties of three and four-point correlation functions of large-charge phonon primaries can be found in \cite{dondi2022spinning} which is also the subject of \Cref{Chapter3}. This work completes a gap in the literature of earlier works that aimed to compute the relevant CFT data~\cite{monin2017semiclassics,jafferis2018conformal,komargodski2021spontaneously,cuomo2021ope,cuomo2021note,Cuomo:2020thesis}.

\begin{figure}[ht]
    \centering
    \includegraphics[scale=0.5]{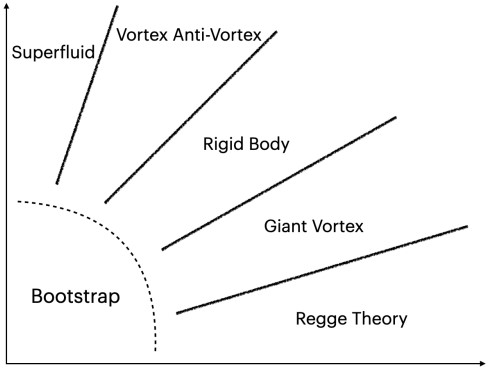}
    \caption{Phases for the ground state on the cylinder.\newline
Figure adopted from \cite{cuomo2023giant}.}
    \label{fig:phasesforgroundstate}
\end{figure}

We anticipate that the superfluid phonon \acrshort{eft} is going to cease working by the time that the spin \(J\) becomes commensurate with the charge \(\mathcal{Q}\), and a new \acrshort{eft} will likely rise. The limitations of the semiclassical methodology and the validity of the phonon primary description have been investigated by Badel et al.~\cite{badel2022identifying} using the small-\(\varepsilon\) expansion. A first endeavour to classify the phases of the system based on the spin \(J\) and the charge \(\mathcal{Q}\) was made by Cuomo et al.~\cite{cuomo2018rotating}. From then on, there have been many developments~\cite{cuomo2020superfluids,cuomo2021large,cuomo2023giant} and it has been shown that the phase diagram for the system is richer than originally anticipated and is depicted in \Cref{fig:phasesforgroundstate}.
There are two extremal limits, the one of large charge \(\mathcal{Q}\) and small spin \(J\) in which the ground state is in the conformal superfluid phase and where the scaling dimension of the lowest operator goes like \cref{eq.largechargescale}, and the one of large spin \(J\) and relatively small charge \(\mathcal{Q}\), which is the large spin \acrshort{eft} of~\cite{alday2007comments,komargodski2013convexity,fitzpatrick2013analytic}, and where we find a Regge like behaviour~\cite{regge1959introduction} with scaling dimension
\begin{equation}
    \Delta_J = J + \dots,
\end{equation}
Already from Cuomo et al.~\cite{cuomo2018rotating} it has been theorised that when \(J \gg \mathcal{Q}^{1/2} \) the system is expressed by one single vortex anti-vortex pair and as the spin continues to increase we pass to the next phases. A new giant vortex phase can be found in \cite{cuomo2023giant} along with an overall nice analysis of the phases. Moreover, it should be noted that the small charge \(\mathcal{Q}\) and small spin \(J\) regime is reachable through the numerical bootstrap technique that was discussed in \(\pqty{\ref{m2}}\). As a side note, the vortex phases have also been studied for parity-violating \acrshort{cfts} in~\cite{cuomo2021large}.

The story gets more complicated when we try to study a large-charge sector with a global non-Abelian symmetry group. The spectrum of the \acrlong{eft} is much richer than before as it has been shown in~\cite{alvarez2017compensating} and in addition to the superfluid phonons of the Abelian \(O(2)\) sector there are also new non-relativistic Goldstone bosons of type II \footnote{The difference between the usual type I Goldstones and the newly found type II will be briefly discussed in \cref{sec.Goldstonetheorem}.}. These type II Goldstones are likely to characterise new primaries operators of large charge that live in several representations of the global symmetry group, but concrete examples are still lacking in the literature. Nonetheless, what is expected is that an analogous situation to the case of the vortices will take place, \emph{i.e.} when a critical number of type II Goldstones appears in the system, new phases will emerge. Up to now, there are no explicit suggestions on how these phases will look like except that they have to be in accordance with ground states that are spatially inhomogeneous as shown in~\cite{hellerman2019note,hellerman2021observables}. It is clear that in the light of a recent work~\cite{banerjee2022subleading} that applies Monte Carlo methods to compute the scaling dimensions of local large charge primaries at the \(O(4)\) \acrshort{wf} fixed point, more work is needed in this direction.

The whole story may differ when we want to examine fermionic \acrshort{cfts}, since it is not a priori clear if they can be described by a superfluid \acrshort{eft} in the spirit of~\cite{hellerman2015cft,monin2017semiclassics}. Although this is the main subject of \Cref{Chapter5} it can be said that there are models like the free fermion found in~\cite{komargodski2021spontaneously} or the \acrshort{gn} model of~\cite{dondi2022fermionic} that the large-charge ground state is that of a \textit{Fermi sphere}, while others exhibit what is known as \acrfull{bcs} \textit{superconductivity}~\cite{cooper1956bound,bardeen1957microscopic} and can still be expressed by a superfluid \acrshort{eft}. Hence, it is easy to imagine that for interacting systems possessing a large charge Fermi sphere ground state, there should exist a \textit{Landau Fermi liquid} \acrshort{eft} in the spirit of Polchinski~\cite{polchinski1992effective} and Shankar~\cite{shankar1994renormalization} that may introduce fresh universal properties like the ones that appeared in the superfluid case. Any proposed Fermi liquid \acrlong{eft} should be simultaneously compatible with conformal symmetries and at the same time exhibit a \acrshort{bcs} instability to account for these fermionic models that acquire a superfluid ground state. It has also been theorised that an entirely new large-charge class may emerge, which is the \textit{non-Fermi liquid} phase. Although this behaviour is familiar in condensed matter theory --- \emph{e.g.} see~\cite{cooper2009anomalous,Sachdev2011} --- it is not presently clear if it can exist in a \acrshort{cft} sector of large global charge. 

As a last note, just a while ago it has been found by Dondi et al.~\cite{dondi2021resurgence} that \acrshort{lce} are, in fact, \textit{asymptotic series} and not \textit{convergent} ones. Even though this is the topic of \Cref{Chapter4} the punchline is that the \acrshort{lce} of the \(O\pqty{N}\) vector model examined in the above paper exhibits a double factorial growth, which is stronger than the one of classical \acrshort{qft}. Hence, non-perturbative corrections play an important role when we try to sum up the series to extrapolate the large-charge result to the small-charge limit. The appropriate mathematical framework that describes the relation between asymptotic series and the non perturbative contributions is \textit{Resurgence}~ \cite{ecalle1981fonctions}.

So, to recapitulate, there are three known large-charge types of behaviour for \acrshort{cfts} in \(d \geq 3\) spacetime dimensions \footnote{The \(d=2\) case is special, and a brief note can be found at the end of the section.} where only large charge is assumed, particularly:
\begin{enumerate}[label=C.\roman*,ref=C.\roman*]
    \item \label{C.1} \textbf{Moduli class:} This is the class least examined in this thesis, except for the free scalar \acrshort{cft}. In this class, the charged operator does not correspond to a meaningful state on the cylinder and the scaling dimension of the lowest charged operator goes like \(\Delta \sim \Qp\) instead of the scaling law of \cref{eq.largechargescale}. There is no \acrshort{eft} description and other methods are required. Notable examples are \(\mathcal{N}=2\) SYM models like~\cite{hellerman2017operator,hellerman2017large,hellerman2019universal,bourget2018limit,grassi2021extremal}.
    \item \label{C.2} \textbf{Superfluid class:} In this class the large charge ground state on the cylinder corresponds to the one about the homogeneous conformal superfluid. These are systems that have an \(O(2)\) internal symmetry that may be a subgroup of a bigger global symmetry group like the \(O\pqty{N}\). Several confirmations of this statement using the small-\(\varepsilon\) expansion can be found in~\cite{antipin2020charging,antipin2020chargingg,antipin2021more}. This class also contains these fermionic models that exhibit \acrshort{bcs} superconductivity, like the models examined in~\cite{chodos2000two,klimenko2012superconducting,zhukovsky2017superconductivity}. In the large charge literature there are only two known examples of fermionic models in the conformal superfluid phase, Gross-Neveu-Yukawa models examined by Antipin et al.~\cite{antipin2022yukawa} using the small-\(\varepsilon\) expansion, and pure fermionic models that were studied by Dondi et al.~\cite{dondi2022fermionic} using the large-\(N\) approximation. Finally, some monopole models with large magnetic charge enjoy some properties of conformal superfluids~\cite{cuomo2021large,de2018large}. 
    \item \label{C.3} \textbf{Fermi sphere class:} The last class is the one that the large charge ground state on the cylinder is that of a fully filled Fermi sphere. This class is apparent in certain fermionic models, and thus far the only known examples are the free fermion of Komargodski et al.~\cite{komargodski2021spontaneously} and \acrshort{gn} type theories that do not support a \acrshort{bcs} superconducting phase~\cite{dondi2022fermionic}.
\end{enumerate}
A special case is \(d=2\), where the \acrlong{lce} has been studied only in~\cite{komargodski2021spontaneously,araujo20222d}. In the work of Komargodski et al.~\cite{komargodski2021spontaneously}, it was demonstrated that the \(U\pqty{1}\) large charge sector decouples from the rest of the system, and thus it cannot control the low-energy dynamics. This is the reason it is not possible to write a low energy \acrshort{eft} as in the case of \(d \geq 3\) and the approach fails. Nonetheless, in~\cite{araujo20222d} it was shown that even if the \acrshort{lce} on its own is not enough, it is fruitful when used in parallel with another controlling parameter and examined in the double scaling limit of the theory.

In this thesis, the phases that appear are the conformal superfluid of \cref{C.2} and the Fermi sphere of \cref{C.3}. So for the rest of the chapter, key ingredients to better understand them, like \acrlong{ssb} and the Goldstone theorem, are laid down. 


\subsection{SSB and Goldstone theorem}\label{sec.Goldstonetheorem}

As stated in \cref{sec.largecharge} the \acrlong{lce} is restricted to systems with global symmetries, but through them, it systematically permits to examine theories that would be inaccessible otherwise. The idea of fixing the charge and restricting the system to the large-charge Hilbert space sector results in a ground state that breaks boost invariance, and the combination of simultaneously breaking boost and charge invariance gives rise to a condensate and a number of Goldstone mode. 

Actually, in~\cite{alvarez2017compensating} it was established that the large-charge domain of the non-Abelian theory contains at the same time a combination of the relativistic and of the non-relativistic configuration of the Goldstone's theorem. The authors show that the spectrum contains one type I or relativistic Goldstone boson which has a dispersion relation that goes like \(\omega \sim \frac{k}{\sqrt{d-1}}\) and also \(N-1\) type II or non-relativistic Goldstones that have a quadratic dispersion \(\omega \sim k^2\). This counting of Goldstones is a verification of the analysis of Nielsen and Chadha~\cite{nielsen1976count}. 

Hence, we see that the breaking of Lorentz invariance by fixing the charge gives rise to a profuse phenomenology, that includes \acrshort{eft}s formulated in the non-relativistic limit, a work initiated by Leutwyler~\cite{leutwyler1994nonrelativistic} back in \(1994\) and fully solved only a few years ago by Watanabe and Murayama~\cite{watanabe2014effective}. 

The subject of how to correctly count and characterise Goldstone bosons both in the relativistic and the non-relativistic limit is a vast and complicated one, as many results have appeared scattered in the literature. The best attempt to gather them up and present them in a clear and comprehensible way is in~\cite{Gaume:2020bmp}.

Hence, before we move to the more specialised parts of the thesis, we will briefly review the Goldstone theorem and the \acrshort{ssb} using the simplest possible derivation, that of the complex scalar field that we first came upon in \cref{eq.U(1)system}, but with a non-trivial ground state. 

\subsubsection{Spontaneous Symmetry Breaking and Gapless States}\label{sec.ssbandgaplessstates}

As we saw in \cref{sec.classicalfield}, for a system expressed by the Lagrangian \(\mathscr{L}\pqty{\phi, \del\phi}\) Noether's theorem dictates that to every continuous symmetry corresponds the conserved current \(\mathcal{J}^\mu\) of \cref{eq.currentconservation}. From \cref{eq.chargeconservationproof} this implies that the charge \(\mathcal{Q}\) of \cref{eq.charge} is time independent
\begin{equation*}
    \dv{t}\mathcal{Q}=0,
\end{equation*}
assuming of course that no sources are present at the boundaries at infinity.

Hence, given some continuous symmetry, a quantum state of the theory has to transform appropriately under the aforesaid symmetry as
\begin{equation}\label{eq.symmetrytransformationofstates}
    \ket{\psi} \longrightarrow e^{i a \mathscr{Q}} \ket{\psi},
\end{equation}
with \(a\) being a real, continuous, arbitrary, not space—dependent parameter of the symmetry. Specifically, in the case that the ground state \(\ket{0}\) of the theory is invariant under the aforementioned symmetry, using \cref{eq.symmetrytransformationofstates} corresponds to 
\begin{equation}
    \ket{0} \longrightarrow e^{i a \mathscr{Q}} \ket{0} = \ket{0},
\end{equation}
which entails that
\begin{equation}\label{eq.groundstateannihilated}
    \mathscr{Q}\ket{0} =0.
\end{equation}
Put differently, if acting with the charge operator \(\mathscr{Q}\) on the ground state, the latter is annihilated then there exists a continuous symmetry related to the said charge that the ground state is invariant under. This is the traditional definition of symmetries in quantum mechanics. 

However, if by any chance the charge does not annihilate the ground state of the theory, \emph{i.e.}
\begin{equation}
    \mathscr{Q}\ket{0} \neq 0,
\end{equation}
it signifies that
\begin{equation}\label{eq.statesa}
    \ket{0} \longrightarrow e^{i a \mathscr{Q}} \ket{0} \equiv \ket{a} \neq \ket{0},
\end{equation}
where the states \(\ket{a}\) are described using the real continuous parameter \(a\) of the initial symmetry transformation and are related to the ground state by it. This is the definition of a broken symmetry. 

The most interesting case arises though when the following equations  
\begin{align}\label{eq.ssbconditions}
    \dv{t}\mathcal{Q}&=0, & \mathscr{Q}\ket{0} &\neq 0,
\end{align}
are concurrently satisfied. This situation represents the forenamed \acrfull{ssb}. This is the circumstance that the charge is still time independent and conserved, yet the ground state of the theory is no longer symmetry invariant. To recapitulate
\begin{align}
   & \pqty{\dv{t}\mathcal{Q}=0, \qquad \mathscr{Q}\ket{0} \neq 0} &\Longrightarrow& &\textrm{\acrshort{ssb}}.
\end{align}
Considering that we retain current conservation, it can be shown using the results of~\cite{federbush1960uniqueness} that the conserved charge commutes with the Hamiltonian \(H\) of the theory
\begin{equation}
    \bqty{H, \mathscr{Q}} =0.
\end{equation}
At this point we want to see how the Hamiltonian acts on the states \(\ket{a}\) given in \cref{eq.statesa}. We get that
\begin{align}
    H \ket{a} & = H e^{i a \mathscr{Q}} \ket{0} = e^{i a \mathscr{Q}} H \ket{0} \nonumber \\
    & = E_0 e^{i a \mathscr{Q}} \ket{0} \nonumber \\
    & = E_0 \ket{a},
\end{align}
where \(E_0\) is the energy of the ground state, and in the first line we used the fact that the Hamiltonian commutes with the charge. Therefore, we perceive that the concurrent satisfaction of \cref{eq.ssbconditions} ensues the existence of a continuous family of states \(\ket{a}\) that are degenerate since they have the same energy \(E_0\) as the ground state of the theory and traversing between them extracts no additional toll. 

In the case of a relativistic \acrshort{qft}, as we will see in \cref{sec.U(1)ssb} the Goldstone modes are related to massless particles. But before seeing an explicit example, let us briefly turn to the general theory. In \(d \geq 3\) spacetime dimensions \footnote{The case of \(d=2\) is peculiar since there can be no \acrshort{ssb} of continuous symmetries. This statement has been proved in~\cite{hohenberg1967existence,mermin1966absence} in the statistical mechanical context, while in the context of \acrshort{qfts} it has been proven in~\cite{coleman1973there}. Also, the case of \(d=3\) is atypical, since in such systems in most cases \acrshort{ssb} is possible at zero temperature but not in finite one.}, there are two possible modes that a continuous symmetry can appear, the \textit{Wigner-Weyl phase} where the symmetries are unbroken and the Nambu-Goldstone phase:
\begin{itemize}[left= 0pt]
    \item \underline{\textbf{\emph{Wigner-Weyl phase}}}

    In this phase, the ground state of the theory \(\ket{0}\) is invariant under the symmetries of the system. This is the case that the conserved charge operators commute with the Hamiltonian and also annihilate the state \(\ket{0}\) as in \cref{eq.groundstateannihilated} so that
    \begin{align}
        \bqty{H, \mathscr{Q}} & = 0, & \mathscr{Q} \ket{0} & = 0. 
    \end{align}
    Furthermore, the conserved current operators \(\mathcal{J}^\mu\) also annihilate the above-mentioned state so that 
    \begin{equation}
        \mathcal{J}^\mu \ket{0} =0. 
    \end{equation}
    Any excited states are generated as in \cref{eq.genericstate} by adding multiple particles to the ground state 
\begin{equation*}
    \ket{k_1,k_2,\dots,k_n} \coloneqq a^\dagger({k}_1) a^\dagger({k}_2) \dots a^\dagger({k}_n) \ket{0}.
\end{equation*}
Hence, in the case of a relativistic system and for a symmetry that exists in the Wigner phase, the particles develop multiplets of the said symmetry and every particle belonging to the same multiplet has the same mass \(m\)\footnote{In the non-relativistic case, instead of the mass, particles share a similar dispersion relation \(\omega\).}.
    \item \underline{\textbf{\emph{Nambu-Goldstone mode}}}

On the other hand, in this mode, as we saw, the ground state \(\ket{0}\) is not symmetry invariant and alternatively, there exists a continuous group of degenerate ground states related by the symmetry in question. On this account, the charge or the current operators do not annihilate the ground state of the theory
\begin{align}
    \mathscr{Q} \ket{0}& \neq 0, & \mathcal{J}^\mu \ket{0} & \neq 0,
\end{align}
and in this spirit the particles do not form multiplets of the said symmetry. In lieu, the \textit{Goldstone theorem} dictates other exciting consequences for this mode:
\begin{enumerate}
    \item In a relativistic system, every spontaneously broken generator of the system is related to a massless scalar field, the \textit{relativistic Goldstone boson}. 
    
    For non-relativistic systems, the counting is trickier, since there is not a one-to-one correspondence between broken generators and the non-relativistic Goldstone bosons. This is the work of Nielsen and Chadha~\cite{nielsen1976count}, with the counting rule appearing in~\cite{watanabe2011number} and fully proven in~\cite{hidaka2013counting, watanabe2014effective}. Therefore, assuming that we have a system with \(m\) spontaneously broken generators, \(n_I\) relativistic or type I Goldstones modes and \(n_{II}\) non-relativistic or type II modes, the correct counting law is
    \begin{equation}
        n_I + 2 n_{II} \geq m.
    \end{equation}
    \item The currents \(\mathcal{J}_a^\mu\) of species \(a\) of the spontaneously broken symmetries generate Goldstone bosons when acting on the ground state 
    \begin{equation}
        \mathcal{J}_a^\mu\pqty{x} \ket{0} \propto \ket{a\pqty{x}},
    \end{equation}
    as we also saw in \cref{eq.statesa}.
\end{enumerate}
\end{itemize}

It is interesting to note that given a continuous group of symmetries \(\mathbb{G}\), it is possible to partially break it down to a subgroup \(\mathbb{H} 	\subset \mathbb{G}\). Now assuming that the action \(S\) of the theory and the Hamiltonian \(H\) are invariant under every symmetry \(\in \mathbb{G}\) but on the other hand, the ground state \(\ket{0}\) is invariant only under the symmetries that belong \(\in \mathbb{H}\), then symmetries in the proper subgroup \(\mathbb{H}\) exist in the Wigner-Weyl phase, even though the rest of the symmetries  \(\in \mathbb{G}/\mathbb{H}\) appear in the Nambu-Goldstone phase.

So having reviewed the Wigner-Weyl phase and the Nambu-Goldstone phase, it is important to note that the \acrlong{lce} lies between the two phases, in the sense that the entire theory is still invariant under a family of global symmetries and so it belongs to the Wigner phase, but in the large-charge sector the ground state satisfies the Nambu-Goldstone condition 
\begin{equation*}
    \mathscr{Q} \ket{0} \neq 0,
\end{equation*}
and can thus be expressed by a mixture of the relativistic and non-relativistic modes of the \acrshort{ssb}.

\subsubsection{An example: \(U\pqty{1}\) \acrlong{ssb} }\label{sec.U(1)ssb}

To illustrate the \acrshort{ssb} process of a global symmetry and the appearance of massless modes, we will examine the easiest possible example, that of a complex scalar field of \cref{eq.U(1)system}. The Lagrangian of the system at hand comes with a quartic interaction term as
\begin{align}\label{eq.U(1)Lagrangianexample}
    \mathscr{L} &= T -V \nonumber \\
    & = - \del_\mu \varphi^* \del^\mu \varphi - \mu^2 \varphi^* \varphi - \frac{g}{2} \pqty{\varphi^* \varphi}^2,
\end{align}
where \(T\) is the kinetic term and \(V\) is the classical potential.

As we have seen in \cref{sec.classicalfield}, the Lagrangian possesses a \(U\pqty{1}\) symmetry
\begin{align*}
    \varphi(x) & \to \varphi' (x) = e^{ia} \varphi (x), & \varphi^* (x) & \to {\varphi^*}'(x)  = e^{-ia} \varphi^* (x),
\end{align*}
as in \cref{eq.U(1)symmetry} which holds true even in the presence of the interaction term. As shown in \cref{sec.classicalfield} the aforesaid symmetry is isomorphic to complex plane rotations given by \cref{eq.O2toU1transformationfields} as
\begin{align*}
    \varphi & = \frac{1}{\sqrt{2}} \left( \phi_1 + i \phi_2 \right), & \varphi^{*} & = \frac{1}{\sqrt{2}} \left( \phi_1 - i \phi_2 \right),
\end{align*}
so that \( U\pqty{1} \simeq O\pqty{2}\). 

The classical potential \(V\) of the theory is 
\begin{equation}\label{eq.classicalpotentialterm}
    V = \mu^2 \varphi^* \varphi + \frac{g}{2} \pqty{\varphi^* \varphi}^2,
\end{equation}
and has to be bounded from below, and hence the coupling constant \(g\) has to be positive definite. But the parameter \(\mu^2\) has no such restriction and can acquire either sign. The theory has an entirely different behaviour depending on the sign of \(\mu\) hence it would be beneficial to examine both regimes.

\(\bullet\) For \(\mu^2 > 0\) we are in the usual case where there exists a unique ground state \(\ket{0}\) and the field \(\varphi\) has a zero \acrfull{vev}
\begin{equation}
   \mel{0}{\varphi}{0} = \expval{\varphi} =0,
\end{equation}
and the potential \(V\) is minimized at 
\begin{equation}
    \pqty{\varphi^* \varphi}_0 = 0.
\end{equation}
The system is invariant under the \(U(1)\) phase symmetry and the excited states of the system are made-up from a family of particles and/or antiparticles that are characterised by the same mass \(m^2 = \mu^2\).

\(\bullet\) On the other hand, for \(\mu^2 < 0\) things change considerably. The minimum of the potential \(V\) now lies in
\begin{equation}\label{eq.ssbminima}
    \pqty{\varphi^* \varphi}_0 = -\frac{\mu^2}{g} \equiv v^2 >0,
\end{equation}
where \(v^2\) is the \acrlong{vev} of the operator \(\varphi^* \varphi\), \emph{i.e.} 
\begin{equation}
    \mel{0}{\varphi^* \varphi}{0} = v^2.
\end{equation}
Thus we have that
\begin{equation}\label{eq.fakemass}
    \mu^2 = - g v^2,
\end{equation}
and using \cref{eq.classicalpotentialterm,eq.fakemass} we can rewrite the classical potential \(V\) as
\begin{align}
    V &= - g v^2 \varphi^* \varphi + \frac{g}{2} \pqty{\varphi^* \varphi}^2 \nonumber \\
    & = \frac{g}{2} \pqty{\varphi^* \varphi - v^2}^2 + \textrm{const}.
\end{align}
\begin{figure}[ht]
    \centering
    \includegraphics[scale=0.30]{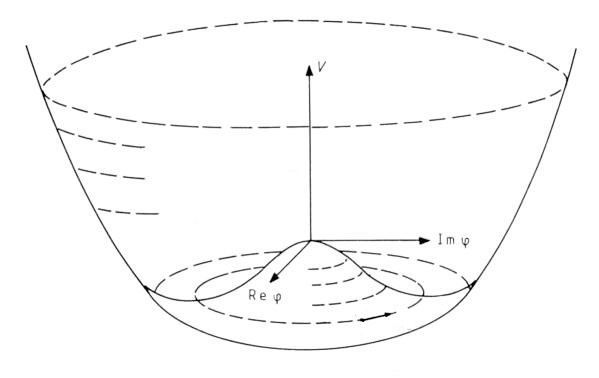}
    \caption{The potential \(V\) of the system exhibits a local maximum at \(\phi =0\) and a circle of degenerate local minima at \(\pqty{\varphi^* \varphi}_0 = v^2\).}
    \label{fig.ringofmimima}
\end{figure}
Now we can analyse the behaviour of \(V\) and we see that instead of a local minimum, the system displays a local maximum at \(\varphi =0\) which is the \(U\pqty{1}\) symmetric point. On the other hand, the system exhibits a circle of degenerate minima that can be found for 
\begin{equation}
    \varphi = v \times e^{i \theta},
\end{equation}
where \(\theta\) is a \(2\pi\)-periodic phase, and we assume that there is no conical singularity present. We can see that none of the minima is invariant under the global \(U\pqty{1}\) symmetry, but they are rather connected by it. 

Therefore, in a semiclassical approximation in perturbation theory the system does not display a unique ground state, but instead a whole family of degenerate ground states that are continuously interconnected by the phase as in \Cref{fig.ringofmimima}.  

\begin{figure}[ht]
    \centering
\includegraphics[scale=0.4]{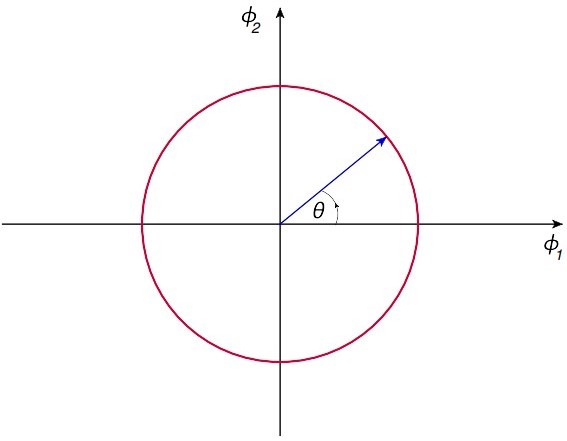}
    \caption{The minima of \(V\) for \(\mu^2 <0\) as seen in the complex \( \pqty{\phi_1, \phi_2}\) plane. The blue line corresponds to the fixed radius \(v\).}
    \label{fig:pji1phi2plane}
\end{figure}

Instead of the above analysis, we can use \cref{eq.O2toU1transformationfields} and project the system into the complex \( \pqty{\phi_1, \phi_2}\) plane. The locus points of the system's minima form a circle with a fixed real radius \(v\) that using \cref{eq.ssbminima} is found to be
\begin{equation}
    \frac{\phi_1^2 + \phi_2^2}{2} = v^2,
\end{equation}
and an undetermined phase \(\theta\) as in \Cref{fig:pji1phi2plane} that needs to be fixed arbitrarily for a ground state to be chosen. 
There is therefore a choice to be made for the phase and some explicit possibilities are
\begin{align}
    \expval{\phi_1} &= \sqrt{2} v, & \expval{\phi_2} & = 0, \label{eq.easiestchoice} \\
    \expval{\phi_1} &=  v, & \expval{\phi_2} & = v, \\
& \vdots & &\vdots \nonumber \\
\expval{\phi_1} &= 0,  & \expval{\phi_2} & = \sqrt{2} v. 
\end{align}
It is clear that any choice of the phase is possible, and it is exactly this choice that embodies the \acrshort{ssb}.

All we have to do now is to choose a vacuum state and examine the spectrum of the system at a semiclassical level. One of the most convenient but nonetheless equivalent choices is to pick \cref{eq.easiestchoice} and from \cref{eq.O2toU1transformationfields} this corresponds to a complex field \(\varphi\) having a real and positive \acrshort{vev}
\begin{equation}
    \expval{\varphi} = + v.
\end{equation}
At this point, we shift \(\varphi\) by its \acrshort{vev} and expand it into real and imaginary parts around the chosen vacuum state as 
\begin{equation}\label{eq.vevplusexpansion}
    \varphi\pqty{x} = v + \frac{\sigma\pqty{x} + i \pi \pqty{x}}{\sqrt{2}},
\end{equation}
where in the above equation \(\sigma\pqty{x}\) and \(\pi \pqty{x}\) are real scalar fields that satisfy
\begin{align}
    \mel{0}{\sigma\pqty{x}}{0} & =0, & \mel{0}{\pi \pqty{x}}{0}& =0.
\end{align}
The whole choice coincides with 
\begin{align}
    \phi_1 &= v + \frac{\sigma\pqty{x}}{\sqrt{2}}, & \phi_2 & = \frac{\pi \pqty{x}}{\sqrt{2}}.
\end{align}
Now we can use \cref{eq.vevplusexpansion} to compute the potential \(V\) and the kinetic terms \(T\) regarding \(\sigma\) and \(\pi\). First we need to compute the quantity \( \varphi^* \varphi - v^2\) as
\begin{align}
     \varphi^* \varphi - v^2 & = v^2 + \sqrt{2} v \sigma + \frac{1}{2} \sigma^2 + \frac{1}{2} \pi^2 - v^2 \nonumber \\
     & = \sqrt{2} v \sigma + \frac{1}{2} \sigma^2 + \frac{1}{2} \pi^2,
\end{align}
so that \(V\) is 
\begin{align}
    V &= \frac{g}{2} \pqty{\varphi^* \varphi - v^2}^2 \nonumber \\
    &= gv^2 \sigma^2 + \frac{gv}{\sqrt{2}} \pqty{\sigma^3 + \sigma \pi^2} + \frac{g}{8} \pqty{\sigma^2 + \pi^2}^2. 
\end{align}
Meanwhile, the kinetic terms can be directly computed using \cref{eq.vevplusexpansion} by bearing in mind that \(v\) is a real number and \(\sigma\) and \(\pi\) are real fields as
\begin{align}
    \del_\mu \varphi^* \del^\mu \varphi = \frac{1}{2} \pqty{\del_\mu \sigma}^2 + \frac{1}{2} \pqty{\del_\mu \pi}^2.
\end{align}
Therefore now we can express the Lagrangian of \cref{eq.U(1)Lagrangianexample} in terms of \(\sigma\) and \(\pi\) 
\begin{equation}\label{eq.ssblagrangian}
    \mathscr{L} = -\frac{1}{2} \pqty{\del_\mu \sigma}^2 - \frac{1}{2} \pqty{\del_\mu \pi}^2 -gv^2 \sigma^2 + \order{\pqty{\sigma,\pi}^3},
\end{equation}
where we neglected the cubic and the quartic order of the Lagrangian. 

From \cref{eq.ssblagrangian} we observe that when we expand around \cref{eq.vevplusexpansion} we end up with a theory where the \(\sigma\) field has acquired a mass \(m_{\sigma}\) while the other field \(\pi\) remains massless. That is
\begin{align}
    m_{\sigma}^2& = 2 g v^2, & m_{\pi} & =0.
\end{align}
A useful note is that  \(\sigma\) and \(\pi\) are independent since the phase symmetry does not link them. 

Beyond the semiclassical approximation, in the full quantum analysis to every order in perturbation theory, the field \(\pi\) remains precisely massless. This is the \textit{gapless Goldstone boson} promised by the Goldstone theorem of \cref{sec.ssbandgaplessstates}. In this case, we started with a continuous global \(U(1)\) phase symmetry that has one generator, and it has been spontaneously broken resulting in one massless Goldstone state.

A final remark is in order: observe that the outcome of the above analysis would be identical for any angle \(\theta = \theta_0\) and it does not work solely for the choice of \cref{eq.easiestchoice} where \(\theta=0\). The way to see this is to pick a different parametrisation of the complex field \(\varphi\) from the beginning. Instead of \cref{eq.vevplusexpansion} we express it as
\begin{equation}\label{eq.differentreparametrisation}
    \varphi\pqty{x} \equiv \bqty{v + \frac{h \pqty{x}}{\sqrt{2}}} e^{i \chi \pqty{x}},
\end{equation}
where again \(h\) and \(\chi\) are real scalars.
Then, if we look at \cref{eq.differentreparametrisation} closely, we perceive that the field \(\pi\) does not contribute to the potential term and hence it does not acquire a mass term. We can insert \cref{eq.differentreparametrisation} into \cref{eq.U(1)Lagrangianexample} and get an expression for the Lagrangian in terms of the  \(h\) and \(\chi\) fields as
\begin{equation}\label{eq.differentformLagragian}
    \mathscr{L} = -\frac{1}{2} \del_\mu h \del^\mu h - \frac{1}{2} \del_\mu \chi \del^\mu \chi -gv^2 h^2 + \dots,
\end{equation}
where again we neglect cubic powers and above. 

The Lagrangian of \cref{eq.differentformLagragian} completely matches the one of \cref{eq.U(1)Lagrangianexample}, where again there is a massive and a massless field with masses
\begin{align}
    m_{h}^2& = 2 g v^2, & m_{\chi} & =0.
\end{align}

%

\chapter[\texorpdfstring%
{\(O(2)\) model as a working example}%
{O(2)}]%
{\(O(2)\) model as a working example}%

\label{Chapter3} 

\epigraph{\itshape “Those who are not shocked when they first come across quantum theory cannot possibly have understood it.”}{Niels Bohr, \textit{Essays on Atomic Physics and Human Knowledge}}

In \Cref{Chapter2} we came upon the importance of \acrlong{cfts} in physics and how they are uniquely determined by the \acrshort{cft} data, \emph{i.e.} the scaling dimension \(\Delta\) of primary operators and the operator product expansion coefficients \( C_{\mathcal{O}_1\mathcal{O}_2\mathcal{O}_3}\). Afterwards, in \cref{sec.largecharge} we explained why the \acrshort{lce} leads to significant diminutions in the computations of these quantities.  

Thus far in the literature, the primary concern has been to compute the conformal dimension of large charge-operators, a work mainly carried out in \cite{hellerman2015cft,alvarez2017compensating,monin2017semiclassics,badel2019epsilon,giombi2021large,antipin2020charging,antipin2020charging2,jack2021anomalous,jack2022scaling} and autonomously confirmed via lattice computations in \cite{banerjee2018conformal,banerjee2019conformal,banerjee2022subleading,singh2022large}. Only a few three and four-point correlation functions have dispersedly turned up to this point in \cite{monin2017semiclassics,jafferis2018conformal,arias2020correlation,cuomo2021note,komargodski2021spontaneously}. Thus, in this chapter that follows closely \cite{dondi2022spinning} we intend to fill in the previously mentioned gap by systematically gathering and studying these three and four-point correlation functions with current insertions for a general \acrshort{cft} in \(d\)-spacetime dimensions that exhibits a global \(O\pqty{2}\) internal symmetry. Hence, we try to assemble and exhibit known results in a self-sufficient and autonomous manner and in a common language, but we even surpass the latest results by calculating for the first time correlators computed not in the large-charge scalar ground state but between spinning phonon states, which are excitations over the said scalar ground state. The results that we present are compatible with the form of correlators with conformal invariance and are in agreement to the ones presented in \cref{sec.stateoperator} and \Cref{sec:constraints}.

The plan of this chapter goes as following: in \cref{sec:O(2)_review}, we start by reviewing the primary elements of the \(O\pqty{2}\) sector, starting from the \acrshort{uv} Lagrangian and briefly revising the \acrfull{lsm} and the \acrfull{nlsm}. Then in \cref{sec.canonicalandpathintegralquantisation} we examine both the canonical and the path integral quantisation of the system, and we compute the elementary two point correlation function both in the large charge scalar ground state \(\braket{\mathcal{Q}}{\mathcal{Q}}\) and in the single phonon state \(\braket{\myatop{\mathcal{Q}}{\ell_2 m_2}}{\myatop{\mathcal{Q}}{\ell_1 m_1}}\). Afterwards, in \cref{sec:ConformalAlgebraAndChargeCorrelators} we display the state-of-the-art results of this chapter, which are correlation functions with insertions of the conserved current \(\mathcal{J}^\mu\) and/or the energy momentum tensor \(T^{\mu \nu}\) sandwiched between single phonon states. Finally, in \cref{sec:HLH} we review correlation functions where a light-charge operator is inserted in an ensemble created by heavy states.


\section[\texorpdfstring%
{Overview of the \(O(2)\) model at large charge}%
{O(2)}]%
{Overview of the \(O(2)\) model at large charge}%
\label{sec:O(2)_review}

So at this point, we want again to study the \(O\pqty{2}\) vector model of \cref{eq.freelagragian} but at fixed charge in \(d\) spacetime dimensions. The original work can be found in \cite{hellerman2015cft} and a more detailed version is in \cite{Gaume:2020bmp}. Here we will present a brief recap of their analysis to derive the \acrshort{nlsm} \acrshort{eft}. 

We start from the \acrshort{uv} fixed point where the theory is free, and we add a relevant deformation that drives the theory away from it. The relevant \acrshort{uv} Lagrangian written in terms of a complex scalar field similar to \cref{eq.U(1)system} is
\begin{equation}
\mathscr{L}_{\textrm{\acrshort{uv}}} = - \del_\mu \varphi^*_{\textrm{\acrshort{uv}}} \del^\mu \varphi_{\textrm{\acrshort{uv}}} - r \varphi^*_{\textrm{\acrshort{uv}}} \varphi_{\textrm{\acrshort{uv}}} - 4u \pqty{\varphi^*_{\textrm{\acrshort{uv}}}\varphi_{\textrm{\acrshort{uv}}}}^2,
\end{equation}
where in the above expression \(r\) is chosen in such a fine-tuned manner that the system flows to an \acrshort{ir} fixed point, which for \(d\leq 3\) is a \acrlong{wf} fixed point studied. It has been shown in \cite{fei2014critical} that an analogous argumentation can be made for \(d >3\) but now rather than to an \acrshort{ir} fixed point, the system flows to a \acrshort{uv} fixed point. A priori, we want to work on the cylinder \(\setR \times \setS^{d-1}\) with a radius \(R_0\) which is conformally equivalent to flat space to make use of the state-operator correspondence that we discussed in \cref{sec.stateoperator}. But only for now, to keep things a little more general, we can consider any pseudo-Riemannian manifold \(\pqty{\mathcal{M}, g_{\mu \nu}}\) that can be foliated as \(\setR \times \Sigma\), where \(\Sigma\) is a Cauchy hypersurface. 

We start by introducing a \acrshort{uv} cut-off \(\Lambda_{\textrm{\acrshort{uv}}}\) that limits the integration units, and then we lower the cut-off by integrating out high-frequency modes. When we reach the \acrshort{ir}, the system is still described by a complex scalar field \(\varphi_{\textrm{\acrshort{ir}}}\) which is related to the \acrshort{uv} field \(\varphi_{\textrm{{\acrshort{uv}}}}\) by an elaborate conversion. 

The field \(\varphi_{\textrm{\acrshort{ir}}}\) can be expressed in terms of an angular and a radial mode as
\begin{equation}
    \varphi_{\textrm{\acrshort{ir}}} = \frac{a}{\sqrt{2}}e^{i b \chi},
\end{equation}
where the field \(a \in \setR^+\), while the field \(\chi\) denotes the phase and is \(2 \pi\) periodic. On the other hand, \(b\) is just a parameter that is there to encapsulate the possibility of a conical singularity. Actually, from this particular parametrisation of the field \(\varphi_{\textrm{\acrshort{ir}}}\) we can see how the model took its name since the internal \(U \pqty{1}\) phase symmetry shifts the \(\chi\) field as
\begin{equation}
    \chi \overset{U \pqty{1}}{\longrightarrow} \chi' = \chi + \textrm{const}.
\end{equation}
The goal is to examine the system at the \acrshort{ir} conformal \acrshort{wf} fixed point but in a domain of large fixed global charge \(\mathcal{Q}\). By fixing the charge and studying the system in the fixed charge sector, the ground state of the said sector is not the same as the ground state of the full theory and this fixing will introduce a \acrshort{ssb} to the system which by virtue of the Goldstone theorem should give rise to a massless Goldstone boson. Assuming that there are no other light \acrfull{dof} we should be able to express the effective Lagrangian of the system in terms of these lightest \acrshort{dof}, which are the Goldstone modes.

In a spirit similar to \cref{sec.U(1)ssb} we can write the effective theory in terms of the fields \(a\) and \(\chi\). Since we are in the \acrshort{ir} fixed point, the corresponding Lagrangian which is known as \acrlong{lsm} has to be scale invariant and can be denoted as 
\begin{equation}\label{eq.lsmlagrangian}
    \mathscr{L}_{\textrm{\acrshort{lsm}}}  = -\frac{1}{2} \del_\mu a \del^\mu a - \frac{1}{2}b^2 a^2 \del_\mu \chi \del^\mu \chi - \frac{\xi R}{2}a^2 - \frac{d-2}{2d} g a^{2d/(d-2)},
\end{equation}
where there are also higher-derivative and higher-curvature terms that are consistent with conformal invariance. In \cref{eq.lsmlagrangian} the term \(\xi\) is the \textit{conformal coupling} which is equal to 
\begin{equation}
    \xi = \frac{(d-2)}{4(d-1)},
\end{equation}
the terms \(b\) and \(g\) are dimensionless Wilsonian coefficients that encrypt the physical data of the modes that were integrated out and \(R\) is the curvature scalar of \(\Sigma\). From dimensional analysis and scale invariance, we can compute the mass dimension of the fields \(a\) and \(\chi\). Since the derivative terms have a mass dimension \(\bqty{\del_\mu} =+1\) it is not hard to find the corresponding mass dimensions for the aforementioned fields, bearing in mind that \(\chi \) is a pure phase hence dimensionless, thus 
\begin{align}\label{eq.massdimensionofachi}
    \bqty{a} &= \frac{d-2}{2}, & \bqty{\chi} &= 0.
\end{align}
In accordance with \cref{eq.massdimensionofachi} a kinetic term of the field \(\chi\) has to come with a term that scales as \(d-2\) hence the \(a^2\) choice. Similarly, the Ricci coupling that has a mass dimension \(\bqty{R} = 2\) has to be coupled with an equivalent term. Finally, the only choice for a potential term that does not contain any dimensionful coupling is to scale as \(a^{2d/(d-2)}\). Doing the full \acrshort{rg} analysis --- see \cite{Gaume:2020bmp} --- it can be shown that the validity of the \acrshort{eft} of \cref{eq.lsmlagrangian} with cut-off scale \(\Lambda\) lies between
\begin{equation}
    \frac{1}{L} \ll \Lambda \ll \frac{\mathcal{Q}^{1/ \pqty{d-1}}}{L},
\end{equation}
where \(L\) is the length scale of \(\Sigma\) and for the cylinder \(\setS^{d-1}\) is its radius \(R_0\).

At this point, we want to compute the Euler-Lagrange \cref{eq.extremalofLangrangian} for the \acrshort{lsm} Lagrangian given in \cref{eq.lsmlagrangian}. We assume that the ground state maintains rotational symmetry, and thus it should exhibit a homogeneous behaviour \( \grad \varphi = 0\).

Then, the \acrlong{eom} for the two fields are
\begin{align}
    \pdv{\mathscr{L}_{\textrm{\acrshort{lsm}}}}{\chi} - \del_t \pqty{\pdv{\mathscr{L}_{\textrm{\acrshort{lsm}}}}{\pqty{\del_t \chi}}} & =  \del_t \pqty{b^2a^2 \dot{\chi}} =0,\label{eq.eomforchifield}\\
    \pdv{\mathscr{L}_{\textrm{\acrshort{lsm}}}}{a} - \del_t \pqty{\pdv{\mathscr{L}_{\textrm{\acrshort{lsm}}}}{\pqty{\del_t a}}} & = b^2 a \dot{\chi}^2 - \xi R a - g a^{\pqty{\frac{2d}{d-2}}-1} - \ddot{a} =0. \label{eq.eomforafield}
\end{align}
We notice that \cref{eq.eomforchifield} is nothing else but the charge conservation \(\del_t \mathcal{J}^t =0\) of \cref{eq.currentconservation} for a homogeneous solution. We can use \cref{eq.charge} to compute the conserved charge \(\mathcal{Q}\) or the charge density \(\rho\) defined as
\begin{equation}\label{eq.chargedensityrho}
    \rho \equiv \frac{\mathcal{Q}}{\textrm{Vol} \pqty{\Sigma}} = b^2 a^2 \dot{\chi}.
\end{equation}
From the homogeneity of the ground state, consistency of the Euler-Lagrange equations and from fixing the charge, the minimum solution for the field \(a\) corresponds to a constant fixed value \(v\) giving rise to a \acrshort{vev}
\begin{equation}\label{eq.veva}
    \expval{a}  = v.
\end{equation}
Using \cref{eq.eomforchifield,eq.chargedensityrho,eq.veva} leads to 
\begin{align}\label{eq.solutionforchi}
    \chi & = \mu t, & \mu & = \frac{\rho}{b^2 v^2}.
\end{align}
We can plug \cref{eq.veva,eq.solutionforchi} back into \cref{eq.eomforafield} and multiply by \(v\). Then we derive an expression for the \acrshort{vev} \(v\pqty{\rho}\) in terms of the charge density 
\begin{equation}\label{eq.potentialwithcentrifugal}
    \frac{\rho^2}{b^2 v^2} - \xi R v^2 - g v^{2d/\pqty{d-2}} =0.
\end{equation}
The last \cref{eq.potentialwithcentrifugal} is reminiscent of the classical potential \(V\) which has now acquired a centrifugal term proportional to \( \propto v^{-2}\) and through that the minimum of the potential is moved away from the origin. For \(R=0\) the equilibrium value of \(v\) sits at
\begin{equation}\label{eq.vevv}
    v \sim \frac{\mathcal{Q}^{1/4}}{\pqty{ \textrm{Vol}\pqty{\Sigma} b \sqrt{g}}^{1/4} }.
\end{equation}

In accordance with \cref{sec.U(1)ssb}, from \cref{eq.lsmlagrangian} we see that the field \(a\) is gapped and taking the charge \(\mathcal{Q}\) large, it is clear from \cref{eq.vevv} that it gains a large \acrshort{vev}. Hence, it should be removed from low energy effective action. 

To achieve that, we should integrate out \(a\) from the \acrshort{lsm} Lagrangian of \cref{eq.lsmlagrangian} which would generate the \acrfull{nlsm} Lagrangian \(\mathscr{L}_{\textrm{\acrshort{nlsm}}} \pqty{\chi}\). But realistically, this cannot happen and the closest we can get is to utilize the saddle-point approximation and remove the field \(a\) using its \acrshort{eom} and the equilibrium value of \cref{eq.vevv}. To determine tree level contributions beyond leading order, we will make use of the symmetries of the system, and write every possible term compatible with conformal invariance and with the correct dimensionality. Hence, we have a Wilsonian effective action in the spirit of \cref{sec.Wilsonianrg} that includes every term consistent with the symmetries of the theory. But working on a sector of large charge leads to significant simplifications. By utilising how terms in the effective action scale regarding the charge \(\mathcal{Q}\) we only keep terms that have a positive \(\mathcal{Q}\) scaling.

The leading term in the \acrshort{nlsm} \acrshort{eft} is generated by using the \acrshort{eom} of the field \(a\) so the Lagrangian reads:
\begin{equation}\label{eq.nlsmlagragian}
    \mathscr{L}_{\textrm{\acrshort{nlsm}}} =  c_1 \pqty{- \del_\mu \chi \del^\mu \chi}^{d/2}+ \dots,
\end{equation}
and \(c_1\) is an unknown Wilsonian coefficient. This is precisely the action that we will use to compute the \acrshort{cft} data of phonon states. 

At this point, it would be interesting to discuss the symmetry breaking pattern of the system. We see that we start with a \acrshort{cft} with the conformal Lie group \(SO\pqty{d,2} \) and a global \(O\pqty{2}\) or \(U\pqty{1}\) symmetry. Then by fixing the charge and having a ground state solution that depends on time, \emph{e.g.} the \(\chi\) field of \cref{eq.solutionforchi}, we spontaneously break boost invariance and time translation invariance and the ground state is no longer invariant under the global symmetry. The breaking of conformal invariance leads to the massive radial mode or gapped Goldstone, and the breaking of the phase symmetry generates the gapless Goldstone boson. Thus, the breaking pattern is
\begin{align*}
    &SO\pqty{d,2} \times O\pqty{2}_{\Qp} &\to&  &SO\pqty{d} \times D \times O\pqty{2}_{\Qp}& &\curly&  & SO\pqty{d} \times D'.
\end{align*}
where \(D\) is the dilatation generator on flat space \footnote{And of time translations on the cylinder.} and \(D' = D + \mu \Qp \) is the ground state's emerging helical symmetry where \(\Qp\) is the generator of the global phase symmetry. In the above symmetry breaking pattern, the first step produces the gapped Goldstone mode, while the second one generates the gapless mode. 

\subsection{Effective field theory}
\label{sec:effect-field-theory}

Although \cref{sec:O(2)_review} followed closely \cite{hellerman2015cft,Gaume:2020bmp}, in~\cite{monin2017semiclassics} an alternative but equivalent approach was presented, using the path integral formulation. Considering a \acrshort{cft} in \(d\)-dimensional flat space which is conformally equivalent to the cylinder, \(\setR \times \setS^{d-1}\) we want to study scalar primary operators \(\Opp\) of large global \(O\pqty{2}\) charge that generate the state \(\ket{\mathcal{Q}}\).

In the cylinder frame, the correlation function of these primary operators \(\Opp\) in the large separation limit \(\tau_\text{out} - \tau_\text{in} \equiv \beta \to \infty\) is denoted as
\begin{equation} \label{eq:corr}
  \braket{ \Qp, \infty}{\Qp , - \infty} = \lim_{\beta \longrightarrow \infty} \mel{\Qp} {e^{-\beta  {H}_{\text{cyl}}}}{\Qp}.
\end{equation}
Using the \acrshort{ccwz} construction Monin et al.~\cite{monin2017semiclassics} have demonstrated that correlation functions of the form of \cref{eq:corr} are described by a weakly couple \acrshort{eft} based on the coset model with the following breaking pattern \footnote{We chose to present the breaking pattern in Euclidean signature as the rest of the chapter will be in this notation.}
\begin{align*}\label{eq:coset}
  &{SO(d+1,1) \times U(1)_Q} &\longrightarrow& &{SO(d) \times U(1)_{D+\mu \Qp}}.
\end{align*}
We observe that the symmetry breaking pattern is exactly the same as the one that we encounter in \cref{sec:O(2)_review} which is thereafter known as the \emph{conformal superfluid} \footnote{In the above analysis the state \(\ket{\Qp}\) is an eigenstate of the charge \(\Qop\ket{\Qp} = \Qp \ket{\Qp}\), hence it is not a superfluid state. The correct assumption is that the saddle point that dominates the correlation function of \cref{eq:corr} is that of a superfluid state.}. 

The Lagrangian is given by \cref{eq.nlsmlagragian} and in Euclidean signature using \cref{eq.Wickrotation} the bulk effective action written in terms of the Goldstone field \(\chi\) of \cref{eq.solutionforchi} is
\begin{equation}\label{eq:actionnlsm}
  S\bqty{\chi}= - c_1 \int\limits_{\setR \times \setS^{d-1}} \dd{\tau}\dd{S}  \pqty{ - \del_\mu \chi \del^\mu \chi }^{d/2} + \text{curvature couplings},
\end{equation}
where \(c_1\) is a Wilsonian parameter related to the \acrshort{uv} theory, $\dd{S} = {R_0}^{d-1} \dd{\Omega}$ and 
\begin{align}\label{eq:chiandmu}
    \chi& = \chi^{\saddle} + \pi\pqty{\tau,\n} \nonumber \\
    & = -i\mu \tau + \pi\pqty{\tau,\n},
\end{align}
where \(\pi\pqty{\tau,\n}\) are the quantum fluctuations on top of the ground state. For fixed and large \(\Qp\) a hierarchy of states is produced,
\begin{equation*}
  \frac{1}{R_0} \ll E \ll \mu \sim \frac{\Qp^{1/(d-1)}}{R_0},
\end{equation*}
where $R_0$ is of course the radius of the cylinder and --- from \cref{eq.solutionforchi,eq.vevv} --- \(\mu\pqty{\Qp}\sim \Qp^{1/\pqty{d-1}}\) is understood as the chemical potential, which is now a function of \(\Qp\). Hence, the action of \cref{eq:actionnlsm} is valid up to the cut-off \(\Lambda\sim \mu \) and \( \pqty{R_0 \mu} \gg 1 \) is the relevant dimensionless ratio which controls the validity of the \acrshort{eft}. 

Now, the power of the \acrshort{lce} is that consecutive terms in the above action are constructed from a combination of \( \norm{\dd \chi}^2  = \pqty{g^{\rho \sigma} \del_\rho \chi \del_{\sigma} \chi}\) and higher-curvature invariant terms and every physical observable is written in terms of a power series in inverse powers of \(\mu\). For example, the overall ground state action, which consists of the bulk action plus the fixed boundary terms evaluated at the saddle point, reads
\begin{equation}\label{eq:S_expansion}
 S^{\saddle} =  \pqty{\frac{\tau_\text{out}-\tau_\text{in}}{R_0}} \Sum_{n=0}^{\infty} \, \alpha_n (R_0\mu)^{d-2n},
\end{equation}
where $\alpha_n$ depends on the Wilsonian parameter \(c_1\) and all the other Wilsonian coefficients that accompany the subsequent terms in the above expansion. Nevertheless, it is possible having an additional controlling parameter like large-\(N\) to compute them directly, \emph{e.g.} see \cite{alvarez2017compensating,de2018large}.


\subsection{Classical treatment}
\label{sec:classical_treatment}

At this point, we want to review the classical treatment of the \acrshort{nlsm} Lagrangian. Inserting \cref{eq:chiandmu} into the Euclidean version of \cref{eq.nlsmlagragian} and expanding up to quadratic order in the fluctuations \(\pi\pqty{\tau,\n}\) while neglecting higher-order terms of the action, the Lagrangian takes the form
\begin{equation}\label{eq:Quadratic_Lagrangian}
\mathscr{L} = - c_1 \mu^d - i c_1 d \mu^{d-1}  \dot{\pi} + c_1 \mu^{d-2} \frac{d(d-1)}{2} \pqty{ \dot\pi^2 + \frac{1}{d-1} (\del_i \pi)^2} +\order{\mu^{d-3}}.
\end{equation}
We can compute the conjugate momentum of the field \(\pi\) using the Euclidean version of \cref{eq.conjugatemomentum} truncated up to  linear order in the field as
\begin{equation}
    \Pi = \eval*{ i \fdv{\mathscr{L}}{\dot{\pi}} }_{\text{lin}} =  c_1 d \mu^{d-1} + i c_1 d(d-1) \mu^{d-2} \dot{\pi}.
\end{equation}
At leading order and neglecting interactions, this choice will provide the canonical Poisson brackets of the free theory as in \cref{sec.canonicalquantisation}. 

It has been shown in \cite{alvarez2017compensating} that it is possible to expand the fields $\pi$ and $\Pi$ into the usual normal mode decomposition in cylindrical coordinates as
\begin{align}
\pi(\tau, \n) & = \pi_0 -  \frac{i \Pi_0 \tau}{c_1 \Omega_d R_0^{d-1}  d(d-1) \mu^{d-2} } \label{eq:pfielddecomposition} \\
    + & \frac{1 }{\sqrt{c_1 R_0^{d-1} d(d-1) \mu^{d-2} }} \mathlarger{\mathlarger{\sum}}_{\ell \geq 1 , m}\pqty{ \frac{a_{\ell m}}{\sqrt{2\omega_{\ell}}} e^{-\omega_{\ell} \tau} Y_{\ell m} (\n)  +  \frac{ a^*_{\ell m} }{ \sqrt{2\omega_{\ell}} } e^{\omega_{\ell} \tau} Y_{\ell m}^* (\n)}\nonumber,  \\
    \Pi(\tau,\n)& = c_1 d \mu^{d-1} + \frac{\Pi_0}{\Omega_d R_0^{d-1}}  \label{eq:Pifielddecomposition}\\
   +& i \sqrt{\frac{c_1 d(d-1)\mu^{d-2}}{R_0^{d-1}}} \mathlarger{\mathlarger{\sum}}_{\ell,m} \pqty{ - a_{\ell m} \sqrt{\frac{\omega_\ell}{2}} e^{-\omega_\ell \tau} Y_{\ell m} (\n) + a^*_{\ell m} \sqrt{\frac{\omega_\ell}{2}} e^{\omega_\ell \tau} Y_{\ell m}^* (\n) } \nonumber, 
\end{align}
and in \cref{eq:pfielddecomposition} the sum in \(\ell\) starts from \(1\) as we have already pulled out the constant zero modes of the fields \(\pi_0\) and \(\Pi_0\), \(\Omega_d = \frac{2 \pi^{d/2}}{\Gamma(d/2)}\) denotes the volume of \(\setS^{d-1}\) and the \(Y_{\ell m}\) are the hyperspherical harmonics defined in \Cref{sec:Ylm-identities}. The dispersion relation for the frequencies \(\omega_\ell\) is computed from the eigenvalues of the angular momentum operator on the \(\setS^{d-1}\) sphere, and it is
\begin{equation}\label{eq:DispersionRelation}
   \omega_\ell= \frac{1}{R_0}\sqrt{\frac{\ell(\ell+d-2)}{(d-1)} } \ .
\end{equation}
Similarly to \cref{sec.canonicalquantisation} the Fourier coefficients $a_{\ell m}$ can be expressed using \cref{eq:pfielddecomposition,eq:Pifielddecomposition} as
\begin{equation}\label{eq:ExtractingTheOscillatorModes}
	a_{\ell m} = \sqrt{ \frac{c_1 d (d-1) \mu^{d-2}}{2 \omega_{\ell} \, R_0^{d-1} } } \int\limits_{\setS^{d-1}} \dd S \bqty{\pi(\tau, \n) \del_\tau \pqty{Y^*_{\ell m}(\n) e^{\omega_{\ell} \tau}} - \del_\tau \pi(\tau, \n) Y^*_{\ell m}(\n) e^{\omega_{\ell} \tau}}.
\end{equation}
Using \cref{eq:ExtractingTheOscillatorModes} and the Poisson bracket of \(\pi\) and \(\Pi\), \emph{i.e.} \(\Bqty{\pi, \Pi} =1 \) generates the relation between the Fourier modes
\begin{equation}
    \Bqty{ a_{\ell m},  a_{\ell' m'}^* } = \delta_{\ell \ell'} \delta_{m m'}.
\end{equation}
Using \cref{eq.currentconservation,eq.charge} the Euclidean charge is found to be
\begin{equation}\label{eq:chargeofnlsm}
    \Qp = \int\limits_{\setS^{d-1}} \dd{S}  \mathcal{J}^{\tau} = \int\limits_{\setS^{d-1}} \dd{S}  \Pi  =c_1 d \Omega_d(\mu R_0)^{d-1} + \Pi_0. 
\end{equation}
We can see that the charge, at leading order, depends on the zero mode \(\Pi_0\) corresponding to the ground state. Thus, the scale \(\mu\) can be related to the ground state charge \(\Qp_0\) --- at \(\Pi_0=0\) --- as
\begin{equation}\label{eq:mu_vs_Q}
  \mu = \left[ \frac{\Qp_0}{c_1 d R_0^{d-1} \Omega_d} \right]^{1/(d-1)} \ .
\end{equation}
This verifies our claim from  \cref{sec:effect-field-theory} that the dimensionless ratio \(\mu R_0 \sim \Qp_0^{1/(d-1)}\) controls the validity of the \acrshort{eft}. Moreover, from \cref{eq:chargeofnlsm} we see that the generic charge \(\Qp\), at leading order in fluctuations, depends on the homogeneous zero mode term \(\Pi_0\) cumulatively
\begin{equation}
   \Qp = \Qp_0 + \Pi_0 .
\end{equation}
Via the state-operator correspondence of \cref{sec.stateoperator}, the scaling dimension of the charged operator \(\Opp\) is related to the cylinder Hamiltonian and can be computed for a generic solution as
\begin{align}\label{eq:scalingdimensionofQ}
\Delta &= R_0 E_{\text{cyl}} \nonumber \\
& = R_0 \int\limits_{\setS^{d-1}} \dd{S} \bqty{ i \Pi \dot{\chi} +\Lp } \nonumber \\
&= R_0 \Bqty{\ \ \int\limits_{\setS^{d-1}} \dd{S} \bqty{ i \Pi \dot{\pi} + \Lp } + \mu {\int\limits_{\setS^{d-1}} \dd{S} \Pi}} \nonumber\\
&= c_1 (d-1) (\mu R_0)^{d} \Omega_d  + \mu R_0 \Pi_0 + c_1 R_0 \mu^{d-2} \frac{d(d-1)}{2} \int\limits_{\setS^{d-1}} \dd{S} \pqty{  \dot{\pi}^2 + \frac{1}{d-1} (\partial_i \pi)^2 }  \nonumber\\
&= \Delta_0 + \frac{\partial \Delta_0}{\partial \Qp_0} \Pi_0 + \frac{1}{2} \frac{\partial^2 \Delta_0}{\partial \Qp_0 \partial \Qp_0} \Pi_0^2 + R_0 \mathlarger{\sum}_{\ell \geq 1 , m} \ \omega_{\ell}  a_{\ell, m}^* a_{\ell m } \ , 
\end{align}
where in the first line we employed \cref{eq.energyandscaling}, in the second line we applied the Euclidean version of \cref{eq.Hamiltoniandensity} \footnote{Remember from \cref{eq.Wickrotation} that \(t \to -i \tau\) thus \(\del_t = i \del_\tau \). Hence, for the Hamiltonian of \cref{eq.Hamiltoniandensity} we have \(H = \Pi \del_t \phi - \mathscr{L}_M \overset{t \to -i \tau}{\longrightarrow} i \Pi \del_\tau \phi + \mathscr{L}_E\).} and the Lagrangian is given by \cref{eq:Quadratic_Lagrangian}, in the third line we utilized \cref{eq:chiandmu} and finally, we have defined 
\begin{align}
  & \Delta_0  = c_1 (d-1) \Omega_d(\mu R_0)^d + \order{(R_0\mu)^{d-2}} \, \label{eq:Delta0} \ , \\ 
  & \frac{\del \Delta_0}{\del \Qp_0}  = R_0 \mu \ , \\
 & \frac{\del^2 \Delta_0}{\del \Qp_0 \del \Qp_0}  = \frac{1}{c_1 d(d-1) \Omega_d (R_0\mu)^{d-2}} \ .
\end{align}
In our definitions \(\Delta_0\) agrees with the classical leading-order contribution to \cref{eq:S_expansion}.

Moreover, notice from the third line of \cref{eq:scalingdimensionofQ} that the Hamiltonian \(H_\chi\) can be written \emph{w.r.t} to the Hamiltonian \(H_\pi\) as
\begin{equation}\label{eq.effectivetimeevolution}
    H_\chi = H_\pi + \mu \Qp,
\end{equation}
and it is shifted by a factor of \(\mu \Qp\) which means that the generator of the effective time evolution for the field \(\pi\) is given by \cref{eq.effectivetimeevolution}, a fact consistent with a superfluid Goldstone fluctuation.


\section{Canonical and path integral quantisation}\label{sec.canonicalandpathintegralquantisation}

Having looked into the classical behaviour of the \(O\pqty{2}\) model at large charge in \cref{sec:O(2)_review}, we now want to examine its quantum behaviour. Similarly to \cref{subsec.Quantisation}, we will start our analysis with the familiar canonical approach in the operator language and then move on to the path integral formulation.


\subsection{Canonical quantization}\label{sec:canonicalquant}

As we explored in \cref{sec.canonicalquantisation} the idea behind canonical quantisation is to promote the classical fields to quantum operators. Furthermore, as we reviewed in \cref{sec.stateoperator}, on the cylinder \(\setR \times \setS^{d-1}\) the proper time coordinate is \(\tau\), hence in order to employ canonical quantisation on the cylinder we foliate our manifold by \(\tau\)-slicing, and to each slice of fixed \(\tau\) we relate the suitable Hilbert space \(\mathscr{H}_{\Qp}\). 

The fields \(\pi\) and \(\Pi\) of \cref{eq:pfielddecomposition,eq:Pifielddecomposition} are promoted to operators, and we impose equal-time commutation relations similar to \cref{eq.commutationfields} so that
\begin{equation}\label{eq:commutatorpiandPi}
    \comm{ {\pi}(\tau,\n)}{ {\Pi}(\tau, \n')} = i \delta_{\setS^{d-1}}(\n,\n'),
\end{equation}
where with \(\delta_{\setS^{d-1}}(\n,\n')\) we denote the delta function on \(\setS^{d-1}\). Therefore, similarly to \cref{eq.fouriercommutators}, we can use \cref{eq:commutatorpiandPi,eq:pfielddecomposition,eq:Pifielddecomposition} to find the following non-zero commutation relations for the Fourier modes \(a_{\ell m } \) and \(a^{\dagger}_{\ell' m'} \) and the zero modes \(\pi_0\) and \(\Pi_0\) as
\begin{align}\label{eq:CanonicalCommutators1}
  \comm{ {\pi}_0}{ {\Pi}_0} &= i, & \comm{ a_{\ell m }}{ a^{\dagger}_{\ell' m'}} = \delta_{\ell\ell'} \delta_{mm'}.
\end{align}
All the rest of the commutators are zero.

Since \cref{eq:CanonicalCommutators1} exhibits ladder commutators, we demand that the vacuum state $\ket{\Qp}$ of the theory to be eliminated by the annihilation operators, and thus it has to satisfy
\begin{equation}\label{eq:QisVacuumOfTheAs}
   a_{\ell m} \ket{\Qp} =  {\Pi}_0 \ket{\Qp} = 0.
\end{equation}
Being in finite volume, the charge operator \(\Qop \) acting on \(\mathscr{H}_\Qp\) is well-defined and is given in terms of the zero
modes only as
\begin{align}\label{eq:chargeoperatordefinition}
   \Qop &= \int\limits_{\setS^{d-1}} \dd{S}  {\Pi}(\tau,\n) = \Qp_0  {\Id} +  {\Pi}_0 \ , &  \Qop\ket{\Qp} &= \Qp_0 \ket{\Qp}.
\end{align}
Consequently, the only mode that carries charge is \(\pi_0\) and thus acting with it on the charged vacuum state \(\ket{\Qp}\) can increase its value. This is more evident by utilizing \cref{eq:CanonicalCommutators1} 
 and computing the subsequent commutators 
\begin{align}\label{eq:howchargechanges}
  \comm{ {\pi}_0}{ \Qop } &= i \ , & \comm{ \Qop}{  a_{\ell m}} &= \comm{ \Qop}{ a_{\ell m}^\dagger } = 0.
\end{align}
Therefore, given the vacuum state \(\ket{\Qp}\) we can apply \(\pi_0\) to create a new state
\begin{equation}
  \ket{ \Qp + q} = e^{i  {\pi}_0 q} \ket{\Qp} = \exp\bqty{\frac{i q}{\Omega_d R_0^{d-1}} \int\limits_{\setS^{d-1}} \dd{S}  {\pi}(\tau,\n)} \ket{\Qp} ,
\end{equation}
that has a charge \(\Qp_0+q\) and scaling dimension \(\Delta_0(\Qp_0 + q)\) \footnote{$\Delta_0 = \Delta_0(\Qp)$ is defined via equations \eqref{eq:mu_vs_Q} and \eqref{eq:Delta0}.}. Given that 
\begin{equation}
    \comm{ a_{\ell m}}{ {\pi}_0} = 0,
\end{equation}
states like this are annihilated by the ladders  \(a_{\ell m}\), but they are not annihilated by \(\Pi_0\) due to \cref{eq:CanonicalCommutators1}, and as a consequence they are
not zero modes of \(\Pi_0\), they do not represent degenerate vacua and lead to no degeneracy in the spectrum, but instead they are gapped
\begin{equation}
  \Delta_0(\Qp_0+q) - \Delta_0(\Qp_0) \sim q (R_0 \mu) \ .
\end{equation}
Considering the fact that the charge operator 
\(\Qop \) acts in a non-trivial manner only on the \(\pi_0\) operator, we can conclude that \(\pi_0\) has to be compact 
\begin{equation}
    \pi_0 \sim \pi_0 + 2\pi  {\Id} \, ,
\end{equation}
which would imply that \(q \in \mathbb{Z}\). Having a mass gap signifies that these states live at the \acrshort{eft} cut-off.

The quantised Hamiltonian \(D\) that is related to the classical quantity of \cref{eq:scalingdimensionofQ} can be expressed in the usual manner as a normal-ordered operator \(\normord{H}\) where in this case the normal ordering is defined for the vacuum state \(\ket{\Qp}\) as
\begin{equation}
    \mel{ \Qp}{ \normord{H}}{ \Qp } = \Delta_0/R_0,
\end{equation}
and a Casimir contribution of the fluctuations of the vacuum as 
\begin{align}\label{eq:Delta1}
   {D}& = R_0 \normord{ {H}} {}+ \Delta_1  {\Id},  &\text{where}&  &\Delta_1 & \coloneqq \frac{1}{2} \Sum_{\ell =1}^{\infty} \Sum_{m = -\ell}^{\ell} \pqty{R_0\omega_\ell}.
\end{align}
The Casimir energy contribution needs to be regulated --- see \Cref{AppendixB} for details --- and it has physical significance. Given that the one-loop Casimir contribution is of order \(\order{\Qp^0}\), it is necessary to keep an eye on every tree-level computation up to this order. For a \acrshort{cft} in \(d=3\) spacetime dimensions, it has been shown in \cite{hellerman2015cft} that 
\begin{equation}
	\Delta_0 = d_{3/2} \Qp^{3/2} + d_{1/2} \Qp^{1/2} + \order{\Qp^{-1/2}},
\end{equation}
where \( d_{3/2}\) and \(d_{1/2}\) depend on the unknown Wilsonian coefficients. For a general spacetime manifold, there are \(\ceil{(d+1)/2} \) terms that scale positively in \(\Qp \). 

At this point, we can go on and compute the commutators between \(D\) \footnote{For the computations we use that \(D\) takes the form:  \begin{equation*}
            D/R_0 = \int\limits_{\setS^{d-1}} \dd{S} \, \normord{T_{\tau\tau}} \ \propto  \mu {\Pi_0} 
        + \mathlarger{\sum_{m,\ell}} \omega_\ell \, {a_{\ell m}} {a^\dagger_{\ell m}} + \frac{\Pi_0^2}{2c_1 \Omega_d R_0^{d-1}  d \mu^{d-2}}  + \order{\varepsilon / \mu^{3}} .
\end{equation*}} and the rest of the operators. Doing so, we find 
\begin{align}\label{eq:CanonicalCommutators2}
  \comm{ {D}}{ a_{\ell m}} &= - R_0 \omega_\ell   a_{\ell m} ,
  &\comm{ {D}}{ a_{\ell m}^\dagger } &=  R_0 \omega_\ell   a_{\ell m}^\dagger, \\
   \comm{ {D}}{ {\pi}_0} &= -i \mathlarger{\mathlarger{\sum}}_{k=1}^\infty \frac{\Delta_0^{(k)}}{(k-1)!} {\Pi}_0^{k-1} ,
  &\comm{ {D}}{ {\Pi}_0} &=0.
\end{align}
Thus from the above commutators, in the usual manner, we can see the ones that give rise to excited states. 

As in \cref{sec.canonicalquantisation} the Fock space is constructed by repeated action of the creation operators as
\begin{equation}\label{eq:generic_state}
a_{\ell_1 m_1}^\dagger \dots a_{\ell_k m_k}^\dagger \ket{\Qp}.
\end{equation}
Since according to \cref{eq:howchargechanges} acting on the vacuum \(\ket{\Qp}\) with creation operators \(a_{\ell m}^\dagger\) does not raise the charge, these states have charge $\Qp_0$ and given \cref{eq:CanonicalCommutators2}, each subsequent application of \(a_{\ell m}^\dagger\) raises the scaling dimension by \(R_0 \omega_{\ell}\) so that
\begin{equation}\label{eq:scalingDimPhonon}
	\Delta = \Delta_0 + \Delta_1+ \mathlarger{\mathlarger{\sum}}_{i =1}^k (R_0 \omega_{\ell_i}).
\end{equation}
These are the \emph{superfluid phonon} states which correspond to spinning primaries \(\Opp_{\ell m}\) labelled by different quantum numbers, \emph{e.g.} spin \(\ell\), but same charge \(\Qp\) as \(\Opp\). 

The only descendant state is the one that that has \(\ell =1\) so it contains at least one creation operator \(a_{1m}^\dagger\). Using \cref{eq:DispersionRelation} we observe that for \(\ell =1\) the dispersion is \(\omega_1 = {1}/{R_0}\) so that the scaling dimension for the operator \( a_{1m}^\dagger \ket{\Qp}\) is 
\begin{equation}
    \Delta = \Delta_0 + \Delta_1 + 1 \ .
\end{equation}
Applying multiple times \(a_{1m}^\dagger\) on the ground state only increases its scaling dimension by one in each consecutive application, therefore according to \cref{sec.primaryfields} this is like acting with a derivative \( \del_\mu\) on the ground state hence the raising operator \(P_\mu\) in this frame is but the Goldstone mode for \(\ell=1\).

Before, we claimed that the new primaries are spinning operators and the reason is the following: there exists a unitary operator \(U \pqty{R}\) with \(R \in SO\pqty{d} \) on \(\mathscr{H}_Q\) that represents the \(R\) rotations --- see Tung~\cite{tung1985group} \(\S \, 7\) for details --- and looking at the transformation properties of the Fourier modes and their decomposition in \cref{eq:pfielddecomposition,eq:Pifielddecomposition}, they transform under \(U \pqty{R}\) as
\begin{align}
{U}(R)  a_{\ell m}^\dagger  {U}^\dagger(R) &= \sum_{m'} D^{\ell}_{mm'}(R^{-1}) \, a_{\ell m'}^\dagger \ , & R & \in SO(d),
\end{align}
where the quantity $D^{\ell}_{mm'}$ is the Wigner's \(D\)-matrix in \(d=3\)-dimensions or equivalently in \(d>3\) it is a generalised finite dimensional irreducible representation of \(SO(d)\). Given that \(SO(d)\) is the group of Euclidean rotations, states defined in \cref{eq:generic_state} correspond to spinning primaries. 

Finally, already in \cref{sec.largecharge} we mentioned that the validity of the superfluid \acrshort{eft} depends on the overall spin \(J\) and that it is not possible to describe every phonon state within it. From our analysis above, it is evident from \cref{eq:scalingDimPhonon} that when \(\ell\) gets too big, the last term of the equation \(R_0 \omega_{\ell}\) becomes comparable with the leading tree-level contribution \(\Delta_0\) which ends the validity of the \acrlong{lce}. Although the leading term in \(\Delta_0\) scales as \(\Qp^{d/\pqty{d-1}}\), there are in total \(\ceil{(d+1)/2} \) terms that scale positively in \(\Qp \) that originate from higher-curvature terms in the Lagrangian, with the last one scaling as \( \Qp^{1/(d-1)}\). Therefore, this sets a cut-off for \(\ell\) since the \acrshort{eft} cannot contain phonon states with 
\begin{equation}\label{eq:ell_cutoff}
\ell_{\text{cutoff}} \sim \Qp^{1/(d-1)}.
\end{equation}
On that account, such systems have a new \acrshort{eft} description --- see \cite{cuomo2023giant} and references therein.

\subsection{Path integral methods}\label{sec:pathIntergral}

Although the canonical quantisation is the easiest framework to discuss the spectrum of the theory and the existence of the charged spinning primary operators \(\Opp_{\ell m}\) which is in accordance with the superfluid hypothesis for a \acrshort{cft} with a global \(O(2)\) phase symmetry, one can show that an equivalent framework is the path integral approach, that has the advantage that it may also incorporate sub-leading corrections originating in interaction terms in \cref{eq:Quadratic_Lagrangian}. In this section, we will review the basic elements of the path integral method and compute the one-loop scaling \(\Delta_1\). 

Instead of the \(\pi\) and \(\Pi\) basis for the Hilbert space \(\mathscr{H}_\Qp\) that we explored in \cref{sec:canonicalquant} we can instead use the alternative basis 
\begin{align}
   {\chi}(\n) \ket{\chi} &= \chi(\n) \ket{\chi}, &  {\Pi}(\n) \ket{\Pi} &= \Pi(\n) \ket{\Pi},
\end{align}
written in terms of eigenstates of the field \(\chi\) and the conjugate momentum \(\Pi\). Following the common prescription, their eigenstates satisfy the usual relation
\begin{equation}\label{eq:bracketchiPi}
  \braket{ \chi}{\Pi} = \exp{i \int\limits_{\setS^{d-1}} \dd{S} \pqty{\chi \Pi}}.
\end{equation}
From \cref{eq:chargeoperatordefinition} and equations thereafter, it is evident that we can, in general, express the vacuum \(\ket{\Qp}\) as a superposition of momentum eigenstates that do not contain the non-trivial zero mode \(\Pi_0\) as
\begin{equation}\label{eq:Qsuperpositionofeigenstates}
  \ket{\Qp} = \mathscr{N}_\Qp \int \DD{\Pi} \delta( \Pi_0 )  \Psi_\Qp(\Pi) \ket{\Pi},
\end{equation}
where \(\mathscr{N}_Q\) is the usual normalisation parameter at fixed charge \(\Qp\) and \(\Psi_\Qp\) is the vacuum wave function. In the infinite separation limit, where \(\tau \to \infty\), correlation functions do not depend on the specifics of \(\Psi_\Qp\) and hence it only has an impact on the overall normalisation, therefore without loss of generality we set its value to \(\Psi_\Qp =1\). We can combine \cref{eq:bracketchiPi,eq:Qsuperpositionofeigenstates} to explore the relation between charge eigenstates \(\ket{\Qp}\) and field eigenstates, \(\ket{\chi}\) which is given by
\begin{align}%
\label{eq:overlap_chiQ}
	\braket{\chi}{\Qp } & = \mathscr{N}_\Qp \Int \DD{\Pi} \delta(\Pi_0) \exp{i \int\limits_{\setS^{d-1}} \dd{S} \pqty{\chi \Pi}} \nonumber \\
 & = \begin{cases}
		\mathscr{N}_\Qp \exp\Bqty{\frac{i \Qp}{\Omega_d R_0^{d-1}} \mathlarger{\int}\limits_{\setS^{d-1}} \dd{S} \chi} & \text{if \(\chi \) is constant,} \\
		0 & \text{otherwise} .
	\end{cases}
\end{align}
As a matter of fact, on a slice of fixed \(\tau\) on the cylinder, integrating over the spherical part allows us to separate the zero modes of the fields as
\begin{equation}
	\chi_0 = \int\limits_{\setS^{d-1}} \dd{S} \chi .
\end{equation}
The bracket of \cref{eq:overlap_chiQ} defines the proper boundary conditions for correlation functions of the form \(\mel{\Qp}{\dots}{\Qp}\), and generalises the quantum mechanical case of \(d=1\) where these boundary conditions are equivalent to open boundary conditions on the segment. Typically, we care about correlation functions, where the vacuum states \(\ket{\Qp}\) are inserted on the cylinder in the limit of infinite separation --- \emph{i.e.} \(\tau = \pm \infty\) --- therefore their effect on the boundary conditions is inconsequential.


\subsubsection{Two-point functions}\label{sec:twoPointFn}

Now, we are ready to formulate the path integral for the states presented in \cref{eq:generic_state}. We will review the amplitude for scalar and spinning primaries, where their insertions on the cylinder are done at infinite time separation.

\subsubsection[\texorpdfstring%
{\(\braket{ \Qp}{\Qp}\) correlator}%
{<Q|Q>}]%
{\(\braket{ \Qp}{\Qp}\) correlator}%
\label{sec:QQ_corr}

The vacuum correlation function of two scalar primaries \(\Opp\) inserted at \(\tau_\text{out} > \tau_\text{in}\) takes the form 
\begin{align} \label{eq:QQcorrelator}   
 \expval{\Opp*(\tau_\text{out})\Opp(\tau_\text{in})} & =  \mel{\Qp}{e^{-\frac{(\tau_\text{out} - \tau_\text{in})}{R_0}   D} }{\Qp} \nonumber \\
 &= \int \DD{\chi_{\text{in}}} \DD{\chi_{\text{out}}} \,  \braket{\Qp}{\chi_{\text{out}}} \braket{\chi_{\text{in}}}{\Qp} \int\limits_{\chi(\tau_\text{in}) = \chi_{\text{in}}}^{\chi(\tau_\text{out}) = \chi_{\text{out}}} \DD{\chi} \, e^{- S \bqty{\chi}} \nonumber \\ 
 &= \abs{\mathscr{N}_Q}^2\int \DD{\chi} \exp\bqty{-\int\limits_{\tau_\text{in}}^{\tau_\text{out}} \dd{\tau} \int\limits_{\setS^{d-1}} \dd{S} \Bqty{\Lp  + \frac{i \Qp}{\Omega_d R_0^{d-1}}    \dot \chi} } \nonumber \\
 & \coloneq \Anew ,
\end{align}
where in the second line we used \cref{eq:overlap_chiQ} and the Lagrangian is given in \cref{eq:Quadratic_Lagrangian}. The path integral above should be seen as the working definition of the correlation function with starting point the bulk action given in \cref{eq:actionnlsm}, without any need to reference the canonically quantized framework any more.

At this point, we want to compute the path integral of \cref{eq:QQcorrelator}. As we have seen in \cref{sec.pathintegral}, generically, the path integral is dominated by the classical contribution of a quantum field. Therefore, we expand our field \(\chi\) into a classical part plus quantum fluctuations as 
\begin{equation}\label{eq:fieldchiexpansion}
\chi \pqty{\tau, \n}= \chi^{\saddle} \pqty{\tau, \n} + \pi \pqty{\tau, \n},
\end{equation}
where \(\chi^{\saddle}(\tau, \n)\) is the classical saddle-point solution to the action's minimisation problem 
\begin{equation}\label{eq:actionminimasation}
  \delta S[\chi] = \int\limits_{\tau_\text{in}}^{\tau_\text{out}} \dd{\tau} \dd{S} \left( - \del_\mu \frac{ \del \mathscr{L}}{\del (\del_\mu \chi)} \right) \delta \chi + \eval*{ \int \dd{S}  \left(  \frac{ \del\mathscr{L}}{\del (\del_\tau \chi)} + \frac{i \Qp}{\Omega_d R_0^{d-1}} \right) \delta  \chi }_{\tau_\text{in}}^{\tau_\text{out}} .
\end{equation}
It is clear from the variation of the action in \cref{eq:actionminimasation} that the first term that corresponds to the bulk contribution has to be smooth for the \acrlong{eom} to be valid. Therefore, the quantity
\begin{equation}
  \frac{\del \mathscr{L}}{\del(\del_\mu \chi)} = c_1 d (-\del_\mu \chi \del^\mu \chi)^{d/2-1} \del^\mu \chi ,
\end{equation}
is required to be free of any divergences. 

In the semiclassical approximation, the classical trajectory that satisfies the boundary conditions and minimises the action is 
\begin{equation}
    \dot{\chi}^{\saddle} = - i \mu,
\end{equation}
which corresponds to the homogeneous saddle point of \cite{hellerman2015cft} and reads
\begin{equation}
    \chi^{\saddle}(\tau, \n) = - i \mu \tau + \pi_0,
\end{equation}
where \(\pi_0\) is a constant parameter. Moreover, the boundary condition in the exponent of \cref{eq:QQcorrelator} fixes the value of \(\mu\) in terms of the conserved current \(\mathcal{J}^{\tau}\) as
\begin{equation}\label{eq:formofthecharge}
    \frac{\Qp}{\Omega_d R_0^{d-1}}  = i\eval{\frac{\del \mathscr{L}}{\del(\del_\tau \chi)}}_{\dot{\chi}^{\saddle} = - i \mu}
     = c_1 d \mu^{d-1},
\end{equation}
and we have already found this relation in the classical treatment of \cref{sec:classical_treatment} in \cref{eq:mu_vs_Q}. As a result, the overall action \(S\bqty{\chi}\) up to quadratic order, written in terms of the bulk action and the boundary term in an expansion around the field configuration of \cref{eq:fieldchiexpansion}, is
\begin{align}\label{eq:Action_quadratic}
\eval{S\bqty{\chi}}_{\chi = \chi^{\saddle} + \pi} &= \Int \dd{\tau} \dd{S} \bqty{\Lp  + \frac{i \Qp}{\Omega_d R_0^{d-1}}    \dot \chi}\mathlarger{\eval}_{\chi = \chi^{\saddle} + \pi}  \\
& =\Delta_0 \frac{\tau_\text{out}-\tau_{\text{in}}}{R_0} + c_1 \mu^{d-2} \frac{d(d-1)}{2} \int\limits_{\tau_\text{in}}^{\tau_\text{out}} \dd \tau \int\limits_{\setS^{d-1}} \dd{S} \pqty{ \dot{\pi}^2 + \frac{1}{(d-1)} (\del_i \pi)^2 }. \nonumber
\end{align}
We observe that the boundary term above and the linear term in \cref{eq:Quadratic_Lagrangian} cancel each other out, therefore the zero mode terms of \cref{eq:scalingdimensionofQ} are gone.

We can choose the normalisation parameter \(\mathscr{N}_\Qp\) of \cref{eq:QQcorrelator} in such a way that the correlation function becomes 
\begin{equation}\label{eq:definitionofA}
  \Anew = R_0^{-2 (\Delta_0 + \Delta_1 +\dots)}  \exp\Bqty{- \frac{(\tau_\text{out} - \tau_\text{in})}{R_0} \bqty{ \Delta_0+\Delta_1 +\dots}} ,
\end{equation}
which is exactly the result of \cref{eq.twopointfunctioncylinder}. Furthermore,  \(\Delta_1\) is the Casimir energy contribution related to the fluctuations \(\pi\) on top of the homogeneous saddle point \(\chi^{\saddle}\) that we first came upon in \cref{eq:Delta1}.

As a final note, as we have seen in \cref{sec.stateoperator}, the action of scalar primary operators inserted at \(\tau = \pm \infty\) \footnote{This is an insertion at \(x_{\text{in}}=0, x_{\text{out}} = \infty  \) on \(\setR^d\).} on the cylinder \(\setR \times \setS^{d-1}\)  projects on the reference states \(\ket{\Qp}\) and \(\bra{\Qp}\) as
\begin{align}\label{eq:operatorsandstates}
  \lim_{\tau_{\text{in}} \to - \infty } \Opp(\tau, \n) \ket{0} \coloneq \ket{\Qp}, &  & \lim_{\tau_{\text{out}} \to \infty } \bra{0}\Opp(\tau, \n)^{\dagger}  \coloneq \bra{\Qp},
\end{align}
where the hermitian conjugation on the cylinder is performed as
\begin{equation}\label{eq.cylinderconjugation}
    \Opp(\tau, \n)^\dagger = \Opp*(-\tau,\n).
\end{equation}
Lastly, the Weyl map to $\setR^d$ has been laid down in \cref{eq.twopointtransform}.


\subsubsection[\texorpdfstring%
{\(\braket{\myatop{\Qp}{\ell_2 m_2}}{\myatop{\Qp}{\ell_1 m_1} } \) correlators}%
{<phonon|phonon>}]{\(\braket{\myatop{\Qp}{\ell_2 m_2}}{\myatop{\Qp}{\ell_1 m_1} } \) correlators}%
\label{sec:phonon2pt}

Following our analysis of the two-point function of scalar primaries \( \Opp\) at large charge, now we want to examine correlation functions of spinning primaries \(\Opp_{\ell m}\). 

The reference states that we are interested in are the single-phonon states \(\ket{\myatop{\Qp}{\ell m}}\) which are the simplest version of \cref{eq:generic_state}, with only one ladder operator \(a^\dagger_{\ell m}\) acting on the charged vacuum \(\ket{\Qp}\) as
\begin{align}\label{eq:one_phonon}
	\ket{\myatop{\Qp}{\ell m}} &=  a_{\ell m}^\dagger \ket{\Qp}, &&  \text{where} && \ket{\myatop{\Qp}{0 0}} =\ket{\Qp}.
\end{align}
By virtue of the Wigner–Eckart theorem \cite{wigner1927einige,eckart1930application} which implies that 
\begin{equation}
    \mel{\Qp}{ a_{\ell_2 m_2} a_{\ell_1 m_1}^\dagger}{\Qp} \sim \delta_{\ell_1 \ell_2} \delta_{m_1 m_2},
\end{equation}
and restricts the form of the correlator, we can compute the two-point correlation function in the canonical quantisation framework, by applying the commutation relations of the ladder operators \(a_{\ell m}\) and \( a_{\ell m}^\dagger \) of \cref{eq:CanonicalCommutators1,eq:CanonicalCommutators2} as
\begin{align}
\braket{ \myatop{\Qp}{\ell_2 m_2}}{\myatop{\Qp}{\ell_1 m_1}} & = \mel{ \Qp}{ a_{\ell_2 m_2} \, e^{ -\frac{ \pqty{\tau_\text{out} - \tau_\text{in}} }{R_0} D }  a_{\ell_1 m_1}^\dagger} {\Qp} \label{eq:spinningtwopointfunction}\\
&= \Anew e^{-(\tau_\text{out} -\tau_\text{in}) \omega_\ell} \delta_{\ell_1 \ell_2} \delta_{m_1 m_2}. \label{eq:one_phonon_correlator_quad}
\end{align}
Using the form of \cref{eq:definitionofA} we observe that 
\begin{equation}
    \small{\exp\Bqty{- \frac{(\tau_\text{out} - \tau_\text{in})}{R_0} \bqty{ \Delta_0+\Delta_1 +\dots}} \times \exp\Bqty{-(\tau_\text{out} -\tau_\text{in}) \omega_\ell} = \exp\Bqty{- \frac{(\tau_\text{out} - \tau_\text{in})}{R_0} \Delta}},
\end{equation}
with \(\Delta \) being the conformal dimension of \cref{eq:scalingDimPhonon} for a single-phonon state. As expected, the conformal dimension of the phonon state increased by \(R_0\omega_\ell\) and the form of the correlator is consistent with the structure of the spinning two-point correlation function in \Cref{sec:constraints}.

Although the aforementioned result holds true up to quadratic order in the Hamiltonian, it cannot enclose any impact from loop corrections, therefore a path integral formulation is necessary. 

The starting point is \cref{eq:ExtractingTheOscillatorModes} in order to write the ladder operators in terms of the fields and insert them in \cref{eq:spinningtwopointfunction}. Following the same logic as in \cref{eq:QQcorrelator}, we find that
\begin{multline}
  \braket{ \myatop{\Qp}{\ell_2 m_2}}{\myatop{\Qp}{\ell_1 m_1}} = { \frac{ c_1 d (d-1) \mu^{d-2}}{2 R_0^{d-1}\sqrt{ \omega_{ \ell_2} \omega_{ \ell_1} } } } \int \dd{S(\n_\text{out})} \int \dd{S(\n_\text{in})} Y^*_{\ell_2 m_2} (\n_\text{out}) \, Y_{\ell_1 m_1} (\n_\text{in}) \\
 \times \Anew \lim_{ \substack{ \tau \to \tau_\text{in} \\ \tau' \to \tau_\text{out}}} \pqty{ \omega_{\ell_2} - \del_{\tau'}{} } \pqty{\omega_{ \ell_1} + \del_{ \tau} {} } \expval{ \pi (\tau', \n_\text{out})\pi (\tau, \n_\text{in})},
\end{multline}
where we have introduced the two-point function of the fluctuations \(\pi\), 
\begin{equation}
\expval{ \pi (\tau', \n_\text{out})  \pi (\tau, \n_\text{in}) }  = \frac{1}{\braket{ \Qp , \tau'}{\Qp , \tau}} \int \DD{\pi} \pi(\tau', \n_\text{out} ) \pi(\tau, \n_\text{in} ) \, e^{- S[\pi]} ,
\end{equation}
and the action \(S[\pi]\) is given in  \cref{eq:Action_quadratic}.

All the information of the spectrum of the theory is included in the full version of the two-point function. Nevertheless, to replicate the outcome of 
\cref{eq:one_phonon_correlator_quad} we can use only the tree-level result --- see \Cref{sec.Goldstonepropagator} for details of the computation --- which was originally derived in \cite{komargodski2021spontaneously} and reads
\begin{equation}\label{eq:Greenfunction}
   \expval{ \pi (\tau', \n_\text{out} ) \pi (\tau, \n_\text{in} )} = \frac{1}{c_1 d(d-1) (\mu R_0)^{d-2} } \pqty{ \Sum_{\ell=1}^\infty \Sum_m e^{-\omega_\ell \abs{\tau'-\tau}} \frac{Y_{\ell m}^*(\n_\text{out}) Y_{\ell m}(\n_\text{in})}{2 R_0 \omega_\ell } - \frac{\abs{ \tau' - \tau} }{2 R_0 \Omega_d} }.
\end{equation}
Via the state operator correspondence, the phonon state \(\ket{\myatop{\Qp}{\ell m}}\) is projected by the action of the symmetric spinning operator inserted at \(\tau =- \infty \) as
\begin{equation}
  \lim_{\tau_{\text{in}} \to - \infty} \Vpp[\Qp][\ell m]\pqty{\tau, \n} \ket{0} \coloneq  \ket*{\myatop{\Qp}{\ell m}}.
\end{equation}
Finally, it is easy to generalise the computation of the two-point function of spinning primaries to systems with more phonon states using the canonical quantisation framework. As a matter of fact, for the two-phonon state \(\ket{\myatop{\Qp}{(\ell m) \otimes (\ell' m')}} \) we have the following result 
\begin{multline}
        \braket{ \myatop{\Qp}{(\ell_2 m_2) \otimes (\ell'_2 m'_2)}}{\myatop{\Qp}{(\ell_1 m_1) \otimes (\ell'_1 m'_1)}} =  \eval{ \Qp}{ a_{\ell_2 m_2 } a_{\ell'_2 m'_2} \, e^{ - \pqty{\tau_\text{out} - \tau_\text{in}}   D / R_0 }  a_{\ell'_1 m'_1}^\dagger  a_{\ell_1 m_1}^\dagger}{\Qp} \\
        = \Anew  e^{- \pqty{\tau_\text{out} -\tau_\text{in}} \pqty{ \omega_{\ell_2} + \omega_{\ell'_2} } } \pqty{ \delta_{\ell_1 \ell_2} \delta_{m_1 m_2} \delta_{\ell'_1 \ell'_2} \delta_{m'_1 m'_2} + \delta_{\ell_1 \ell'_2} \delta_{m_1 m'_2} \delta_{\ell'_1 \ell_2} \delta_{m'_1 m_2} }.
\end{multline}
For every consecutive phonon excitation that is added to the system, the energy of the relative state increases by \(\omega R_0\) in accordance with \cref{eq:generic_state} and also, there appears a summation over all permutations of the Kronecker deltas. As long as no state contains an \(\ell=1\) quantum number, they correspond to spinning primaries in some reducible representation. For instance, in \(d = 3\) we have
\begin{equation}
    \ell \otimes \ell' = (\ell + \ell') \oplus (\ell + \ell'-2) \oplus \dots \oplus \abs{\ell-\ell'}.
\end{equation}


\subsection{One-loop scaling dimension and regularization}\label{sec:loopCorrections}

In the spirit of \cref{sec.thermalQFT} it is possible to set up our system on the thermal circle \(\setS^1_\beta \times \setS^{d-1}\) where we have compactified the temporal dimension in a circle of circumference \(\beta\). Then taking the limit \(\beta \to \infty\) which is the zero temperature limit, we recover the initial predictions of our theory.

We want to compute the one-loop scaling dimension \(\Delta_1\). On the thermal circle \(\setS^1_\beta \times \setS^{d-1}\) this corresponds to
\begin{equation}
    \Delta_1 = -\lim_{\beta \rightarrow \infty} \pdv{\beta} \log \mathcal{Z}_0,
\end{equation}
where \(\mathcal{Z}_0\pqty{\beta}\) is the free partition function related to the generating functional of \cref{eq.generatingfunctional}.
This computation amounts to the evaluation of the functional determinant 
\begin{equation}\label{eq:functionaldeterminantdelta1}
     \frac{1}{2\beta} \ln\det\bqty{- \del_\tau^2 -\Laplacian_{\setS^{d-1}}^2}  =  \frac{1}{2} \Sum_{\ell >0} M_\ell (R \omega_\ell),
\end{equation}
where \(\omega_\ell\) are the eigenvalues of the Laplacian with degeneracy \(M_\ell\) and \(\omega_n = 2\pi n /\beta\) are the Matsubara frequencies that we have already summed over ---  \cite[see][Appendix B]{monin2017semiclassics} for details. The above result matches the expression that we have found in \cref{eq:Delta1}. This sum is divergent and has to be regularised. 

So for now, we will work explicitly in \(d=3\) and we will use a heat kernel regularisation, \emph{e.g.} see Monin \cite{monin2016partition}. A more general approach using a smooth cut-off regulator is found in \Cref{sec:regularisation}.

For now, we start with the expression
\begin{equation}
    I = \Sum_{\ell =1}^\infty \pqty{2 \ell + 1} \sqrt{\ell \pqty{\ell + 1}},
\end{equation}
which matches exactly  \cref{eq:functionaldeterminantdelta1} in three dimensions, and we extract a \(\frac{1}{2\sqrt{2}}\) prefactor that we will reinstate later. We rewrite it as 
\begin{equation}
    I = 2 \Sum_{\ell =1}^\infty \pqty{ \ell + \frac{1}{2}} \sqrt{\pqty{\ell + \frac{1}{2}}^2 - \frac{1}{4}}.
\end{equation}
The above result is the energy for a boson that is conformally coupled plus a mass term \(m^2 = - \frac{1}{4}\). We can now split this into a convergent and a divergent part as
\begin{align}
    I &= 2 \Sum_{\ell =1}^\infty \bqty{\pqty{ \ell + \frac{1}{2}} \sqrt{\pqty{\ell + \frac{1}{2}}^2 - \frac{1}{4}} - \pqty{\pqty{\ell + \frac{1}{2}}^2 - \frac{1}{8}}}  \nonumber \\
    &+ 2 \Sum_{\ell =1}^\infty \bqty{\pqty{\ell + \frac{1}{2}}^2 - \frac{1}{8} \pqty{\ell + \frac{1}{2}}^0} \\
    & = I_{\text{conv}} + I_{\text{div}}. \nonumber
\end{align}
We can use the Hurwitz zeta function 
\begin{equation}
    \zeta\pqty{s,a} = \Sum_{n=0}^{\infty} \frac{1}{\pqty{n+a}^s},
\end{equation}
to compute the divergent part as
\begin{equation}
    I_{\text{div}} = 2 \zeta\pqty{2,\frac{1}{2}} - \frac{1}{4} \zeta\pqty{0,\frac{1}{2}} - \frac{1}{4} = - \frac{1}{4}.
\end{equation}
The convergent part can be evaluated numerically as \(I_{\text{conv}} = -0.01509 \). Now if we multiply our combined results by the prefactor \(\frac{1}{2\sqrt{2}}\) we have the one-loop scaling dimension
\begin{equation} \label{eq:scalinggoldstone}
    \Delta_1 = -0.0937256. 
\end{equation}
As stated before, since this contribution comes at order \(\order{\Qp^0}\) and it gets no other corrections, it is a universal prediction of the theory and a feature of all superfluid \acrshort{eft}s.


\section{Correlators with current insertions}\label{sec:ConformalAlgebraAndChargeCorrelators}

As we saw in \cref{sec.canonicalandpathintegralquantisation} there is a complete agreement in the predictions of the canonical and the path integral quantisation as long as we use the tree-level result for the two-point correlation function of the fluctuations \(\pi\). Any higher-loop corrections are suppressed by powers of \(\Qp\), therefore working at large charge has the advantage that physical data at the fixed point can be encapsulated by a free theory. Hence, this permits the computation of n-point functions of the strongly coupled \acrshort{cft} with the use of the operator algebra \eqref{eq:CanonicalCommutators1} only.

On that account, in this section, we compute three and four-point correlation functions of large charge primaries with insertions of the conserved current \(\mathcal{J}^\mu\) and/or the energy-momentum tensor \(T^{\mu \nu}\). Although some of these correlation functions have already been computed in \cite{monin2017semiclassics,komargodski2021spontaneously,jafferis2018conformal,Cuomo:2020thesis} for the case where \(\ell=0\) which corresponds to scalar primaries \(\Opp\), we go beyond the state-of-the-art, and we compute them for spinning primary operators \(\Vpp[\Qp][\ell m]\). 

In our analysis, the results that we present have to be seen as an expansion in the charge \(\Qp\) with only the classical contribution and the leading-order quantum correction present. Furthermore, the form of our correlators  is in agreement with the ones that are presented in \cref{sec.stateoperator} and \Cref{sec:constraints}.

\subsection{Conserved currents and Ward identities in the EFT}\label{sec:currents}

The classical currents of the \acrshort{nlsm} \acrshort{eft} of \cref{eq:actionnlsm} are computed using the Euclidean version of \cref{eq.currentconservation,eq.ThetaEM} and are found to be
\begin{align}
    \mathcal{J}_\mu &= c_1 d (-\del_\mu \chi \del^\mu \chi)^{d/2-1} \del_\mu \chi ,\label{eq:currentnlsm}\\
    T_{\mu\nu} &= c_1 \Bqty{ d  (-\del_\mu \chi \del^\mu \chi)^{d/2-1} \del_\mu \chi \del_\nu \chi + g_{\mu\nu}  (-\del_\mu \chi \del^\mu \chi)^{d/2} } \label{eq:EMtensornlsm}.
\end{align}
We can write them as an expansion in \(\chi = \chi^{\saddle} + \pi\pqty{\tau,\n} \) up to \(\order{\pi^2}\) as
\begin{subequations}\label{eq:ExpansionsOfTsAndQsInTheField}
\begin{align}
  & \mathcal{J}_\tau = - i \frac{\Qp}{\Omega_d R_0^{d-1}} \Bqty{ 1 + \frac{i}{\mu}  (d-1)  \dot{\pi} - \frac{  (d-2) (d-1) }{2 \mu^2} \bqty{ \dot{\pi}^2 + \frac{ ( \del_i \pi )^2 }{R_0^2 (d-1)} } + \order{\mu^{-3}} } \label{eq:Jtau}, \\
  & \mathcal{J}_i = \frac{\Qp}{\Omega_d R_0^{d-1}} \Bqty{\frac{1}{\mu R_0} \del_i \pi + \frac{i }{\mu} \frac{(d-2)}{\mu R_0} \dot{\pi} \del_i \pi + \order{\mu^{-3}}},\label{eq:Ji} \\
  & T_{\tau\tau} = - \frac{ \Delta_0 }{ \Omega_d R_0^{d}} \Bqty{1 + i \frac{ d}{ \mu}  \dot{\pi} - \frac{ d (d-1) }{ 2 \mu^2} \bqty{ \dot{\pi}^2 + \frac{(d-3) \, (\del_i \pi)^2 }{R_0^2 (d-1)^2} } + \order{\mu^{-3}} }, \\
   & T_{\tau i} = - i \frac{\Delta_0}{\Omega_d R_0^{d}} \Bqty{ \frac{1}{\mu R_0} \frac{d}{d-1} \partial_i \pi + \frac{i}{\mu} \frac{d}{\mu R_0} \dot{\pi}\partial_i \pi + \order{\mu^{-3}}  }, \\
  & 
  \begin{multlined}[][.7\linewidth]
  T_{ij} = \frac{ \Delta_0 }{ \Omega_d R^{d} } \frac{ h_{ij}}{(d-1)} \Bqty{ 1 + i \frac{ d }{ \mu } \dot{\pi} - \frac{ d (d-1) }{ 2 \mu^2} \bqty{ \dot{\pi}^2 + \frac{ (\del_i \pi)^2 }{R_0^2 (d-1)} }  +\order{\mu^{-3}} } \\
    + \frac{ \Delta_0 }{   \Omega_d R^{d}} \frac{1}{(\mu R_0)^2}\frac{d}{(d-1)}  \Bqty{ \del_i \pi \del_j \pi + \order{\mu^{-3}} },
    \end{multlined}
\end{align}
\end{subequations}
where we denote $h_{ij}$ the metric of the $d-1$-sphere. Since the saddle point $\chi^{\saddle}$ is homogeneous, it is clear from \cref{eq:EMtensornlsm,eq:currentnlsm} that at leading order \(\mathcal{J}_i = T_{\tau i} = 0\).

As already stated, we will study correlation functions of the above currents in the canonical quantisation framework of \cref{sec:canonicalquant} which is adequate for leading-order results. From \cref{eq.charge,eq.Hamiltoniandensity} we know that by integrating the temporal components \(\mathcal{J}_\tau\) or \(T_{\tau\tau}\) over the spatial hypersurface \(\setS^{d-1}\) generates the operators
\begin{align}
    \Qop & = \Int\limits_{\setS^{d-1}} \dd{S} \mathcal{J}_{\tau} \ , &
    D &= R_0 \Int\limits_{\setS^{d-1}} \dd{S} T_{\tau \tau}.
\end{align}
By inserting these operators at any time \(\tau\) on the cylinder, it is possible to measure the charge or the scaling dimension of an operator \(\mathscr{O}_i\) inserted at \(\tau_i \neq \tau\). This is a direct consequence of the Ward identities
\begin{align}
\expval{\Qop(\tau) \mathlarger{\prod_{i}} \mathscr{O}_i(\tau_i, \n_i) } &=  \Sum_{ \tau_i < \tau} \Qp_i \expval{\mathlarger{\prod}_{i} \mathscr{O}_i(\tau_i, \n_i ) },  \label{eq:WardIdentitiesQ}\\
\expval{  {D}(\tau) \mathlarger{\prod}_{i} \mathscr{O}_i(\tau_i, \n_i) } &=  \Sum_{ \tau_i < \tau} \Delta_i \expval{ \mathlarger{\prod}_{i} \mathscr{O}_i(\tau_i, \n_i ) }, \label{eq:WardIdentitiesD}
\end{align}
which are a generalisation of the Noether's theorem of current conservation for quantum operators. Since these identities are true at every order in a loop expansion, they can be employed to constrain the form of correlation functions with the current insertions of \cref{eq:ExpansionsOfTsAndQsInTheField}.


\subsection[\texorpdfstring%
{\(\mel{\myatop{\Qp}{\ell_2 m_2}}{\mathcal{J}}{\myatop{\Qp}{\ell_1 m_1} } \)}%
{<QJQ>}]{\(\mel{\myatop{\Qp}{\ell_2 m_2}}{ \mathcal{J} } {\myatop{\Qp}{\ell_1 m_1} } \) correlators}%
\label{sec:VJV}
We start by computing the three-point correlation functions between two spinning primary \footnote{For \(\ell=1\) these are not primaries but descendants, so we do not consider this special case.} operators \(\Vpp[\Qp][\ell m] \ket{0}=  a^\dagger_{\ell m} \ket{\Qp}\) inserted at \(\tau_\text{in} , \tau_\text{out}\) and the current \(\mathcal{J}_\mu(\tau,x)\), inserted at a time \(\tau_\text{in} < \tau < \tau_\text{out}\). These are found to be
\begin{align}
  &\begin{aligned}\
  &\expval{\Vpp* \mathcal{J}_{\tau}(\tau,\n) \Vpp} = - i \frac{\Qp}{\Omega_d R_0^{d-1}} \Bigg\{ \Anew[\Delta_\Qp + R_0 \omega_{\ell_2}] \delta_{\ell_1 \ell_2}  \delta_{m_1 m_2}  \\
  & + \Anew[\Delta_\Qp + R_0 \omega_{\ell_1}][\Delta_\Qp + R_0 \omega_{\ell_2}][\tau] (d-1)(d-2) \Omega_d \frac{ R_0 \sqrt{ \omega_{ \ell_2} \omega_{ \ell_1} } }{ 2 d  \Delta_0 } \bqty{ Y_{\ell_2 m_2}^* (\n) Y_{\ell_1 m_1} (\n) - \frac{ \partial_i Y_{\ell_2 m_2}^* (\n) \, \partial_i Y_{\ell_1 m_1} (\n) }{ R_0^2 (d-1) \omega_{\ell_2} \omega_{\ell_1} } }  \Bigg\} , 
  \end{aligned} \label{eq:3pointspinningandJ}\\
  &\expval{\Vpp* \mathcal{J}_i(\tau,\n) \Vpp} =  i \frac{ Q (d-2)}{2 \Delta_0 R_0^{d-1} d } \Anew[ \Delta_\Qp + R_0 \omega_{ \ell_{1}} ][ \Delta_\Qp + R_0 \omega_{ \ell_{2}} ][ \tau]  \bqty{ \sqrt{ \frac{ \omega_{ \ell_2} }{ \omega_{ \ell_1} } } Y_{\ell_2 m_2}^* (\n) \del_i Y_{\ell_1 m_1} (\n) - (1 \leftrightarrow 2 )^* },
\end{align}
and we introduced two new notations 
\begin{align}
    \Anew[ \Delta_\Qp + R_0 \omega_{\ell_2}]  &\coloneqq \Anew  e^{- (\tau_\text{out} - \tau_\text{in}) \, \omega_{ \ell_2} } ,\\
     \Anew[\Delta_1][\Delta_2][\tau] &\coloneqq e^{- \Delta_2 ( \tau_\text{out} - \tau)/R_0} e^{- \Delta_1 (\tau - \tau_\text{in})/R_0},
\end{align}
that generalise the quantity \(\Anew = \Anew[\Delta]\) that we originally defined in \cref{eq.twopointfunctioncylinder}. We also have to mention that the case of $\ell_i=0$  in \cref{eq:3pointspinningandJ} was originally computed in \cite{monin2017semiclassics}.

By integrating \(\mathcal{J}_\tau\) over the spatial hypersurface \(\setS^{d-1}\) we derive the charge, and therefore we can use the integral version of the Ward identity of \cref{eq:WardIdentitiesQ}, where we integrate the three-point function of \cref{eq:3pointspinningandJ} over the \(d-1\)-sphere to derive
\begin{equation}\label{eq:wardconfirmation}
    \Int\limits_{\setS^{d-1}} \dd{S(\n)} \expval{\Vpp* \mathcal{J}_{\tau}(\tau,\n) \Vpp} = - i \Qp  \Anew[\Delta_\Qp + R_0 \omega_{\ell_2}] \delta_{\ell_1 \ell_2}  \delta_{m_1 m_2},
\end{equation}
which is precisely what was predicted by the Ward identity, giving us a nice consistency check.

At this point, we want to analyse the structure of the correlation function with the \(\mathcal{J}_\tau\) insertion. From \cref{eq:Jtau} it is easy to notice that the current is a combination of a homogeneous classical part that is also time independent and quantum fluctuations.  

From the first line of \cref{eq:Jtau} which corresponds to the classical contribution, we observe that it is actually proportional to the two point function of two spinning primaries
\begin{equation*}
	\braket{ \myatop{\Qp}{\ell_2 m_2}} {\myatop{\Qp}{\ell_1 m_1}} = \Anew[\Delta_\Qp + R_0 \omega_{\ell_2}] \delta_{\ell_1 \ell_2}  \delta_{m_1 m_2},
\end{equation*}
that we have seen in \cref{eq:one_phonon_correlator_quad}.

On the other hand, the quantum part is not homogeneous and being on the sphere it can be written in terms of spherical harmonics, but should have the same tensor structure as the left-hand side of \cref{eq:Jtau}. Furthermore, since the Ward identities hold at all orders, the integral of this part over the sphere should vanish, which was independently confirmed in  \cref{eq:wardconfirmation}.

In general, any physical observable can be split into a classical part and quantum fluctuations. Therefore, in this spirit, if we analyse the correlation function with the \(\mathcal{J}_i\) insertion, we see that due to \cref{eq:Ji}, there is no classical contribution and the only thing that remains is the piece of the inhomogeneous quantum fluctuations. For \(\ell_i = 0\) the ground state is $\ket{\Qp}$ and the Ward identities demand that to every order
\begin{equation}
    \mel{\Qp}{\mathcal{J}_i}{\Qp} = 0.
\end{equation}
From \cref{eq:3pointspinningandJ} we can also read off the \acrshort{ope} coefficient
\begin{equation}
    C_{\Vpp[-\Qp][\ell m] \mathcal{J}_\tau \Vpp[\Qp][\ell m]} =\frac{ \expval{\Vpp[-\Qp][\ell m] \mathcal{J}_{\tau}(\tau,\n) \Vpp[\Qp][\ell m]} }{ \expval{\Vpp[-\Qp][\ell m]  \Vpp[\Qp][\ell m] }} = - i \frac{\Qp}{\Omega_d R_0^{d-1}}.
\end{equation}
We can generalise our computation to higher-phonon states, \emph{e.g.} for two phonons we get
\begin{align}
& \mel{ \myatop{\Qp}{(\ell_2 m_2) \otimes (\ell'_2 m'_2)}}{\mathcal{J}_\tau (\tau, \n)}{\myatop{\Qp}{(\ell_1 m_1) \otimes (\ell'_1 m'_1)} } = - i \frac{\Qp }{ \Omega_d R_0^{d-1}} \Anew[\Delta_\Qp + R_0 \omega_{\ell_2} + R_0 \omega_{\ell'_2}] \nonumber \\
 & \times \Bigg\{ \pqty{ \delta_{\ell_1 \ell_2} \delta_{m_1 m_2} \delta_{\ell'_1 \ell'_2} \delta_{m'_1 m'_2} + \delta_{\ell_1 \ell'_2} \delta_{m_1 m'_2} \delta_{\ell'_1 \ell_2} \delta_{m'_1 m_2} } \nonumber \\
 &+ \Omega_d \frac{ (d-2) (d-1) }{2d \, \Delta_0 } \Bigg[ \frac{ R_0 \sqrt{ \omega_{\ell'_2} \omega_{\ell'_1}} }{ e^{ (\tau - \tau_\text{in}) ( \omega_{\ell'_1} - \omega_{\ell'_2}) } } \bqty{  Y_{\ell'_1 m'_1}(\n) Y_{\ell'_2 m'_2}^*(\n) - \frac{ \partial_i Y_{\ell'_1 m'_1}(\n) \partial_i Y_{\ell'_2 m'_2}^*(\n) }{R_0^2 (d-1) {\omega_{\ell'_2} \omega_{\ell'_1}} } }  \delta_{\ell_2 \ell_1} \delta_{m_2 m_1} \nonumber \\
 &+ \frac{ R_0 \sqrt{ \omega_{\ell'_2} \omega_{\ell_1}} }{ e^{ (\tau - \tau_\text{in}) ( \omega_{\ell_1} - \omega_{\ell'_2} ) } } \bqty{ Y_{\ell_1 m_1} (\n) Y_{\ell'_2m'_2}^*(\n) - \frac{ \partial_i Y_{\ell_1 m_1}(\n) \partial_i Y_{\ell'_2 m'_2}^*(\n) }{R_0^2 (d-1) {\omega_{\ell'_2} \omega_{\ell_1}} } } \delta_{\ell_2 \ell'_1} \delta_{m_2 m'_1} \nonumber \\
 &+ \frac{ R_0 \sqrt{ \omega_{\ell_2} \omega_{\ell'_1}} }{ e^{ (\tau - \tau_\text{in}) ( \omega_{\ell'_1} - \omega_{\ell_2} ) } } \bqty{ Y_{\ell'_1 m'_1}(\n) Y_{\ell_2 m_2}^*(\n) - \frac{ \partial_i Y_{\ell'_1 m'_1}(\n) \partial_i Y_{\ell_2 m_2}^*(\n) }{R_0^2 (d-1) {\omega_{\ell_2} \omega_{\ell'_1}} } } \delta_{\ell'_2 \ell_1} \delta_{m'_2 m_1} \nonumber \\
 &+ \frac{ R_0 \sqrt{ \omega_{\ell_2} \omega_{\ell_1}} }{ e^{ (\tau - \tau_\text{in}) ( \omega_{\ell_1} - \omega_{\ell_2} ) } } \bqty{  Y_{\ell_1 m_1} (\n) Y_{\ell_2 m_2}^* (\n) - \frac{ \partial_i Y_{\ell_1 m_1}(\n) \partial_i Y_{\ell_2 m_2}^*(\n) }{R_0^2 (d-1) {\omega_{\ell_2} \omega_{\ell_1}} } } \delta_{\ell'_2 \ell'_1} \delta_{m'_2 m'_1} \Bigg] \Bigg\} .
\end{align}
In all our computations thus far we neglected any linear terms of \(\pi\) that appear in \cref{eq:ExpansionsOfTsAndQsInTheField}. Although this poses no problem for our computations, in the special case that we want to compute correlation functions where the two spinning primaries have a different number of phonons this term cannot be omitted as can be seen in the following example of correlation functions with an insertion of the current sandwiched between a scalar and a one-phonon primary operators 
\begin{align}
    \mel{ \Opp*}{\mathcal{J}_\tau (\tau, \n)} {\myatop{\Qp}{\ell m} }
    &= -\frac{ \Qp (d-1)}{ \Omega_d R_0^{d-1} } \sqrt{ \frac{ \Omega_d }{ 2d } \frac{ R_0 \omega_{\ell} }{ \Delta_0 } } \Anew[\Delta_\Qp + R_0 \omega_{\ell}][\Delta_\Qp ][\tau] Y_{\ell m} (\n) ,  \\
    \mel{ \Opp*} {\mathcal{J}_i (\tau, \n)}{ \myatop{\Qp}{\ell m} }
    &= \frac{ \Qp}{ \Omega_d R_0^{d-1} } \sqrt{ \frac{ R_0 \, \Omega_d }{ 2d \, \Delta_0 R_0 \omega_{\ell} } } \, \Anew[\Delta_\Qp + R_0 \omega_{\ell}][\Delta_\Qp ][\tau] \, \partial_i Y_{\ell m} (\n). 
\end{align}
Similar relations hold true for every correlation function with a \(\mathcal{J}\) or \(T\) insertion. From this point on, we will not reference this special case again.


\subsection[\texorpdfstring%
{\(\mel{\myatop{\Qp}{\ell_2 m_2}}{\mathcal{J} \mathcal{J}}{\myatop{\Qp}{\ell_1 m_1}}\)}%
{<OJJO>}]%
{\(\mel{\myatop{\Qp}{\ell_2 m_2}}{\mathcal{J} \mathcal{J}}{\myatop{\Qp}{\ell_1 m_1}}\) correlators}%
\label{sec:VJJV}
The next correlation functions that we will examine are four-point functions between two spinning primary operators \(\Vpp[\Qp][\ell m]\) that are inserted at \(\tau_\text{in}\) and \(\tau_\text{out}\) and two currents \(\mathcal{J}_\mu\) inserted at times \(\tau > \tau' \) so that~\( \tau_\text{out} > \tau > \tau' > \tau_\text{in}\). \newline 
We start by inserting two \(\mathcal{J}_\tau\) as
\begin{align}
        &\expval{\Vpp* \mathcal{J}_\tau(\tau,\n) \mathcal{J}_\tau(\tau',\n') \Vpp } = - \Anew[\Delta_\Qp + R_0 \omega_{ \ell_2}] \frac{ \Qp^2 }{ \Omega_d^2 R_0^{2d-2}} \, \delta_{\ell_1 \ell_2} \delta_{m_1m_2} \nonumber\\
        &\shoveright{ \times \Bigg\{ 1 + \frac{ (d-1)^2  }{ 2 d \, \Delta_1 } \sum_{ \ell} e^{-|\tau -\tau'| \omega_{\ell} } R_0 \omega_{\ell} \frac{ (d+2\ell -2)}{ (d-2) } C^{ \frac{d }{2} -1}_{\ell} (\n \cdot \n') \Bigg\} } \nonumber \\
       & {+ \Bigg\{ \Anew[\Delta_\Qp + R_0 \omega_{ \ell_1}][\Delta_\Qp + R_0 \omega_{ \ell_2}][\tau] \frac{ \Qp^2 (d-1)^2 }{ 2 \Omega_d  R_0^{2d-2} d  } \frac{ R_0\sqrt{ \omega_{ \ell_1} \omega_{ \ell_2} } }{ \Delta_0 } \,\Bigg( - \frac{ Y_{\ell_2 m_2}^* (\n) Y_{\ell_1 m_1} (\n') }{ e^{- (\tau-\tau') \omega_{\ell_1}} } } \nonumber \\
        &+ \frac{ (d-2) }{ (d-1)} \bqty{\frac{ \del_i Y_{ \ell_1 m_1} (\n) \del_i Y_{\ell_2 m_2}^* (\n) }{ (d-1)\, R_0^2 {\omega_{ \ell_1} \omega_{ \ell_2}} } - Y_{\ell_1 m_1} (\n) Y_{\ell_2 m_2}^* (\n) } \Bigg)  +  \pqty{(\tau, \n) \leftrightarrow (\tau', \n') }  \Bigg\} ,
\end{align}
and in the above expression, we have introduced the Gegenbauer polynomials \(C^{{d}/{2}-1}_{\ell}\), which are defined as
\begin{equation}\label{eq:DefGegenbauerPolynomials}
 C^{d/2-1}_{\ell}(\n\cdot \n') = \frac{(d-2)\Omega_d}{d+2\ell-2} \sum_{m} Y_{\ell m }^* (\n) Y_{\ell m} (\n') .
\end{equation}
The case of $\ell_i=0$ was initially computed in~\cite{Cuomo:2020thesis}.

Again, a nice consistency check can be provided employing the Ward identity of \cref{eq:WardIdentitiesQ}. If we integrate the four-point function over the spatial hypersurface at the fixed temporal slice \((\tau, \n)\) we should recover the charge \(\Qp\) times an expression that does not depend on \(\tau\). Indeed, doing the computation, we find that
\begin{equation}
        \Int\limits_{\setS^{d-1}} \dd{S(\n)} \expval{\Vpp* \mathcal{J}_\tau(\tau,\n) \mathcal{J}_\tau(\tau',\n') \Vpp } = -i \Qp \expval{\Vpp* \mathcal{J}_{\tau}(\tau',\n') \Vpp} ,
\end{equation}
where we recover the expression of \cref{eq:3pointspinningandJ}.

The rest of the components of the \(\mel{\myatop{\Qp}{\ell_2 m_2}}{\mathcal{J} \mathcal{J}}{\myatop{\Qp}{\ell_1 m_1}}\) four-point function are
\begin{align}
  &\expval{\Vpp* \mathcal{J}_\tau(\tau, \n) \mathcal{J}_i (\tau', \n') \Vpp } = 0 , \\
  &\begin{multlined}[][.9\linewidth]
  \expval{\Vpp* \mathcal{J}_i (\tau, \n) \mathcal{J}_j (\tau',\n') \Vpp } = \Anew[\Delta_\Qp + R_0 \omega_{\ell_1}][\Delta_\Qp + R_0 \omega_{\ell_2}][\tau] \frac{ \Qp^2 }{2d \Omega_d R_0^{2d -2} \Delta_0  } \\
  \times \Bigg[ \del_i \del'_j \Sum_{ \ell} \frac{ e^{-|\tau -\tau'| \omega_{\ell} } }{ R_0 \omega_{\ell} } \frac{ (d+2\ell -2)}{ (d-2) \Omega_d} C^{\frac{d }{2} - 1}_{\ell} (\n \cdot \n') \, \delta_{ \ell_2 \ell_1} \delta_{m_2 m_1} \\
   + \frac{\del_j Y_{\ell_2 m_2}^* (\n') \del_i Y_{\ell_1 m_1} (\n) }{ e^{ (\tau - \tau') \omega_{\ell_2}} R_0 \sqrt{ \omega_{ \ell_1} \omega_{ \ell_2} } } 
  + \frac{ \del_i Y_{\ell_2 m_2}^* (\n) \del_j Y_{\ell_1 m_1}(\n') }{ e^{- (\tau- \tau') \omega_{ \ell_1}} R_0 \sqrt{\omega_{\ell_1} \omega_{\ell_2}}} \Bigg] .
\end{multlined}
\end{align}
We notice that for the four-point function with two insertions of the spatial current \(\mathcal{J}_i\) the leading order classical contribution vanishes, but since this correlation function is not protected by the symmetries, there are sub-leading terms that appear.

For \(\ell_i = 0 \) the aforementioned correlation functions simplify significantly,
\begin{align}
 &\begin{multlined}[][.9\linewidth]
    \expval{ \Opp* \mathcal{J}_\tau(\tau,\n) \mathcal{J}_\tau(\tau',\n') \Opp } = -   \Anew\frac{ \Qp}{ (\Omega_d R_0^{d -1})^2}  \\
    \times \bigg[ \Qp + \frac{(d-1) }{ 2 \mu } \Sum_{\ell} \omega_{ \ell}e^{-|\tau -\tau'| \omega_{\ell} }  \frac{ (d+2\ell -2)}{ (d-2)} C^{\frac{d}{2}-1}_{\ell} (\n \cdot \n') \bigg] ,
\end{multlined}\\
    &\expval{ \Opp* \mathcal{J}_i(\tau,\n) \mathcal{J}_j(\tau', \n') \Opp } =  \frac{ \Qp \Anew }{ 2 \mu \Omega_d (d-1) R_0^{2d-1}} \del_i \del'_j  \Sum_{\ell} \frac{ (d+2\ell -2) \, C^{ \frac{d }{2}-1}_{\ell} (\n \cdot \n') }{ (d-2) \Omega_d  \omega_{\ell} e^{ |\tau -\tau'| \omega_{ \ell} } } ,\\
    &\expval{ \Opp* \mathcal{J}_\tau(\tau, \n) \mathcal{J}_i (\tau', \n') \Opp } = 0 .
\end{align}
Homogeneity of the ground state of two scalar primaries \(\Opp\) ensures that the \(\mel{\Qp}{\mathcal{J}_i\mathcal{J}_j}{\Qp}\) correlator vanishes exactly due to rotational invariance. Moreover, these correlators satisfy \cref{eq:WardIdentitiesQ} as expected.


\subsection[\texorpdfstring%
{\(\mel{\myatop{\Qp}{\ell_2 m_2}}{T}{ \myatop{\Qp}{\ell_1 m_1}}\)}%
{<QTQ>}]%
{\(\mel{\myatop{\Qp}{\ell_2 m_2}}{T}{\myatop{\Qp}{\ell_1 m_1} }\) correlators}%
\label{sec:VTV}
Finally, we compute the three-point correlation functions between two spinning primary  operators \(\Vpp[\Qp][\ell m]\) inserted at \(\tau_\text{in} , \tau_\text{out}\) and the stress-energy tensor \(T\), inserted at a time \(\tau_\text{in} < \tau < \tau_\text{out}\).

We start by examining the \(T_{\tau\tau}\) component
\begin{multline}\label{eq:3pt_Ttautau}
    \expval{\Vpp* T_{\tau\tau} (\tau, \n)  \Vpp } = - \Anew[\Delta_\Qp + R_0 \omega_{\ell_{1}}][\Delta_\Qp + R_0 \omega_{\ell_{2}}][\tau] \frac{1 }{ \Omega_d R_0^d}  \Bigg\{ (\Delta_0+ \Delta_1) \delta_{\ell_2 \ell_1} \delta_{m_2 m_1} \\
    + \frac{ \Omega_d }{2} R_0\sqrt{ \omega_{ \ell_1} \omega_{ \ell_2} } \, \bigg[ (d-1) Y_{ \ell_2 m_2}^* (\n) Y_{\ell_1 m_1} (\n) 
    - \frac{(d-3) }{ (d-1)} \frac{ \del_i Y_{\ell_2 m_2}^* (\n) \del_i Y_{\ell_1 m_1} (\n) }{ R_0^2 { \omega_{ \ell_1} \omega_{\ell_2}}} \bigg] \Bigg\} .
\end{multline}
It is useful to note that at the limit of infinite separation, \emph{i.e.} \(\tau_\text{out}, \tau_\text{in} \to \pm \infty\) and when \(\ell_1=\ell_2\) so that the scaling dimension of the two spinning primaries is the same, the correlation function is completely independent of \(\tau\) as in \cref{eq.threepointnotau}. 

Moreover, integrating \cref{eq:3pt_Ttautau} over the spatial hypersurface \(\setS^{d-1}\) should satisfy the Ward identity of \cref{eq:WardIdentitiesD} and be \(\tau\) independent 
\begin{equation}
        \Int\limits_{\setS^{d-1}} \dd{S(\n)} \expval{\Vpp* T_{\tau\tau} (\tau, \n)  \Vpp } = - \Anew[ \Delta_\Qp + R_0 \omega_{ \ell}] \frac{1}{R_0} \pqty{ \Delta_0 + \Delta_1 + R_0 \omega_{\ell_2}} \, \delta_{\ell_1 \ell_2} \delta_{m_1 m_2} .
\end{equation}
The rest of the components of the correlation function \(\mel{\myatop{\Qp}{\ell_2 m_2}}{T}{\myatop{\Qp}{\ell_1 m_1} }\) are
\begin{align}
   &\expval{\Vpp*  T_{\tau i}(\tau, \n) \Vpp } = \Anew[\Delta_\Qp + R_0 \omega_{\ell_{1}}][\Delta_\Qp + R_0 \omega_{\ell_{2}}][\tau] \frac{1}{ 2 R_0^{d}}\bigg[ \sqrt{ \frac{ \omega_{ \ell_2} }{ \omega_{ \ell_1} } } Y_{\ell_2 m_2}^* (\n) \del_i Y_{\ell_1 m_1} (\n) - (1 \leftrightarrow 2 )^* \bigg],\label{eq:Tticorrelator} \\ 
  &\begin{multlined}[][.9\linewidth]
    \expval{\Vpp*  T_{ij}(\tau, \n) \Vpp } = \Anew[ \Delta_\Qp + R_0 \omega_{ \ell_1}][ \Delta_\Qp + R_0 \omega_{ \ell_2}][\tau] \frac{ 1 }{ (d-1)  \Omega_d R_0^d} \\
    \times \Bigg\{  h_{ij} \bqty{ \pqty{ \Delta_0 + \Delta_1 }  \delta_{ \ell_2 \ell_1} \delta_{m_2 m_1} 
     + \frac{\Omega_d R_0 \sqrt{\omega_{\ell_1} \omega_{\ell_2}} }{2} 
     \times \pqty{ (d-1)  Y_{ \ell_2 m_2}^* (\n) Y_{\ell_1 m_1} (\n) 
    - \frac{ \del_i Y_{ \ell_2 m_2}^* (\n) \del_i  Y_{\ell_1 m_1} (\n) }{ R_0^2 \omega_{ \ell_1 } \omega_{ \ell_2 } } } }  \\
    + R_0\sqrt{ \omega_{ \ell_1} \omega_{ \ell_2} } \Omega_d \frac{ \del_{ ( i} Y_{ \ell_2 m_2 }^* (\n)  \del_{j) }  Y_{\ell_1  m_1} (\n) }{ R_0^2 \omega_{ \ell_1} \omega_{ \ell_2}} \Bigg\} .
\end{multlined}
\end{align}
An important observation and a consistency check to our results is the following: from \cref{eq.tracelessEM} we know that the trace of the energy-momentum tensor is zero for a theory with conformal invariance. Therefore, computing the three-point correlation function with an insertion of the trace of \(T= T_{\tau \tau} + h^{ij} T_{ij}\) should be zero independent of the boundary states. That being the case 
\begin{equation}
    \expval{\Vpp* \pqty{T_{\tau \tau}(\tau, \n) + h^{ij} T_{ij}(\tau, \n)} \Vpp } = \expval{\Vpp* T_{\tau \tau}(\tau, \n)  \Vpp } + \expval{\Vpp* h^{ij} T_{ij}(\tau, \n) \Vpp }. 
\end{equation}
Since we have already computed \(\expval{\Vpp* T_{\tau \tau}(\tau, \n)  \Vpp }\) we just have to compute the correlator
\begin{multline}\label{eq.hijTij}
    \expval{\Vpp* h^{ij} T_{ij}(\tau, \n) \Vpp } = \Anew[ \Delta_\Qp + R_0 \omega_{ \ell_1}][ \Delta_\Qp + R_0 \omega_{ \ell_2}][\tau] \frac{ 1 }{ (d-1) \, \Omega_d R_0^d} \Bigg\{ (d-1) \left( \Delta_0 + \Delta_1\right)  \delta_{ \ell_2 \ell_1} \delta_{m_2 m_1}  \\
    + \frac{\Omega_d R_0 \sqrt{\omega_{\ell_1} \omega_{\ell_2}} }{2 } \bigg( (d-1)^2 \, Y_{ \ell_2 m_2}^* (\n) Y_{\ell_1 m_1} (\n) - (d-3) \, \frac{ \del_i Y_{ \ell_2 m_2}^* (\n) \del_i  Y_{\ell_1 m_1} (\n) }{ R_0^2 \omega_{ \ell_1 } \omega_{ \ell_2 } } \bigg) \Bigg\} ,
\end{multline}
therefore, adding together the results of \cref{eq:3pt_Ttautau,eq.hijTij} we see that they cancel each other out as predicted. \newline
In the case that \(\ell_i = 0 \) the above correlators simplify as
\begin{align}
  &\expval{ \Opp* T_{\tau\tau}(\tau, \n)  \Opp } = - \Anew \frac{ \Delta_0 + \Delta_1 }{ \Omega_d R_0^d} , \\
  &\expval{ \Opp* T_{\tau i}(\tau, \n) \Opp } = 0 , \\
  &\expval{\Opp* T_{ij}(\tau, \n) \Opp} = \frac{\Anew }{(d-1)} \frac{\Delta_0 + \Delta_1 }{ \Omega_d R_0^d} h_{ij} .
\end{align}
The correlation function with an insertion of \(T_{\tau i}\) has to be zero due to the rotational symmetry of the scalar ground state, which was already obvious from the fact that \cref{eq:Tticorrelator} did not contain a leading-order contribution. Furthermore, due to the homogeneity of the ground state, the three-point function with a single insertion of \(T_{ij}\) has to be analogous to the metric \(h_{ij}\) of the spatial hypersurface, which in this case is the \(\setS^{d-1}\) sphere. 

Finally, the four-point correlation function \(\mel{\myatop{\Qp}{\ell_2 m_2}}{T T}{ \myatop{\Qp}{\ell_1 m_1}}\) has in total six components that correspond to insertions of the energy-momentum tensors at \(\tau>\tau'\) between spinning primaries \(\Vpp[\Qp][\ell m]\) at \(\tau_\text{out},\tau_\text{in}\) such that \(\tau_\text{out} > \tau > \tau' > \tau_\text{in}\). Moreover, there are six components that belong to the correlation function \(\mel{\myatop{\Qp}{\ell_2 m_2}}{T\mathcal{J}}{\myatop{\Qp}{\ell_1 m_1}}\) with a single insertion of the conserved current \(\mathcal{J}_\mu\) and one insertion of the energy-momentum tensor \(T\) at times \(\tau > \tau'\).

Since these results are very lengthy and disrupt the flow of the thesis and at the same time they do not pose any additional computational challenge, they are moved to \Cref{sec:TTandTJcorrelators}.


\section{Heavy--light--heavy correlators}\label{sec:HLH}

In this final part of the chapter, we are interested in computing three and four-point correlation functions of spinning charged primary operators \(\Vpp[\Qp][\ell m]\) with the insertion of “light” charged spinning primary operators \(\mathscr{O}^q\) with \(q \ll \Qp\) in the middle. As long as the inserted operators have \(q \ll \Qp\), they do not affect the validity of the large-charge saddle point that we examined in \cref{sec.canonicalandpathintegralquantisation} but rather act as probes around it. \newline
Therefore, to complete our study of large-charge spinning correlators, we will shortly review the procedure of their computation that has originally appeared in~\cite{jafferis2018conformal,monin2017semiclassics,Cuomo:2020thesis}.

Laying down the groundwork, we start by noticing that in the validity of the \acrshort{eft} every operator should be constructed with regard to the Goldstone field \(\chi\)~\cite{monin2017semiclassics} by matching their quantum numbers. Therefore, assuming that the operator at hand has a small charge \(q\), a scaling dimension \(\delta\) and transforms in some representation of spin \(\ell\) it takes the form 
\begin{equation}
    \Opp[q][\delta][\ell m] = c^{(1)}_{\delta,\ell,q} \Proj^{\nu_1 \dots \nu_\ell}_{\qquad \ell m} \del_{\nu_1} \chi \dots \del_{\nu_\ell} \chi   \pqty{ \del \chi }^{\delta-\ell} e^{iq\chi} + \dots,
\end{equation}
where with \(c^{(1)}_{\delta,\ell,q}\) we denote a Wilsonian coefficient that does not depend on the charge, and cannot be determined in the scope of the \acrshort{eft}. Moreover, \(\Proj^{\nu_1 \dots \nu_\ell}_{\qquad \ell m}{}\) represents the change from Cartesian to spherical basis and is properly defined in \cref{eq:ProjectorToSphericalBasis}.

For the case that \(\ell=0\) we get
\begin{equation}\label{eq:lightchargeprimary}
    \Opp[q][\delta] = c^{(1)}_{\delta,q} \pqty{ \del \chi }^{\delta} e^{iq\chi} + \dots,
\end{equation}
where we use the shorthand notation
\begin{align}
  \Opp[q][\delta] &= \Opp[q][\delta][0 0] , & c^{(1)}_{\delta,q} &= c^{(1)}_{\delta,0,q} ,
\end{align}


\subsection[\texorpdfstring%
{\(\expval{ \Vpp[-\Qp-q][\ell_2 m_2]  \Opp[q][\delta] \Vpp }\)}%
{<QqQ>}]%
{The \(\expval{ \Vpp[-\Qp-q][\ell_2 m_2]  \Opp[q][\delta] \Vpp }\) correlator}%
\label{sec:VqV}

We start by computing the three-point function 
\begin{equation}
    \expval{ \Vpp[-\Qp-q][\ell_2 m_2] (\tau_\text{out})  \Opp[q][\delta](\tau_c, \n_c) \Vpp (\tau_\text{in}) },
\end{equation}
with a light scalar primary inserted at \(\tau_c\). The hermitian conjugation on the cylinder is similar to \cref{eq.cylinderconjugation} and reads
\begin{equation}
    \left[ \Opp[q][\delta] (\tau , \n) \right]^\dagger = \Opp*[q][\delta] (-\tau , \n) .
\end{equation}
From the general form of the three-point functions and the symmetries of the system the classical contribution should look like
\begin{equation}\label{eq:formof3pointwithsmallcharge}
        \expval{ \Vpp[-\Qp-q][\ell_2 m_2] (\tau_\text{out}) \,  \Opp[q][\delta](\tau_c, \n_c) \, \Vpp (\tau_\text{in}) } = \mathcal{C}^\delta_{\Qp +q, q, \Qp} \delta_{\ell_1\ell_2} \delta_{m_1 m_2}   e^{- \omega_{ \ell_2} (\tau_\text{out} - \tau_\text{in})} e^{- \Delta_{\Qp +q} \frac{ (\tau_\text{out} -\tau_c) }{R_0} } e^{- \Delta_{ \Qp} \frac{ ( \tau_c -\tau_\text{out}) }{R_0} } ,
\end{equation}
which is in accord with the form of \cref{eq:limit3pointCFT}. Using dimensional analysis, we see that a factor of \(R_0^\delta\) has to be extracted from the \acrshort{ope} coefficient, since it is originating in the insertion of \(\Opp[q][\delta]\) so that
\begin{equation}
R_0^{-\delta} \, \tilde{ \mathcal{C }}^\delta_{ \Qp +q ,q ,\Qp} = \mathcal{C }^\delta_{ \Qp +q ,q ,\Qp} .
\end{equation}
Our goal is to replicate the result of \cref{eq:formof3pointwithsmallcharge} in the semiclassical path-integral framework.

Inserting the small charge \(q\) operator in the system is similar to adding a source term in the action which may affect the saddle-point solution and changes the \acrlong{eom} to
\begin{align}
    \nabla_\mu \mathcal{J}^\mu &= \nabla_\mu \pdv{ S}{\pqty{\del_\mu \chi }} = \frac{i q   \delta (\tau - \tau_c) \delta (\n - \n_c) }{ \sqrt{g} }, &\begin{dcases} \mathcal{J}^\mu(\tau_\text{in}= -\infty, \n_1) = \frac{ \delta^{\mu}_0 \Qp }{ R_0^{d-1} \Omega_d} \\
    \mathcal{J}^\mu(\tau_\text{out}= +\infty, \n_2) = \frac{ \delta^{\mu}_0 ( \Qp + q) }{ R_0^{d-1} \Omega_d}
    \end{dcases} .
\end{align}
In the usual manner, when we have an inhomogeneous \acrfull{pde} with a source, we can write the solution as a combination of the homogeneous solution of the \acrshort{pde} where the source is zero and a particular solution which is identical to the Green's function. 

The new boundary conditions change the form of the \(l=0\) term but for small enough \(q\) to leading order the solution remains the same as for the homogeneous case, up to the additional particular solution to account for the inhomogeneity,
\begin{equation}\label{eq:newhomogeneousolution}
        \chi(\tau , \n) = - i \mu \tau + \pi_0 + \pi (\tau,\n) + q p \pqty{\tau, \n}.
\end{equation}
The particular solution \(p\) has to satisfy the inhomogeneous equation
\begin{equation}
         - c_1 \mu^{d-2} {d(d-1)} \pqty{\del_\tau^2 + \frac{1}{\pqty{d-1}} \Laplacian_{\setS^{d-1}} }p(\tau,\n) =  \frac{i q   \delta (\tau - \tau_c) \delta (\n - \n_c) }{ \sqrt{g} } .
\end{equation}
and has the solution
\begin{equation}
         p(\tau,\n) = -\frac{  i \, q }{ c_1 d (d-1) (\mu R_0)^{d-2} } \pqty{ - \frac{\abs{\tau - \tau_c}}{2 R_0 \Omega_d} \theta (\tau-\tau_c) + \mathlarger{\mathlarger{\sum}}_{\ell=1}^{\infty} \mathlarger{\mathlarger{\sum}}_m  e^{-\omega_\ell\abs{\tau-\tau_c}} \frac{ Y_{\ell m}(\n_c)^* Y_{\ell m} (\n)  }{ 2 R_0 \omega_\ell} } ,
\end{equation}
where \(\theta(\tau-\tau_c)\) is required to satisfy the different boundaries at \(\tau_{\text{in},\text{out}} = \mp \infty\) and also
\begin{equation}
    \mu \propto \Qp^{\frac{1}{(d-1)}} .
\end{equation}
This is indeed identical to the propagator in \cref{eq:Greenfunction}.

At \(\tau_\text{in} \to -\infty\) and \(\tau_\text{out} \to \infty\) the source only affects the overall normalisation of the fields, so they retain the form of the unperturbed case in~\cref{eq:pfielddecomposition,eq:Pifielddecomposition}. Thus, the reference states are
\begin{align}\label{eq:referencestates}
\lim_{\tau_{\text{out}} \to \infty } \bra{0}\Vpp[-\Qp-q][\ell_2 m_2] \pqty{\tau, \n}  & \coloneq \bra{\myatop{\Qp+q}{\ell_2 m_2}} = \bra{\Qp+q}  a_{\ell_2 m_2}, \\
 \lim_{\tau_{\text{in}} \to - \infty} \Vpp[\Qp][\ell_1 m_1]\pqty{\tau, \n} \ket{0} & \coloneq  \ket{\myatop{\Qp}{\ell_1 m_1}} = a^\dagger_{\ell_1 m_1} \ket{\Qp}.
\end{align}
Using the above results we can compute the correlation function to leading order as
\begin{equation}
    \begin{aligned}
    \expval{\Vpp[-\Qp-q][\ell_2 m_2]  \Opp[q][\delta](\tau_c, \n_c) \Vpp } &= \mel{\Qp+q}{  a_{\ell_2 m_2}  \Opp[q][\delta](\tau_c, \n_c)    a_{\ell_1 m_1}^{\dagger} }{\Qp}  \\
    &= c^{(1)}_{\delta,q} \mel{\Qp+q}{   a_{\ell_2 m_2}  \pqty{ \del \chi}^{\delta} e^{iq \chi(\tau_c, \n_c) }    a_{\ell_1 m_1}^{\dagger}}{\Qp}\\
    &= c^{(1)}_{\delta,q} \mu^\delta \mel{\Qp+q}   {a_{\ell_2 m_2}  e^{iq \chi(\tau_c, \n_c) }    a_{\ell_1 m_1}^{\dagger} }{\Qp} + \dots \\
    &= c^{(1)}_{\delta,q} \mu^\delta  e^{\mu q \tau_c} \mel{\Qp+q}{a_{\ell_2 m_2}   e^{iq   \pi (\tau_c, \n_c) + iq^2   p(\tau_c, \n_c) }    a_{\ell_1 m_1}^{\dagger}}{\Qp} + \dots \ , 
    \end{aligned}
\end{equation}
where in the first line we used \cref{eq:referencestates}, in the second line we used the form of the light scalar primary of \cref{eq:lightchargeprimary} and in the last line we used \cref{eq:newhomogeneousolution}. The leading term can further be expanded in powers of \(q \ll 1\) so that
\begin{align}
        \expval{ \Vpp[-\Qp-q][\ell_2 m_2] \,    \Opp[q][\delta](\tau_c, \n_c) \, \Vpp  } &= \frac{ c^{(1)}_{\delta,q} \mu^\delta }{  e^{ \mu q ( \tau_\text{out} - \tau_c ) } } \Bqty{\mel{\Qp}{a_{\ell_2 m_2}   a_{\ell_1 m_2}^{\dagger}}{\Qp} + \order{ q} } \nonumber \\
        &= (R_0\mu)^\delta \frac{ c^{(1)}_{\delta,q} }{ R_0^\delta } \delta_{\ell_2 \ell_1} \delta_{m_2 m_1} \, e^{- \Delta_{\Qp +q} \frac{ (\tau_\text{out} -\tau_c) }{R_0} - \Delta_{\Qp } \frac{ (\tau_c -\tau_\text{in}) }{R_0} }  e^{- (\tau_\text{out}-\tau_\text{in}) \omega_{ \ell_2} } \nonumber \\
        & =  \frac{ c^{(1)}_{\delta,q} (R_0\mu)^\delta}{ R_0^\delta } \delta_{\ell_2 \ell_1} \delta_{m_2 m_1} \Anew[\Delta_\Qp \, + \omega_{\ell}][\Delta_{\Qp + q } \, + \omega_{\ell} ][\tau_c] + \dots \, .\label{eq:VqV}
\end{align}
The above result can be generalised if we instead insert a light spinning operator in some representation of spin \(\ell\) as
\begin{equation}\label{eq:VVV}  
        \mel{\Qp +q}{a_{\ell_2 m_2} \Opp[q][\delta][\ell m] (\tau_c, \n_c)   a_{\ell_1 m_1 }^\dagger}{\Qp} = \frac{  c^{(1)}_{\delta,\ell,q} (R_0\mu)^\delta}{ R_0^\delta } \braket{ \ell_2 m_2  ; \ell, m}{\ell_1 m_1 } \Anew[\Delta_\Qp + \omega_{ \ell_1}][\Delta_{\Qp + q }+ \omega_{ \ell_2}][\tau_c]  + \dots \, ,
\end{equation}
and we denote with \( \braket{\ell_2 m_2 ; \ell, m}{\ell_1 m_1 } \) the relevant Clebsch-Gordan coefficient. Quantum corrections to \cref{eq:VqV,eq:VVV} have been computed in~\cite{jafferis2018conformal}.

Taking the special case that \(\ell_i=0\), the light operator \( \Opp[q][\Delta][\ell m] \) is sandwiched between two large-charge scalar primaries and since the overall charge should be conserved, we deduce that one of the scalar operators has charge \(-\Qp -q\). Moreover, due to rotational invariance of the homogeneous ground state, the only light operators that we can insert with no-trivial results are for \(\ell = 0\) as
\begin{equation}
   \mel{\Qp+q}{\Opp[q][\delta][\ell m] (\tau_c , \n_c)}{\Qp} \propto  \delta_{\ell,0} \, \mu^\delta e^{- \Delta_{Q} \frac{ (\tau_\text{out} -\tau_\text{in}) }{R_0} } e^{ \mu q ( \tau_c -\tau_\text{out})} = \mu^\delta \Anew[\Delta_\Qp ][\Delta_{\Qp+q} ][\tau_c] \delta_{\ell 0} \, .
\end{equation}
The above correlation function was initially derived in~\cite{monin2017semiclassics,cuomo2021note}. The \acrshort{ope} coefficient has also been computed to be
\begin{equation}\label{eq:OPElight}
   \tilde{\mathcal{C }}^\delta_{ \Qp +q ,q ,\Qp} =  c^{(1)}_{\delta,q} (R_0\mu )^\delta \bqty{ 1 - \frac{ \frac{q^2}{2 } \mathlarger{\sum}_{ \ell ,m} \frac{ Y_{\ell m}^* (\n_c) Y_{\ell m} (\n_c) }{ R_0 \omega_\ell } }{ c_1 d (d-1) (R_0\mu)^{d-2} } + \dots } + \dots \, .
\end{equation}
In the case of \(d=3\) the dominant correction is the first one, since it is solely suppressed by \(\mu \sim \sqrt{\Qp}\) and it can be computed via zeta function regularization utilising that
\begin{equation}
   \mathlarger{\sum}_{\ell,m} \frac{ Y_{\ell m}^* (\n_c) Y_{\ell m} (\n_c) }{ \omega_\ell } = \frac{1}{\Omega_d} \mathlarger{\sum}_{\ell} \frac{ M_{\ell} }{ \omega_{ \ell}} = \frac{\sqrt{d-1} R_0 \zeta_{\setS^{d-1}}(1/2)}{\Omega_d},   
\end{equation}
where \(M_\ell\) is the degeneracy of \(\omega_\ell\) and we used the results of \Cref{sec:Ylm-identities}. So, at \(d=3\) the \acrshort{ope} coefficient of \cref{eq:OPElight} are found to be \cite{cuomo2021note}
\begin{equation}
\tilde{\mathcal{C }}^\delta_{ \Qp +q ,q ,\Qp} \propto 
\pqty{\Qp}^{\frac\delta2} \bqty{ 1 + 0.0164523 \times \frac{ q^2 \sqrt{12\pi} }{\sqrt{c_1 \Qp} } + \dots } + \dots .
\end{equation}


\subsection[\texorpdfstring%
{\(\expval{\Vpp* \Opp[-q][\delta] \Opp[q][\delta] \Vpp }\)}%
{<QqqQ>}]%
{The \(\expval{\Vpp* \Opp[-q][\delta] \Opp[q][\delta] \Vpp }\) correlator}%
\label{sec:VqqV}

Finally, we generalise the result of \cref{sec:VqV} by computing the four-point function
\begin{equation}
    \expval{ \Vpp[-\Qp-q_d-q_c][\ell_2 m_2] (\tau_2) \, \Opp[q_d ][\delta_d] (\tau_d, \n_d) \, \Opp[q_c][\delta_c] (\tau_c, \n_c) \, \Vpp[ \Qp][\ell_1 m_1] (\tau_1) } ,
\end{equation}
where \(q_d \sim q_c \ll \Qp\). These form of operators were initially presented in \cite{jafferis2018conformal}.

The new \acrshort{eom} with the double scalar insertions are rather similar to the \acrlong{eom} of the three-point correlation function with just one light scalar insertion
\begin{align}
    \nabla_\mu \mathcal{J}^\mu &= \frac{i q_d \delta (\tau - \tau_d) \delta (\n - \n_d) }{ \sqrt{g}} + \frac{i q_c  \delta (\tau - \tau_c) \delta (\n - \n_c) }{ \sqrt{g}} ,
     &\begin{cases}
     \mathcal{J}^\mu(\infty, \n) = \frac{ \delta^{\mu}_0 (\Qp + q_d + q_c) }{ R_0^{d-1} \Omega} , \\ \mathcal{J}^\mu(-\infty, \n) = \frac{ \delta^{\mu}_0 \Qp }{ R_0^{d-1} \Omega} .
     \end{cases}
\end{align}
At leading order the four-point function reads
\begin{align}
    \expval{ \Vpp[-\Qp- q_d- q_c][\ell_2 m_2]  \, \Opp[q_d][\delta_d]  \, \Opp[q_c][\delta_c]  \, \Vpp[\Qp][\ell_1 m_1] } &=  \frac{ c^{(1)}_{\delta_d,q_d} c^{(1)}_{\delta_c,q_c} }{ \mu^{- \delta_d - \delta_c} e^{ \mu (q_d+ q_c) \tau_\text{out}} } \mel{\Qp}{a_{\ell_2 m_2} e^{iq_d \chi } e^{iq_c \chi }    a_{\ell_2 m_2}^{\dagger}}{\Qp} + \dots \nonumber \\
    & = \frac{ c^{(1)}_{\delta_d,q_d} c^{(1)}_{\delta_c,q_c} }{ \mu^{-\delta_d -\delta_c} } e^{- \mu q_d (\tau_\text{out} - \tau_d) - \mu q_c (\tau_\text{out} - \tau_c)  } \mel{\Qp}{a_{\ell_2 m_2} e^{- \frac{(\tau_\text{out}-\tau_\text{in})}{R_0}   D }   a_{\ell_2 m_2}^{\dagger}}{\Qp} + \dots \nonumber \\
    &\begin{multlined}[][.6\linewidth]
    = (R_0\mu)^{\delta_d +\delta_c} \frac{ c^{(1)}_{\delta_d,q_d} c^{(1)}_{\delta_c,q_c} }{ R_0^{ \delta_d +\delta_c} }  \delta_{\ell_2 \ell_1} \delta_{m_2 m_1} e^{ - (\tau_\text{out}-\tau_\text{in}) \omega_{ \ell_2} } \\
    \times e^{- \Delta_{\Qp +q_d+q_c} \frac{ (\tau_\text{out} -\tau_c) }{R_0} - \Delta_{\Qp+q_c} \frac{ (\tau_d -\tau_c) }{R_0} - \Delta_{\Qp } \frac{ (\tau_c -\tau_\text{in}) }{R_0} } + \dots
    \end{multlined}
\end{align}
For the special case that \(q=q_c=-q_d\) the correlation function becomes
\begin{equation}
    \expval{ \Vpp[-\Qp][\ell_2 m_2]  \,\Opp[-q][\delta] \, \Opp[q ][\delta]\, \Vpp[\Qp][\ell_1 m_1] } = (R_0\mu)^{2 \delta} \frac{ \abs{c^{(1)}_{\delta ,q} }^2 }{R_0^{2\delta} } e^{- \Delta_{\Qp} \frac{ (\tau_\text{out} - \tau_\text{in}) }{R_0} - q \frac{ \del \Delta_\Qp }{\del \Qp }  \frac{ (\tau_d - \tau_c) }{R_0} - (\tau_\text{out} -\tau_\text{in}) \omega_{ \ell_2} } \delta_{ \ell_2 \ell_1} \delta_{m_2 m_1} + \dots 
\end{equation}
Also, next-to-leading order corrections of the above correlation function have appeared in~\cite{jafferis2018conformal}.

\bigskip

As a final remark for this chapter, the analysis that we performed holds true for the case of the \(O\pqty{2}\) model in \(d\) spacetime dimensions at large fixed charge, or more generally for a \acrshort{cft} that has the homogeneous \(O\pqty{2}\) sector as a part of a larger symmetry group like the \(O\pqty{N}\). But once we want to move away from the homogeneous solution and study the full non-Abelian sector, things get more involved. The first thing that we notice is that the fixing parameter should not be just a set of charges, but a whole representation. For example, studying the homogeneous sector \cite{alvarez2017compensating} of \(O\pqty{2}\) corresponds to studying the completely symmetric representation. But other representations beyond the completely symmetric can be reached in one of the following ways: 
\begin{enumerate}
    \item Exciting the type II Goldstones \cite{nielsen1976count,watanabe2013massive,watanabe2020counting} that we mentioned in \cref{sec.Goldstonetheorem} and are charged under the global symmetry.
    \item Start from the beginning from an inhomogeneous ground state, that corresponds to a separate saddle point. 
\end{enumerate}
It has been theorised that both approaches should produce the same results in the correct limit, but each of them has its own technical challenges. To start with, the type II Goldstones come in at order \(1/\mu\)~\cite{alvarez2017compensating}, which also explains why they played no part in the computations of this chapter, therefore to capture their contribution, there is the requirement that more sub-leading terms shall be added to the \acrshort{eft}~\cite{Gaume:2020bmp}. On the other hand, besides the relatively easy case of the \(O(4)\) model, there is no other known inhomogeneous saddle in the literature, and even there an analytic expression is only accessible in a very special limit~\cite{banerjee2019conformal}. Finally, inhomogeneous ground states break the \(SO(d)\) rotational invariance that has proven so crucial in our analysis, therefore the computation of correlators around the inhomogeneous saddle for both tree-level and quantum fluctuations will be much more technically involved.

%

\chapter[\texorpdfstring%
{Resurgence and the \(O(N)\) vector model}%
{O(N)}]%
{Resurgence and the \(O(N)\) vector model}%
\label{Chapter4}

\epigraph{\itshape “It doesn't matter how beautiful your theory is, it doesn't matter how smart you are. If it doesn't agree with experiment, it's wrong.”}{Richard P. Feynman}

One of the most impressive results of the early days work in the large-charge expansion is that it seems to work for small charges too, a result that is quite astounding as it is. Generally, we expect that the semiclassical expansion should work only for systems that exhibit numerous \acrshort{dof}. This detail was initially discovered during the comparison of the results for the conformal dimension \(\Delta_{\Qp}\) of the lowest charged operator \(\Opp\) in the large charge expansion and the lattice simulations for the cases of the  \(O(2)\) model~\cite{banerjee2018conformal} and then again for the \(O(4)\) model~\cite{banerjee2019conformal}. It was pointed out that considering only a minimal number of terms in the effective action was enough to replicate the results of the lattice with high precision.

From the viewpoint of the \acrshort{eft}, it cannot be explained why the results in the large-charge limit can be extrapolated to the small charge regime. Nevertheless, by adding another controlling parameter, \emph{e.g.} large \(N\), to the theory beyond the large charge \(\Qp\), we can move beyond the validity of the large-charge \acrshort{eft} and try to explain this behaviour. A neat example is the \(O(2N)\) vector model at large charge and large \(N\)~\cite{alvarez2019large,giombi2021large}. When we find ourselves in the double-scaling limit defined as \(\Qp\to \infty, \, \,  N\to \infty\), with their ratio \(\Qp/(2N) = \Qb \) being constant, we are in a position to make exact predictions at leading order in \(N\) for every value of the charge \(\Qb\). 

Therefore, based on the work of Alvarez-Gaume et al. \cite{alvarez2019large}, in this chapter we go a step further and demonstrate that the \acrshort{lce} studied in the double-scaling regime is, in fact, asymptotic, and this characteristic is connected to the asymptotic nature of the Seeley--DeWitt expansion~\cite{de1963dynamical,seeley1967complex} of the heat-kernel trace and the associated zeta function on the two-sphere \(\setS^2\)~\cite{CANDELAS1984397,dowker2005barnes}.  Generically, it was asserted by Dyson~\cite{PhysRev.85.631} that asymptotic series are usually encountered in perturbative solutions of quantum mechanical problems. This characteristic implies the existence of non-perturbative phenomena in the underlying theory, and the contemporary approach to the topic is called resurgence asymptotics or, for simplicity, resurgence. Present day reviews include ~\cite{Dorigoni_2019,Aniceto_2019} and the interested reader is referred there for a list of applications both in physics and in mathematics, and also for further references.

The plan of this chapter is as follows: in  \cref{sec:asymptotics}, we start by reviewing the \(O(2N)\) vector model in \(d=3\) spacetime dimensions at the \acrshort{wf} fixed point, and then we move on to study its asymptotic behaviour. To do so, we employ the resurgent methodology, and we illustrate which non-perturbative corrections appear when we go to the double-scaling limit of the \acrshort{lce}. More specifically, in \cref{sec:torus} we study the system on the torus \(\setT^2\) and in \cref{sec:perturbative-sphere} we repeat our analysis for the sphere \(\setS^2\), writing our results in terms of both the usual perturbative series and the newly found exponentially suppressed non-perturbative contributions. Then in \cref{sec:worldline}, we start in \cref{sec:path-integral} by reformulating our analysis and developing a geometrical picture where we reforge the heat-kernel in the form of a path integral of a quantum mechanical particle  moving along closed geodesics. In \cref{sec:torus-path-integral} we study again the case of the torus, while in \cref{sec:sphere-path-integral} we discuss the sphere and interpret the previously found exponential contributions in the form of worldline instantons of a particle that has a mass which is equal to the chemical potential $\mu$ and is moving along the aforementioned geodesics. Therefore, in \cref{sec:resurgence}, we put together the outcome of the two prior sections, and we derive the precise form of the grand potential that is suitable for every charge value \(\Qp\). Then we can extrapolate our large charge results to the small charge regime, to find out that they match with great accuracy, a fact that is also numerically substantiated. Finally, in  \cref{sec:lessons-from-large-N}, we conjecture that the geometrical interpretation of the wordline geodesics is robust enough to work for the generic case of finite \( N\), which is beyond the double-scaling limit. This fact lends credibility towards the general validity of our analysis. By making the assumption that the qualitative characteristics of the geometrical interpretation also carry on to finite  \( N\), we can utilise our knowledge of the effects that originate from the leading exponential terms to obtain some constraints on the form of the perturbative expansion, although in finite \(N\) the Wilsonian coefficients are not reachable within the validity of the \acrshort{eft}. In the end, we find out that only a few number of terms are more than enough to fully describe the theory and match the small charge analysis with an excellent precision as estimated by the lattice computations.


\section{Asymptotics at large charge}%
\label{sec:asymptotics}

Our goal is to apply the resurgence methodology to study the asymptotics
of the \(O(2N)\) vector model in the limit that both the charge \(\Qp\) and \(N\) are large, which we call the double-scaling limit.

More specifically, after a brief introduction to the model in an abstract compact pseudo-Riemannian manifold, we examine the system for the case of the torus, \(\setS^1 \times \setT^2\) and the sphere, \(\setS^1 \times \setS^2\) writing our results in terms of both the usual perturbation series but also the non-perturbative corrections which are exponentially suppressed.

\subsection[\texorpdfstring%
{\(O(2N)\) model}%
{O(2N)}]%
{\(O(2N)\) model}%
\label{sec:O2N}

We start our analysis by considering the Landau--Ginzburg \(O(2N)\) model with real scalar fields in \(d=3\) spacetime dimensions on the Euclidean manifold \(\pqty{\mathcal{M} = \setR \times \Sigma}\), where \(\Sigma\) is a compact pseudo-Riemannian two-dimensional surface of volume \(V\). The generating functional \footnote{Remember from \cref{eq.partitionfunctionandgenerating} that the generating functional \(\mathcal{Z}\bqty{J=0}\) at \(\setR \times \Sigma\) and the partition function \(\mathcal{Z}\pqty{\beta}\) at \(\setS^1 \times \Sigma\) are the same.} \(\mathcal{Z}\) at \(J=0\) is
\begin{equation}
    \mathcal{Z}\bqty{J=0} \coloneqq \mathcal{Z} = \int \mathcal{D}\phi_i \, \exp\pqty{-S[\phi_i]},
\end{equation}
where the action of the model up to \(\bqty{m}=3\) is given by
\begin{equation}
  S[\phi_i] = \mathlarger{\mathlarger{\sum}}_{i=1}^{2N}  \mathlarger{\int} \dd{\tau} \dd{\Sigma} \bqty{\frac{1}{2} g^{\mu \nu} \del_\mu \phi_i \del_\nu \phi_i + \frac{r}{2} \phi_i \phi_i  + \frac{u}{4}  \pqty{\phi_i \phi_i}^2 + \frac{v}{8}  \pqty{\phi_i \phi_i}^3}.
\end{equation}
The coupling \(r\) contains the conformal coupling \(\xi\), hence the action is Weyl invariant at the limit that $u=v=0$. In the spirit of \cref{eq.O2toU1transformationfields} we pair the fields in complex scalars as

\begin{align}
    \varphi_i & = \frac{1}{\sqrt{2}} \left( \phi_i + i \phi_{N+i} \right), & \varphi^{*}_i & = \frac{1}{\sqrt{2}} \left( \phi_i - i \phi_{N+i} \right).
\end{align}
We note that the \(O(2N)\) symmetry is not evident any more, while a \(U(N) \subset O(2N)\) symmetry emerges. In terms of the complex scalar fields, the above action reads 
\begin{equation}\label{eq:UV-Hamiltonian}
  S[\varphi_i] = \mathlarger{\mathlarger{\sum}}_{i=1}^{N}  \mathlarger{\int} \dd{\tau} \dd{\Sigma} \bqty{g^{\mu \nu} \del_\mu \varphi_i^* \del_\nu \varphi_i + r \varphi_i^* \varphi_i  + \frac{u}{2}  \pqty{\varphi_i^* \varphi_i}^2 + \frac{v}{4}  \pqty{\varphi_i^* \varphi_i}^3}.
\end{equation}
At this point we introduce two new fields \cite{zinn2021quantum}, a Lagrange multiplier \(\lambda\) and an auxiliary field \(\rho\) and we impose the constraint
\begin{equation}
    \rho = \varphi^*_i \varphi_i,
\end{equation}
by the following integral over the Lagrange multiplier \(\lambda\) using the identity
\begin{equation}
    1 = \int \dd{\rho} \delta\pqty{\varphi^*_i \varphi_i - \rho} = \frac{1}{2 \pi i} \int \dd{\rho} \dd{\lambda} \exp\pqty{\lambda \pqty{\varphi^*_i \varphi_i -\rho}},
\end{equation}
where we note that in the complex plane of \(\lambda\) the above contour integral runs parallel to the imaginary axis. This transformation corresponds to a different representation \cite{domb2000phase,brezin1976field} of the generating functional 
\begin{equation}
    \mathcal{Z} = \mathlarger{\int} \mathcal{D} \varphi_i \, \mathcal{D} \rho \, \mathcal{D} \lambda \, \exp\pqty{- S\bqty{\varphi_i, \rho, \lambda}},
\end{equation}
where the action is 
\begin{equation}
    S\bqty{\varphi_i, \rho, \lambda} = \mathlarger{\mathlarger{\sum}}_{i=1}^{N}  \mathlarger{\int} \dd{\tau} \dd{\Sigma} \bqty{g^{\mu \nu} \del_\mu \varphi_i^* \del_\nu \varphi_i + \pqty{r + \lambda} \varphi_i^* \varphi_i  -\rho \lambda + \frac{u\rho^2}{2}  + \frac{v\rho }{4} \pqty{\varphi_i^* \varphi_i}^2}.
\end{equation}
The above-stated integral is Gaussian in \(\rho\) and we can integrate it out from the action. This leads to a new effective \footnote{A \(\pqty{\varphi^* \varphi}^4\) term is dropped as irrelevant in the \acrshort{ir}.} action 
\begin{align}
\mathcal{Z} &= \mathlarger{\int} \mathcal{D} \varphi_i  \, \mathcal{D} \lambda \, \exp\pqty{- S\bqty{\varphi_i, \lambda}}, \\
S\bqty{\varphi_i, \lambda} &= \mathlarger{\mathlarger{\sum}}_{i=1}^{N} \mathlarger{\int} \dd{\tau} \dd{\Sigma} \bqty{g^{\mu \nu} \del_\mu \varphi_i^* \del_\nu \varphi_i + \pqty{r + \lambda} \varphi_i^* \varphi_i  - \frac{\lambda^2}{2u} + \frac{v}{4u} \lambda \pqty{\varphi^* \varphi}^2}. \label{eq:O2Neffectiveaction} 
\end{align}
The former Lagrange multiplier $\lambda$ is now promoted to a real field via the above Hubbard--Stratonovich transformation~\cite{hubbard1959calculation,stratonovich1957method}.

If we fine-tune \(r\) to the value of the conformal coupling \cite{orlando2022following,moser2022convexity}, and we set 
\( v /u \to 0, \ \text{when} \ u \to \infty\) the theory flows from a free \acrshort{uv} fixed point at \(u=v=0\) to a strongly coupled \acrshort{ir} conformal point, the \acrlong{wf} fixed point.


\subsubsection{Large charge at large N}%
\label{sec:largeN}

So at this point, we want to shortly review the results of~\cite{alvarez2019large} --- where the interested reader is referred to for a much more extended discussion. We start with \cref{eq:O2Neffectiveaction} and by fine-tuning the relevant parameters we reach the \acrshort{wf} fixed point, where the action reads
\begin{equation}
	S[\varphi_i, \lambda] = \mathlarger{\mathlarger{\sum}}_{i=1}^{N} \mathlarger{\int} \dd{t}\dd{\Sigma} \bqty{ \del_\mu\varphi^*_i\del^\mu\varphi_i +(\xi R+\lambda)\varphi^*_i\varphi_i },
\end{equation}
and \(R\) is  Ricci curvature scalar of \(\Sigma\) while \(\xi =  1/8\) is the conformal coupling. Note that the \(N\) fields have the same coupling to the field \(\lambda\), therefore they are not independent and the symmetry of the system is \(U(N)\) instead of \(U(1)^N\).

The Noether charges of \cref{eq.charge} associated to the subgroup \(U(1)^N \subset U(N)  \subset O(2N)\) are
\begin{equation}
	\hat{\Qp}_i = \mathlarger{\int} \dd{\Sigma} \mathcal{J}^{\tau}_i = i \mathlarger{\int} \dd{\Sigma} \bqty{ \dot{\varphi}^*_i \varphi_i - \varphi^*_i \dot{\varphi}_i },
\end{equation}
and act as rotations on the complex fields. Moreover, recall that our manifold \(\Sigma\) is compact, and therefore the surface terms are zero.

We compactify our temporal dimension on a circle \(\setS^1_{\beta}\) of circumference \(\beta\) and when \(\beta \to \infty\) we recover the usual \(\setR\). We want to compute the fixed-charge partition function on \(\setS^1_\beta \times \Sigma\) which is formally defined as
\begin{align}
\mathcal{Z}_c( \beta, \Qp_i) &= \Tr\bqty{e^{- \beta H } \mathlarger{\prod}_{i=1}^N \delta\pqty{\hat{\Qp}_i-\Qp_i} } \label{eq:Definitioncanonicalpartion} \\
&= \mathlarger{\int}\limits_{-\pi}^{\pi}  \mathlarger{\prod}_{i=1}^N \, \frac{\dd{\theta_i}}{2\pi} \, \mathlarger{\prod}_{i=1}^N \, e^{i \theta_i \Qp_i}   \, \Tr\bqty{ e^{-\beta H} \, \mathlarger{\prod}_{i=1}^N \, e^{- i \theta_i \hat{\Qp}_i}}.
\end{align}
To get from the first to the second line, when we exponentiated the delta function we utilised the fact that the charges commute, \emph{i.e.} \(\comm{\Qp_i}{\Qp_j} = 0\), \(\forall \,i, \, j \) since they are all Abelian. Moreover, charge quantisation indicates that the eigenvalues of $\hat{\Qp}_i$ are integers, and hence we used a Fourier transform on the fundamental domain \(\theta_i \in \bqty{-\pi,\pi}\). 

\Cref{eq:Definitioncanonicalpartion} is the partition function of the canonical ensemble, while  
\begin{equation}\label{eq:grandcanonical}
\mathcal{Z}_{gc}(\beta, \mu_i )= \Tr\bqty{ e^{- \beta H} \, \mathlarger{\prod}_{i=1}^N \, e^{ \pqty{ \mu_i \hat{\Qp}_i} \beta}  }_{\mu_i = - i \theta_i /\beta} \, ,
\end{equation} 
is the grand-canonical partition function with imaginary chemical potentials \(\mu_i = -  i \theta_i/\beta\).

In our case, it suffices to fix the charge that corresponds to the Cartan generator that rotates the field \(\varphi_N\) which means that we confine ourselves to the completely symmetric representation of rank \(\Qp\) --- for details about the fixing of the charge for the \(O(2N)\) model the interested reader is referred to \cite{antipin2020charging} and also~Section~4.1 in~\cite{Gaume:2020bmp}. Therefore, we have
\begin{align}
\mathcal{Z}_c \pqty{\Qp, \beta} & = \Tr\bqty{e^{-\beta H} \delta\pqty{\hat{\Qp} - \Qp}} \nonumber\\
& = \mathlarger{\int}\limits_{-\pi}^{\pi} \frac{\dd{\theta}}{2 \pi} e^{i \theta \Qp} \Tr\bqty{e^{- \beta H - i \theta \hat{\Qp}} }.
\end{align}
Given the fact that the trace above contains the current \(\mathcal{J}^\tau\) that depends on the momenta, summing over the momenta is not that trivial. To remedy that, we introduce a constant background field along the temporal direction for the now gauged \(U(1)\) symmetry so that the trace can be written as
\begin{equation}
    \Tr\bqty{e^{- \beta H - i \theta \hat{\Qp}} } = \mathlarger{\int}\limits_{\text{PBC}} \, \DD{\varphi} \, \DD{\lambda} \, e^{-S_\theta[\varphi,\lambda]} \, ,
\end{equation}
where PBC means periodic boundary conditions, \emph{i.e.} \(\varphi_N \pqty{\beta, \n} = \varphi_N \pqty{0, \n}\). \newline
The modified action \(S_\theta\) reads
\begin{equation}\label{eq:gaugedaction}
	S_\theta[\varphi,\lambda] = \mathlarger{\int} \dd{\tau}\dd{\Sigma} \bqty{\mathlarger{\mathlarger{\sum}}_{i=1}^{N-1} \del_\mu\varphi^*_i \del^\mu\varphi_i +D_\mu\varphi_N^* D^\mu\varphi_N + \pqty{\xi R+\lambda} \mathlarger{\mathlarger{\sum}}_{i=1}^{N}\varphi^*_i \varphi_i } \, ,
\end{equation}
with the covariant derivative \(D_\mu\), defined as
\begin{equation}
	D_\mu \varphi_N  =
    \begin{cases}
		\pqty{\del_\tau +i \frac{\theta}{\beta}}\varphi_N & \text{if } \mu =0, \\
		\del_i\varphi_N & \text{otherwise}.
     \end{cases}
\end{equation}
We observe that \cref{eq:gaugedaction} is quadratic in the \(N-1\) fields \(\varphi_i\) and we can integrate them out to yield an effective action for the
Lagrange multiplier \(\lambda\) and the field \(\varphi_N\) as
\begin{equation}\label{eq:partitionoflambdaandphi}
  \mathcal{Z}_c(\Qp, \beta) = \mathlarger{\int}\limits_{- \pi}^{\pi} \frac{\dd{\theta}}{2\pi} \, e^{i\theta \Qp} \, \DD{\varphi_N} \, \DD{\lambda} \, e^{-S_\theta[\varphi_N,\lambda]},
\end{equation}
where the action reads
\begin{equation}\label{eq:actionoflambdaandphi}
  S_{\theta}[\varphi_N,\lambda] = \pqty{N-1} \Tr\bqty{\log\pqty{ -\del_\tau^2 - \mathlarger{\Laplacian} + \xi R+ \lambda }}
  +\mathlarger{\int} \dd{\tau} \dd{\Sigma} \bqty{D_\mu\varphi_N^* D^\mu\varphi_N + \pqty{\xi R+\lambda}\varphi^*_N \varphi_N }.
\end{equation}
At this point it is useful to take a look at the inverse propagator for the field \(\varphi_N\). In the usual manner of \cref{sec.pathintegral}, we integrate the action by part, and we pass to Fourier space where we find that 
\begin{equation}\label{eq.inversepropagator}
    \Delta_F^{-1} = \Pmqty{0 & \pqty{\omega- \frac{\theta}{\beta}}^2 + p^2 +\pqty{\xi R + \lambda} \\ \pqty{\omega- \frac{\theta}{\beta}}^2 + p^2 +\pqty{\xi R + \lambda}  & 0 }. 
\end{equation}
The zeros of the inverse propagator \(\det\pqty{\Delta_F^{-1}} =0\) lie at 
\begin{equation}
    \omega^2 + \pqty{\sqrt{\xi R + \lambda} \pm \mu}^2 + \pqty{1 \pm \frac{\mu}{\sqrt{\xi R + \lambda}}}p^2 \mp \frac{\mu}{4 \pqty{\xi R + \lambda}^{3/2}} p^4 + \dots = 0,
\end{equation}
where we used that the chemical potential is \(\mu = -i \theta / \beta \). The matrix of \cref{eq.inversepropagator} is invertible, except for \(\mu^2 = \xi R + \lambda \) where it becomes singular, and the field \(\varphi_N\) will exhibit a non-trivial zero mode. What we see here is a manifestation of the Goldstone theorem of \cref{sec.Goldstonetheorem}, in the sense that a non-zero \acrshort{vev} which is the zero mode indicates a \acrshort{ssb} of the global symmetry and the existence of massless modes in the spectrum. 

Going back to \cref{eq:partitionoflambdaandphi}, the path integral in the partition function localises at the saddle point that minimises the action \(S_{\Qp} = - i \theta \Qp + S_{\theta} \bqty{\varphi_N, \lambda}\) with respect to \(\theta\) and the zero modes of the two fields. Therefore, knowing that there can be zero modes, we decompose the fields into \acrshort{vev}s plus fluctuations
\begin{align}
	\varphi_N &= \frac{A}{\sqrt{2}} + u, & \expval{u} & =0,  \label{eq:vevforvarphi}\\
 \lambda&= \mu^2 - \xi R +\hat \lambda = m^2 + \hat \lambda, & \expval{\hat{\lambda}} & = 0 . \label{eq:vevforlambda}
\end{align}
In \cref{eq:vevforlambda} we introduced the parameter \(m\), which is the mass relative to the conformal Laplacian \(\mathlarger{\Laplacian}{} - \xi R\), while \(\mu\) can be interpreted as the mass related to the Laplace--Beltrami operator \(\mathlarger{\Laplacian} = \nabla \cdot \nabla\).

Adding \cref{eq:vevforvarphi,eq:vevforlambda} into \cref{eq:actionoflambdaandphi} the action reads
\begin{equation}
	S_{\theta} = (N-1)\Tr\bqty{\log \pqty{-\del_\tau^2{}-\mathlarger{\Laplacian}{} + \mu^2 + \hat \lambda }} + \mathlarger{\int} \dd{t} \dd{\Sigma} \bqty{D_\mu u^* D^\mu u + \frac{A^2\theta^2}{2\beta^2} + (\mu^2+\hat\lambda)\left|\frac{A}{\sqrt{2}}+u \right|^2 },
\end{equation}
We can perform the quadratic path integral --- for which see \cite{alvarez2019large} for details --- in order to get an effective action written in terms of non-local terms for the fluctuations \(\hat{\lambda}\).

Nonetheless, the saddle point equations are derived by minimising \(S_{\Qp} = - i \theta \Qp + S_{\theta}\) at the zero of the fluctuations \(\hat{\lambda} =0\). Then the action of the saddle point reads
\begin{equation}\label{eq:SQ}
  S_\Qp^{\saddle} = -i\theta Q +\pqty{N-1}\Tr\bqty{\log \pqty{ -\del_\tau^2{}- \mathlarger{\Laplacian}{} + \mu^2 }} + \frac{V\beta A^2}{2}\pqty{\frac{\theta^2}{\beta^2} + \mu^2}.
\end{equation}
We minimise the action \(S_\Qp^{\saddle}\) with respect to \(\mu{}, A\) and \(\theta\) as
\begin{equation}\label{eq;saddlepoint}
	\begin{cases}
	\pdv{S_\Qp^{\saddle}}{\theta}	= & -i\Qp \,+ \, \frac{\theta}{\beta}VA^2 = 0,\\
	\pdv{S_\Qp^{\saddle}}{\mu}	= & \pqty{N-1}\pdv{\mu} \Tr\bqty{\log \pqty{ -\del_\tau^2{}- \mathlarger{\Laplacian}{} + \mu^2 }} + \beta V A^2\mu =0,\\
	\pdv{S_\Qp^{\saddle}}{A}= & V\beta \pqty{\frac{\theta^2}{\beta^2} + \mu^2} A= 0.
	\end{cases}
\end{equation}
The first equation relates the charge \(\Qp\) with the \acrshort{vev} \(A\) while the second equation connects \(\mu\) with the quantum effects that appear in the functional determinant. At large charge, \(\mu\) is the controlling parameter that we will use in our asymptotic expansion.
Finally, the third equation reveals that a non-zero \acrshort{vev} is attainable only if \(\theta^2 = - \mu^2 \beta^2\). We can rewrite \cref{eq;saddlepoint} as
\begin{equation}\label{eq:usefulsaddlepoint}
\begin{cases}
   & \mu  = -i \theta / \beta 	\iff \theta = i \mu \beta, \\
   & i \Qp  = VA^2 \frac{\theta}{\beta} 	\iff \Qp = \mu V A^2, \\
   & VA\mu^2  = - \frac{\pqty{N-1}}{\beta} \pdv{\mu} \Tr\bqty{\log\pqty{ -\del_\tau^2{}- \mathlarger{\Laplacian}{} + \mu^2 }} \iff \Qp = - \frac{\pqty{N-1}}{\beta} \pdv{\mu} \Tr\bqty{\log\pqty{ -\del_\tau^2{}- \mathlarger{\Laplacian}{} + \mu^2 }}.
   \end{cases}
\end{equation}
By taking the double-scaling limit
\begin{align}\label{eq:doublescaling}
	\Qp &\to \infty, & N&\to \infty, & \frac{\Qp}{2N}=\mathbb{Q} \ \text{fixed}, 
\end{align}
only the first two terms of \cref{eq:SQ} dominate, and the canonical free energy of \cref{eq.5.2.6}, at the saddle, becomes
\begin{align}
    F_c^{\saddle}\pqty{\Qp} & = -\frac{1}{\beta} \log\pqty{\mathcal{Z\pqty{\Qp}}} = -\frac{1}{\beta} \log\bqty{\exp{- \pqty{-i \theta \Qp + N\log\bqty{\det\pqty{-\del_\tau^2{}-\mathlarger{\Laplacian}{}+\mu^2} }}}} +\order{N^0} \nonumber \\
    & = - \frac{i \theta \Qp}{\beta} + \frac{N}{\beta} \log\bqty{\det\pqty{-\del_\tau^2{}-\mathlarger{\Laplacian}{}+\mu^2}} +\order{N^0} \nonumber \\
    & = \mu \Qp + \frac{N}{\beta} \log\bqty{\det\pqty{-\del_\tau^2{}-\mathlarger{\Laplacian}{}+\mu^2}} +\order{N^0},
\end{align}
where we used \cref{eq:usefulsaddlepoint} and also the fact that \(\Tr(\log M) =\log(\det M)\). We can express the free energy in terms of the modified charge \(\Qb\) as
\begin{equation}
	F_c^{\saddle}(\mathbb{Q}) = 2N\bqty{ \mu \mathbb{Q} + \frac{1}{2\beta} \log\bqty{\det\pqty{-\del_\tau^2{}-\mathlarger{\Laplacian}{}+\mu^2} }}+\order{N^0}.
\end{equation}
From \cref{eq:usefulsaddlepoint,eq:doublescaling} we can see that $\mu$ is related to $\mathbb{Q}$ via
\begin{equation}
  \Qb \overset{\Qb = \Qp/2N}{=\joinrel=\joinrel=} - \frac{1}{2\beta} \pdv{\mu} \log\bqty{\det\pqty{-\del_\tau^2{}-\mathlarger{\Laplacian}{}+\mu^2} } .
\end{equation}
We recognise the grand potential in the form of the functional determinant \cite{Moshe_2003,zinn2021quantum,kapusta_gale_2006}, as
\begin{align}\label{eq:definitiongrandpotential}
  \omega(\mu) &= - \frac{1}{2\beta}\log\bqty{\det\pqty{-\del_\tau^2{}-\mathlarger{\Laplacian}{}+\mu^2}}, & \Qb & =  \pdv{\mu} \omega\pqty{\mu},
\end{align}
and we can identify the canonical free energy as the Legendre transformation of the grand potential \footnote{To simplify the notation we are using the form of the grand potential and the canonical free energy per degree of freedom, a fact that is possible since we are only keeping results at leading order in \(N\). Instead, we could have used the standard quantities \(F^{\saddle}(\Qp) = \sup_{\mu} ( \mu \Qp - \Omega(\mu))\).} which reads
\begin{equation}\label{eq:canonicalfreeenergylegendre}
  f^{\saddle}(\Qb) \coloneqq \frac{F^{\saddle}(\Qb)}{2N} =\sup_{\mu }(\mu \Qb - \omega(\mu)).
\end{equation}
The functional determinant can be computed in \(\zeta\)-function regularisation \cite{elizalde1994zeta,elizalde2012ten,kirsten2001spectral}, which is also convenient when we consider different manifolds
\footnote{In general, if we consider a differential operator \(\mathscr{O} + m^2\) that has eigenvalues  \(\lambda_n +m^2 \) with \(\lambda_n >0\) on a manifold \(\mathcal{M}\), the zeta function for the  aforementioned operator on that manifold is defined as
\begin{equation}\label{eq:definitionzetafunction}
    \zeta_{\mathscr{O}} (s| \mathcal{M},m) = \mathlarger{\sum}_{n} \pqty{\lambda_n{} +m^2}^{-s} \equiv \Tr\bqty{\mathscr{O}+m^2}^{-s},
\end{equation}
which is a generalisation of the Hurwitz zeta function. Furthermore, using the identity \(\log{x} = - \eval{\dv{x^{-s}}{s}}_{s=0}\) we get
\begin{equation}
\Tr\bqty{\log\pqty{\mathscr{O}+m^2}} = -\eval{\dv{s}\Tr\bqty{\pqty{\mathscr{O}{}+m^2}^{-s}}}_{s=0} =  - \eval{ \dv{s} \zeta_{\mathscr{O}}(s | \mathcal{M},m)}_{s=0} \ .
\end{equation}}
. Therefore, we have
\begin{equation}\label{eq:logdetandzeta}
  \log\bqty{\det\pqty{-\del_\tau^2{}-\mathlarger{\Laplacian}{}+\mu^2}}  = - \eval{ \dv{s} \zeta(s | S^1 \times \Sigma, \mu)}_{s=0},
\end{equation}
and \(\zeta(s| S^1 \times \Sigma, \mu)\) is the spectral zeta function. In our work, we are going to utilise the Mellin representation, which is defined via the integral form
\begin{equation} \label{eq:Mellinrepres}
  \zeta(s | \mathcal{M}, \mu) = \frac{1}{\Gamma(s)}  \Int\limits_0^\infty \frac{\dd{t}}{t} \, t^s \, \Tr\bqty{e^{\pqty{\del_\tau^2{} + \mathlarger{\Laplacian}{} - \mu^2} t} }.
\end{equation}
If we integrate the zeta function of \cref{eq:Mellinrepres} we get
\begin{align}
     \dv{s} \zeta(s | \mathcal{M}, \mu) &=  \dv{s} \Bqty{\frac{1}{\Gamma(s)}  \mathlarger{\int}\limits_0^\infty \frac{\dd{t}}{t} \, t^s \, \Tr\bqty{e^{\pqty{\del_\tau^2{} + \mathlarger{\Laplacian}{} - \mu^2} t} }} \nonumber \\
    & = \mathlarger{\int}\limits_0^\infty \frac{\dd{t}}{t} \, \Tr\bqty{e^{\pqty{\del_\tau^2{} + \mathlarger{\Laplacian}{} - \mu^2} t} }.\label{eq:integratedMellin}
\end{align}
Hence, if we insert \cref{eq:integratedMellin} into \cref{eq:logdetandzeta} the functional determinant reads
\begin{equation}\label{eq:Schwinger-representation}
     \log\bqty{\det\pqty{-\del_\tau^2{}-\mathlarger{\Laplacian}{}+\mu^2}}   = -  \mathlarger{\int}\limits_0^\infty \frac{\dd{t}}{t} \, \Tr\bqty{e^{\pqty{\del_\tau^2{} + \mathlarger{\Laplacian}{} - \mu^2} t} }.
\end{equation}
In our analysis, we focus on product manifolds \(\pqty{\mathcal{M} = \setS^1 \times \Sigma}\) and therefore, we can separate the temporal part \(\setS^1\) \footnote{For the case of the fermions in \cref{Chapter5} this is not that trivial and a more careful approach in the spirit of \cite{kapusta_gale_2006} has to be followed.}. The result for the heat-kernel trace is
\begin{align}\label{eq:heatkernettime}
  \Tr\bqty{ e^\pqty{\del_\tau^2{} t}} & = \Sum_{n \in \setZ} e^{-\frac{4 \pi^2 n^2}{\beta^2} t} = \theta_3\pqty{0, e^{-t \pqty{2 \pi  / \beta}^2}} \nonumber \\
  & = \frac{\beta}{\sqrt{4 \pi  t}}   \Sum_{k \in \setZ} e^{- \frac{k^2 \beta^2}{4 t} }  = \frac{\beta}{\sqrt{4 \pi  t}} \pqty{1 +  \mathlarger{\mathlarger{\sideset{}{'}\sum}}_{k \, \in \, \Bqty{\setZ \setminus \Bqty{0}}} e^{- \frac{k^2 \beta^2}{4 t} } } , 
\end{align}
and we have already pulled out the contribution of the zero mode out of the above summation. In the first line of \cref{eq:heatkernettime} we used that the trace of the operator is a summation upon the Matsubara frequencies \(\omega_n = {2 \pi n}/{\beta}\) and then that the sum is precisely the Jacobi theta function
\begin{equation}\label{eq:Jacobitheta3}
    \theta_3 \pqty{z, q} \coloneqq \Sum\limits_{n = -\infty}^{\infty} q^{n^2} e^{2niz}.
\end{equation}
To get to the second line, we made use of the following special form of the Poisson resummation that we will often use in the following sections: 
\begin{equation}\label{eq:Poissonsummation}
\Sum_{n \in \setZ} \exp\bqty{- a \pqty{M n^2 + bn + c}} = \sqrt{\frac{\pi}{Ma}} \,\Sum_{k \in \setZ} \exp\bqty{-\pqty{\frac{\pi^2}{aM}}k^2 - i \pqty{\frac{\pi b}{M}}k - a \pqty{c- \frac{b^2}{4M}}}.
\end{equation}
For the zero temperature limit \(\beta \to \infty\) that coincides with \(\setS^1 \to \setR\) the sum is dominated by the zero mode, therefore 
\begin{equation}
  \Tr\bqty{ e^\pqty{\del_\tau^2{} t}} =  \frac{\beta}{\sqrt{4 \pi  t}} \pqty{1 +  \mathlarger{\mathlarger{\sideset{}{'}\sum}}_{k \, \in \, \Bqty{\setZ \setminus \Bqty{0}}} e^{- \frac{k^2 \beta^2}{4 t} } }   \underset{\beta \to \infty}{\longrightarrow} \frac{\beta}{\sqrt{4 \pi  t}}.
\end{equation}
At this point, the functional determinant can be written as
\begin{align}\label{functionaldeterminantandzetafunction}
   \log\bqty{\det\pqty{-\del_\tau^2{}-\mathlarger{\Laplacian}{}+\mu^2}}   & = -  \mathlarger{\int}\limits_0^\infty \frac{\dd{t}}{t} \, e^{-\mu^2 t} \Tr\bqty{ e^\pqty{\del_\tau^2{} t}} \Tr\bqty{ e^\pqty{\mathlarger{\Laplacian}{} t}} \nonumber\\
   & \overset{\beta \to \infty}{=\joinrel=\joinrel=} - \frac{\beta}{\sqrt{4 \pi  }}\mathlarger{\int}\limits_0^\infty \frac{\dd{t}}{t} t^{-1/2} e^{-\mu^2 t} \Tr\bqty{ e^\pqty{\mathlarger{\Laplacian}{} t}} \nonumber \\
   & \beta \zeta\pqty*{-\tfrac{1}{2} | \Sigma, \mu}.
\end{align}
and we remember that \(\Gamma(-1/2) = -2 \sqrt{\pi}\) and the definition of \cref{eq:Mellinrepres}. From \cref{eq:definitiongrandpotential} and using the previous results, the grand potential and the charge can be written as
\begin{align}
  \label{eq:grand-potential-zeta}
  \omega(\mu) & = -\frac{1}{2} \zeta(-\tfrac{1}{2} | \Sigma, \mu), & \Qb & = \pdv{\mu} \omega(\mu) = -\frac{\mu}{2} \zeta(\tfrac{1}{2} | \Sigma, \mu),
\end{align}
where we used that
\begin{align}
    \pdv{\mu} \omega(\mu) & =\pdv{\mu}\pqty{-\frac{1}{2} \zeta(-\tfrac{1}{2} | \Sigma, \mu)} 
 = -\frac{1}{2} \pdv{\mu}\bqty{-\frac{1}{\sqrt{4 \pi  }}\mathlarger{\int}\limits_0^\infty \frac{\dd{t}}{t} t^{-1/2} e^{-\mu^2 t} \Tr\bqty{ e^\pqty{\mathlarger{\Laplacian}{} t}}} \nonumber \\
    & = -\frac{\mu}{2 \sqrt{\pi}}\mathlarger{\int}\limits_0^\infty \frac{\dd{t}}{t} t^{1/2} e^{-\mu^2 t} \Tr\bqty{ e^\pqty{\mathlarger{\Laplacian}{} t}} =  -\frac{\mu}{2} \zeta(\tfrac{1}{2} | \Sigma, \mu) ,
\end{align}
and \(\Gamma(1/2) = \sqrt{\pi}\). This equation also relates the charge \(\Qb\) with \(\mu\). Finally, using \cref{eq:canonicalfreeenergylegendre}, the canonical free energy at the saddle reads
\begin{equation}\label{eq;freenergy-saddle}
  F^{\saddle}(\Qb) = 2N \bqty{ \mu \Qb+\frac{1}{2}  \zeta(-\tfrac{1}{2} | \Sigma, \mu) } \overset{f^{\saddle}(\Qb)=\frac{F^{\saddle}(\Qb)}{2N} }{\Longrightarrow} f^{\saddle} \pqty{\Qb} = \mu \Qb+\frac{1}{2}  \zeta(-\tfrac{1}{2} | \Sigma, \mu).  
\end{equation}
In the limit that \(N \to \infty\), the above-mentioned expressions are completely exact \(\forall \hspace{0.5em} \Qb = \Qp/N  \), and therefore we can interpret them as the semiclassical resummation of an infinite series of \(1/N\) corrections. 

The localisation of the integral in \cref{functionaldeterminantandzetafunction} around \(t = 0\) is attainable for large enough values of \(\mu\) and as a consequence, the large-charge problem that we face is reduced to a classical problem, which is the heat-kernel's Weyl asymptotic expansion. It has been shown \cite{de1963dynamical,seeley1967complex} that the expansion can be expressed in terms of Seeley--DeWitt coefficients, as
\begin{equation}\label{eq:Seeleyexpansion}
  \Tr\bqty{ e^\pqty{\mathlarger{\Laplacian}{} t}} \sim \frac{V}{4 \pi t} \pqty{ 1 + \frac{R}{12} t + \dots } .
\end{equation}
In the next sections, we will study the cases of the torus \(\Sigma = \setT^2\) and the two-sphere \(\Sigma = \setS^2\). The case of the torus \(\setT^2\) is chosen as a probe, since being flat the results simplify a lot and the asymptotic expansion includes only the leading order term, since every other term is related to the curvature \(R\) and so vanish. On the other hand, the case of the sphere \(\setS^2\) is chosen to make use of the state-operator correspondence of \cref{sec.stateoperator} and be able to compute the scaling dimension of the lowest charged operator \(\Opp\). In the sphere, the Weyl expansion is indeed asymptotic and for that reason, it can be studied in the framework of resurgence analysis.


\subsection{The torus}
\label{sec:torus}

We start our analysis by studying the case of the torus \(\Sigma = \setT^2\), which is relatively straightforward as the canonical free energy is exact, but nonetheless it demonstrates a couple of qualitative properties that are universal and useful for the subsequent sections.  

For the square torus with sides of length \(L\), the Ricci curvature is zero, \emph{i.e.} \(R=0\), and so via \cref{eq:Seeleyexpansion} only the first Seeley--DeWitt coefficient remains. On that account, the heat trace becomes
\begin{equation}\label{eq:LOT2}
    \Tr\bqty{ e^\pqty{\mathlarger{\Laplacian}_{\setT^2}{} t}} \sim \frac{L^2}{4 \pi t}  + \order{e^{-1/t}},
\end{equation}
while the zeta function of \cref{eq:Mellinrepres} on the torus reads
 \begin{equation}
     \zeta(s| \setT^2, \mu) =  \frac{L^2 \mu^{2(1-s)}}{4 \pi(s-1)}  + \order{e^{-\mu}}.
\end{equation} 
Ergo, we can compute the grand potential \(\omega\pqty{\mu}\), the charge \(\Qb\), and the canonical free energy \(f^{\saddle}\pqty{\Qb}\) of \cref{eq:grand-potential-zeta,eq;freenergy-saddle} as
\begin{align}
\omega(\mu) &= - \frac{1}{2} \zeta(-\tfrac{1}{2}| \setT^2, \mu)  = \frac{L^2 \mu^3}{12 \pi}, \\
\Qb & = -\frac{\mu}{2} \zeta(\tfrac{1}{2}| \setT^2, \mu) = \frac{L^2 \mu^2}{4 \pi} , \\
f^{\saddle}(\Qb) &= \mu \Qb + \frac{1}{2} \zeta(-\tfrac{1}{2}| \setT^2, \mu) =
\frac{4 \sqrt{\pi}}{3L} \Qb^{3/2}.
\end{align}
We see that solving \(\mu\) in terms of the charge \(\Qb\) agrees with the form of \cref{eq:formofthecharge}, with the exception that being in the double-scaling limit we have the power to compute the Wilsonian coefficients \cite{alvarez2019large}.

Although the above solutions are exact in perturbation theory at leading order in the charge \(\Qb\), we can enhance our analysis in the following manner: the  \(\order{e^{-1/t}}\) corrections of~\cref{eq:LOT2} are known and can be expressed in a closed form, using that the spectrum of the Laplacian on the torus \(\setT^2\) is 
\begin{equation}\label{eq:spectrumontorus}
\spec(\mathlarger{\Laplacian}_{\setT^2}) = \left\{ - \frac{4 \pi^2}{L^2} (k_1^2 + k_2^2) \middle| k_1, k_2 \in \setZ \right\}.
\end{equation}
Then, using \cref{eq:spectrumontorus}
the trace of the heat-kernel reads
\begin{align}
   \Tr\bqty{ e^\pqty{\mathlarger{\Laplacian}_{\setT^2}{} t}}  & = \mathlarger{\mathlarger{\sum}}\limits_{k_1 \in \setZ} \, \mathlarger{\mathlarger{\sum}}\limits_{k_2 \in \setZ} e^{-\frac{4 \pi^2 \pqty{k_1^2 + k_2^2}}{L^2}  t } \label{eq;heatkerneltracetorus} \\
   &= \bqty{ \theta_3\pqty{0, e^{- t\pqty{2 \pi / L}^2}}}^2,
\end{align}
where to get to the second line we used the definition of the Jacobi theta function in \cref{eq:Jacobitheta3} and we sum over each momentum eigenvalue \(k_i\) separately. We observe that the heat-kernel on the torus is exactly a square of the theta function, which verifies our claim that the final result can be written in an exact form. 

Since for \(\mu\) large the integral localises at \(t \rightarrow 0^+\), we can apply the Poisson summation formula of \cref{eq:Poissonsummation} in \cref{eq;heatkerneltracetorus} and in the appropriate limit, the heat-kernel reads
\begin{equation} \label{eq:Torus-trace-Poisson}
  \Tr\bqty{ e^\pqty{\mathlarger{\Laplacian}_{\setT^2}{} t}}  = \frac{L^2}{4 \pi t} \pqty{1 + \mathlarger{\sideset{}{'}\sum}\limits_{ \bm{k} \in \Bqty{ \setZ^2 \setminus \Bqty{0}}} e^{- \frac{\norm{\bm{k}}^2 L^2}{4 t} }} \ ,
\end{equation}
where we denote \(\norm{ \bm{k}}^2 = k_1^2 + k_2^2\), and as before, we have already pulled out the contribution of the zero mode in the summation.

Besides \cref{eq:Torus-trace-Poisson} being exact, moreover, it is valid for finite \(t\) as well, and as a consequence, we can use it to compute the sub-leading terms in \cref{eq:Mellinrepres} in the limit of large charge, \emph{i.e.} $\mu \rightarrow \infty$ as
\begin{equation}\label{eq;heatkernelsubleadingterms}
  \zeta(s| \setT^2, \mu) = \frac{L^2 \mu^{2\pqty{1-s} }}{4 \pi ( s -1)} + \frac{L^2}{2 \pi} \mathlarger{\mathlarger{\sideset{}{'}\sum}}_{\bm{k}} \frac{2^{2-s}}{\Gamma(s)} \pqty{\frac{\norm{\bm{k}}L}{\mu} }^{s-1} K_{1-s}(\norm{\bm{k}} \mu L) ,
\end{equation}
where \(K_n(z)\) denotes the modified Bessel function of the second kind \cite{abramowitz1988handbook,arfken1999mathematical} given in integral formula as
\begin{equation}
    K_n(z) = \frac{\Gamma\pqty{n + \frac{1}{2}} (2z)^n}{\sqrt{\pi}} \mathlarger{\int}\limits_{0}^{\infty} \dd{t} \frac{\cos{t}}{\pqty{t^2 + z^2}^{n + 1/2}}.
\end{equation}
We can use \cref{eq;heatkernelsubleadingterms} to compute the sub-leading corrections to the grand potential and the canonical free energy in a closed form as
\begin{align}
  \omega(\mu) &= - \frac{1}{2}  \zeta(-\tfrac{1}{2} | \setT^2, \mu) = \frac{L^2 \mu^3}{12 \pi} \pqty{ 1 + \mathlarger{\mathlarger{\sideset{}{'}\sum}}\limits_{\bm{k}} \frac{e^{-\norm{\bm{k}} \mu L }}{\norm{\bm{k}}^2 \mu^2 L^2 } \pqty{1 + \frac{1}{\norm{\bm{k}} \mu L} }  }, \label{eq:torus-grand-potential} \\
  f^{\saddle}(\Qb) &= \sup_\mu( \mu \Qb - \omega(\mu)) = \frac{4 \sqrt{\pi}}{3 L} \Qb^{3/2} \pqty{1 - \mathlarger{\mathlarger{\sideset{}{'}\sum}}\limits_{\bm{k}} \frac{e^{- \norm{\bm{k}} \sqrt{4 \pi \Qb}}}{8 \norm{\bm{k}}^2 \pi \Qb} + \dots  } \label{eq:torus-free-energy}.
\end{align}
The charge \(\Qp\) is computed via the grand potential \(\omega\pqty{\mu}\) in terms of \(\mu\) and the corresponding relation of the two conjugate variables is recursive, therefore, even though the expressions for \(\omega(\mu)\) and \(\Qb\) are exact, there are more terms present in the expression of the free energy that are related to the fact that \(\Qb\) is a recursive function of \(\mu\). Nevertheless,  extrapolating this result to the small-charge regime up to charge \(\Qb = 1\), the  contribution to the free energy stemming from these farther exponentially suppressed terms is \(\order{10^{-3}}\).

To complete our analysis, we also want to compute the small-charge regime of the grand potential, which corresponds to small-\(\mu\). In order to do so, we will apply the binomial theorem for  \(\mu < 2 \pi/L\) and express \(\omega(\mu)\) in terms of \(\mu\) in a series expansion
\begin{equation}
  \begin{aligned}
    \omega(\mu) &= -\frac{1}{2} \zeta(-\tfrac{1}{2}| \setT^2, \mu) = \eval{ -\frac{1}{2} \mathlarger{\mathlarger{\sum}}_{\bm{k} \in \setZ^2} \pqty{\frac{4 \pi^2}{L^2} \norm{\bm{k}} + \mu^2}^{-s}}_{s=-1/2} \\
    &= - \frac{1}{2} \bqty{ \mu + \frac{2 \pi}{L}  \mathlarger{\mathlarger{\sideset{}{'}\sum}}_{\bm{k} \in \setZ^2} \mathlarger{\mathlarger{\sum}}_{n=0}^\infty \binom{1/2}{n} \pqty{\frac{L \mu}{2 \pi} }^{2n} \norm{\bm{k}}^{1/2 - n}} \\
    &=   - \frac{1}{2} \bqty{ \mu + \frac{2 \pi}{L} \mathlarger{\mathlarger{\sum}}_{n=0}^\infty \binom{1/2}{n} \pqty{\frac{L \mu}{2 \pi} }^{2n} \zeta(-\tfrac{1}{2} + n | \setT^2, 0) },
  \end{aligned}
\end{equation}
and in the first line we used the definition of the zeta function of \cref{eq:definitionzetafunction} and then the binomial expansion. The zeta function defined on the torus is expressed by the Chowla--Selberg formula~\cite{chowla1949epstein}
\begin{equation}
  \zeta(s|\setT^2, 0) = 2 \zeta(2s) + \frac{2^{2s} \sqrt{\pi}}{\Gamma(s)} \Gamma(s-\tfrac{1}{2}) \zeta(2s -1 ) + \frac{2^{s+5/2} \pi^s}{\Gamma(s)} \mathlarger{\mathlarger{\sum}}_{n=1}^\infty n^{s-1/2} \sigma_{1-2s}(n) K_{1/2-s}(n \pi),
\end{equation}
where with \(\sigma\) we denote the divisor function
\begin{equation}
  \sigma_k(n) = \mathlarger{\mathlarger{\sum}}_{d | n} d^k .
\end{equation}
Thereby, we get a convergent sum, and the first few terms read
\begin{equation}
  \omega(\mu) = \frac{0.64443\dots}{L} -\frac{1}{2}  \mu + 0.20064\dots \mu ^2 L  + 0.00816\dots \mu ^4 L^3 + \dots .%
\end{equation}

Finally, we can analyse the form of the grand potential in \cref{eq:torus-grand-potential} to point out some universal properties. In general, it will exhibit the following form: it will be a perturbative expansion in \(\mu\) \footnote{Although in the case of the torus, this expansion ends after a single term.} and then there is a series of exponentially suppressed terms which are regulated by the parameter \(\mu L\) that is dimensionless, and \(L\) would be the scale of the manifold \(\Sigma\). Similarly, from \cref{eq:torus-free-energy} we see that the canonical free energy is expressed as a double expansion based on two parameters: \(1/\Qb\) and \(e^{-\sqrt{4 \pi \Qb}}\). 

These structures are called \emph{trans-series} and their appearance in perturbative problems comes naturally. Generically, considering a problem that exhibits some small parameter \(z\), a trans-series solution to that problem in the limit that \( z \to 0\) acquires the general form \cite{mariño_2015}:
\begin{align}\label{eq:trans-series}
	\Phi(\sigma_k,z) &= \Sum_{k} \sigma_k e^{-A_k/z^{1/\beta_k}} z^{-b_k/\beta_k} \Phi^{(k)}(z) \nonumber\\
& = \Phi^{(0)}(z) +  \Sum_{k \neq 0} \sigma_k e^{-A_k/z^{1/\beta_k}} z^{-b_k/\beta_k} \Phi^{(k{}>{}0)}(z),
\end{align} 
where we pulled the zero-mode out of the sum and all \(\Phi^{(k)}(z)\) in \cref{eq:trans-series} are asymptotic series. Specifically, \(\Phi^{(0)}\) is the formal solution to the problem. Since these are formal solutions, these expressions are coherent only when an appropriate prescription is provided for the proper summation of the series, as we will see in the case of the sphere \(\setS^2\).

We have to emphasise that \emph{resurgent trans-series} are specified by a particular set of relations between the parameters \(A_k, \beta_k , b_k\) of \cref{eq:trans-series} and the asymptotic series \(\Phi^{(k)}\). Due to this set of relations, it is feasible to specify the value of \(\sigma_k \in \mathbb{C}\) which is known as the \emph{trans-series parameter} in such a way that a suitable summation of the aforementioned trans-series will generate an unambiguous function and \(\Phi^{(0)}\) will be its leading perturbative asymptotic expansion. This sort of function is called a \emph{resurgent function} and for details the interested reader is referred to \cite{Dorigoni_2019, edgar2009transseries}. 

To summarise, the form of the heat trace on the torus \(\setT^2\) is of the form of \cref{eq:trans-series} but in a trivial manner, all the \(\Phi^{(k )}\) are precisely one-loop exact. Therefore, there is in no emergence of ambiguities for the asymptotic series and the resurgent function solution matches the trans-series representation, that is the Jacobi theta function \(\theta_3\). In this case, of the sphere \(\Sigma = \setS^2\), this is not true any more, as we will see in \cref{sec:perturbative-sphere}.

\subsection{The sphere}
\label{sec:perturbative-sphere}

Having studied the torus \(\setT^2\), now we move to the sphere \(\setS^2\) of radius \(R_0\). Similarly to the case of the torus, the heat-kernel in the small-\(t\) regime can be expressed as an asymptotic series written in terms of the Seeley--DeWitt coefficients for the two-sphere and can be reassembled into a trans-series, but contrary to the torus, this is not trivial, and it is not Borel resumable, so we need to complement the leading-order perturbative expansion with non-perturbative exponentially suppressed corrections to give it meaning. 

As we will see in \cref{sec:path-integral}, these exponentials have a nice geometrical interpretation in the form of \acrfull{wl} instantons, and the non-perturbative ambiguities that arise are associated with tachyonic instabilities.

\subsubsection{Seeley--DeWitt coefficients.} \label{sec.Seeley--DeWitt_coefficients.}

As in the case of the torus, we want to compute the grand potential \(\omega(\mu)\), the charge \(\Qb\) and the canonical free energy \(f^{\saddle} (\Qb)\) using \cref{eq:grand-potential-zeta,eq;freenergy-saddle}. To do so, we need to know the heat-kernel trace for the case of sphere \(\setS^2\), in order to calculate the zeta function \(\zeta(-\tfrac{1}{2}| \setS^2, \mu)\).

So, the first thing that we need is the spectrum of the Laplacian on \(\setS^2\), which reads
\begin{equation}
\spec\pqty{\mathlarger{\Laplacian}_{\setS^2}} = \left\{ -\frac{\ell \pqty{\ell +1}}{R_0^2} \middle| \ell \in \setN \right\} ,
\end{equation}
and each eigenvalue is \(M_{\ell} =(2 \ell + 1)\) degenerate. It would be more convenient to use the trace of the conformal Laplacian \(\mathlarger{\mathlarger{\Laplacian}}_{\setS^2} - \xi R\) \footnote{Remember that for the manifold \(\mathcal{M} = \setS^1 \times \setS^2\) we get that \(\xi= \frac{(d-2)}{4(d-1)}\) and \(R= 2/R_0^2\)} that we saw in \cref{sec:largeN} which is related to the mass parameter \(m\).

Similarly with the torus, we can apply the Poisson summation formula to rewrite the trace \cite{marklof2004selberg} as
\begin{equation}
	\begin{aligned}
		\Tr\bqty{ e^{ \pqty*{ \mathlarger{\mathlarger{\Laplacian}}_{\setS^2} - \frac{1}{4R_0^2}  } t} } &=  \mathlarger{\mathlarger{\sum}}_{\ell = 0}^\infty (2\ell +1)  e^{- t\pqty{\ell + \frac{1}{2} }^2 /R_0^2 } = \mathlarger{\mathlarger{\sum}}_{\ell = - \infty}^{\infty} \abs{ \ell + \frac{1}{2}} e^{- {t} \left( \ell + \frac{1}{2} \right)^2 /R_0^2} \\
		&= \Sum_{k \in \setZ} (-1)^k \Int\limits_\setR \dd{\rho} \abs{\rho}   e^{-\rho^2 t/R_0^2 + 2 \pi i k \rho },
	\end{aligned}
\end{equation}
where to get from the first to the second line we used that \(\rho = \ell + 1/2\) and the Poisson summation using the relevant Fourier transform. Now
we extract the zero mode and then expand the exponential
\begin{align}
   \Tr\bqty{ e^{ \pqty*{ \mathlarger{\mathlarger{\Laplacian}}_{\setS^2} - \frac{1}{4R_0^2}  } t} }&= \Int\limits_\setR \dd{\rho} \abs{\rho} e^{-\rho^2 t/R_0^2} + \sideset{}{'} \Sum_{k \in \Bqty{\setZ - \setminus \Bqty{0}}} (-1)^k \Int\limits_\setR \dd{\rho} \abs{\rho}   e^{-\rho^2 t/R_0^2 + 2 \pi i k \rho } \nonumber\\
    &= \Int\limits_\setR \dd{\rho} \abs{\rho} e^{-\rho^2 t/R_0^2} + \sideset{}{'} \Sum_{k \in \Bqty{\setZ - \setminus \Bqty{0}}} (-1)^k \Int\limits_\setR \dd{\rho} \abs{\rho}   e^{-\rho^2 t/R_0^2} \cos\pqty{2 \pi k \rho} \nonumber\\
    & = \frac{R_0^2}{t} + 2\sideset{}{'} \Sum_{k \in \Bqty{\setZ - \setminus \Bqty{0}}} (-1)^k \Int\limits_{0}^{\infty} \dd{\rho} \abs{\rho}   e^{-\rho^2 t/R_0^2} \cos\pqty{2 \pi k \rho} \nonumber\\
    &= \frac{R_0^2}{t} + \sideset{}{'} \Sum_{k \in \Bqty{\setZ - \setminus \Bqty{0}}} (-1)^k \bqty{ \frac{R_0^2}{t} - \frac{2 \abs{k} \pi R_0^3}{t^{3/2}} \operatorname{F}\left( \tfrac{\pi R_0 \abs{k}}{\sqrt{t}} \right)  } \label{eq:poissonresumsphere},
\end{align}
where in the last line of our calculation we introduced the Dawson's function \(\operatorname{F}(z)\) which is correlated to the imaginary error function \(\operatorname{erfi}(z)\) as
\begin{equation}\label{eq:dawsonfunctionerror}
  \operatorname{F}(z) = e^{-z^2} \int_0^z \dd{t} e^{-t^2} = \frac{\sqrt{\pi}}{2} e^{-z^2 } \operatorname{erfi}(z) .
\end{equation}
In the case that the values of the argument of the Dawson function are small enough, we are able to utilise its asymptotic expansion. Since we are interested in the small-\(t\) limit, we take the large-\(z\) expansion which is
\begin{equation}
  \operatorname{F}(z) \sim \Sum_{n=0}^\infty \frac{(2n - 1)!!}{2^{n+1}} \pqty{\frac{1}{z} }^{2n + 1},
  \label{eq:Dawson}
\end{equation}
and for our case we get
\begin{align}\label{eq:Dawsonexpanded}
   \operatorname{F}\left( \tfrac{\pi R_0 \abs{k}}{\sqrt{t}} \right) &=  \Sum_{n=0}^\infty \frac{(2n - 1)!!}{2^{n+1}} \pqty{\frac{t^{1/2}}{\pi R_0 \abs{k}} }^{2n + 1} \nonumber \\
   &= \frac{t^{1/2}}{\pi R_0 \abs{k}} \Sum_{n=0}^\infty \frac{(2n - 1)!!}{2^{n+1} \pqty{\pi R_0 \abs{k}}^{2n}} t^n. 
\end{align}
Using \cref{eq:poissonresumsphere,eq:Dawsonexpanded} and after formally manipulating the expression, we obtain the leading asymptotic form of the heat-kernel trace on the two-sphere 
\begin{align}
  \Tr\bqty{ e^{ \pqty*{ \mathlarger{\mathlarger{\Laplacian}}_{\setS^2} - \frac{1}{4R_0^2}  } t} } & \sim \frac{R_0^2}{t} - \Sum_{n=1}^\infty \frac{(-1)^n(1-2^{1-2n})}{n! R_0^{2n-2}} B_{2n} t^{n-1} \nonumber \\
  & \equiv \frac{R_0^2}{t}  \Sum_{n=0}^\infty a_n \pqty{\frac{t}{R_0^2}}^n,
  \label{eq:Phi0}
\end{align}
where \(a_n = \bqty{{(-1)^{n+1}(1-2^{1-2n}) B_{2n}}}\, /\pqty{{n!}} \), we have inserted the zero mode again in the summation, and also we introduced the Bernoulli numbers \(B_{2n}\) which can be written with respect to the Riemann zeta function \cite{arfken1999mathematical} as
\begin{equation}\label{eq:Bernoulliandzeta}
    B_{2n} = (-1)^{n+1} \frac{2 (2n)!}{(2\pi)^{2n}} \zeta(2n).
\end{equation}
The expression of \cref{eq:Phi0} was already mentioned in~\cite{Cahn1975}, and it was based on the previous work of Mulholland in~\cite{mulholland1928asymptotic}.
It is not hard to see that the above series is asymptotic given the fact that at large \(n\) the Seeley--DeWitt coefficients  \(a_n\) are \(n!\) divergent
\begin{align}
	    a_n &= \frac{(-1)^{n+1}(1-2^{1-2n})}{n!} B_{2n} && \overset{Eq. \pqty{\ref{eq:Bernoulliandzeta}}}{\implies} && a_n \sim\frac{2}{ \sqrt{\pi} } \frac{n^{-1/2}}{\pi^{2n} } n!.
	\label{eq:S2coeff}
\end{align}
The divergence of the Seeley--DeWitt coefficients could have been predicted based on the fact that the expansion of the Dawson's function in \cref{eq:Dawson} is also asymptotic.

The expansion in \cref{eq:Phi0} is just a formal solution and, as we discussed in \cref{sec:torus} an appropriate prescription needs to be provided for the summation. Therefore, we assume that if we can complete this series into a resurgent function, a suitable summation of the aforementioned trans-series will generate an unambiguous result.
Hence, the starting point is to correctly identify the structure of the non-perturbative terms in the general form of \cref{eq:trans-series}, and the result of \cref{eq:Phi0} is the perturbative part that can be identified with \(\Phi^{(0)}\). Following \cite{mariño_2015}, the general form of the resurgent trans-series is taken to be
\begin{align}
\Phi(\sigma_k,z) =&\Phi^{(0)}(z) +  \Sum_{k \neq 0} \sigma_k e^{-A_k/z^{1/\beta_k}} z^{-b_k/\beta_k} \Phi^{(k{}>{}0)}(z) , & \Phi^{(k{}>{}0)}(z) &\sim \Sum_{\ell = 0}^\infty a^{(k)}_\ell z^{\ell/\beta_k},
\end{align}
and we denoted \(z = t / R_0^2\).

As we noted, there is a particular set of relations between
the parameters \(A_k, \beta_k, b_k\) and the asymptotic series \(\Phi^{k}\) and more specifically, the coefficients \(a^{(k)}_\ell \in \Phi^{(k{}>{}0)}\) and the above-mentioned parameters are encrypted in the large-order form of the perturbative part
\begin{equation}\label{eq:GenericLargeOrder}
	a_n \sim \Sum_k \, \frac{S_k}{2 \pi i} \, \frac{\beta_k}{A_k^{n \beta_k + b_k}} \, \Sum_{\ell = 0}^\infty \, a_\ell^{(k)} \, A_k^{\ell} \, \Gamma( \beta_k n + b_k - \ell),
\end{equation}
and \(S_k\) stands for the Stokes constants.\footnote{For the torus \(\setT^2\) the large-order form is trivially realized. } The formula of \cref{eq:GenericLargeOrder} is one of the biggest accomplishments in the framework of the resurgence analysis concerning resurgent functions \cite{Dorigoni_2019}.

In the problem at hand, we have complete knowledge of the \(a_n\) in the sense that we can use the following identities
\begin{align}
    \Sum_{k \neq 0} \frac{\pqty{-1}^k}{k^{2n}} & = 2 \Sum_{k=1}^{\infty} \frac{\pqty{-1}^k}{k^{2n}} = 2 \zeta\pqty{2n} \pqty{2^{1-2n} -1}, \\
    \zeta\pqty{2n} & = \pqty{-1}^{n+1} \frac{\pqty{2 \pi}^{2n} B_{2n}}{2 {\pqty{2n}!}},
\end{align}
which we can combine to derive
\begin{equation}
	\Sum_{k\neq 0} \frac{(-1)^k}{k^{2n}} = \frac{\pqty{-1}^{n+1} \pqty{2 \pi}^{2n}}{(2n)!}  \pqty{2^{1-2n}-1}  B_{2n},
\end{equation}
to rewrite \(a_n\) in a suggestive form that we can compare with the generic form of \cref{eq:GenericLargeOrder} as
\begin{equation}\label{eq:suggestiveforman}
a_n = - \Sum_{k\neq0} (-1)^k  \frac{\pqty{2n}!}{n! \, 2^{2n} \, (\pi k)^{2n} } 
 = \frac{1}{\sqrt{\pi}}  \Sum_{k\neq0} (-1)^{k+1} \frac{\Gamma(n+ \tfrac{1}{2})}{(\pi k)^{2n}},
\end{equation}
where we used that
\begin{equation}\label{eq:Gamman+1/2}
    \Gamma(n+ \tfrac{1}{2}) = \frac{\pqty{2n}! \sqrt{\pi}}{n! \, \, 2^{2n}}.
\end{equation}
Comparing \cref{eq:GenericLargeOrder,eq:suggestiveforman} we get the following values for the parameters
\begin{align}\label{eq:valuesofcoefficients}
	\beta_k &= 1, & b_k&= \frac{1}{2}, & A_k &= (\pi k)^2, & \frac{S_k}{2\pi i } a_0^{(k)} &= (-1)^{k+1}  \abs{k} \sqrt{\pi} , & a^{(k)}_{>0} &= 0.
\end{align}
We observe that \(a^{(k)}_{>0} = 0\), and as a consequence, we deduce that the series are truncated to a single term around every exponential. Hence, using \cref{eq:Phi0} along with \cref{eq:trans-series,eq:valuesofcoefficients} indicates that a trans-series representation of the heat-kernel trace has to include the terms
\begin{equation}
	\label{eq:non-perturbative-heat}
 \Tr\bqty{ e^{ \pqty*{ \mathlarger{\mathlarger{\Laplacian}}_{\setS^2} - \frac{1}{4R_0^2}  } t} }\,\supset\, 	 2 i\left( \frac{\pi R_0^2}{t} \right)^\frac{3}{2} \Sum_{k \neq 0} \sigma_k \,(-1)^{k+1} \, \abs{k} \, e^{- (k\pi R_0)^2/t}.
\end{equation}
We need to point out that \cref{eq:non-perturbative-heat} is unambiguously defined up to a \(k\)-dependent complex term \(\sigma_k\) which is the trans-series parameter that we came upon in \cref{eq:trans-series}. The large-order analysis of \(\Phi^{(0)}\) cannot specify the value of \(\sigma_k\), and this is because no matter the choice of the parameter, \(\Phi^{(0)}\) will always be the perturbative asymptotic solution of the above trans-series.

\subsubsection{Grand potential and free energy.}

Having analysed the asymptotic behaviour of the heat-kernel trace, now we move our attention to the grand potential \(\omega(\mu)\) and the canonical free energy \(f^{\saddle}(\Qb)\) which are also asymptotic series.

Since they are associated to the heat-kernel trace through the Mellin transform of the zeta function, they are both higher factorial divergent quantities. This appears to be a characteristic of the model at hand in the double-scaling limit of \cref{eq:doublescaling}, and given that the canonical free energy is related to the scaling dimension via the state-operator correspondence, we can assert that \emph{the \acrshort{lce} of the conformal dimension of the lowest charged operator is itself asymptotic and the relevant coefficients in the asymptotic series diverge like \((2n)!\)}. In \cref{sec:lessons-from-large-N}  we will theorise that this feature is a generic characteristic of \acrlong{lce}.

We can carry out an analogous large-order analysis for the case of the large-\(\mu\) expansion of the operator \(-\mathlarger{\mathlarger{\Laplacian}}_{\setS^2} + \mu^2\) on the two-sphere with the Mellin zeta function,
\begin{equation}
	\zeta(s | \setS^2 , \mu) = \frac{1}{\Gamma(s)}\Int\limits_0^\infty \dd{t} \, t^{s-1} e^{-\mu^2 t} 
\Tr\bqty{e^{ \mathlarger{\mathlarger{\Laplacian}}_{\setS^2} t} }.
\end{equation}
As in \cref{sec.Seeley--DeWitt_coefficients.} , it would be more convenient to utilise the conformal Laplacian \(\mathlarger{\mathlarger{\Laplacian}} -\xi R\) instead, which corresponds to a shift in the mass $\mu^2 \rightarrow {\mu^2}'= \mu^2 - 1/(4R_0^2) \equiv m^2$.
By doing so, and using the Weyl's asymptotic formula \footnote{\Cref{eq:Seeleyexpansion} is but the very few terms of this generic expansion.} \cite{weyl1911asymptotische,rosenberg1997laplacian,vassilevich2003heat} which reads
\begin{equation}
\Tr\bqty{e^{\mathlarger{\mathlarger{\Laplacian}}_{\Sigma} t } } = \Sum_{n=0}^{\infty} \, K_n \, t^{\frac{n}{2}-1},
\end{equation}
for the case of the two-sphere \(\setS^2\), we derive~\cite{alvarez2019large} the following relation
\begin{equation}  
    \zeta(s | \setS^2, m)
    = R_0^2 m^{2(1-s)} \Sum_{n=0}^\infty \, a_n \, \frac{\Gamma( n+s-1) }{\Gamma(s)} \, \frac{1}{(m R_0)^{2n}},
\end{equation}
which, as we anticipated, is an expansion in the limit \(m \sim \mu \rightarrow \infty\) and where \(a_n\) are the Seeley--DeWitt coefficients of the two-sphere \(\setS^2\) that we have already computed in \cref{eq:S2coeff}. Furthermore, the additional gamma function generates a farther \(n!\) increase in the large-order divergence. By setting \(s=-1/2\) we reclaim the expression for the grand potential
\begin{align}
  \label{eq:large-m-grand-potential}
    \omega(m) &= - \frac{1}{2} \zeta \pqty{ \left. - \frac{1}{2} \right| \setS^2 , m^2 } =  R_0^2 m^3 \Sum_{n=0}^\infty \, \frac{ \omega_n}{(m R_0)^{2n}} \nonumber \\
    & =   \frac{1}{3} R_0^2 m^3 - \frac{1}{24} m  + \frac{7}{1920} \frac{1}{m R_0^2} + \dots,
\end{align}
 and where 
\begin{equation}
    \omega_n = -\frac{a_n \Gamma\pqty{n - \frac{3}{2}}}{2 \Gamma\pqty{-1/2}} =  \frac{a_n \Gamma\pqty{n - \frac{3}{2}}}{4 \sqrt{\pi}},
\end{equation}
are the grand potential's coefficients, which we can also express in a closed form using \cref{eq:suggestiveforman}, as
\begin{equation}
  \omega_n =  \frac{1}{4 \pi}  \Sum_{k \neq 0} \, \frac{(-1)^{k+1}}{(\pi k)^{2n}} \, \Gamma\left( n + \frac{1}{2} \right) \, \Gamma \left( n - \frac{3}{2} \right).
  \label{eq:gran-pot-coeff}
\end{equation}
The appearance of the double gamma function in \cref{eq:gran-pot-coeff} makes the comparison with the generic large-order behaviour of the  trans-series coefficients of \cref{eq:GenericLargeOrder} a bit more complicated. Nevertheless, by using the following identity
\begin{equation}
  \begin{aligned}
	  2^{2n} \Gamma(n+\tfrac{1}{2}) \Gamma(n - \tfrac{3}{2}) &=   \sqrt{\frac{\pi}{2} }  \Sum_{k=0}^{\infty} \gamma_k \Gamma\left( 2n - \frac{3}{2} - k \right)  \\
  &= \sqrt{\frac{\pi}{2} } \left[ 8 \Gamma(2n - \tfrac{3}{2}) + 15 \Gamma(2n - \tfrac{5}{2}) + \frac{105}{16}  \Gamma(2n - \tfrac{7}{2}) + \dots \right],
\end{aligned}
\end{equation}
where we can compute the \(\gamma_k\) coefficients recursively, we are now able to compare the two expressions and match the corresponding parameters, remembering that \(z= 1/(mR_0)^2\), to obtain
\begin{align}
	\beta &= 2, & b_k&= -\frac{3}{2}, & A_k &= 2\pi k, & \frac{S_k}{2 \pi i} a_0^{(k)} &= \frac{(-1)^{k+1}}{4\sqrt{2\pi}} \frac{\gamma_0}{(2\pi \abs{k} )^\frac{3}{2}}  , & a^{(k)}_{\ell >0} &= \frac{\gamma_\ell}{\gamma_0}\frac{1}{(2\pi \abs{k})^\ell \gamma_0},
\end{align}
in which case the non-perturbative corrections to \(\omega\pqty{m}\) read
\begin{equation}
  \label{eq:non-perturbative-grand}
\omega(m )   \supset \sqrt{R_0 m^3}   \frac{(-1)^k}{( 2 \pi \abs{k})^{\frac{3}{2}}}  e^{-(2 \pi \abs{k} ) R_0 m}  \Sum_{\ell=0}^\infty \left( \frac{\gamma_\ell}{\gamma_0 } \right) \frac{1}{(2\pi \abs{k} m R_0)^{\ell}}.
\end{equation}
Observe that the structure of the non-perturbative terms is similar to the torus \(\setT^2\), in the sense that there are a series of exponentially suppressed terms which are regulated by the parameter \(2 \pi R_0 m\) that is dimensionless, where \(2 \pi R_0\) would be the length of the manifold.

Finally, we note that the \(\gamma_\ell\) coefficients are factorially growing, and they alter in sign. They appear in the Henkel's expansion of the modified Bessel function of the second kind
\begin{align}
K_2(z) &\sim \sqrt{\frac{\pi}{2z}} ea^{-z} \Sum_{\ell =0}^\infty \left( \frac{\gamma_\ell}{\gamma_0} \right) \frac{1}{z^{\ell}}, & \textrm{as}& &   z \rightarrow \infty,
\end{align} 
and this fact will play in important role in our discussion of the Borel resummation in the next section.

\begin{figure}
  \centering
  \includegraphics[width=.6\textwidth]{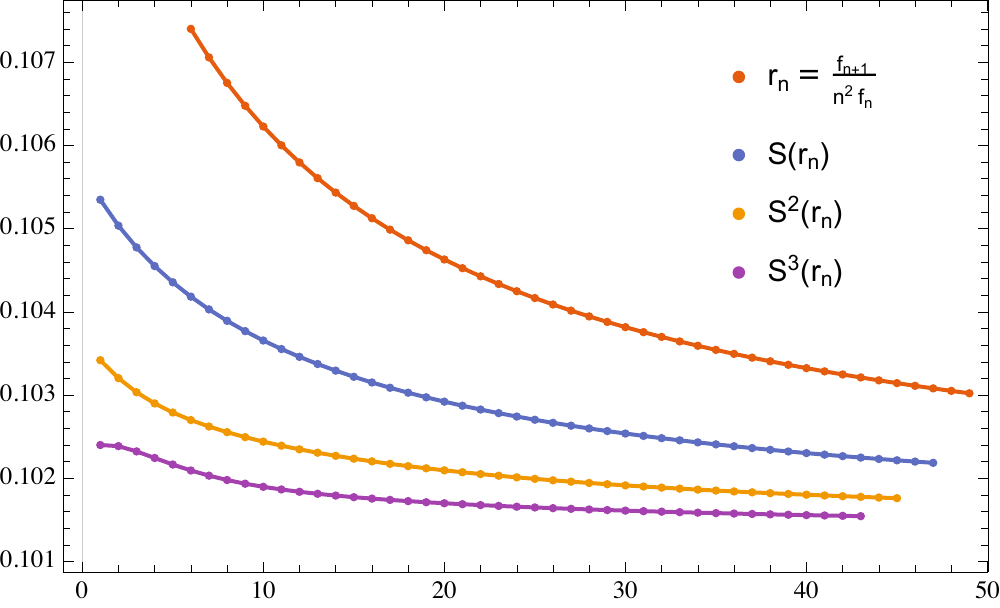}
  \caption{Ratio $r_n = n^{-2} f_{n+1} / f_n $ for the coefficients in the expansion of \(f^{\saddle} (\Qb)\), along with the initial three Shank transforms~\cite{shanks1955non} written in terms of \(n\). The overall convergence at the limit of large \(n\) towards a constant value of order \(\order{1}\) signifies a double-factorial behaviour $f_n \sim (n!)^2$, as anticipated from the outcome of the non-perturbative contributions. }
  \label{fig:fn-coefficients}
\end{figure}

Having understood the non-perturbative asymptotic behaviour of \(\omega(m)\), we now want to study the canonical free energy. The relevant Legendre transform has to be computed order by order in the charge \(\Qb \) initiating the computation from the perturbative contribution
\begin{align}
  \label{eq:Legendre}
  \Qb &= \dv{\mu} \omega(\mu) & \implies &&  R_0 m(\Qb) &=  {\Qb}^{1/2} - \frac{1}{24} {\Qb}^{-1/2} + \frac{43}{5760 }\Qb^{-3/2} + \dots\\
  f^{\saddle}(\Qb) &= \mu \Qb - \omega(\mu) & \implies &&  f^{\saddle}(\Qb) &= \frac{2}{3 R_0} \Qb^{3/2} + \frac{1}{6 R_0} \Qb^{1/2} - \frac{7}{720 R_0} \Qb^{-1/2} + \dots
\label{eq:sphere-free-energy}
\end{align}
The result is an asymptotic series and within the scope of this analysis it suffices to examine solely the first non-perturbative terms that appear in the free energy  \(f^{\saddle}(\Qb)\) and therefore in the suppressed critical exponents. These are derived through the leading-order approximation of \cref{eq:Legendre} and they offer a very high level of precision when compared with the small-charge result. 
Doing so, the non-perturbative corrections read
\begin{equation}
	f^{\saddle}(\Qb) \, \supset \, \frac{\Qb^{3/4}}{R_0}  \frac{(-1)^k}{( 2 \pi \abs{k})^{\frac{3}{2}}}  e^{-(2 \pi \abs{k} ) \sqrt{\Qb}} + \dots \, ,
\end{equation}
which stand for a \(2n!\) factorial divergence of the perturbative terms of the free energy, or in equal grounds a \(2n!\) factorial divergence of the critical exponents, as is depicted in \Cref{fig:fn-coefficients}.

As a final note, from the \acrshort{eft} perspective, the corresponding \((2n)!\) divergence in the free energy is a tree-level and not a quantum effect. By identifying \(\omega (\mu)\) with a \(S_{\text{eff}}\) ---
as we will do in \cref{sec:lessons-from-large-N} --- the Wilsonian coefficients create a divergent series which is to be set side by side with the \(n!\) divergence that we commonly anticipate in \acrshort{qft} that originates in the proliferation of the Feynman diagrams in a perturbative expansion. Therefore, in the limit under consideration, the classically derived divergence is more dominant than the one originating in quantum effects.

By having the precise expressions for every term in the perturbative series, it is feasible to extrapolate the computed result to an arbitrarily small charge if we accomplish to resume the related trans-series into a resurgent trans-series function.
To fulfil this goal, in the next section we will employ the Borel resummation technique.

\subsection{Borel resummation}\label{sec:Borelresum}

In the previous section, we have deduced the generic form of the non-perturbative corrections that are related to the factorially divergent series that we have conjectured as the asymptotics of a resurgent trans-series function. Nonetheless, we still have to give a meaning to the series that we started with, and the Borel resummation \footnote{In \Cref{sec:borel-transform} there is a short review of the Borel transform methodology.} is a method that attains exactly this goal, by systematically including the non-perturbative corrections that we have computed.

We will start our analysis by examining the Borel transformation of the heat-kernel trace expression that we found in \cref{eq:Phi0}, namely
\begin{align*}
    \Tr\bqty{ e^{ \pqty*{ \mathlarger{\mathlarger{\Laplacian}}_{\setS^2} - \frac{1}{4R_0^2}  } t} } & =\frac{R_0^2}{t} \, \Sum_{n=0}^\infty \, a_n \, \pqty{\frac{t}{R_0^2}}^n & a_n & = \frac{(-1)^{n+1}(1-2^{1-2n})}{n!} B_{2n}.
\end{align*}
In general, the Borel transform acts as follows
\begin{align} \label{eq:general-Borel}
\Phi(z) &\sim \Sum_{n=0}^{\infty} \, a_n z^n &\longrightarrow &&  \mathcal{B}\left\{\Phi\right\}(\zeta) &= \Sum_{n=0}^{\infty} \, \frac{a_n}{\Gamma(\beta n + b)} \zeta^n,
\end{align}
For the two-sphere \(\Sigma = \setS^2\), we know from \cref{sec.Seeley--DeWitt_coefficients.} the values of \(\beta = 1\) and \(b = 1/2\), and thus by setting \(z = t/R_0^2\) and neglecting the prefactor \(R_0^2/t\) which we will reintroduce at the end of our analysis to avoid working with a series with negative powers, and mapping \(\zeta \rightarrow \zeta^2\) to get a Borel transform without any branch cuts we get a closed-form expression for the Borel transformation that reads
\begin{equation}
	\mathcal{B}\{\Phi^{(0)} \} (\zeta)=   \Sum_{n=0}^\infty \frac{a_n}{\Gamma(n+1/2)} \zeta^{2n} =  \frac{1}{\sqrt{\pi}} \frac{\zeta}{\sin \zeta} ,
\end{equation}
and to get the above result we used \cref{eq:Gamman+1/2} the Taylor expansion of
\begin{equation}
  \frac{1}{\sin(z)} = 2 \Sum_{n=0}^\infty  B_{2n} \frac{(-1)^n(1-2^{2n-1})}{(2n)!} z^{2n-1}.
\end{equation}
We appropriately Borel resum our expression --- see \Cref{sec:borel-transform} for details --- and we get
\begin{align}
 \mathcal{S} \{ \Phi^{(0)} \} (z) &=  \frac{2}{\sqrt{z}}  \Int\limits_0^\infty \dd{\zeta} \,e^{-\zeta^2/z} 	\mathcal{B}\{\Phi^{(0)} \} (\zeta) \nonumber \\
 &= \frac{2}{\sqrt{\pi z}} \Int\limits_0^\infty \dd{\zeta} \, \frac{\zeta\,e^{-\zeta^2 / z}}{\sin \zeta}.
 \label{eq:Kernel-resummation}
\end{align}
The last expression is exactly the integral representation of the heat-kernel trace in the two-sphere $\setS^2$ that was initially computed in~\cite{perrin1928etude} and was retrieved here in the form of a Borel integral. \newline
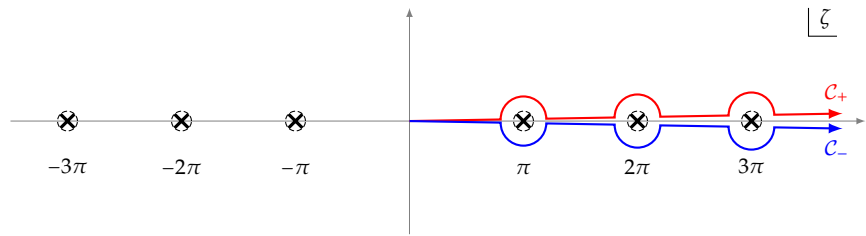
\begin{figure}[h!]
  \centering
  \begin{footnotesize}
    \begin{tikzpicture}[x=1.5cm,y=1.5cm,
      pole/.pic={
        \tikzset{scale=sin 5}
        \clip [preaction={draw, dash pattern=on 2pt off 1pt}] circle [radius=1];
        \draw [very thick] (-1,1) -- (1,-1) (-1,-1) -- (1,1);
      }]

      \draw [help lines, -latex] (-3.5,0) -- (4,0);
      \draw [help lines, -latex]  (0,-1) -- (0,1);
      \foreach \x/\y in {1/{},2/2,3/3}
      {
        \pic at (\x, 0) {pole};
        \node[label={\(\y \pi\)}] at (\x,-.6){};
        \pic at (-\x, 0) {pole};
        \node[label={\(-\y \pi\)}] at (-\x,-.6){};
      }

      \draw[thick, red,-latex, rotate=1] (0,0) -- (8/10, 0) arc (180:0:1/5) -- (12/10, 0) -- (18/10, 0)  arc (180:0:1/5) -- (22/10, 0) -- (28/10, 0)  arc (180:0:1/5) -- (32/10, 0) -- (38/10, 0);
      \node[red] at (3.75,.25) {\(\mathcal{C}_+\)};

      \draw[thick, blue,-latex, rotate=-1] (0,0) -- (8/10, 0) arc (180:360:2/10) -- (12/10, 0) -- (18/10, 0)  arc (180:360:2/10) -- (22/10, 0) -- (28/10, 0)  arc (180:360:2/10) -- (32/10, 0) -- (38/10, 0);
      \node[blue] at (3.75,-.25) {\(\mathcal{C}_-\)};
      
      \draw (3.5,1) -- (3.5,.75) -- (3.75,.75);
      \node at (3.65,.9) {\(\zeta\)};
    \end{tikzpicture}

  \end{footnotesize}
  \caption{The structure of the poles and the choices of the integration contours \(\mathcal{C}_\pm\) for \(\mathcal{S}_\pm \{ \Phi^{(0)} \} (t) \) on \(\setS^2\). The corresponding choices dissent by the residues at \(\zeta = k \pi \), \(k = 1, 2, \dots \).}
  \label{fig:lateral-Borel-sphere}
\end{figure}
Nonetheless, the Borel integral of \cref{eq:Kernel-resummation} is ill-defined since the integrand exhibits simple poles along the integration path for the values of \(\zeta = k \pi, \, \, k  \in \setZ^+\), which signifies that the above series is not Borel summable and the integration ray $\theta = 0$ is a Stokes line.

Therefore, we define two lateral Borel transforms \(\mathcal{S}_\pm \{ \Phi^{(0)} \} (t)\), the integration contours of which pass over the poles or under the poles respectively, as in \Cref{fig:lateral-Borel-sphere}. The above action introduces the subsequent ambiguity to the sum
\begin{equation}
\begin{aligned}
	 [\mathcal{S}_+ - \mathcal{S}_{-} ] \{ \Phi^{(0)} \} (z) &= - (2 \pi i)  \Sum_{k=1}^\infty \Res\limits_{\zeta = k \pi} \left( \frac{2}{\sqrt{\pi z}}  \frac{\zeta\,e^{-\zeta^2 /z}}{\sin \zeta} \right) \\ 
    &= 2 i z \pqty{\frac{\pi}{z} }^{3/2}  \Sum_{k\neq0}^\infty (-)^{k+1}  \abs{k} \,  e^{- k^2 \pi^2/z } ,
\end{aligned}
\end{equation}
which by reinstating the prefactor \(R_0^2/t\) agrees completely with the expression found in~\cref{eq:non-perturbative-heat}. Every single term matches with a pole on the Borel plane along the positive real axis.
But, no matter how we orient the contour, the integration path deviates from the real axis and thus the integral obtains an imaginary contribution. Hence, the still ambiguous trans-series that is associated to the heat-kernel trace reads
\begin{equation}
 \Tr\bqty{ e^{ \pqty*{ \mathlarger{\mathlarger{\Laplacian}}_{\setS^2} - \frac{1}{4R_0^2}  } t} } =  \frac{2 R_0^3}{\sqrt{\pi} t^{3/2}} \Int\limits_{\mathcal{C}_\pm} \dd{\zeta}  \frac{\zeta\, e^{-\zeta^2 R_0^2/t}}{\sin \zeta} + 2i \left( \frac{\pi R_0^2}{t} \right)^{\frac{3}{2}} \Sum_{k\neq 0} \sigma_k^{\pm} (-1)^{k+1}  \abs{k}  e^{- {\pqty{k \pi R_0}^2}/{t}}.
\label{eq:heat-trans-series}
\end{equation}
A similar analysis can be performed for the grand potential, again starting by examining the perturbative coefficients \(\omega_n\) in \cref{eq:gran-pot-coeff}.
Having said that, an elegant closed form is not possible for the Borel transformed coefficient \(\omega^{(0)}\). Therefore, rather of doing that, we implement the Mellin transform to the Borel resumed quantity of \cref{eq:Kernel-resummation} that corresponds to the integral representation of the heat-kernel trace, and we find
\begin{equation}
\begin{aligned}
	\zeta_\pm(s | \setS^2 , m^2 ) 
	&=  \frac{R_0^{2s}}{ \Gamma(s)} \Int\limits_0^\infty \dd{z} z^{s-2} e^{-m^2R_0^2 z} \left[  \mathcal{S}_{\pm} \{ \Phi^{(0)}\} (z) - 1 - \frac{z}{12} \right] +   \frac{ R_0^2 m^{2-2s} }{s-1} + \frac{m^{-2s}}{12} .
	\end{aligned}
\end{equation}
Given that the Mellin integral above has already been analytically continued in order to convergent for \(s = -1/2\),  allows us to swap the order of integration and therefore obtain
\begin{equation}
  \label{eq:resurged-grand-potential}
	\begin{aligned}
	\omega_{\pm}(m) &= - \frac{1}{2} \zeta \left(\left. - \frac{1}{2} \right| \setS^2 , m^2 \right) \\
	&=   \frac{1}{3} R_0^2 m^{3} - \frac{m}{24}  + \frac{m^2 R_0}{\pi}  \Int\limits_{\mathcal{C}^\pm} \frac{ \dd \zeta}{\zeta^2} \left( \frac{\zeta}{\sin \zeta} - 1 - \frac{\zeta^2}{6} \right) K_2(2 m R_0 \zeta).\\
    \end{aligned}
\end{equation}
From the expression of \cref{eq:resurged-grand-potential} we are now able to identify the Borel resummation of the leading series $\omega^{(0)}$ with the subsequent integral 
\begin{equation}
	\mathcal{S}_{\pm}\{\omega^{(0)}\} =  \Int\limits_{\mathcal{C}^\pm} \frac{ \dd \zeta}{\zeta^2} \left( \frac{\zeta}{\sin \zeta} - 1 - \frac{\zeta^2}{6} \right) K_2(2 m R_0 \zeta),
\end{equation}
and for \(\zeta = k \pi\) exhibits a discontinuity at, 
\begin{equation}
  \label{eq:grand-potential-residues}
[	\mathcal{S}_{+}-\mathcal{S}_-]\{\omega^{(0)}\} = \Sum_{k=1}^\infty \frac{(-1)^k}{k^2 \pi^2} K_2(2 \pi k mR_0),
\end{equation}
which agrees completely with the previous results that we had computed in~\cref{eq:non-perturbative-grand}. During our analysis of the heat-kernel trace, we deduced that the non-perturbative corrections are semiclassically exact, in the sense that they only consist of a single term. 
The same logic also applies to the case of the grand potential, as we still possess a single term, yet the non-perturbative contributions include a Bessel function rather than of the ordinary instanton-like exponentials that are common in problems in \acrshort{qft}. 

The non-perturbative ambiguities stemming from the Borel summation in both the case of the grand potential and for the free energy are connected to the ambiguities that appear in the heat-kernel trace in~\cref{eq:heat-trans-series}.
Nevertheless, they can be lifted in several ways:
\begin{itemize}[left= 0pt]
\item The first way, is to impose the reality condition of the heat-kernel trace in~\cref{eq:heat-trans-series}  for \( t \in \setR^+\).
  A priori, this methodology does not guarantee to determine completely the coefficients \(\sigma_k\), yet for our problem it happens to be enough --- see \Cref{sec:Dawson} for details --- and thus we find \(\sigma_k^{\pm} = \pm 1/2\), which hints that \(S_k = 1, \, \forall \, k\).
  The heat-kernel trace then becomes
	\begin{equation}
	\begin{aligned}
	\Tr\bqty{ e^{ \pqty*{ \mathlarger{\mathlarger{\Laplacian}}_{\setS^2} - \frac{1}{4R_0^2}  } t} }  &=  \frac{2}{\sqrt{\pi}} \left( \frac{R_0^2}{t} \right)^{\frac{3}{2}} \Int\limits_{\mathcal{C}_\pm} \dd{\zeta}  \frac{\zeta\, e^{-\zeta^2 R_0^2/t}}{\sin \zeta}  \pm i \left( \frac{\pi R_0^2}{t} \right)^{\frac{3}{2}} \Sum_{k\neq 0} (-1)^{k+1}  \abs{k}  e^{- \pqty{k \pi R_0}^2/{t}} \\
	 &=  \frac{2}{\sqrt{\pi}} \left( \frac{R_0^2}{t} \right)^{\frac{3}{2}} \PV{\bqty{  \Int\limits_{\mathcal{C}_\pm} \dd{\zeta}  \frac{\zeta\, e^{-\zeta^2 R_0^2/t}}{\sin \zeta} }} ,
	 \label{eq:final-heat-trace}
	 \end{aligned}
	\end{equation}
    and the above result is both unambiguous and real, although that it does not look so. This is usually true for many systems that involve \acrfull{ode}~\cite{aniceto2015nonperturbative}. Considering that the heat-kernel trace solves the non-linear partial differential heat equation at coincident points, it is still intriguing to examine how its trans-series structure can be understood from a linear Dawson's \acrshort{ode}.
    \item The second way that we will further explore in the next section, is to deduce a path integral expression of the heat-kernel trace, where the trans-series form emerges naturally from the semiclassical expansion around the non-trivial saddles. For the usual integrals --- see ~\cite{cherman2015decoding} for details ---  such a prescription is enough because it yields the Lefschetz thimble decomposition of the corresponding integral. The same authors demonstrated that for path integrals, unstable saddles also play a part in the cancellation of foresaid ambiguities. Therefore, in \cref{sec:worldline} we will move to this direction and display that an analogous phenomenon is true also in the heat-kernel's trace path integral formulation.
\end{itemize}

\section{Worldline interpretation}%
\label{sec:worldline}

As we saw in \cref{subsec.Quantisation} propagators of elliptic operators can be represented as quantum mechanical path integrals. Actually, in his early days, Feynman initially developed the path integral methodology for non-relativistic systems and after two years he started publishing his well-known set of papers that established relativistic \acrshort{qft} and Feynman diagrams came into being. Nevertheless, he simultaneously established a representation of the quantum electrodynamics \(S\)-matrix written as a relativistic particle path integral in the Appendix of \cite{feynman1950mathematical,feynman1951operator}. This is the particle path integral formalism or in other words the worldline formalism --- see~\cite{schubert2001perturbative} and relevant references within --- where we construct a suitable quantum mechanical particle path integral, to compute the determinant of the aforementioned elliptic operator. 

The methodology has been applied successfully to \acrshort{qft} amplitudes and also to computing effective actions around classical field backgrounds. On that base, heat-kernel traces can be expressed as worldline integrals that consist of a free particle that moves on the curved manifold at hand. Nonetheless, the definition of quantum mechanics on a curved space is not that trivial a matter, and it is known from the time of DeWitt~\cite{dewitt1957dynamical} that it is actually plagued with several ambiguities that are associated to the issue of properly defining the path integration measure on a curved space. Many of these complications have been worked out recently and there is now a perturbative definition of the previously mentioned path integrals --- see \cite{de1995loop} --- that matches the leading Seeley--DeWitt coefficients for generic manifolds~\cite{bastianelli2006path}.

In the following sections, we will apply the worldline methodology to demonstrate that the trans-series function of \cref{eq:final-heat-trace} can be produced as a saddle-point approximation in the limit that \(t \rightarrow 0^+\) of a properly defined quantum mechanical path integral representing the heat-kernel trace. This corresponds to a completely geometrical representation of the non-perturbative corrections and ambiguities that appeared in the previous resurgent analysis. Since the asymptotic character and the relevant ambiguities of the grand potential \(\omega\) and the free energy \(f^{\saddle}(\Qb)\) --- therefore of the scaling dimension \(\Delta_{\Qb}\) --- originate from the similar nature of the heat-kernel trace, the geometrical structure of the latter will carry over to them.


\subsection{The heat-kernel as a path integral}
\label{sec:path-integral}

Although in field theory we are mostly used that path integrals are integrals over fields and not integrals over particles, this does not have to always be the case. Therefore, the first step towards a proper representation of the functional determinant of the operator \(-\del_\tau^2{}-\mathlarger{\Laplacian}{}+\mu^2 \) as a particle path integral is Schwinger's representation of \cref{eq:Schwinger-representation} which reads
\begin{equation*}
     \log\bqty{\det\pqty{-\del_\tau^2{}-\mathlarger{\Laplacian}{}+\mu^2}}   = -  \Int\limits_0^\infty \frac{\dd{t}}{t} \, \Tr\bqty{e^{\pqty{\del_\tau^2{} + \mathlarger{\Laplacian}{} - \mu^2} t} },
\end{equation*}
and its relation to the grand potential \(\omega(\mu)\) is given by~\cref{eq:definitiongrandpotential}.

As explained in detail in \cref{sec:largeN}, on the manifold \(\setS^1_{\beta} \times \Sigma\) \footnote{Taking the limit \(\beta \to \infty\) corresponds to \(\setR \times \Sigma\).} the trace factorises reducing the whole problem to the study of the heat-kernel trace on \(\Sigma\)
\begin{equation}
      \log\bqty{\det\pqty{-\del_\tau^2{}-\mathlarger{\Laplacian}{}+\mu^2}}   = -  \Int\limits_0^\infty \frac{\dd{t}}{t} \, \frac{1}{\sqrt{4 \pi t}} e^{- \mu^2 t} \, \Tr\bqty{e^{ \mathlarger{\Laplacian}{}  t} }.
\end{equation}
The main idea is to match the heat-kernel trace \( \Tr\bqty{e^{ \mathlarger{\Laplacian}{}  t} }\) to the partition function \(\mathcal{Z}(\beta) = \Tr\bqty{e^{-\beta H}}\). Making the comparison, this corresponds to a particle that has an inverse temperature \(\beta =t\) and whose Hamiltonian is given by \(H = - \mathlarger{\mathlarger{\Laplacian}}\), that is a free quantum particle moving on the manifold $\Sigma$ ~\cite{dewitt2003global,camporesi1990harmonic,bastianelli2006path}.
By choosing a map \(x^\mu\) on \(\Sigma\) the classical action of the aforementioned free particle is 
\begin{equation}\label{eq:actionparticle}
  S[x] = \frac{1}{4} \Int_0^t \dd{\tau} g_{\mu \nu}(x) \dot x^\mu(\tau) \dot x^\nu(\tau),
\end{equation}
where \(g_{\mu\nu}\) stands for the metric on $\Sigma$ and $x^\mu:(0,t) \to  \Sigma$ is the worldline defined by the motion of the particle. The heat-kernel trace is therefore associated with a path integral over closed loops which reads
\begin{equation}
  \Tr\bqty{e^{-\beta H}} \overset{H = - \mathlarger{\mathlarger{\Laplacian}}}{\underset{\beta =t}{ =\joinrel=\joinrel=}} \Tr \left[ e^{ \mathlarger{\Laplacian}{} t} \right]  \equiv  \mel{x}{e^{\mathlarger{\mathlarger{\Laplacian}}{} t} }{x}  = \Int\displaylimits_{x(0)=\Bar{x}}^{x(t) = \Bar{x}} \DD{x^\mu} \, e^{-S[x]}.
  \label{eq:working-definiton}
\end{equation}
From this point on, we will consider \cref{eq:working-definiton} as the working definition of the heat-kernel trace in the particle path integral representation.
This representation of the heat-kernel is derived by applying the Feynman--Kac formula, which is however rather involved due to the intrinsic diffeomorphism invariance and also the emerging ordering ambiguities that are generated by the curvature terms when we quantise the Hamiltonian of the system. 
Nonetheless, these problems have been worked out for the case of the semiclassical expansion around the loop \(x_{\rm cl}^\mu(\tau) = 0\) in \cref{eq:working-definiton}.

In the current analysis, we exploit the fact that all of these alterations are subleading in the expansion in the limit that \(t \rightarrow 0^+\), hence it is indeed a semiclassical expansion and by rescaling the worldline time as \(\tau \rightarrow \tau' =  t \tau \), the action of \cref{eq:actionparticle} can be recast as
\begin{equation}
  S[x] = \frac{1}{4t} \Int_0^1 \dd{\tau} g_{\mu \nu}(x) \dot x^\mu \dot x^\nu ,
\end{equation}
and the usual \(\hbar\) expansion of quantum mechanics equals the small-\(t\) expansion of the above heat-kernel where the path integral localises around the saddle points of the action, and we may express it as a perturbative expansion in powers of \(t\).

The \acrshort{eom} of \cref{eq.extremalofLangrangian} can be found by varying the above action and therefore the Euler--Lagrange equations in our case are the geodesics 
\begin{equation}
  \ddot x_{\text{cl}}^\mu + \Gamma^\mu_{\nu \rho}(x) \dot x_{\text{cl}}^\nu \dot x_{\text{cl}}^\rho = 0,
\end{equation}
where \( \Gamma^\mu_{\nu \rho}\) are the Christoffel symbols~\cite{CarrollSean2014Saga}. We observe that the localisation of the heat-kernel trace path integral takes place as a sum over every closed geodesic \(\gamma\) on the manifold \(\Sigma\) and that these non-trivial geodesics are the exact equivalent of the worldline instantons that appear in~\cite{dunne2005worldline,dunne2006worldline} which regulate the non-perturbative corrections of Euler--Heisenberg-type Lagrangians~\cite{dunne2005heisenberg}.

From general calculus arguments we expect that every one of these saddles will exhibit its individual perturbative series in \(t\), weighted by a factor of \(e^{-\ell(\gamma)^2/(4t) }\), where \(\ell(\gamma)\) is the length of the closed geodesics, and therefore the semi-classical expansion reads
\begin{equation}
   \Tr \left[ e^{ \mathlarger{\Laplacian}{} t} \right] \,  = \, t^{-b_0} \, \Sum_{n=0}^\infty \, a_n^{(0)} \, t^{n} + \sideset{}{'} \Sum_{\text{\(\gamma \in \) closed geodesics}} e^{-\frac{\ell(\gamma)^2}{4 t} } t^{-b_\gamma} \, \Sum_{n=0}^\infty \, a_n^{(\gamma)} \, t^{n},
  \label{eq:semiclassical}
\end{equation}
and the second summation is over the non-trivial geodesics, and also the coefficient \(b_\gamma\) is geometrically dependent.
The series \(a_n^{(\gamma)}\) are, in general, anticipated to be factorially growing, due to the usual arguments about the proliferation of Feynman graphs. 

The resemblance of the structure of \cref{eq:semiclassical} to the corresponding one of the general trans-series in~\cref{eq:trans-series} is not an accident, as the latter were introduced to agree with semiclassical expansions, where they naturally appear. Even so, there is a conceptual disagreement regarding the resurgent analysis that was carried out in \cref{sec:asymptotics} as resurgence is independent of the existence of a non-perturbative definition of the observable that we want to study. That is why there is no generic geometric interpretation for the trans-series structure, and there exist ambiguities that can't be fixed immediately. For the problem at hand, we have already demonstrated that the relevant ambiguities can be lifted by imposing the reality condition of the heat-kernel. 
In the next sections, we will show that using the above path integral, we can reproduce the outcome of \cref{eq:final-heat-trace} without any further physical input.


\subsection{The torus}
\label{sec:torus-path-integral}

As in \cref{sec:torus} we begin our analysis by examining the case of the torus \(\setT^2\) that acts as a probe before moving on to the sphere \(\setS^2\), where we will attempt to replicate the result of \cref{eq:Torus-trace-Poisson} using the worldline framework.

We start by considering a square torus with sides of length \(L\), where the corresponding metric reads%
\begin{equation}
  \dd s^2 = g_{ij} \dd{x^i}  \dd{x^j} = (\dd{x^1})^2 + (\dd{x^2})^2.
\end{equation}
Given that the torus is flat, there are no subtleties arising from the curvature of the manifold that would affect our definition in \cref{eq:working-definiton} and the corresponding heat-kernel trace is expressed by the path integral
\begin{equation}\label{eq:pathintegraltorus}
  \Tr\bqty{e^{\mathlarger{\Laplacian}{} t}}   =  \Int\displaylimits_{x(t)=x(0)} \DD{x} e^{- \frac{1}{4t} \Int\limits_0^1 \dd{\tau} ( (\dot x^1)^2 + (\dot x^2)^2) },
\end{equation}
that localises for \(t \to 0^+\) and thus it is possible to calculate it semiclassically using the usual saddle-point approximation methods. 

\begin{figure}[h!]
  \hfill
  \begin{tikzpicture}[scale=.85]
    \coordinate (Origin)   at (0,0);
    \coordinate (XAxisMin) at (-1,0);
    \coordinate (XAxisMax) at (6.5,0);
    \coordinate (YAxisMin) at (0,-1);
    \coordinate (YAxisMax) at (0,3.5);
    \draw [thin,-latex] (XAxisMin) -- (XAxisMax);%
    \draw [thin,-latex] (YAxisMin) -- (YAxisMax);%

    \clip (-.6,-.6) rectangle (6.6,3.6); %

    \draw[thin,gray] (-2,-2) grid[step=1] (7,7);

    \coordinate (Xone) at (1,0);
    \coordinate (Xtwo) at (0,1);

    \filldraw[fill=gray, fill opacity=0.3, draw=black] (Origin)
    rectangle ($(Xone)+(Xtwo)$);

    \draw[thick, red, -latex] (0,0) -- (5,2);
    \draw[thick, blue, -latex] (0,0) -- (1,3);
  \end{tikzpicture} \hfill
  \includegraphics[width=.45\textwidth]{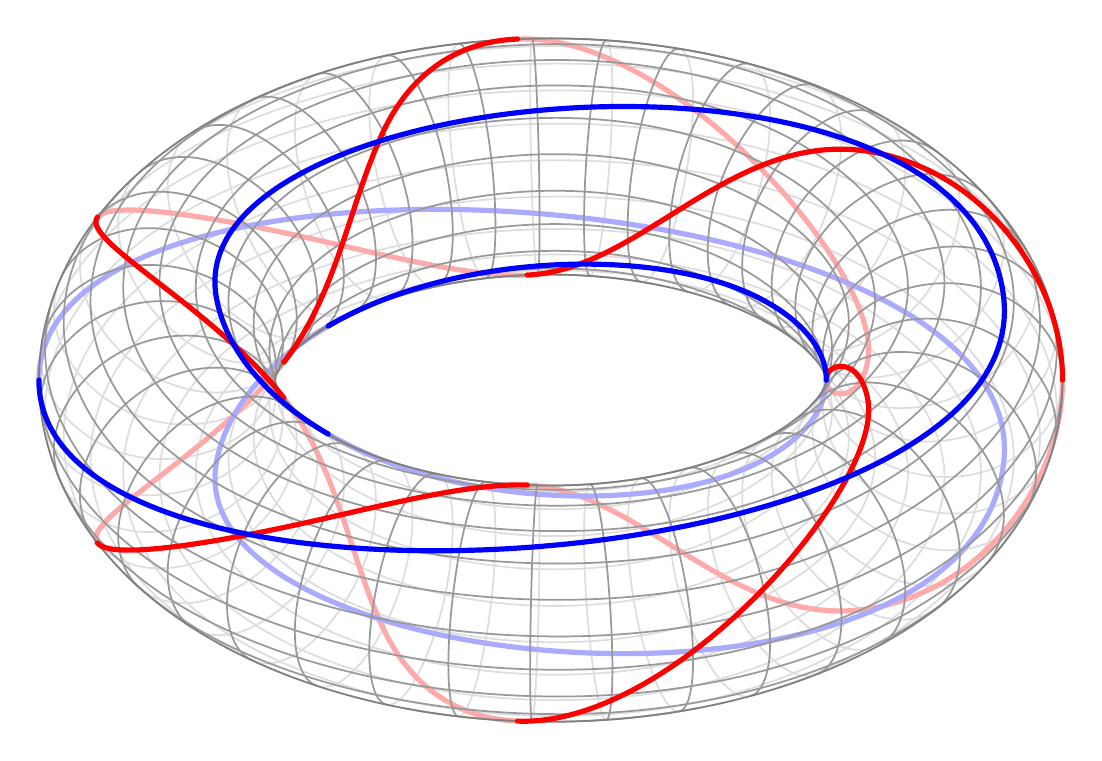}
  \hfill
  \caption{Non-trivial closed geodesics depicted on the torus and labelled by the integers \((1,3)\) for the blue line and \((5,2)\) for the red line as sections in \(\setR^2\) and on a three-dimensional representation.}
  \label{fig:torus-geodesics}
\end{figure}
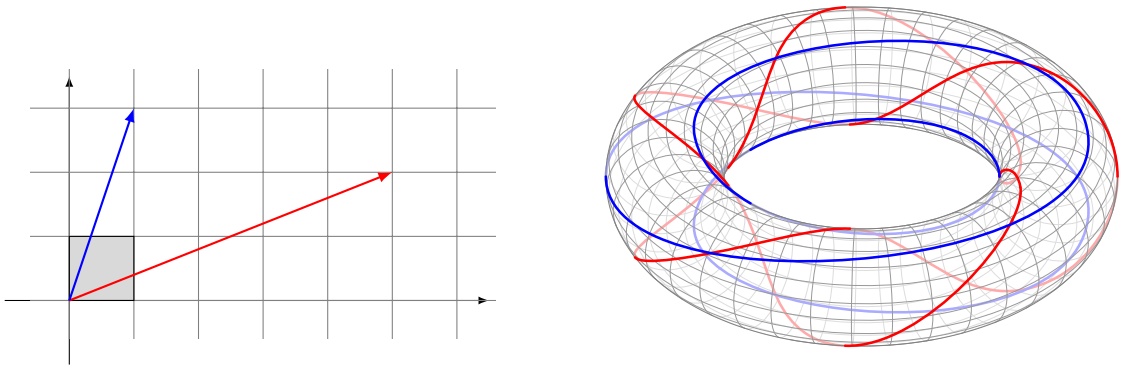

We express the torus as \(\setR^2\) and we identify \(x^i \simeq x^i + L\).
This way, by fixing any point, for example the origin, we end up with a lattice \(\setZ^2 \) of identical points and the corresponding closed geodesics that pass through these points are straight lines that connect the point in the origin to any other point that exists in the lattice as in \Cref{fig:torus-geodesics}.
This indicates that the closed geodesics can be expressed by pairs of integer numbers \((k_1, k_2)\) that also include the trivial zero length geodesic.
The corresponding length \(\ell\) of the closed geodesics can be computed by a simple trigonometric identity and it reads
\begin{equation}
  \ell(k_1, k_2) = L \sqrt{k_1^2 + k_2^2}.
\end{equation}
In the usual manner, the field \(x\) can be split into two parts, the first part is the classical solution \(X^i_{\text{cl}}\) and the second part are the quantum fluctuations \(h^i(\tau)\),
\begin{equation}
  x^i(\tau ) = X^i_{\text{cl}}(\tau) + h^i(\tau) = k^i L \tau + h^i(\tau) .
\end{equation}
Given that the torus \(\setT^2\) is flat, the corresponding action will be quadratic, and therefore we are able to separate the two parts
\begin{equation}
  S[x] = S[X_{\text{cl}}] + S[h] = \frac{L^2 (k_1^2 + k_2^2)}{4 t} + \frac{1}{4t} \Int\limits_0^1 \dd{\tau} \left[ (\dot h^1)^2 + (\dot h^2)^2 \right] .
\end{equation}
And then, using \cref{eq:pathintegraltorus} we can express the path integral as
\begin{align}\label{eq:halffinishedtorus}
\Tr\bqty{e^{\mathlarger{\Laplacian}{} t}}   & =  \Int\limits_{\setT^2} \dd{x}  \Int\displaylimits_{x(1)=x(0)=x} \DD{x} e^{S[x] } \nonumber \\
  & =L^2 \Sum_{\bm{k} \in \setZ^2} e^{-\frac{L^2 (k_1^2 + k_2^2)}{4 t}} \Int\displaylimits_{h(t) = h(0) = 0} \DD{h} e^{-S[h]},
\end{align}
and in the first line we have included the integral over the fixed point \(x\) that every geodesic passes through because due to translation invariance, the specific point is not relevant as any choice would be equivalent, and therefore it needs to be integrated over.
The fluctuation part of the path integral is Gaussian and can be evaluated up to a normalisation parameter \(\mathcal{N}\). We will choose the usual worldline normalisation \(\mathcal{N} = \pqty{4\pi t}^{-\frac{d}{2}}\), where \(d\) are the components of the path integral --- see \cite{SchubertWorld,strassler1992field} and \Cref{sec;pathintegralnormalisation} for details of the calculation --- to get 
\begin{equation}
 \Int\displaylimits_{h(t) = h(0) = 0} \DD{h} e^{- \frac{1}{4t} \int_0^1 \dd \tau \dot{h}^2} = \frac{1}{\sqrt{4\pi t}}.
 \label{eq:normalisation-factor}
\end{equation}
The final result is produced by combining \cref{eq:halffinishedtorus,eq:normalisation-factor} and reads
\begin{equation}
 \Tr\bqty{e^{\mathlarger{\Laplacian}{} t}}  = \frac{L^2}{4\pi t} \Sum_{\bm{k} \in \setZ^2} e^{- \frac{L^2 \norm{\bm{k}}^2}{4t}} ,
\end{equation}
where again we denote \(\norm{ \bm{k}}^2 = k_1^2 + k_2^2\) and it is in perfect agreement with \cref{eq:Torus-trace-Poisson} that was computed using the Poisson resummation.

In the literature, this correspondence is referred to as spectrum-geodesic duality for compact manifolds. For every eigenvalue of the Laplace operator there exists a corresponding closed geodesic, and the interested reader is referred to ~\cite{camporesi1990harmonic,grosche2013path} for a discussion.

\subsection{The sphere}
\label{sec:sphere-path-integral}

We now move on to the case of the sphere \( \setS^2\) where, a priori, generic ambiguities generated by the curvature of the manifold are involved. Nevertheless, since these are subleading, they will not affect our semiclassical analysis. We start by employing a generalisation of \cref{eq:working-definiton} which reads
\begin{equation}
  \mel{y}{e^{ \mathlarger{\mathlarger{\Laplacian}}{} t }}{x} = \Int\displaylimits_{x(0)=x}^{x(t) = y} \DD{x^\mu}  \exp\pqty{- \frac{1}{4t } \Int\limits_0^1 \dd \tau \, g_{\mu\nu}(x) \dot{x}^\mu \dot{x}^\nu}.
\label{eq:path-integral-def}
\end{equation}
We want to go to polar coordinates, where  \(x^\mu = (\theta,\phi)\) and \(\theta \, \mathlarger{\mathlarger{\in}} \, \bqty{0, \pi}\) is the polar angle while \(\phi \, \mathlarger{\mathlarger{\in}} \,  \bqty{0, 2\pi}\) is the azimuthal angle and the volume element is expressed as
\begin{equation}
    \DD{x^\mu} = \sin(\theta) \, \DD{\theta} \, \DD{\phi}.
\end{equation}
From now on, \(\theta\) and \(\phi\) will be our worldline fields, and on the two-sphere \(\setS^2\) the action and the \acrshort{eom} are given by
\begin{align}
  S\bqty{\theta, \phi} &= \frac{R_0^2}{4t} \Int_0^1 \dd{\tau} \left[ \dot{\theta}^2 + \sin^2 \theta \dot \phi^2 \right],  \\
  &\begin{cases}
      \ddot{\phi} + 2 \cot(\theta) \dot{\theta} \dot{\phi} = 0, \\
      \ddot{\theta} - \dot{\phi}^2 \sin(2 \theta) = 0. 
    \end{cases}
	\label{eq:action-EOM}
\end{align}	
We are looking for the classical solution of \cref{eq:action-EOM} and as we want to compute a heat-kernel trace we know from both \cref{sec.thermalQFT,eq:working-definiton} that we have to look at coincident endpoints \(x = y = \Bar{x}\). It is rather straightforward to solve the above \acrshort{eom} for both fields that satisfy our boundary conditions \footnote{The action of \cref{eq:path-integral-def} written in those coordinates is clearly \emph{not} rotationally invariant for \(\theta\) since there exists two singular points which are the poles. Nevertheless, this problem comes out only at a higher order in \(t\), therefore the leading order result in the limit \(t \rightarrow 0^+\) is all right.}, and although that the polar field has a unique classical solution which is the equator, for the azimuthal field we can easily think of a situation that for two points \(x^{\text{in}}, x^\text{f}\), the final point \(x^\text{f}\) coincides with \(x^{\text{in}}\) after \(k\) wrappings around the equator. 

Therefore, we deduce that there exists an infinite number of winding geodesics solving the \acrshort{eom} and the classical trajectory between the two points is parametrised as
\begin{align}
	\theta_{\rm cl}(\tau) &= \pi/2 , &	\phi_{\rm cl}(\tau) &= 2 \pi k \tau , & k &\in \setZ.
\end{align}	
We can split the worldline fields into classical solutions and fluctuations, and we introduce the fluctuations \(h_\theta, h_\phi\) around the above classical solutions, that satisfy Dirichlet boundary conditions. We can then express the heat-kernel trace at coincident points as
\begin{equation}
 \mel{x}{e^{ \mathlarger{\mathlarger{\Laplacian}}{}  t }}{x} = e^{- \frac{(2\pi k R_0)^2}{4 t} } \Int\displaylimits_{h_i(t) = h_i(0) = 0}  \DD{h_\theta} \, \DD{h_\phi} \, \exp\pqty{  - \frac{R_0^2}{4t} \Int\limits_0^1 \dd{\tau} \bqty{ \dot{h}_\theta^2 - (2\pi k)^2 h_\theta^2 + \dot{h}_\phi^2 + \order{h^3} }   },
\end{equation}
where we have learnt in \cref{sec.U(1)ssb} that this is the quadratic action of two free fields, one massless and one massive. The higher-order terms, \emph{i.e.} \(\order{h^3}\), which correspond to interactions, contribute to higher order in the parameter \(t\) and therefore they are ignored. 

The first integral over the massless field \(h_\phi\) can be easily computed and actually reproduces our normalization of \cref{eq:normalisation-factor} with the succeeding substitution \(t \rightarrow t /R_0^2\).

\begin{figure}[ht]
  \hfill
  \includegraphics[width=.3\textwidth]{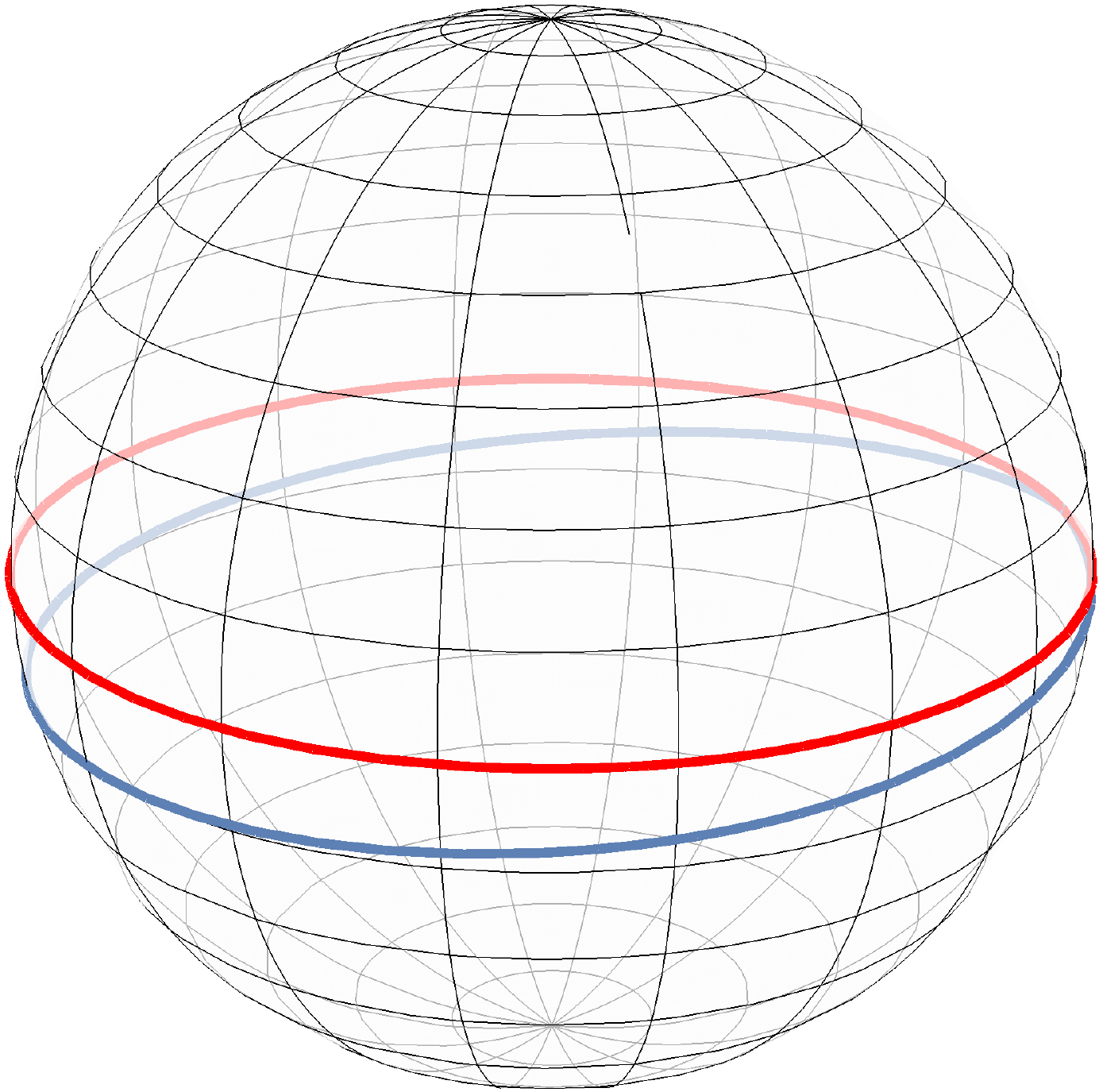}
  \hfill
    \includegraphics[width=.3\textwidth]{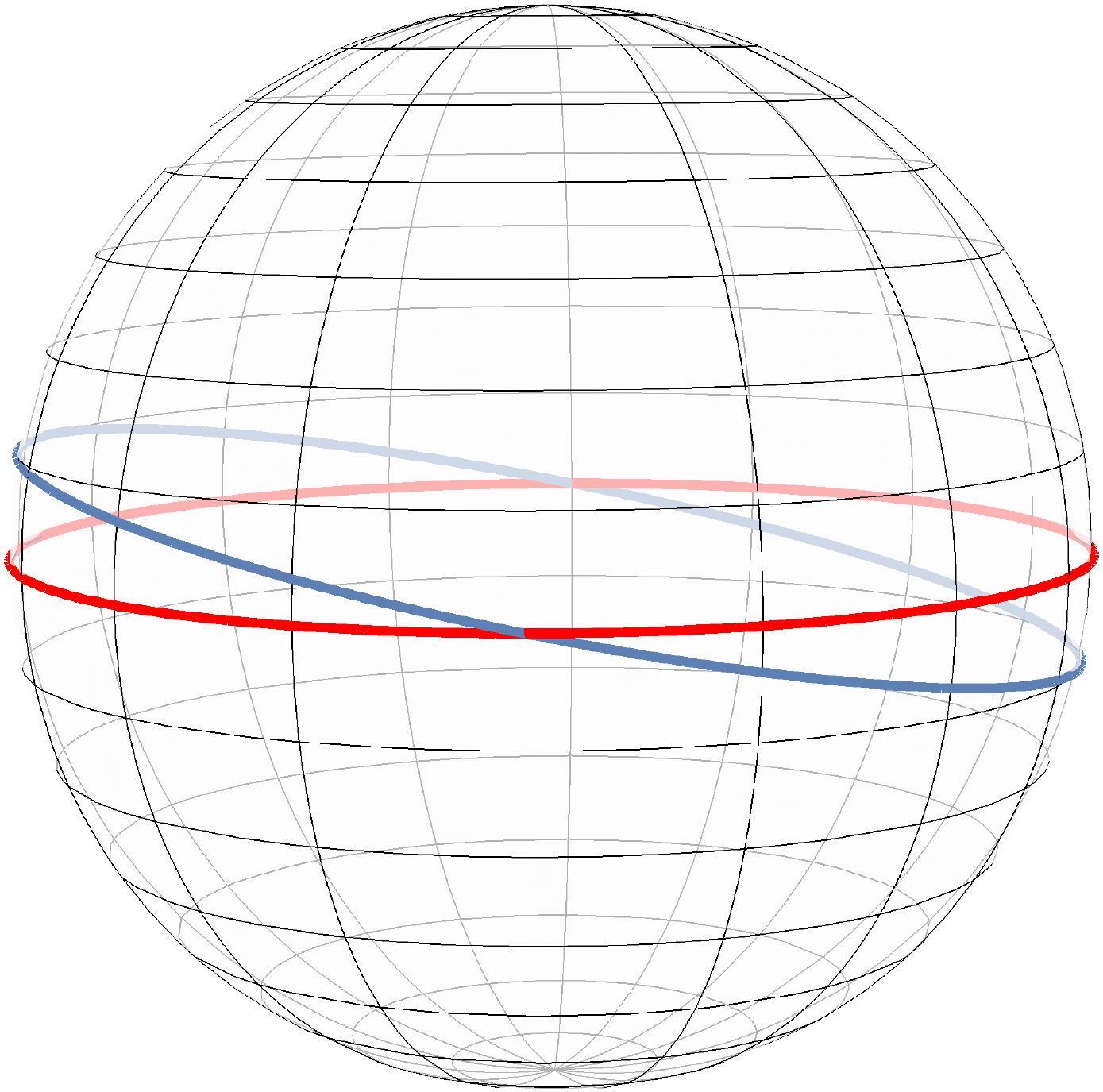}
  \hfill
  \includegraphics[width=.3\textwidth]{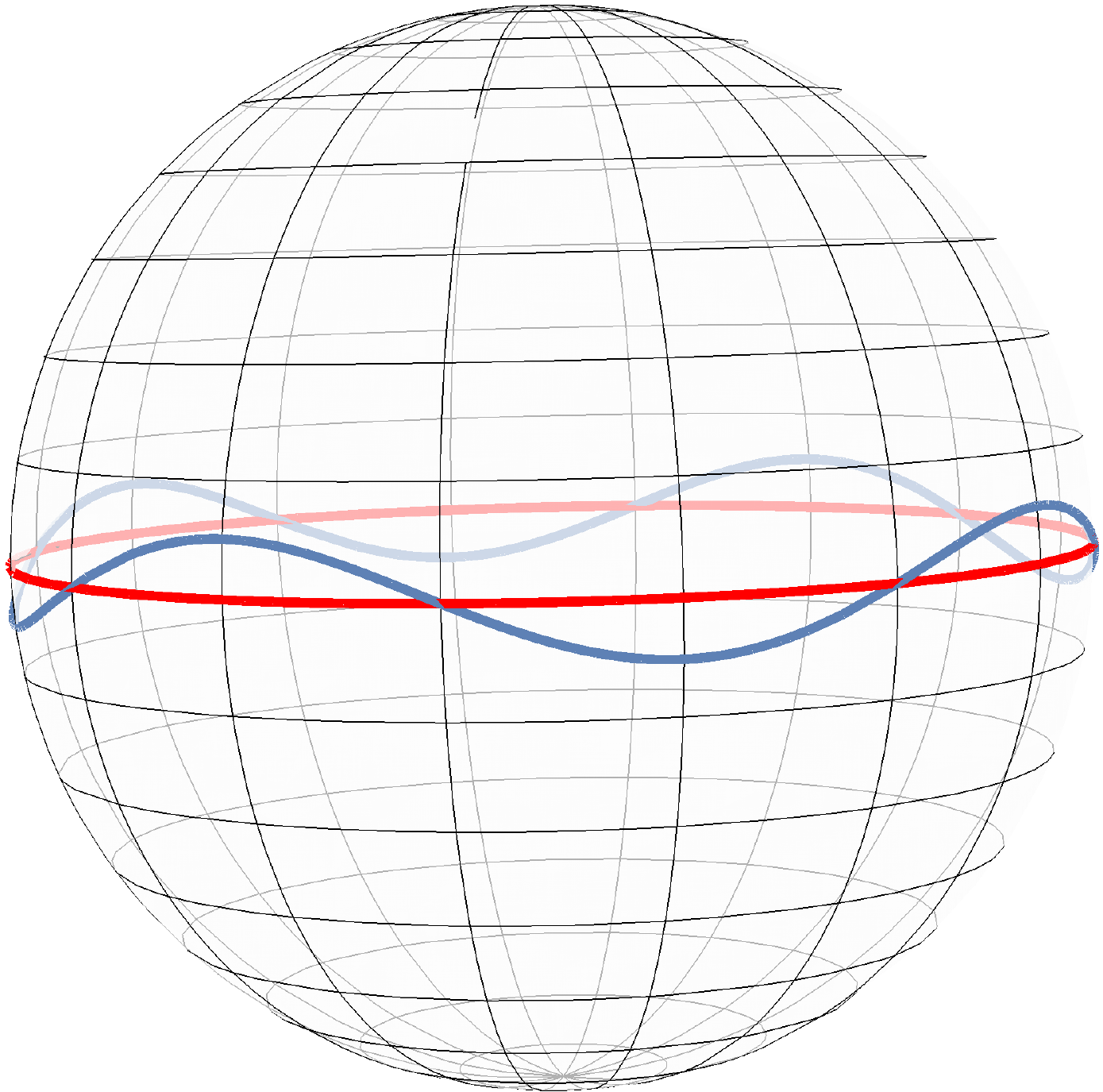}
  \caption{For \(k=1\) which corresponds to a single winding around the equator (red line), the unstable mode \(n=1\) is depicted on the left with blue, the zero mode \(n=2\) is in the middle and a massive mode \(n=8\) is on the right.}
  \label{fig:unstable-mode-sphere}
\end{figure}

On the contrary, the integral over the massive field \(h_\theta\) contains both a zero-mode and also multiple negative modes.
This can be seen in the following manner: by expanding the fluctuations in a complete orthonormal basis in the usual mode decomposition of the eigenfunctions that satisfies the following differential equation
\begin{align}
	h_\theta(\tau) &= \Sum_{n=1} c_{n} h_\theta^n (\tau) ,& - \frac{R_0^2}{2t} \left[ \partial_\tau^2 +(2\pi k)^2 \right] h_\theta^n(\tau) &= \lambda_n  h_\theta^n(\tau),
\end{align}
where the eigenbasis for the above eigenfunctions and eigenvalues is explicitly
\begin{align}\label{eq:zeroandtachyonicmodes}
	h_\theta^n &= \sqrt{2} \sin ( \pi n \tau), & \lambda_n &= \frac{\pi^2R_0^2}{2t} \pqty{n^2 - 4 k^2}.
\end{align}
Similarly, we can express the measure of integration on the space of the fluctuations \(h_\theta\) regarding the Fourier modes as
\begin{equation}
	\Int \DD{h_\theta} \equiv \mathlarger{\mathlarger{\prod}}_{n=1}^\infty \Int \frac{\dd{c_n}}{\sqrt{2\pi}}.
\end{equation}
From \cref{eq:zeroandtachyonicmodes} we observe that the zero mode lies at \(n = 2k\) and we need to treat it separately, while at the same time for \(n < 2k\) there are \(2k-1\) modes which we will denote \(h_\theta^{n<2k}\) that are tachyonic. Unlike the torus case of before, where we found that winding geodesics were topologically stable saddles, these winding geodesics are clearly not topologically stable since, as we can see in \Cref{fig:unstable-mode-sphere} they can be contracted to a point.

The aforementioned zero mode stands for a rigid rotation of the sphere as seen in \Cref{fig:unstable-mode-sphere}, that happens to be a symmetry of the action. Actually, we can easily observe that
\begin{equation}
  \begin{cases}
    \theta^\alpha_{\rm cl} = \frac{\pi}{2} + \alpha \sin ( 2 \pi k \tau), \\
    \phi_{\rm cl} = 2\pi k \tau,
  \end{cases}
\end{equation}
corresponds to a set of solutions of the \acrlong{eom} (\ref{eq:action-EOM}) at leading order in the coefficient \(\alpha \in (0,\pi)\).
That is precisely the fluctuation \(h_\theta^{2k}\), and applying the rules of instanton calculus we can exchange the integral from over the Fourier mode amplitude \(c_{2k}\) with an integral over the coefficient $\alpha$, which changes the integration measure as
\begin{equation}
	\Int \frac{\dd{c_{2k}}}{\sqrt{2\pi}} = \sqrt{\frac{1}{2}} \Int \frac{\dd{\alpha}}{\sqrt{2\pi}} =  \frac{\sqrt{ \pi }}{2}.
\end{equation}
The following functional determinant is obtained by the integral over the rest of the modes
\begin{equation}
\sideset{}{'}\det \left( - \frac{R_0^2}{2t} \left( \partial_\tau^2 + (2\pi k)^2 \right)  \right)^{-\frac{1}{2}} = \frac{\sqrt{2} \pi \abs{k} R_0}{ \sqrt{t}} \det \left( - \frac{R_0^2}{2 t} \partial_\tau^2 \right)^{-\frac{1}{2}} \sideset{}{'}\det \left(  \text{Id} + \frac{4\pi^2 k^2}{ \partial_\tau^2}    \right)^{-\frac{1}{2}} ,
\end{equation}
where we have to mention that no multiplicative anomaly is generated by the splitting of the determinant \cite{monin2016partition}. What we did was to simultaneously divide and multiply the above equation with the \(n = 2k\) eigenvalue of \(\partial_\tau^2\) since the first determinant on the right corresponds to our previous normalisation. The leftover determinant does not require regularisation and is evaluated to be
\begin{equation}\label{eq:determinantregularised}
	\sideset{}{'}\det \left(  \text{Id} + \frac{4\pi^2 k^2}{\partial_\tau^2}    \right)^{-\frac{1}{2}}  = e^{\frac{i\pi \nu_{q}}{2}} \left( \mathlarger{\mathlarger{\prod}}_{\substack{n =0\\n \neq 2k}}^{\infty} \abs{ 1 - \frac{4k^2}{n^2}}\right)^{-\frac{1}{2}} = \sqrt{2} \,e^{\frac{i\pi \nu_{q}}{2} },
\end{equation}
where for \(k>0\) we evaluated the infinite product in the following manner
\begin{equation}
	\mathlarger{\mathlarger{\prod}}_{n \neq 2k } \abs{ 1 - \frac{4k^2}{n^2}} = \mathlarger{\mathlarger{\prod}}_{n=1}^{2k-1} \left( \frac{4k^2}{n^2} - 1 \right) \mathlarger{\mathlarger{\prod}}_{n=2k+1}^{\infty} \left( 1 - \frac{4k^2}{n^2} \right) = \frac{ \Gamma(4k)}{2k \Gamma(2k)^2 } \frac{\Gamma(2k+1)^2}{\Gamma(4k+1)} = \frac{1}{2}.
	\label{eq:func-det}
\end{equation}
In \cref{eq:determinantregularised} the Morse index \(\nu_q\) was introduced, whose existence is natural in the context of functional determinants that have \(q\) negative modes \cite{horvathy2011maslov}. It can be understood as an analogue of the intersection numbers that arise in the Lefschetz thimble decomposition in the case of ordinary integrals.

In the context of our analysis, it is because the analytic continuation of Gaussian integrals with negative modes features a two-fold ambiguity 
\begin{equation}
	\Int \frac{\dd c_n}{\sqrt{2\pi}}\, e^{ \frac{1}{2} \lambda_n c_n^2 } = e^{\pm i \frac{\pi}{2}}  \frac{1}{\sqrt{\lambda_n}}.
	\label{eq:morse}
\end{equation}
In the previous computation of the determinant in \cref{eq:func-det} we chose to factor out the \(q = 2k-1\) individual phases so that every term in the above infinite product is positive definite.
Each phase is independently chosen for each negative mode and the same continuation procedure is selected for every one of them, to get
\begin{equation}
e^{\frac{i \pi \nu_q}{2}} = (\pm i)^{2k-1} = \mp i (-1)^k.
\end{equation}
We can gather and put all of the above results together, so that at the end we get the semiclassical saddle-point approximation of the heat-kernel trace that reads
\begin{equation}\label{eq:pathIntHeatTrace}
	\Tr[ e^{\mathlarger{\mathlarger{\Laplacian}}{} t} ] = \frac{R_0^2}{t} \pqty{1 + \order{t}} \pm i \left( \frac{\pi R_0^2}{t} \right)^{\frac{3}{2}} \sideset{}{'}\Sum_{k \in \Bqty{\setZ - \setminus \Bqty{0}}} (-1)^{k+1} \abs{k} e^{- \frac{k^2\pi^2 R_0^2}{t} } \pqty{1 + \order{t}}.
\end{equation}
We observe that this results is in perfect agreement with the unambiguous result of \cref{eq:final-heat-trace}.

This methodology does not let us calculate subleading order corrections in every sector.
Nevertheless, there exists a methodology that calculates the Seeley--DeWitt coefficients only in the perturbative sector using a diagrammatic expansion, and the interested reader is referred to ~\cite{bastianelli2017quantum}.
As far as we know, up to this point there was no attempt of reproducing any fluctuations of the non-trivial sectors.

Although this goes beyond the current analysis, the equivalence between the formalisms that we computed allows us to come to the following intriguing conclusions 
\begin{itemize}
	\item The structure of the trans-series concerning the heat-kernel trace on the two-sphere $\setS^2$, and therefore the trans-series structure of the scaling dimension \(\Delta_\Qb\) of charged operators in the \acrshort{lce} for the case of the \(O(2N)\) model, is completely specified by geometrical arguments. 
	
 \item Moreover, the saddle points that drive the factorial growth in the \acrshort{lce} do \emph{not} have to be stable. The non-perturbative structure of the model is commonly guided by topological considerations, but resurgent asymptotics can also be driven by saddles that are in the same topological class, like the example of the equator on \(\setS^2\). This fact was already explored for two-dimensional field theories in~\cite{cherman2015decoding}. 
 
\item The existence of negative modes and the relevant choice of the proper continuation in \cref{eq:morse} that give rise to the non-trivial Morse index corresponds to a geometric understanding of the Borel ambiguity computed in the framework of resurgence analysis in \cref{eq:final-heat-trace}. Selecting different phases is like choosing different paths to avoid the pole singularities in the Borel plane in every way possible. At the very end, this is associated with the Lefschetz thimble integral decomposition~\cite{witten2011analytic}.
\end{itemize}


\section{Comparison with the small charge expansion}
\label{sec:resurgence}

Up to now, we have explored the large-charge limit of the \(O(2N)\) vector model and its asymptotics using the resurgence methodology.
But, as we stated before, being in the double-scaling limit of the theory, we are in a position to make exact predictions at leading order
in \(N\) for every value of the charge \(\Qb\), therefore the small-charge regime is within reach, and we can express the grand potential \(\omega\) as a convergent expansion in terms of the chemical potential \(\mu\). 

To do so, we start from the expression of the grand potential in \cref{eq:grand-potential-zeta} written in terms of the zeta-function on the sphere \(\setS^2\) but in the limit \(\mu \to 0\) we use a binomial expansion instead of an asymptotic --- for details see \cite{alvarez2019large} \S \, 3.3 and Appendix B --- which now reads
\begin{align}
  \omega(\mu) & = - \frac{1}{2} \zeta(-\tfrac{1}{2}| \setS^2, \mu) = - \frac{1}{2} \eval{ \Sum_{l=0}^\infty (2l + 1) \left( \frac{l(l+1)}{R_0^2} + \mu^2 \right)^{-s}}_{s = -1/2} \nonumber \\
 & = -  R_0^{-2s} \eval{ \Sum_{k=0}^\infty \binom{-s}{k} \zeta(2s+2k-1, \tfrac{1}{2}) \pqty{\mu^2 R_0^2 - \frac{1}{4}}^k}_{s=-1/2},
  \label{eq:pot-small-q}
\end{align}
where with \(\zeta(s,a)\) we denote the Hurwitz zeta function
\begin{equation}
  \zeta(s,a) = \Sum_{n=0}^\infty ( n + a)^{-s},
\end{equation}
and we also observe that it would have been more convenient if in the previous expansion we had used the conformal mass~\(m^2 = \mu^2 - {1}/{4 R_0^2}\) instead. By using the following expression of the zeta function
\begin{equation}
    \zeta\pqty{2n, \frac{1}{2}} = \pqty{2^{2n} -1} \zeta\pqty{2n} = \pqty{-1}^{n+1} \pqty{2^{2n}-1} \frac{B_{2n} \pqty{2 \pi}^{2n}}{2 \pqty{2n}!},
\end{equation}
we can rewrite \cref{eq:pot-small-q} in terms of \(m\) and the Bernoulli numbers as
\begin{equation}\label{eq:smallchargegrandpotential}
  \omega(m) = R_0 m^2 \Sum_{k=0}^\infty (-1)^k  \binom{1/2}{k+1} \frac{(2\pi)^{2k} (2^{2k}-1) B_{2k}}{2 (2k)!} (R_0 m)^{2k}.
\end{equation}
The above expression is a convergent expansion and not an asymptotic, and its radius of convergence is \(\abs{R_0m}  < 1/2\). From the relation between the conformal mass \(m\) and the mass \(\mu\) we see that the previous mentioned radius of convergence corresponds to \(\mu^2 =0\). In this special case, the mode \(\ell=0\) in \cref{eq:pot-small-q} becomes a zero mode and should be treated separately
\begin{align}
    \zeta\pqty{s | \setS^2, 0} & = 2 \Sum_{k=0}^{\infty} \Sum_{\ell =1}^{\infty} \binom{-s}{k} \pqty{\ell + \frac{1}{2}}^{-2s -2k+1} \pqty{-\frac{1}{4}}^k \nonumber \\
    & = \Sum_{k=0}^{\infty} \pqty{-1}^k \binom{-s}{k} 2^{1-2k} \zeta\pqty{2s+2k-1, \frac{3}{2}},
\end{align}
which was first derived in \cite{carletti1994minakshisundaram}. The expression converges rapidly, and it is possible to numerically evaluate it so that
\begin{equation}
    \frac{1}{2\sqrt{2}} \zeta\pqty*{-\frac{1}{2} | \setS^2, 0} = - 0.09372546 \dots
\end{equation}
which is precisely the contribution to the conformal dimension related to the Casimir energy of the Goldstone modes that we also found in \cref{eq:scalinggoldstone} in \cref{sec:loopCorrections}.

But most importantly, from the aforementioned radius of convergence, we deduce that the singularity that determines it, lies at the  value \(m^2 = -1/(4R_0^2)\) which is negative. This is why the expansions in the two distinguished limits \(m^2 \to 0^+\) and \(m^2 \to \infty\) can be smoothly interpolated for all values of \(m^2 >0\). It is therefore possible to express the Legendre relation between \(\mu\) and the charge \(\Qb\) order by order and the very first few are
\begin{align}
 R_0  f^{\saddle}(\Qb) &= \frac{\Qb}{2} +\frac{4 \Qb^2}{\pi ^2}+\frac{16 \left(\pi ^2-12\right) \Qb^3}{3 \pi ^4} + \frac{16 \left(384-48 \pi ^2+\pi ^4\right) \Qb^4}{3 \pi ^6} \dots \label{eq:smallchargeenergy}\\
 R_0 \mu &= \frac{1}{2} +\frac{8 \Qb}{\pi ^2} +\frac{16 \left(\pi ^2-12\right) \Qb^2}{\pi ^4} + \frac{64 \left(384-48 \pi ^2+\pi ^4\right) \Qb^3}{3 \pi ^6} + \dots \, ,
\end{align}
We note that the canonical free energy in the small-charge limit is a convergent expansion as well, and we can estimate its radius of convergence to \(\abs{\Qb} \lessapprox 0.28\dots\). Once more, the leading singularity exists at negative values of the charge, which allows the small-charge regime and the large-charge regime to be related without obstruction. 

\begin{figure}[ht]
  \centering
  \hfill
  \includegraphics[width=.45\textwidth]{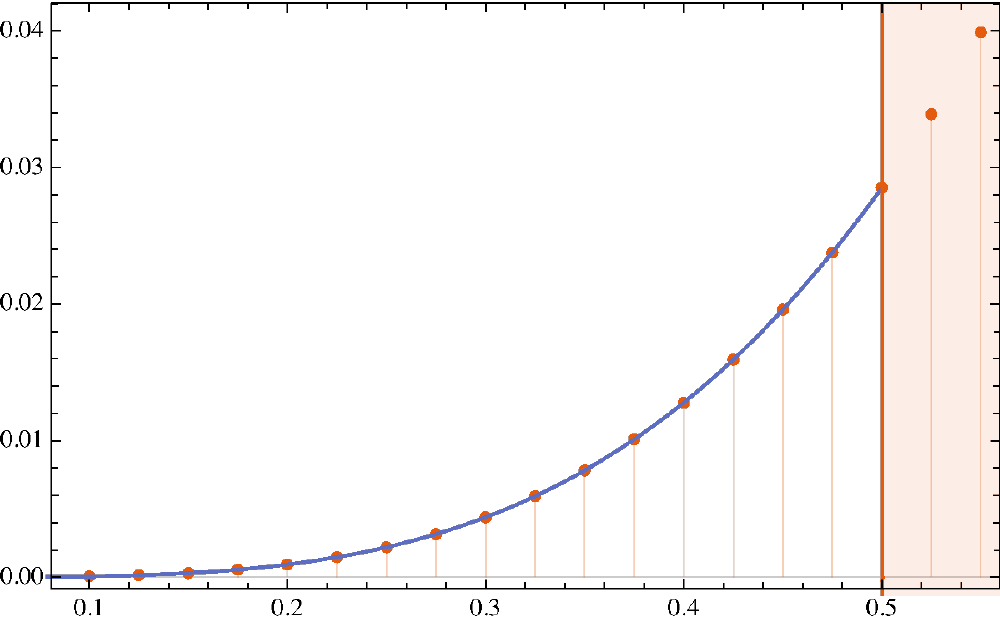}
  \hfill
  \includegraphics[width=.45\textwidth]{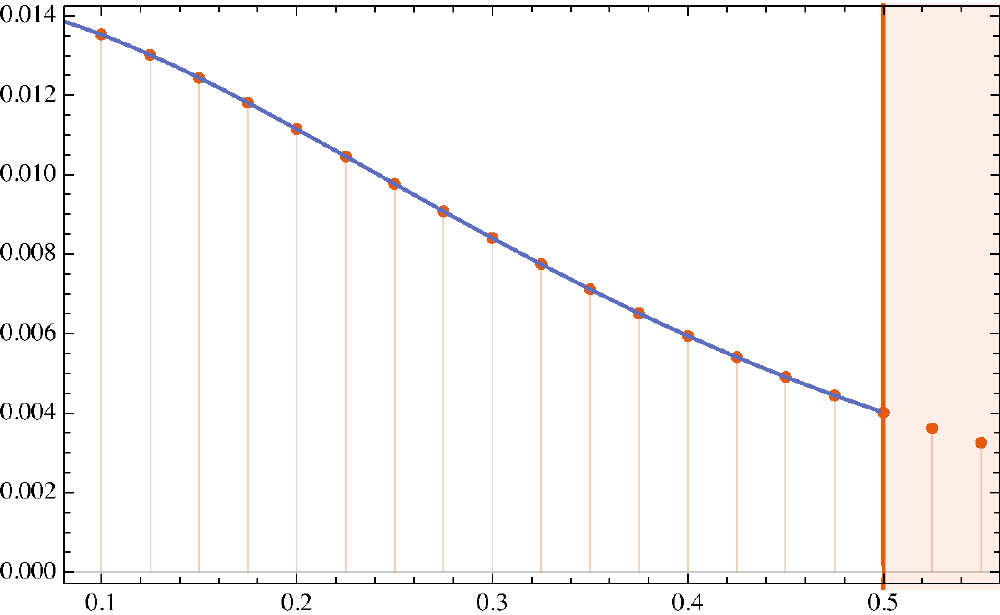}  
  \caption{We present the real and the imaginary component of the lateral Borel summation for the case of the grand potential depicted by dots 
 and compared to the small-charge limit and the exponential corrections coming from worldline instantons, which are the continuous line, as function of \(m R_0 \). The two methodologies agree completely at the level of resolution of our numerical calculation.}
  \label{fig:resurgence-vs-convergent}
\end{figure}

Now, we are in a position to compare the small-charge regime result of \cref{eq:smallchargegrandpotential} with the expression we derived for the grand potential \(\omega\) of
\cref{eq:resurged-grand-potential} in the large-charge asymptotic regime that we can express as
\begin{equation}
  \omega(m) = \PV \left[\frac{R_0 m^2}{\pi} {\Int_0^\infty \dd{\zeta} \frac{K_{2}(2 m R_0 \zeta)}{\zeta\sin(\zeta)}  }\right].
\end{equation}
For that matter, we tried to compare both the real and the imaginary parts of the lateral Borel summation of \(\omega\) with the values that we get from the small-charge expansion which is convergent and the exponential contributions of the worldline computation. The results are depicted in \cref{fig:resurgence-vs-convergent} and we observe that the aforementioned approaches are in perfect agreement with each other at the point of validity of our numerical simulation. For instance, if we pick the value \(m R_0 = 0.4\) which for the charge is \(\Qb \simeq 0.187\dots \) we compute that the small-charge expansion and the asymptotic resurgent expression agree up to eight digits 
\begin{align}
  \eval{R_0 \omega(mR_0=0.4)}_{\text{small charge}} &= 0.012\, 777\,296\,63 \dots \\
  \eval{R_0 \omega(mR_0=0.4)}_{\text{resurgence}} &= 0.012\,777\,297\,69 \dots
\end{align}
As a final note, the factor that limits the computation for the resurgence analysis is computer time, since for every additional digit the computer needs ten times more than what it took for the previous one. The relative error between the two computations if of order \(8 \times 10^{-8}\), that is smaller by six orders of magnitude than the exponential contribution \(e^{-2 \pi \times 0.4} \approx 8 \times 10^{-2}\). This outcome is a forceful indication that we have considered every non-perturbative correction in our analysis.

\section[\texorpdfstring%
{Lessons from large \(N\)}%
{<N>}]%
{Lessons from large \(N\)}%
\label{sec:lessons-from-large-N}

In this final section, our goal is to use the precise results that we derived while working in the double scaling limit~\cite{alvarez2019large} to extract some generic properties of the \acrshort{lce}.

When not in the double scaling limit, we have to use the \acrshort{eft} prescription that is explained in detail in \Cref{Chapter3} and especially in \cref{sec:O(2)_review}. For this section, the important result is that the bulk effective action can be written as an expansion in terms of the Goldstone field \(\chi\) as in \cref{eq:actionnlsm} 
\begin{equation*}
  S\bqty{\chi}= - c_1 \int\limits_{\setR \times \setS^{d-1}} \dd{\tau}\dd{S}  \pqty{ - \del_\mu \chi \del^\mu \chi }^{d/2} + \text{curvature couplings},
\end{equation*}
while at the same time physical observables are expressed as a series in inverse powers of the charge \(\Qp\). For example, the overall ground state action evaluated at the saddle point reads
\begin{align}
    &S^{\saddle} =  \pqty{\frac{\tau_\text{out}-\tau_\text{in}}{R_0}} \Sum_{n=0}^{\infty} \alpha_n (R_0\mu)^{d-2n},  &&\text{where}& \mu \sim \frac{\Qp^{1/(d-1)}}{R_0}.
\end{align}
Therefore, the energy of the ground state on the two sphere \(\setS^2\), which is related to the scaling dimension \(\Delta_\Qp\) of the lowest charged primary operator via the state-operator correspondence of \cref{eq.energyandscaling}, is
\begin{align}\label{eq:groundstateenergylargeNlargeQ}
    E & = -\frac{1}{\beta} \log\mathcal{Z} \simeq - \frac{1}{\beta} \log\pqty{e^{-S^{\saddle}}} \nonumber \\
   & =  \frac{1}{R_0} \Sum_{r=0}^{\infty} \alpha_n \, \Qp^{\frac{d-2n}{d-1}}
     = \frac{1}{R_0} \pqty{\alpha_0 \Qp^{3/2} + \alpha_1 \Qp^{1/2} + \dots},
\end{align}
where we used that \(\tau_\text{out}-\tau_\text{in} = \beta\) and we specialised to \(d=3\) spacetime dimensions. The coefficients \(\alpha_r\) are inaccessible within the \acrshort{eft} since they are related to the Wilsonian parameters that connect the \acrshort{ir} \acrshort{eft} with the initial \acrshort{uv} theory, and they have to be inserted in the \acrshort{eft} as an external input. They can be computed either in a double-scaling limit~\cite{orlando2019safe,badel2019epsilon,badel2020feynman,cuomo2021note,watanabe2021accessing,arias2019large,arias2020correlation,arias2020uv,antipin2020charging,antipin2021untangling,antipin2020charging2,jack2021anomalous,jack2022scaling}  or via lattice simulations \cite{banerjee2018conformal,banerjee2019conformal}.
In the double-scaling limit analysis that we performed, the grand potential corresponds to the effective action evaluated at the saddle point while the canonical free energy corresponds to the energy of the ground state
\begin{align}\label{eq:grandpotentialandfreeenergy}
	&\Omega(\mu) = - \frac{1}{\beta} \eval{S\bqty{\chi}}_{\chi = - i\mu \tau}, & F^{\saddle}_c(\Qp) = E.
\end{align}
Therefore, combining the results of \cref{eq:sphere-free-energy,eq:doublescaling,eq:canonicalfreeenergylegendre,eq:groundstateenergylargeNlargeQ} we can deduce that the coefficients \(\alpha_r\) of the \(O(N)\) vector model at large \(N\) and large charge in the \acrshort{wf} fixed point are
\begin{align}
  \alpha_0 &= \frac{2}{3 (2N)^{1/2}}, & \alpha_1 &= \frac{(2N)^{1/2}}{6} , & \alpha_2 &= -\frac{7 (2N)^{3/2}}{720} _,
\end{align}
and so forth.

Even though the above coefficients cannot be calculated within the validity of the \acrshort{eft}, we can still apply our previous non-perturbative analysis to make some statements about their general large-order behaviour.
In order to do so, we will make the subsequent assumptions that are valid for any \(N\):
\begin{enumerate}
\item the large-charge expansion is a priori asymptotic;
\item there is a first non-trivial saddle of a worldline path integral for a particle with a mass \(\mu\) that can be derived as the leading singularity in the Borel plane.
\end{enumerate}
The later assumption is because we are examining a \acrshort{cft} with no intrinsic scales, and therefore the sole dimensionful parameter in our system is related to the density of the fixed charge.%
\footnote{{The above proportionality factor is determined by the mass of the lightest \acrshort{dof} in the spectrum. In the case that \(N>1\) we can farther conjecture that this is a gapped Goldstone mode~\cite{cuomo2021gapped} and its mass \(\mu\)  is set by the symmetries of the system, and it is not possible to be renormalized by any loop effects~\cite{brauner2006spontaneous,nicolis2013implications}. Then we would have got \(\varpi = 1\) which matches with the result of the double-scaling limit.}}

This indicates that we should expect that the conformal dimension takes the general form of a double expansion in inverse powers of the charge \(1/\Qp\) and in \(e^{-2 \pi {\varpi} R_0 \mu}\)as
\begin{equation}
  \label{eq:dimensions-double-series}
  \Delta_{\Qp} = \Qp^{3/2} \Sum_{n} \, \alpha_n \, \frac{1}{\Qp^n}  + C_1 \, \Qp^{\kappa_1} \, e^{-3 \pi {\varpi} \alpha_0 \sqrt{\Qp}} \, \Sum_{n} \, \alpha^{(1)}_n \,  \frac{1}{\Qp^{n/2}}  + \dots,
\end{equation}
and in the above we denote by \(C_1\),  \(\kappa_1\) and \({\varpi}\) some constant parameters of the underlying theory, while we  utilised the fact that the chemical potential \(\mu\) and the charge \(\Qp\) are Legendre dual variables
\begin{equation}
  \label{eq:chemical-potential-charge-any-N}
  \mu = \frac{1}{R_0} \fdv{\Delta_\Qp}{\Qp} = \frac{3}{2 R_0} \alpha_0 \Qp^{1/2} + \dots
\end{equation}
In \cref{eq:dimensions-double-series} the subsequent series is an expansion in \(1/\Qp^{1/2}\), which is in total agreement with the exact expression that we have found in \cref{eq:non-perturbative-grand}.\footnote{This above trans-series structure, that has an expansion in $\Qp^{1/2}$ in the non-perturbative limit, is also observed in the case of Large-$N$ asymptotics of matrix models~\cite{marino2008nonperturbative,ahmed2017transmutation}.}

From \cref{eq:dimensions-double-series} we observe that parameters that appear in the non-perturbative part of the expansion are related to the large-order behaviour of the preceding perturbative series in \(1/\Qp\). Therefore, we can improve our conjecture about the 
structure of the exponential term to a general prediction about the large-order character of the \(\alpha_n\).

As a general rule, in the case that at large \(n\), the parameters \(\alpha_n\), are divergent 
\begin{equation}
  \alpha_n \sim \frac{(\beta n)!}{A^n} \, , 
\end{equation}
then we acquire an optimal truncation of the above perturbative series at a value that we denote \(N^*\) which corresponds to a minimum of \(\alpha_n \Qp^{-n}\), which for the case at hand reads
\begin{equation}
  N^* \approx \frac{1}{\beta} \abs{A \Qp}^{1/\beta},
\end{equation}
and the relevant error order is 
\begin{equation}\label{eq:error}
  \epsilon(\Qp) \sim e^{-(A \Qp)^{1/\beta}} .
\end{equation}

In our case, we flip this logic. Assuming that the structure of the leading non-perturbative terms remain exactly the same for any value of \(N\), we can use \cref{eq:dimensions-double-series,eq:error} to read the value of \(A\) and \(\beta\) as
\begin{align}
  \beta &= 2, & A &= 9 \pi^2 {\varpi}^2 \alpha_0^2,
\end{align}
which verifies that the asymptotic growth is $(2n)!$:
\begin{equation}
  \alpha_n \sim \frac{ (2n)!}{(3 \pi {\varpi} \alpha_0)^{n}  } ,
\end{equation}
and this is optimally truncated to the value
\begin{equation}
  N^* \approx \frac{3 \pi {\varpi} \alpha_0}{2} \Qp^{1/2} .
\end{equation}
The non-perturbative effects of the theory obtain their leading contribution through the aforementioned \((2n)!\) semiclassical divergence.
From instanton calculus --- see \cite{dondi2021resurgence} for details --- we know that the usual quantum corrections have a \(n!\) growth and their effect is of order \(e^{-\Qp^{3/2}}\). Also note that there is an interplay between the small-\(n\) and large-\(n\) parameters. In fact, the non-perturbative expansion is related to the large-\(n\) character of \(\alpha_n\) via the resurgence analysis and is connected to the small-\(n\) constants via the \acrlong{eom} in \cref{eq:chemical-potential-charge-any-N}.
This is precisely the reason we can express the optimal truncation in terms of the lowest parameter \(\alpha_0\).

The above analysis tries to illuminate the lattice results for the \(O(2)\) and the \(O(4)\) model~\cite{banerjee2018conformal,banerjee2019conformal}.
It was observed in these two papers that the \acrshort{lce} stays perfect for small values of the charge \(\Qp\) as well, and also that only a small number of terms are adequate to calculate the conformal dimensions of the lowest charged operators.
On general grounds, we expect \(\alpha_0\) to be of \(\order{1}\) and lattice results for the \(O(2)\) and \(O(4)\) model yield \(\alpha_0 \sim 0.337(3)\) and \(\alpha_0 \sim 0.301(3)\) respectively.

With the above generic assumptions about the large-order behaviour, our analysis predicts that the generic optimal truncation value is for \(N^* = \order*{\sqrt{\Qp}}\) which comes with an error that is of order \(\order*{e^{-\pi \sqrt{\Qp}}}\), and these predictions are in perfect agreement with the numerical outcome. In~\cite{banerjee2018conformal,banerjee2019conformal} it has been found that only the first two to three terms in the relevant expansion are enough to replicate the lattice results with excellent precision for charges up to \(\Qp = \order{10}\).
For a charge \(\Qp = 1\), the relevant error is of order \(\order{10^{-2}}\) and for a charge \(\Qp=11\), the error becomes of order \(\order{10^{-5}}\), which have to be compared to \(e^{-\pi } \approx 4 \times 10^{-2}\) and \(e^{-\pi \sqrt{11}} \approx 3 \times 10^{-5}\).

\bigskip

As a final remark for this chapter, we have applied the resurgence methodology for the purpose of analysing and expanding previously derived results of the \acrshort{lce} for the case of the \(O(N)\) model in \(d=3\) spacetime dimensions at the \acrshort{wf} conformal fixed point~\cite{alvarez2019large}, utilising results that were obtained in the double-scaling limit of the theory, defined as \(\Qp\to \infty, \, \,  N\to \infty\), with their ratio \(\Qp/(2N) = \Qb \) being kept constant. We have investigated two distinct cases for the system defined on the manifold \(\setR \times \Sigma\) : either \(\Sigma = \setT^2\) is the two-torus, or \(\Sigma=\setS^2\) is the two-sphere. In the latter case, we can apply the state-operator correspondence (\S \,\ref{sec.stateoperator}) to compute the scaling dimension \(\Delta_\Qb\) of the lowest charged primary operator \(\Opp\) from the ground-state energy, which corresponds to the canonical free energy in the double-scaling limit. We calculated the usual perturbative series as well as the exponentially suppressed non-perturbative contributions for both cases, while for the sphere \(\setS^2\), resurgence analysis alone fails to provide an unambiguous result for the non-perturbative corrections. This ambiguity can be resolved in two ways, either by utilising the resurgence methodology for the Dawson's function, or using a geometric interpretation in terms of the worldline instantons. 

The second procedure does not a priori depend on large \(N\) given the fact that the final result is a finite-volume effect that is connected to the geometric structure of the compactification manifold. Therefore, we obtain a nice geometric understanding of both the non-perturbative contributions and of the Borel ambiguities and also the picture is robust enough to let us go beyond the double-scaling limit and suggest a precise form for the grand potential \(\omega\) that holds true for all values of the charge \(\Qb\). We were able to verify our proposition numerically with excellent precision, and we were in a position to conjecture that the \acrshort{lce} is always asymptotic, even in finite \(N\), with an optimal truncation \(N^* = \order*{\sqrt{\Qp}}\) and an error of order \(\epsilon(\Qp) = \order*{e^{- \sqrt{\Qp}}}\) which is in agreement with the lattice simulations~\cite{banerjee2018conformal,banerjee2019conformal}. The non-perturbative corrections that we compute originate from the fact that the \acrshort{eft} is an asymptotic expansion on its own.

%

\chapter[\texorpdfstring%
{Fermionic Models at large \(N\)}%
{<N>}]%
{Fermionic Models at large \(N\)}%

\label{Chapter5} 

\epigraph{\itshape “The art of doing mathematics consists in finding that special case which contains all the germs of generality.”}{David Hilbert}

Although most of the work in the \acrshort{lce} is centred around bosonic theories, a few attempts confronting the topic of relativistic fermionic theories at large charge have been made~\cite{delacretaz2022thermalization,komargodski2021spontaneously,antipin2022yukawa} \footnote{Meanwhile, the non-relativistc \acrshort{cft} describing the case of the unitary Fermi gas has been examined at large charge in~\cite{favrod2018large,kravec2019nonrelativistic,kravec2019spinful,orlando2021near,hellerman2020droplet,pellizzani2022operator,hellerman2022nonrelativistic}}. As an interesting turn of events, it was discovered in \cite{komargodski2021spontaneously} that the free fermion at large-charge does not fall within the conformal superfluid class, but the conformal dimension of the lowest operator of large-charge scales as \(\Qp^{3/2}\) and the ground state is identified with a Fermi surface. 

In this chapter, following closely Dondi et al.~\cite{dondi2022fermionic}, we try to bridge this gap, and we consistently analyse various fermionic models in \(d=3\) spacetime dimensions in Euclidean signature at large charge \(\Qp\) and large \(N\)~\cite{alvarez2019large}. These models all exhibit a four-fermion interaction term, and we will utilise the standard large-\(N\) methodology that we also employed in~\cref{sec:O2N} to perform a Hubbard–Stratonovich transformation via which we rewrite the four-fermion interaction term as a Yukawa-type term, by introducing a complex scalar Stratonovich field. If the field remains auxiliary, the conformal phase of the theory lies in the \acrshort{uv}, and is accessible only through the large-\(N\) expansion. On the other hand, by introducing kinetic terms for the Stratonovich field in \(d=3\) dimensions, we get the \acrshort{uv} completion of the model, and the conformal phase of the theory lies in the \acrshort{ir}~\cite{Moshe_2003}. No matter if we make the Stratonovich field dynamical or not, the large-charge primary operator \(\Opp\) is contained in the spectrum of the \acrshort{cft} and depending on the nature of the initial interaction, we have found two distinct qualitative behaviours :
\begin{enumerate}
    \item First there is the \acrlong{gn}-type of behaviour and the free fermion falls into this category. In the \acrshort{gn} case at large charge, we observed that there is no \acrshort{ssb} for the \(U(1)_B\) Baryon symmetry and strictly in the \(N \to \infty\) limit the large-\(N\) physics is described in terms of an approximate Fermi sphere. It is not yet clear if the Fermi surface remains when subleading corrections in \(N\) are considered, but we note that this large-charge sector has no \(\Qp^0\) universal contribution corresponding to the Casimir energy of fluctuations, since there are no Goldstone modes.
    \item The second is the \acrlong{njl}-type of behaviour. The \acrshort{njl} model and its generalisations exhibit simultaneously a \(U(1)_B\) baryon symmetry and a \(U(1)_A\) axial symmetry, and in specific large-charge sectors the \(U(1)_A\) can be spontaneously broken. In that case, the large-charge ground state coincides with the conformal superfluid paradigm, but the scaling dimension of the lowest charged primary operator of the theory has different numerical coefficients than the bosonic case~\cite{alvarez2019large}, indicating that the fermionic \acrshort{cft} lies in a different universality class. Meanwhile, we specifically verified that the spectrum of fluctuations over the large-charge ground state contains the anticipated conformal Goldstone mode and therefore, the conformal dimension exhibits the universal \(\Qp^0\) term corresponding to the Casimir energy of the fluctuations that we found and computed in \Cref{Chapter3,Chapter4}. Finally, working in large-\(N\), we can access the small-charge limit of the theory where the conformal dimension of the lowest charge scalar operator is in accordance with the usual perturbative result for the free bosonic scalar operator of mass dimension one and charge two whose energy we computed in \cref{eq:smallchargeenergy}.
\end{enumerate}
We note that both the \acrlong{gn} model and \acrlong{njl} model are anticipated to exhibit interacting fixed points for any spacetime dimension \(2 < d < 4\). Even though we will not use the small-\(\varepsilon\) expansion in the current chapter, nor we will examine the characteristics of the models in any spacetime dimension apart from \(d=3\), many features of the Lagrangians of the \acrshort{gn} and \acrshort{njl} models can be realised if we dimensionally reduce them from \(d=4\). 

As a final note, for every fermionic model that supports a large-charge superfluid ground state, there is a physically intuitive way to comprehend the existence of a bosonic condensate. For instance, for the \acrshort{njl} model we can carry out a \acrfull{pg} transformation~\cite{pauli1957conservation,gursey1958relation}, defined as
\begin{align}
        \Psi &\mapsto \frac{ 1}{2} \left[ (1-\Gamma_5) \Psi + (1+ \Gamma_5) C_4 \bar\Psi^T  \right] ,&
        \bar\Psi &\mapsto \frac{ 1}{2} \left[ \bar\Psi (1+\Gamma_5) - \Psi^T C_4 (1- \Gamma_5) \right] ,
\end{align}
to derive a model that exhibits a Cooper-type interaction~\cite{kleinert1998two,ebert2016competition}.
Every computation can be executed in the context of the Cooper model, with the same results as before. In the Cooper-pair context, it is evident that the nature of the condensate is that of Cooper pairs that describe a superconductor. The attractive interaction of the system results in a Cooper instability, and we now have a system described by condensing bosons at large charge, and that is why the results are so similar to the \( O(N) \) vector model.

The plan of the chapter is as follows: in \cref{sec:models} we present all the fermionic models that we want to examine and their \acrshort{uv} completions. In \cref{sec:symmetry} we work on \(\setS^1_{\beta} \times \setT^2\) which taken at \(\beta \to \infty\) and simultaneously, at large volume coincides with the flat space, and we examine the existence or not of a bosonic condensate in the large-charge sector for the different models. Meanwhile, in \cref{sec:fluctuations} we explicitly compute the spectrum of the fluctuations of the \acrshort{gn} and the \acrshort{njl} model by examining the one-loop propagator. Finally, in \cref{sec:conformalDim}, working on \(\setS^1_{\beta} \times \setS^2\), we can use the state-operator correspondence, and we compute the scaling dimension of the lowest charged scalar operator \(\Opp\) for the different models, in the large-charge and the small-charge limit.

\section{The Models}%
\label{sec:models}

We study fermionic models in \(d=3\) spacetime dimensions with Euclidean metric that exhibit a second-order phase transition. In our analysis, we will use the large-\(N\) approximation to take advantage of the simplifications that appear in this limit, and we will concentrate our attention on a few specific cases with a small symmetry group on top of the \(SO(2N)\) symmetry.

All the models that we examine are derived by deforming the \acrshort{cft} of the free fermion by adding a four-fermion interaction term with an irrelevant coupling \(g\). Assuming that the models feature a fundamental scale \(\Lambda\) at the \acrshort{uv} limit, then in the case that the temperature and the density of the system are zero the coupling \(g\) acquires a critical value \(g_c^{-1} \sim \Lambda\) and there is a \acrshort{cft} representing the critical point between the two phases in which certain symmetries exhibit \acrshort{ssb}. We aim to examine a critical limit like that, but for finite charge density. 
Based on the analysis of \Cref{sec.renormalisable}, the presence of the irrelevant coupling \(g\) renders these kinds of models non-renormalisable in the usual power-counting sense, therefore taking the limit \(\Lambda \rightarrow \infty\) is not well-defined. Nevertheless, it was shown~\cite{gross1974dynamical,parisi1975theory} that they can be renormalised using the large-\(N\) expansion~\cite{Moshe_2003} and the \acrshort{rg} flow connects the conformal phase at the critical point \(g_c\) in the \acrshort{uv} limit with the free fermion \acrshort{cft} found in the \acrshort{ir} limit. There is strong evidence that the aforementioned conformal phases hold out in the finite-\(N\) limit; however, an appropriate \acrshort{rg} analysis demands some sort of \acrshort{uv} completion, which is commonly achieved by introducing additional scalar \acrshort{dof} that interact with the fermions through some Yukawa interaction~\cite{zinn1991four}, with the corresponding Yukawa coupling being relevant in \(d < 4\). Therefore, these completed models are free in the \acrshort{uv} limit while they are strongly interacting in the \acrshort{ir} limit, with the \acrshort{cft} living there being the same as the four-fermion models. The corresponding \acrshort{cft} gets weakly coupled in \(d=4-\varepsilon\), permitting perturbative calculations of the conformal data, and such a computation at large charge can be found in~\cite{antipin2022yukawa}. In the strict large-\(N\) limit, it is enough to look at the minimal models containing just fermionic matter, and therefore we shall only shortly mention the complete version in the \acrshort{uv}.
 
The explicit models that we will study are the \acrfull{gn} and the chiral \acrlong{gn} or \acrfull{njl} model along with its \( SU(2) \times SU(2) \) generalisation. Normally, they are studied either for finite \(N\) in \(d = 4 - \varepsilon\) and \(d = 2 + \varepsilon\)  spacetime dimensions, or for \(2< d < 4\) using the \(1/N\) expansion. Obviously, in \(d=3\) there is no natural notion of chirality, and the usual solution to that is to dimensionally reduce the \(4d\)-model utilising four-component fermions. In \(d=3\) spacetime dimensions, a four-component fermion exists in some reducible representation, so in reality we double the number of flavours in the flavour group, of which the “chiral” symmetry is a part. A system that contains \(N\) free and massless Dirac fermions in \(d = 3+1\) exhibits a \(O(2N)\) global symmetry, since every Dirac particle is decomposable to two Majorana fermions. By reducing to \(d=3\), every \(4d\)-Majorana fermion is decomposed into two irreducible Majorana particles, and the emerging symmetry group is \(O(4N)\). This is precisely the symmetry of the kinetic part in the action of the models that we examine, which is, in general, broken by the four-fermion interaction terms. Nevertheless, by using the previous mentioned four-dimensional reducible representation in \(d=3\) we can introduce a notion of chirality and properly define a \(\Gamma_5\) matrix as in \cref{sec:RedRep}. In this spirit, throughout the chapter we mention axial and chiral symmetries, despite the fact that in \(d=3\) they are dimensionally reduced to standard global symmetries.

The equivalence between the results obtained at fixed \(N\) in the \(4 - \varepsilon\) and \(2 + \varepsilon\) expansions and the large-\(N\) expansion for generic spacetime dimensions has been shown, and this also admits the case of \(d = 3\)~\cite{fei2016yukawa}. Finally, the conformal phases found in large-\(N\) are strongly believed~\cite{chandrasekharan2013quantum} to exist in \(d=3\) and finite \(N\) as well.

\subsection{Gross-Neveu model}

For three-dimensional fermionic theories with an even number \(2N\) of fermion fields \(\psi_{i=1...2N}\) it is convenient to introduce a reducible representation of the Clifford algebra in the following way:
 \begin{align}
 	\Gamma_\mu &= \sigma_3 \otimes \gamma_\mu = \begin{pmatrix}
 	\gamma_\mu & 0 \\ 0 & - \gamma_\mu 
 	\end{pmatrix}, &   \Psi_i &\equiv  \begin{pmatrix}
 	\psi_i \\ \psi_{i+N}
 \end{pmatrix}, & \bar \Psi_i &= \Psi_i^\dagger \Gamma_3, &
 i &= 1, ... , N.
 \end{align}
Therefore, the Lagrangian of the \acrlong{gn} model~\cite{gross1974dynamical} written in terms of the above reducible representation reads
\begin{equation}\label{eq:L-GN}
  \Lp = \Sum_{i=1}^{N} \, \bar \Psi_i  \Gamma^\mu \del_\mu \Psi_i - \frac{ g}{N}  \left( \Sum_{i =1}^N \, \bar \Psi_i \Psi_i \right)^2,
\end{equation}
and the global symmetry of the model is explicitly \(O(2N)\times O(2N)\). Furthermore, we note that there is an Abelian \(U(1)_B\) diagonal subgroup that is related to the transformation
\begin{equation}
   U(1)_B : \Psi_i \to e^{i\alpha} \Psi_i.
\end{equation}
Following the large-\(N\) Hubbard--Stratonovich methodology of \cref{sec:O2N}, we introduce an auxiliary field and a Lagrange multiplier \(\sigma\), and after integrating out the auxiliary field we are left with an action where the multiplier has been promoted to a Stratonovich field which replaces the fermionic bilinear \(\Bar{\Psi} \Psi\). The corresponding Lagrangian reads  
\begin{equation}\label{eq:GNmodelauxiliary}
  \Lp =  \Sum_{i=1}^{N} \, \bar \Psi_i \left( \Gamma^\mu \del_\mu + \sigma \right)  \Psi_i  + \frac{N}{4g} \sigma^2 .
\end{equation}
In the limit that \(g \to \infty\) we can reach the critical point of the theory, where we neglect the \(\sigma^2\) term. This is a second-order phase transition that separates broken and unbroken phases of \( \mathbb{Z}_2 \) chiral symmetry which is realised as 
\begin{equation}
\mathbb{Z}_2 : \Psi \rightarrow -\Gamma_\mu \Psi .
\end{equation}
The proper \acrshort{uv} completion of the \acrshort{gn} model at finite \(N\) is the \acrfull{gny} model~\cite{Moshe_2003} derived by promoting the auxiliary Stratonovich field \(\sigma\) to a dynamical field resulting in the Lagrangian
\begin{equation}
    \Lp =  \Sum_{i=1}^{N} \, \bar \Psi_i \left( \Gamma^\mu \del_\mu + \sigma \right)  \Psi_i  + \frac{1}{2 g_Y} \partial_\mu \sigma \partial_\mu \sigma.
\end{equation}
It is easy to see that the coupling \(g_Y\) is relevant and in the \acrshort{ir} limit it grows large, so that the critical action of the \acrshort{gn} model and of the \acrshort{gny} model at the critical point are the same.

\subsection{Nambu–Jona–Lasinio -- type models}

The chiral \acrlong{gn} or \acrlong{njl} model~\cite{nambu1961dynamical,nambu1961dynamical2} is a classic four-fermion interaction model that exhibits a continuous chiral symmetry in \(d=4\) spacetime dimensions with Lagrangian
\begin{equation}\label{eq:L-NJL}
  \Lp =  \Sum_{i=1}^{N} \, \bar \Psi_i  \Gamma^\mu \del_\mu \Psi_i - \frac{ g}{N}  \bqty{  \pqty{ \Sum_{i=1}^{N} \, \bar \Psi_i \Psi_i }^2 - \pqty{\Sum_{i=1}^{N} \, \bar \Psi_i \Gamma_5 \Psi_i }^2  } \, ,
\end{equation}
and we note that the $\bar{\Psi} \Gamma_5 \Psi$ fermionic bilinear has a  $Sp(2N)$-invariance, and thus the symmetry group of the system is decreased regarding the \acrlong{gn} model of \cref{eq:L-GN} to
\begin{equation*}
    [O(2N)\times O(2N)] \cap Sp(2N) = U(N).
\end{equation*}
In addition to the \(U(1)_B\) baryon symmetry of the \acrshort{gn} model, an additional \(U(1)_A\) axial symmetry manifests from the combination of the different quartic interactions as in \cref{eq:L-NJL}:
\begin{equation}
  \Psi_i \to e^{i \alpha \Gamma_5} \Psi_i .
\end{equation}
Therefore, the total internal symmetry of the model \footnote{We neglect the Poincaré or the conformal symmetries of the model, and we only examine the internal symmetry. Furthermore, we can always use that \(U(N) \simeq SU(N) \times U(1)\). A brief sketch goes as follows: for \(A \in U(N)\) : \(A = \det A \times \frac{A}{\det A}\). For  \(\det A = e^{i \chi} \in U(1), \ \chi \in \setR\), then \(\mathbb{M} =(A/ \det A) \) has \(\det \mathbb{M} =1\), hence  \(\mathbb{M} \in SU(N)\).} reads
\begin{equation}
    U(N) \times U(1)_A \simeq SU(N) \times U(1)_B \times U(1)_A. 
\end{equation}
At the critical point, the conformal version of the \acrshort{njl} model sits between two phases in which the \(U(1)_A\) axial is either spontaneously broken or not.
In the first phase, a Goldstone mode appears, in contrast to the \acrshort{gn} case where the chiral symmetry is discrete. 

The generalisation of the \acrshort{njl} model was studied~\cite{nambu1961dynamical,nambu1961dynamical2} by examining a system with two-flavour fermions \(\Psi_{i,f}, \, f = 1, 2\) and the corresponding Lagrangian is
\begin{equation}
  \Lp =  \Sum_{i = 1}^{N} \Sum_{f = 1}^2  \, \bar\Psi_{i,f}  \Gamma^\mu \del_\mu \Psi_{i,f} - \frac{ g}{N}  \left[  \left( \Sum_{i = 1}^{N} \Sum_{f = 1}^2 \, \bar \Psi_{i,f} \Psi_{i,f} \right)^2 - \Sum_{a=1}^3 \left( \Sum_{i = 1}^{N} \Sum_{f = 1}^2 \, \bar \Psi_{i,f} \Gamma_5 \sigma^a_{fg} \Psi_{i,g} \right)^2  \right] ,
\end{equation}
and \(\sigma^a\) denote the Pauli matrices. The symmetry group of the previous \acrshort{njl} model got enhanced into \(U(1)_B \times SU(2)_L \times SU(2)_R \), while the total internal symmetry of the system is \(U(N) \times SU(2)_L \times SU(2)_R\) and we note that the \(U(1)_A \) axial symmetry is no longer present and is replaced by the \(SU(2)_L \times SU(2)_R\) symmetry group which acts on the fermions as
\begin{align}
  \Psi_{i,f} &\to e^{i \frac{1 + \Gamma_{5}}{2}  \omega_a^L \sigma^a_{fg}} \Psi_{i,g}, & \text{and} &&   \Psi_{i,f} &\to e^{i \frac{1 - \Gamma_{5}}{2} \omega_a^R \sigma^a_{f g}} \Psi_{i,g}  .
\end{align}
The above symmetry exists due to the pseudo-real nature of \( SU(2) \), for which there is no completely symmetric symbol \(d_{abc}\).

In our analysis, we will investigate the \(U(1)_B \times U(1)_A \) and \(U(1)_B \times SU(2)_L \times SU(2)_R \) \acrshort{njl} models --- which in a shorthand notation we refer to as  \(U(1)\)-\acrshort{njl} and \(SU(2)\)-\acrshort{njl} --- using the large-\(N\) expansion, utilising the unbroken \( SU(N) \) symmetry that exists in both cases.

In the case of finite \(N\), these models are well-defined through their \acrshort{uv} completions \footnote{
For simplicity, we suppress all global symmetry indices from this point on.} and accessible using perturbation theory in \(d=2+\varepsilon\) and \(d=4-\varepsilon\). The \acrshort{uv} completed version of the \(U(1)\)-\acrshort{njl} reads
\begin{equation}\label{eq:NJLUV}
  \Lp =  \bar \Psi \bqty{ \Gamma^\mu \del_\mu + \Phi \pqty{\frac{1 + \Gamma_5}{2} }  + \Phi^{*} \pqty{ \frac{1 - \Gamma_5}{2} } } \Psi  + \frac{1}{g_Y} \del_{\mu} \Phi^{*} \del_{\mu} \Phi,
\end{equation}
where the \( U(1)_A \) symmetry manifests as
\begin{align}
  \Psi &\to e^{i \alpha \Gamma_5} \Psi,  & \Phi &\to e^{-2 i \alpha} \Phi .
\end{align}
Similarly to the \acrshort{gn} case, the completed version of the \(U(1)\)-\acrshort{njl} model has an  \acrshort{ir} fixed point in the limit \( g_Y \rightarrow \infty \), which formally generates the critical action that we also employ in the strict large-\(N\) limit. Then, the dynamical field \(\Phi\) of \cref{eq:NJLUV} is identified as the initially auxiliary Stratonovich field that replaced the complex fermionic bilinear  \( \bar{\Psi} \Psi + \bar{\Psi} \Gamma_5 \Psi \).

For the case of the \( SU(2) \)-\acrshort{njl} model, there exists a similar completion if we introduce a group of real fields \(\sigma, \pi_{a = 1,2,3} \), in terms of which the \acrshort{uv} Lagrangian reads
\begin{equation}
  \Lp =   \bar \Psi \pqty{ \Gamma^\mu \del_\mu  + \sigma + i  \pi_a \sigma^a  \Gamma_5 
 } \Psi + \frac{1}{2 g_Y} \pqty{\del_{\mu} \sigma \del_{\mu} \sigma + \del_{\mu} \pi^a \del_{\mu} \pi^a},   
\end{equation}
and in this instance, the \(SU(2)_L \times SU(2)_R \) symmetry is infinitesimally realised as
\begin{align}
  \delta_{L,R} \Psi &= i \left(\frac{1 \pm \Gamma_{5}}{2} \right) \omega_{a} \sigma^a \Psi, &
    \begin{cases}
        \delta_{L,R} \sigma = \pm \omega_a \pi_a ,\\
        \delta_{L,R} \pi_a = \mp \omega_{a} \sigma +  \epsilon_{abc} \pi_b \omega_c.
    \end{cases}
\end{align}
In a similar manner, the symmetry manifests in terms of the field \(\Phi = \sigma + \pi_a \sigma^a\),   as
\begin{align}
 \begin{cases}
   \Psi \to e^{i \frac{1 + \Gamma_5}{2} \omega_a \sigma^a} \Psi ,  \\
   \Phi \to  \Phi e^{ - i \omega_a \sigma^a}  ,
 \end{cases}
 &&
    \begin{cases}
      \Psi \to e^{i \frac{1 - \Gamma_5}{2} \omega'_a \sigma^a} \Psi , \\
      \Phi \to  e^{  i \omega'_a \sigma^a}  \Phi.
    \end{cases}
\end{align}
This model is known as iso\acrshort{njl}.

\subsection{Cooper model}

We have observed that all the models that we have examined thus far contain a fermion-antifermion interaction term. But, in condensed matter physics, to study superconductivity which is occurring through Cooper pairs we need to look at difermion interaction terms. Therefore, fermionic models at large-\(N\) that are aimed at studying superconductivity at finite \(U(1)_B\)-charge density, additionally to \acrshort{gn} or \acrshort{njl}-types of interactions~\cite{ebert2005competition,ebert2016competition}, should also contain the following interaction term
\begin{equation}
    (4f)_{Cp} = \frac{g}{N} \bar \Psi C \bar \Psi^T \, \Psi^T C \Psi .
\end{equation}
For our purposes, we care about the model that contains only the Cooper pair term
\begin{equation}
  \Lp =  \bar \Psi \Gamma^\mu \del_\mu \Psi + \frac{g}{N} \bar \Psi C_4 \bar \Psi^T \, \Psi^T C_4 \Psi,
 \end{equation}
where \(C_4\) is the charge-conjugation matrix defined in \cref{eq:chargeconjugationmatrix} in~\Cref{sec:RedRep}.

At the critical point --- antithetically to what occurs in the \acrshort{gn} model --- the Cooper model at zero temperature and finite density admits a non-trivial solution to the gap equation,  bringing about a superconducting phase. 

It can be shown --- see \cref{sec:Pauli-Gursey} for details --- that the Cooper model is dual to the \acrshort{njl} model~\cite{ebert2016competition} and this becomes apparent by employing the \acrlong{pg} transformation~\cite{pauli1957conservation,gursey1958relation,kleinert1998two},
\begin{align}
        &\Psi \mapsto \frac{ 1}{2} \left[ (1-\Gamma_5) \Psi + (1+ \Gamma_5) C_4 \bar\Psi^T  \right] , &
        &\bar\Psi \mapsto \frac{ 1}{2} \left[ \bar\Psi (1+\Gamma_5) - \Psi^T C_4 (1- \Gamma_5) \right] .
\end{align}
Therefore, before we move to the next sections, some comments on the Cooper model:
\begin{itemize}[left= 0pt]
    \item The \acrlong{pg} transformation is a linear involution and hence it only impacts the path integral measure up to a trivial rescaling.
    \item The Cooper and the \acrshort{njl} model, both enjoy a \(U(1)_A \times U(1)_B\) symmetry. 
    Their duality maps the \(U(1)_B\)-chemical potential of the Cooper model to the $U(1)_A$-chemical potential of the \acrshort{njl} model, and the other way around. Therefore, any results found using the \acrshort{njl} model at finite \(U(1)_A\) hold true for the Cooper model at finite \(U(1)_B\).
    \item At the critical point, any quantities found in the Cooper model have to agree with the same quantities computed in the \acrshort{njl} model up to a  \acrshort{pg} transformation, which we specifically checked in the large-\(N\) limit in leading-\(N\) for the case of the ground state energy at both models.
\end{itemize}
%


\section[\texorpdfstring%
{Symmetry breaking at large \(N\)}%
{<N>}]%
{Symmetry breaking at large \(N\)}%
\label{sec:symmetry}

As it will be obvious in the next sections, whether the \acrshort{lce} results in simplifications in the computation of the \acrshort{cft} data in the large-charge sector depends strongly on the occurrence or not of a condensate. On that account, we shall first examine the symmetry breaking of the \acrshort{gn} and the \acrshort{njl}-type of models.

In any case, irrespective of the presence of \acrlong{ssb} in the underlying theory, it is possible to compute the conformal dimension of specific primary operators from finite-density ground states on the cylinder, using the state-operator correspondence of \cref{sec.stateoperator}. In short, a charged primary operator \(\Opp\) with a conformal dimension \(\Delta_{\Qp}\) is related to a state \(\ket{\Qp}\) with charge \(\Qp\) on the cylinder of radius \(R_0\) as in \cref{eq:operatorsandstates}. The energy of the charged state is computed using \cref{eq.energyandscaling} as
\begin{equation}
	E(\Qp) = \Delta_{\Qp} / R_0 .
\end{equation}
As we discussed in \cref{sec:lessons-from-large-N}, the energy of the charged ground state corresponds to the canonical free energy at the saddle point as \( F^{\saddle}_c(\Qp) = E(\Qp)\). Therefore, we start by considering the thermal \acrshort{cft} on $\setS^1_\beta \times \Sigma$ and we study the canonical partition function
\begin{equation}
    \mathcal{Z}_c\pqty{\Qp,\mu} = e^{-\mu \beta \Qp} \Tr\bqty{ e^{- \beta H - \mu \hat{\Qp}} } \, ,
\end{equation}
where we notice the grand-canonical partition function of \cref{eq:grandcanonical} which reads
\begin{equation}
\mathcal{Z}_{gc}(\beta, \mu )= \Tr\bqty{ e^{- \beta H - \mu \hat{\Qp}} } \, ,
\end{equation} 
where \(\mu\) is the chemical potential. We can follow the analysis of \cref{sec:largeN} and the bottom line is that for the models that we will examine, in the large-\(N\) limit, the grand-canonical partition function \(\mathcal{Z}_{gc}(\beta, \mu )\) acquires a path integral representation that can be calculated precisely order-by-order and reads
\begin{equation}
    \mathcal{Z}_{gc}(\beta, \mu ) \xrightarrow[N \rightarrow \infty]{\beta \rightarrow \infty} e^{ \beta \Omega(\mu)},
\end{equation}
where $ \Omega$ is the thermodynamical grand potential, which reads
\begin{equation}\label{eq:grandpotentialgeneral}
    \Omega(\mu) = -\frac{1}{\beta} S^{\saddle}_{\text{eff}, \,  (0)} \, ,
\end{equation}
in accordance with \cref{eq:grandpotentialandfreeenergy}. For fermionic theories in the critical point \(g \to \infty\) this coincides with
\begin{equation}
   \Omega(\mu)  =  \frac{N}{\beta} \Tr\bqty{\log\pqty{G^{\mu}}^{-1}} \, ,
\end{equation}
where \(G^{\mu}\) is the fermionic Green's function at fixed chemical potential. This is similar to the expression of the grand potential of \cref{eq:definitiongrandpotential} but since we are dealing with Grassmann variables the functional determinant has the opposite sign than in the bosonic case \footnote{An additional difference is that in the above definition, the grand potential is not defined per degree of freedom.}.

Using this expression, we can compute the charge \(\Qp\) and the canonical free energy at the saddle point as
\begin{align}
{\Qp} &=  \pdv{\Omega}{\mu}, & {F_c^{\saddle}(\Qp)} &= -\frac{1}{\beta} \log \mathcal{Z}_c = \sup_{\mu} \pqty{\mu \Qp - \Omega(\mu)}.
\label{eq:this_is_basic}
\end{align}
In this section we work on the manifold \(\setS^1_\beta \times \setT^2\), therefore we will perform our analysis on the torus \(\setT^2\) which produces the leading-order result of  \(\setS^1_\beta \times \setS^2\) in the macroscopic limit \(R_0 \rightarrow \infty\). We are going to utilise the same expressions to represent energy and charge density, and we shall normalise our results by the volume \(V\) of the torus. In the following section, we will not differentiate between flat space and torus.

\subsection{Gross--Neveu model}\label{sec:GN-det}
 
We start by studying the \acrlong{gn} model for \(2N\) Dirac fermions in \(d=3\) spacetime dimensions in Euclidean signature at a finite temperature \(1/\beta\) and a finite \(U(1)_B\)-chemical potential \(\mu\). Utilising the reducible representation presented in~\Cref{sec:RedRep} and expressing the Lagrangian in terms of the Stratonovich field \(\sigma\) of \cref{eq:GNmodelauxiliary}, the corresponding action reads
\begin{equation}
S \, \bqty{ \sigma, \Psi} = \Int\limits_{\setS^1_\beta \times \setT^2} \dd{\tau} \dd[2]{x}\left[  \bar{\Psi} \pqty{ \Gamma^\mu \partial_\mu - \mu \Gamma_3 + \sigma} \Psi + \frac{N}{4 g} \sigma^2   \right].
\end{equation}
The above model features a fundamental cut-off scale \(\Lambda\) and for general values of the coefficients \((g,\beta,\mu)\) the Stratonovich field gets a \acrshort{vev} and the discrete parity symmetry \(\Psi \rightarrow - \Gamma_5 \Psi\) is spontaneously broken. Given some general postulates, the aforementioned \acrshort{vev} is homogeneous \(\expval{\sigma} = \sigma_0\) and the Stratonovich field can be expanded in the classical \acrshort{vev} plus quantum fluctuations around it as \(\sigma = \sigma_0 + \hat\sigma/\sqrt{N}\). Then we can perform the integral over the Grassmann variables and write an effective action for the field \(\sigma\) around the above saddle point as 
\begin{equation}\label{eq:Seff_GN}
S^{\saddle}_{\text{eff}} = N \left\{  \beta V \frac{\sigma_0^2}{4g} - \Tr \bqty{\log \big(G^{(\mu)}\big)^{-1} }   \right\} + \frac{1}{2} \left[ \Tr \bqty{G^{(\mu)} \hat\sigma G^{(\mu)} \hat\sigma} + \frac{1}{4g}  \Int_{\setS^1_\beta \times \mathbb{R}^2} \dd{\tau} \dd[2]{x} \hat{\sigma}^2 \right] + \mathcal{O}(N^{-1}),
\end{equation}
and with \(G^{(\mu)}\) we denote the fermionic Green's function at finite chemical potential\footnote{We will utilise the notation $X = (\tau,\vec{x})$ for points on $\setS^1_\beta \times \mathbb{R}^2$.}
\begin{equation}
    G^{(\mu)}(X,Y) = \mel{ X} {(\Gamma^\mu \partial_\mu - \mu \Gamma_3 + \sigma_0 )^{-1}}{ Y}.
    \label{eq:GN_prop}
\end{equation}
%

\subsubsection{Leading-order action and gap equation} 

We can write the expression of the grand potential \(\Omega(\mu)\) of \cref{eq:grandpotentialgeneral} in momentum-space from the leading-order part of the effective action of \cref{eq:Seff_GN}, using the form of the Fourier transforms and the fermionic Matsubara sums given in \Cref{sec:loop} as \footnote{See \cite{kapusta_gale_2006} for details of the computation. Also, the expression is normalised by the volume \(V\).}
\begin{equation}\label{eq:GN_grandpotential}
\frac{ \Omega(\mu)}{N} \coloneqq  -\frac{\sigma_0^2}{4 g} + 2 \Int\limits^\Lambda \frac{\dd[2]{p}}{(2\pi)^2} \Bqty{ \omega_p + \frac{1}{\beta} \log \pqty{ 1 + e^{-\beta (\omega_p +\mu)} } + (\mu \leftrightarrow - \mu) },
\end{equation}
where we assume that \(\sigma_0 , \, \mu \ll \Lambda\) and, moreover, we have introduced the notation \(\omega_p^2 = p^2 + \sigma_0^2\). 

To compute the condensate \(\sigma_0\) we have to solve the gap equation
\begin{equation}
    \fdv{S^{\saddle}_{\text{eff}, \,  (0)}}{\sigma_0} =0,
\end{equation}
and to properly read off the symmetry phases~\cite{Moshe_2003} we introduce the coupling at criticality \(g_c^{-1} = \Lambda/\pi\). For non-zero values of the condensate \(\sigma_0\) the gap equation becomes
\begin{equation}
0 =  \left(\frac{1}{g} - \frac{1}{g_c}\right) - \frac{1}{\pi}  \left( \sigma_0- \frac{1}{\beta} \log (1 + e^{\beta (\sigma_0+\mu)}) - \frac{1}{\beta} \log (1 + e^{\beta (\sigma_0-\mu)})    \right).
\end{equation}
Therefore, solving for \(\sigma_0\) we get the closed form expression
\begin{equation}\label{eq:gapsigma0}
	e^{\beta \sigma_0} = \frac{1}{2} \Bqty{  e^{\beta \pi \pqty{ \frac{1}{g_c} - \frac{1}{g} }} - 2 \cosh \beta \mu + \sqrt{ \pqty{e^{\beta \pi \pqty{ \frac{1}{g_c} - \frac{1}{g} }} - 2 \cosh \beta \mu }^2 -4 }   }.
\end{equation}
To find the exact value of \(\sigma_0\) we separate the expression \eqref{eq:gapsigma0} to the following phases
\begin{enumerate}[left = 0pt]
    \item \underline{For \(g > g_c\)} 
    
    The only non-trivial solution is found at the limit of zero temperature and chemical potential, \emph{i.e.} \(\beta \to \infty , \, \mu =0\) and reads
\begin{equation}
   \eval*{\sigma_0}_{\mu,\beta^{-1} = 0} = \pi \left( \frac{1}{g_c} - \frac{1}{g} \right).
\end{equation}
The above solution survives also at the limit of zero temperature and finite chemical potential if only \(\mu < \mu_c = \sigma_0\big|_{\mu,\beta^{-1} = 0}\). For the case of \( \mu > \mu_c\) we observe parity restoration.
\item \underline{For \(g = g_c\)} 

The value \(g = g_c\) is the quantum critical point of the \acrshort{gn} model at large-\(N\) and signifies a second-order phase transition between the two phases of the theory: the one with parity and the one where parity is broken. At the critical point, there is no non-trivial solution for the condensate \(\sigma_0\) for any value of the chemical potential \(\mu\). 
\end{enumerate}
Therefore, at the critical point for finite chemical potential \(\mu\), we have that \(\sigma_0 = 0\) and the ground state of the corresponding \acrshort{cft} at the zero temperature limit \(\beta \to \infty\) is that of a filled Fermi sphere where the fermions are massless. The grand potential for the above configuration is computed using \cref{eq:GN_grandpotential} and reads
\begin{align}
\frac{\Omega(\mu)}{N} & =  2 \Int\limits_{\mu<|p|<\Lambda} \frac{\dd^2 p}{(2\pi)^2} \, \abs{p} +2 \mu \Int\limits_{|p|<\mu} \frac{\dd^2 p}{(2\pi)^2} \nonumber \\
& =  \frac{\Lambda^3}{3 \pi} + \frac{\mu^3}{6\pi}.
\end{align}
We can use the result of the grand potential and \cref{eq:this_is_basic} to compute the normalised \(U(1)_B\) charge \(\Qp\) and the normalised canonical energy of the Fermi-sphere ground state as
\begin{align}
	\frac{\Qp}{N} &= \frac{\mu^2}{2\pi},  & \frac{F^{\saddle}_c \pqty{\Qp}}{N} &= \frac{1}{3\pi} \pqty{2 \pi \frac{\Qp}{N} }^{3/2} .
\label{eq:densities_GN}
\end{align}
As we saw in \cref{Chapter4}, the computation on the torus \(\setT^2\) provides the leading order result in \(\Qp\) for the ground state energy of the Fermi-sphere in the large-charge expansion on the sphere \(\setS^2\), and therefore for the conformal dimension \(\Delta_{FS}\) of the Fermi sphere operator \(\Opp_{FS}\)~\cite{komargodski2021spontaneously} which is the lightest charged primary in the theory. As a matter of fact, we have checked that at leading order, the result for the \acrshort{gn} \acrshort{cft} is precisely the same as the free fermion \acrshort{cft}.

We should point out that the absence of \acrlong{ssb} at leading order in the large-\(N\) limit in the \acrlong{gn} model at large charge has two possible explanations:
\begin{enumerate}
\item We have an interacting theory that at large-\(N\), the large-charge ground state is not that of a superfluid but of an exact Fermi surface, that would make this theory to be the first non-free example with such a behaviour. If this is the case, there should be a transition between Fermi sphere physics at large-\(N\) and superfluid physics at small-\(N\) at some value \(N^{*}\) to accommodate the emergent supersymmetry at \(N=1/2\)~\cite{hellerman2015cft,fei2016yukawa,antipin2022yukawa}\footnote{As we label
\(N\), the \acrlong{gn} model is correctly defined for half-integer values of \(N\) in \(d=3\).}.
\item The Fermi surface at finite values of \(N\) is never precisely free, and differently to the superfluid \acrshort{eft}s examined in~\cite{hellerman2015cft,alvarez2017compensating}, any interactions taking place in the Fermi surface at low energies are not automatically suppressed. If that is the case, the \acrshort{ssb} is an exponentially suppressed effect and simply inaccessible in \(1/N\) perturbation theory. This possibility would not require a finite-\(N\) transition, and it would be in accordance with the emergent supersymmetry~\cite{fei2016yukawa} and the massive Goldstino~\cite[see][\S~4]{hellerman2015cft} at \(N=1/2\).
\end{enumerate} 
Since we have not yet analysed the four-fermion interaction about the Fermi surface ground state, we cannot say for sure which is the correct possibility.

\subsection{Nambu–Jona–Lasinio model}\label{sec:NJL-det}

The next model we will study is the \( U(1) \)-\acrshort{njl} model with a finite \(U(1)_A\)-chemical potential \(\mu\), written in terms of the auxiliary Stratonovich field \(\Phi\) replacing the complex fermionic bilinear \(\Bar{\Psi} \Psi + \Bar{\Psi} \Gamma_5 \Psi \) with action
\begin{equation}
\label{eq:U(1)-NJL-chem}
  S \bqty{\Phi, \Psi} =  \Int\limits_{\setS^1_\beta \times \setT^2} \dd{\tau} \dd[2]{x} \bqty{ \bar \Psi \pqty{ \Gamma^\mu \partial_\mu - \mu \Gamma_3 \Gamma_5 + \Phi P_+ + \bar{\Phi} P_- } \Psi + \frac{N}{4 g} \abs{\Phi}^2 },
\end{equation}
where in the preceding expression we introduced the chiral projectors
\begin{equation}
    P_{\pm} = \frac{1 \pm \Gamma_5}{2}.
\end{equation} 
The above chemical potential \(\mu\) is sourcing a finite charge density for the \(U(1)_A\) axial symmetry
\begin{align}
	\Psi &\to e^{i\alpha \Gamma_5}\Psi, & \Phi &\to e^{-2i\alpha}\Phi.
\end{align}
We will proceed as in the \acrshort{gn} case of \cref{sec:GN-det}, and therefore we can assume that the Stratonovich field \(\Phi\) will acquire a \acrshort{vev} that will spontaneously break the \(U(1)_A\) symmetry, and moreover that the aforementioned \acrshort{vev} is homogeneous \(\expval{\Phi} = \Phi_0\). Then the normalised thermodynamic grand potential \eqref{eq:grandpotentialgeneral} is computed from the leading-\(N\) term of the effective action and reads
\begin{align}
\frac{\Omega(\mu)}{N} =  -\frac{\abs{\Phi_0}^2}{4 g} + \Int\limits^\Lambda \frac{\dd[2]{p}}{(2\pi)^2} \left\{ \Omega_+ + \Omega_- + \frac{2}{\beta} \log \left( 1 + e^{-\beta \Omega_+} \right) + \frac{2}{\beta} \log \left( 1 + e^{-\beta \Omega_-} \right) \right\},
\end{align}
and with \(\Omega_{\pm}\) we denote the one-particle on-shell energies defined as
\begin{equation} \label{eq:disp_NJL}
    \Omega_{\pm}^2 \coloneqq \abs{\Phi_0}^2 + (\abs{p} \pm \mu)^2.
\end{equation}
The difference with the \acrlong{gn} model analysed in \cref{sec:GN-det} is that it is not possible to get a Fermi sphere configuration if \(\Phi_0 \neq 0\), given that the one particle on-shell energies are always positive definite \(\Omega_{\pm} \geq 0\). Therefore, for zero temperature, \emph{i.e.} \(\beta \to \infty\), we can drop the thermal logarithm terms, in which case the grand potential reads
\begin{equation}
\label{eq:NJL-FermionSpectrum}
\lim_{\beta \rightarrow \infty}	\frac{\Omega(\mu)}{N} = - \frac{\abs{\Phi_0}^2}{4 g} + \Int\limits^{\Lambda} \frac{\dd[2]{p}}{(2\pi)^2} \bqty{\sqrt{\pqty{\abs{p}+\mu}^2 + \abs{\Phi_0}^2} + \sqrt{\pqty{\abs{p}-\mu}^2 + \abs{\Phi_0}^2} }.
\end{equation}
To explicitly calculate the values of the condensate \(\Phi_0\) we have to solve the gap equation and as before we also include the critical coupling \(g_c^{-1} = \Lambda/\pi\). Thus, the gap equation in the zero temperature limit is
\begin{equation}
0  = \frac{1}{2}\left( \frac{1}{g} - \frac{1}{g_c} \right) + \frac{1}{2\pi^2} \left[\sqrt{ \abs{\Phi_0}^2 + \mu^2} - \mu  \arctanh\pqty{ \frac{\mu}{\sqrt{ \abs{\Phi_0}^2 + \mu^2 }}}  \right], 
\end{equation}
and given that it depends solely on $\abs{\Phi_0}$, we infer that the \acrshort{vev} is real and positive, \emph{i.e.} \(\expval{\Phi} = \Phi_0 > 0\). In contrast to the \acrshort{gn} case, at the critical point \(g= g_c\) the gap equation always features a non-trivial solution for finite values of the chemical potential \(\mu\), which is 
\begin{align}\label{eq:NJL_vev_flat}
  &\Phi_0 = \mu \sqrt{\kappa_0^2-1} , & &\text{where}& &\kappa_0 \tanh \kappa_0 =1,
\end{align}
and we can numerically evaluate it as \(\Phi_0/\mu = 0.6627\dots\).

The above solution verifies our postulate the that the finite chemical potential ground state will spontaneously break the \(U(1)_A\) axial symmetry by providing a \acrshort{vev} to the Stratonovich field \(\Phi\), which in turn will play the part of the large order parameter. We also note that symmetry cannot be restored in any value apart from zero chemical potential, \emph{i.e.} \(\mu=0\), since conformal symmetry forbids the existence of any new scale that would separate the broken and the unbroken phase. 

\begin{figure}[ht]
  \centering
    \begin{footnotesize}
    \begin{tikzpicture}
      \node at (0,0) {\includegraphics[width=.75\textwidth]{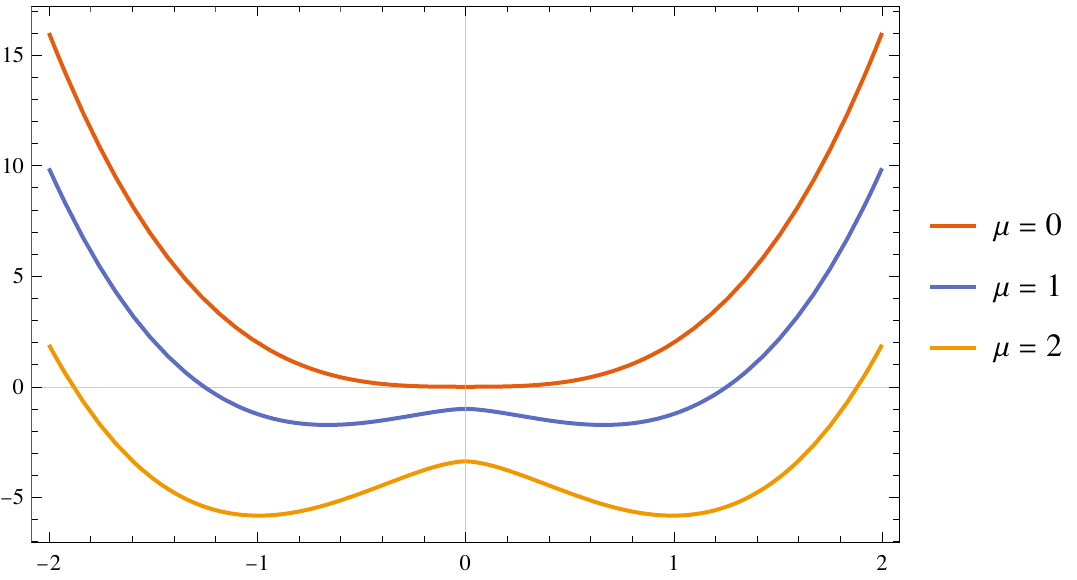}};
      \node at (-6.25,3.25) {\(\Omega^2\)};
      \node at (4.5,-3) {\(\Phi_0\)};
    \end{tikzpicture}
  \end{footnotesize}
	\caption{Plot of leading order in \(N\) of the canonical potential for the \acrshort{njl} model with $\mathfrak{Im}\pqty{\Phi} =0$ for different values of \(\mu\). For \(\mu >0 \) the minima at $\Phi\neq 0$ signals \acrshort{ssb}. }
 \label{fig:minima}
\end{figure}

We can make the situation more clear by computing the renormalized potential \(\Omega\) explicitly for some generic constant configuration of the auxiliary field \(\Phi\).
We will use the minimal subtraction scheme, given that the divergent part of the grand potential is independent of the chemical potential \(\mu\), 
\begin{equation}
  \begin{aligned}
	\frac{\Omega(\mu)}{N} - \frac{\Omega(0)}{N}  &= \Int\frac{\dd^2 p}{(2\pi)^2}\left[\Omega_+ + \Omega_- - 2\sqrt{p^2+ \abs{\Phi}^2}\right]\\
	&= \frac{1}{6\pi}\left[3 \abs{\Phi}^2\mu \arctanh\pqty{\frac{\mu}{\sqrt{\abs{\Phi}^2+\mu^2}}} +\pqty{\mu^2 - 2\abs{\Phi}^2}\sqrt{\abs{\Phi}^2 + \mu^2} + 2\abs{\Phi}^3\right],
  \end{aligned}
\end{equation}
and then add once more the renormalized value \(\Omega\). For \(\mu =0\) the one-particle on-shell energy reads \(\Omega^2_{\pm} = \abs{\Phi}^2 + p^2  \) which is precisely the same expression as \(\omega_p\) in the \acrshort{gn} model, and the regularised integral expression reads
\begin{equation}
   \Omega(0) = -\frac{N \abs{\Phi}^3}{3\pi}.
\end{equation}
Thus, the renormalised potential is
\begin{equation}
	\frac{\Omega(\mu)}{N} = \frac{1}{6\pi}\left[3 \abs{\Phi}^2\mu \arctanh\pqty{\frac{\mu}{\sqrt{\abs{\Phi}^2+\mu^2}}} + \pqty{\mu^2 - 2\abs{\Phi}^2} \sqrt{\abs{\Phi}^2 + \mu^2} \right], 
\end{equation}
and we observe that it is \(U(1)_A\)-invariant as anticipated, and at the same time it features a \(\setS^1\)-worth of vacua for the value of \(\abs{\Phi} = \mu \sqrt{\kappa_0^2 -1}\) as is depicted in~\Cref{fig:minima}.
The ground state of this model is that of the conformal superfluid and using \cref{eq:this_is_basic} we can again compute the \(U(1)_A\)-charge and the canonical free energy as 
\begin{align} \label{eq:flat_space_NJL}
	\frac{\Qp}{N} &= \frac{\kappa_0^3 \mu^2}{2\pi} , & \frac{F_c^{\saddle}(\Qp)}{N} &= \frac{1}{3\pi \kappa_0^{3/2}} \pqty{2\pi \frac{\Qp}{N}}^{3/2}.
\end{align}
Again, we note that this computation on the torus \(\setT^2\) gives the leading order result in \(\Qp\) for the ground state energy of the conformal superfluid at large charge on the sphere \(\setS^2\), and therefore for the conformal dimension \(\Delta_{SF}\) of the superfluid large charge operator \(\Opp_{SF}\). This configuration cannot support a Fermi sphere solution, and hence the whole charge is held in the superfluid.

As a final note, we might try to repeat the above computation for a finite \(U(1)_B\)-charge instead, however the eigenvalues of \cref{eq:disp_NJL} are replaced by
\begin{equation}
    \Omega_{\pm} = \sqrt{p^2 + \abs{\Phi_0}^2} \pm \mu,
\end{equation}
which are precisely the same eigenvalues that appear in the \acrshort{gn} model. Therefore, we conclude that for this choice of chemical potential there is no superfluid sector, and instead the ground state at large charge is described by a Fermi sphere with the same charge and energy as in \cref{eq:densities_GN}.

\subsection[\texorpdfstring%
{\(SU(2)\)-NJL model}%
{SU(2)}]%
{\(SU(2)\)-NJL model}%
\label{sec:su2su2-det}

The final model that we will examine is the \(SU(2)\)-\acrshort{njl} model. Now, we can choose from multiple charge densities to decide which we will source. We will study the model for finite values of the chemical potential for the \(\sigma^3\) and \(\Gamma_5 \sigma^3\) generators. The corresponding critical action reads
\begin{equation}
\label{eq:SU(2)-NJL-chem}
  S \bqty{\sigma, \pi_a, \Psi} =  \Int\limits_{\setS^1_\beta \times \setT^2} \dd{\tau} \dd[2]{x} \left[ \bar \Psi \left( \Gamma^\mu \partial_\mu + \sigma + i \pi_a \sigma^a \Gamma_5 - \left\{
        \begin{aligned} &\mu_V \Gamma_3 \sigma^3 \\
                        &\mu_A \Gamma_3 \Gamma_5 \sigma^3 \end{aligned}
                    \right\}
        \right) \Psi  \right].
\end{equation}
The grand potential is once more computed from the leading term of the effective action and now reads
\begin{align}
\frac{\Omega^{V,A}}{N} =  2 \Int\limits^\Lambda \frac{\dd[2]{p}}{(2\pi)^2} \left\{ \Omega_+^{V,A} + \Omega_-^{V,A} + \frac{2}{\beta} \log \left( 1 + e^{-\beta \Omega_+^{V,A}} \right) + \frac{2}{\beta} \log \left( 1 + e^{-\beta \Omega_+^{V,A}} \right) \right\},
\end{align}
where in a similar manner as in \cref{sec:NJL-det} we introduce the one-particle on-shell energies 
\begin{align}
\Omega^{V}_{\pm} &= \sqrt{|\Phi_2|^2 + \left(\sqrt{(|p| + |\Phi_1|} \pm \mu_V \right)^2 }, &
\Omega^{A}_{\pm} &= \sqrt{|\Phi_1|^2 + \left(\sqrt{(|p| + |\Phi_2|} \pm \mu_A \right)^2 },
\end{align}
where in terms of real fields \(\sigma, \pi_a\) we have \(|\Phi_1|^2 = \sigma^2 + \pi_3^2, \, |\Phi_2|^2 = \pi_1^2 + \pi_2^2\). Independently of the choice of the sourcing, there is no configuration for the gap equation at zero temperature for which the field combinations \(\Phi_1 , \Phi_2\) acquire a \acrshort{vev} simultaneously. It turns out that when \(\mu_V\) is switched on, the only solution that can be found is
\begin{align}
    \abs{\Phi_{1,0}}& = 0, &  \abs{\Phi_{2,0}} & = \mu_{V,A} \sqrt{\kappa_0^2 -1},
\end{align}
where \(\kappa_0\) is the same as in \cref{eq:NJL_vev_flat} .

The same analysis holds true when \(\mu_A\) is switched on instead, with the only difference being that the \acrshort{vev}s for \(\Phi_1, \Phi_2\) are exchanged. 

Given that the one-particle on-shell energies are always positive, \emph{i.e.} \(\Omega_{\pm}^{V,A} \geq 0\), we conclude that no Fermi sphere can arise in the \(\beta \to \infty\) limit, and the large charge ground state configuration is that of the conformal superfluid, similar to the \(U(1)\)-\acrshort{njl} model.
The only difference with the previous result is an overall factor of two, that appears in the grand potential and the canonical free energy.

\section{Spectrum of fluctuations}
\label{sec:fluctuations}

In \cref{sec:symmetry} we determined the large-charge ground state in flat space for the \acrshort{gn} and the \(U(1)\)-\acrshort{njl} and \(SU(2)\)-\acrshort{njl} models. In this section, we aim to analyse the fluctuations' spectrum on top of the aforesaid ground state. 

As a preliminary assessment, we expect that the models that exhibit \acrshort{ssb} and the ground state is that of a conformal superfluid, follow the pattern of the large charge \(O(N)\) vector model so that the spectrum contains one type I Goldstone mode with the following dispersion relation
\begin{equation}
    \omega_I = \frac{p}{\sqrt{2}} + \dots
\end{equation}
and a massive particle of order \(\mu\)~\cite{hellerman2015cft}. \newline
On the contrary, in the case of the \acrlong{gn} model, where the large charge ground state is that of a Fermi sphere and there is no \acrshort{ssb}, the conformal superfluid paradigm does not apply, and we do not expect the appearance of a conformal Goldstone mode. 

\subsection{Gross-Neveu model}
\label{sec:gn-fluctuations}

First, we examine the \acrshort{gn} model, and we anticipate the quantum fluctuations on top of the Fermi sphere ground state to be both of bosonic and of fermionic nature. The former are generated from the Stratonovich field \(\sigma\), while the latter are particle hole excitations present in the free fermionic theory~\cite{polchinski1992effective}. 

To understand the impact of the fluctuations, we should go to the \acrfull{nlo} term in the effective action \(S^{\saddle}_{\text{eff}, \, (2)}\) \eqref{eq:Seff_GN} at the critical point where \(\sigma_0 =0\). Then the whole analysis is reduced to the computation of \( \Tr \bqty{G^{(\mu)} \sigma G^{(\mu)} \sigma}\), 
which is simpler to perform in momentum space \footnote{Our notation is \(P = (\omega_n , \vec{p})\) for momenta on $\setS^1_\beta \times \mathbb{R}^2$, where $\omega_n = \frac{\pqty{2n+1} \pi}{\beta}$ are the Matsubara frequency for fermions.}, where it reads 
\begin{align}
	\Tr \bqty{ G^{(\mu)} \sigma G^{(\mu)} \sigma} &= - \SumInt \dd[2]{p}\, \sigma(-P) \sigma(P)  \SumInt \frac{\dd[2]{k}}{ (2\pi)^2 \beta}  \Tr\bqty{G^{(\mu)}(K) G^{(-\mu)}(P-K)} ,
 \label{eq:doubletrace}
\end{align}
where we used that \(G^{(\mu)}(X,Y) = -G^{(-\mu)}(Y,X) \) and the fermionic propagator \eqref{eq:GN_prop} at the critical point in momentum space for finite charge density which reads
\begin{align}
	G^{(\mu)}(P)&= \frac{i \Gamma_\mu \tilde{P}^\mu}{\tilde{P}^2}, & \tilde{P}& = (\omega_n - i \mu, \vec{p}).
\end{align}
The one-loop integral of \cref{eq:doubletrace} can be depicted in the following Feynman diagram 
\begin{equation}
\vcenter{\hbox{\includegraphics[scale=1]{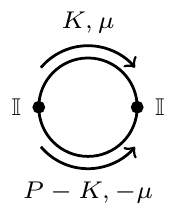}}} \, = 2 P^2 I_2 -4 I_1,
\end{equation}
and sums up to the computation of two master integrals \(I_1,I_2\), where details of the computation can be found in \Cref{sec:loop}. At the limit that \(\beta \to \infty\) this reads
\begin{equation}
    \SumInt \frac{\dd[2]{k}}{ (2\pi)^2 \beta}  \Tr\bqty{G^{(\mu)}(K) G^{(-\mu)}(P-K)} =  i^2 \bqty{\frac{\sqrt{P^2}}{4} + \frac{\mu}{\pi}} .
\end{equation}
Therefore using \cref{eq:Seff_GN} we obtain the quadratic action of fluctuations, which is
\begin{equation}
 	S^{\saddle}_{\text{eff}, \, (2)} = \frac{1}{2} \Tr\bqty{G^{(\mu)\sigma G^{(\mu)} \sigma }} = \frac{1}{2} \SumInt \dd^2 p \, \sigma(-P) \sigma(P) \left[ \frac{\sqrt{P^2}}{4} + \frac{\mu}{\pi} \right] .
 \end{equation}
This action is non-local and does not actually correspond to stable bosonic excitations on top of the fermionic ground state. As a matter of fact, the \(\mu/ \pi\) is a decay constant term of \(\order{\mu}\) rather than a mass term, and a similar case has been found in~\cite{hands1993four}.

\subsection{Nambu–Jona–Lasinio model}
\label{sec:njl-fluctuations}

In \cref{sec:NJL-det} we observed that fixing the \(U(1)_A\)-charge generates a \acrshort{vev} for the Stratonovich field \(\Phi\) and from the determinant of the grand potential in the zero temperature limit in \cref{eq:NJL-FermionSpectrum}, we note that every fermion gets a mass that is equal to
\begin{equation}
	m_F^2 = \mu^2 + \Phi_0^2 = \kappa_0^2\mu^2,
\end{equation}
where we used the explicit value of the \acrshort{vev} for \(\Phi_0\) from \cref{eq:NJL_vev_flat}.
This fact indicates that the \(U(1)_B\) symmetry remains unbroken, while the axial symmetry is broken. 

Therefore, the \(U(1)\)-\acrshort{njl} is a suitable candidate to verify the predictions of the superfluid \acrshort{eft} by explicitly computing the \acrshort{nlo} corrections in the functional determinant, which are subleading in \(N\). 

Following \cref{sec.U(1)ssb}, as in~\cref{eq.vevplusexpansion} we can shift the \(\Phi\) field by its \acrshort{vev} and expand the fluctuation into real and imaginary parts \(\hat \Phi = \hat\sigma + i \hat\pi\) around the chosen vacuum state \(\expval{\Phi} = \Phi_0\) as 
\begin{equation}
    \Phi = \Phi_0 + \frac{\hat{\Phi}}{\sqrt{N}} = \Phi_0 + \frac{\hat\sigma + i \hat\pi}{\sqrt{N}},
\end{equation}
and then express the Lagrangian of \cref{eq:U(1)-NJL-chem} in the critical limit in terms of the vacuum state plus the quantum fluctuations on top
\begin{equation}\label{eq:langrangianfluctuationsnjl}
	\Lp_{\Phi_0} =  \bar\Psi \Gamma_\mu \partial^\mu \Psi + \Phi_0\bar\Psi\Psi - \mu \bar\Psi\Gamma_3\Gamma_5\Psi + \frac{1}{ \sqrt{N}} \pqty{ \hat\sigma \bar\Psi\Psi+ i\hat\pi \bar\Psi\Gamma_5\Psi } .
\end{equation}
To determine the fluctuations' spectrum over the ground state, we have to compute the inverse propagator \(G^{-1} (P)\) of $\hat\Phi$ at one fermion loop.

We can read the fermion propagator from \cref{eq:langrangianfluctuationsnjl} and we can write it in momentum space using the notation of \Cref{sec:loop} where it explicitly reads
\begin{align}\label{eq:propagatornjl}
	\Delta^{(\mu,\Phi_0)}( P) & = (-i\slashed{P} +\Phi_0 -\mu\Gamma_3\Gamma_5)^{-1} \nonumber \\
	& = \frac{
    \left( \omega^2 + k^2 + \Phi_0^2 - \mu^2 + 2 \mu (i \omega \Gamma_3 + \Phi_0 ) \Gamma_3 \Gamma_5 \right) 
    }{\left(\omega^2 + \Phi_0^2 + (\mu + k)^2 \right)\left(\omega^2 + \Phi_0^2 + (\mu - k)^2 \right)} \left( i\slashed{P} +\Phi_0 -\mu\Gamma_3 \Gamma_5\right) ,
\end{align}
and we use the notation \({P} = (\omega, \vec{p})\) for momenta on \(\setS^1_\beta \times \setR^2\), and \(\slashed{P} = \Gamma^\mu P_\mu\). We note that due to the absence of a Fermi surface ground state, we can work straight away in the zero temperature limit \(\beta \rightarrow \infty\), therefore \(\omega\) are the continuous Matsubara frequencies, \emph{i.e.} \(\omega_n \overset{\beta \to \infty} {\longrightarrow} \omega\). 

From the explicit form of \cref{eq:propagatornjl} we observe that the fermionic propagator is antisymmetric \(\Delta^{\pqty{\mu,\Phi_0}}( -P) = -\Delta^{\pqty{-\mu,-\Phi_0}}( P)\) and utilising it we can compute the following Feynman graphs in momentum space which will provide the inverse propagator for the scalar
fluctuations 
\begin{align}
\vcenter{\hbox{\includegraphics[scale=1]{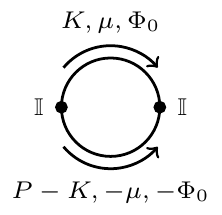}}} \, &=	G^{-1}_{\sigma\sigma}(P) = -\Int \frac{\dd^3k}{(2\pi)^3} \Tr\left[\Delta^{(\mu,\Phi_0)}(K)\Delta^{(-\mu,-\Phi_0)}(P-K) \right],\\
\vcenter{\hbox{\includegraphics[scale=1]{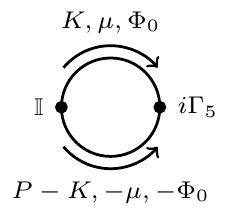}}} \, &=	G^{-1}_{\sigma\pi}(P) = -i\Int \frac{\dd^3k}{(2\pi)^3}\Tr\left[\Delta^{(\mu,\Phi_0)}(K)\Gamma_5 \Delta^{(-\mu,-\Phi_0)}(P-K) \right],\\
\vcenter{\hbox{\includegraphics[scale=1]{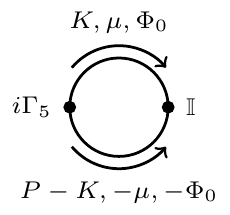}}} \, &=	G^{-1}_{\pi\sigma}(P) = -i\Int \frac{\dd^3k}{(2\pi)^3}\Tr\left[ \Gamma_5 \Delta^{(\mu,\Phi_0)}(K) \Delta^{(-\mu,-\Phi_0)}(P-K) \right] ,\\
\vcenter{\hbox{\includegraphics[scale=1]{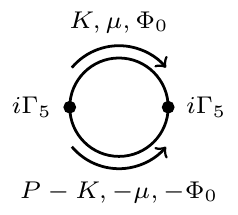}}} \, &=	G^{-1}_{\pi\pi}(P) = \Int \frac{\dd^3k}{(2\pi)^3}\Tr\left[\Gamma_5 \Delta^{(\mu,\Phi_0)}(K)\Gamma_5 \Delta^{(-\mu,-\Phi_0)}(P-K) \right].
\end{align}
Every integral above can be expanded in the regime of interest which is \((P/\mu)\ll1\).

The zero order of the expansion equals the value of \(P=0\) and needs regularisation. We choose to regularise it by subtracting the \(\mu=0\) result, since the divergence is independent of \(\mu\). The analytical computation can be found in \cref{sec:njl-computations} and the final result reads
\begin{equation}
	G^{-1}(P)\Big|_{\mathcal{O}(0)} =  \begin{pmatrix}
		G^{-1}_{\sigma\sigma}(0) & G^{-1}_{\sigma\pi}(0)\\
		G^{-1}_{\pi\sigma}(0) & G^{-1}_{\pi\pi}(0) 
	\end{pmatrix} = \frac{\kappa_0\mu}{\pi}
    \begin{pmatrix}
		1 & 0\\
		0 & 0 
	\end{pmatrix},
\end{equation}
where \(\kappa_0\) is given in \cref{eq:NJL_vev_flat} and the result is consistent with the existence of a massive and of a massless mode. 

Higher orders in the expansion have no need for regularisation, and we can straightforwardly compute them. The linear order in \(P\) reads
\begin{equation}
	G^{-1}(P) \Big|_{\mathcal{O}(P/\mu)} =  \frac{\kappa_0 \omega}{2\pi } \begin{pmatrix}
	    0 & -1 \\  1 & 0
	\end{pmatrix},
\end{equation}
while the quadratic order is
\begin{equation}
	G^{-1}(P) \Big|_{\mathcal{O}(P^2/\mu^2)} = \begin{pmatrix}
        \frac{(2 \kappa_0^2 -1) \omega^2 }{ 12 \pi \kappa_0 (\kappa_0^2 -1) \mu} + \frac{ (3\kappa_0^6 - 2\kappa_0^4 - 2 \kappa_0^2 + 2) p^2 }{ 24 \pi \kappa_0^3 (\kappa^2 -1) \mu}  & 0  \\
        0   &  \frac{ \kappa_0 \omega^2 }{ 4 \pi(\kappa_0^2 -1) \mu} + \frac{ \kappa_0^3 p^2 }{ 8\pi (\kappa_0^2 -1 ) \mu }
    \end{pmatrix} .
\end{equation}
As usual, we can compute the dispersion relations from the zeros of the inverse propagator
\begin{equation}
	G^{-1}(P) =  \begin{pmatrix}
        \frac{\kappa_0\mu}{\pi}+ \frac{2 \kappa_0^2 \left(2 \kappa_0^2-1\right) \omega ^2+\left(3 \kappa_0^6-2 \kappa_0^4-2 \kappa_0^2+2\right) p^2}{24 \pi  \kappa_0^3 \left(\kappa_0^2-1\right) \mu } &  - \frac{\kappa_0}{2\pi } \omega \\
     \frac{\kappa_0}{2\pi } \omega & \frac{2 \kappa_0 \omega^2 + \kappa_0^3 p^2 }{8 \pi ( \kappa_0^2 -1) \mu }
	\end{pmatrix} + \mathcal{O}(P^3/\mu^3),
\end{equation}
therefore for the two modes they read
\begin{align}\label{eq:U(1)-spectrum}
    \omega_{1}^2 &= - \frac{1}{2}p^2 +\dots ,\\
    \omega_{2}^2 &= - 12 \frac{ \left( \kappa_0^2 -1\right) \kappa_0^4 }{ \left( 2 \kappa_0^2-1 \right) } \mu^2 - \frac{\left(5 \kappa_0^6-5 \kappa_0^4-\kappa_0^2+2\right) }{ 2\kappa_0^2 (2 \kappa_0^2 -1) } p^2 +\dots,
\end{align}
From \cref{eq:U(1)-spectrum}, we can identify the conformal Goldstone \(\omega_I = \omega_1\) and we also observe the radial massive mode of mass order \(\order{\mu}\). This result confirms our analysis and demonstrates that the conformal superfluid \acrshort{eft} is applicable in this case.

Finally, we have repeated this same analysis for the case of the \(SU(2)\)-\acrshort{njl} model for finite values of the chemical potentials \(\mu_{A,V}\). We discovered precisely the same spectrum as in \cref{eq:U(1)-spectrum} with the addition of two additional degenerate gapped modes, with the following dispersion relation
\begin{equation}
    \omega ^2 = - 4  \kappa_0^2  \mu^2 - \frac{ \left( \kappa_0 ^2-1 \right) p^2}{ \kappa_0^2 } + \dots 
\end{equation}
We conclude that the conformal superfluid \acrshort{eft} describes both the \(U(1)\)-\acrshort{njl} and the \(SU(2)\)-\acrshort{njl} model, and the spectrum of their quantum fluctuations agrees completely with the expectant predictions considering their symmetry breaking pattern, as we will discuss in \cref{sec:SBB-patterns}.

\subsection{Symmetry breaking patterns and Goldstones}
\label{sec:SBB-patterns}

The analysis of \cref{sec:njl-fluctuations} indicates that the ground state of both the \(U(1)\)-\acrshort{njl} and the \(SU(2)\)-\acrshort{njl} models is described by the conformal superfluid \acrshort{eft} and the spectrum of neither of the theories contains a Goldstone mode of type II with a quadratic dispersion relation. This may come as a surprise, given that naively we might think that the \(SU(2)\)-\acrshort{njl} is a suitable candidate in which the inhomogeneous sectors of the \(O(4)\) model can be further studied~\cite{hellerman2019note,hellerman2021observables,banerjee2019conformal,banerjee2022subleading}. Nevertheless, the obtained result is totally consistent with the known counting rules of Goldstone modes.

The actions of \cref{eq:U(1)-NJL-chem,eq:SU(2)-NJL-chem} at the conformal point and for zero chemical potential \(\mu\) feature the following total symmetry
\begin{equation}
  SO(4,1) \times SU(N) \times U(1)_B \times 
  \begin{cases}
    U(1)_A  & \text{($U(1)$-\acrshort{njl})},\\
    SU(2)_L \times SU(2)_R & \text{($SU(2)$-\acrshort{njl}).} 
  \end{cases}
\end{equation}
By fixing the charge and introducing the axial chemical potentials $\mu_A$, the symmetry is reduced to
\begin{equation}\label{eq:leftover-symm}
D \times SO(3) \times SU(N) \times U(1)_B \times
\begin{cases}
U(1)_A & \text{($U(1)$-\acrshort{njl})},\\
U(1)^{(3)}_{A} \times U(1)^{(3)}_B & \text{($SU(2)$-\acrshort{njl}),}
\end{cases}
\end{equation}
where \(D\) is the dilatation generator on flat space and of time translations on the cylinder, and \(U(1)_{A,B}^{(3)}\) are the global phase symmetries generated by \(\Gamma_5 \sigma^3\) and \(\sigma^3\) respectively which are the Cartans of \(SU(2)_L \times SU(2)_R\). So, we observe that by introducing the axial chemical potentials, we have explicitly broken the \(SU(2)_L \times SU(2)_R\) symmetry down to \(U(1)^{(3)}_{A} \times U(1)^{(3)}_B\). 

Seeing the breaking as a two-step procedure~\cite{Gaume:2020bmp}, at this point the ground state solution at large-\(N\) spontaneously breaks a linear combination of \(D\) and of the global phase symmetry \(U(1)_A\) for both models. This is the same symmetry breaking pattern that we discussed in \cref{sec:O(2)_review} and agrees with the typical process of symmetry breaking for a non-Lorentz invariant theory~\cite{nielsen1976count,watanabe2011number,watanabe2013massive,hidaka2013counting} that we discussed in~\cref{sec.ssbandgaplessstates}. Assuming that we have a system with \(m\) spontaneously broken generators, \(n_I\) relativistic Goldstones modes with linear dispersion relations and \(n_{II}\) non-relativistic modes with quadratic dispersion relations, the correct counting law~\cite{watanabe2020counting} is
\begin{equation}
    n_I + 2 n_{II} = m.
\end{equation}
Given that the \(U(1)\)-\acrshort{njl} and \(SU(2)\)-\acrshort{njl} models have exactly \(m=1\) only one type I Goldstone mode with linear dispersion relation can exist which is the conformal superfluid phonon. As we saw at the end of \cref{sec:njl-fluctuations}, the difference is that the \(SU(2)\)-\acrshort{njl} contains two additional gapped modes that can be merged into a complex scalar field that is charged under the remaining unbroken \(U(1)_B^{(3)}\) symmetry.

\section{Conformal dimensions and local CFT spectrum}%
\label{sec:conformalDim}

In this final section, we want to compute the scaling dimension of the lowest primary operator of large charge \(\Qp\) for the \acrlong{gn} and the \(U(1)\)-\acrshort{njl} model. 

In our analysis thus far, we have examined the aforementioned models on the torus \(\setT^2\) of volume \(V\), but to be able to compute conformal dimensions of charged operators, we should be able to use the state-operator correspondence \eqref{eq.energyandscaling} and study the systems of interest on the cylinder \(\setR \times \setS^2\). This can be accomplished by using a Weyl transformation --- see \Cref{sec:cylinder_spinor} for details --- so that the action for the free fermion at fixed charge on \(\setR \times \setS^2\) reads
\begin{align}
	S &= \Int\limits_{\setR \times \setS^2} \bqty{  \bar{\Psi} \pqty{ \slashed{D} - \mu \Gamma_\tau + \sigma} \Psi  }, & \text{with} && \Qp &= \Int\limits_{\setS^2} \bar{\Psi} \Gamma_\tau \Psi,
\end{align}
where with \(\slashed{D}\) we denote the Dirac operator \footnote{The form of the Dirac operator along with the notations used can be found on~\Cref{sec:app_fermions} and \(\Gamma_{\tau} = \Gamma_3\) in our notation.} on the manifold \(\setR \times \setS^2\) .

\subsection{Gross-Neveu model}\label{sec:scalingGN}

We start with the \acrshort{gn} model at large-\(N\), and the grand potential \(\Omega(\mu)\) evaluated on \(\setS_{\beta}^1 \times \setS^2\) reads 
\begin{align}
\frac{\Omega(\mu)}{N} = -\frac{\sigma_0^2}{4g} + \frac{2}{(4 \pi R_0^2)}  \Sum_{j = \frac{1}{2}} \, \pqty{2j+1} \bqty{  \sqrt{\omega_j^2 + \sigma_0^2} + \text{thermal contributions} },
\end{align}
and \(\omega_j = (j+1/2)/R_0\) denote the eigenvalues of the Dirac operator on the two-sphere \(\setS^2\) and the expression is analogous to \cref{eq:GN_grandpotential} on the torus \(\setT^2\).

Solving the gap equation for the condensate \(\sigma_0\), we can easily deduce that there is no non-trivial solution where \(\sigma_0 \neq 0\) at zero temperature \(\beta \to \infty\) and at the critical point \(g = g_c\) for any value of \(\mu \), precisely like the torus \footnote{The sums need to be regulated using a smooth cut-off, \emph{e.g.} \(e^{-\omega_j/\Lambda}\), to preserve diffeomorphism- invariance.}. Therefore, at the limit \(\beta \to \infty\) the grand potential reads
\begin{equation}
	\frac{\Omega(\mu)}{N} =  \frac{1}{2\pi R_0^2} \bqty{ \Sum_{\omega_j > \mu} \pqty{2j+1}\omega_j + \mu \Sum_{\omega_j < \mu } \pqty{2j +1} },
\end{equation}
describing a Fermi-sphere ground state where the fermions are massless. Using \cref{eq:this_is_basic}, we compute the charge and the canonical free energy that corresponds to the energy of the Fermi's sphere ground state as
\begin{align}
\frac{\Qp}{N} &= \frac{1}{2\pi R_0^2} \floor{\mu R_0} (   \floor{\mu R_0} +1  ), & \frac{F^{\saddle}_c(\Qp)}{N} &= \frac{1}{6 \pi R_0^3}   \floor{\mu R_0} (  \floor{\mu R_0} +1  ) ( 2 \floor{\mu R_0} +1 ),
\end{align}
and we use the floor function to point out that the energy levels on the cylinder are discrete. The limit \(R_0 \to \infty\) replicates the torus result in \cref{sec:GN-det}.

The above Fermi sphere ground state corresponds to the charged scalar primary \(\Opp_{FS}\) that was first discussed in~\cite{komargodski2021spontaneously} in relation to the free fermionic \acrshort{cft}, and it is parity-even. Actually, it turns out that \(\Opp_{FS}\) is the lowest charged primary operator even for the interacting \acrshort{cft} of the \acrlong{gn} model because the Stratonovich field \(\sigma\) does not condense. The same result is true also for the \(U(1)\)-\acrshort{njl} model when we fix the \(U(1)_B\) charge. The charge \(\Qp\) and the conformal dimension \(\Delta_{FS} (\Qp)\) of \(\Opp_{FS}\) are
\begin{align}
	\frac{\Qp}{2N} &=\floor{\mu R_0} (  \floor{\mu R_0} +1  ), &
 \frac{\Delta_{FS} (\Qp)}{2N} &=  \frac{1}{3}  \floor{\mu R_0} (  \floor{\mu R_0} +1  ) ( 2 \floor{\mu R_0} +1 ) =  \frac{ \Qp}{ 6N} \sqrt{ \frac{ 2 \Qp}{N} + 1},
 \label{eq:fermi_sphere_scaling}
\end{align}
where we normalized \(\Qp,\,\text{and} \, \Delta_{FS}(\Qp)\) by \(2N\), which is the total number of \(3d\) Dirac fermions and the conformal dimension \(\Delta_{FS}(\Qp)\) is depicted in~\cref{fig:CGN_scaling} in terms of \(\Qp\).

The asymptotics of the scaling dimension at the limit \(\Qp/ 2N \to \infty\) can be consistently computed so that the first orders read
\begin{equation}
\frac{\Delta_{FS} (\Qp)}{N} =
\frac{2}{3} \pqty{\frac{\Qp}{2N}}^{3/2} + \frac{1}{12} \pqty{\frac{\Qp}{2N}}^{1/2} - \frac{1}{192} \pqty{\frac{\Qp}{2N}}^{-1/2} + \order{\pqty{\frac{\Qp}{2N}}^{-3/2}} , 
\end{equation}
As we discussed before, this large-charge sector does not contain Goldstone excitations that describe new primary operators with \(\sim \order{1}\) gap. As we saw in \cref{sec:gn-fluctuations}, fluctuations of the Stratonovich field \(\sigma\) cannot systematically be used for describing new primaries with gap of order \(\sim \order{1}\) from the Fermi-surface ground state. It rather contains particle-hole type excitations that create new --- in general fermionic and with spin --- primaries with the same charge \(\Qp\) and gap \(\delta \Delta\sim \order{1}\). Given the fact that this is a strictly large-\(N\) analysis, it is not clear if this effect persists as well for finite values of \(N\).

\subsection{Nambu–Jona–Lasinio model}\label{sec:conformalNJL}

Next, we want to compute the conformal dimension of the lowest charged operator for the \(U(1)\)-\acrshort{njl} model. The ground state of the model at fixed \(U(1)_A\) charge corresponds to a conformal superfluid, and therefore studying the \acrshort{cft} on \(\setS^1_\beta \times \setS^2\) via the state-operator correspondence we compute the conformal dimension \(\Delta_{SF}(\Qp)\) of the scalar primary \(\Opp_{SF}\). In our analysis, we focus explicitly on the \(U(1)\)-\acrshort{njl} model, but we note that the result holds true for the \(SU(2)\)-\acrshort{njl} of \cref{sec:su2su2-det} at leading order in \(N\), by replacing \(N\) by \(2N\).

\begin{figure}[ht]
\centering
\includegraphics[scale=0.6]{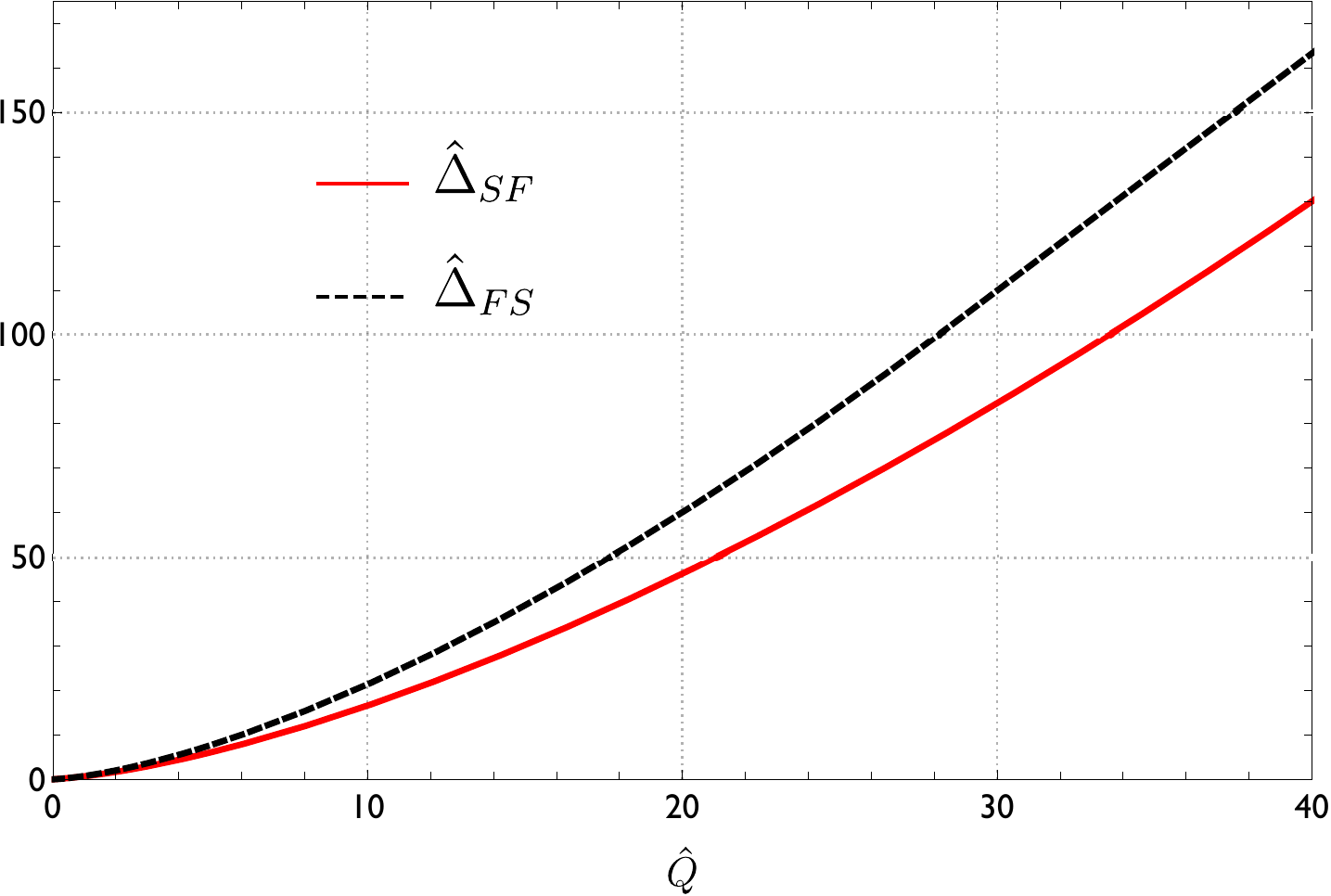}
\caption{Conformal dimension ${\Delta}_{FS}$ of the primary operator $\Opp_{FS}$ in the \acrshort{gn} model and the \(U(1)\)-\acrshort{njl} with fixed \(U(1)_B\) charge compared to the conformal dimension \({\Delta}_{SF}\) of the charged superfluid primary operator \(\Opp_{SF}\) in the \acrshort{njl} model at finite \(U(1)_A\)-charge.}
\label{fig:CGN_scaling}
\end{figure}

At the critical point, and at the limit \(\beta \to \infty\), the grand potential reads
\begin{align}
\frac{\Omega(\mu)}{N} &=  \frac{1}{(4 \pi R_0^2)}  \Sum_{j = \frac{1}{2}}^\infty (2j+1) \left\{  \Omega_+ + \Omega_-  \right\}, & \Omega_{\pm}^2 &= \abs{\Phi}^2 + (\omega_j \pm \mu)^2,
\end{align}
where \(\omega_j = (j+1/2)/R_0\) denote the eigenvalues of the Dirac operator on the two-sphere \(\setS^2\). As in \cref{sec:scalingGN}, we use \cref{eq:this_is_basic} to compute the charge and the canonical free energy, and using \cref{eq.energyandscaling} the latter corresponds to the conformal dimension of the scalar operator \(\Opp_{SF}\) as
\begin{align}
\frac{\Qp}{2N} &= \frac{1}{2}\Sum_{j=\frac{1}{2}}^\infty (2j+1)\left\{ \frac{\omega_j+\mu}{\Omega_+} - \frac{\omega_j-\mu}{\Omega_-}  \right\}, \label{eq:njlcharge} \\
\frac{\Delta_{SF}(\Qp)}{2N} &= - \frac{R_0}{2} \Sum_{j=\frac{1}{2}}^\infty (2j+1) \left\{ \Omega_+ + \Omega_- \right\}  + (\mu R_0) \frac{\Qp}{2N}.
\label{eq:NJL_scalingdim}
\end{align}
Furthermore, we need to evaluate the Stratonovich field \(\Phi\) on the solution \(\Phi_0\) of the gap equation,
\begin{equation}
G \coloneqq \frac{1}{2 R_0} \left. \Sum_{j=\frac{1}{2}}^\infty (2j+1) \left\{ \frac{1}{\Omega_+} + \frac{1}{\Omega_-}  \right\} \right|_{\Phi= \Phi_0} = 0.
\label{eq:gap_NJL}
\end{equation}
In the above equations, the expression of the charge is finite, but on the other hand, the expressions for the gap equation and the conformal dimension need to be regularised. One way to accomplish that, is to remove the leading divergence in the sums and then add them back using a zeta-function regularisation. This way, we obtain the following regulated forms
\begin{align}
G^{\text{reg}} &= \frac{1}{2} \Sum_{j = \frac{1}{2}}^\infty  \left\{ (2j+1) \frac{1}{R_0}\left[\frac{1}{\Omega_+} + \frac{1}{\Omega_-} \right] - 4   \right\}  + 2 \zeta(0), \\
\frac{\Delta^{\text{reg}}}{2N} &= - \frac{1}{2} \Sum_{j=\frac{1}{2}}^\infty \left\{ (2j+1) R_0 \left[ \Omega_+ + \Omega_- \right] - 4 R_0^2\omega_j^2 -2 R_0^2\Phi^2   \right\} - R_0^2 \Phi^2 \zeta(0) + (\mu R_0) \frac{\Qp}{2N}.
\end{align}
Given that the infinite sum is convergent now, we can numerically compute the preceding regulated expressions for different values of \(\Qp\) and the result is depicted in~\Cref{fig:CGN_scaling} along with the conformal dimension \(\hat{\Delta}_{FS}(\Qp) = \Delta_{FS}(\Qp)/2N\) for the primary operator \(\Opp_{FS}\) that appears in the \acrlong{gn} and the \(U(1)\)-\acrshort{njl} models with fixed \(U(1)_B\) charge.

We can also analytically compute the first terms in the asymptotic expansion of the conformal dimension \(\Delta_{SF}\) in the limits of large charge, \emph{i.e.} \(\frac{\Qp}{N} \gg 1\) and small charge, \emph{i.e.} \(\frac{\Qp}{N} \ll 1\). This will be done in the final two sections.

\subsubsection{Large-\(\Qp\) limit}%
\label{sec:largeQ}

For finite \(U(1)_A\) charge, we can deduce that there are three dimensionful quantities in our problem, \(\mu\), \(\Phi_0\) and \(R_0\) respectively, nevertheless in the equations of \cref{sec:conformalNJL} the only two dimensionless ratios that appear are \(R_0 \mu,\, R_0 \Phi_0\). Since in this section we care about the large-\(\Qp\) limit, the dominant scale is the chemical potential \(\mu\), satisfying \(R_0 \mu \gg 1\).

Therefore, we can write an expansion in inverse powers of \(\mu\) for \cref{eq:NJL_vev_flat}, that is the solution of the gap equation, which reads
\begin{equation}
	R_0 \Phi_0 = \sqrt{\kappa_0^2-1}\left(\mu R_0 + \frac{\kappa_1}{\mu R_0}+ \frac{\kappa_2}{(\mu R_0)^3}+\dots\right).
 \label{eq:ansatz}
\end{equation}
To determine the parameters \(\kappa_i\) we have to solve the gap \cref{eq:gap_NJL}, which we split into regular and divergent contribution as
\begin{align}
G & =	G_r + G_d = 0, \label{eq:gapfunctionsplit}\\
G_r &=  \Sum_{\ell=1} \ell \left[\frac{1}{\sqrt{(\ell + \mu R_0)^2+ (\Phi_0 R_0)^2}} + \frac{1}{\sqrt{(\ell - \mu R_0)^2+ (\Phi_0 R_0)^2}} - 2 \frac{1}{\sqrt{\ell ^2+ (\Phi_0 R_0)^2}} \right] ,\\
G_d &=  2 \Sum_{\ell=1} \ell  \frac{1}{\sqrt{\ell ^2+ (\Phi_0 R_0)^2}},
\end{align}
and we changed the sum in \cref{eq:gap_NJL} as \(j = \ell - 1/2\).

We can compute the regular part using the Euler-Maclaurin formula, and we write it as an expansion in \(\mu R_0 \gg 1\) which then reads
\begin{equation}
	G_r = 2 \mu R_0 \left[ - \kappa_0 + \sqrt{\kappa_0^2-1} + \arccoth(\kappa_0)\right] + \frac{1}{6 \mu R_0}\left[-\frac{1}{\kappa_0}-12\kappa_0\kappa_1 +\frac{1+12(\kappa_0^2-1)\kappa_1}{\sqrt{\kappa_0^2-1}}  \right]+\dots
\end{equation}
On the other hand, the divergent contribution can be computed using zeta-function regularisation \footnote{We explicitly checked that the derived  expansion agrees with the cutoff-independent contribution in smooth cut-off regularisation.} as
\begin{equation}\label{eq:divergentintegralnjl}
	G_d = 2 \eval*{ \Sum_{\ell=1}^\infty \ell (\ell^2+R_0^2\Phi_0^2)^{-s}}_{s=1/2}= \eval*{ \frac{2}{\Gamma(s)}\Int\limits_0^\infty \frac{\dd{t}}{t}t^s\Sum_{\ell=1}^\infty \ell e^{-(\ell^2+R_0^2\Phi_0^2) t} }_{s=1/2}.
\end{equation}
From \cref{eq:NJL_vev_flat} we can easily deduce that for \(\mu R_0 \gg 1\), we also have \(\Phi_0 R_0 \gg 1\). Then it is clear how to proceed, since we have performed a similar computation in \cref{sec:largeN}. The integral in \cref{eq:divergentintegralnjl} localises at \(t=0\)  for long enough values of \(\Phi_0 R_0\) and using the following expansion for the sum
\begin{equation}
\Sum_{\ell=1}^\infty \, \ell \, e^{-\ell^2t} = \frac{e^{-t}}{12} \, \left(2t + 5 +\frac{6}{t} + \dots \right),
\end{equation}
the problem reduces to the following expression
\begin{equation}
	G_d = -2 \Phi_0 R_0 - \frac{1}{6\Phi_0 R_0} - \frac{1}{120 (\Phi_0 R_0)^3} + \dots.
\end{equation}
Inserting the ansatz of \cref{eq:ansatz} in the divergent expression, it is possible to solve \cref{eq:gapfunctionsplit} order-by-order in \(R_0 \mu\) and then solve for the parameters \(\kappa_i\). Then, we observe that the solution for \(\kappa_0\) is given by \cref{eq:NJL_vev_flat} and, in fact, appears in every subsequent coefficient \(\kappa_{i >0}\)
\begin{align}
	\kappa_0 \tanh \kappa_0 &= 1, &
	\kappa_1 &= -\frac{1}{12 \kappa_0^2}, &
	\kappa_2 &= \frac{33-16 \kappa_0^2}{1440 \kappa_0^6}, & &\dots
 \label{eq:coeffs}
\end{align}
We repeat the same procedure to calculate the divergent sum appearing in the conformal dimension of \cref{eq:NJL_scalingdim}, therefore we divide it into regular and divergent part as
\begin{align}
	\Delta_r &= -2N\Sum_{\ell = 1} \ell
	\left[ \sqrt{(\ell +\mu R_0)^2 + \Phi_0^2R_0^2} + \sqrt{(\ell -\mu R_0)^2 +\Phi_0^2R_0^2} - 2\sqrt{\ell^2 + \Phi_0^2R_0^2}\right],\\
	\Delta_d &= - 4N \Sum_{\ell=1}^\infty \ell \sqrt{\ell^2 + \Phi_0^2R_0^2}.
\end{align}
Again, the regular part can be computed by applying the Euler-Maclaurin formula
\begin{equation}
	\Delta_r = -\frac{2 N}{3} (R_0\mu)^3 \pqty{ 3(\kappa_0^2-1) \arccoth{\kappa_0} + 3\kappa_0 - 2 \kappa_0^3 + 2 (\kappa_0^2-1)^{\frac{3}{2}} }+\dots,
\end{equation}
while the divergent part is computed again using the Mellin representation of zeta-function, and reads
\begin{equation}
	\Delta_d =  N \pqty{ \frac{4(\Phi_0 R_0)^3}{3} + \frac{\Phi_0 R_0}{3} - \frac{1}{60 \Phi_0 R_0} + \dots } .
\end{equation}
Finally, if we invert order-by-order the charge equation~\eqref{eq:njlcharge} we can acquire a relation for the chemical potential as a function of the charge, \emph{i.e.} \(\mu = \mu (\Qp)\) and since the summation is convergent we just use the Euler--Maclaurin formula. 

Eventually, by combining the results of \cref{eq:ansatz,eq:coeffs} along with the relation \(\mu = \mu (\Qp)\), we deduce the asymptotic expansion for the conformal dimension \(\Delta_{SF} (\Qp\) as:
\begin{equation}
  \frac{\Delta_{SF}(\Qp)}{2N} = \frac{2}{3} \pqty{\frac{\Qp}{2N \kappa_0}}^{3/2} + \frac{1}{6} \pqty{\frac{\Qp}{2N \kappa_0}}^{1/2} + \frac{11-6\kappa_0^2}{720\kappa_0^{2}} \pqty{\frac{\Qp}{2N \kappa_0}}^{-1/2} + \dots
\end{equation}
Observe that the leading term in the expansion matches the expression that we had computed in the torus for the energy of the ground state~\eqref{eq:flat_space_NJL}. So, we come to the same conclusion as in \cref{Chapter4}, that the subleading terms in the computation in the large-charge expansion in \(\setS^2\) are due to an expansion in the curvature and the leading term relies solely on the manifold's volume, which for \(\setS^2\) is \(V= 4\pi r_0^2\).

On a relevant note, the conformal Goldstone mode that we computed in~\cref{sec:fluctuations} is independent of \(N\) and contributes with a universal term to the conformal dimension at order $N^0$, $\Qp^0$. We can compute it numerically and its value agrees completely with the result in the \(O(2)\) and \(O(N)\) models that we saw in \cref{Chapter3,Chapter4} respectively and is given in \cref{eq:scalingON}.

\subsubsection{Small-\(\Qp\) limit}%
\label{sec:smallQ}

Finally, we want to examine the small charge limit, $\Qp/N \ll 1$. For zero values of the charge, it is apparent that the free energy vanishes due to conformal invariance, given the fact that it corresponds to the scaling dimension of the identity operator. In our representation with regard to the scalar field \(\Phi\), we have to consider its conformal coupling \(\xi\) to the curved
background. For the cylinder \(\setR \times \setS^2\) this corresponds to a mass term \(m = 1/(2R_0)\) \footnote{A similar term we have found in \cref{Chapter4} regarding the conformal Laplacian, which was shifted by a mass term \( m^2 =1/4R_0^2\).} that for us arises from the chemical potential~\cite{alvarez2019large}.

By computing the free energy and the charge for \(\mu =1/(2R_0)\), they both vanish and also the gap equation on \(\setS^2\) is satisfied with a zero \acrshort{vev}, $\Phi_0=0$. Therefore, the value of \(\mu =1/(2R_0)\) is the correct one to perform the small-charge expansion around, and for convenience, we write the chemical potential as this critical value plus a small deviation, \emph{i.e.} \(\mu = 1/ (2R_0) + \hat{\mu}\), and then we expand every expression in \cref{eq:NJL_scalingdim} around the small \(\hat \mu\). When \(\hat{\mu} = 0\) we have symmetry restoration, \(\Phi_0\) does not get a \acrshort{vev} and using this we can write an ansatz for \(\Phi_0\) as
\begin{equation}
	\hat \mu = \mu_2 \Phi_0^2 R_0 + \mu_4 \Phi_0^4  R_0^3 + \dots 
 \label{eq:ansatz_2}
\end{equation}
We can compute the charge by expanding in the limit \(\Phi_0 R_0 \ll 1\), and we find
\begin{equation}
	\frac{\Qp}{2N} = \frac{\pi^2}{4} (\Phi_0R_0)^2 - \frac{\pi^2}{16}(\pi^2 - 16\mu_2) (\Phi_0R_0)^4 
 + \frac{\pi^2}{48} \left( \pi^4 + 12 \pi^2 (\mu_2^2 - 2 \mu_2) + 48 \mu_4   \right) (\Phi_0 R_0)^6 + \dots,
\end{equation}
which obviously vanishes for \(\Phi_0 = 0\). To determine the coefficients \(\mu_{i}\) in the ansatz of \cref{eq:ansatz_2} we have to solve order-by-order in small $\hat{\mu}$ the gap~\cref{eq:gap_NJL}. Thus, we separate it again into a divergent and a convergent contribution --- by first removing the expression for $\hat\mu = 0$ and then adding it again --- with each of them having a proper expansion in the limit $\Phi_0 R_0 \ll 1$. The convergent part reads   
\begin{equation}
	G_r  = \frac{\pi^2}{2} \mu_2 \pqty{\Phi_0R_0}^2 - \frac{\pi^2}{4}\pqty{\pi^2\mu_2 - 4 \mu_2^2 - 2\mu_4} \pqty{\Phi_0R_0}^4+\dots,
\end{equation}
while for the divergent part we use the zeta-function regularisation to get 
\begin{align}
	G_d &= \Sum_{\ell = 0}^\infty (2\ell+1)\frac{1}{\sqrt{(\ell +\frac{1}{2})^2+ (\Phi_0 R_0)^2}} = 2 \Sum_{\ell=0} (\ell+\frac{1}{2}) \Sum_{k=0} \binom{-\sfrac{1}{2}}{k}(\Phi_0R_0)^{2k}(\ell+\frac{1}{2})^{2(-1/2-k)} \nonumber \\
  &= 2 \Sum_{k=0} \binom{-\sfrac{1}{2}}{k}(\Phi_0R_0)^{2k} \zeta(2k; \frac{1}{2}) = -\frac{\pi^2 (\Phi_0R_0)^2}{2} + \frac{\pi^4 (\Phi_0R_0)^4}{8} + \dots.	
\end{align}
To determine all the parameters $\mu_i$, we solve the gap \cref{eq:gapfunctionsplit} order-by-order and we find
\begin{align}
	\mu_2 &= 1, & \mu_4 &=\frac{\pi^2-8}{4}, &&\dots
\end{align}
Similarly to the large-\(\Qp\) expansion, we use the same methodology to evaluate the divergent part that appears in the conformal dimension~\eqref{eq:NJL_scalingdim} by dividing it into convergent and divergent parts
\begin{align}
  &\begin{multlined}[][\arraycolsep]
    \Delta_r = -2N\Sum_{\ell = 1} \ell 
    \left[ \sqrt{\pqty{\ell + \frac{1}{2} +\hat\mu R_0}^2+\pqty{\Phi_0 R_0}^2} + \sqrt{\pqty{\ell - \frac{1}{2} - \hat\mu R_0}^2+\pqty{\Phi_0 R_0}^2} \right.\\
    - \left.\sqrt{\pqty{\ell + \frac{1}{2}}^2+\pqty{\Phi_0 R_0}^2} + \sqrt{\pqty{\ell - \frac{1}{2}}^2+\pqty{\Phi_0 R_0}^2} \right],
  \end{multlined}
  \\
&\Delta_d = - 4N \Sum_{\ell=1}^\infty \pqty{\ell+ \frac{1}{2}} \sqrt{\pqty{\ell+\frac{1}{2}}^2+ \pqty{\Phi_0 R_0}^2}.
\end{align}
The regular part is expanded for $\Phi_0 R_0 \ll 1$ and reads
\begin{equation}
\Delta_r = N \pqty{ \frac{\pi^2}{2}\mu_2 (\Phi_0 R_0)^4 + \dots } \, ,
\end{equation}
while for the divergent part we use zeta-function regularisation to obtain
\begin{align}
\Delta_d &= -4 N \Sum_{\ell = 0}^\infty \pqty{\ell+ \frac{1}{2}} \Sum_{k=0} \binom{\sfrac{1}{2}}{k}\pqty{\Phi_0 R_0}^{2k} \pqty{\ell+\frac{1}{2}}^{2(1/2-k)} =- 4 N \Sum_{k=0} \binom{\sfrac{1}{2}}{k}\pqty{\Phi_0 R_0}^{2k} \zeta(2k-2; \frac{1}{2}) \nonumber\\
&=  N \frac{\pi^2}{4}\pqty{R_0 \Phi_0}^4+\dots
\end{align}
Finally, by simply inverting $\Qp= \Qp(\Phi_0)$:
\begin{equation}
	R_0\Phi_0 = \frac{2}{\pi} \pqty{\frac{\Qp}{2N}}^{1/2}+\frac{\pi^2-16}{\pi^3} \pqty{\frac{\Qp}{2N} }^{3/2}+\dots \, ,
\end{equation}
we can compute the scaling dimension in the small-\(\Qp\) limit as
\begin{equation}
	\frac{\Delta(\Qp)}{2N}  = \frac{1}{2}\left(\frac{\Qp}{2N}\right) +\frac{2}{\pi^2}\left(\frac{\Qp}{2N} \right)^2+ \dots
\end{equation}
Given that $\Phi$ has charge two, the leading-order contribution $\Delta(\Qp) = \Qp/2$ comes to no surprise, as it is the anticipated result for the charged operator $\Phi^{\Qp/2}$ in the free-field limit. Moreover, at leading order, the result is \(N\) independent, so we conclude that the leading order solution will apply to the \(SU(2)\)-\acrshort{njl} case as well.

\bigskip

As a final remark, in this chapter we examined various fermionic models with four-fermion interaction terms in \(d=3\) spacetime dimensions at large charge and large \(N\), and specifically, we studied the \acrlong{gn} model, the \acrlong{njl} model and its \(SU(2)_L \times SU(2)_R\) generalisation. 

As a result of the fermionic nature of our models, we observe that the ground state at large-charge exhibits two distinct physical descriptions, contingent upon whether the examined model possesses a symmetry that is raised to an axial symmetry in \(d=4\) or not. If this does not happen, which is the case of the \acrlong{gn} model, then the sector of fixed large charge is described in terms of a Fermi surface. On the contrary, if such a symmetry exists, fixing the axial charge leads to a conformal superfluid description of the ground state similar to the bosonic cases studied in detail in the literature~\cite{hellerman2015cft,monin2017semiclassics,alvarez2019large,Gaume:2020bmp}. The \acrlong{njl}-type of models exhibit such a behaviour and there is a physically intuitive way that explains it. For the case of the \(U(1)\)-\acrshort{njl} model, we can perform a \acrlong{pg} transformation and map it to the Cooper model. The existence of an attractive interaction gives rise to a Cooper instability, and the formation of Cooper pairs produces a bosonic condensate, leading to the Fermi's sphere breakdown. 

Finally, in the models that the large-charge ground state coincides with the conformal superfluid paradigm, the scaling dimension of the lowest charged primary operator of the theory has different numerical coefficients than the bosonic case~\cite{alvarez2019large}, indicating that the fermionic \acrshort{cft} lies in a different universality class. Meanwhile, we specifically verified that the spectrum of fluctuations over the large-charge ground state contains the anticipated conformal Goldstone mode and therefore, the conformal dimension exhibits the universal \(\Qp^0\) contribution corresponding to the Casimir energy of the fluctuations.


\appendix 



\chapter{Preliminaries for QFTs, CFTs and LCE } 

\label{AppendixA} 

\section{Elements of Differential Geometry }\label{sec.differentialgeometry}

This section follows closely \cite{wald2010general}. One of the first things we want to define in a precise mathematical formulation is the concept of \textit{manifolds}. Vaguely speaking, a manifold is a space that looks locally like \(\mathbb{R}^n\) but may be curved and have different global properties. The physical reason to define a manifold as explained in \cite{CarrollSean2014Saga} is our inability to define a global reference frame throughout the whole spacetime. A precise definition of manifolds is useful because, through that, we can define \textit{tangent vectors} as \textit{directional derivatives} acting on some arbitrary functions defined on that manifold.

\textbf{\large{Definition(\textit{Manifold})}}: An \(C^\infty\) \(d\)-dimensional real Manifold \(\mathcal{M}\) is simply a set along with a collection of subsets \({\lbrace{\mathcal{U}_\alpha}\rbrace}\) that should satisfy certain properties:
\begin{enumerate}
\item Each point  \(p \in \mathcal{M}\) belongs to at least one subset $\mathcal{U}_\alpha$ , while the collection of subsets ${\lbrace{\mathcal{U}_\alpha}\rbrace}$ covers $\mathcal{M}$ ; $\mathcal{M}$ = $\bigcup_{\substack{\alpha}}$ $\mathcal{U}_\alpha$.
\item For each $\alpha$, exists a one-to-one, onto, map $\phi_\alpha :\mathcal{U}_\alpha \rightarrow \phi_\alpha (\mathcal{U}_\alpha) $ where $\phi_\alpha (\mathcal{U}_\alpha) $ is an open subset of $\mathbb{R}^d$.
\item Finally, the sets should be smoothly sewn together. If any two sets $\mathcal{U}_\alpha$ and $\mathcal{U}_\beta$ overlap in such a way that $\mathcal{U}_\alpha \bigcap \mathcal{U}_\beta \neq \varnothing $ then, there should be a map $\phi_\alpha \circ {\phi_\beta }^{-1}$ that should take points from $\phi_\beta[\mathcal{U}_\alpha \cap \mathcal{U}_\beta] \subset \phi_\beta (\mathcal{U}_\beta) \subset \mathbb{R}^d $ to certain points in $\phi_\alpha[\mathcal{U}_\alpha \cap \mathcal{U}_\beta] \subset \phi_\alpha(\mathcal{U}_\alpha) \subset \mathbb{R}^d $. Moreover, there is the requirement that these subsets of $\mathbb{R}^d$ are open, and that the map is infinitely continuously differentiable.
\end{enumerate}
The most usual word used by physicists for maps $\phi_\alpha$ is coordinate system.

By having a definition of manifolds, we want to move on to vectors. We are mostly used from Euclidean Geometry to add vectors together. But when someone takes under consideration curved geometries, such concepts are no longer valid because at this point we lose the ability to move vectors uniquely around a manifold while leaving them “unchanged”. Fortunately, there is a limit where the vector-space structure notion can be recovered. It's the limit of \textit{tangent vectors} or “infinitesimal displacements” about a point. We note that for certain manifolds, for example the sphere, that can be conceived as objects embedded in $\mathbb{R}^d$, it is not hard to imagine the tangent vector at a point $p$ by drawing the tangent plane in which this vector belongs to.
This can be generalized to every manifold embedded into a \(d\)-dimensional Euclidean space, but since not always a manifold can be embedded in a higher dimensional space, we want to avoid defining our tangent vectors in a way that refers to the embedding of our manifold in $\mathbb{R}^d$, and we want to give a definition that is independent of coordinates. 

In $\mathbb{R}^d$ exists a one-to-one correspondence between vectors and directional derivatives. In such a way we define the \textit{tangent vector}, by considering only the intrinsic structure of the manifold without thinking about the embeddings, as a \textit{directional derivative}. The main concept is that directional derivatives are characterized by their linearity and “Leibniz rule” behaviour. \\
Let \(f : \mathcal{M} \rightarrow \mathbb{R}\) be a smooth function on \(\mathcal{M}\). Then we define \(\mathcal{F}\) to be the collection of all smooth functions on \(\mathcal{M} \ (C^{\infty} (\mathcal{M}) \rightarrow \mathbb{R})\).\\
\textbf{\large{Definition(\textit{Tangent vector})}}: We define a tangent vector $V$ at a point $p \in \mathcal{M}$ to be a map $V : \mathcal{F} \rightarrow \mathbb{R}$ which satisfies the following properties

\begin{itemize}
    \item Linearity : \(
      V(af+bg)= aV(f)+bV(g),  \quad \forall \quad f,g \in \mathcal{F} ,\quad a,b \in \mathbb{R}\),
    \item Leibniz rule : $V(fg)=f(p)V(g)+g(p)V(f)$.
\end{itemize}
Furthermore, the set of all maps from $C^\infty(\mathcal{M}) \rightarrow \mathbb{R}$ to a point $p \in \mathcal{M}$ defines the tangent space $\mathcal{T}_p(\mathcal{M})$ which is a \(d\)-dimensional vector space, isomorphic to $\mathbb{R}^d$. We can define a set of basis vectors for any tangent space to be given by $\partial_\mu(p)=\frac{\partial}{\partial x^\mu}(p) \in \mathcal{T}_p(\mathcal{M}) $ , where $x^\mu$ is the local coordinate system, so any vector $V \in \mathcal{T}_p(\mathcal{M})$ can be written as :
\begin{equation}
  V=V^\mu \partial_\mu.
\end{equation}
For completeness, we can define a vector field as a set of vectors with exactly one at each point in spacetime.


\section{Curved space QFT}\label{sec.curvedQFT}

In principle, it is easy to generalize theories from flat to curved space. The only thing we should do is write them in a coordinate-invariant way, and make sure they stay valid when spacetime is curved. This section is inspired by Gibbons~\cite{hawking1980general} \S 13 and \cite{Wald:1995yp}.

A general quantum field theory on an arbitrary fixed background consists of the following :
\begin{enumerate}
    \item A \textit{Hilbert space} $\mathscr{H}$\\
    \item A classical spacetime manifold $(\mathcal{M}, g_{\mu\nu})$\\
    \item Field operators $\big\{ \phi \big\}$ that act on $\mathscr{H}$ and are defined on $\mathcal{M}$ as tensorial (\textit{bosonic}) distributions. (We can, of course, include fermions).\\
    \item The fields $\big\{ \phi \big\}$ satisfy wave equations which, in the case of no mutual or self interactions, are linear between the particles.\\
    \item The fields satisfy a \textit{“field algebra”}. (commutation relations for the operators).\\
    \item There exist rules on how \textit{Fock} bases on $\mathscr{H}$ or physical observables can be constructed.\\
    \item “Renormalisation” or “regularisation” procedures.\\
\end{enumerate}
So first thing first, we assume that every physical quantity corresponds to a self adjoint linear map acting on the Hilbert space (observable). Then physical systems can either be pure states or mixed states that are defined through density matrices $(\rho)$. Density matrices are embedded with an extra degree of uncertainty beyond that of quantum mechanics. Then our manifold should be able to pose a Cauchy problem so  $(\mathcal{M}, g_{\mu\nu})$ should be globally hyperbolic and in general time orientable. Space orientability is useful but not necessary. The action on a curved space for a scalar field is given by :\\
\begin{equation}\label{5.1.1}
    S= -\frac{1}{2} \int d^d(x) {\sqrt{-g}}(g^{\mu\nu} \nabla_{\mu}\phi \nabla_{\nu}\phi +m^2 \phi +\xi R \phi^2)
\end{equation}\\
The equation of motion for this scalar field is :\\
\begin{equation}\label{5.1.2}
   \Box \phi -m^2 \phi -\xi R \phi = 0    
\end{equation}\\
$\xi$ is a direct coupling between the field and the curvature scalar $R$. For simplicity, from this point on we assume that we have minimal coupling so we take : \\
\begin{equation}\label{5.1.3}
    \xi = 0.
\end{equation}\\
Another popular choice would be to adopt the \textit{conformal coupling} so in that case we would have that :\\
\begin{equation}
    \xi = \frac{(d-2)}{4(d-1)}.
\end{equation}\\
If we assume that we have a manifold $(\mathcal{M}, g_{\mu\nu})$ which is globally hyperbolic it can be proved that it can be \textit{foliated} into a set of Cauchy hypersurfaces $\Sigma$ in such a way that :\\
\begin{equation}
    \mathcal{M}= \mathbb{R} \times \Sigma,
\end{equation}\\
where $\mathbb{R}$ is the time direction and $\Sigma$ is a $d-1$ Riemannian manifold. In that case we proceed like in flat space. Denoting that $x^{\mu}=(x^0 , \vec{x})$ we can now write the action as follows : \\
\begin{equation}
    S = \int d^d x \Lp.
\end{equation}\\
Where $\Lp$ is the Lagrangian density. The conjugate momentum is as usual :\\
\begin{equation}
    \pi(\vec{x})=\frac{\partial \Lp }{\partial(\nabla_0 \phi)} = -\sqrt{-g}g^{\mu 0}\nabla_{\mu}\phi= - \sqrt{h} \eta^{\mu} \nabla_{\mu}\phi(\vec{x}).
\end{equation}\\
Where $h_{ij}$ is the induced metric on $\Sigma$ and $\eta^{\mu}$ is the vector normal to $\Sigma$. The quantization procedure continues by the imposition of canonical commutation relations such that :\\
\begin{equation}
    \begin{split}
        & \Big[ \phi(\vec{x}), \pi(\vec{x})  \Big] = i \hbar \delta^{d-1}(\vec{x},\vec{y}), \\
        & \Big[ \phi(\vec{x}),\phi(\vec{y}) \Big] = 0, \\
        & \Big[ \pi(\vec{x}) , \pi(\vec{y}) \Big] = 0. \\
    \end{split}
\end{equation}\\
And we define $\delta^{d-1}(\vec{x},\vec{y})$ as :\\
\begin{equation}
    \int_{\Sigma} d^{d-1} \vec{x} f(\vec{y}) \delta^{d-1}(\vec{x},\vec{y}) = f(\vec{x}).
\end{equation}\\
Up to this step, we assumed we were on a local system, so the global spacetime structure was irrelevant for the commutators. By the time that we wish to represent these commutators utilizing operators acting on a Hilbert space, the structure of the background spacetime metric enters the definition of the inner product and becomes important. The inner product is then expressed as :\\
\begin{equation}\label{5.1.10}
    (f,g) = \int_{\Sigma} d{\Sigma_{\mu}}{\mathcal{J}}^{\mu}(f,g) .
\end{equation}\\
Where :\\
\begin{equation}\label{5.1.11}
    {\mathcal{J}}^{\mu}(f,g)= - i \sqrt{-g}g^{\mu\nu}(f^{*}\nabla_{\nu}g-(\nabla_{\nu}f^{*})g).
\end{equation}\\
Implying the EOM $\nabla_\mu \mathcal{J}^{\mu} = 0$ so the current is conserved and the inner product $(f,g)$ is independent of the choice of the spacelike slice $\Sigma$. Using eqs. \ref{5.1.10} and eqs. \ref{5.1.11} we can clearly see that :\\
 \begin{equation}\label{5.1.12}
     (f,g)^* = -(f^*,g^*) = (g,f).
 \end{equation}\\
 Because : \\
 \begin{equation}\label{5.1.13}
     (j^{\mu}(f,g))^{*} = i \sqrt{-g}g^{\mu\nu}(f\nabla_\nu g^* - (\nabla_\nu f)g^*)= -i\sqrt{-g}g^{\mu\nu}(g^*\nabla_\nu f - f\nabla_\nu g^*) = j^{\mu}(g,f)
 \end{equation}\\
 Furthermore from \ref{5.1.12} we got that for $g = f^*, -(f^*,f) = (f^*,f) $. So clearly $(f,f^*) = 0$ so the inner product is not positive definite. At this point in the flat space we would introduce a set of positive and negative frequency modes and then expand the field operator $\phi$ in terms of these modes. But since in general there is no timelike killing vector, for an abstract spacetime, there is no natural a priori notion of positive frequency. In order to overcome this problem we decompose the space of solutions into a positive part and its conjugate 
 \begin{equation}
     S = S_p \oplus S_{p}^*
 \end{equation}
 And we demand that
 \begin{align}\label{5.1.15}
     &(f,f) > 0 \qquad \forall f  \in S_p, & 
     &(f,g^*) = 0 \qquad \forall  f,g  \in S_p
 \end{align}
 Now we can define a set of creation and annihilation operators for $f$ as
 \begin{align}\label{5.1.16}
     a(f) &= (f,\phi), & a^{\dagger}(f) &= -a(f^*) = -(f^*,\phi).
 \end{align}
 Using the canonical commutation relations we can find that these operators satisfy the following algebra
 \begin{align}\label{5.1.17}
     [a(f),a^{\dagger}(g)] &= (f,g),& [a(f),a(g)] &= [a^{\dagger}(f),a^{\dagger}(g)] = 0.
 \end{align}
 So we can find a vacuum state annihilated by the annihilation operators:\\
 \begin{equation}\label{5.1.18}
     a(f)\vert 0\rangle = 0 \phantom{X} \forall f \in S_p
 \end{equation}\\
 And for this vacuum state we can construct a Fock space by repeated action of creation operators of the form :\\
 \begin{equation}\label{5.1.19}
     a^{\dagger}(f_{i_1})... a^{\dagger}(f_{i_\kappa})\vert 0 \ \rangle,
 \end{equation}\\
 for all n and all $f_{i_\kappa} \in S_p$ \\
 From this point on, given a positive norm subspace we can construct an orthonormal basis $f_n\in S_p$. Using that we rewrite the creation and annihilation operators in that base as :\\
 \begin{equation}\label{5.1.20}
     a_n=(f_n,\phi), \phantom{X} a^{\dagger}_n= (-f_{n}^*,\phi), \phantom{X}a_n a_m^{\dagger}- a_m^{\dagger}a_n = \delta_{nm}
 \end{equation}\\
 With that in mind we can expand our field operator in the usual mode decomposition :\\
 \begin{equation}\label{5.1.21}
    \phi = \Sum\limits_n(a_n f_n + a_{n}^{\dagger} f_{n}^{*})
 \end{equation}\\
 In this basis we can write the vacuum state as $\vert 0_f \rangle$ and we demand :\\
 \begin{equation}\label{5.1.22}
     a_n \vert 0_f \rangle = 0, \phantom{X} \forall n
 \end{equation}\\
 The usual $n$-particle basis of states is :\\
 \begin{equation}\label{5.1.23}
     a_{n_1}^{\dagger}...a_{n_\kappa}^{\dagger}\vert 0_f \rangle.
 \end{equation}\\
 And so a state with $\omega_n$ excitations can be written as :\\
 \begin{equation}\label{5.1.24}
     \vert \omega_n \rangle = \frac{1}{\sqrt{\omega_n!}}{(a_n)}^{\omega_n}\vert 0_f \rangle
 \end{equation}\\
 We can even define a number counting operator as :\\
 \begin{equation}\label{5.1.25}
     N_{f_n}=a_{n}^{\dagger}a_n
 \end{equation}\\
 We should note that we are careful, and we should remember that these are defined regarding the orthonormal basis $f_n$. For later use for a stationary spacetime, we resolve $\phi$ into normal modes as:\\
 \begin{equation}\label{5.1.26}
     \phi (t,\Vec{x}) = \Sum\limits_n\Bigg( \frac{a_n \overline{f}_n(\Vec{x})}{\sqrt{2\omega_n}}e^{-i\omega_n t} + \frac{a_n^{\dagger} \overline{f}_n^*(\Vec{x})}{\sqrt{2\omega_n}}e^{i\omega_n t}\Bigg)
 \end{equation}\\ 
 where $f_n$ obeys:\\
 \begin{equation}\label{5.1.27}
     \nabla^{2}_f \overline{f}_n + \partial_\mu g_{00} h^{\mu\nu}\partial_\nu \overline{f}_n + m^2 \sqrt{g_{00}} \overline{f}_n = \Lp\overline{f}_n = -\omega_n^2\overline{f}_n
 \end{equation}\\
 Where the first term is the Laplace-Beltrami operator.


\section{KMS}\label{sec.KMS}

The expectation value for any operator \(\mathscr{O}\) in thermal equilibrium at a temperature \(T = 1/{\beta}\) is

\begin{equation}\label{5.3.9}
    {\langle \mathscr{O} \rangle}_\beta = \frac{1}{Z} \text{tr}{\big(e^{-\beta H}\mathscr{O}\big)}.
\end{equation}
\\
In some cases, this definition cannot be defined properly, for example in most \acrshort{qfts} the average of a single operator is normalized to zero. Yet, a generalization is possible by considering the expectation value of a couple of time dependant operators, $\mathscr{\Tilde{O}}(t)$ and $\mathscr{O}(0)$ as

\begin{equation}
    \big\langle  \mathscr{\Tilde{O}}(t) \mathscr{O}(0)  \big\rangle = \big\langle  e^{itH} \mathscr{\Tilde{O}} e^{-itH} \mathscr{O}  \big\rangle.
\end{equation}
Using \cref{eq.Wickrotation}and applying \cref{5.3.9} we get

\begin{align}
    {\langle  \mathscr{\Tilde{O}}_{-i\beta} \mathscr{O}  \rangle}_{\beta} & = \frac{1}{Z}\text{tr}\left({e^{-\beta H}\left(e^{\beta H} \mathscr{\Tilde{O}} e^{-\beta H}\right)\mathscr{O}}\right) \nonumber \\
    & =  \frac{1}{Z}\text{tr}\left( \mathscr{\Tilde{O}} e^{-\beta H} \mathscr{O}\right) = \frac{1}{Z}\text{tr}\left(e^{-\beta H} \mathscr{O} \mathscr{\Tilde{O}} \right)\\
    & = {\langle  \mathscr{O} \mathscr{\Tilde{O}}\rangle}_{\beta}, \nonumber
\end{align}
where the cyclicity of the trace has been used and also $\mathscr{\Tilde{O}}$ and $\mathscr{O}$ are taken to be bounded. Hence, we have at an expression that only contains finite quantities, and it is known as the \textit{KMS} condition

\begin{equation}\label{KMS}
    {\big\langle \mathscr{\Tilde{O}}_{-i\beta} \mathscr{O}\big\rangle}_{\beta} = {\big\langle  \mathscr{O}\mathscr{\Tilde{O}} \big\rangle}_{\beta}.
\end{equation}
\\
Thermal states have to obey the KMS condition, which is a nice measure to check if a system is thermal or not. 


\section{Classification of renormalisability and running of the mass. }\label{sec.renormalisable}

To classify theories, we need to know the mass dimension of their coupling constants. We will briefly show how this works in the case of the \(\phi^4\) theory with the Lagrangian of \cref{eq.phi^4langragian}
\begin{equation*}
    \mathscr{L} = -\frac{1}{2} \del^\mu \phi \del_\mu \phi - \frac{1}{2} m^2 \phi^2 - \frac{g}{4!} \phi^4.
\end{equation*}
We start with the following mass dimensions: \(\bqty{E} = 1 = \bqty{t^{-1}} = \bqty{x^{-1}}\). Then, given the form of the Lagrangian, we can compute
\begin{align}
    0 & = \bqty{S}, & S = \Int \dd[d]{x} \Lp, \\
    d & =\bqty{\Lp}, \\
    1 & = \bqty{\pdv{x^\mu}}, \\
    \frac{d-2}{2} & = \frac{1}{2} \pqty{\bqty{\Lp}- 2\bqty{\pdv{x^\mu}}} = \bqty{\phi},\\
    2 & = d- (d-2) = \bqty{\Lp} - 2 \bqty{\phi} = \bqty{m^2}, \\
    d-p\pqty{\frac{d-2}{2}} & = \bqty{\Lp} - p \bqty{\phi} = \bqty{g_p}.\label{eq:massdimensioncoupling}
\end{align}

Therefore, theories are classified as follows
\begin{enumerate}
    \item Non-renormalisable. \newline
    At least one \(g_p\) in the theory has \(\bqty{g_p}<0\). Any \(n\)-point function contains \(\infty\) many divergences.
    \item Super-renormalisable \newline
    All \(\bqty{g_p} > 0\), finite number of divergent diagrams. 
    \item Renormalisable \newline
    No \(g_p\) in the theory has \(\bqty{g_p}<0\) and there is at least one \(g_p\) that satisfies \(\bqty{g_p} =0\). Then there is a finite number of \(n\)-point functions that contain \(\infty\) many divergencies. 
\end{enumerate}

\subsection{Running of the mass}

From Källén–Lehmann spectral representation --- \cite[see][]{Kallen:1952zz,Lehmann1954berEV} --- the exact propagator has an isolated pole at \( p^2 = - m_{\textrm{phys}}^2 \), e.g.
\begin{equation}
    \eval{\pqty{G^\pqty{2}\pqty{p,-p}}^{-1}}_{p^2 = - m_{\textrm{phys}}^2} = \eval{\Gamma^\pqty{2}\pqty{p,-p}}_{p^2 = - m_{\textrm{phys}}^2}=0.
\end{equation}
From \cref{eq.exactpropagator} this means that
\begin{align}\label{eq.physicalmass}
   0& = - m_{\textrm{phys}}^2 + m^2 - \underbrace{\eval{\Pi_{\textrm{\acrshort{ms}}}\pqty{p^2}}_{p^2 = - m_{\textrm{phys}}^2}}_{\order{g^2}} \Rightarrow \nonumber \\
    m_{\textrm{phys}}^2 & = m^2 - \eval{\Pi_{\textrm{\acrshort{ms}}}\pqty{p^2}}_{p^2 = - m_{\textrm{phys}}^2}  + \order{g^4}
\end{align}
The physical mass cannot depend on \(\mu\) thus
\begin{equation}\label{eq.physicalmassdepend}
    0 = \pdv{\log m_\textrm{phys}}{\log \mu}.
\end{equation}
But the \( \eval{\Pi_{\textrm{\acrshort{ms}}}\pqty{p^2}}_{p^2 = - m_{\textrm{phys}}^2}\) is given by \cref{eq.selfenergyresult} and it depends both on \(m^2\) and on \(\mu\), hence using \cref{eq.physicalmass,eq.physicalmassdepend} we can derive the anomalous mass dimension \(\frac{\mu}{m} \pdv{m}{\mu} =\pdv{\log m}{\log \mu} \) of \cref{eq.gammamfunctiong}.


\section{CFT}

\subsection{Infinitesimal generators}

\paragraph{Form of special conformal transformation}
We show how special conformal transformation can be seen as inversion-translation-inversion. We start with an inversion,
\begin{equation}
x{'}^\mu = \frac{x^\mu}{x^2}.
\end{equation}
Then we perform a translation
\begin{equation}
	x{''}^\mu = x{'}^\mu - a^\mu.
\end{equation}
Finally we perform another inversion, such that
\begin{equation}
x^{*\mu} = \frac{{(x{''})}^\mu}{(x{''})^2}.
\end{equation}
Combining the above equations we get
\begin{align}
 x^{*\mu} & = \frac{x{'}^\mu - a^\mu}{{(x{''})}^2} 
 = \frac{\frac{x^\mu}{x^2} - a^\mu}{{(x{''})}^2}
= \frac{{x^\mu} - a^\mu {x^2}}{{x^2}{(x{''})}^2} \nonumber\\
& = \frac{{x^\mu} - a^\mu {x^2}}{{x^2}\Big(\frac{x^\mu}{x^2}- a^\mu\Big)\Big(\frac{x_\mu}{x^2}- a_\mu\Big)} 
 = \frac{{x^\mu} - a^\mu {x^2}}{{x^2}\Big(\frac{x^\mu x_\mu}{x^4} - \frac{x^\mu a_\mu}{x^2}- \frac{x_\mu a^\mu}{x^2} + a^\mu a_\mu \Big)} \nonumber\\
&= \frac{{x^\mu} - a^\mu {x^2}}{1-2ax+a^2 x^2},
\end{align}
where we used that $x^2={x^\mu}{x_\mu}$, and also that $a^\mu x_\mu = a_\mu x^\mu = ax$ and that $a^\mu a_\mu = a^2$. 

\paragraph{Infinitesimal generators}

\begin{enumerate} 
	\item \emph{We work out the infinitesimal generator starting from the finite form of the scale transformation, 
	\begin{equation}
			x'^\mu = \alpha x^\mu.
\end{equation}} \newline
We start by parameterizing $\alpha = e^a, \ a \in \mathcal{R}$, so that
\begin{equation}
		{x'}^{\mu} =  e^a x^\mu.
\end{equation}
Then we Taylor expand to first order, i.e. $\order{a^2}$, to obtain the infinitesimal transformation
\begin{equation}
		{x'}^{\mu} = x^\mu + a x^\mu.
\end{equation}
Then we find the generator as
\begin{equation}\label{Generator}
	G_a = - i \bigg( \frac{\delta x^\mu}{\delta \epsilon_a}\partial_\mu \bigg)+ i \frac{\delta \Phi}{\delta \epsilon_a}
\end{equation}
In which case we get $\frac{\delta x^\mu}{\delta a} = x^\mu $, so that 
\begin{equation}
		G_a = -i x^\mu {\partial}_\mu \equiv D
\end{equation}
which is the generator of dilatation/scale transformations.
		
\item \emph{We work out the infinitesimal generator starting from the finite form of the SCT, 
	\begin{equation}\label{sct}
			x'^\mu = \frac{x^\mu - b^\mu \vec{x}^2}{1- 2\vec{b}\cdot \vec{x} + b^2 \vec{x}^2 }.
\end{equation}}
We start by writing this as 
\begin{equation}
{x'}^{\mu} = \Big( x^\mu - b^\mu x^2 \Big) \frac{1}{1- \Big( 2bx -b^2 x^2 \Big)}.
\end{equation}
This allows us to utilise the identity $ \frac{1}{1-c} = 1+c, \, \abs{c} \ll 1$  to get
\begin{align}
{x'}^{\mu} & = \Big( x^\mu - b^\mu x^2 \Big) \Big( 1 + 2bx -b^2x^2 \Big) + \order{b^3} \nonumber \\
& = \Big( x^\mu - b^\mu x^2 \Big) \Big(1 +2bx \Big) + \order{b^2} \nonumber \\
& = x^\mu + 2 (bx) x^\mu - b^\mu x^2 + \order{b^2}.
\end{align}
This is the infinitesimal transformation of the SCT. The same result can be obtained by Taylor expansion. Then, we can use again \cref{Generator}, and $ \frac{\delta x^\mu}{\delta b^\nu} = 2 x_\mu {\delta}^{\mu}_{\ \nu} x^\mu$ , $ \frac{\delta x^\mu}{\delta b^\nu} = - x^2{\delta}^{\mu}_{\ \nu} $ to get
\begin{align}
G_a & = -i \Big( 2 x_\mu {\delta}^{\mu}_{\ \nu} x^\mu \partial_\mu - x^2 {\delta}^{\mu}_{\ \nu} \partial_\mu \Big) \nonumber \\
		& = -i \Big( 2 x_\nu x^\mu \partial_\mu - x^2 \partial_\nu \Big).
\end{align}
Since these are dummy indices, we can exchange $ \nu \iff \mu$ to get
\begin{equation}
		K_\mu = -i \Big( 2 x_\mu x^\nu \partial_\nu - x^2 \partial_\mu \Big).
\end{equation}
\end{enumerate}
	

\paragraph{Scale factor of the SCT}

\emph{We work out explicitly the scale factor $\Omega(x)$ of the SCT given in \cref{sct}.} 
We have that 
\begin{equation}\label{scalefactror}
	\Omega(x) \eta_{\mu \nu} = \eta_{\rho \sigma} \frac{\partial x^{\rho}}{\partial x^{' \mu}}\frac{\partial x^{\sigma}}{\partial x ^{' \nu}}
\end{equation}
and the desired transformation is 
\begin{equation}
	{x'}^{\mu} = \frac{x^\mu - b^\mu x^2}{1 - 2bx + b^2 x^2},
\end{equation}
which can be rewritten as 
\begin{equation}
	x^\mu = {x'}^{\mu} \Big( 1 -2bx + b^2x^2 \Big) + b^\mu x^2.
\end{equation}
Then, it is easy to derive that 
\begin{equation}\label{rho}
	\frac{\partial x^\rho}{\partial {x'}^{\mu}} = \delta^{\rho}_{\ \mu} \Big( 1 - 2bx + b^2 x^2 \Big).
\end{equation}
And similarly for sigma
\begin{equation}\label{sigma}
	\frac{\partial x^\sigma}{\partial x^{' \nu}} = \delta^{\sigma}_{\ \nu} \Big( 1 - 2bx + b^2 x^2 \Big).
\end{equation}
Then combining \cref{rho,sigma,scalefactror} we get
\begin{align}
	\Omega(x) \eta_{\mu \nu} & = \eta_{\rho \sigma} \delta^{\rho}_{\ \mu} \delta^{\sigma}_{\ \nu} \Big( 1 - 2bx + b^2 x^2 \Big)  \Big( 1 - 2bx + b^2 x^2 \Big) \nonumber \\ 
	& = \eta_{\mu \nu} {\Big( 1 - 2bx + b^2 x^2 \Big)}^2.
\end{align}
And hence, the scale factor for the SCT is
\begin{equation}
	\Omega(x) = {\Big( 1 - 2bx + b^2 x^2 \Big)}^2.
\end{equation}
	
\paragraph{Commutation rules of the conformal algebra.}

\emph{We explicitly verify the commutation rules of the generators of the conformal algebra involving the new generators $D$ and $K_\mu$.}
\begin{enumerate}
\item 
\begin{align}
[D , P_\mu] & = [-i x^\nu \del_\nu, -i \partial_\mu]
		= i^2 x^\nu \partial_\nu \partial_\mu - i^2 \partial_\mu x^\nu \partial_\nu \nonumber \\
		& = - i^2 \delta_{\mu}^{\ \nu} \partial_\nu = -i^2 \partial_\mu \nonumber\\
		& = i \Big( -i \partial_\mu \Big) = i P_\mu
\end{align}
\item
\begin{align}
		[D, K_\mu] & = [-i x^\rho \partial_\rho,-2i x_\mu x^\nu \partial_\nu + i x^2 \partial_\mu ] \nonumber\\
		& = [-i x^\rho \partial_\rho,-2i x_\mu x^\nu \partial_\nu] + [-i x^\rho \partial_\rho,i x^2 \partial_\mu] \nonumber \\
		& = \bigg( 2 i^2 x^\rho \partial_\rho \big( x_\mu x^\nu \big) \partial_\nu - 2i^2 x_\mu x^\nu \partial_\nu x^\rho \partial_\rho \bigg) + \bigg(-i^2 x^\rho \partial_\rho \big( x_\sigma x^\sigma \big) \partial_\mu - (-i^2) x^2 \partial_\mu x^\rho \partial_\rho \bigg) \nonumber \\
		& = \Big(-2 x^\rho \eta_{\rho \mu}x^\nu \partial_\nu - 2 x^\rho x_\mu \delta_{\rho}^{\ \nu} \partial_\nu + 2 x_\mu x^\nu \delta_{\nu}^{\ \rho}\partial_\rho \Big) \nonumber \\
		& + \Big( x^\rho \eta_{\rho \sigma} x^{\sigma}\partial_{\mu} + x^\rho x_\sigma \delta_{\rho}^{\ \sigma}\partial_{\mu} - x^2 \delta_{\mu}^{\ \rho} \partial_\rho \Big) \nonumber \\
		& = \Big( -2 x_\mu x^\nu \partial_\nu - 2 x_\mu x^\nu \partial_\nu + 2 x_\mu x^\nu \partial_\nu \Big) + \Big( x_\sigma x^\sigma \partial_\mu + x_\rho x^\rho \partial_\mu - x^2 \partial_\mu \Big) \nonumber \\
		& = - 2 x_\mu x^\nu \partial_\nu + x^2 \partial_\mu = -i K_\mu.
\end{align}
Since
\begin{equation}
		-i K_\mu = -i (-i) (2 x_\mu x^\nu \partial_\nu -x^2 \partial_\mu) = -2 x_\mu x^\nu \partial_\nu + x^2 \partial_\mu.
\end{equation}
\item
\begin{align}
[K_\mu,P_\nu] & = [-2i x_\mu x^\sigma \partial_\sigma + i x^2 \partial_\mu, -i \partial_\nu] \nonumber \\
		& = [-2i x_\mu x^\sigma \partial_\sigma, -i \partial_\nu]+[i x^2 \partial_\mu, -i \partial_\nu] \nonumber \\
		& = \Big(2i^2 x_\mu x^\sigma \partial_\sigma \partial_\nu - (-2i) (-i)\partial_\nu (x_\mu x^\sigma )\partial_\sigma \Big) + \Big( -i^2 x^2 \partial_\mu \partial_\nu - (-i^2) \partial_\nu(x^2)\partial_\mu \Big) \nonumber \\
		& = -2i^2 \delta_{\nu \mu} x^\sigma \partial_\sigma - 2i^2 x_\mu \delta_{\nu}^{\ \sigma}\partial_{\sigma} + i^2 \partial_\nu (x_\rho x^\rho)\partial_\mu \nonumber \\
		& = 2 \delta_{\nu \mu} x^\sigma \partial_\sigma + 2 x_\mu \partial_\nu + i^2 \delta_{\nu \rho} x^\rho \partial_\mu + i^2 x_\rho \delta_{\nu}^{\ \rho}\partial_\mu \nonumber \\
		& = 2 \delta_{\nu \mu} x^\sigma \partial_\sigma + 2 x_\mu \partial_\nu - x_\nu \partial_\mu - x_\nu \partial_\mu \nonumber \\
		& = 2 \delta_{\nu \mu} x^\sigma \partial_\sigma + 2x_\mu \partial_\nu - 2 x_\nu \partial_\mu \nonumber \\
		& = 2 \Big( \delta_{\mu \nu} x^\sigma \partial_\sigma + x_\mu \partial_\nu - x_\nu \partial_\mu \Big) = 2i \Big( \eta_{\mu \nu}D - J_{\mu \nu}\Big).
\end{align}
Since
\begin{align}
		2i \Big( \eta_{\mu \nu}D - J_{\mu \nu}\Big) & = 2i \eta_{\mu \nu} D - 2i J_{\mu \nu}  \nonumber \\
		& = 2i \eta_{\mu \nu} (-i x^\mu \partial_\mu) - 2i (i) (x_\mu \partial_\nu - x_\nu \partial_\mu) \\
		& = 2 \eta_{\mu \nu}x^\mu \partial_\mu + 2 x_\mu \partial_\nu - 2x_\nu \partial_\mu.
\end{align}
\item
\begin{align}
[K_\mu, J_{\nu \rho}] & = [-2i x_\mu x^\sigma \partial_\sigma + i x^2 \partial_\mu, i x_\nu \partial_\rho - i x_\rho \partial_\nu] \nonumber\\
		& = [-2i x_\mu x^\sigma \partial_\sigma,i x_\nu \partial_\rho]+ [-2i x_\mu x^\sigma \partial_\sigma,- i x_\rho \partial_\nu] +[i x^2 \partial_\mu,i x_\nu \partial_\rho] + [i x^2 \partial_\mu,- i x_\rho \partial_\nu] \nonumber \\
		& = \Big( -2i(i)x_\mu x^\sigma \partial_\sigma x_\nu \partial_\rho - (-2i)(i) x_\nu \partial_\rho (x_\mu x^\sigma) \partial_\sigma \Big) \nonumber\\
		& + \Big( -2i(-i) x_\mu x^\sigma \partial_\sigma x_\rho \partial_\nu - (-2i)(-i)x_\rho \partial_\nu (x_\mu x^\sigma) \partial_\sigma \Big) \nonumber \\
		& + \Big( i^2 x^2 \partial_\mu x_\nu \partial_\rho - i^2 x_\nu \partial_\rho (x^2) \partial_\mu \Big) \nonumber \\
		& + \Big(-i^2 x^2 \partial_\mu x_\rho \partial_\nu - (-i) (i) x_\rho \partial_\nu (x^2) \partial_\mu \Big) \nonumber \\
		& = -2i^2 x_\mu x^\sigma \delta_{\sigma \nu} \partial_\rho + 2i^2 x_\nu \delta_{\rho \mu} x^\sigma \partial_\sigma + 2i^2 x_\nu x_\mu \delta_{\rho}^{\ \sigma} \partial_\sigma \nonumber \\
		& + 2i^2 x_\mu x^\sigma \delta_{\sigma \rho} \partial_\nu - 2i^2 x_\rho \delta_{\nu \mu} x^\sigma \partial_\sigma - 2i^2 x_\rho x_\mu \delta_{\nu}^{\ \sigma} \partial_\sigma \nonumber \\
		& + i^2 x^2 \delta_{\mu \nu} \partial_\rho - i^2 x_\nu \delta_{\rho \sigma}x^\sigma \partial_\mu -i^2 x_\nu x_\sigma \delta_{\rho}^{\ \sigma}\partial_\mu \nonumber \\
		& - i^2 x^2 \delta_{\mu \rho} \partial_\nu + i^2 x_\rho \delta_{\nu \sigma}x^\sigma \partial_\mu + i^2 x_\rho x_\sigma \delta_{\nu}^{\ \sigma}\partial_\mu \nonumber \\
		& = 2i^2 x_\nu \delta_{\rho \mu} x^\sigma \partial_\sigma - 2i^2 x_\rho \delta_{\nu \mu} x^\sigma \partial_\sigma +i^2 x^2 \delta_{\mu \nu} \partial_\rho - i^2 x^2 \delta_{\mu \rho} \partial_\nu \nonumber \\
		& = i \Big( 2i \delta_{\rho \mu} x_\nu  x^\sigma \partial_\sigma - i \delta_{\rho \mu} x^2 \partial_\nu - 2i \delta_{\mu \nu} x_\rho x^\sigma \partial_\sigma + i \delta_{\mu \nu} x^2 \partial_\rho \Big) \nonumber \\
		& = i \Big( -i \delta_{\mu \nu} \big( 2 x_\rho x^\sigma \partial_\sigma - x^2 \partial_\rho \big) + i \delta_{\rho \mu} \big( 2 x_\nu x_\sigma \partial_\sigma - x^2 \partial_\nu \big) \Big) \nonumber \\
		& = i \Big( \delta_{\mu \nu} K_\rho - \delta_{\mu \rho} K_\nu \Big).
\end{align}
\end{enumerate}
	
\paragraph{Representations of the conformal group}
\begin{enumerate} 
	\item \emph{We verify the commutation relations of \cref{reducedalgebra}}

To do so, we will use the following
\begin{align}
		\tilde{\Delta} & = -i x^\mu \partial_\mu ,\\
		\kappa_\mu & = -i ( 2 x_\mu x^\nu \partial_\nu - x^2 \partial_\mu ), \\
		S_{\mu \nu} & = i ( x_\mu \partial_\nu - x_\nu \partial_\mu ).
\end{align}
Using these, we get:
\begin{enumerate}
\item 
\begin{equation}
	\begin{split}
			[\tilde{\Delta},S_{\mu \nu}] & = [-i x^\rho \partial_\rho, ix_\mu \partial_\nu - i x_\nu \partial_\mu]\\
			& = [-i x^\rho \partial_\rho, ix_\mu \partial_\nu] + [-i x^\rho \partial_\rho, - i x_\nu \partial_\mu]\\
			& = \Big( - i^2 x^\rho \partial_\rho \big( x_\mu \partial_\nu \big) + i^2 x_\mu \partial_\nu \big( x^\rho \partial_\rho \big) \Big) + \Big( i^2 x^\rho \partial_\rho \big( x_\nu \partial_\mu \big) - i^2 x_\nu \partial_\mu \big( x^\rho \partial_\rho \big) \Big)\\
			& = - i^2 x^\rho \delta_{\rho\mu}\partial_\nu + i^2 x_\mu \delta_{\nu}^{\ \ \rho} \partial_\rho + i^2 x^\rho \delta_{\rho \nu}\partial_\mu -i^2 x_\nu \delta_{\mu}^{\ \ \rho}\partial_\rho\\
			& = -i^2 x_\mu \partial_\nu + i^2 x_\mu \partial_\nu + i^2 x_\nu \partial_\mu - i^2 x_\nu \partial_\mu\\
			& = 0 .
	\end{split}
\end{equation}
			
			\item 
			\begin{equation}
			\begin{split}
			[\tilde{\Delta},\kappa_\mu] & = [-i x^\rho \partial_\rho,-2i x_\mu x^\nu \partial_\nu + i x^2 \partial_\mu ]\\
			& = [-i x^\rho \partial_\rho,-2i x_\mu x^\nu \partial_\nu] + [-i x^\rho \partial_\rho,i x^2 \partial_\mu]\\
			& = \big( 2 i^2 x^\rho \partial_\rho \big( x_\mu x^\nu \big) \partial_\nu - 2i^2 x_\mu x^\nu \partial_\nu x^\rho \partial_\rho \big) + \big(-i^2 x^\rho \partial_\rho \big( x_\sigma x^\sigma \big) \partial_\mu - (-i^2) x^2 \partial_\mu x^\rho \partial_\rho \big) \\
			& = \big(-2 x^\rho \delta_{\rho \mu}x^\nu \partial_\nu - 2 x^\rho x_\mu \delta_{\rho}^{\ \ \nu} \partial_\nu + 2 x_\mu x^\nu \delta_{\nu}^{\ \ \rho}\partial_\rho \big)\\
			& + \big( x^\rho \delta_{\rho \sigma} x^{\sigma}\partial_{\mu} + x^\rho x_\sigma \delta_{\rho}^{\  \ \sigma}\partial_{\mu} - x^2 \delta_{\mu}^{\ \ \rho} \partial_\rho \big)\\
			& = \big( -2 x_\mu x^\nu \partial_\nu - 2 x_\mu x^\nu \partial_\nu + 2 x_\mu x^\nu \partial_\nu \big) + \big( x_\sigma x^\sigma \partial_\mu + x_\rho x^\rho \partial_\mu - x^2 \partial_\mu \big)\\
			& = - 2 x_\mu x^\nu \partial_\nu + x^2 \partial_\mu\\
			& = -i \kappa_\mu.
			\end{split}
			\end{equation}
			
			\item
			\begin{equation}
			\begin{split}
			[\kappa_\nu, \kappa_\mu] & = [-2i x_\nu x^\rho \partial_\rho + i x^2 \partial_\nu, -2i x_\mu x^\sigma \partial_\sigma + i x^2 \partial_\mu]\\
			& = [-2i x_\nu x^\rho \partial_\rho, -2i x_\mu x^\sigma \partial_\sigma] + [-2i x_\nu x^\rho \partial_\rho, i x^2 \partial_\mu]\\
			& + [i x^2 \partial_\nu,-2i x_\mu x^\sigma \partial_\sigma] + [i x^2 \partial_\nu, i x^2 \partial_\mu]\\
			& = \Big(4i^2 x_\nu x^\rho \partial_\rho (x_\mu x^\sigma)\partial_\sigma - 4i^2 x_\mu x^\sigma\partial_\sigma ( x_\nu x^\rho) \partial_\rho \Big)\\
			& + \Big(-2i^2 x_\nu x^\rho \partial_\rho (x_\omega x^\omega)\partial_\mu + 2i^2 x^2 \partial_\mu (x_\nu x^\rho) \partial_\rho \Big)\\
			& + \Big( -2i^2 x^2 \partial_\nu (x_\mu x^\sigma) \partial_\sigma + 2i^2 x_\mu x^\sigma \partial_\sigma (x_\zeta x^\zeta) \partial_\nu \Big)\\
			& + \Big( i^2 x^2 \partial_\nu ( x_\xi x^\xi) \partial_\mu - i^2 x^2 \partial_\mu (x_\gamma x^\gamma) \partial_\nu \Big)\\
			& = \Big( 4i^2 x_\nu x^\rho \delta_{\rho \mu} x^\sigma \partial_\sigma + 4i^2 x_\nu x^\rho x_\mu \delta_{\rho}^{\ \ \sigma}\partial_\sigma - 4i^2 x_\mu x^\sigma \delta_{\sigma \nu} x^\rho \partial_\rho - 4i^2 x_\mu x^\sigma x_\nu \delta_{\sigma}^{\ \ \rho}\partial_\rho \Big)\\
			& + \Big( -2i^2 x_\nu x^\rho \delta_{\rho \omega} x^\omega \partial_\mu - 2i^2 x_\nu x^\rho x_\omega \delta_{\rho}^{\ \ \omega}\partial_\mu + 2i^2 x^2 \delta_{\mu \nu }x^\rho \partial_\rho + 2i^2 x^2 x_\nu \delta_{\mu}^{\ \ \rho} \partial_\rho \Big)\\
			& + \Big( -2i^2 x^2 \delta_{\nu \mu}x^\sigma \partial_\sigma - 2i^2 x^2 x_\mu \delta_{\nu}^{\ \ \sigma}\partial_\sigma + 2i^2 x_\mu x^\sigma \delta_{\sigma \zeta}x^\zeta \partial_\nu + 2i^2 x_\mu x^\sigma x_\zeta \delta_{\sigma}^{\ \ \zeta}\partial_\nu \Big)\\
			& + \Big( i^2 x^2 x_\xi \delta_{\nu}^{\ \ \xi}\partial_\mu + i^2 x^2 x^\xi \delta_{\xi \nu}\partial_\mu - i^2 x^2 \delta_{\mu \gamma}x^\gamma \partial_\nu -i^2 x^2 x_\gamma \delta_{\mu}^{\ \ \gamma}\partial_\nu \Big)\\
			& = -4 x_\nu x_\mu x^\sigma \partial_\sigma - 4 x_\nu x^\sigma x_\mu \partial_\sigma + 4 x_\mu x_\nu x^\rho \partial_\rho + 4 x_\mu x^\sigma x_\nu \partial_\sigma + 2 x_\nu x^2 \partial_\mu\\
			& + 2 x_\nu x^\omega x_\omega \partial_\mu - 2 x^2 \delta_{\mu \nu}x^\rho \partial_\rho - 2 x^2 x_\nu \partial_\mu + 2 x^2 \delta_{\nu \mu} x^\sigma \partial_\sigma + 2 x^2 x_\mu \partial_\nu\\
			& - 2 x_\mu x^\sigma x_\sigma \partial_\nu - 2 x_\mu x^\sigma x_\sigma \partial_\nu - x^2 x_\nu \partial_\mu - x^2 x_\nu \partial_\mu + x^2 x_\mu \partial_\nu + x^2 x_\mu \partial_\nu \\
			& = 0 .
			\end{split}
			\end{equation}
			
			\item
			\begin{equation}
			\begin{split}
			[\kappa_\mu,S_{\nu \rho}]& = [-2i x_\mu x^\sigma \partial_\sigma + i x^2 \partial_\mu, i x_\nu \partial_\rho - i x_\rho \partial_\nu]\\
			& = [-2i x_\mu x^\sigma \partial_\sigma,i x_\nu \partial_\rho]+ [-2i x_\mu x^\sigma \partial_\sigma,- i x_\rho \partial_\nu] +[i x^2 \partial_\mu,i x_\nu \partial_\rho] + [i x^2 \partial_\mu,- i x_\rho \partial_\nu]\\
			& = \Big( -2i(i)x_\mu x^\sigma \partial_\sigma x_\nu \partial_\rho - (-2i)(i) x_\nu \partial_\rho (x_\mu x^\sigma) \partial_\sigma \Big)\\
			& + \Big( -2i(-i) x_\mu x^\sigma \partial_\sigma x_\rho \partial_\nu - (-2i)(-i)x_\rho \partial_\nu (x_\mu x^\sigma) \partial_\sigma \Big)\\
			& + \Big( i^2 x^2 \partial_\mu x_\nu \partial_\rho - i^2 x_\nu \partial_\rho (x^2) \partial_\mu \Big)\\
			& + \Big(-i^2 x^2 \partial_\mu x_\rho \partial_\nu - (-i) (i) x_\rho \partial_\nu (x^2) \partial_\mu \Big)\\
			& = -2i^2 x_\mu x^\sigma \delta_{\sigma \nu} \partial_\rho + 2i^2 x_\nu \delta_{\rho \mu} x^\sigma \partial_\sigma + 2i^2 x_\nu x_\mu \delta_{\rho}^{\ \ \sigma} \partial_\sigma\\
			& + 2i^2 x_\mu x^\sigma \delta_{\sigma \rho} \partial_\nu - 2i^2 x_\rho \delta_{\nu \mu} x^\sigma \partial_\sigma - 2i^2 x_\rho x_\mu \delta_{\nu}^{\ \ \sigma} \partial_\sigma\\
			& + i^2 x^2 \delta_{\mu \nu} \partial_\rho - i^2 x_\nu \delta_{\rho \sigma}x^\sigma \partial_\mu -i^2 x_\nu x_\sigma \delta_{\rho}^{\ \ \sigma}\partial_\mu\\
			& - i^2 x^2 \delta_{\mu \rho} \partial_\nu + i^2 x_\rho \delta_{\nu \sigma}x^\sigma \partial_\mu + i^2 x_\rho x_\sigma \delta_{\nu}^{\ \ \sigma}\partial_\mu\\
			& = 2i^2 x_\nu \delta_{\rho \mu} x^\sigma \partial_\sigma - 2i^2 x_\rho \delta_{\nu \mu} x^\sigma \partial_\sigma +i^2 x^2 \delta_{\mu \nu} \partial_\rho - i^2 x^2 \delta_{\mu \rho} \partial_\nu\\
			& = i \Big( 2i \delta_{\rho \mu} x_\nu  x^\sigma \partial_\sigma - i \delta_{\rho \mu} x^2 \partial_\nu - 2i \delta_{\mu \nu} x_\rho x^\sigma \partial_\sigma + i \delta_{\mu \nu} x^2 \partial_\rho \Big)\\
			& = i \Big( -i \delta_{\mu \nu} \big( 2 x_\rho x^\sigma \partial_\sigma - x^2 \partial_\rho \big) + i \delta_{\rho \mu} \big( 2 x_\nu x_\sigma \partial_\sigma - x^2 \partial_\nu \big) \Big)\\
			& = i \Big( \delta_{\mu \nu} \kappa_\rho - \delta_{\mu \rho} \kappa_\nu \Big)\\
			& = i \Big( \eta_{\mu \nu} \kappa_\rho - \eta_{\mu \rho} \kappa_\nu \Big).
			\end{split}
			\end{equation}
			
			\item
			\begin{equation}
			\begin{split}
			[S_{\mu \nu}, S_{\rho \sigma}] & = [i x_\mu \partial_\nu - i x_\nu \partial_\mu, i x_\rho \partial_\sigma - i x_\sigma \partial_\rho]\\
			& = [i x_\mu \partial_\nu,i x_\rho \partial_\sigma] + [i x_\mu \partial_\nu,- i x_\sigma \partial_\rho] + [-i x_\nu \partial_\mu,i x_\rho \partial_\sigma] + [- i x_\nu \partial_\mu,- i x_\sigma \partial_\rho]\\
			& = \Big( i^2 x_\mu \partial_\nu (x_\rho) \partial_\sigma - i^2 x_\rho \partial_\sigma(x_\mu) \partial_\nu \Big) + \Big( -i^2 x_\mu \partial_\nu (x_\sigma) \partial_\rho + i^2 x_\sigma \partial_\rho (x_\mu) \partial_\nu \Big)\\
			& + \Big( -i^2 x_\nu \partial_\mu (x_\rho) \partial_\sigma + i^2 x_\rho \partial_\sigma (x_\nu) \partial_\mu \Big) + \Big( i^2 x_\nu \partial_\mu (x_\sigma) \partial_\rho - i^2 x_\sigma \partial_\rho ( x_\nu) \partial_\mu \Big)\\
			& = - x_\mu \delta_{\nu \rho} \partial_\sigma + x_\rho \delta_{\sigma \mu} \partial_\nu + x_\mu \delta_{\nu \sigma}\partial_\rho - x_\sigma \delta_{\rho \mu} \partial_\nu\\
			& + x_\nu \delta_{\mu \rho} \partial_\sigma - x_\rho \delta_{\sigma \nu} \partial_\mu - x_\nu \delta_{\mu \sigma}\partial_\rho + x_\sigma \delta_{\rho \nu}\partial_\mu\\
			& = \eta_{\nu \rho}\big( x_\sigma \partial_\mu - x_\mu \partial_\sigma \big) + \eta_{\sigma \mu} \big( x_\rho \partial_\nu - x_\nu \partial_\rho \big)\\
			& + \eta_{\nu \sigma} \big( x_\mu \partial_\rho - x_\rho \partial_\mu \big) + \eta_{\mu \rho} \big( x_\nu \partial_\sigma - x_\sigma \partial_\nu \big)\\
			& = i \eta_{\nu \rho} i \big( x_\mu \partial_\sigma - x_\sigma \partial_\mu \big) + \dots\\
			& = i \Big( \eta_{\nu \rho} S_{\mu \sigma} + \eta_{\mu \sigma}S_{\nu \rho} - \eta_{\mu \rho} S_{\nu \sigma} - \eta_{\nu \sigma}S_{\mu \rho} \Big).
			\end{split}
			\end{equation}
		\end{enumerate}
		So we can sum up the algebra as: 
		\begin{equation}
		\begin{split}
		[\tilde{\Delta},S_{\mu \nu}] & = 0,\\
		[\tilde{\Delta},\kappa_\mu] & = -i \kappa_\mu,\\
		[\kappa_\nu, \kappa_\mu] & = 0,\\
		[\kappa_\mu,S_{\nu \rho}]& =i \Big( \eta_{\mu \nu} \kappa_\rho - \eta_{\mu \rho} \kappa_\nu \Big),\\
		[S_{\mu \nu}, S_{\rho \sigma}] & = i \Big( \eta_{\nu \rho} S_{\mu \sigma} + \eta_{\mu \sigma}S_{\nu \rho} - \eta_{\mu \rho} S_{\nu \sigma} - \eta_{\nu \sigma}S_{\mu \rho} \Big).
		\end{split}
		\end{equation}
		
\item \emph{We want to verify \cref{eq:eq:Hausedorf}}

We will use the Hausdorff formula 
	\begin{equation}
		e^{-A}Be^A = B + [B,A] + \frac{1}{2!} [[B,A],A]+ \dots
\end{equation}
\begin{enumerate}
\item We will start with the dilatation
\begin{equation}
			e^{i x^\rho P_\rho}D e^{-i x^\rho P_\rho},
\end{equation}
where $ A = -i x^\rho P_\rho$ and $ B = D$. Then, using \cref{Hausdorff} we get
\begin{equation}
\begin{split}
			e^{i x^\rho P_\rho}D e^{-i x^\rho P_\rho} & = D + [D,-i x^\rho P_\rho ] + \frac{1}{2}[[D,-i x^\rho P_\rho], - x^\sigma P_\sigma]\\
			& = D - i x^\rho [D, P_\rho] - \frac{1}{2}i x^\rho [[D, P_\rho],- i x^\sigma P_\sigma]\\
			& = D - i x^\rho i P_\rho + \frac{1}{2}i^2 x^\rho x^\sigma [[D, P_\rho],P_\sigma]\\
			& = D - i x^\rho i P_\rho - \frac{1}{2}i x^\rho x^\sigma [P_\rho, P_\sigma]\\
			& = D + x^\rho P_\rho = D + x^\nu P_\nu.
\end{split}
\end{equation}
And we used that $ [D, P_\rho] = i P_\rho$ and also that $[P_\mu, P_\nu]= 0$.
\item For the SCT we have the following
\begin{equation}
			e^{i x^\rho P_\rho} K_\mu e^{-i x^\rho P_\rho},
\end{equation}
where $ A = -i x^\rho P_\rho$ and $ B = K_\mu$. Then, \cref{Hausdorff} we derive
\begin{equation}
	\begin{split}
			e^{i x^\rho P_\rho} K_\mu e^{-i x^\rho P_\rho} & = K_\mu + [K_\mu, - i x^\rho P_\rho] + \frac{1}{2}[[K_\mu, i x^\rho P_\rho], -i x^\sigma P_\sigma]\\
			& = K_\mu - i x^\rho [K_\mu, P_\rho] + \frac{1}{2}i^2 x^\rho x^\sigma [K_\mu, P_\rho], P_\sigma]\\
			& = K_\mu -  i x^\rho \Big( 2i \eta_{\mu \rho} D - 2i J_{\mu \rho} \Big) - \frac{1}{2}x^\rho x^\sigma [2i \eta_{\mu \rho} D - 2i J_{\mu \rho}, P_\sigma]\\
			& = K_\mu + 2 x^\rho \eta _{\mu \rho} D - 2 x^\rho J_{\mu \rho}  - \frac{1}{2}x^\rho x^\sigma [2i \eta_{\mu \rho} D, P_\sigma] - \frac{1}{2}x^\rho x^\sigma [- 2i J_{\mu \rho}, P_\sigma]\\
			& = K_\mu + 2 x_\mu D - 2 x^\rho J_{\mu \rho} - i x^\rho x^\sigma [D, P_\sigma] + i x^\rho x^\sigma [J_{\mu \rho}, P_\sigma]\\
			& = K_\mu + 2 x_\mu D - 2 x^\rho J_{\mu \rho} i x^\rho x^\sigma \eta_{\mu \rho} i P_\sigma - i x^\rho x^\sigma [P_\sigma, J_{\mu \rho}]\\
			& = K_\mu + 2 x_\mu D - 2 x^\rho J_{\mu \rho} + x_\mu x^\sigma P_\sigma - i x^\rho x^\sigma \Big( i \eta_{\sigma \mu} P_\rho - i \eta_{\sigma \rho} P_\mu \Big)\\
			& = K_\mu + 2 x_\mu D - 2 x^\rho J_{\mu \rho} + x_\mu x^\sigma P_\sigma + x^\rho x^\sigma \eta_{\sigma \mu} P_\rho - x^\rho x^\sigma \eta_{\sigma \rho} P_\mu\\
			& = K_\mu + 2 x_\mu D - 2 x^\rho J_{\mu \rho} + 2 x_\mu x^\nu P_\nu - x^2 P_\mu\\
			& = K_\mu + 2 x_\mu D - 2 x^\nu J_{\mu \nu} + 2 x_\mu x^\nu P_\nu - x^2 P_\mu.
	\end{split}
\end{equation}
Here we used that $ [D, P_\rho] = i P_\rho$ and that $ [P_\rho, J_{\mu \nu}] =  i \big( \eta_{\rho \mu} P_\nu - \eta_{\rho \nu} P_\mu \big)$, and we also renamed certain dummy indices.
\end{enumerate}
\item \emph{We write down the full transformation rules of the fields under $D$ and $K_\mu$.}
To derive the full transformation rules, we use the above results, but we replace $ D = \tilde{\Delta}$, $J_{\mu \nu} = S_{\mu \nu}$ and finally $K_\mu = \kappa_\mu$. Then we get the following
\begin{itemize}
\item 
\begin{equation}
			D \Phi (x) = \tilde{\Delta} \Phi (x) - i x^\nu \partial_\nu \Phi (x) = \Big( - i x^\nu \partial_\nu + \tilde{\Delta}   \Big) \Phi (x),
\end{equation}
\item
\begin{equation}
			K_\mu \Phi (x) = \{ \kappa_\mu + 2 x_\mu \tilde{\Delta} - 2 x^\nu S_{\mu \nu} + -2i x_\mu x^\nu \partial_\nu +i x^2 \partial_\mu \} \Phi (x).
\end{equation}
\end{itemize}
\end{enumerate}

\subsection{The energy-momentum tensor}\label{sec.conformalEM}

\begin{enumerate} 
\item \emph{We want to verify that $\partial_\mu \partial_\lambda \partial_\rho X^{\lambda\rho\mu\nu}=0$ for $X^{\lambda\rho\mu\nu}$.}

where 
\begin{equation}\label{Chi}
		X^{\lambda \rho \mu \nu }= \frac{2}{d-2}\Bigg( \eta^{\lambda \rho} \sigma_{+}^{\mu \nu} - \eta^{\lambda \mu} \sigma_{+}^{\rho \nu} - \eta^{\lambda \nu} \sigma_{+}^{\mu \rho} + \eta^{\mu \nu} \sigma_{+}^{\lambda \rho} - \frac{1}{d-1} \big( \eta^{\lambda \rho}\eta^{\mu \nu}-\eta^{\lambda \mu}\eta^{\rho \nu} \big) \sigma^{a}_{+ a}  \Bigg).
\end{equation}
We have to calculate the following
\begin{enumerate}
\item $ \partial_\mu \partial_\lambda \partial_\rho \eta^{\lambda \rho} \sigma_{+}^{\mu \nu} = \partial_\mu \partial_\lambda \partial^\lambda \sigma_{+}^{\mu \nu} = \square \partial_\mu \sigma_{+}^{\mu \nu} = \square \sigma_{+, \  \mu}^{\mu \nu}.$
\item $ \partial_\mu \partial_\lambda \partial_\rho \eta^{\lambda \mu} \sigma_{+}^{\rho \nu} = \partial^\lambda \partial_\lambda \partial_\rho \sigma_{+}^{\rho \nu} = \square \partial_\rho \sigma_{+}^{\rho \nu} = \square \sigma_{+, \  \rho}^{\rho \nu} = \square \sigma_{+, \  \mu}^{\mu \nu}.$
\item $ \partial_\mu \partial_\lambda \partial_\rho \eta^{\lambda \nu} \sigma_{+}^{\mu \rho} = \partial_\mu \partial^\nu \partial_\rho \sigma_{+}^{\mu \rho} = \partial^\nu \sigma_{+, \ \mu \rho}^{\mu \rho}.$
\item $ \partial_\mu \partial_\lambda \partial_\rho \eta^{\mu \nu} \sigma_{+}^{\lambda \rho} = \partial_\rho \partial^\nu \partial_\lambda \sigma_{+}^{\lambda \rho} = \partial^\nu \sigma_{+, \ \lambda \rho}^{\lambda \rho}= .\partial^\nu \sigma_{+, \ \mu \rho}^{\mu \rho}.$
\end{enumerate}
We observe that all the indices are summed over with only $\nu$ being a free index, so the rest are dummy indices that we have the freedom to rename. Furthermore, we have to calculate the following
\begin{enumerate}
	\item $ \partial_\mu \partial_\lambda \partial_\rho \eta^{\lambda \rho}\eta^{\mu \nu}\sigma^{a}_{+ a} = \partial^\nu \partial^\rho \partial_\rho \sigma^{a}_{+ a} = \partial^\nu \square \sigma^{a}_{+ a}.$
	\item $ \partial_\mu \partial_\lambda \partial_\rho \eta^{\lambda \mu}\eta^{\rho \nu}\sigma^{a}_{+ a} = \partial^\lambda \partial^\nu \partial_\lambda \sigma^{a}_{+ a} = \partial^\nu \square \sigma^{a}_{+ a}.$
\end{enumerate}
So it if we plug these into $\partial_\mu \partial_\lambda \partial_\rho X^{\lambda \rho \mu \nu} $ we get
\begin{equation}
	\begin{split}
		\partial_\mu \partial_\lambda \partial_\rho X^{\lambda \rho \mu \nu } & = \frac{2}{d-2}\Bigg( \square \sigma_{+, \  \mu}^{\mu \nu} - \square \sigma_{+, \  \mu}^{\mu \nu} - \partial^\nu \sigma_{+, \ \mu \rho}^{\mu \rho} + \partial^\nu \sigma_{+, \ \mu \rho}^{\mu \rho} - \frac{1}{d-1} \big( \partial^\nu \square  - \partial^\nu \square \big)\sigma^{a}_{+ a} \Bigg)\\
		& = 0.
\end{split}
\end{equation}
Which is the requested result. 
\item \emph{We want to show that the term $\frac{1}{2}\partial_\lambda \partial_\rho X^{\lambda\rho\mu\nu}$ is symmetric under $\mu \leftrightarrow \nu$.}

To show that it is symmetric under $\mu \leftrightarrow \nu$, it is enough to show that $\partial_\lambda \partial_\rho X^{\lambda\rho\mu\nu} - \partial_\lambda \partial_\rho X^{\lambda\rho\nu\mu} = 0 $. 
The starting point is the following equation~\cite{Di_Francesco_1997}
\begin{equation}
	X^{\lambda\rho\mu\nu} - X^{\lambda\rho\nu\mu} = \frac{2}{(d-2)(d-1)} \sigma^{a}_{+ a} \big( \eta^{\lambda \mu}\eta^{\rho \nu} - \eta^{\lambda \nu}\eta^{\rho \mu} \big).
\end{equation}
Then, it is not hard to show that 
\begin{equation}
\begin{split}
		\partial_\lambda \partial_\rho \Big( X^{\lambda\rho\mu\nu} - X^{\lambda\rho\nu\mu} \Big) & \simeq \partial_\lambda \partial_\rho \eta^{\lambda \mu}\eta^{\rho \nu} \sigma^{a}_{+ a} - \partial_\lambda \partial_\rho \eta^{\lambda \nu}\eta^{\rho \mu}  \sigma^{a}_{+ a}\\
		& = \partial^\mu \partial^\nu \sigma^{a}_{+ a} - \partial^\nu \partial^\mu \sigma^{a}_{+ a}\\
		& = 0.
	\end{split}
\end{equation}
This proves the desired result.
\item 
\emph{We want to show that $T^{\mu\nu}$ is indeed traceless.}

We want to show that the modified energy-momentum tensor
\begin{equation}\label{energy-momentum}
		T^{\mu \nu} = T^{\mu \nu}_c + \partial_\rho B^{\rho \mu \nu} + \frac{1}{2}\partial_\lambda \partial_\rho X^{\lambda \rho \mu \nu},
\end{equation}
is traceless, in other words that
\begin{equation}
	T^\mu_{\  \mu} = 0.
\end{equation}
We start by multiplying \cref{energy-momentum} by $\eta_{\mu \nu}$, in which case we obtain
\begin{equation}\label{situation}
	\begin{split}
		\eta_{\mu \nu} T^{\mu \nu} & = \eta_{\mu \nu} \Big( T^{\mu \nu}_c + \partial_\rho B^{\rho \mu \nu} + \frac{1}{2}\partial_\lambda \partial_\rho X^{\lambda \rho \mu \nu} \Big)\\
		& = T^{\mu}_{c \ \mu} + \partial_\rho B^{\rho \mu}_{\ \ \ \mu} + \frac{1}{2}\partial_\lambda \partial_\rho X^{\lambda \rho \mu}_{\ \ \ \ \ \mu}.
	\end{split}
\end{equation}
We need to calculate $\partial_\rho B^{\rho \mu}_{\ \ \ \mu}$ and also $\frac{1}{2}\partial_\lambda \partial_\rho X^{\lambda \rho \mu}_{\ \ \ \ \ \mu}$.
\begin{itemize}
\item We start with $\partial_\rho B^{\rho \mu}_{\ \ \ \mu}$. By definition, we know that
\begin{equation}
	B^{\rho \mu \nu} = \frac{i}{2} \Bigg( \frac{\partial \Lp}{\partial (\partial_\rho \Phi)} S^{\nu \mu} \Phi + \frac{\partial \Lp}{\partial (\partial_\mu \Phi)} S^{\rho \nu} \Phi + \frac{\partial \Lp}{\partial (\partial_\nu \Phi)} S^{\rho \mu} \Phi \Bigg).
\end{equation}
We multiply with $\eta_{\mu \nu}$ to get
\begin{equation}
	\begin{split}
			B^{\rho \mu}_{\ \ \ \mu} = \eta_{\mu \nu} B^{\rho \mu \nu} & = \frac{i}{2} \Bigg( \frac{\partial \Lp}{\partial (\partial_\rho \Phi)} S^{\mu}_{\ \mu} \Phi + \frac{\partial \Lp}{\partial (\partial^\nu \Phi)} S^{\rho \nu} \Phi + \frac{\partial \Lp}{\partial (\partial^\mu \Phi)} S^{\rho \mu} \Phi \Bigg)\\
			& = \frac{i}{2} \Bigg( \frac{\partial \Lp}{\partial (\partial^\nu \Phi)} S^{\rho \nu} \Phi + \frac{\partial \Lp}{\partial (\partial^\mu \Phi)} S^{\rho \mu} \Phi \Bigg)\\
			& = \frac{i}{2} \Bigg( \frac{\partial \Lp}{\partial (\partial^\mu \Phi)} S^{\rho \mu} \Phi + \frac{\partial \Lp}{\partial (\partial^\mu \Phi)} S^{\rho \mu} \Phi \Bigg)\\
			& = - i \Bigg( \frac{\partial \Lp}{\partial (\partial^\mu \Phi)} S^{\mu \rho} \Phi \Bigg),
	\end{split}
\end{equation}
where in the second line we used that $S^{\mu \nu} $ is traceless, in the third line the fact that the only free index is $\rho$ so we have the freedom to rename the dummy indices, and finally in the forth line we used that by definition $S^{\mu \nu } = - S^{\nu \mu}$. So, it is easy to compute that
\begin{equation}\label{requirement1}
			\partial_\rho B^{\rho \mu}_{\ \ \ \mu} = - i \partial_\rho \Bigg( \frac{\partial \Lp}{\partial (\partial^\mu \Phi)} S^{\mu \rho} \Phi \Bigg).
\end{equation}
\item The case of $\frac{1}{2}\partial_\lambda \partial_\rho X^{\lambda \rho \mu}_{\ \ \ \ \ \mu}$ is more lengthy. By definition we know that
\begin{equation}
			X^{\lambda \rho \mu \nu }= \frac{2}{d-2}\Bigg( \eta^{\lambda \rho} \sigma_{+}^{\mu \nu} - \eta^{\lambda \mu} \sigma_{+}^{\rho \nu} - \eta^{\lambda \nu} \sigma_{+}^{\mu \rho} + \eta^{\mu \nu} \sigma_{+}^{\lambda \rho} - \frac{1}{d-1} \big( \eta^{\lambda \rho}\eta^{\mu \nu}-\eta^{\lambda \mu}\eta^{\rho \nu} \big) \sigma^{a}_{+ a}  \Bigg).
\end{equation}
So we need to compute the following
\begin{itemize}
\item $\partial_\lambda \partial_\rho (\eta^{\lambda \rho} \sigma^{\mu \nu}_{+})= \partial_\lambda \partial^\lambda  \sigma^{\mu \nu}_{+} = \square \sigma^{\mu \nu}_{+} $.
\item $\partial_\lambda \partial_\rho (-\eta^{\lambda \mu} \sigma^{\rho \nu}_{+})= - \partial^\mu \partial_\rho  \sigma^{\rho \nu}_{+}=- \partial^\mu \partial_\lambda  \sigma^{\lambda \nu}_{+} $.
\item $\partial_\lambda \partial_\rho (- \eta^{\lambda \nu} \sigma^{\rho \mu}_{+})= - \partial_\rho \partial^\nu  \sigma^{\rho \mu}_{+}=- \partial^\nu \partial_\lambda  \sigma^{\lambda \mu}_{+} $.
\item $\partial_\lambda \partial_\rho (\eta^{\mu \nu} \sigma^{\lambda \rho}_{+})= \eta^{\mu \nu} \partial_\lambda \partial_\rho \sigma^{\lambda \rho}_{+}  $.
\end{itemize}
And we renamed some dummy indices for future convenience. Now we should contract with $\eta_{\mu \nu}$, to derive
\begin{itemize}
\item $\eta_{\mu \nu}\partial_\lambda \partial_\rho (\eta^{\lambda \rho} \sigma^{\mu \nu}_{+}) = \eta_{\mu \nu} \square \sigma^{\mu \nu}_{+} =  \square \sigma^{\mu}_{+ \ \mu}$.
\item $\eta_{\mu \nu} \Big( -\eta^{\lambda \mu} \sigma^{\rho \nu}_{+}  (- \eta^{\lambda \nu} \sigma^{\rho \mu}_{+} \Big) = - \eta_{\mu \nu} \partial^\mu \partial_\lambda  \sigma^{\lambda \nu}_{+} - \eta_{\mu \nu}\partial^\nu \partial_\lambda  \sigma^{\lambda \mu}_{+} = -2 \partial_\lambda \partial_\rho \sigma^{\lambda \rho}_{+}$, where we renamed the dummy indices.
\item $\eta_{\mu \nu} \partial_\lambda \partial_\rho (\eta^{\mu \nu} \sigma^{\lambda \rho}_{+})= \eta_{\mu \nu} \eta^{\mu \nu} \partial_\lambda \partial_\rho \sigma^{\lambda \rho}_{+} = d \partial_\lambda \partial_\rho \sigma^{\lambda \rho}_{+}$.
\end{itemize}
Hence,
\begin{equation}
\begin{split}
\frac{2}{d-2} \eta_{\mu \nu} \partial_\lambda \partial_\rho & \Big( \eta^{\lambda \rho} \sigma_{+}^{\mu \nu} - \eta^{\lambda \mu} \sigma_{+}^{\rho \nu} - \eta^{\lambda \nu} \sigma_{+}^{\mu \rho} + \eta^{\mu \nu} \sigma_{+}^{\lambda \rho} \Big)\\
& = \frac{2}{d-2} \Big( \square \sigma^{\mu}_{+ \ \mu}  -2 \partial_\lambda \partial_\rho \sigma^{\lambda \rho}_{+} + d \partial_\lambda \partial_\rho \sigma^{\lambda \rho}_{+} \Big)\\
& = 2 \partial_\lambda \partial_\rho \sigma^{\lambda \rho}_{+} + \frac{2}{d-2} \square \sigma^{\mu}_{+ \ \mu}.
\end{split}
\end{equation}
Now we need to compute the following
\begin{itemize}
\item $\eta_{\mu \nu}\partial_\lambda \partial_\rho \big( \eta^{\lambda \rho}\eta^{\mu \nu} \sigma^{\mu}_{+ \ \mu} \big) = d \eta^{\lambda \rho} \partial_\lambda \partial_\rho \sigma^{\mu}_{+ \ \mu} = d \partial_\lambda \partial^\lambda  \sigma^{\mu}_{+ \ \mu} =d \square  \sigma^{\mu}_{+ \ \mu}  $.
\item $\eta_{\mu \nu}\partial_\lambda \partial_\rho \big( \eta^{\lambda \mu}\eta^{\rho \nu} \sigma^{\mu}_{+ \ \mu} \big) = \eta_{\mu \nu}\eta^{\lambda \mu}\eta^{\rho \nu} \partial_\lambda \partial_\rho\sigma^{\mu}_{+ \ \mu} = \delta^{\lambda}_{\ \ \nu} \eta^{\rho \nu} \partial_\lambda \partial_\rho\sigma^{\mu}_{+ \ \mu} = \eta^{\lambda \rho} \partial_\lambda \partial_\rho\sigma^{\mu}_{+ \ \mu} = \square \sigma^{\mu}_{+ \ \mu}  $.
\end{itemize}
Therefore, 
\begin{equation}
-\frac{2}{(d-1)(d-2)}\eta_{\mu \nu}\partial_\lambda \partial_\rho \Bigg( \Big( \eta^{\lambda \rho}\eta^{\mu \nu} - \eta^{\lambda \mu}\eta^{\rho \nu} \Big) \sigma^{\mu}_{+ \ \mu} \Bigg) =-\frac{2}{d-2}\square \sigma^{\mu}_{+ \ \mu},
\end{equation}
So finally,
\begin{equation}
\begin{split}
\eta_{\mu \nu} \partial_\lambda \partial_\rho X^{\lambda \rho \mu \nu } = 2 \partial_\lambda \partial_\rho \sigma^{\lambda \rho}_{+} & = \partial_\lambda \partial_\rho (\sigma^{\lambda \rho} + \sigma^{\rho \lambda})\\
& = 2 \partial_\lambda \partial_\rho \sigma^{\lambda \rho} = 2 \partial_\rho \Big( \partial_\lambda \sigma^{\lambda \rho} \Big)\\
& = 2 \partial_\rho V^\rho.
\end{split}
\end{equation}
In the first line we used the definition of $\sigma^{\lambda \rho}_{+}$, and to go from the first line to the second, we used that fact that due to the symmetry of $\partial_\lambda \partial_\rho$, the antisymmetric part of $\sigma^{\rho \lambda}$ is zero and only the symmetric part remains. This gives us the desired result
\begin{equation}
\eta_{\mu \nu} \frac{1}{2} \partial_\lambda \partial_\rho X^{\lambda \rho \mu \nu } = \partial_\mu V^\mu.
\end{equation}
The $V^\mu$ is known as the virial and by definition, it is given by
\begin{equation}
\begin{split}
	V^\mu & = \frac{\partial \Lp}{\partial \big( \partial^\rho \Phi \big)} \big( \eta^{\mu \rho} \Delta + i S^{\mu \rho} \big) \Phi\\
	& =  \frac{\partial \Lp} {\partial \big( \partial_\mu \Phi \big)}\Delta \Phi + i \frac{\partial \Lp}{\partial \big( \partial^\rho \Phi \big)}  S^{\mu \rho} \Phi.
\end{split}
\end{equation}
Then, 
\begin{equation}\label{requirement2}
	\partial_\mu V^\mu = \Delta \partial_\mu \Big( \frac{\partial \Lp} {\partial \big( \partial_\mu \Phi \big)} \Phi \Big) + i \partial_\mu \Big( \frac{\partial \Lp}{\partial \big( \partial^\rho \Phi \big)}  S^{\mu \rho} \Phi \Big).
\end{equation}
\end{itemize}
The final requirement comes from the hypothesis that the current is conserved, \emph{i.e.} $\partial_\mu j^\mu_D = 0$, which gives 	
\begin{equation}\label{requirement3}
\eta_{\mu \nu} T^{\mu \nu}_c = - \Delta \partial_\mu \Big( \frac{\partial \Lp} {\partial \big( \partial_\mu \Phi \big)} \Phi \Big).
\end{equation}
If we plug \cref{requirement1,requirement2,requirement3} into \cref{situation}, we get
\begin{equation}
\begin{split}
		\eta_{\mu \nu} T^{\mu \nu} & = T^{\mu}_{c \ \mu} + \partial_\rho B^{\rho \mu}_{\ \ \ \mu} + \frac{1}{2}\partial_\lambda \partial_\rho X^{\lambda \rho \mu}_{\ \ \ \ \ \mu}\\
		& = - \Delta \partial_\mu \Bigg( \frac{\partial \Lp} {\partial \big( \partial_\mu \Phi \big)} \Phi \Bigg) - i \partial_\rho \Bigg( \frac{\partial \Lp}{\partial (\partial^\mu \Phi)} S^{\mu \rho} \Phi \Bigg)\\
		& + \Delta \partial_\mu \Bigg( \frac{\partial \Lp} {\partial \big( \partial_\mu \Phi \big)} \Phi \Bigg) + i \partial_\mu \Bigg( \frac{\partial \Lp}{\partial \big( \partial^\rho \Phi \big)}  S^{\mu \rho} \Phi \Bigg)\\
		& = - i \partial_\rho \Bigg( \frac{\partial \Lp}{\partial (\partial^\mu \Phi)} S^{\mu \rho} \Phi \Bigg) + i \partial_\rho \delta^\rho_{\ \mu} \Bigg( \frac{\partial \Lp}{\partial \big( \partial^\rho \Phi \big)}  S^{\mu \rho} \Phi \Bigg)\\
		& = - i \partial_\rho \Bigg( \frac{\partial \Lp}{\partial (\partial^\mu \Phi)} S^{\mu \rho} \Phi \Bigg)  +  i \partial_\rho \Bigg( \frac{\partial \Lp}{\partial (\partial^\mu \Phi)} S^{\mu \rho} \Phi \Bigg) =0
\end{split}
\end{equation}
And we showed that the modified energy-momentum tensor is indeed traceless, as long as we have current conservation. 
\end{enumerate}


\subsection{Correlation functions of primary operators}
We will start with the two-point function to show explicitly what happens, so that afterwards we can expand our techniques to the three- and four-point functions. The starting point is how an $n$-point function transforms
\begin{equation}\label{generaltransform}
		\langle \phi_1(x_1) \phi_2(x_2) \dots \phi_n(x_n) \rangle = {\abs{\frac{\partial x'}{\partial x}}}_{x=x_1}^{\Delta_1 / d}{\abs{\frac{\partial x'}{\partial x}}}_{x=x_2}^{\Delta_2 / d} \dots  {\abs{\frac{\partial x'}{\partial x}}}_{x=x_n}^{\Delta_n / d} \langle \phi_1(x_1') \phi_2(x_2') \dots \phi_n(x_n') \rangle. 
\end{equation} 
We want to check this for the four possible symmetries of CFT : translations, i.e. $x' = x+a $, Lorentz transformations, i.e $x'=\Lambda x$, dilatations, i.e. $x'=\lambda x$ and SCTs, i.e. $x'= \frac{x-bx^2}{1-2bx+b^2x^2}$. 
\begin{enumerate}
	\item We start with translations. From \cref{generaltransform} we derive that
\begin{equation}
\begin{split}
\langle \phi_1(x^\mu_1) \phi_2(x_2^\nu) \rangle & = \delta_{\ \rho}^\mu \delta_{\ \sigma}^\nu \langle \phi_1(x^\rho_1+ a^\rho) \phi_2(x_2^\sigma + a^\sigma) \rangle\\
& = \langle \phi_1(x^\mu_1 + a^\mu) \phi_2(x_2^\nu+ a^\nu) \rangle.
\end{split}
\end{equation}
From translation invariance we deduce that if we define our two-point function as the kernel $G(x_1,x_2) \equiv \langle \phi_1(x_1) \phi_2(x_2) \rangle$, then
\begin{equation}
\begin{split}
			G(x_1,x_2)& =G(x_1+a,x_2+a)\\
			&= G(x_1+a-x_2-a,x_2+a-x_2-a)\\
			& = G(x_1-x_2,0).
\end{split}
\end{equation}
Thus the propagator depends on wholly in the difference between $x_1 \ \& \ x_2$.
\item From the Lorentz transformations we get that
\begin{equation}
\begin{split}
\langle \phi_1(x^\mu_1) \phi_2(x_2^\nu) \rangle & =\abs{\Lambda^\mu_{\ \rho}}^{\Delta_1 } \abs{\Lambda^\nu_{\ \sigma}}^{\Delta_2 } \langle \phi_1(\Lambda^\mu_{\ \rho}x^\rho_1) \phi_2(\Lambda^\nu_{\ \sigma}x_2^\sigma) \rangle \\
& = \big(\det{\Lambda^\mu_{\ \rho}}\big)^{\Delta_1 } \big(\det{\Lambda^\nu_{\ \sigma}}\big)^{\Delta_2 } \langle \phi_1(\Lambda^\mu_{\ \rho}x^\rho_1) \phi_2(\Lambda^\nu_{\ \sigma}x_2^\sigma) \rangle \\
& = \langle \phi_1(\Lambda^\mu_{\ \rho}x^\rho_1) \phi_2(\Lambda^\nu_{\ \sigma}x_2^\sigma) \rangle,
\end{split}
\end{equation}
since $\det\Lambda = 1$. 
Combining the first two transformations, we know of one quantity that depends on the difference between two points, and it is also Lorentz invariant. This is the absolute value of the interval between two different spacetime points, i.e. $\abs{x_1 -x_2} \equiv \sqrt{\eta_{\mu \nu}\big(x_1^\mu - x_2^\mu\big)\big(x_1^\nu-x_2^\nu\big)}$.

It is obvious that this quantity depends on the difference between $x_1 \ \& \ x_2$. It is not hard to see that if we pick a different frame, let's say $x'=\Lambda x$ we have,
\begin{equation}
\begin{split}
			\abs{x_1' -x_2'} & = \sqrt{\eta_{\mu \nu}\big(x_1^{'\mu} - x_2^{'\mu} \big)\big(x_1^{'\nu} -x_2^{'\nu} \big)} \\
			& = \sqrt{\eta_{\mu \nu} \Lambda^\mu_{\ \rho}\Lambda^\nu_{\ \sigma} \big(x_1^\rho - x_2^\rho \big)\big(x_1^\sigma -x_2^\sigma \big)}\\
			& = \sqrt{\eta_{\rho \sigma}  \big(x_1^\rho - x_2^\rho \big)\big(x_1^\sigma -x_2^\sigma \big)}\\
			& = \abs{x_1 -x_2}.
	\end{split}
\end{equation}
From here we can write that the most general form of the two-point function up to this point and up to trivial coefficients is
\begin{equation}\label{generalform}
	\langle \phi_1(x_1) \phi_2(x_2) \rangle = C_{12} {\abs{x_1 -x_2}}^\alpha,
\end{equation}
An important note that has to be made here, is that the result for the first two symmetries is the same, independently of the order of the function, \emph{i.e.} the result holds for two-point functions and for $n$-point functions. This has to do with the fact that the Jacobian equals one. As we will see, this is not the case for dilatations and SCTs as these behave differently for two-point functions, three-point functions, etc, and it is these two transformations that put such firm constraints in the form of the two and three point functions.
\item From the dilatation invariance we have that:
\begin{equation}\label{1}
\begin{split}
\langle \phi_1(x_1) \phi_2(x_2)) \rangle & = \lambda^{\Delta_1}\lambda^{\Delta_2}\langle \phi_1(\lambda x_1) \phi_2(\lambda x_2) \rangle \\
& = \lambda^{{\Delta_1}+{\Delta_2}}\langle \phi_1(\lambda x_1) \phi_2(\lambda x_2) \rangle.
\end{split}
\end{equation}
But, we can use explicitly \cref{generalform} to find that for $x' = \lambda x$
\begin{equation}\label{2}
\begin{split}
\langle \phi_1(\lambda x_1) \phi_2(\lambda x_2) \rangle & = C_{12} {\abs{\lambda x_1 - \lambda x_2}}^\alpha \\
& = C_{12}{\sqrt{\eta_{\mu \nu}\big(\lambda x_1^{\mu} - \lambda x_2^{\mu} \big)\big(\lambda x_1^{\nu} - \lambda x_2^{\nu} \big)}}^\alpha \\
& = \lambda^\alpha C_{12} \sqrt{\eta_{\mu \nu}\big(x_1^{\mu} - x_2^{\mu} \big)\big(x_1^{\nu} -x_2^{\nu} \big)}\\
& = \lambda^\alpha C_{12} {\abs{x_1 -x_2}}^\alpha\\
& = \lambda^\alpha \langle \phi_1(x_1) \phi_2(x_2) \rangle.
\end{split}
\end{equation}
But now, using \cref{1,2}, for the equality to stand, we see that the following condition must hold
\begin{equation}
\alpha = - \big( {\Delta_1}+{\Delta_2} \big).
\end{equation}
And so, we can write the two-point function as:
\begin{equation}
			\langle \phi_1(x_1) \phi_2(x_2) \rangle = \frac{C_{12}}{{\abs{x_1 -x_2}}^{{\Delta_1}+{\Delta_2}}}.
\end{equation}
\item Applying the SCT and using that $\abs{\frac{\partial x'}{\partial x}} = \frac{1}{{(1-2bx+b^2 x^2)}^d} = \frac{1}{{\gamma_i}^d}$, we get the following
\begin{equation}\label{SCTapply}
	\begin{split}
	\langle \phi_1(x_1) \phi_2(x_2) \rangle & = {\bigg( \frac{1}{{(1-2bx+b^2 x^2)}^d} \bigg)}_{x=x_1}^{\Delta_1 /d} {\bigg( \frac{1}{{(1-2bx+b^2 x^2)}^d} \bigg)}_{x=x_2}^{\Delta_2 /d} \langle \phi_1(x^{'}_1)\phi_2(x^{'}_2) \rangle\\
	& = \frac{1}{{\gamma_1}^{\Delta_1}}\frac{1}{{\gamma_2}^{\Delta_2}}\frac{C_{12}}{{\abs{x^{'}_1 -x^{'}_2}}^{{\Delta_1}+{\Delta_2}}}. 
	\end{split}
\end{equation}
At this point, we should calculate how the absolute value of the interval between the two points changes under SCT.
\begin{equation}\label{SCTscale}
	\begin{split}
	\abs{x^{'}_1 - x^{'}_2} & = \sqrt{\eta_{\mu \nu}\big(x_1^{'\mu} - x_2^{'\mu} \big)\big(x_1^{'\nu} -x_2^{'\nu} \big)}\\
	& = \sqrt{x_1^{'\mu}x_{1 \ \mu}^{'} - x_1^{'\mu}x_{2 \ \mu}^{'} - x_2^{'\mu}x_{1 \ \mu}^{'}+x_2^{'\mu} x_{2 \ \mu}^{'}}\\
	& = \sqrt{\frac{\big( x_1^\mu -b^\mu x_1^{\ 2} \big)}{\gamma_1} \frac{\big( x_{1\mu} -b_\mu x_1^{\ 2} \big)}{\gamma_1} + \frac{\big( x_2^\mu -b^\mu x_2^{\ 2} \big)}{\gamma_2} \frac{\big( x_{2\mu} -b_\mu x_2^{\ 2} \big)}{\gamma_2} - \frac{\omega}{\gamma_1 \gamma_2}}\\
	& = \sqrt{\frac{x_1^{\ 2} (1-2bx_1 + b^2x_1^{ \ 2})}{\gamma_1 (1-2bx_1 + b^2x_1^{\ 2})} + \frac{x_2^{\ 2} (1-2bx_2 + b^2x_2^{\ 2})}{\gamma_2 (1-2bx_2 + b^2 x_2^{\ 2})}  -\frac{\omega}{\gamma_1 \gamma_2} }\\
	& = \sqrt{\frac{x_1^{\ 2}}{\gamma_1} + \frac{x_2^{\ 2}}{\gamma_2}-\frac{\omega}{\gamma_1 \gamma_2}}\\
	& = \sqrt{\frac{x_1^{\ 2} \gamma_2 + x_2^{\ 2} \gamma_1 -\omega }{\gamma_1 \gamma_2}}\\
	& = \sqrt{\frac{x_1^{\ 2} + x_2^{\ 2} - 2 x_1 x_2}{\gamma_1 \gamma_2}}\\
	& = \frac{\abs{x_1 -x_2}}{\sqrt{\gamma_1 \gamma_2}},
\end{split}
\end{equation}
where we set that $\omega = 2 \big( x_1^\mu -b^\mu x_1^{\ 2} \big) \big( x_{2\mu} -b_\mu x_2^{\ 2} \big) $. Hence, we can find that 5
\begin{equation}\label{SCTtransformation}
	\frac{1}{{\abs{x^{'}_1 - x^{'}_2}}^{\Delta_1 + \Delta_2}}= \frac{1}{{\abs{x_1 - x_2}}^{\Delta_1 + \Delta_2}} \gamma_1^{\frac{\Delta_1 + \Delta_2}{2}} \gamma_2^{\frac{\Delta_1 + \Delta_2}{2}} 
\end{equation}
Hence, plugging \cref{SCTtransformation} into \cref{SCTapply} we find that
\begin{equation}
	\langle \phi_1(x_1) \phi_2(x_2) \rangle = \frac{C_{12}}{{\abs{x_1 - x_2}}^{\Delta_1 + \Delta_2}} \gamma_1^{\frac{\Delta_1 + \Delta_2}{2}} \gamma_2^{\frac{\Delta_1 + \Delta_2}{2}}\frac{1}{{\gamma_1}^{\Delta_1}}\frac{1}{{\gamma_2}^{\Delta_2}}.
\end{equation}
Thus, we have the following matching conditions
\begin{itemize}
\item $\frac{\Delta_1 + \Delta_2}{2} - \Delta_1 = 0$,
\item $\frac{\Delta_1 + \Delta_2}{2} - \Delta_2 = 0$,
\end{itemize}
which have the unique solution that 
\begin{equation}
			\Delta_1 = \Delta_2 = \Delta.
\end{equation}
\end{enumerate}
The reason that we underwent so much trouble, is that by doing the full analysis in the simplest case, it is much easier to proceed for the three- and afterwards the four-point functions. 

So, in the same spirit as before, we should start with \cref{generaltransform}, but with three fields now, and repeat all the steps. But, we have observed that the translation and Lorentz invariance apply the same to all $n$ point functions, and thus we can deduce that the three point function is determined by the absolute value of the interval between the three points, in some power, in other words
\begin{equation}\label{general3point}
		\langle \phi_1(x_1) \phi_2(x_2) \phi_3(x_3) \rangle = C_{123} {\abs{x_1-x_2}}^\alpha {\abs{x_2-x_3}}^b {\abs{x_1-x_3}}^c,
\end{equation}
where from now on we will use $\abs{x_{12}} = \abs{x_1 - x_2}$ for shortness. Let's see now what happens under dilatations and SCTs.
\begin{enumerate}
\item From dilatation, we have
\begin{equation}
			\begin{split}
			\langle \phi_1(x_1) \phi_2(x_2) \phi_3(x_3) \rangle & = \lambda^{\Delta_1 + \Delta_2 + \Delta_3} \langle \phi_1(\lambda x_1) \phi_2(\lambda x_2) \phi_3(\lambda x_3) \rangle \\
			& = \lambda^{\Delta_1 + \Delta_2 + \Delta_3} C_{123} {\abs{\lambda x_{12}}}^\alpha {\abs{\lambda x_{23}}}^b {\abs{\lambda x_{13}}}^c\\
			& = \lambda^{\Delta_1 + \Delta_2 + \Delta_3+\alpha+b+c} C_{123} {\abs{ x_{12}}}^\alpha {\abs{ x_{23}}}^b {\abs{ x_{13}}}^c\\
			& =\lambda^{\Delta_1 + \Delta_2 + \Delta_3+\alpha+b+c} \langle \phi_1(x_1) \phi_2(x_2) \phi_3(x_3) \rangle.
\end{split}
\end{equation}
Thus, we obtain the constraint that
\begin{equation}\label{constraint3}
\Delta_1 + \Delta_2 + \Delta_3+\alpha+b+c=0 \Leftrightarrow \Delta_1 + \Delta_2 + \Delta_3= -\alpha - b -c.
\end{equation}
\item Under SCT, and following \cref{SCTapply}, we have that
\begin{equation}
\begin{split}
\langle \phi_1(x_1) \phi_2(x_2) \phi_3(x_3) \rangle & = \frac{1}{{\gamma_1}^{\Delta_1}}\frac{1}{{\gamma_2}^{\Delta_2}}\frac{1}{{\gamma_3}^{\Delta_3}} C_{123} {\abs{ x^{'}_{12}}}^\alpha {\abs{ x^{'}_{23}}}^b {\abs{ x^{'}_{13}}}^c\\
& = \frac{C_{123}}{{\gamma_1}^{\Delta_1}{\gamma_2}^{\Delta_2}{\gamma_3}^{\Delta_3}} \frac{{\abs{ x_{12}}}^\alpha}{{(\gamma_1 \gamma_2)}^{\alpha /2 }} \frac{{\abs{ x_{23}}}^b}{{(\gamma_2 \gamma_3)}^{b /2 }} \frac{{\abs{ x_{13}}}^c}{{(\gamma_1 \gamma_3)}^{c /2 }}.
\end{split}
\end{equation}
and to go from the second to the third line, we used \cref{SCTscale}. From this, we find the following three constraints
\begin{gather}
\Delta_1 + \alpha/2 + c/2 =0,\\
\Delta_2 + \alpha /2 + b/2 = 0,\\
\Delta_3 + b/2 +c/2 =0.
\end{gather}
We can solve for $\alpha, b, c$ to get
\begin{gather}
\alpha = \Delta_3 - \Delta_1 - \Delta_2,\\
b = \Delta_1 - \Delta_2 -\Delta_3,\\
c = \Delta_2 - \Delta_1 - \Delta_3.
\end{gather}
\end{enumerate}
We can check explicitly that this unique set of solutions satisfies \cref{constraint3}. Using these, the final form of the three-point function is
\begin{equation}
\langle \phi_1(x_1) \phi_2(x_2) \phi_3(x_3) \rangle  = \frac{C_{123}}{\abs{ x_{12}}^{\Delta_1 + \Delta_2 - \Delta_3} \abs{ x_{23}}^{\Delta_2 + \Delta_3 - \Delta_1} \abs{ x_{13}}^{\Delta_1 + \Delta_3 - \Delta_2} }.
\end{equation}
Since we have normalized the fields for the two-point function, we cannot normalize them again, hence $C_{123}$ is an important part of the CFT. So, we can see that if we know the scaling dimension of the fields (which can be calculated through the two-point function), the only unknown part of the three-point function are the OPE coefficients. So, once more, we observe that the conformal transformation imposes some strong constraints on the form of the three-point function. 

For the four-point function, again we can use the same logic as for the three-point function, and by using translation and Lorentz invariance, we can write it in the form
\begin{equation}
		\langle \phi_1(x_1) \phi_2(x_2) \phi_3(x_3) \phi_4(x_4) \rangle = C_{1234} \abs{ x_{12}}^\alpha \abs{ x_{13}}^b \abs{ x_{14}}^c \abs{ x_{23}}^d \abs{ x_{24}}^\epsilon \abs{ x_{34}}^z.
\end{equation}
As we will soon see, the coefficients in front are not as innocent as the ones of the two and three-point functions. For now, we can concentrate on the dilatation invariance, which is the same as the three point function but with more coefficients. Thus,
\begin{equation}
		\langle \phi_1(x_1) \phi_2(x_2) \phi_3(x_3) \phi_4(x_4) \rangle = \\
		\lambda^{\Delta_1 + \Delta_2 + \Delta_3 + \Delta_4 + \alpha +b + c+ d + \epsilon +z} \langle \phi_1(x_1) \phi_2(x_2) \phi_3(x_3) \phi_4(x_4) \rangle,
\end{equation}
which gives the constraint 
\begin{equation}
		\Delta_1 + \Delta_2 + \Delta_3 + \Delta_4 = -\big( \alpha +b + c+ d + \epsilon +z \big).
\end{equation}
From the SCT invariance, and using \cref{SCTscale} we have
\begin{multline}
		\langle \phi_1(x_1) \phi_2(x_2) \phi_3(x_3) \phi_4(x_4) \rangle = \\
		\frac{C_{1234}}{{\gamma_1}^{\Delta_1}{\gamma_2}^{\Delta_2}{\gamma_3}^{\Delta_3}{\gamma_4}^{\Delta_4}} \frac{{\abs{ x_{12}}}^\alpha}{{(\gamma_1 \gamma_2)}^{\alpha /2 }} \frac{{\abs{ x_{13}}}^b}{{(\gamma_1 \gamma_3)}^{b /2 }} \frac{{\abs{ x_{14}}}^c}{{(\gamma_1 \gamma_4)}^{c /2 }} \frac{{\abs{ x_{23}}}^d}{{(\gamma_2 \gamma_3)}^{d /2 }} \frac{{\abs{ x_{24}}}^d}{{(\gamma_2 \gamma_4)}^{\epsilon /2 }}\frac{{\abs{ x_{34}}}^d}{{(\gamma_3 \gamma_4)}^{z /2 }}.
\end{multline}
By matching the gamma, we derive the following conditions
\begin{gather}\label{set1}
		\alpha + b+c = - 2 \Delta_1,\\\label{set2}
		\alpha + d + \epsilon = - 2 \Delta_2,\\\label{set3}
		b+d+z = - 2 \Delta_3,\\ \label{set4}
		c+\epsilon+ z = - 2 \Delta_4.
\end{gather}
But this set of equations is impossible to be solved as we have six unknowns for four equations. But for the four-point function, there is a catch. When having four points and more, it is possible to create certain coefficients that preserve the CFT symmetries. The trick is to use the absolute value of the interval of two points, which is a priori Lorentz and translation invariant. Thus, the catch is to find the correct combination that will preserve dilatation and SCT invariance. It is clear that this is not the case for two and three point functions. The two-point function is too simple, as it contains only one interval. For the three-point function it is impossible to create any coefficient that is invariant under dilatation, e.g. for $\chi = \frac{\abs{x_{12}}\abs{x_{23}}}{\abs{x_{13}}}$ scales like $\lambda$ under dilatations, $\zeta = \frac{\abs{x_{12}}}{\abs{x_{13}}}$ is invariant under dilatations but it is not invariant under SCT, etc. But for the four point function we can create some cross ratios, which are
\begin{equation}
		\chi_1 = \frac{\abs{x_{12}}\abs{x_{34}}}{\abs{x_{13}}\abs{x_{24}}}, \ \ \ \chi_2 = \frac{\abs{x_{12}}\abs{x_{34}}}{\abs{x_{23}}\abs{x_{14}}}.
\end{equation}
The dilatation invariance is obvious. What is interesting is that in order for $\chi_1 \ \& \ \chi_2$ to be SCT invariant, we gain two more equations or, more precisely, constraints (remember from our analysis above, we had more unknowns than equations, which meant that we had some freedom in the scaling). Thus, from $\chi_1 \ \& \ \chi_2$ if we SCT transform them as usual, we derive that 
\begin{gather}
		b + \epsilon = \alpha + z,\\
		d + c = \alpha + z.
\end{gather}
These two, combined with equations \cref{set1,set4} lead to
\begin{gather}
		2\alpha + z - \epsilon +c = - 2 \Delta_1\\
		2\alpha +z -c +\epsilon = -2 \Delta_2\\
		2\alpha + 3z - c - \epsilon = - 2 \Delta_3\\
		c + \epsilon + z = - 2 \Delta_4
\end{gather}
Now it is obvious that we have the correct number of unknowns and equations, thus by solving the system we get the following results
\begin{gather}
		\alpha = -\frac{2 \Delta_1}{3}-\frac{2 \Delta_2}{3}+\frac{ \Delta_3}{3}+ \frac{ \Delta_4}{3}\\
		b= -\frac{2 \Delta_1}{3}     - \frac{2 \Delta_3}{3} + \frac{ \Delta_4}{3}+ \frac{ \Delta_2}{3}\\
		c=   -\frac{2 \Delta_1}{3} - \frac{2 \Delta_4}{3} + \frac{ \Delta_2}{3}   + \frac{\Delta_3}{3}           \\
		d=   -\frac{2 \Delta_2}{3} - \frac{2 \Delta_3}{3} + \frac{ \Delta_4}{3} + \frac{ \Delta_1}{3}\\
		\epsilon = -\frac{2 \Delta_2}{3} -\frac{2 \Delta_4}{3} + \frac{\Delta_1}{3} + \frac{ \Delta_3}{3}\\
		z= -\frac{2 \Delta_3}{3} - \frac{2 \Delta_4}{3} + \frac{ \Delta_1}{3} + \frac{\Delta_2}{3}.
\end{gather}
In short notation, these can be written as 
\begin{equation}
		\Delta/3 - \Delta_i -\Delta_j , \ \ \ \Delta= \Sum_{i=1}^4 \Delta_i.
\end{equation}
Hence, the four-point function can be written in terms of conformal blocks
\begin{equation}
		\langle \phi_1(x_1) \phi_2(x_2) \phi_3(x_3) \phi_4(x_4) \rangle = \mathcal{F}\big( \chi_1, \chi_2 \big) \mathlarger{\mathlarger{\prod}}_{i < j}^4 x_{ij}^{\Delta/3 - \Delta_i -\Delta_j},
\end{equation}
where $\mathcal{F}\big( \chi_1, \chi_2 \big)$ is a function of all possible cross ratios.


\section{Constraints from conformal symmetry}
\label{sec:constraints}

To use the state-operator correspondence, we worked in the cylinder frame in the limit of infinite separation, and the two of the insertions are taken to be at $\tau = \pm \infty$.
For operators with spin, it is more convenient to work in the spherical tensor basis.

An object that transforms into an irreducible representation of \(SO(d)\) is in the usual Cartesian basis expressed by a completely symmetric and traceless tensor \(T_{\nu_1 \dots \nu_{\ell}}\).
But in the spherical basis, the pair \(\ell, m\) labels the same object.
To pass over we apply the operator
\(\Proj{^{\nu_1 \dots \nu_\ell}_{\ell m}}{}\), that we can write as an integral on the sphere as
\begin{equation}\label{eq:ProjectorToSphericalBasis}
  \Proj{^{\nu_1 \dots \nu_\ell}_{\ell m}}{} = k_{d, \ell}  \Int \dd{\Omega} n^{\nu_1} \dots n^{\nu_{\ell}} Y^*_{\ell m}(\n)  \, ,
\end{equation}
where \(k_{d,\ell}\) is a normalisation factor that arises by requiring that \(\Proj{}\) squares to one,
\begin{equation}
  \abs{\Proj_{\ell m}{}}^2 = \delta_{\mu_1 \nu_1} \dots \delta_{\mu_{\ell} \nu_{\ell}} \pqty{ \Proj{^{\nu_1 \dots \nu_\ell}_{\ell m}}}^{*} \Proj{^{\mu_1 \dots \mu_\ell}_{\ell m}} = 1,
\end{equation}
which reads
\begin{equation}
  k_{d,\ell} = \sqrt{\frac{2^{\ell} }{ \Omega_d} \frac{ \Gamma \left( \frac{d}{2} + \ell\right) }{ \ell! \,\, \Gamma \left( \frac{d}{2} \right) } } \, .
\end{equation}
The simplest non-trivial example is for the vector \(V_{\mu}\) in \(d =3 \), that is mapped to \(V_{1m}\) with components
\begin{equation}
  \begin{pmatrix}
    V_{1,-1} \\
    V_{1,0} \\
    V_{1,1}
  \end{pmatrix}
  = 
  \begin{pmatrix}
    - \frac{1}{\sqrt{2}} \pqty*{ V_1 + i V_2 } \\
    V_3 \\
    \frac{1}{\sqrt{2}} \pqty*{ V_1 - i V_2 }
  \end{pmatrix}.
\end{equation}
We know that the two-point correlation function of two primary operators is non-zero only when they feature the same scaling dimension \(\Delta\), and also they transform in conjugate representations. 
On the cylinder frame, in the limit \(\tau_\text{out} - \tau_\text{in} \gg 1\) we get, up to a normalization
\begin{equation}\label{eq:TwoPointCylinder}
  \expval{ \Opp*[q][\Delta][\ell \bar m](\tau_\text{out}, \n_2) \Opp[q][\Delta][\ell m](\tau_\text{in}, \n_1) } = e^{-(\tau_\text{out} - \tau_\text{in}) \Delta/R_0}  I^{\ell}_{m \bar m}(\n_2) \coloneq \Anew I^{\ell}_{m \bar m}(\n_2),
\end{equation}
and we used that in the infinite-separation limit the unit vector evaluated at the direction of the separation is
\begin{equation}
  \n = \frac{x-y}{\abs{x-y}} = \frac{  e^{\tau_\text{out}/R_0} \n_2 - e^{\tau_\text{in}/R_0} \n_1 }{|e^{\tau_\text{out}/R_0}\n_2 - e^{\tau_\text{in}/R_0} \n_1|} \overset{\tau_{\text{out},\text{in}} \to \pm \infty}{\longrightarrow} \n_2 .
\end{equation}
Where \(I^{\ell}_{m \bar m}\) is the intertwiner between the two representations

\begin{equation}
  I^{\ell}_{m \bar m}(\n) = \delta_{m \bar m} - \frac{2 \Omega_d}{M_{\ell}} Y^{*}_{\ell \bar m}(\n) Y_{\ell m}(\n) .
\end{equation}
For three point functions, in the limit of large separation $\tau_{\text{in},\text{out}} \to \mp \infty$ on the cylinder the result does not depend on the scaling dimension of the middle operator \(\Delta_c\)
\begin{equation}\label{eq:limit3pointCFT}
         \expval{ \Opp[2][\Delta_2] \Opp[c][\Delta] \Opp[1][\Delta_1] } \longrightarrow \mathcal{C}_{\mathscr{O}_1c\mathscr{O}_2} e^{- \Delta_2 ( \tau_\text{out} - \tau)} e^{-\Delta_1(\tau - \tau_\text{in})} =  \Anew[\Delta_1][\Delta_2][\tau] \mathcal{C}_{\mathscr{O}_1c\mathscr{O}_2}.
\end{equation}
For correlators including operators with spin, the scalar part does not change and is the same as in \cref{eq.generalform3pointfunction,eq.formof3pointfunction} but it is supplemented by an appropriate tensor structure~\cite{osborn1994implications,costa2011spinning}.
For instance, in the scalar--scalar--spin-\(\ell\) correlation function we need to multiply by \((V^{(ijk)} \cdot t)^{\ell}\), where
\begin{equation}
    V^{(ijk)} = \frac{\abs{x_{ki}} \abs{x_{kj}}}{\abs{x_{ij}}} \pqty{ \frac{x_{ki}}{\abs{x_{ki}}^2} - \frac{x_{kj}}{\abs{x_{kj}}^2} } ,
\end{equation}
and  $t$ is a supplementary vector which squares to zero, $t^2=0$, to guarantee the tracelessness of \(V^{(ijk)}\).
This item has an especially simple expression as a spherical tensor.
Observing that \(\Proj{^{\nu_1 \dots \nu_\ell}_{\ell m}}{}\) is antisymmetric and also traceless by construction, we do not need to deduct any traces, and we just need to calculate one integral
\begin{multline}
    V^{(ijk)}_{\ell m} = \Proj{^{\mu_1 \dots \mu_\ell}_{\ell m}} {} V^{(ijk)}_{\mu_1} \dots V^{(ijk)}_{\mu_\ell} = k_{\ell , d} \int \dd{\Omega} Y^*_{\ell m}(\n) \pqty{\n \cdot V^{(ijk)}}^\ell \\
    = \frac{1}{k_{\ell, d}} \frac{ \abs{ \abs{x_{kj}}^2 x_{ki} - \abs{x_{ki}}^2 x_{kj}}^{\ell }}{ \abs{x_{ij}}^{\ell} \abs{x_{ki}}^{\ell} \abs{x_{kj}}^{\ell} }  Y^*_{\ell m} \pqty{ \frac{  \abs{x_{kj}}^2 x_{ki} - \abs{x_{ki}}^2 x_{kj}  }{  \abs{ \abs{x_{kj}}^2 x_{ki} - \abs{x_{ki}}^2 x_{kj}} } } .
\end{multline}
In the infinite separation limit, writing \(x_i = R_0 e^{\tau_\text{out}/R_0} \n_2\), \(x_j = R_0 e^{\tau_\text{in}/R_0} \n_1\),  \(x_k = R_0 e^{\tau/R_0} \n\),  we deduce that
\begin{equation}
    V^{(ijk)}_{\ell m} =  \frac{1}{k_{\ell, d}} Y^{*}_{\ell m}(\n) \pqty*{1 + \order{e^{-(\tau_\text{out} - \tau )/R_0}}} ,
\end{equation}
as expected.


\chapter[\texorpdfstring%
{Preliminaries for the \(O(2)\) model}%
{O(2)}]%
{Preliminaries for the \(O(2)\) model}%

\label{AppendixB}


\section{Hyperspherical harmonics and their properties}
\label{sec:Ylm-identities}

We collect useful formulas related to spherical harmonics in \(d\) dimensions~\cite{avery2017hyperspherical}.

The hyperspherical harmonic \(Y_{\ell m}\) is the eigenfunction of the Laplacian on \(S^{d-1}\)
\begin{equation}
  -\Laplacian_{S^{d-1}} Y_{\ell m} (\n) = \ell ( \ell + d - 2 ) Y_{\ell m} (\n) ,
\end{equation}
where \(\ell = 0, 1, \dots \) and \(m \) is a \( d - 2\) components vector that satisfies
\begin{equation}
l \ge m_1 \ge m_2 \ge \dots \ge m_{d-3} \ge \abs{m_{d-2}}.	
\end{equation}
We note that the lowest component \(m_{d-2}\) is related to the usual \(SO(3)\) quantum number and this is the sole component that can feature a negative sign.
We denote \(m^*\) the vector which has the sign of \(m_{d-2}\) flipped.
This is present in the conjugation property
\begin{equation}
	Y_{\ell m }^* = (-1)^{m_{d-2}} Y_{\ell m^*}  .
\end{equation}
The above eigenvalue is independent of \(m\), and has the multiplicity
\begin{equation}\label{eq:laplacian_degenerancy}
  M_{\ell} = \frac{(d + 2 \ell - 2) \Gamma(d + \ell - 2)}{\Gamma(\ell + 1) \Gamma(d - 1)} \, .
\end{equation}
Given that the Laplacian is self-adjoint, the \(Y_{\ell m}\) form an orthonormal basis for \(L^2(S^{d - 1})\)
\begin{equation}
  \scalar{ Y_{\ell m}, Y_{\ell' m'}} = \Int_{S^{d -1}} \dd{\Omega} Y_{\ell m}(\n) Y^{*}_{\ell'  m' }(\n) = \delta_{\ell \ell'} \delta_{m m'} \, .
\end{equation}
We define the resized versions of the volume element and of the eigenvalues of the Laplacian as
\begin{align}
  \dd{S} &= R_0^{d-1} \dd{\Omega}, & \omega_{\ell}^2 &= \frac{\ell (\ell + d - 2)}{(d-1) R_0^2}. 
\end{align}
Some helpful identities that are obtained summing over the indices \(m\) read
\begin{align}\label{eq:Y_orthogonality}
  \Sum_m Y_{\ell m}(\n) Y^{*}_{\ell m}(\n) &= \frac{M_{\ell}}{\Omega_d} \, ,\\
  \Sum_m Y_{\ell m}(\n) \del_i Y^{*}_{\ell m}(\n) &= 0 \, ,\\
  \Sum_m \del_i Y_{\ell m}(\n) \del_j Y^{*}_{\ell m}(\n) &= \frac{M_{\ell}}{\Omega_d} ( R_0 \omega_{\ell} )^2 h_{ij}(\n) \, ,
\end{align}
where with $\Omega_d = \frac{2 \pi^{d/2}}{\Gamma(d/2)} $ we denote the volume of the \(d - 1\) sphere. 

Sums involving the eigenvalues $\omega_\ell$ can be written as
\begin{equation}
  \Sum_{\ell, m} \omega_{\ell}^s Y_{\ell m}(\n) Y^{*}_{\ell m}(\n)  = \frac{\Sigma(s)}{\Omega_d R_0^s}= \frac{\zeta( - \sfrac{s}{2} \mid S^{d-1} ) }{(d - 1)^{s/2} R_0^s \Omega_d} \, ,
\end{equation}
and the $\Lambda$-independent part of $\Sigma(s)$ is associated to the zeta-function on the sphere~\cite{monin2017semiclassics}. 
\begin{equation}
  \zeta(s \mid S^{d-1}) = \Tr[ (-\Laplacian_{S^{d-1}}{})^s] \, .
\end{equation}
For the special case of \(s =1\) we retrieve the Casimir energy of a free scalar 
\begin{equation}
  \Sum_{\ell, m} \omega_{\ell} Y_{\ell m}(\n) Y^{*}_{\ell m}(\n) = \frac{\Sigma(1)}{\Omega_D R_0} = \frac{2 \Delta_1}{\Omega_d R_0} \ .
\end{equation}

In a similar manner, sums having open derivative indexes can be calculated as
\begin{equation}
  \Sum_{\ell, m} \omega_{\ell}^s \del_i Y_{\ell m}(\n) \del_j Y^{*}_{\ell m}(\n) = \frac{\Sigma(s+2)}{\Omega_d R_0^s} h_{ij} =  \frac{\zeta(-\sfrac{s}{2} - 1 \mid S^{d - 1})}{(d-1)^{s/2 + 1}  R_0^s \Omega_d} h_{ij} \, ,
\end{equation}
and, for \(s = -1\)
\begin{equation}
  \Sum_{\ell,m} \frac{1}{\omega_{\ell}} \del_i Y_{\ell m}(\n) \del_j Y^{*}_{\ell m}(\n) = \frac{R_0 \Sigma(1)}{\Omega_d} h_{ij} = \frac{2 R_0 \Delta_1}{\Omega_d} h_{ij} \, .
\end{equation}
Finally, monomials can be expressed in terms of Gegenbauer polynomials
\begin{equation}
    (\n \cdot \n')^\ell = \frac{ \ell!}{2^\ell} \Sum_{s=0}^{\floor{\frac{\ell}{2}}} \frac{ \left( \frac{d}{2} - 1 + \ell -2s\right) \Gamma \left( \frac{d}{2} -1 \right)}{ s! \, \Gamma \left( \frac{d}{2} + \ell -s \right)} C^{d/2 -1}_{\ell-2s} (\n \cdot \n') .
\end{equation}
Additionally, the Gegenbauer polynomials exhibit an addition property of the form
\begin{equation}\label{eq:Gegenbauer_addition}
C_{\ell_a}^{d/2-1}(\n\cdot \n') C_{\ell_b}^{d/2-1}(\n\cdot \n') =\Sum_{k=0}^{\min(\ell_a, \ell_b)} \braket{k}{\ell_a \ell_b} C^{d/2-1}_{\ell_a + \ell_b -2k}(\n \cdot \n') ,
\end{equation}
and the coefficients $\braket{ k}{\ell_a \ell_b}$  read
\begin{multline}
 \braket{k}{\ell_a \ell_b } = \left( \frac{d}{2} -1 - 2 k + \ell_a + \ell_b \right) \frac{\Gamma(\ell_a + \ell_b +1 -2k ) }{\Gamma\left( \frac{d}{2} -1 \right)^2 \Gamma(\ell_a + \ell_b -2k + d-2)} \\
\times \frac{\Gamma\left( \frac{d}{2} + k -1 \right) 
\Gamma(\ell_a + \ell_b - k + d-2) \Gamma\left( \ell_a - k + \frac{d}{2} -1 \right) \Gamma\left( \ell_b - k + \frac{d}{2} -1 \right) }{ \Gamma(k+1)  \Gamma\left( \ell_a + \ell_b -k +\frac{d}{2} \right) \Gamma(\ell_a - k +1 ).\Gamma(\ell_b -k +1) } .
\end{multline}
The above is a $SO(d)$ generalisation of angular momentum addition in $d=3$ spacetime dimensions.
%


\section{The Goldstone propagator}\label{sec.Goldstonepropagator}

We consider the eigenvalue problem
\begin{equation}
    \left(- \partial_\tau^2 - \frac{1}{(d-1)R_0^2} \Delta_{\setS^{d-1}} \right) \phi_n(\tau,x) = \lambda_n \phi_n(\tau,x)
\end{equation}
The solutions read
\begin{align}
    &\left\{ e^{\pm i\alpha \tau} Y_\ell^m(x) \,\Big| \,  \left(\alpha^2 + \omega_\ell^2 \right) ; \,\, \ell \in \mathbb{N} ,\, \alpha \in \mathbb{R}^+ \right\} ,&  &\omega_\ell = \frac{1}{R_0} \sqrt{\frac{\ell(\ell +d-2)}{d-1}}
\end{align}
The eigenset can be formally expressed on the cylinder where we can impose Dirichlet or Neumann boundary conditions
\begin{align}
 D: & \left\{ \sqrt{\frac{2}{\beta R_0^{d-1}}} \sin\left[ \frac{n\pi}{\beta} \left( \tau + \frac{\beta}{2} \right) \right] Y_\ell^m(x) \,\Big| \,  \left(\frac{n^2\pi^2}{\beta^2} + \omega_\ell^2 \right) ; \,\, \ell \in \mathbb{N} ,\, \alpha \in \mathbb{R}^+ \right\}  \\
 N: & \left\{  \sqrt{\frac{2}{\beta R_0^{d-1}}} \cos\left[ \frac{n\pi}{\beta} \left( \tau + \frac{\beta}{2} \right) \right] Y_\ell^m(x) \,\Big| \,  \left(\frac{n^2\pi^2}{\beta^2}  + \omega_\ell^2 \right) ; \,\, \ell \in \mathbb{N} ,\, \alpha \in \mathbb{R}^+ \right\}  
\end{align}
We have chosen such a normalisation such that both in both the boundary conditions we get
\begin{align}
&R_0^{d-1} \Int\limits_{-\beta/2}^{\beta/2} \dd \tau \Int \dd \Omega_d \, \phi_n^* \phi_m = \delta_{nm}, &  &\Sum_n \phi_n^*(\tau_1, x_1) \phi_n(\tau_2,x_2) = \delta_{I \times \setS^{d-1}}
\end{align}
We use the spectrum to compute the Green's function as usual
\begin{align}
&G(\tau_\text{in}, x_1 | \tau_\text{out} , x_2 ) = \Sum_n' \frac{\phi_n(\tau_\text{in},x_1)^* \phi_n(\tau_\text{out} , x_2)}{\lambda_n}, \\
&\left( -\partial_\tau^2 - \frac{1}{(d-1)R_0^2} \Delta_{\setS^{d-1}} \right) G = \delta_{I \times \setS^{d-1}} - \Sum_{n \,\in\, \rm z.m.} \phi_n^* \phi_n 
\end{align}
There are two zero modes
\begin{align}
&\frac{1}{\sqrt{\beta}} Y_0^0, & &\tau \sqrt{\frac{12}{\beta^3}} Y_0^0
\end{align}
which do not satisfy Dirichlet boundary conditions, while the first one satisfies Neumann boundary conditions. Excluding this zero modes produces the same double summation range for both boundary condition which can be split as
\begin{equation}
    \Sum_{(\ell,m),n\neq 0} \cdots \,\, = \underbrace{\Sum_{(\ell,m) \neq 0} \Sum_{n >0} \cdots \,}_{G^{(D,N)}_I} + \underbrace{|Y_0^0|^2 \Sum_{n\geq 1} \cdots}_{G^{(D,N)}_{II}} \ .
\end{equation}
Starting from
\begin{align}
     G^{(D)} &= \frac{2}{\beta} \underset{\tiny{n=m=n\neq 0}}{\Sum_{\ell,m}  \Sum_{n\in\mathbb{Z}}} \frac{\sin\left(\frac{\pi n }{\beta} \left(\tau_\text{in} + \frac\beta2\right) \right) \sin\left(\frac{\pi n }{\beta} \left(\tau_\text{out} + \frac\beta2\right) \right) }{\left(\frac{\pi^2 n^2 }{\beta^2} + \omega_\ell^2 \right)}  Y_\ell^m(x_1)^* Y_\ell^m(x_2) \\
     G^{(N)}  &= \frac{2}{\beta} \underset{\tiny{n=m=n\neq 0}}{\Sum_{\ell,m}  \Sum_{n\in\mathbb{Z}}} \frac{\cos\left(\frac{\pi n }{\beta} \left(\tau_\text{in} + \frac\beta2\right) \right) \cos\left(\frac{\pi n }{\beta} \left(\tau_\text{out} + \frac\beta2\right) \right) - \frac{1}{2} \delta_{n0} }{\left(\frac{\pi^2 n^2 }{\beta^2} + \omega_\ell^2 \right)}  Y_\ell^m(x_1)^* Y_\ell^m(x_2) \\
\end{align}
The propagator on the $I$ part can be computed separately as
\begin{align}
G^{(D,N)}_I &= \frac{1}{\Omega_d R_0^{d-1}} \frac{2\beta}{\pi^2} \Sum_{n \geq 1} \frac{1}{n^2} \left[ \cos\left( \frac{n\pi}{\beta}(\tau_\text{out}-\tau_\text{in}) \right) \mp \cos\left( \frac{n \pi}{\beta}(\tau_\text{in}+\tau_\text{out}+\beta) \right)  \right] \\
&= \frac{1}{\Omega_d R_0^{d-1}}\begin{cases}
\frac{\beta}{2}\left[ 1 - \frac{2|\tau_\text{out}-\tau_\text{in}|}{\beta} - \frac{4\tau_\text{out}\tau_\text{in}}{\beta^2} \right] \\
\frac{\beta}{6}\left[ 1 - \frac{6|\tau_\text{out}-\tau_\text{in}|}{\beta} + \frac{6(\tau_\text{out}^2 + \tau_\text{in}^2)}{\beta^2} \right]
\end{cases}
\end{align}
where we utilised that
\begin{align}
 &\Sum_{k\geq 1} \frac{\cos(kx)}{k^2} = \frac{\pi^2}{6} - \frac{\pi |x|}{2} + \frac{x^2}{4}, &  &-2\pi \leq x \leq 2\pi .
\end{align}
The double sum part reads
\begin{align}
  G^{(D,N)}_{II} &= \frac{1}{R_0^{d-1}} \frac{2\beta}{\pi^2 } \Sum_{{(\ell,m)\neq 0}} \left\{ \Sum_{n\geq 1} \frac{\cos\left(\frac{\pi n }{\beta} \left(\tau_\text{out} - \tau_\text{in}\right) \right) }{\left(n^2 + \frac{\beta^2}{\pi^2 } \omega_\ell^2 \right)} \mp \Sum_{n\geq1} \frac{\cos\left(\frac{\pi n }{\beta} \left(\tau_\text{in} + \tau_\text{out}+ \beta \right) \right) }{ \left( n^2 + \frac{\beta^2}{\pi^2 } \omega_\ell^2 \right) } + \begin{cases} 0 \\ \frac{\pi^2}{\beta^2 \omega_\ell^2}  \end{cases} \right\} Y_\ell^m(x_1)^* Y_\ell^m(x_2) \\
  &= \frac{1}{R_0^{d-1}} \Sum_{\ell \geq 1, m} \frac{Y_\ell^m(x_1)^* Y_\ell^m(x_2)}{\omega_\ell^2} \left\{ \frac{ \omega_\ell  \cosh \left[ \omega_\ell (\beta - | \tau_\text{out}-\tau_\text{in} |) \right] \mp \omega_\ell \cosh \left[ \omega_\ell (\tau_\text{in}+\tau_\text{out}) \right] }{\sinh \left(\beta \omega_\ell \right) } \right\}
\end{align}
where we utilised
\begin{align}
    &\Sum_{k\geq 1} \frac{\cos(kx)}{k^2+a^2} = \frac{\pi}{2a} \frac{\cosh\big(a(\pi-|x|)\big)}{\sinh(a\pi)} -\frac{1}{2a^2}, & &-2\pi \leq x \leq 2\pi.
\end{align}
We can also consider the $\setS^1$ propagator, for which only the constant zero mode has to be removed, as in the Neumann case of the segment. This reads
\begin{align}
G^{\setS^1} &= \left (\sum_{\ell \geq 1 , m , n} +\sum_{\ell = 0 , m , n \neq 0}\right) e^{ i \frac{n \pi}{\beta} (\tau_\text{out}-\tau_\text{in})} \frac{1}{\beta R_0^{d-1}}\frac{ Y_\ell^m(x_1)^* Y_\ell^m(x_2) }{\frac{n^2 \pi^2}{\beta^2} + \omega_\ell^2} \\
&= \Sum_{\ell \geq 1 , m , n \geq 0} \left\{ \cos \left(\frac{n \pi}{\beta} (\tau_\text{out}-\tau_\text{in}) \right) - \frac{1}{2} \delta_{n0} \right\} \frac{2}{\beta R_0^{d-1}}\frac{ Y_\ell^m(x_1)^* Y_\ell^m(x_2) }{\frac{n^2 \pi^2}{\beta^2} + \omega_\ell^2}  + \frac{2\beta}{ \pi^2 R_0^{d-1} \Omega_d} \Sum_{n \geq1} \frac{1}{n^2} \cos \left( \frac{n\pi}{\beta}(\tau_\text{out}-\tau_\text{in}) \right) \\
&= \frac{1}{2} \left( G^{(D)} + G^{(N)} \right) \\
&=  \frac{1}{R_0^{d-1}} \sum_{\ell \geq 1, m} \frac{Y_\ell^m(x_1)^* Y_\ell^m(x_2)}{\omega_\ell^2} \left\{ \frac{ \omega_\ell  \cosh \left[ \omega_\ell (\beta - | \tau_\text{out}-\tau_\text{in} |) \right] }{\sinh \left(\beta \omega_\ell \right) } \right\} + \frac{2 \beta}{ \Omega_d R_0^{d-1}} \left[ \frac{1}{6} - \frac{1}{2\beta} |\tau_\text{out} - \tau_\text{in}| + \frac{1}{4\beta^2}(\tau_\text{out}-\tau_\text{in})^2\right]
\end{align}
By taking the limit $\beta \to \infty$ and neglecting the constant contribution $\sim \beta$ we get
\begin{align}
    G^{\mathbb{R}} &= \frac{1}{R_0^{d-1}} \left\{ - \frac{1}{\Omega^d} | \tau_\text{out}-\tau_\text{in} | + \Sum_{\ell =1,m} \frac{Y_\ell^m(x_1)^* Y_\ell^m(x_2)}{\omega_\ell} e^{-|\tau_\text{out}-\tau_\text{in}| \omega_\ell} \right\}  \\
    &= \frac{1}{\Omega_d R_0^{d-1}} \left\{ - | \tau_\text{out}-\tau_\text{in}| + \Sum_{\ell=1} \frac{d+2\ell-2}{(d-2)} \frac{e^{-|\tau_\text{out}-\tau_\text{in}| \omega_\ell}}{\omega_\ell} C^{d/2-1}_{\ell}(\hat{x}_1 \cdot \hat{x_2} )  \right\}
\end{align}
The result is fine up to an overall normalisation, and on $\mathbb{R}$ eigenfunctions are normalised as follows
\begin{align}
&\Int \dd \Omega_d Y_\ell^m(\hat{x}_1)^* Y_{\ell'}^{m'}(\hat{x}_2) \Int \frac{\dd \omega}{2\pi} e^{i \omega(\tau_\text{out}-\tau_\text{in})} = \delta(\tau_\text{out}-\tau_\text{in}) \delta_{mm'}\delta_{\ell\ell'}, & &\lambda_{\ell,\omega}= \omega^2 + \omega_\ell^2
\end{align}
There is no need to remove the zero modes any more, thus
\begin{align}
    G^{\mathbb{R}} &= \frac{1}{R_0^{d-1}} \Int \frac{\dd \omega}{2\pi} e^{i \omega|\tau_\text{out}-\tau_\text{in}|} \Sum_{\ell =0} \frac{Y_\ell^m(\hat{x}_1)^* Y_{\ell}^{m}(\hat{x}_2)}{\omega^2 + \omega_\ell^2} \\
    &= \frac{1}{R_0^{d-1}} \Sum_{\ell =1} e^{- \omega_\ell |\tau_\text{out}-\tau_\text{in}|} \frac{Y_\ell^m(\hat{x}_1)^* Y_{\ell}^{m}(\hat{x}_2)}{2\omega_\ell} - \frac{1}{2R_0^{d-1} \Omega_d} |\tau_\text{out}-\tau_\text{in}|
\end{align}
The $\pi$ fluctuation correlation function reads in our normalisation
\begin{equation}
    \langle \pi(\tau_\text{out},y) \pi(\tau_\text{in} , x) \rangle = \frac{1}{c_1 d(d-1) (\mu R_0)^{d-2}} \left\{ \Sum_{\ell=1} e^{-\omega_\ell|\tau_\text{out}-\tau_\text{in}|} \frac{Y_\ell^m(x) Y_\ell^m(y)^*}{2R_0\omega_\ell} - \frac{1}{2\Omega_d R_0} |\tau_\text{out}-\tau_\text{in}| \right\}
\end{equation}
%


\section{Regularisation}\label{sec:regularisation}

For future convenience, we introduce the regularization
\begin{equation}\label{eq:regulated_sphere_sum}
\Sigma(s) = \lim_{\Lambda \to \infty} \Sum_{\ell >0} M_\ell (R_0 \omega_\ell)^{s} e^{-\omega_\ell^2/ \Lambda^2}.
\end{equation}
The one-loop scaling dimension reads $\Delta_1 = \frac{1}{2} \Sigma(1)$.
for various spacetime dimensions.
The use of a momentum-dependent regulator is natural.



\section{Correlation insertions} \label{sec:TTandTJcorrelators}

\subsection[\texorpdfstring%
{$\mel{\myatop{\Qp}{\ell_2 m_2}}{T T}{\myatop{\Qp}{\ell_1 m_1}}$}%
{<QTTQ>}]%
{$\mel{\myatop{\Qp}{\ell_2 m_2}}{T T}{\myatop{\Qp}{\ell_1 m_1}}$ correlators.}%
\label{sec:VTTV}

We compute the correlators with two insertions of the stress-energy tensor~\cite{dondi2022spinning} at $\tau>\tau'$ between spinning operators $\Vpp[Q][\ell m]$ at $\tau_\text{in},\tau_\text{out}$ such that $\tau_\text{out} > \tau > \tau' > \tau_\text{in}$.
There are a total of six correlators which read
\begin{align}
	& \expval{\Vpp* T_{\tau\tau}(\tau, \n) T_{\tau\tau}(\tau', \n') \Vpp } = \Anew[\Delta_\Qp + R_0 \omega_{ \ell_1}][\Delta_\Qp + R_0 \omega_{ \ell_2}][\tau] \frac{ \Delta_0 }{ \Omega_d^2 R_0^{2d}} \nonumber \\
	&\Bigg\{  \bigg[ { \Delta_0} + {2 \Delta_1} + \frac{ d }{ 2} \Sum_{\ell} e^{-|\tau -\tau'| \omega_{\ell} } R_0 \omega_{\ell} \frac{ (d+2\ell -2)}{ d-2 } C^{ \frac{d }{2} -1}_{\ell} (\n \cdot \n') \bigg] \delta_{\ell_1 \ell_2} \delta_{m_1 m_2} \nonumber \\
	&+ \frac{ d \Omega_d}{ 2} R_0 \sqrt{ \omega_{\ell_1} \omega_{\ell_2} } \bigg[   Y^*_{\ell_2 m_2} (\n) Y_{\ell_1 m_1} (\n') \, e^{ (\tau -\tau') \omega_{ \ell_1} } + Y^*_{\ell_2 m_2} (\n') Y_{\ell_1 m_1} (\n) \, e^{- (\tau -\tau') \omega_{ \ell_2} } \bigg] \Bigg\} \nonumber \\
	&+ \Bigg\{ \, \Anew[\Delta_\Qp + R_0 \omega_{ \ell_1}][\Delta_\Qp + R_0 \omega_{ \ell_2}][\tau] \frac{\Omega_d \Delta_0 R_0\sqrt{ \omega_{\ell_1} \omega_{\ell_2} } }{ 2 \Omega_d^2 R_0^{2d} } \bigg( \, (d-1) \, Y_{\ell_1 m_1} (\n) Y^*_{\ell_2 m_2} (\n) \nonumber\\
   & - \frac{(d-3)}{(d-1)} \frac{ \del_i Y_{\ell_1 m_1} (\n) \del_i Y^*_{\ell_2 m_2} (\n) }{ R_0^2 {\omega_{ \ell_1} \omega_{ \ell_2}} } \bigg)\, + \Big((\tau, \n) \leftrightarrow (\tau', \n') \Big)  \Bigg\} .
\end{align}
This correlator is symmetric under $(\tau, \n) \leftrightarrow (\tau', \n')$ and the $\ell=0$ case has already appeared in~\cite{komargodski2021spontaneously}.
\begin{align}
         &\expval{\Vpp* T_{ij} (\tau, \n)  T_{kn} (\tau', \n') \Vpp } = \Anew[\Delta_\Qp + R_0 \omega_{ \ell_1}][\Delta_\Qp + R_0 \omega_{ \ell_2}][\tau] \frac{ \Delta_0 }{ (d-1)^2 \Omega_d^2 R_0^{2d} } \nonumber \\
          &\Bigg\{  \bigg[ \Delta_0 + 2 \Delta_1  + \frac{ d }{ 2} \sum_{\ell} e^{-|\tau -\tau'| \omega_{\ell} } R_0 \omega_{\ell} \frac{ (d+2\ell -2)}{ d-2} C^{ \frac{d }{2} -1}_{\ell} (\n \cdot \n') \bigg] h_{ij} h_{kn}   \delta_{\ell_2 \ell_1} \delta_{m_2 m_1} \nonumber \\
         &+ \frac{ d \, \Omega_d }{ 2} R_0 \sqrt{ \omega_{\ell_2} \omega_{ \ell_1} } \left( Y_{\ell_2  m_2}^* (\n) Y_{\ell_1 m_1} (\n') \, e^{ (\tau - \tau') \omega_{ \ell_1}} + Y_{\ell_2  m_2}^* (\n') Y_{\ell_1 m_1} (\n) \, e^{ -(\tau -\tau') \omega_{ \ell_2} } \right) h_{ij} h_{kn}  \Bigg\} \nonumber \\
       & + \Bigg\{ \Anew[\Delta_\Qp + R_0 \omega_{ \ell_1}][\Delta_\Qp + R_0 \omega_{ \ell_2}][\tau] \frac{\Omega_d \Delta_0 R_0 \sqrt{ \omega_{ \ell_1 } \omega_{\ell_2} } }{ 2 (d-1) \, \Omega_d^2 R_0^{2d} } \bigg[ 2 \frac{ \del_{( i}  Y_{ \ell_2  m_2}^* (\n) \del_{j)} Y_{\ell_1 m_1} (\n) }{ R_0^2 (d-1) \, { \omega_{ \ell_1} \omega_{ \ell_2} } } + Y_{\ell_2 m_2}^* (\n)   Y_{\ell_1 m_1} (\n) h_{ij} \nonumber \\
       & - \frac{ \del_i Y_{\ell_2 m_2}^* (\n) \del_i Y_{\ell_1 m_1} (\n) }{ R_0^2 (d-1) \, { \omega_{\ell_1} \omega_{\ell_2}} } h_{ij} \bigg] \, h_{kn} +  \Big((\tau, \n, {ij} ) \leftrightarrow (\tau', \n', {kn} ) \Big)  \Bigg\} .
\end{align}
This correlator is symmetric under $(\tau, \n, ij ) \leftrightarrow (\tau', \n', kn )$.
\begin{align}
        &\expval{\Vpp* T_{\tau i} (\tau,\n)  T_{\tau j} (\tau',\n') \Vpp } = - \Anew[ \Delta_\Qp + R_0 \omega_{ \ell_1}][ \Delta_\Qp + R_0 \omega_{ \ell_2}][\tau] \frac{\Delta_0 d }{ 2 (d-1)^2 \Omega_d R_0^{2d} } \nonumber \\
        &\Bigg\{ \del_i \del'_j \Sum_{\ell} \frac{e^{-|\tau -\tau'|  \omega_{\ell} } }{R_0 \omega_{\ell} } \frac{ (d+2\ell -2)}{ (d-2) \Omega_d} C^{ \frac{d }{2} -1}_{\ell} (\n \cdot \n')  \, \delta_{\ell_2 \ell_1}\delta_{m_2 m_1} + \frac{ \partial_i Y_{\ell_2 m_2}^* (\n) \, \partial_j' Y_{\ell_1 m_1} (\n') }{ R_0 \sqrt{ \omega_{ \ell_2} \omega_{ \ell_1} }  e^{ - (\tau - \tau') \omega_{ \ell_1}} } \nonumber \\
       & + \frac{ \partial_j' Y_{\ell_2 m_2}^* (\n') \, \partial_i Y_{\ell_1 m_1} (\n) }{ R_0 \sqrt{ \omega_{ \ell_2} \omega_{ \ell_1} }  e^{ (\tau - \tau') \omega_{\ell_2}} } \, \Bigg\} .
\end{align}
This correlator is symmetric under $(\tau, \n, i ) \leftrightarrow (\tau', \n', j )$.
\begin{align}
        &\expval{\Vpp* T_{\tau i} (\tau, \n)  T_{\tau\tau} (\tau', \n') \Vpp } = - \Anew[\Delta_\Qp + R_0 \omega_{ \ell_1}][\Delta_\Qp + R_0 \omega_{ \ell_2}][\tau] \frac{ \Delta_0 d }{2 \Omega_d R_0^{2d}} \frac{1}{(d-1)} \nonumber \\
       & \Bigg\{ \del_i \sum_{\ell} e^{- |\tau -\tau'| \omega_{\ell} } \frac{ (d+2\ell -2)}{ (d-2) \Omega_d} C^{ \frac{d }{2} -1}_{\ell} (\n \cdot \n') \, \delta_{ \ell_2 \ell_1} \delta_{m_2 m_1} + \sqrt{ \frac{ \omega_{\ell_2} }{ \omega_{\ell_1} } } \frac{ Y_{\ell_2 m_2}^* (\n') \del_i Y_{\ell_1 m_1} (\n) }{ e^{ (\tau -\tau') \omega_{ \ell_2} }} \nonumber  \\
       & - \sqrt{ \frac{ \omega_{\ell_1} }{ \omega_{\ell_2} } } \frac{ \del_i Y_{\ell_2 m_2}^* (\n) \, Y_{\ell_1 m_1} (\n') }{ e^{ -(\tau -\tau') \omega_{\ell_1}}} + \frac{ (d-1) }{ d } \bigg[ \sqrt{ \frac{ \omega_{\ell_2} }{ \omega_{\ell_1} } } Y_{\ell_2 m_2}^* (\n)  \del_i Y_{\ell_1 m_1} (\n) - (1 \leftrightarrow 2 )^* \bigg] \, \Bigg\} .
\end{align}
Since $T_{\tau i}$ vanishes on the homogeneous ground-state, this correlator only receives a second-order contribution from the linear parts and the quadratic term of $T_{\tau i}$. 

Moving to  $T_{\tau i} T_{j k}$ we find
\begin{align}
    &\expval{\Vpp* T_{\tau i} (\tau, \n)  T_{j k} (\tau', \n') \Vpp } = \Anew[\Delta_\Qp + R_0 \omega_{\ell_{1}}][\Delta_\Qp + R_0 \omega_{\ell_{2}}][\tau] \frac{\Delta_0 d }{2 \Omega_d R_0^{2d} } \frac{ h_{jk} }{ (d-1)^2} \nonumber \\
    &\Bigg\{ \partial_i \sum_{\ell} e^{- |\tau-\tau'| \omega_{\ell}} \frac{ (d+ 2\ell -2) }{ (d-2) \Omega_d} C^{\frac{ d}{2} -1}_\ell (\n \cdot \n') \, \delta_{\ell_2 \ell_1} \delta_{m_2 m_1} + \sqrt{ \frac{ \omega_{ \ell_2} }{ \omega_{ \ell_1}} } \frac{ Y_{\ell_2 m_2}^* (\n') \, \partial_i Y_{\ell_1 m_1} (\n) }{ e^{ (\tau-\tau') \omega_{\ell_2} } } \nonumber \\
    & - \sqrt{ \frac{ \omega_{ \ell_1} }{ \omega_{ \ell_2}} } \frac{ Y_{\ell_1 m_1} (\n') \, \partial_i Y_{\ell_2 m_2}^* (\n) }{ e^{- (\tau-\tau') \omega_{\ell_1} } } + \frac{ (d-1) }{ d} \bigg[ \sqrt{ \frac{ \omega_{ \ell_2} }{ \omega_{ \ell_1} } } Y_{\ell_2 m_2}^* (\n) \del_i Y_{\ell_1 m_1} (\n) - (1 \leftrightarrow 2)^* \bigg] \Bigg\} .
\end{align}
Again, besides the linear terms, only the quadratic term of $T_{\tau i}$ contribute at second order.
In addition, the correlator $\expval{\Vpp* T_{\tau i} (\tau, \n) \, h^{jk}(\n') T_{j k} (\tau', \n') \Vpp }$ differs only by a minus sign from the former result with an insertion of $T_{\tau\tau} (\tau', \n')$, as enforced by conformal invariance.
\begin{align}
    \expval{\Vpp* T_{\tau \tau} (\tau, \n)  T_{ij} (\tau', \n') \Vpp } =  - \Anew[\Delta_\Qp + R_0 \omega_{ \ell_1}][\Delta_\Qp + R_0 \omega_{ \ell_2}][\tau] \frac{ \Delta_0 }{ \Omega_d^2 R_0^{2d}} \frac{ h_{ij} }{ (d-1)}  \nonumber \\
    \Bigg\{ \bigg[ \Delta_0 + 2 \Delta_1 + \frac{ d \, \Omega_d }{ 2} \Sum_{\ell} R_0 \omega_{\ell} e^{- |\tau - \tau'| \omega_{\ell} } \frac{ (d+ 2\ell -2) }{ (d-2) \Omega_d} C^{ \frac{d }{2} -1}_{\ell} (\n \cdot \n') \bigg] \delta_{\ell_2 \ell_1} \delta_{m_2 m_1} \nonumber \\
    + \frac{ d \, \Omega_d  }{2} R_0 \sqrt{ \omega_{ \ell_2} \omega_{ \ell_1} } \bigg[  Y_{\ell_2 m_2}^* (\n) Y_{\ell_1 m_1} (\n')  e^{ (\tau - \tau') \omega_{ \ell_1} } + Y_{\ell_2 m_2}^* (\n') Y_{\ell_1 m_1} (\n) e^{-(\tau - \tau') \omega_{ \ell_2} } \bigg] \Bigg\} \nonumber \\
    - \Anew[\Delta_\Qp + R_0 \omega_{\ell_{1}}][\Delta_\Qp + R_0 \omega_{\ell_{2}}][\tau] \frac{ \Delta_0 R_0 \sqrt{ \omega_{ \ell_1} \omega_{ \ell_2} } }{ 2\Omega_d R_0^{2d}} h_{ij} \Bigg\{ Y_{ \ell_2 m_2}^* (\n) Y_{\ell_1 m_1} (\n) - \frac{(d-3) }{ (d-1)^2} \frac{ \del_i Y_{\ell_2 m_2}^* (\n) \del_i Y_{\ell_1 m_1} (\n) }{ R_0^2 { \omega_{ \ell_1} \omega_{\ell_2}}} \Bigg\} \nonumber \\
     - \Anew[ \Delta_\Qp + R_0 \omega_{ \ell_1}][ \Delta_\Qp + R_0 \omega_{ \ell_2}][\tau'] \frac{ \Delta_0 R_0 \sqrt{ \omega_{ \ell_1} \omega_{ \ell_2}} }{2 \Omega_d R_0^{2d}} \Bigg\{ \, h_{ij} \Big[ \, Y_{ \ell_2 m_2}^* (\n') Y_{\ell_1 m_1} (\n') - \frac{ \del_i Y_{ \ell_2 m_2}^* (\n') \del_i  Y_{\ell_1 m_1} (\n') }{ (d-1) \, R_0^2 \omega_{ \ell_1 } \omega_{ \ell_2 } } \Big] \nonumber \\
    + 2 \frac{ \del_{ ( i} Y_{ \ell_2 m_2 }^* (\n')  \del_{j) }  Y_{\ell_1  m_1} (\n') }{ (d-1) \, R_0^2 \omega_{ \ell_1} \omega_{ \ell_2}} \Bigg\} .
\end{align}
The above correlator is not symmetric in $(\tau, \n) \leftrightarrow (\tau', \n')$, however, by conformal invariance, the correlator $\expval{\Vpp* T_{\tau \tau} (\tau, \n) h^{ij}(\n) T_{ij} (\tau', \n') \Vpp }$ is symmetric in $(\tau, \n) \leftrightarrow (\tau', \n')$.

For the special case $\ell =0$, the above correlators read
\begin{align}
 &\begin{multlined}[][.9\linewidth]\expval{ \Opp* T_{\tau\tau}(\tau, \n) T_{\tau\tau} (\tau', \n') \Opp} =  \Anew  \frac{ \Delta_0 }{\Omega_d^2 R_0^{2d}} \bigg[ \Delta_0  + 2 \Delta_1 \\
	 + \frac{ d  }{2} \Sum_{\ell} e^{-|\tau -\tau'| \omega_{\ell} } R_0 \omega_{\ell} \frac{ (d+2\ell -2)}{ d-2} C^{ \frac{d }{2} -1}_{\ell} (\n \cdot \n') \bigg] ,
  \end{multlined} \\
    &\begin{multlined}[][.9\linewidth]
    \expval{ \Opp* T_{ij}(\tau, \n) T_{kn}(\tau', \n') \Opp } =  \Anew \frac{\Delta_0 }{ \Omega_d^2 R_0^{2d} } \frac{  h_{ij} h_{kn} }{(d-1)^2} \bigg[ \Delta_0 +  2\Delta_1 \\
     + \frac{ d }{2 } \Sum_{\ell} \frac{ R_0 \omega_{\ell} }{ e^{ |\tau -\tau'| \omega_{\ell} } } \frac{ (d+2\ell -2)}{ d-2} C^{ \frac{d }{2} -1}_{\ell} (\n \cdot \n') \bigg] ,
    \end{multlined} \\
    &
    \expval{\Opp* T_{\tau i} (\tau, \n) T_{\tau j} (\tau', \n') \Opp } = - \frac{ \Anew \Delta_0 d }{ 2 (d-1)^2 \Omega_d^2 R_0^{2d} } \del_i \del'_j \Sum_{\ell}  \frac{ e^{- |\tau -\tau'| \omega_{\ell} } }{ R_0 \omega_{\ell} } \frac{ (d+2\ell -2)}{d-2} C^{ \frac{d }{2} -1}_{\ell} (\n \cdot \n') ,
     \\
    &
    \expval{ \Opp* T_{\tau i}(\tau, \n) T_{\tau\tau}(\tau', \n') \Opp } = - \frac{ \Anew \Delta_0 d }{ 2 (d-1) \Omega_d^2 R_0^{2d} } \del_i \Sum_{\ell} e^{- |\tau -\tau'| \omega_{\ell} } \frac{ (d+2\ell -2)}{ d-2} C^{ \frac{d }{2} -1}_{\ell} (\n \cdot \n') ,
\\
&
    \expval{\Opp* T_{\tau i} (\tau, \n)  T_{j k} (\tau', \n') \Opp } = \frac{ \Anew \Delta_0 d \, h_{jk} }{2 (d-1)^2 \Omega_d^2 R_0^{2d} } \partial_i \Sum_{\ell} e^{- |\tau-\tau'| \omega_{\ell}} \frac{ (d+ 2\ell -2) }{ d-2} C^{\frac{ d}{2} -1}_\ell (\n \cdot \n') ,
\\
&\begin{multlined}[][.9\linewidth]
    \expval{\Opp* T_{\tau \tau} (\tau, \n)  T_{ij} (\tau', \n') \Opp } = - \Anew \frac{ \Delta_0 }{ \Omega_d^2 R_0^{2d}} \frac{ h_{ij} }{ (d-1)} \bigg[ \Delta_0 + 2 \Delta_1 \\
      + \frac{ d  }{ 2} \Sum_{\ell} R_0 \omega_{\ell} e^{- |\tau - \tau'| \omega_{\ell} } \frac{ (d+ 2\ell -2) }{ d-2} C^{ \frac{d }{2} -1}_{\ell} (\n \cdot \n') \bigg] .
\end{multlined}
\end{align}
The $\ell_1 = \ell_2 = 0$ correlator with insertions of $T_{\tau i} T_{\tau \tau}$ was computed in the macroscopic limit $R \rightarrow \infty$ in~\cite{komargodski2021spontaneously}.

\subsection[\texorpdfstring%
{$\mel{\myatop{\Qp}{\ell_2 m_2}}{T\mathcal{J}}{ \myatop{\Qp}{\ell_1 m_1}}$}%
{<QTJQ>}]%
{$\mel{\myatop{\Qp}{\ell_2 m_2}}{T\mathcal{J}}{ \myatop{\Qp}{\ell_1 m_1}}$ correlators.}%
\label{sec:VTJV}

We consider correlation functions with one insertion of the stress-energy tensor and one insertion of the $O(2)$-current at times $\tau > \tau'$ between spinning operators $\Vpp[\Qp][\ell m]$ at $\tau_\text{in}, \tau_\text{out}$ so that we have the ordering $\tau_\text{out} > \tau > \tau' > \tau_\text{in}$. There are in total six correlators involving which read
\begin{align}
\begin{multlined}[][.9\linewidth]
        \expval{\Vpp* T_{\tau i}(\tau, \n) \mathcal{J}_\tau(\tau', \n') \Vpp } =  - i \Anew[ \Delta_\Qp + R_0 \omega_{ \ell_1}][ \Delta_\Qp + R_0 \omega_{ \ell_2}][\tau] \frac{ \Qp }{2 \Omega_d R_0^{2d-1}} \\
        \Bigg\{  \del_i \Sum_{\ell} e^{- |\tau -\tau'| \omega_{\ell} }  \frac{ (d+2\ell -2)}{ (d-2) \Omega_d} C^{ \frac{d }{2} -1}_{\ell} (\n \cdot \n') \, \delta_{\ell_2 \ell_1} \delta_{m_2 m_1}  + \sqrt{ \frac{ \omega_{\ell_2} }{ \omega_{\ell_1} } } \frac{ Y_{\ell_2 m_2}^* (\n') \del_i Y_{\ell_1 m_1} (\n) }{ e^{ (\tau - \tau') \omega_{\ell_2}} } \\
         - \sqrt{ \frac{ \omega_{ \ell_1} }{ \omega_{ \ell_2} } } \frac{ \del_i Y_{\ell_2 m_2}^* (\n) Y_{\ell_1 m_1} (\n') }{ e^{- (\tau - \tau') \omega_{\ell_1}} } + \bigg( \sqrt{ \frac{ \omega_{\ell_2} }{ \omega_{\ell_1} } }   Y_{\ell_2 m_2}^* (\n) \, \del_i Y_{\ell_1 m_1} (\n) - (1 \leftrightarrow 2)^* \bigg) \, \Bigg\} .
\end{multlined}
\end{align}
$T_{\tau i}$ vanishes on the homogeneous ground state and thus only receives quadratic contributions from the linear terms and the quadratic term of $T_{\tau i}$. 
\begin{align}
\begin{multlined}[][.9\linewidth]
        \expval{\Vpp* \mathcal{J}_i (\tau, \n) T_{\tau\tau} (\tau', \n') \Vpp} = - i \Anew[\Delta_\Qp + R_0 \omega_{ \ell_1}][\Delta_\Qp + R_0 \omega_{ \ell_2}][\tau] \frac{ \Qp }{2 \Omega_d R_0^{2d-1}} \\
        \Bigg[ \delta_{\ell_2 \ell_1} \delta_{m_2 m_1} \del_i \Sum_{\ell} e^{- |\tau -\tau'| \omega_{\ell} } \frac{ (d+2\ell -2)}{ (d-2) \Omega_d} C^{ \frac{d }{2} -1}_{\ell} (\n \cdot \n') + \sqrt{ \frac{ \omega_{\ell_2} }{ \omega_{\ell_1} } }  \frac{ Y_{\ell_2 m_2}^* (\n') \, \del_i Y_{\ell_1 m_1} (\n) }{ e^{ (\tau -\tau') \omega_{\ell_2}} } \\
        - \sqrt{ \frac{ \omega_{\ell_1} }{ \omega_{ \ell_2} }} \frac{ \del_i Y_{\ell_2 m_2}^* (\n) \, Y_{\ell_1 m_1} (\n') }{ e^{ -(\tau -\tau') \omega_{ \ell_1}} } + \frac{ (d-2)}{ D} \bigg( \sqrt{ \frac{ \omega_{ \ell_2} }{ \omega_{ \ell_1} } } Y_{\ell_2 m_2}^* (\n) \del_i Y_{\ell_1 m_1} (\n) - (1 \leftrightarrow 2)^* \bigg) \Bigg] .
\end{multlined}
\end{align}
This correlator is related to the previous one which is clear from the expansions in \cref{eq:ExpansionsOfTsAndQsInTheField}.
\begin{align}\label{eq:TtautauJtauCorrelator}
\begin{multlined}[][.9\linewidth]
        \expval{\Vpp* T_{\tau \tau}(\tau, \n) \mathcal{J}_\tau(\tau', \n') \Vpp } = i \Anew[\Delta_\Qp + R_0 \omega_{\ell_{1}}][\Delta_\Qp + R_0 \omega_{ \ell_{2}} ][\tau] \frac{ \Qp (d-1) }{2 \Omega_d R_0^{2d -1}} R_0 \sqrt{\omega_{ \ell_2} \omega_{ \ell_1} } \\
        \Bigg\{ \, \bigg[ \frac{ Y_{\ell_2 m_2}^* (\n') Y_{\ell_1 m_1} (\n) }{ e^{ (\tau-\tau') \omega_{\ell_2}} }+ \frac{ Y_{\ell_2 m_2}^* (\n) Y_{\ell_1 m_1} (\n') }{ e^{- (\tau-\tau') \omega_{\ell_1}} } \bigg] + \Sum_{\ell} \frac{ R_0 \omega_\ell \, e^{- |\tau - \tau'| \omega_\ell}}{  R_0 \sqrt{\omega_{ \ell_2} \omega_{ \ell_1} }}   \frac{ (d+ 2\ell -2) }{ (d-2) \Omega_d} C_\ell^{ \frac{ d}{2} -1} (\n \cdot \n') \\
        +  \frac{ (d-2)}{d} \Big[ Y_{\ell_2 m_2}^* (\n') Y_{\ell_1 m_1} (\n') - \frac{ \partial_i Y_{\ell_2 m_2}^* (\n') \, \partial_i Y_{\ell_1 m_1} (\n') }{ R_0^2 (d-1) \omega_{\ell_2} \omega_{\ell_1} } \Big] 
        +\frac{2 }{ (d-1)} \Bigg[ \frac{1 }{\Omega_d} \Big( \frac{ \Delta_0 + \Delta_1 }{R_0\sqrt{ \omega_{ \ell_1} \omega_{ \ell_2} }} \Big) \delta_{\ell_2 \ell_1} \delta_{m_2 m_1} \\
        + \frac{ 1}{2} \bigg( (d-1) Y_{ \ell_2 m_2}^* (\n) Y_{\ell_1 m_1} (\n) 
        - \frac{(d-3) }{ (d-1)} \frac{ \del_i Y_{\ell_2 m_2}^* (\n) \del_i Y_{\ell_1 m_1} (\n) }{ R_0^2 { \omega_{ \ell_1} \omega_{\ell_2}}} \bigg) \Bigg]  \Bigg\} .
\end{multlined}
\end{align}
Here, the quadratic term from $\mathcal{J}_\tau$ vanishes after integration over $\n'$, whereas the quadratic term from $T_{\tau\tau}$ remains finite after integration over $\n$. This is so because it has to correct the energy by $R_0\omega_{\ell_2}$~\cite{dondi2022spinning}.
\begin{align}
\begin{multlined}[][.9\linewidth]
        \expval{\Vpp* T_{\tau i}(\tau, \n) \mathcal{J}_j (\tau', \n') \Vpp } = -i \Anew[\Delta_\Qp + R_0 \omega_{\ell_{1}}][\Delta_\Qp + R_0 \omega_{ \ell_{2}} ][\tau] \frac{ \Qp  }{2 \Omega_d R_0^{2d-1} } \frac{ 1}{ (d-1)} \\
        \Bigg\{ \partial_i \partial'_j \Sum_{\ell} \frac{ e^{- |\tau -\tau'| \omega_\ell} }{ R_0 \omega_{\ell} } \frac{ (d+2\ell -2) }{ (d-2) \Omega_d } C_\ell^{ \frac{d}{2} -1} (\n\cdot \n') \, \delta_{\ell_2 \ell_1} \delta_{m_2 m_1} + \frac{ \partial'_j Y_{\ell_2 m_2}^* (\n') \partial_i Y_{\ell_1 m_1} (\n) }{ R_0 \sqrt{ \omega_{ \ell_1} \omega_{ \ell_2}} \, e^{(\tau -\tau') \omega_{ \ell_2}} }\\
        + \frac{ \partial_i Y_{\ell_2 m_2}^* (\n) \partial'_j Y_{\ell_1 m_1} (\n') }{ R_0 \sqrt{ \omega_{ \ell_1} \omega_{ \ell_2}} \, e^{-(\tau -\tau') \omega_{ \ell_1}} } \Bigg\} .
\end{multlined}
\end{align}
Both $T_{\tau i}$ and $\mathcal{J}_i$ vanish on the homogeneous ground state and therefore the sole quadratic contribution originates from the two linear terms.
\begin{align}
\begin{multlined}[][.9\linewidth]
        \expval{\Vpp* T_{ij}(\tau, \n) \mathcal{J}_\tau (\tau', \n') \Vpp } = - i \Anew[\Delta_\Qp + R_0 \omega_{ \ell_{1}} ][ \Delta_\Qp + R_0 \omega_{ \ell_{2}} ][\tau] \frac{ \Qp }{ \Omega_d R_0^{2d-1} }  \\
        \Bigg\{ h_{ij} \bigg[ \frac{ \left( \Delta_0 + \Delta_1 \right) }{ \Omega_d (d-1)} \delta_{ \ell_2 \ell_1} \delta_{m_2 m_1} + \frac{ 1}{2} \Sum_{\ell}  e^{- |\tau -\tau'| \omega_\ell}  R_0 \omega_{\ell}  \frac{ (d+2\ell -2) }{ (d-2) \Omega_d } C_\ell^{\frac{d}{2} -1} (\n\cdot \n') \, \delta_{\ell_2 \ell_1} \delta_{m_2 m_1} \\
        + \frac{ R_0 \sqrt{\omega_{\ell_2} \omega_{\ell_2}} }{ 2} \bigg( \frac{Y_{\ell_2 m_2}^* (\n') Y_{\ell_1 m_1} (\n)}{e^{(\tau - \tau') \omega_{\ell_2} } } + \frac{ Y_{\ell_2 m_2}^* (\n) Y_{\ell_1 m_1} (\n') }{ e^{- (\tau - \tau') \omega_{\ell_1} } } + \Big[ 1 + \frac{ (d-2)}{d} \Big] Y_{ \ell_2 m_2}^* (\n) Y_{\ell_1 m_1} (\n) \\
        - \Big[ 1 + \frac{ (d-2)}{d} \Big] \frac{ \del_i Y_{ \ell_2 m_2}^* (\n) \del_i  Y_{\ell_1 m_1} (\n) }{ (d-1) \, R_0^2 \omega_{ \ell_1 } \omega_{ \ell_2 } } \, \bigg)  \bigg] \, + \frac{ R_0\sqrt{ \omega_{ \ell_1} \omega_{ \ell_2} } }{ (d-1)} \frac{ \del_{ ( i} Y_{ \ell_2 m_2 }^* (\n)  \del_{j) }  Y_{\ell_1  m_1} (\n) }{ R_0^2 \omega_{ \ell_1} \omega_{ \ell_2}}  \Bigg\} .
\end{multlined}
\end{align}
The above correlator is associated to the $T\mathcal{J}$ correlator in \cref{eq:TtautauJtauCorrelator} because $h^{ij} T_{ij} = -T_{\tau\tau}$, due to conformal invariance.
\begin{align}
\begin{multlined}
        \expval{\Vpp* \mathcal{J}_i (\tau, \n) T_{jk} (\tau', \n') \Vpp } = i \Anew[ \Delta_\Qp + R_0 \omega_{ \ell_{1}} ][ \Delta_\Qp + R_0 \omega_{ \ell_{2}} ][ \tau]  \frac{ \Qp }{2 \Omega_d R_0^{2d-1} } \frac{ h_{jk} }{ (d-1)}  \\
        \Bigg\{\partial_i \Sum_{\ell} e^{- |\tau-\tau'| \omega_{\ell}} \frac{ (d+ 2\ell -2)}{ (d-2) \Omega_d} C_{\ell}^{\frac{d}{2} -1} (\n \cdot \n') \, \delta_{\ell_2 \ell_1} \delta_{m_2 m_1} + \sqrt{ \frac{ \omega_{\ell_2} }{ \omega_{\ell_1} } } \frac{ Y_{\ell_2 m_2}^* (\n') \, \partial_i Y_{\ell_1 m_1} (\n) }{ e^{ (\tau-\tau') \omega_{\ell_2}} }  \\
         - \sqrt{ \frac{ \omega_{\ell_1} }{ \omega_{\ell_2} } } \frac{ Y_{\ell_1 m_1} (\n') \, \partial_i Y_{\ell_2 m_2}^*  (\n) }{ e^{- (\tau-\tau') \omega_{\ell_1}} } + \frac{ (d-2) }{ d} \bigg[ \sqrt{ \frac{ \omega_{ \ell_2} }{ \omega_{ \ell_1} } } Y_{\ell_2 m_2}^* (\n) \del_i Y_{\ell_1 m_1} (\n) - (1 \leftrightarrow 2 )^* \, \bigg] \, \Bigg\} .
\end{multlined}
\end{align}
The above correlator is analogous to $h_{jk}$ since the quadratic term in the expansion of $T_{jk}$ only appears at cubic order, but this is no longer true once we include higher-order corrections.
%


\chapter[\texorpdfstring%
{Preliminaries for the \(O(N)\) model and Resurgence}%
{O(N)}]%
{Preliminaries for the \(O(N)\) model and Resurgence}%

\label{AppendixC} 

Most of the sections of the Appendix are based on~\cite{dondi2021resurgence} and on an unpublished set of notes by Nicola Dondi.

\section{The Borel transform}
\label{sec:borel-transform}

The \emph{Borel transform} is an operation acting in the space of power series in the following manner
\begin{align}
\Phi(z) &\sim \Sum_{n=0}^{\infty} a_n z^n &\longrightarrow &&  \mathcal{B}\left\{\Phi\right\}(\zeta) &= \Sum_{n=0}^{\infty} \frac{a_n}{\Gamma(\beta n + b)} \zeta^n,
\end{align}
where we presume a generic large order behaviour of $a_k$ as in \cref{eq:GenericLargeOrder} with $b \equiv \max\{ b_k\},\, \beta \equiv \max \{ \beta_k\}$. This assumption guarantees that the series defined by $\mathcal{B}\left\{ \Phi \right\}$ is convergent in a disc displaced at the beginning of the $\zeta$-plane, also referred as \emph{Borel plane}.

The analytic properties of the Borel transforms can be straightaway deduced from \cref{eq:GenericLargeOrder} and for example
\begin{align}
	\frac{a_n}{\Gamma(\beta n + b)} &\sim \frac{1}{A^{\beta n +b}} & \text{then} && \mathcal{B}\{ \Phi\} (\zeta ) &\xrightarrow{\zeta \rightarrow A^\beta} \frac{A^{-b}}{1-\zeta\, / A^{\beta}} + \text{regular}.
	\label{eq:small-example}
\end{align}
Then we can define the \emph{Borel resummation} of $\Phi$ as 
\begin{equation}
\mathcal{S}\{ \Phi\} (z) = \frac{1}{ \beta} \Int\limits_0^\infty \frac{\dd{\zeta}}{\zeta}  \left( \frac{\zeta}{z} \right)^{\frac{b}{\beta}} e^{- \left(\zeta/z \right)^{1/\beta}} \mathcal{B}\{\Phi\}(\zeta).
\label{eq:def-Borel}
\end{equation}
Utilising the definition of the Gamma function, it is clear that $\mathcal{S}\left\{ \Phi \right\}(z) \sim \Phi(z)$ as $z \rightarrow 0^+$. Nevertheless, if the integral is well-defined, it specifies a function calculable for all values of $z$, which stands for the “sum'' of the divergent series $\Phi$.
This function is without ambiguities unless $\mathcal{B}\{\Phi\}$ exhibits singularities along the integration line. In this case, we need to define a directional summation $\mathcal{S}_\theta$ by integrating along the ray with angle $\theta$ in the Borel plane: the expression in \cref{eq:def-Borel} matches to the case $\theta= 0$. 

A ray $\theta$ along which $\mathcal{B}\{\Phi\}$ features singularities is a \emph{Stokes ray} and there are ambiguities in the Borel resummation.
We then define lateral Borel summations $\mathcal{S}_{\theta^{\pm}}$ by deforming the contour of integration to avoid the poles. This represents the existence of a branch cut for $\operatorname{Arg}(z) = \theta$ for the Borel resummation, with the discontinuity calculated as
\begin{equation}
[\mathcal{S}_{\theta^+} - \mathcal{S}_{\theta^-} ]\{\Phi\} = - [ \mathcal{S}_{\theta^-} \circ \Disc_\theta ] \{ \Phi\}.
\end{equation}
The above discontinuity is strictly non-perturbative, as we can see from the example in \cref{eq:small-example} where $\theta = 0$ 
\begin{equation}
[\mathcal{S}_{0^+} - \mathcal{S}_{0^-} ]\{\Phi\}  = - \frac{2\pi i }{\beta } z^{- b/\beta} e^{-A/z^{1/\beta}} = - \Disc_0 \{ \Phi\}.
\end{equation}
For a non-Borel summable perturbative series $\Phi^{(0)}$, the expression $\Disc\{ \Phi^{(0)}\} $ supplies the structure of the non-perturbative terms that we have to add to the trans-series expression in \cref{eq:trans-series}, and the large-order behaviour in \cref{eq:GenericLargeOrder} results applying Cauchy's integral representation.  
The lateral Borel summation of the general trans-series is determined, as
\begin{equation}
\mathcal{S}_{\theta^\pm} \{	\Phi \} (\sigma_k,z) = \mathcal{S}_{\theta^\pm} \{ \Phi^{(0)} \} (z) + \Sum_{k} \sigma_k^\pm e^{- A_k /z^{1/\beta_k}} z^{-b_k / \beta_k} \, \mathcal{S}_{\theta^\pm} \{\Phi^{(k)} \} (z).
	\label{eq:Borel-trans-series}
\end{equation} 
This will not determine a unique resurgent function for $\text{Arg}(z) = \theta$ unless we fix some way the ambiguity associated with the integration path.
This can be achieved by the imposition of extra conditions on the lateral Borel summation, that picks out the trans-series parameters $\sigma_k^\pm$ in such a manner that the ambiguity is lifted. When this is accomplishable, we achieved “semiclassical decoding'' in the language of~\cite{marino2021testing}.

\section{Trans-series representation of the Dawson's function}
\label{sec:Dawson}

We construct the trans-series representation of the Dawson's mathematical function that we found in \cref{sec:perturbative-sphere}.
We will exhibit that no non-perturbative ambiguities remain by the time that we enforce the reality condition for the heat-kernel trace of $\setS^2$.
The methodology is standard and follows the one used in Euler's and Riccati's \acrshort{ode}~\cite{Dorigoni_2019}.
The Dawson's function constitutes the unique solution to the Cauchy problem
\begin{align}
	\dv{F}{z} + 2z F(z) &= 1, & F(0) &= 0.
\end{align}
It is apparent that $z = \infty$ is a critical point of the \acrshort{ode}.
We may try to seek a solution regarding an asymptotic series around this point, that turns out to be
\begin{align}
F(z) &\sim \Sum_{k=0}^\infty a_n  \frac{1}{z^{2k+1}} =  \Sum_{k=0}^\infty \frac{(2k-1)!!}{2^{k+1}} \frac{1}{z^{2k+1}} & \text{for } z &\rightarrow \infty.
\label{eq:Dawson-asymptotic}
\end{align}
The above solution has two problems: first is that it is an asymptotic series, so it is only formally defined; and second that there is no constant of integration that we can fix using the initial condition.

Both of these issues have a common resolution. We define a Borel transform as follows
\begin{equation}
	\mathcal{B}\{ \Phi\}(\zeta) =  \Sum_{k=0}^\infty \frac{a_n}{\Gamma \left( k + \frac{1}{2} \right)} \zeta^{k+1}=\frac{1}{2\sqrt{\pi}} \frac{\zeta}{1-\zeta}.
\end{equation}
The above series is not Borel summable due to the singularity at $\zeta =1$, and thus we have to compute the proper lateral Borel summations as
\begin{equation}
	\mathcal{S}_{\pm} \{ \Phi\} (z) = \Int\limits_{\pm} \dd{\zeta} \zeta^{-\frac{3}{2}} e^{-z^2 \zeta} \mathcal{B}\{ \Phi\}(\zeta).
\end{equation}
We can compute the difference between the lateral Borel summation, which is
\begin{equation}
\left[\mathcal{S}_{+} - \mathcal{S}_-\right]\{ \Phi\}(z)   =  - \Disc \{ \Phi \}(z) = - \frac{2\pi i}{2\sqrt{\pi}} e^{-z^2}.
\end{equation}
But this quantity solves to the homogeneous \acrshort{ode}.
Therefore, we can then express the general solution to the \acrshort{ode} as a trans-series placed at $z = \infty$  with a single trans-series parameter $\sigma$ in the following manner
\begin{equation}
	\Xi (z,\sigma) =  \Sum_{k=0}^\infty \frac{(2k+1)!!}{2^{k+1}} \frac{1}{z^{2k+1}} + \sigma e^{-z^2}.
\end{equation}
Its lateral Borel summation are
\begin{equation}
	\mathcal{S}_{\pm} \left\{ \Xi \right\} (z,\sigma_{\pm}) = \frac{1}{2\sqrt{\pi}} \Int\limits_{\pm} \frac{\dd{\zeta}}{\sqrt{\zeta}} \frac{e^{-z^2 \zeta}}{1-\zeta} + \sigma_{\pm } e^{-z^2},
\end{equation}
that can be shown to solve the inhomogeneous \acrshort{ode}.

To specify the trans-series parameter, we can enforce the outcome to be real for real $z$, which fixes 
\begin{equation}
	\Im \sigma_{\pm} =\pm \frac{ \pi}{2 \sqrt{\pi}},
\end{equation}
in which case the coefficients of the non-perturbative exponentials are strictly imaginary. 
The real part of $\sigma_{\pm}$ remains unfixed and equals the integration constant that had disappeared when we had tried a power series ansatz for the solution.
A generic boundary condition fixes its value as
\begin{align}
	F(0) &= c, & \implies && \Re \sigma_{\mp} &= \frac{\sqrt{\pi}}{2} e^{-c^2} \operatorname{erfi}(c).
\end{align}
For $F(0) = 0$ we receive $\Re \sigma_{\pm} = 0$, thus the reality condition is adequate to determine the whole non-perturbative ambiguity. 
Also observe that this fixed representation of the trans-series resummation matches with the definition of the Dawson's function:
\begin{equation}
	\mathcal{S}_{\pm} \left\{ \Xi \right\} (z,\sigma_{\pm}) = \frac{1}{2 \sqrt{\pi}} \operatorname{P.V.} \Int\limits_{0}^\infty \dd{\zeta} \frac{e^{-z^2 \zeta}}{\sqrt{\zeta}(1-\zeta)} = \frac{\sqrt{\pi}}{2}e^{-z^2} \operatorname{erfi}(z) \equiv F(z), \quad z >0.
\end{equation}
Given that the factorial growth of the Dawson's function drives the ones we found in the heat-kernel trace, the grand potential and the canonical free energy, we can reason that a reality condition is adequate to fix all of the above non-perturbative ambiguities.


\subsection{Path integral normalisation}\label{sec;pathintegralnormalisation}

We are interested in the following path integral 
\begin{equation}
	I = \Int_{DBC} \mathcal{D} x \, e^{- \frac{m}{2} \Int_0^t \dd \tau \dot{x}^2 },
\end{equation}
which can be evaluated as
\begin{equation}
	I =  \lim_{ \substack{N\to \infty, \epsilon\to 0 \\ N \epsilon = t} }  \left(\frac{m}{2 \pi  \epsilon} \right)^\frac{Nd}{2} \Int \mathlarger{\mathlarger{\prod}}_{i=1}^{N-1} \dd^d x_i
         \exp \left\{ -\frac{m}{2 \epsilon} \Sum_{i=0}^{N-1} (x_{i+1} - x_i )^2 \right\},
\end{equation}
and we define
\begin{equation}
	K(\epsilon , x_{i+1} , x_i) = \left( \frac{m}{2\pi \epsilon} \right)^\frac{d}{2} \, e^{- \frac{m}{2\epsilon} ( x_{i+1} - x_i )^2 }.
\end{equation}
Then for \(i=1\) we find the following relation
\begin{align}
	\Int \dd^d x_1 \, e^{ - \frac{m}{2\epsilon} \left[ (x_{2} - x_1)^2 + (x_1 - x_{0})^2 \right]} &= \Int \dd^d x_1 \, e^{ - \frac{m}{2\epsilon} \left[ x_2^2 + x_0^2 + 2 x_1^2 - 2 x_1 (x_2 + x_0 )\right]} \nonumber\\
	&= \left( \frac{\pi \epsilon }{ m}\right)^{\frac{d}{2}} \, e^{- \frac{m}{2\epsilon} \left[ x_2^2 + x_0^2 - \frac{1}{2}(x_2 + x_0)^2 \right]} \nonumber\\
 &= \left( \frac{\pi \epsilon }{ m}\right)^{\frac{d}{2}} \, e^{- \frac{m}{4\epsilon}(x_2 - x_0)^2 },
\end{align}
and in terms of the \(K\) we get
\begin{equation}
	\Int \dd^d x_1 K(\epsilon , x_2, x_1 ) K(\epsilon , x_1, x_0 ) = \left( \frac{m}{4 \pi \epsilon} \right)^\frac{d}{2} e^{- \frac{m}{4\epsilon} (x_2 - x_0)^2} \equiv K(2\epsilon , x_2 , x_0).
\end{equation}
Now we can repeat the same computation but for \(i=2\) using the prior results
\begin{align}
\Int \dd^d x_2\, e^{- \frac{m}{4\epsilon} \left[ 2(x_3-x_2)^2 + (x_2 - x_0)^2 \right] } &= \Int \dd^d x_2 \, e^{- \frac{m}{4\epsilon} \left[ 2x_3^2 + x_0^2 + 3 x_2^2 - 2 x_2 (2x_3 +x_0) \right] } \\
&= \left( \frac{4 \pi \epsilon}{3 m} \right)^{\frac{d}{2}} \, e^{- \frac{m}{4 \epsilon} \left[ 2x_3^2 + x_0^2 - \frac{1}{3}(2x_3 + x_0)^2 \right] } \\
& = \left( \frac{4 \pi \epsilon}{3 m} \right)^{\frac{d}{2}} \, e^{- \frac{m}{6 \epsilon} (x_3 - x_0)^2 }.
\end{align}
And we can repeat the same thing for \(K\) as
\begin{equation}
	\Int \dd^d x_2 K(\epsilon , x_3, x_2 ) K(2\epsilon , x_2, x_0 ) = \left( \frac{m}{6 \pi \epsilon} \right)^\frac{d}{2} e^{- \frac{m}{6\epsilon} (x_2 - x_0)^2} \equiv K(3\epsilon , x_3 , x_0)
\end{equation}
By repeating the same procedure indefinitely and taking the appropriate limits we get
\begin{align}
	I &=  \lim_{ \substack{N\to \infty, \epsilon\to 0 \\ N \epsilon = t} }  \Int \left[ \mathlarger{\prod}_{i=1}^{N-1} \dd^d x_i \right] K(\epsilon, x_N , x_{N-1} ) \mathlarger{\prod}_{i=1}^{N-1} 
    K(\epsilon , x_{i} , x_{i-1}) \nonumber\\
    &=  \lim_{ \substack{N\to \infty, \epsilon\to 0 \\ N \epsilon = t} } K(N\epsilon , x_N , x_0) \\
    &= \left( \frac{m}{2\pi t} \right)^\frac{d}{2} e^{ - \frac{m}{2 t} (y-x)^2 }
\end{align}
To compare with the mode decomposition,  with $\zeta$-function regularisation we change variable in the discretised version as $x_k = x + \frac{k \epsilon}{t} (y-x) + h_k$ from which we see that $h_0 = h_N = 0$. The Jacobian is unity $\dd^d x_k = \dd^d h_k$ and $x_{k+1} - x_k = h_{k+1} - h_k + \frac{\epsilon}{t} (y-x)$ so that we get
\begin{align}
	I &=  \lim_{ \substack{N\to \infty, \epsilon\to 0 \\ N \epsilon = t} }  \left(\frac{m}{2 \pi  \epsilon} \right)^\frac{Nd}{2} \Int \mathlarger{\prod}_{i=1}^{N-1} \dd^d x_i
	\exp \left\{ -\frac{m}{2 \epsilon} \Sum_{i=0}^{N-1}\left[ \frac{\epsilon^2}{t^2} (y-x)^2 + (h_{k+1} - h_k)^2  + \frac{2\epsilon}{t} (h_{k+1} - h_k)(y-x)\right] \right\} \\
	&=  e^{- \frac{m}{2t} (y-x)^2} \lim_{ \substack{N\to \infty, \epsilon\to 0 \\ N \epsilon = t} }  \left(\frac{m}{2 \pi  \epsilon} \right)^\frac{Nd}{2} \Int \mathlarger{\prod}_{i=1}^{N-1} \dd^d h_k
	\exp \left\{ -\frac{m}{2 \epsilon} \Sum_{i=0}^{N-1}\left[  (h_{k+1} - h_k)^2 \right] \right\} 
\end{align}
The linear term in $h_{k+1} - h_k$ telescopes to $0$ as $h_0 = h_N = 0$. We then divide out such terms defining the normalisation
\begin{align}
	I &= e^{- \frac{m}{2t} (y-x)^2} \mathcal{N}, & 	\mathcal{N} &= \Int_{\rm DBC} \mathcal{D} h \, e^{- \frac{m}{2} \Int\limits_0^t \dd \tau \dot{h}^2}
\end{align}
We have observed that in time slicing we have that 
\begin{equation}
\mathcal{N} = \left( \frac{m}{2\pi t} \right)^\frac{d}{2}
\end{equation}

\section{Optimal truncation in the double-scaling limit}
\label{sec:optimal-truncation}

In the double-scaling limit, we know every series coefficients of $\omega$, thus we can calculate the Borel transform and evaluate the canonical free energy to any given precision. This is not doable in the generic \acrshort{eft} arising at finite values of $N$, where we need to use numerical computations on the lattice to get the first few coefficients. Here we present what sort of results we should expect for the optimal truncation utilising the results in the double-scaling limit. More sophisticated investigation such as Borel-Pad\'e and conformal mappings~\cite{costin2019resurgent} can produce  significant improvements on the optimal truncation approximations.

\begin{figure}
  \centering
  \includegraphics[width=.48\textwidth]{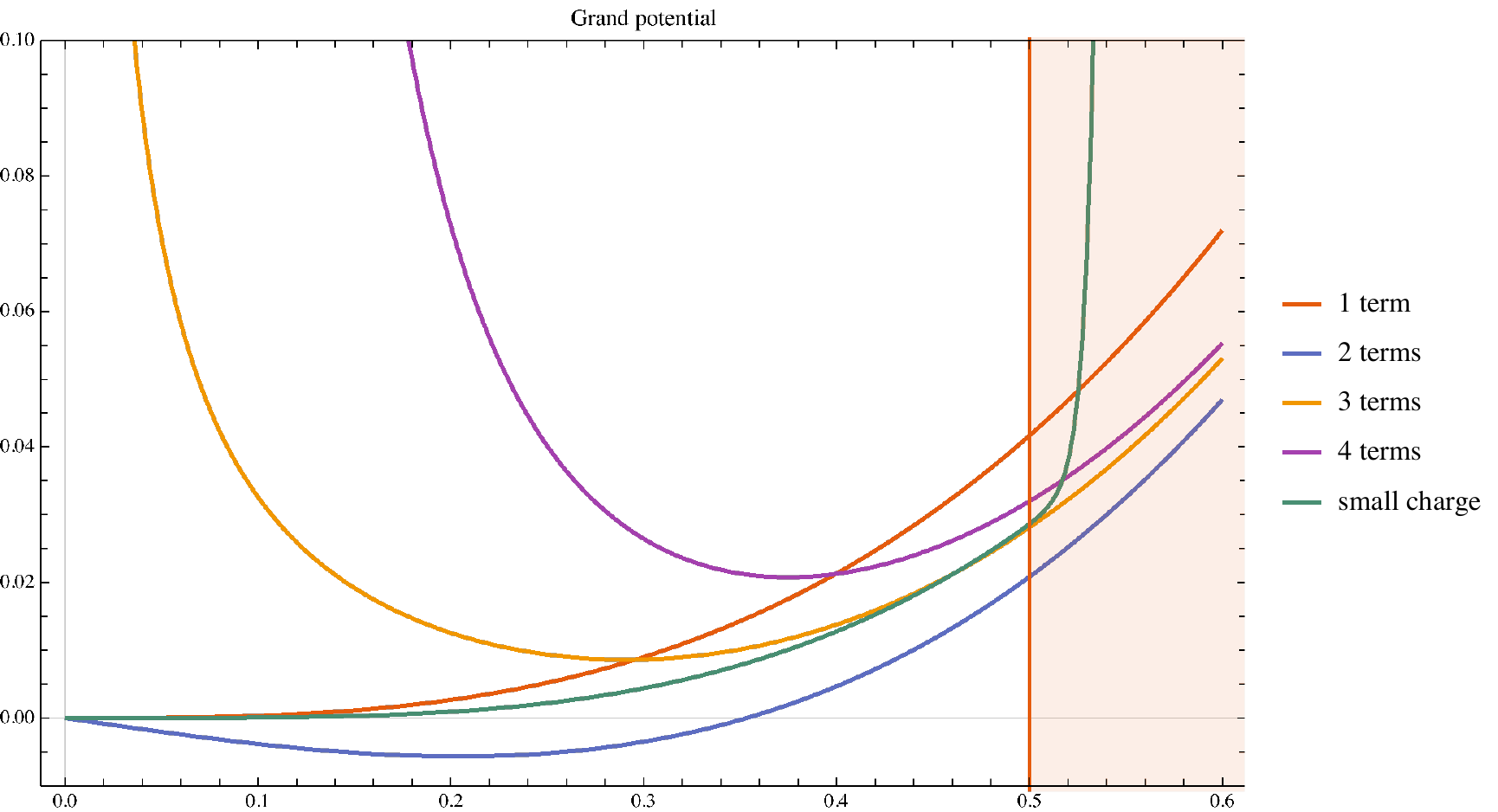} \hfill
  \includegraphics[width=.48\textwidth]{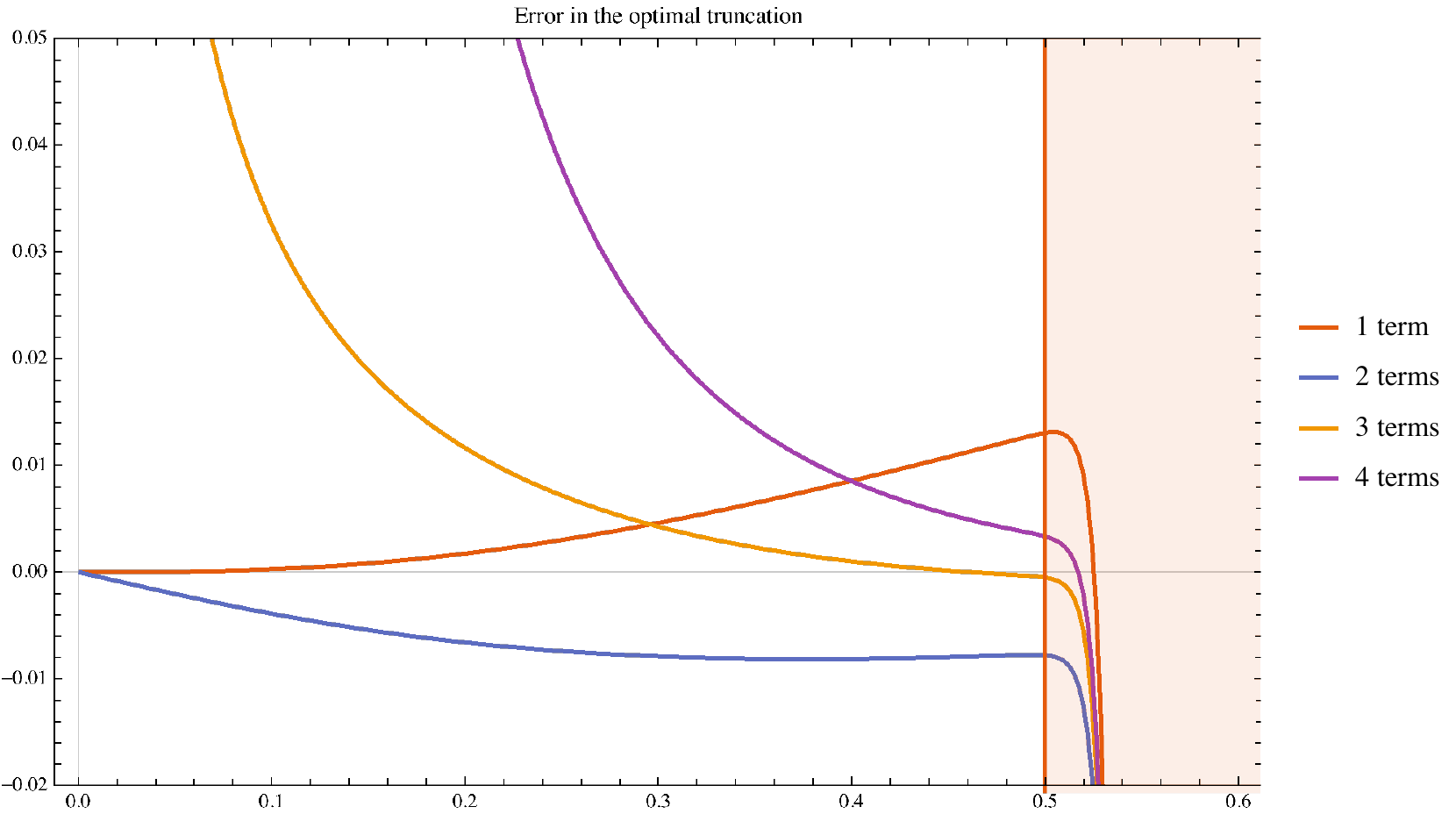}
  \caption{Left: Grand potential $\omega$ as function of $m$ in the small charge expansion and for different numbers of terms. The small-charge expansion breaks down in the red-shaded region. Right: Error in the truncation w.r.t. the exact small-charge expansion.}
  \label{fig:asymptotic-vs-convergent}
\end{figure}

We consider the grand potential \(\omega(m)\) in \cref{eq:large-m-grand-potential}
\begin{equation}
  \omega(m) = R_0^2 m^3 \Sum_{n=0}^\infty \frac{(-1)^n \pqty{1 - 2^{1 - 2n}} B_{2n} \Gamma(\frac{3}{2}-n )}{2 \sqrt{\pi} \Gamma(n+1)} \frac{1}{(R_0m)^{2n}} =  R_0^2 m^3  \Sum_{n=0}^\infty \omega_n \frac{1}{(R_0m)^{2n}}.
\end{equation}
In the thesis and in \cite{dondi2021resurgence} we exhibit that if for large values of \(n\) the constants \(\omega_n\) grow like
\begin{equation}
  \omega_n \sim (\beta n)! A^{-n},
\end{equation}
then the series has an optimal truncation given by the value of \(n\) for which \(\omega_n (R_0m)^{-2n}\) has a saddle:
\begin{equation}
  \label{eq:optimal-truncation}
  N^* \approx \frac{1}{\beta} \abs{A (R_0m)^2}^{1/\beta},
\end{equation}
and the relevant error in the truncation is of order 
\begin{equation}
  \epsilon(m) \sim e^{-(A R_0^2 m^2)^{1/\beta}} .
\end{equation}
Here, for large \(n\) we have
\begin{equation}
  \omega_n \sim \frac{ (2n )! (4 \pi^2)^{-n} }{\sqrt{\pi} n^{5/2}} ,
\end{equation}
so we match \(A = 4 \pi^2\), \(\beta = 2\) and therefore the optimal truncation is for the value
\begin{equation}
  N^* \approx \pi R_0 m .
\end{equation}
Utilising that for large values of  \(\Qb \), the coefficient \(m\) scales as \(R_0 m \sim \sqrt{\Qb} \ \) , we discover that in the expansion of the grand potential, and of the canonical free energy, the optimal truncation is found at the value 
\begin{equation}
  N^* \approx \pi \sqrt{\Qb} . 
\end{equation}
In \Cref{fig:asymptotic-vs-convergent} we compare the asymptotic expansion truncated at the \(N\)-th term with the convergent small-charge expansion.
Empirically, we note that the optimal truncation for \(m < 0.5\)  is at \(N = 3\) terms until we hit the limit where the convergent expansion cannot be trusted any longer.
This is quite close to the asymptotic estimate above that would give \(N^* \lessapprox 3\) in this regime.

\chapter{Preliminaries for Fermionic models} 

\label{AppendixD} 

\section{3d Fermions}\label{sec:app_fermions}

In this appendix following~\cite{dondi2022fermionic}, we gather some background material for fermionic theories in \(d=2+1\) and \(d=3\) spacetime dimensions to make the present work as self-contained as possible.

\subsection{Gamma matrices in the Dirac convention in 3d} \label{sec:notation}
The gamma matrices in \(d=2+1\) and \(d=3\) dimensions are built out of the Pauli matrices
 \begin{align}\label{Pauli}
 	\sigma_1 & = \begin{pmatrix}
 		0 & 1 \\ 1 & 0
 	\end{pmatrix}, & 
 	\sigma_2 &= \begin{pmatrix}
 		0 & -i \\ i & 0
 	\end{pmatrix}, &
 	\sigma_3 &=  \begin{pmatrix}
 		1 & 0 \\ 0 & -1
 	\end{pmatrix},
 \end{align}
 in the following manner
 \begin{align}\label{eq:GammaMatricesConvention}
 d&=2+1:	& &\gamma_0 = i\sigma_3 , & &\gamma_{1,2} = \sigma_{1,2} \\
 d&=3:	 & &\gamma_\mu = \sigma_\mu, & &\mu = 1,2,3.
 \end{align}
And the Clifford algebra in our convention reads
\begin{equation}
    \{ \gamma_\mu , \gamma_\nu \} = 2 \eta_{\mu \nu}.
\end{equation}
We work in the signature $\eta_{\mu\nu} = (-1,1,1)$ for $2+1$ dimensional Minkowski spacetime and in this signature $\gamma_0$ is anti-Hermitian, while spatial $\gamma$s are Hermitian.\\
Furthermore, the gamma matrices satisfy
\begin{align}
(\gamma_i)^2 & = - (\gamma_0)^2 = 1, & \gamma_0 \gamma_\mu \gamma_0 &= (\gamma_\mu)^\dagger.
\end{align}
Complex (Dirac) spinors $\psi$ transform in the usual representation of $SO(1,2),\, SO(3)$ generated by these gamma matrices.\\
The Dirac conjugate in our notation is

\begin{align}
   d& =2+1: & \bar{\psi}& = \psi^\dagger \gamma_0, \\
   d& =3: &\bar{\psi} &=\psi^\dagger \gamma_3.
\end{align}
The continuation from Minkowski to Euclidean spacetime is achieved as follows

\begin{align}
    t  & \rightarrow -i \tau, &\partial_t &\rightarrow i \partial_\tau, &\gamma_0  &\rightarrow i \gamma_3, & \gamma_i &\rightarrow \gamma_i.
\end{align}
Given the above transformation rules, the massive Dirac action is continued in the following manner:

\begin{align}
 &i \underbrace{\Int \dd t \dd^2 x \, \left [ \bar{\psi} ( i \gamma^\mu \del_\mu + i m )\psi\right] }_{S_M} &\rightarrow& &-\underbrace{ \Int \dd^3 x  \left[  \bar{\psi} (\gamma^\mu \del_\mu + m) \psi \right ] }_{S_E} .
 \end{align}
 

 \subsection[\texorpdfstring%
{Spinors on $\setS^1_\beta \times \setS^2$}%
{<S1S2>}]%
{Spinors on $\setS^1_\beta \times \setS^2$}%
\label{sec:cylinder_spinor}

\paragraph{Spinors in spherical coordinates.}
~\\
We follow closely the treatment sketched in \cite{borokhov2003topological}.\\
The Hermitian Dirac operator on $\mathbb{R}^3$ is
\begin{equation}
     i \gamma^\mu \del_\mu = - \vec{\sigma} \cdot \vec{p}
\end{equation}
with momentum operator $p_\mu = - i \partial_\mu$ and $\vec{\sigma} = \sigma_i$ the Pauli matrices of \cref{Pauli}.\\
We define the generalized angular momentum and total angular momentum as
\begin{align}
	\vec{L} &= \vec{r} \times \vec{p} , & \vec{J} &= \vec{L} + \frac{\vec{\sigma}}{2} ,& [ \vec{L}, \vec{r} ] &= [\vec{J}, \vec{r}] = 0.
\end{align}
These are both Hermitian operators. The eigenfunctions of $\vec{L}^2$ are ordinary spherical harmonics:

\begin{align}
	\vec{L}^2 Y_{\ell m} &= \ell (\ell+1 ) Y_{\ell m} , & L_z Y_{\ell m} &= m Y_{\ell m}, & \ell &= 0,1,2... & m &= -\ell , ... \ell.
\end{align}
Using these, we can build simultaneous eigenfunctions of $\{ \vec{J}^2 ,  J_z, \vec{L}^2 , \vec{S}^2 \}$. These are spinor spherical harmonics:

\begin{align}
	\phi^+_{ j m_j} &= \begin{pmatrix}
	\sqrt{\frac{\ell+m+1}{2\ell+1}} Y_{\ell m} \\ \sqrt{\frac{\ell-m}{2\ell+1}} Y_{\ell m+1} 
\end{pmatrix}, &
\phi^-_{ j m_j} &= \begin{pmatrix}
- \sqrt{\frac{\ell-m}{2\ell+1}}  Y_{\ell m} \\ \sqrt{\frac{\ell+m+1}{2\ell+1}} Y_{\ell m+1} 
\end{pmatrix}.
\end{align}
The wave functions $\phi^{\pm}$ correspond respectively to the cases $j = \ell \pm 1/2$ and $m_j = m \pm 1/2$. These have the following quantum numbers:
\begin{align}
	&\begin{cases}
		\vec{L}^2 \phi^\pm_{j m_j} &= \ell(\ell+1) \phi^\pm_{j  m_j}\\
				\vec{J}^2 \phi^\pm_{j m_j} &= j(j+1) \phi^\pm_{j  m_j}\\
						J_z \phi^\pm_{j m_j} &= m_j \phi^\pm_{j  m_j}
	\end{cases} & \begin{cases}
	j =  \frac{1}{2} ,  \frac{3}{2} , \frac{5}{2} ... \\
	m_j = -j ... j
\end{cases}
\end{align}
and are $(2j+1)$-degenerate. Any spinor in $\mathbb{R}^3$ can be decomposed in this orthonormal basis. 
It is convenient to introduce the radial $\gamma$ matrix $\gamma_r = \vec{\gamma} \cdot \hat{r}$. The Dirac operator can then be written as
\begin{equation}
	i \gamma^\mu \del_\mu = i \sigma_r \left\{  \frac{\partial}{\partial r} - \frac{1}{r} \left[  \vec{J}^2 - \vec{L}^2 - \frac{3}{4}  \right]    \right\} ,
\end{equation}
and is diagonal in the $\phi^\pm$ basis.

\paragraph{Weyl map to the cylinder.}
~\\
We perform a Weyl transformation to the cylinder by letting

\begin{align}
    r&=e^{\tau}, & \eta_{\mu\nu}&=R_0e^{2 \tau} g_{\mu\nu}, & \psi_{\mathbb{R}^3}&=e^{-\tau} \psi_{\mathbb{R} \times \setS^2}.
\end{align}
If we foliate $\mathbb{R}^3$ radially we can define the Dirac conjugate as $\psi^\dagger = \bar{\psi} \sigma_r$. Then the free Dirac action on $\mathbb{R}^3$ reads
\begin{align}
	S &= \Int\limits_{\mathbb{R}^3}\, \bar{\psi}\slashed{\partial}\psi= \Int\limits_{\mathbb{R} \times \setS^2} \bar{\psi} \slashed{D} \psi, & \slashed{D} &= \gamma_r  \left\{  \frac{\partial}{\partial \tau} - \frac{1}{R_0} \left[ \hat{J}^2 - \hat{L}^2 + \frac{1}{4}  \right] \right\}.
\end{align}
The eigenfunctions on the cylinder are
\begin{equation}
	\Psi^\pm_{n j m_j}(\tau , \hat{x}) = e^{-i \omega_n \tau} \phi_{j m_j}^{\pm}(\hat{x}),
\end{equation}
where $\hat{x}$ is a point on $\setS^2$. We will make use of the following relations when computing functional determinant on $\setS_\beta^1 \times \setS^2$:
\begin{align}
	\Int\limits_{\mathbb{R} \times \setS^2} \, (\Psi^{\pm}_{j m_j})^\dagger  \Psi^{\pm}_{j' m_j'} &= \delta_{jj'} \delta_{m_j m_j'} \begin{pmatrix}
		1 & 0\\ 0 & 1
	\end{pmatrix}, \\
	\Int_{\mathbb{R} \times \setS^2} \, (\Psi^{\pm}_{j m_j})^\dagger \gamma_r \Psi^{\pm}_{j' m_j'} &= \delta_{jj'} \delta_{m_j m_j'} \begin{pmatrix}
		0 & -1\\ - 1 & 0
	\end{pmatrix}, \\
  	\Int_{\mathbb{R} \times \setS^2} \, (\Psi^{\pm}_{j m_j})^\dagger (i \slashed{D}) \Psi^{\pm}_{j' m_j'} &= \delta_{jj'} \delta_{m_j m_j'} \begin{pmatrix}
		0 & \omega_n -i \omega_j\\  \omega_n+ i \omega_j & 0
	\end{pmatrix},
\end{align}
where we introduced
\begin{align}
	\omega_n &= \frac{(2n+1)\pi}{\beta}, & \omega_j &= \frac{1}{R_0} \left( j + \frac{1}{2} \right),
\end{align}
which are the Matsubara frequencies and the eigenvalues of the Dirac operator on the sphere, respectively.


\subsection{Reducible Representation} 
\label{sec:RedRep}

For 3$d$ fermionic theories with an even number $2N$ of fermion fields $\psi_{a=1...2N}$ it is convenient to introduce a reducible representation of the Clifford algebra as follows:
 \begin{align}
 	\Gamma_\mu &= \sigma_3 \otimes \gamma_\mu = \begin{pmatrix}
 	\gamma_\mu & 0 \\ 0 & - \gamma_\mu 
 	\end{pmatrix}, &   \Psi_a &\equiv  \begin{pmatrix}
 	\psi_a \\ \psi_{a+N}
 \end{pmatrix}, & a = 1, ... , N.
 \end{align}
then we can pick
\begin{equation}
  \Gamma_5 = \sigma_1 \otimes 1 =
  \begin{pmatrix}
    & 1 \\ 1 
  \end{pmatrix}.
\end{equation}
The charge conjugation matrix is 
\begin{equation}\label{eq:chargeconjugationmatrix}
    C_4 = \Gamma_2 = \sigma_3 \otimes C = \begin{pmatrix}
     \sigma_2 &  \\
       & -\sigma_2
    \end{pmatrix} ,
\end{equation}
and satisfies
\begin{align}
    &C_4 = C_4^{-1} = C_4^\dagger = - C_4^T = - C_4^*, \quad &C_4 \Gamma_\mu C_4 = - (\Gamma_\mu)^T .
\end{align}
In addition, it holds that
\begin{align}
    \{ \Gamma_5 ,C_4\} =0 .
\end{align}
In terms of spinors the reducible four-dimensional representation consists of two two-dimensional irreducible spinors,
\begin{align}
    &\Psi = \big( \psi_{1} , \psi_{2} \big)^T , \quad &\bar \Psi = \Psi^\dagger \Gamma_3 =  \big( \psi_{1}^\dagger \gamma_3 , - \psi_{2}^\dagger \gamma_3 \big) = \big( \bar{\psi}_{1} , - \bar{\psi}_{2} \big) ,
\end{align}
for $a=1,\dots,N$. \\
As a concrete example, the action of the $U(1)$-\acrshort{njl} model in terms of this reducible representation can be written as
\begin{equation}
  S =  \Int \dd^3x \left( \bar \Psi \Gamma^\mu \del_\mu \Psi - \frac{ g}{N} \left( (\bar \Psi \Psi)^2 - ( \bar \Psi \Gamma_5 \Psi)^2 \right) \right) .
\end{equation}


\section[\texorpdfstring%
{\(U(1)\) Pauli--Gürsey transformation}%
{<U1>}]%
{\(U(1)\) Pauli--Gürsey transformation}%

\label{sec:Pauli-Gursey}

We consider the following transformation
\begin{equation}
    \begin{aligned}
        &\Psi \mapsto \frac{ 1}{2} \left[ (1-\Gamma_5) \Psi + (1+ \Gamma_5) C_4 \bar\Psi^T  \right] , \\
        &\bar\Psi \mapsto \frac{ 1}{2} \left[ \bar\Psi (1+\Gamma_5) - \Psi^T C_4 (1- \Gamma_5) \right] .
    \end{aligned}
\end{equation}
Note that the explicit expression of the transformation relies upon the chosen convention for the gamma matrices.
This transformation is an involution, as it maps $\Psi_a$ to $\Psi_a$ after applying it twice.
Under the \acrshort{pg} transformation, the kinetic term remains invariant,
\begin{align}
   & \Int \dd^3 x \, \bar \Psi \Gamma^\mu \partial_\mu \partial \Psi &\mapsto& &\Int \dd^3 x \, \bar \Psi  \Gamma^\mu \partial_\mu \partial \Psi ,
\end{align}
while the Cooper \acrshort{bcs} interaction term is related to the \(U(1)\)-\acrshort{njl} interaction term,
\begin{equation}
    \begin{aligned}
        &&- \bar \Psi C_4 \bar \Psi^T \, \Psi^T C_4 \Psi \mapsto \bar \Psi (1+ \Gamma_5) \Psi \, \bar \Psi (1- \Gamma_5) \Psi .
    \end{aligned}
\end{equation}
The converse statement also applies, given that the \acrshort{pg} transformation is an involution.

Eventually, the \acrshort{pg} transformation maps the fermion number chemical potential into the axial chemical potential and vice versa,
\begin{align}
        \bar \Psi \Gamma_3 \mu \Psi &\mapsto \bar \Psi ( - \Gamma_3 \Gamma_5 \mu ) \Psi , & \bar \Psi \Gamma_3 \Gamma_5 \mu \Psi &\mapsto \bar \Psi ( - \Gamma_3 \mu ) \Psi .
    \end{align}
In total, we get the following map for the chiral \acrshort{gn} model:
\begin{equation}
    \begin{aligned}
        S &= \Int \dd^3x \left( \bar \Psi ( \Gamma^\mu \del_\mu - \mu \Gamma_3 \Gamma_5 ) \Psi - \frac{ g}{N} \left( (\bar \Psi \Psi)^2 - ( \bar \Psi \Gamma_5 \Psi)^2  \right) \right) \\
        &= \Int \dd^3x \left( \bar \Psi ( \Gamma^\mu \del_\mu - \mu \Gamma_3 \Gamma_5 ) \Psi - \frac{ g}{N} \bar \Psi ( 1 + \Gamma_5 ) \Psi \, \bar \Psi ( 1 - \Gamma_5 ) \Psi \right) \\
        &\mapsto \Int \dd^3x \left( \bar \Psi ( \Gamma^\mu \del_\mu + \mu \Gamma_3 ) \Psi + \frac{ g}{N} \bar \Psi C_4 \bar \Psi^T \, \Psi^T C_4 \Psi \right) .
    \end{aligned}
\end{equation}
For wholeness, the Cooper model written with regard to the Stratonovich fields reads
\begin{equation}
  \Lp =   \bar \Psi \Gamma^\mu \del_\mu \Psi + i \frac{\Phi}{2} \bar \Psi C_4 \bar \Psi^T + i \frac{ \Phi^*}{ 2} \Psi^T C_4 \Psi + \frac{ N}{4g} \Phi^{*} \Phi .
\end{equation}

\section{Finite-density loop integrals and Matsubara sums}
\label{sec:loop}

\subsubsection[\texorpdfstring%
{Fourier transforms on $\setS^1_\beta \times \mathbb{R}^2$}%
{<S1R>}]%
{Fourier transforms on $\setS^1_\beta \times \mathbb{R}^2$}%
	
We designate a point on $\setS^1_\beta \times \mathbb{R}^2$ as $X = (\tau, x)$ and momenta as $P = (\omega_n , \vec{p})$ where $\omega_n = \pi (2n+1)/\beta$ are fermionic Matsubara frequencies. The Fourier transforms in our normalisation conventions read
\begin{align}
\delta(X-X') &=\SumInt \frac{\dd^d p}{\beta (2\pi)^d} e^{- i P \cdot (X-X')} & \delta_{nn'} \delta(p-p') &= \Int \frac{\dd \tau \dd^d x}{\beta (2\pi)^d} e^{- i X \cdot (P-P')} \\
f(X) &=\SumInt \frac{\dd^d p}{\sqrt{\beta(2\pi)^d}} e^{- i P \cdot X} f(P), & f(P) &= \Int \frac{\dd \tau \dd^d x}{\sqrt{\beta (2\pi)^d}}  e^{i P \cdot X} f(X).
\end{align}
%

\subsubsection{Matsubara sums}

The master formula for fermionic Matsubara sums that we encounter in the computation of the Dirac determinant is
\begin{equation}
\Sum_{n \in \mathbb{Z}} \log \left[ \frac{(2n+1)^2 \pi^2 + A^2}{(2n+1)^2 \pi^2 + 1 } \right] = A + 2 \log \left( 1 + e^{- A} \right).
\end{equation}
%

\subsection[\texorpdfstring%
{GN scalar integrals at finite $\mu,\beta$}%
{<mubeta>}]%
{GN scalar integrals at finite $\mu,\beta$}%

We collect the scalar integral utilized to obtain the results in \cref{sec:gn-fluctuations}.
Following~\cite{laine2016basics}, one-loop integrals at finite temperature and chemical potential can all be derived from the massive scalar integral
\begin{equation}
	\Int \frac{\dd^d k}{(2\pi)^d} \frac{1}{[k^2+m^2]^\alpha} = \frac{1}{(4\pi)^{\frac{d}{2}}} \frac{\Gamma(\alpha-d/2)}{\Gamma(\alpha)} (m^2)^{-\alpha + \frac{d}{2}} .
\end{equation}
Recalling that $\tilde{K} = ( \omega_n - I \mu , \vec{k})$ the first scalar integral we used can be computed as
\begin{align}
	I_1 &=  \SumInt \frac{\dd^d k}{\beta (2\pi)^d} \frac{1}{\tilde{K}^2} \nonumber\\
	&= \frac{\Gamma(1-d/2)}{(4\pi)^\frac{d}{2}} \Sum_{n \in \mathbb{Z}} \frac{1}{[(\omega_n - i \mu)^2]^{1 - \frac{d}{2}}} \nonumber\\
	&=\frac{\Gamma(1-d/2)}{(4\pi)^\frac{d}{2}} \left( \frac{2\pi}{\beta} \right)^{-2 + d} \Sum_{n \in \mathbb{Z}} \frac{1}{\left[ \left( n + \frac{1}{2} - i \bar{\mu} \right)^2 \right]^{1 - \frac{d}{2}} }\nonumber\\
	&=\frac{\Gamma(1-d/2)}{(4\pi)^\frac{d}{2}} \left( \frac{2\pi}{\beta} \right)^{-2 + d} \left\{    \zeta\left( 2 - d , \frac{1}{2} - i \bar{\mu} \right) +   \zeta\left( 2 - d , \frac{1}{2} + i \bar{\mu} \right)   \right\},
\end{align}
where we normalise bare quantities as $\bar{\mu} = \beta \mu/(2\pi)$ etc. At \(\beta \to \infty\) this becomes
\begin{equation}
	\lim_{\beta \rightarrow \infty}  I_1 = - \frac{\mu}{4\pi} .
\end{equation}
The $I_2$ integral has three scales: $\beta,\mu, P$ where $P$ is an external momentum. It can be computed similarly to $I_1$ where we use a Feynman parametrization to merge the propagators
\begin{align}
	I_2 &=\SumInt \frac{\dd^2 k \dd^2 q}{\beta (2\pi)^2} \frac{\delta(K+Q-P)}{\tilde{K}^2 (\tilde{Q}^\dagger)^2 } \nonumber\\
	&= \Int_0^1 \dd x \SumInt \frac{\dd^2 k }{\beta (2\pi)^2} \frac{1}{ \left[ k^2 + \left\{   x(1-x) p^2 + (1-x)(\omega_n - i \mu)^2 + x (\omega_m - \omega_n + i \mu)^2   \right\} \right]^2 }\nonumber \\
	&= \frac{\Gamma(2-d/2)}{\beta (4\pi)^{\frac{d}{2}}} \left( \frac{2\pi}{\beta} \right)^{d-4} \Int\limits_0^1 \dd x \Sum_{n \in \mathbb{Z}}  \frac{1}{ \left[ (n + \frac{1}{2}  - i \bar{\mu} - x \bar{\omega}_m )^2 + x(1-x) (\bar{p}^2 + \bar{\omega}^2)   \right]^{2- \frac{d}{2}} }\nonumber\\
	&= \frac{\Gamma(2-d/2)}{\beta (4\pi)^{\frac{d}{2}}} \left( \frac{2\pi}{\beta} \right)^{d-4} \Int\limits_0^1 \dd x \Sum_{n \in \mathbb{Z}}  \frac{1}{ \left[ (n +A)^2 +B  \right]^{2- \frac{d}{2}} }\nonumber\\
	&= \frac{\Gamma(2-d/2)}{\beta (4\pi)^{\frac{d}{2}}} \left( \frac{2\pi}{\beta} \right)^{d-4} \Int\limits_0^1 \dd x \left\{ \frac{1}{[A^2 + B]^{2-\frac{d}{2}}} +F(2-d/2 ; A,B) +F(2-d/2 ; -A,B) \right\},
\end{align}
where we introduced 
\begin{equation}
	A = \frac{1}{2} - i \mu  - x \omega_m, \quad B = x(1-x) [\bar{p}^2 +\bar{\omega}_m^2 ].
\end{equation}
The functions $F$ are special $\zeta$-functions found in \cite{elizalde2012ten}. At \(\beta \to \infty\) we find
\begin{equation}
	\lim_{\beta \rightarrow \infty} I_2 = \frac{1}{8\sqrt{\omega_m^2 + p^2}} = \frac{1}{8 \sqrt{P^2}} .
\end{equation}
%

\subsection{NJL loop integrals} \label{sec:njl-computations}

We want to compute the following integrals:
\begin{align}
	G^{-1}_{\sigma\sigma}(P) &= -\Int \frac{\dd^3k}{(2\pi)^3} \Tr\left[\Delta^{(\mu,\sigma)}(K)\Delta^{(-\mu,-\sigma)}(P-K) \right],\\
	G^{-1}_{\sigma\pi}(P) &= -i\Int \frac{\dd^3k}{(2\pi)^3}\Tr\left[\Delta^{(\mu,\sigma)}(K)\Gamma_5 \Delta^{(-\mu,-\sigma)}(P-K) \right],\\
	G^{-1}_{\pi\sigma}(P) &= -i\Int \frac{\dd^3k}{(2\pi)^3}\Tr\left[ \Gamma_5 \Delta^{(\mu,\sigma)}(K) \Delta^{(-\mu,-\sigma)}(P-K) \right] ,\\
	G^{-1}_{\pi\pi}(P) &= \Int \frac{\dd^3k}{(2\pi)^3}\Tr\left[\Gamma_5 \Delta^{(\mu,\sigma)}(K)\Gamma_5 \Delta^{(-\mu,-\sigma)}(P-K) \right],
\end{align}
where the fermion propagator is given by
\begin{equation}
\begin{split}
	\Delta^{(\mu,\sigma)}( P) & = (-i\slashed{P} +\Phi_0 -\mu\Gamma_3\Gamma_5)^{-1}\\
	& = \frac{
    \left( \omega^2 + k^2 + \Phi_0^2 - \mu^2 + 2 \mu (i \omega \Gamma_3 + \Phi_0 ) \Gamma_3 \Gamma_5 \right) 
    }{\left(\omega^2 + \Phi_0^2 + (\mu + k)^2 \right)\left(\omega^2 + \Phi_0^2 + (\mu - k)^2 \right)} \left( i\slashed{P} +\Phi_0 -\mu\Gamma_3 \Gamma_5\right) .
\end{split}
\end{equation}

\subsubsection{Zeroth order in $P /\mu$}
At zeroth order in $P/\mu$ these integrals are now
\footnotesize\begin{align}
&G^{-1}_{\sigma\sigma}(P)|_{\mathcal{O}(0)}  =\Int \frac{\dd^2 k \dd k_0}{(2 \pi)^3}   \Bigg[ 4 \Phi_0^2 \left(\frac{1}{\left((k+\mu )^2+k_0^2+\Phi_0 ^2\right)^2}+\frac{1}{\left((k-\mu )^2+k_0^2+\Phi_0 ^2\right)^2}\right) 	-\frac{2}{(k-\mu )^2+k_0^2+\Phi_0 ^2}-\frac{2}{(k+\mu )^2+k_0^2+\Phi ^2} \Bigg],\\
 &G^{-1}_{\sigma\pi}(P)|_{\mathcal{O}(0)}  =0,\\
 &G^{-1}_{\pi\sigma}(P)|_{\mathcal{O}(0)}  =0,\\
 &G^{-1}_{\pi\pi}(P)|_{\mathcal{O}(0)}  =- \Int \frac{\dd^2 k \dd k_0}{(2 \pi)^3} \left[\frac{2}{(k+\mu )^2+k_0^2+\Phi_0 ^2}+\frac{2}{(k-\mu )^2+k_0^2+\Phi_0 ^2}\right].
\end{align}
\normalsize We then perform the residue integral over $k_0$,
\begin{align}
     G^{-1}_{\sigma\sigma}(P)|_{\mathcal{O}(0)} & =- \Int \frac{\dd^2 k }{(2 \pi)^3} \left[\frac{2 \pi  (k-\mu )^2}{\left((k-\mu )^2+\Phi_0 ^2\right)^{3/2}}+\frac{2 \pi  (k+\mu )^2}{\left((k+\mu )^2+\Phi_0 ^2\right)^{3/2}} \right], \\
	G^{-1}_{\pi\pi}(P)|_{\mathcal{O}(0)} & = \Int \frac{\dd^2 k }{(2 \pi)^3} \left[2 \pi  \left(\frac{1}{\sqrt{(k+\mu )^2+\Phi_0 ^2}}+\frac{1}{\sqrt{(k-\mu )^2+\Phi_0 ^2}}\right) \right].
\end{align}
The remaining integrals are divergent. The divergence is however independent of $\mu$, so we can simply subtract the expression for $\mu=0$ to regularize them:
\begin{multline}
  G^{-1}_{\sigma\sigma}(P)|_{\mathcal{O}(0)} = \Int \frac{\dd^2 k }{(2 \pi)^3} \left[ \frac{4 \pi  k^2}{\left(k^2+\Phi_0 ^2\right){}^{3/2}}-\frac{2 \pi  (k-\mu )^2}{\left((k-\mu )^2+\Phi _0^2\right){}^{3/2}}-\frac{2 \pi  (k+\mu )^2}{\left((k+\mu )^2+\Phi _0^2\right){}^{3/2}} \right] 
                           \\ - \Int \frac{\dd^2 k }{(2 \pi)^3} \frac{4 \pi  k^2}{\left(k^2+\Phi_0 ^2\right){}^{3/2}}
\end{multline}
\begin{multline}                           
	G^{-1}_{\pi\pi}(P)|_{\mathcal{O}(0)}  = \Int \frac{\dd^2 k }{(2 \pi)^3} \left[-\frac{ 4 \pi}{ \sqrt{ k^2+ \Phi_0 ^2}} +\frac{2 \pi}{ \sqrt{ (k-\mu )^2+\Phi_0 ^2}} +\frac{2 \pi}{ \sqrt{ (k+\mu )^2+\Phi_0^2}} \right] \\+ 4 \pi \Int \frac{\dd^2 k }{(2 \pi)^3} \frac{1}{\sqrt{k^2+\Phi_0^2}}.
\end{multline}
The divergent integrals after regulation give
\begin{align}
    &\Int \frac{\dd^2 k }{(2 \pi)^3} \frac{4 \pi  k^2}{\left(k^2+\Phi_0 ^2\right){}^{3/2}}  = \frac{8}{(2 \pi)^3} \pi^2 \Int \dd k \ k \frac{k^2}{\left(k^2+\Phi _0^2\right){}^{3/2}} \equiv - \frac{16}{(2 \pi)^3} \pi^2 \Phi_0,\\
     &\Int \frac{\dd^2 k }{(2 \pi)^3} \frac{4 \pi}{\sqrt{k^2+\Phi_0^2}} \equiv - \frac{8}{(2 \pi)^3} \pi^2 \Phi_0.
\end{align}
We can then perform the spatial integral over the momentum $k$. We get
\begin{align}
    G^{-1}_{\sigma\sigma}(P)|_{\mathcal{O}(0)} & = \frac{8 \pi ^2}{(2 \pi)^3} \left(2 \sqrt{\mu ^2+\Phi_0 ^2}-\mu  \arctanh\left(\tfrac{\mu }{\sqrt{\mu ^2+\Phi_0 ^2}}\right)\right), \\
    G^{-1}_{\pi\pi}(P)|_{\mathcal{O}(0)} & = \frac{8 \pi ^2}{(2 \pi)^3} \left(\sqrt{\mu ^2+\Phi_0 ^2}-\mu  \arctanh\left(\tfrac{\mu }{\sqrt{\mu ^2+\Phi_0 ^2}}\right)\right).
\end{align}
Finally we can use the \acrshort{eom},
\begin{align}
	\Phi_0 &=\sqrt{\kappa_0^2 -1}\mu, &
	\arctanh*(\tfrac{1}{\kappa_0}) &=\kappa_0
\end{align}
to find the final result
\begin{align}
    G^{-1}_{\sigma\sigma}(P)|_{\mathcal{O}(0)} & = \frac{\kappa_0 \pi}{\mu}, & 
    G^{-1}_{\pi\pi}(P)|_{\mathcal{O}(0)} & =0.
\end{align}

\subsubsection{First order in $P/ \mu$}

At order 1 in $P/\mu$ the following two integrals are an odd function of $k_0$ and $k_1,k_2$, hence under integration it follows that
\begin{align}
F^{-1}_{\sigma\sigma}(P)|_{\mathcal{O}(P/ \mu)} & = 0,\\ G^{-1}_{\pi\pi}(P)|_{\order{P/\mu}} & =0.
\end{align}
The remaining two are computed as follows:
\begin{align}
    G^{-1}_{\sigma\pi}(P) |_{\order{P/\mu}}& = \Int \frac{\dd^2 k \dd k_0}{(2\pi)^3} \frac{4 \mu ^2 \frac{\omega}{\mu} \left(-3 k^4+2 k^2 \left(-k_0^2+\mu ^2-\Phi_0 ^2\right)+\left(k_0^2+\mu ^2+\Phi_0 ^2\right)^2\right)}{\left((k-\mu )^2+k_0^2+\Phi_0 ^2\right)^2 \left((k+\mu )^2+k_0^2+\Phi_0 ^2\right)^2}, \\
    G^{-1}_{\pi\sigma}(P) |_{\order{P/\mu}} & = -\Int \frac{\dd^2 k \dd k_0}{(2\pi)^3}\frac{4 \mu ^2 \frac{\omega}{\mu}    \left(-3 k^4+2 k^2 \left(-k_0^2+\mu ^2-\Phi_0 ^2\right)+\left(k_0^2+\mu ^2+\Phi_0 ^2\right)^2\right)}{\left((k-\mu )^2+k_0^2+\Phi_0 ^2\right)^2 \left((k+\mu )^2+k_0^2+\Phi_0 ^2\right)^2}.
\end{align}
We perform the integral over $k_0$ to obtain
\begin{align}
    G^{-1}_{\sigma\pi}(P) |_{\order{P/\mu}}& = \pi \Int \frac{\dd^2 k}{(2 \pi)^3}   \left(\frac{\mu  (\mu -k)}{\left((k-\mu )^2+\Phi_0 ^2\right)^{3/2}}+\frac{\mu  (k+\mu )}{\left((k+\mu )^2+\Phi_0 ^2\right)^{3/2}}\right) \frac{\omega}{\mu} ,\\
    G^{-1}_{\pi\sigma}(P) |_{\order{P/\mu}} & = \pi \Int \frac{\dd^2 k}{(2 \pi)^3}    \left(\frac{\mu  (k-\mu )}{\left((k-\mu )^2+\Phi_0 ^2\right)^{3/2}}-\frac{\mu  (k+\mu )}{\left((k+\mu )^2+\Phi_0 ^2\right)^{3/2}}\right)\frac{\omega}{\mu} .
\end{align}
We can then perform the spatial integral over the momentum $k_1$ and $k_2$ 
\begin{align}
     G^{-1}_{\sigma\pi}(P) |_{\order{P/\mu}}& = \frac{2 \pi ^2}{(2 \pi)^3} \mu \frac{\omega}{\mu}   \log \left(\tfrac{2 \mu  \left(\mu -\sqrt{\mu ^2+\Phi_0 ^2}\right)}{\Phi_0 ^2}+1\right) ,\\
     G^{-1}_{\pi\sigma}(P) |_{\order{P/\mu}} & = \frac{2 \pi ^2}{(2 \pi)^3} \mu \frac{\omega}{\mu}  \log \left(\tfrac{2 \mu  \left(\sqrt{\mu ^2+\Phi_0 ^2}+\mu \right)}{\Phi_0 ^2}+1\right) .
\end{align}
Finally using the \acrshort{eom} and simplifying we end up with
\begin{align}
    G^{-1}_{\sigma\pi}(P) |_{\order{P/\mu}}& =  -\frac{\kappa_0  \omega }{2 \pi }, &  G^{-1}_{\pi\sigma}(P) |_{\order{P/\mu}} & = \frac{\kappa_0  \omega }{2 \pi }.
\end{align}

\subsubsection{Second order in $P^2 / \mu^2$}

Next we consider the quadratic part in $P$. By rotational invariance, at second order in $\mathcal{O}(P^2/\mu^2)$, the integrand must have the form
\begin{equation}
	A(k)\omega^2+ B(k)\omega(P\cdot k) + C(k)P^2+ D(k) (k\cdot P)^2.
\end{equation}
The $B(k)$ piece does not contribute as it is an odd function of $k_1,k_2$. Similarly, the cross-term in $(k\cdot P)^2$ will not contribute. The part of the integrand that contributes is thus
\begin{equation}
	A(k)\omega^2+  C(k)P^2+ D(k) (k_1^2p_1^2+k_2^2p_2^2).
\end{equation} 
Given the symmetry under the exchange $1 \leftrightarrow 2$, this is a function of $p_1^2+p_2^2$.\\
We will split the computation into two parts, one with $\omega$ and the other with $P$.\\
We start with the $\omega$ part. After performing the $k_0$ residue integral the two remaining integrals are
\begin{align}
    I^{\omega}_{\sigma \sigma} & = \frac{1}{2} \Int \frac{\dd^2 k}{(2 \pi)^3} \pi  \mu ^2 \left (\frac{\omega}{\mu} \right)^2 \left(\frac{(k-\mu )^2}{\left((k-\mu )^2+\Phi_0 ^2\right)^{5/2}}+\frac{(k+\mu )^2}{\left((k+\mu )^2+\Phi_0 ^2\right)^{5/2}}\right) , \\
    I^{\omega}_{\pi \pi} & = \Int \frac{\dd^2 k}{(2 \pi)^3} \frac{1}{2} \pi  \mu ^2 \left (\frac{\omega}{\mu} \right)^2 \left(\frac{1}{\left((k+\mu )^2+\Phi_0 ^2\right)^{3/2}}+\frac{1}{\left((k-\mu )^2+\Phi_0 ^2\right)^{3/2}}\right).
\end{align}
We can then perform the integrals over $k_1$ and $k_2$ to get
\begin{align}
     I^{\omega}_{\sigma \sigma} & = \frac{1}{(2 \pi)^3} \left (\frac{\omega}{\mu} \right)^2 \frac{2 \pi ^2 \mu ^2  \left(\mu ^2+2 \Phi_0 ^2\right)}{3 \Phi_0 ^2 \sqrt{\mu ^2+\Phi_0 ^2}}, &
     I^{\omega}_{\pi \pi}  & = \frac{1}{(2 \pi)^3} \left (\frac{\omega}{\mu} \right)^2\frac{2 \pi ^2 \mu ^2  \sqrt{\mu ^2+\Phi_0 ^2}}{\Phi_0 ^2}.
\end{align}
By using the \acrshort{eom} and simplifying we finally get
\begin{align}
   I^{\omega}_{\sigma \sigma}& = \frac{\omega ^2-2 \kappa_0 ^2 \omega ^2}{12 \pi  \kappa_0  \mu -12 \pi  \kappa_0 ^3 \mu }, &
   I^{\omega}_{\pi \pi}  & = -\frac{\kappa_0  \omega ^2}{4 \pi  \mu -4 \pi  \kappa_0 ^2 \mu }.
\end{align}
Similarly we can repeat the same procedure for the parts that depend on $P, \ p_1, \ p_2$. First we perform the residue integral over $k_0$ and then the integrals over $k_1, \ k_2$. The result is
\begin{align}
    I^P_{\sigma \sigma} & = \frac{1}{2} \frac{\pi ^2 \mu}{(2 \pi)^3} \left( \frac{p}{\mu} \right)^2 \left(2 \arctanh\left(\tfrac{\mu }{\sqrt{\mu ^2+\Phi_0 ^2}}\right)+\frac{2 \mu  \left(\mu ^4+\Phi_0 ^4\right)}{3 \Phi_0 ^2 \left(\mu ^2+\Phi_0 ^2\right)^{3/2}}\right),\\
    I^P_{\pi \pi} & = \frac{\pi ^2 \mu}{(2 \pi)^3}  \left( \frac{p}{\mu} \right)^2 \left(\frac{\mu  \sqrt{\mu ^2+\sigma ^2}}{\Phi_0 ^2}+\arctanh\left(\tfrac{\mu }{\sqrt{\mu ^2+\Phi_0 ^2}}\right)\right).
\end{align}
Again, by using the \acrshort{eom} and simplifying we finally get
\begin{align}
   I^P_{\sigma \sigma} & =\frac{\left(3 \kappa_0 ^6-2 \kappa_0 ^4-2 \kappa_0 ^2+2\right) p^2}{24 \pi  \kappa_0 ^3 \left(\kappa_0 ^2-1\right) \mu }, &
   I^P_{\pi \pi}  & =-\frac{\kappa_0 ^3 p^2}{8 \pi  \mu -8 \pi  \kappa_0 ^2 \mu }.
\end{align}
So finally we can put it all together to get
\begin{align}
    G^{-1}_{\sigma\sigma}(P)|_{\mathcal{O}(P^2/ \mu^2)} & = \frac{\omega ^2-2 \kappa_0 ^2 \omega ^2}{12 \pi  \kappa_0  \mu -12 \pi  \kappa_0 ^3 \mu } + \frac{\left(3 \kappa_0 ^6-2 \kappa_0 ^4-2 \kappa_0 ^2+2\right) p^2}{24 \pi  \kappa_0 ^3 \left(\kappa_0 ^2-1\right) \mu },\\
    G^{-1}_{\pi\pi}(P)|_{\order{P^2/\mu^2}} & =   -\frac{\kappa_0  \omega ^2}{4 \pi  \mu -4 \pi  \kappa_0 ^2 \mu } -\frac{\kappa_0 ^3 p^2}{8 \pi  \mu -8 \pi  \kappa_0 ^2 \mu }.
\end{align}

\section{Resurgence Analysis of the eigenvalues of the Dirac Operator}

Taking as \(\Sigma = \mathbb{S}^2\)  we want to perform a similar analysis to \cref{sec:perturbative-sphere} but now for the eigenvalues of the Dirac operator. After some formal manipulation, see \cref{sec:cylinder_spinor} the spectrum of Laplacian for spin-\(1/2\) particles reads
\begin{equation}
    \textnormal{spec} \pqty{\Delta_{\mathbb{S}^2}^{1/2}} = \Bqty{ -\frac{(\ell + 1)^2}{R_0^2} \bigg| \ell \in \mathbb{N} }.
\end{equation}
and each eigenvalue is degenerate with multiplicity \(M_{\ell} = 4 \pqty{\ell + 1}\).

Applying the Poisson summation formula, the trace reads 
\begin{align}
\Tr\bqty{e^{\Delta_{\mathbb{S}^2}^{1/2}t }} & = 4 \Sum_{\ell = 0}^{\infty} (\ell +1) e^{-t (\ell + 1)^2/{R_0^2}} = 2 \Sum_{\ell = - \infty}^{\infty} \abs{\ell +1} e^{-t {(\ell + 1)^2}/ {R_0^2} } \\
   & = 2 \Sum_{- \infty}^{\infty} \Int_{\mathbb{R}} \dd \rho \abs{\rho} e^{- \rho^2 t / R_0^2 +2 \pi i k \rho},
\end{align}
Now we are able to extract the zero mode out of the summation and afterward expand the exponential as
\begin{align}
\Tr\bqty{e^{\Delta_{\mathbb{S}^2}^{1/2}t }} & = 2  \Int_{\mathbb{R}} \dd \rho |\rho| e^{- \rho^2 t /R_0^2 } + 2 \Sum\limits_{k \in \Bqty{\mathbb{Z} - \setminus \Bqty{0}}} \Int_{\mathbb{R}} \dd \rho |\rho| e^{- \rho^2 t / R_0^2 } \cos{(2 \pi k  \rho)} \nonumber \\
& = 2  \Int_{\mathbb{R}} \dd \rho |\rho| e^{- \rho^2 t /R_0^2 } + 4 \Sum\limits_{k \in \Bqty{\mathbb{Z} - \setminus \Bqty{0}}} \Int_{0}^{\infty} \dd \rho |\rho| e^{- \rho^2 t /R_0^2 } \cos{(2 \pi k  \rho)} \nonumber \\
& = \frac{2 R_0^2}{t} + 4 \Sum\limits_{k \in \Bqty{\mathbb{Z} - \setminus \Bqty{0}}} \Int_{0}^{\infty} \dd \rho |\rho| e^{- \rho^2 t /R_0^2 } \cos{(2 \pi k  \rho)} \nonumber \\
& = \frac{2 R_0^2}{t} +  2 \Sum\limits_{k \in \Bqty{\mathbb{Z} - \setminus \Bqty{0}}}  \bqty{ \frac{R_0^2}{t} - \frac{2 \abs{k} \pi R_0^3}{t^{3/2}} \operatorname{F}\left( \tfrac{\pi R_0 \abs{k}}{\sqrt{t}} \right)  }.
\end{align}
In the last line we have introduced the Dawson's function \(\operatorname{F}(z)\) which is related to the error function as in \cref{eq:dawsonfunctionerror} and for small values of its argument we can use its asymptotic expansion as in \cref{eq:Dawson}. Following \cref{sec:perturbative-sphere} the leading asymptotic form of the heat-kernel trace reads
\begin{equation}\label{eq:heat-kerneltracedirac}
\Tr\bqty{e^{\Delta_{\mathbb{S}^2}^{1/2}t }} \sim \frac{2R_0^2}{t} -2 \Sum_{n=1}^\infty \frac{(-1)^{n+1}  B_{2n} \ t^{n-1}}{n! \ R_0^{2n-2}} \equiv \frac{2R_0^2}{t}  \Sum_{n=0}^\infty a_n \pqty{\frac{t}{R_0^2}}^n,
\end{equation}
where \(B_{2n}\) are the Bernoulli numbers. As in \cref{sec:perturbative-sphere} we assume into the completion of a resurgent trans-series. Upon using the identities
\begin{gather}
    \Sum_{k \neq 0} \frac{1}{k^{2n}} = 2 \zeta(2n)=  (-1)^{n+1} \frac{(2 \pi)^{2n}}{(2n)!} B_{2n} ,\\
    (-1)^{n+1} B_{2n} = (2n)! \Sum_{k \neq 0} \frac{1}{k^{2n}} \frac{1}{(2 \pi)^{2n}},
\end{gather}
and \cref{eq:Gamman+1/2} we can write \(a_n\) in the suggestive form
\begin{equation}\label{eq:alphadirac}
	a_n = - \frac{(2n)!}{n! \ 2^{2n} \pi^{2n}} \Sum_{k \neq 0} \frac{1}{k^{2n}} = - \frac{1}{\sqrt{\pi}}  \Sum_{k \neq 0} \frac{\Gamma(n+ \tfrac{1}{2})}{(\pi k)^{2n}}.
\end{equation}
Comparing \cref{eq:alphadirac} with \cref{eq:GenericLargeOrder}, we find that
\begin{align}
	\beta &= 1, & b_k&= \frac{1}{2}, & A_k &= (\pi k)^2, & \frac{S_k}{2\pi i } a_0^{(k)} &= -  \abs{k} \sqrt{\pi} , & a^{(k)}_{>0} &= 0.
\end{align}
Therefore, a trans-series representation of the heat-kernel trace contains the terms

\begin{equation}
 \Tr\bqty{e^{\Delta_{\mathbb{S}^2}^{1/2}t }} \,\supset\, -	4 i \pqty{ \frac{\pi R_0^2}{t} }^\frac{3}{2} \Sum_{k \neq 0} \sigma_k^{\pm} \, | k |\, e^{- (k\pi R_0)^2/t}.
\end{equation}

\subsubsection{Borel resummation}

At this point, we will follow closely \cref{sec:Borelresum}. We begin by Borel transforming the heat-kernel trace of \cref{eq:heat-kerneltracedirac} --- see \Cref{sec:borel-transform} for details. In our case, we get
\begin{equation}
    \mathcal{B}\{\Phi^{(0)} \} (\zeta)=   \Sum_{n=0}^\infty \frac{a_n}{\Gamma(n+1/2)} \zeta^{2n} =  \Sum_{n=0}^\infty \frac{(-1)^n B_{2n} }{n! \Gamma(n+1/2)} \zeta^{2n}
\end{equation}
Now, by using the following identities 
\begin{gather}
    \Gamma(n + \frac{1}{2}) = \frac{(2n)! \sqrt{\pi}}{2^{2n} n!},\\
    \cot{x} = \Sum_{n=0}^{\infty} \frac{1}{x} \frac{(-1)^n 2^{2n} B_{2n}}{(2n)!} x^{2n}.
\end{gather}
we derive
\begin{equation}
    \mathcal{B}\{\Phi^{(0)} \} (\zeta)=  \frac{\zeta \cot{\zeta}}{\sqrt{\pi}}.
\end{equation}
The appropriate Borel resummation in the direction $\theta = 0$ is then 

\begin{equation}
 \mathcal{S} \{ \Phi^{(0)} \} (z)=   \Int_0^\infty \frac{\dd{\zeta}^2}{\zeta^2} \bigg(\frac{\zeta^2}{z}\bigg)^{1/2}\,e^{-\zeta^2/z} 	\mathcal{B}\{\Phi^{(0)} \} (\zeta) = \frac{2}{\sqrt{\pi z}} \Int_0^\infty \dd \zeta\, \zeta\,e^{-\zeta^2 / z} \cot{\zeta}
\end{equation}
This is the integral representation of the heat-kernel trace on $\mathbb{S}^2$ for fermions, found originally in	\cite{Miele_1997} but now retrieved as a Borel integral. The above integral is ill-defined given that $\cot{\zeta} $ has simple poles along the integration path for $\zeta = k \pi, \ k = \in \mathbb{Z}^+$, which signifies that the above series is not Borel summable and therefore $\theta = 0$ is a Stokes line.
To remedy that, we define lateral Borel transforms $\mathcal{S}_+ \{ \Phi^{(0)} \} (t)$,the integration contours pass of which pass over/under the poles
\begin{equation}
\begin{aligned}
	 [\mathcal{S}_+ - \mathcal{S}_{-} ] \{ \Phi^{(0)} \} (z) &= - (2 \pi i)  \Sum_{k=1}^\infty \Res_{\zeta = k \pi} \left( \frac{2}{\sqrt{\pi z}}  \zeta\,e^{-\zeta^2 / z} \cot{\zeta} \right) \\ 
    &= - \frac{2 \pi i}{\sqrt{\pi z}} \Sum_{k \neq 0}^{\infty} \pi |k|  e^{- (k \pi)^2 / z},
\end{aligned}
\end{equation}
where we used that 

\begin{align}
    2 \Sum_{k=1}^{\infty} &= \Sum_{k \neq 0}^{\infty}, &
    \lim_{x \rightarrow 0} x \cot{x} &= 1.
\end{align}
Reinstating the prefactor $\frac{2R_0^2}{t}$ and setting $z= \frac{t}{R_0^2}$ we finally get
\begin{equation}
    [\mathcal{S}_+ - \mathcal{S}_{-} ] \{ \Phi^{(0)} \} (z) = - 4i \bigg( \frac{\pi R_0^2}{t} \bigg)^{3/2} \Sum_{k \neq 0}^{\infty} |k| e^{- \frac{k^2 \pi^2 R_0^2}{t}}.
\end{equation}
Therefore, the trans-series associated to the heat-kernel trace is
\begin{equation}
\Tr\bqty{e^{\Delta_{\mathbb{S}^2}^{1/2}t }} = \frac{4 R_0^3}{\sqrt{\pi} t^{3/2}} \Int_{\mathcal{C}_\pm} \dd{\zeta}  \zeta\, e^{-\zeta^2 R_0^2/t} \cot{\zeta } - 4i \left( \frac{\pi R_0^2}{t} \right)^{\frac{3}{2}} \Sum_{k\neq 0} \sigma_k^{\pm}  \abs{k}  e^{- \frac{k^2 \pi^2 R_0^2}{t}}.
\end{equation}


\printbibliography[heading=bibintoc]



\begin{declaration}
\phantomsection
\addchaptertocentry{\authorshipname} 
\setlength\parindent{0pt} 

{\noindent\huge\bfseries\authorshipname\par\vspace{10pt}}
\vspace*{5mm}
\textbf{Last name, first name:} \lname, \fname\\[3mm]
\textbf{Matriculation number:} \matrnumber\\[1.2cm]

I declare herewith that this thesis is my own work and that I have not used any sources other than
those stated. I have indicated the adoption of quotations as well as thoughts taken from other authors
as such in the thesis. I am aware that the Senate pursuant to Article 36 paragraph 1 litera r of the
University Act of September 5th, 1996 and Article 69 of the University Statute of June 7th, 2011 is
authorized to revoke the doctoral degree awarded on the basis of this thesis.
For the purposes of evaluation and verification of compliance with the declaration of originality and the
regulations governing plagiarism, I hereby grant the University of Bern the right to process my
personal data and to perform the acts of use this requires, in particular, to reproduce the written thesis
and to store it permanently in a database, and to use said database, or to make said database
available, to enable comparison with theses submitted by others.
\\[1.5cm]

Bern, \today \\[4mm]

\includegraphics[scale = 0.20]{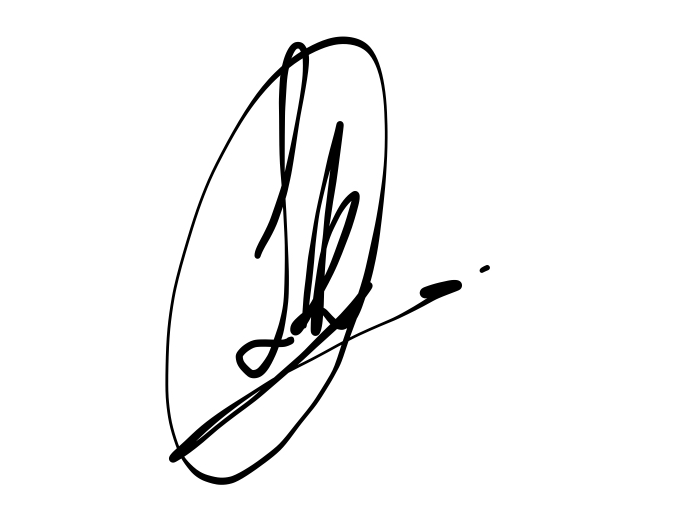}\\[3mm]
\authorname

\end{declaration}

\end{document}